\documentclass[12pt]{article}
\usepackage{amssymb}
\usepackage{amsmath}
\usepackage{graphics}
\usepackage{graphicx}
\usepackage{psfrag}
\usepackage[off]{auto-pst-pdf}
\usepackage{wrapfig}
\usepackage[mathscr]{eucal}
\usepackage{textgreek}
\usepackage{tikz}
\usetikzlibrary{cd}
\usepackage{youngtab}
\usepackage{tikz-cd}
\usepackage{subcaption}
\usepackage{comment} 

\usepackage{hyperref}
\usepackage[numbers,sort&compress]{natbib}
\usepackage{hypernat}
\hypersetup{linkbordercolor=.8 .8 1,
            citebordercolor=.7 1 .7}

\DeclareFontFamily{U}{euc}{}
\DeclareFontShape{U}{euc}{m}{n}{<-6>eurm5<6-8>eurm7<8->eurm10}{}%
\DeclareSymbolFont{AMSc}{U}{euc}{m}{n} 
\DeclareMathSymbol{\upsi}{\mathord}{AMSc}{"20}
\DeclareMathSymbol{\ulambda}{\mathord}{AMSc}{"15}
\DeclareMathSymbol{\usigma}{\mathord}{AMSc}{"1B}
\DeclareMathSymbol{\utau}{\mathord}{AMSc}{"1C}
\DeclareMathSymbol{\uups}{\mathord}{AMSc}{"1D}
\DeclareMathSymbol{\uepsilon}{\mathord}{AMSc}{"F}
\DeclareMathSymbol{\ueta}{\mathord}{AMSc}{"11}
\DeclareMathSymbol{\uvarphi}{\mathord}{AMSc}{"27} 

\catcode`\@=11

\def\@citex[#1]#2{%
\if@filesw \immediate \write \@auxout {\string \citation {#2}}\fi
\@tempcntb\m@ne \let\@h@ld\relax \def\@citea{}%
\@cite{%
  \@for \@citeb:=#2\do {%
    \@ifundefined {b@\@citeb}%
      {\@h@ld\@citea\@tempcntb\m@ne{\bf ?}%
      \@warning {Citation `\@citeb ' on page \thepage \space undefined}}%
      {\@tempcnta\@tempcntb \advance\@tempcnta\@ne%
      \@tempcntb\number\csname b@\@citeb \endcsname \relax%
      \ifnum\@tempcnta=\@tempcntb 
        \ifx\@h@ld\relax%
          \edef \@h@ld{\@citea\csname b@\@citeb\endcsname}%
        \else%
          \edef\@h@ld{\ifmmode{-}\else--\fi\csname b@\@citeb\endcsname}%
        \fi%
      \else
        \@h@ld\@citea\csname b@\@citeb \endcsname%
        \let\@h@ld\relax%
      \fi}%
    \def\@citea{,\penalty\@highpenalty\,}%
  }\@h@ld
}{#1}}

\def\@citeb#1#2{{[#1]\if@tempswa , #2\fi}}
\def\@citeu#1#2{{$^{#1}$\if@tempswa , #2\fi }}
\def\@citep#1#2{{#1\if@tempswa , #2\fi}}

\def\bcites{         
        \catcode`\@=11
        \let\@cite=\@citeb
        \catcode`\@=12
}

\def\upcites{         
        \catcode`\@=11
        \let\@cite=\@citeu
        \catcode`\@=12
}

\def\plaincites{      
        \catcode`\@=11
        \let\@cite=\@citep
        \catcode`\@=12
}

\newcount\hour
\newcount\minute
\newtoks\amorpm
\hour=\time\divide\hour by 60
\minute=\time{\multiply\hour by 60 \global\advance\minute by-\hour}
\edef\standardtime{{\ifnum\hour<12 \global\amorpm={am}%
        \else\global\amorpm={pm}\advance\hour by-12 \fi
        \ifnum\hour=0 \hour=12 \fi
        \number\hour:\ifnum\minute<10 0\fi\number\minute\the\amorpm}}
\edef\militarytime{\number\hour:\ifnum\minute<10 0\fi\number\minute}

\def\draftlabel#1{{\@bsphack\if@filesw {\let\thepage\relax
   \xdef\@gtempa{\write\@auxout{\string
      \newlabel{#1}{{\@currentlabel}{\thepage}}}}}\@gtempa
   \if@nobreak \ifvmode\nobreak\fi\fi\fi\@esphack}
        \gdef\@eqnlabel{#1}}
\def\@eqnlabel{}
\def\@vacuum{}
\def\marginnote#1{}
\def\draftmarginnote#1{\marginpar{\raggedright\scriptsize\tt#1}}
\def\draft{
        \pagestyle{plain}
        \overfullrule=2pt
        \oddsidemargin -.5truein
        \def\@oddhead{\sl \phantom{\today\quad\militarytime} \hfil
        \smash{\Large\sl DRAFT} \hfil \today\quad\militarytime}
        \let\@evenhead\@oddhead
        \let\label=\draftlabel
        \let\marginnote=\draftmarginnote
        \def\ps@empty{\let\@mkboth\@gobbletwo
        \def\@oddfoot{\hfil \smash{\Large\sl DRAFT} \hfil}
        \let\@evenfoot\@oddhead}
        \def\@eqnnum{(\theequation)\rlap{\kern\marginparsep\tt\@eqnlabel}%
        \global\let\@eqnlabel\@vacuum}  }

\def\section{\@startsection {section}{1}{\z@}{3.ex plus 1ex minus
 .2ex}{2.ex plus .2ex}{\large\bf}}
\def\subsection{\@startsection{subsection}{2}{\z@}{2.75ex plus 1ex minus
 .2ex}{1.5ex plus .2ex}{\bf}}

\def\abstract{\if@twocolumn
\section*{Abstract}
\else 
\begin{center}
{\bf Abstract\vspace{-.5em}\vspace{0pt}}
\end{center}
\quotation
\fi}

\catcode`\@=12

\newcommand{\beq}{\begin{equation}}
\newcommand{\eeq}{\end{equation}}
\newcommand{\beqa}{\begin{eqnarray}}
\newcommand{\eeqa}{\end{eqnarray}}
\newcommand{\dd}{{\rm d}}

\newcommand{\Z}{{\mathbb Z}}

\newcommand{\Q}{{\mathbb Q}}

\newcommand{\R}{{\mathbb R}}

\newcommand{\C}{{\mathbb C}}

\newcommand{\PP}{{\mathbb P}}
\newcommand{\e}{\,{\rm e}}
\newcommand{\CP}{\mathbb{CP}}

\newcommand{\be}{\begin{equation}}
\newcommand{\ee}{\end{equation}}
\newcommand{\bea}{\begin{eqnarray}}
\newcommand{\eea}{\end{eqnarray}}

\def\to{\rightarrow}

\def\lae{\mathrel{\mathop{\smash{\lower .5 ex \hbox{$\stackrel<\sim$}}}}}
\def\lae{\mathrel{\mathop{\smash{\lower .5 ex \hbox{$\stackrel>\sim$}}}}}

\def\l:{\mathopen{:}\,}
\def\r:{\,\mathclose{:}}

\catcode`\@=11
\def\theequation{\arabic{equation}}
\catcode`\@=12

\bcites

\catcode`\@=11
\def\theequation{\thesection.\arabic{equation}}
\@addtoreset{equation}{section}
\@addtoreset{footnote}{section}
\@addtoreset{footnote}{subsection}
\catcode`\@=12

\newcommand{\bepsilon}{\overline{\epsilon}}
\newcommand{\bPhi}{\overline{\Phi}}

\newcommand{\bphi}{\overline{\phi}}
\newcommand{\bpsi}{\overline{\psi}}

\newcommand{\bi}{\overline{\imath}}
\newcommand{\bj}{\overline{\jmath}}
\newcommand{\bl}{\overline{l}}

\newcommand{\bz}{\overline{z}}
\newcommand{\bchi}{\overline{\chi}}
\newcommand{\blambda}{\overline{\lambda}}
\newcommand{\bsigma}{\overline{\sigma}}
\newcommand{\bSigma}{\overline{\Sigma}}
\newcommand{\bareta}{\overline{\eta}}
\newcommand{\barxi}{\overline{\xi}}
\newcommand{\bartial}{\overline{\partial}}

\newcommand{\lrD}{\overleftarrow{D}\!\!\!\!\!
\overrightarrow{\vbox to 8.2pt{}}}
\newcommand{\bD}{\overline{D}}

\newcommand{\wh}{\widehat}
\newcommand{\wt}{\widetilde}

\newcommand{\nn}{\nonumber}

\newcommand{\ttt}{\mathfrak{t}}
\newcommand{\im}{{\rm i}}

\newcommand{\half}{{1\over 2}}
\newcommand{\ch}{{\rm ch}}

\newcommand{\sDelta}{\mathit{\Delta}}

\newcommand{\Arg}{{\rm Arg}}

\newcommand{\Rs}{\utau}
\newcommand{\Is}{\uups}

\newcommand{\cs}{\mbox{\textsigma}}
\newcommand{\rs}{\mbox{\texttau}}
\newcommand{\is}{\mbox{\textupsilon}}

\newcommand{\rmP}{{\rm P}}

\newcommand{\Sym}{{\rm Sym}}
\newcommand{\sym}{S}
\newcommand{\GL}{{\rm GL}}
\newcommand{\SL}{{\rm SL}}
\newcommand{\Hom}{{\rm Hom}}
\newcommand{\Ext}{{\rm Ext}}
\newcommand{\rmH}{{\rm H}}
\newcommand{\Db}{{\rm D}^b}
\newcommand{\bgamma}{\mbox{\large $\gamma$}}
\newcommand{\sgamma}{\mbox{$\gamma$}}
\newcommand{\fund}{{{\rm F}}}
\newcommand{\sfund}{\texttt{F}}
\newcommand{\Symt}{\Sym^2\fund}
\newcommand{\symt}{S^2\fund}
\newcommand{\brane}{\mathfrak{B}}
\newcommand{\bV}{\mathscr{V}}
\newcommand{\rmV}{{\rm V}}
\newcommand{\AN}{\textit{\textbf{A}}}
\newcommand{\AO}{A}
\newcommand{\rmGRR}{{\rm GRR}}

\newcommand{\Kos}{\mathfrak{K}}
\newcommand{\Lc}{\mathfrak{LC}}
\newcommand{\Lb}{\mathfrak{L}}
\newcommand{\elll}{h}
\newcommand{\elln}{n}

\newcommand{\Complex}{\mathfrak{C}}

\newcommand{\Gr}{\!\textit{\textbf{G}}}
\newcommand{\fb}{\!\textit{\textbf{E}}}
\newcommand{\Pf}{\!\textit{\textbf{P}}}

\newcommand{\pathGtoP}{{\mathcal P}^{\,{}^{\downarrow}}}
\newcommand{\pathPtoG}{{\mathcal P}^{\,{}^{\uparrow}}}

\newcommand{\bfD}{{\bf D}}
\newcommand{\bfS}{{\bf S}}

\newcommand{\wtq}{\widetilde{q}}

\newcommand{\textw}{\raisebox{0.5mm}{\scalebox{.8}[.8]{$\bigwedge$}}}
\newcommand{\cktt}{\texttt{R}}

\newcommand{\BBB}{\mathfrak{B}}

\begin{document}

\begin{titlepage}

\begin{center}

\vspace*{0.5cm}
       {\large \bf Grade Restriction Rule For D-Brane Transport In \\[0.1cm]
         Non-Abelian Gauged Linear Sigma Models}
\vskip 0.5cm       
{Richard Eager$,\!\!{}^{{}^{1,2}}\,$\, 
Kentaro Hori$,\!\!{}^{{}^{2}}\,$\,
Johanna Knapp${}^{{}^{3}}\,$\,
and\, Mauricio Romo${}^{{}^{4}}\,$}\\
\vskip 0.5cm
$\!{}^{{}^{1\,}}${\small\sl Department of Physics, University of Maryland,
  College Park, MD 20742, USA}\\[0.1cm]
$\!{}^{{}^{2\,}}${\small\sl Kavli Institute for the Physics and Mathematics
  of the Universe (WPI), University of Tokyo\\[-0.05cm]
  Kashiwa, Chiba 277-8583, Japan}\\[0.1cm]
$\!{}^{{}^{3\,}}${\small\sl School of Mathematics and Statistics,
  The University of Melbourne\\[-0.05cm]
  Parkville, VIC, 3010, Australia}\\[0.1cm]
$\!{}^{{}^{4\,}}${\small\sl Center for Mathematics and Interdisciplinary Sciences,
  Fudan University and\\[-0.07cm]
  Shanghai Institute for Mathematics and Interdisciplinary Sciences
  (SIMS)\\[-0.05cm]
  Shanghai, 200433, China}

\end{center}
\vskip 0.2cm
\begin{abstract}
We study the low energy behaviour of B-type D-branes
in non-Abelian gauged linear sigma models.
The main problem is to determine the grade restriction rule
for D-brane transport from one phase to another along paths.
We analyze the convergence condition of
the integral for the hemisphere partition function.
We perform a thorough analysis in a particular model with two phases,
one weakly coupled and one strongly coupled, both
corresponding to Calabi-Yau threefolds.
General methods are developed in this study
and are applied to other models successfully.
The integrals for a set of special
Chan-Paton representations exhibit critical behaviour around
each singular point on the phase boundary.
From each of these points emanate ``no entry zones'' where
a convergent contour is lost or is forced to traverse Gamma function poles.
Knowledge of the special representations and no entry zones is strong enough
to uniquely determine the grade restriction rule.
In the course of this study, we encounter a problem in Weyl invariance of
convergent contours, which forces us to modify the grade restriction criterion.
The special representations play an important r\^ole in the
monodromy around the singular point.
We also obtain exact results on boundary conditions in supersymmetric
QCDs and apply them to understand the low energy behaviour of D-branes in
the strong coupling phase.
\end{abstract}

\end{titlepage}

\pagestyle{empty}

{\footnotesize

\tableofcontents

}

\setcounter{page}{0}

\newpage

\setcounter{page}{1}
\pagestyle{plain}
\section{Introduction}

Gauged linear sigma models (GLSMs) \cite{Wittenphases}
provide a simple and global picture of the moduli spaces of two-dimensional
(2,2) superconformal field theories.
The space of Fayet-Iliopoulos (FI) parameters is decomposed
by the pattern of gauge symmetry breaking into chambers, called \emph{phases},
which correspond to low energy theories of various characters
such as non-linear sigma models with Calabi-Yau target spaces and
Landau-Ginzburg orbifolds.
The FI parameter has its superpartner, the theta parameter,
and the theory is well behaved except on a locus of complex
codimension one in the FI-theta parameter space.
In particular, different phases can be connected by paths that circumvent
the singular locus, and a class of observables 
can be related by continuation along the paths.
A method to relate D-branes in different phases
was introduced in \cite{HHP} for models with Abelian gauge groups.
The main purpose of the present work is to extend the method to models with
non-Abelian gauge groups.

Our focus will be 
D-branes that preserve the B-type half of the (2,2) supersymmetry \cite{OOY},
called D-branes of type B or B-branes in short,
in GLSMs of Calabi-Yau type.
The chiral sector of B-branes is the main tool to study their
low energy behaviour.
It is invariant under the renormalization group flow and
can be studied using the high energy description.
It is also invariant under deformation of the twisted chiral parameter
of the theory, which is the FI-theta parameter in GLSMs. 
In particular, it is constant along a path in the parameter space.
If the path connects different phases,
B-branes in one phase can be transported to the other phase
without changing the chiral sector,
and the transport depends only on the homotopy class of the path.
The rule of transport is what is determined in \cite{HHP} in Abelian models,
which we would like to study further.
The category of B-branes is an important invariant
obtained by applying the idea of chiral ring to the boundary.
B-branes are equivalent in the chiral sector when
they are isomorphic in the category.
The transport of B-branes along a path between different phases
determines an equivalence of the corresponding categories in the two phases.
The transport along a loop around the singular locus determines
an autoequivalence of the category at the base point of the loop.

An important notion in GLSMs is \emph{regularity}, that is,
the absence of non-compact flat directions in the effective potential
$U_{\!\it eff}(\sigma)$ for the scalar superpartner $\sigma$
of the gauge potential. 
The singular locus of the FI-theta parameter space is nothing but
the locus at which such a flat direction is present \cite{Wittenphases}.
The basic idea of \cite{HHP} is to require that the D-branes respect
the regularity of the bulk theory.
Part of the data to specify a D-brane is the representation
of the gauge group placed at the boundary, the Chan-Paton representation,
and its charge $q$ contributes linearly to the effective
potential $V_{\!{\it eff}, q}(\sigma)$ at the boundary.
Another element is the boundary condition on $\sigma$ specified by a
Lagrangian submanifold $\bgamma$ of the complexified Lie algebra
of the gauge group.
The regularity requirement is that
$V_{\!{\it eff}, q}(\sigma)$ is bounded below on $\bgamma$.
For the transport along a path between two phases,
there must be a continuous family of such $\bgamma$ over the path.
This imposes a constraint on the Chan-Paton representations
that can be included in the branes, called the \emph{grade restriction rule}.
It depends on the homotopy class of the paths
which we call the \emph{window}.
It turns out that this requirement uniquely determines the rule of D-brane
transport.

A possible way to extend this rule to non-Abelian GLSMs is provided by
the hemisphere partition function \cite{SuTe,HoOk,HR} which computes the
central charge of the D-brane \cite{BaPl}.
It is expressed as an integral of Mellin-Barnes type
in the Cartan subalgebra $\ttt_{\C}$
of the complexified Lie algebra of the gauge group,
with the contour $\bgamma$ corresponding to
the boundary condition on the scalar $\sigma$.
It was observed in \cite{HR} that, in theories with Abelian gauge groups,
the integrand for the charge $q$ part of the brane asymptotes
to $\e^{-2\pi V_{{\it eff},q}(\usigma)}$ and existence of
a family of convergent contours $\bgamma$ over paths between phases
agrees with the grade restriction rule of \cite{HHP}.
It was proposed in \cite{HR} that the grade restriction rule in a general GLSM
can be obtained from the condition of convergence of
the hemisphere partition function.
It was successfully applied in a simple non-Abelian model.
We would like to explore the proposal in more general GLSMs.

A distinctive feature of non-Abelian GLSMs is the possibility of
\emph{strongly coupled phases} where continuous parts of the gauge symmetry
remain unbroken.
To understand the low energy behaviour of the theory in such a phase is
difficult in general but sometimes possible.
The first case studied is the \emph{R\o dland model} with $U(2)$ gauge group
\cite{HoTo}.
It has one weakly coupled phase where the gauge symmetry is completely broken
and one strongly coupled phase where the $SU(2)$ subgroup remains unbroken,
with three singular points in between.
The theory in the weakly coupled phase reduces in the standard way to
the sigma model with a Calabi-Yau target $X$.
In the strongly coupled phase, after understanding the low energy dynamics of
$SU(2)$ gauge theory, it was found that the theory also reduces to
the sigma model with another Calabi-Yau target $Y$.
From the way $X$ and $Y$ appear,
the two phases are called the \emph{Grassmannian phase}
and the \emph{Pfaffian phase}, respectively.
After \cite{HoTo}, several models with strongly coupled phases
have been studied and found to have interesting low energy behaviour
\cite{Hduality,HoKn,GeJo,HoKn2,CKS,KnSh,KnMc}.
One of them is the \emph{Hosono-Takagi model} with gauge group
$(U(1)\times O(2))/\{(\pm 1, \pm {\bf 1}_2)\}$,
with a very similar structure of phases as the R\o dland model but
also with interesting differences.
When we study D-branes in non-Abelian GLSMs, 
we cannot avoid this representative class of models.

The categories of B-branes in a class of theories
are known or expected to be equivalent to categories studied in mathematics.
For example, in the sigma model with a
compact K\"ahler target $X$, the category is known \cite{HHP,BondalRosly,Block}
to be equivalent to the derived category of coherent sheaves on $X$.
Therefore, the equivalences of categories between phases in GLSMs
have a clear mathematical meaning and hence have attracted the interest
of mathematicians. Some of such equivalences had been proven independently or
inspired by GLSMs. The work of \cite{HHP} was directly motivated by
the mathematical proof by Orlov \cite{OrlovCY/LG} of the equivalence for
the Calabi-Yau/Landau-Ginzburg correspondence \cite{Wittenphases}.
The result of \cite{HHP} was formulated mathematically \cite{Segal}
and is extended significantly in \cite{DHL,BFK} to 
the theories in which the gauge group can be non-Abelian
and the matter fields take values in a class of non-linear spaces,
though the phases are weakly coupled.
Some explicit examples including the ``simple model'' mentioned above
are worked out in \cite{DS,Donovan}.
The derived equivalence of the two Calabi-Yau manifolds $X$ and $Y$
that appear in the R\o dland model is proved in \cite{BC,KuzGrPf,ADS}.
In particular, the proof by Addington, Donovan and Segal \cite{ADS}
is given in a form that looks similar to the grade restriction rule.
The proof is extended in \cite{Rennemo} to the case of
Hosono-Takagi model.

We take the R\o dland model
as our primary example to study the grade restriction rule
using the hemisphere partition function.
The analysis was conducted in several stages,
starting with the test of convergence with na\"ive choices of contours
and then performing analytical and numerical studies back and forth,
which eventually led us to a decisive conclusion.
Through this analysis, we developed general methods to determine
the grade restriction rule. These are applied successfully to other systems
with non-Abelian gauge groups including the Hosono-Takagi model.
A key discovery is the local behaviour of the integrand
around the singular points in the FI-theta parameter space,
which we now highlight.

Let us look at the contribution to the hemisphere partition function of
an irreducible representation $\rmV$ in the Chan-Paton factor,
called the hemisphere integral for $\rmV$.
Its integrand asymptotes to $\e^{-\AN_\rmV(\usigma)}$ for some
explicit real valued function $\AN_\rmV(\usigma)$ on $\ttt_{\C}$
 which scales as
 $\AN_\rmV(\lambda\usigma)=\lambda\AN_\rmV(\usigma)$ for $\lambda>0$,
 called the \emph{growth function},
except along the half-hyperplanes of $\ttt_{\C}$
on which the poles of the integrand lie, called the \emph{walls}.
(The growth function for the charge $q$ representation of an Abelian gauge group
is nothing but the effective boundary potential
$2\pi V_{{\it eff},q}(\usigma)$ of \cite{HHP}.)
For the transport along a path between phases,
the problem is to find a family of contours $\bgamma$ over the path
on which $\AN_\rmV$ is bounded below from $0$ while staying away from the walls
all the way.
The discovery is that a set of \emph{special representations} $\{\rmV_s\}$
is associated to each singular point $t_*$ in the FI-theta parameter space.
Their growth functions $\AN_{\rmV_s}$ identically vanish on branches of
the non-compact flat directions of $U_{\it eff}(\sigma)$ at $t_*$
and the region $\AN_{\rmV_s}\geq 0$ exhibits a critical behaviour around $t_*$.
Also, there is a \emph{no entry zone} extending from $t_*$:
as the FI-theta parameter enters there from outside,
a convergent contour $\bgamma$ goes out of the admissible region
$\AN_{\rmV_s}\geq 0$ or is forced to hit the walls.
Its exact location can be determined:
The boundary corresponding to blocking of the contour deformation by a wall
is where the hypersurface $\AN_{\rmV_s}=0$ on the wall is singular.
The information of special representations and no entry zones
is strong enough to uniquely determine the grade restriction rule.
The special representations play an important r\^ole
also in the monodromies along the loops around the singular points.

Another important discovery is that it is not always possible to
find a family of \emph{Weyl invariant} convergent contours $\bgamma$.
This forces us to abandon the interpretation of the contour $\bgamma$
as the data to specify the boundary condition on the vector multiplet
put forward in \cite{HR}.
A revised interpretation is presented:
existence of a family of convergent contours,
whether Weyl invariant or not, is the criterion for
the boundary interaction with the standard Neumann boundary condition
to respect the regularity.

Understanding the low energy behaviour of D-branes in a strongly coupled phase
is a new challenge which we face in non-Abelian GLSMs.
In this paper, we attack this important problem
in the Pfaffian phase of the R\o dland model
by a combination of several approaches.
First of all, regularity is a non-trivial issue in the entire phase,
since the classical potential has a non-compact flat direction
due to the unbroken $SU(2)$ gauge symmetry
which is lifted by quantum corrections.
It turns out that there is a non-trivial condition from regularity:
among the irreducible representations
$\sym^h\fund(i)=\Sym^h\fund\otimes \det^{\otimes i}\fund$
of the gauge group $U(2)$, where $\fund$ is the fundamental doublet,
only those with $h=0,1,2$ are acceptable 
as Chan-Paton representations.
This may be regarded as the grade restriction rule
of the strongly coupled phase.
The hemisphere partition function can be employed to learn about
the topology of the low energy image of a given brane in gauge theory.
In particular, we find branes whose partition functions match with the central
charges of D0-branes and D6-branes in the Calabi-Yau manifold $Y$.
In order to obtain more precise information, we take the most fundamental
approach.
That is, we study boundary $SU(2)$ dynamics and apply the result to
the strongly coupled phase,
extending what is done in \cite{HoTo} in the bulk. 
Recall that the $SU(2)$ supersymmetric QCD with seven fundamentals
with a mass matrix of rank four flows to the free theory of the baryons
made of the three massless quarks.
We identify a brane in the QCD,
which we denote by $\brane^{SU(2)}_{3,7}$,
that flows to the Neumann boundary condition on the baryons.
This enables us to prove that the candidates for the D0 and D6
branes indeed flow to the D0 and D6 branes in $Y$.
Furthermore, we show that gauge theory representatives of general branes
in the Calabi-Yau $Y$ are provided by a functor constructed in \cite{ADS},
up to a shift, by observing that it
can be described in terms of our QCD brane $\brane^{SU(2)}_{3,7}$.

To determine the rule of D-brane transport from the grade restriction rule,
we need to show that, in each phase,
the renormalization group flow induces a one-to-one correspondence
between the chiral equivalence classes of grade restricted branes in
gauge theory and branes in the low energy theory,
more precisely, an equivalence between the categories thereof.
The problem includes a purely mathematical component
which had already been solved in \cite{ADS} for one of the windows
of the R\o dland model (and in \cite{DonovanPfGr} for the rest of the windows
after the input from the present work --- see the end of this introduction).
However, we need to examine whether the mathematical construction is
relevant to the physics of the renormalization group flow.
We find that it is indeed the case.
During the process, we also decided to make everything as explicit as possible.
The effort turned out to be fruitful.
It helps us to understand the flow itself better
and enables us to describe the transport and monodromy in ways that
had not been realized.
The most important is the \emph{grade restriction process}
in which a given gauge theory brane is replaced by a grade restricted
brane which has the same low energy behaviour in the phase.
This is done by a sequence of brane replacement processes in each of which
a given brane is replaced by another by binding a brane which is empty
in the phase. Therefore, the construction of such \emph{empty branes}
is important.

Let us describe the empty branes in the two phases of the R\o dland model.
In the Grassmannian phase,
they are obtained by finding matrix factorization completions of
the twisted Lascoux complexes
which are used in \cite{DS,Donovan} for a similar purpose in simpler models
with vanishing superpotential.
Furthermore, the resulting empty branes must be bound together
in forms that can be used in the brane replacement process
to switch grade restricted representatives,
from one window to another.
This yields three series of
empty branes in the Grassmannian phase, denoted by $\Lb^h(i)$,
with $h=0,1,2$ and $i\in\Z$.
In the Pfaffian phase, empty branes are obtained from the Koszul complexes
which were used for similar purpose in Abelian GLSMs \cite{HHP}.
After matrix factorization completion and tensor product with representations
of the gauge group, we obtain three series of empty branes in the
Pfaffian phase, denoted by $\Kos\otimes\sym^h\fund(i)$,
with $h=0,1,2$ and $i\in\Z$.
One important property of these branes is that
each has the special representations at one of the singular points
as the edge components.
These branes are not empty in the opposite phases.
The branes $\Kos\otimes \sym^h\fund(i)$ descend to D6-branes of rank
$h+1$ in the Grassmannian phase while
$\Lb^h(i)$ descend to D6-branes of rank $5,4,1$ for $h=0,1,2$ in the Pfaffian
phase. In fact, the latter are the branes which we prove to
flow to D6-branes in $Y$ using the understanding of boundary $SU(2)$ dynamics.

The empty branes can be used to describe the monodromies along loops
around singular points with base points inside phases,
in the same way as in Abelian models \cite{HHP}. 
Such a monodromy can be regarded as the composition of the D-brane transport
from one phase to another through one window and the backward transport
through another, and requires a switch of grade restricted representatives
in the phase where the loop turns around.
This is done using the empty brane in that phase
which has the special representation at the singular point
as the edge components.
The effect is the Seidel-Thomas spherical twist by the low energy image of
this brane in the phase of the base point.
For example, for loops around the singular points of the
R\o dland model, the monodromies are the twists by vector bundles of
ranks $1,2,3$ on $X$ if the base point is in the Grassmannian phase
and by vector bundles of
ranks $5,4,1$ on $Y$ if the base point is in the Pfaffian phase.

Another notable result, obtained in the R\o dland model,
is the presentation of the D-brane transport from
the Pfaffian phase to the Grassmannian phase
as Fourier-Mukai transforms.
This is possible due to the identification
of gauge theory branes that reduce to D0-branes in the Pfaffian phase.
The transport of D0-branes can be done in families and this
tells us the candidates for the integral kernel for each window, up to
tensor product with line bundle on $Y$.
The ambiguity is fixed by using the information
from the hemisphere partition function.
We find that the Fourier-Mukai transform
found by Borisov and C\u ald\u araru \cite{BC}
is the same as the transport through one of the windows
composed with the tensor products with certain line bundles in $X$ and $Y$.

The rest of the paper is organized as follows.

In Section~\ref{sec:Problem}, we describe the main problem 
and make some preparations.
After a brief introduction to B-branes in GLSMs and
a review of the grade restriction proposal of \cite{HHP} in Abelian models,
we state the problem we would like to attack in this paper.
We then introduce the hemisphere partition function \cite{SuTe,HoOk,HR}
and explain how it may be used to find the grade restriction rule in general.
As a simple application, we determine the grade restriction rule for
a class of supersymmetric QCDs that will be relavant
in the strongly coupled phases.

Sections \ref{sec:RodlandGRR}-\ref{sec:transport} are devoted to
the R\o dland model.
In Section~\ref{sec:RodlandGRR}, we determine the grade restriction rule.
We start by describing the walls and properties of the growth function.
We find contours $\bgamma_+$ and $\bgamma_-$ in the weakly coupled and
strongly coupled phases respectively:
$\bgamma_+$ works for all representations of the gauge group
while $\bgamma_-$ works for grade restricted representations
in the strongly coupled phase.
We find four classes of windows between the two phases.
Two of them are represented by straight paths at constant values of the theta
parameter
and the other two are represented by curved paths with varying theta values. 
We first examine the convergence of the hemisphere integral on the real locus
and find that the set of allowed representations is too small to produce
low energy branes in either of the two phases.
We then discuss how to enlarge the set and
find a contour $\bgamma_0$ that allows sets which may be large enough.
It can be used for transport through the two windows of straight paths.
We next consider the other two windows of curved paths
and describe the results of a numerical analysis
outlined in Appendix~\ref{subapp:curved}.
In Section~\ref{subsec:catastroph}, motivated by the numerical analysis,
we study the local behaviour of the growth function near the singular points
and find the special representations $\{\rmV_s\}$ as described above,
as well as the beginning of the no-entry zones.
In Section~\ref{subsec:noentry}, we study
the no-entry zones which are found numerically in Appendix~\ref{subapp:genpaths}
to extend indefinitely away from the singular points.
We explain the way to find their exact location
from the singularity of $\AN_{\rmV_s}$ on the walls.
We then find that the information of special representations and
no-entry zones uniquely determines the grade restriction rule
for each of the windows.
We check that this also works in Abelian models studied in \cite{HHP}.
We then discuss the issue of Weyl invariance and conclude
with the revision of the grade restriction criterion.

In Section~\ref{sec:Grassmannian}, we study the low energy behaviour of
D-branes in the Grassmannian phase. Applying the method of \cite{HHP},
we determine the sigma model image of a general gauge theory brane
and also propose a way to lift sigma model branes to gauge theory.
We write down the Chern character formula of the sigma model image
of a gauge theory brane
and find the relation between the hemisphere partition function
and the large volume central charge on the Calabi-Yau $X$.
As a preparation for further discussion, we make a digression
to explain some mathematical background, including
basic facts on homological algebra, Borel-Weil-Bott theorem and
invariant theory. This also enables us to prove that
the proposed lift from sigma model to gauge theory is always possible.
We then proceed to the construction of empty branes in this phase.
We introduce a triangular grid to label the representations of $U(2)$,
which is used here to describe the twisted Lascoux complexes and their
matrix factorization completions but also will be used for other purposes
such as depicting the grade restricted subsets.
We describe the grade restriction process and
outline the construction of the empty branes $\Lb^h(i)$ mentioned above,
with their use in the window switch.

In Section~\ref{sec:Pfaffian}, we study the low energy behaviour of D-branes
in the Pfaffian phase.
We first make remarks on the empty branes in this phase and
change of R-charge frame.
We then compute the hemisphere partition function and propose its relation to
the large volume central charge on the Calabi-Yau $Y$.
We construct gauge theory branes whose partition functions
match with the central charge of D0-branes.
We also find that the partition functions of $\Lb^h(i)$
agree with the central charges of D6-branes. 
These are the candidates for the D0 and D6-branes.
We next move on to the study of boundary conditions in
the $SU(2)$ supersymmetric QCD and obtain the result
on the brane $\brane^{SU(2)}_{3,7}$ mentioned above.
We then study the low energy behaviour of the candidates for D0 and D6-branes
and prove, using the result on $\brane^{SU(2)}_{3,7}$,
that they indeed flow to D0 and D6-branes.
We next discuss the brane replacement process
where the higher Ext vanishing condition plays an important r\^ole.
Finally, we argue that the flow from the category of grade restricted branes
to the category of branes in $Y$ is an equivalence.
The key player is the lift of branes from $Y$ to the gauge theory
which we find combining the result of \cite{ADS} and the boundary flow
in the $SU(2)$ QCD.

In Section~\ref{sec:transport}, 
we study the rule of D-brane transport along paths
between the two phases as well as the monodromies along loops
in the moduli space.
We start by describing how to use the grade restriction rule to
determine the transport.
We then compute the transport at the level of D-brane charges
using the hemisphere partition function.
We next work out the actual transport through four windows
for some particular branes.
The first examples are the D6-branes for which we
work out the transport in both directions, from $X$ to $Y$ and from $Y$ to $X$.
The next examples concern transport of D0-branes in $Y$.
We find that they are transported to vector bundles of rank $\pm 4$ on $X$
or two term complexes of vector bundles on $X$ of effective rank $\pm 1$,
depending on the window.
As mentioned above, with the aid of the hemisphere partition function,
this exercise enables us to describe the transport
from $Y$ to $X$ through each window as a Fourier-Mukai transform.
We next determine the monodromies along loops around the singular points.
The results are the Seidel-Thomas spherical twists as described above.
We find commutativity of monodromies around two points.
We then determine the monodomies along loops around the two large volume limits
in the moduli space.
We have a shift $[-2]$ in one of them and we find that it can be understood
by the flow of grading proposed by Douglas \cite{Douglas}.
We end with the confirmation that the homotopy equivalence
relations among paths and loops in the moduli space are reflected in
the transport and monodromy.

In Section~\ref{sec:other}, we apply the methods
developed in Section~\ref{sec:RodlandGRR}
to analyze the convergence condition
in other one parameter models with strongly coupled phases.
The models discussed are:
the dual of the R\o dland model with gauge group
$(U(1)\times USp(4))/\{(\pm 1,\pm{\bf 1}_4)\}$
(rank 3),
the Hosono-Takagi model (with rank 2 gauge group
$(U(1)\times O(2))/\{(\pm 1,\pm{\bf 1}_2)\}$ as already mentioned),
the dual of Hosono-Takagi model with gauge group
$(U(1)\times SO(4))/\{(\pm 1,\pm{\bf 1}_4)\}$ (rank 3), and
generalizations of the R\o dland model
with the same gauge group but with different numbers of fields.
We focus on the grade restriction rule and leave the other aspects
for future study. However, we do construct empty branes in one of
the two phases, and this is enough to determine the monodromies along loops
around the singular points with the base point in the opposite phase.
In the dual R\o dland model, the result reproduces the monodromies
with a base point in the Pfaffian phase in the original R\o dland model.
This produces an explicit correspondence between D-branes under
the Seiberg-like duality \cite{Hduality}.
In the Hosono-Takagi model and its dual, the monodromy obtained in one model
provides a prediction for the monodromy in the other.

In Appendix~\ref{app:SUSY}, we discuss boundary interactions and boundary
conditions that preserve the B-type supersymmetry in
2d (2,2) supersymmetric gauge theories and non-linear sigma models.
This is mostly a review to fix our conventions,
but we provide a streamlined description for the most general situation
which has not been written up in the past.
We include a mathematical result on adjoint invariant Lagrangian
submanifolds of the complexified Lie algebra of a compact Lie group.

In Appendix~\ref{app:Cat}, we describe the categories of D-branes of type B in
non-linear sigma models and gauge theories.
We introduce operators $\bfD$ and $\bfS$ that help us to prove
that the morphisms are described by the zero modes of a set of variables.
We then introduce some categories studied in mathematics,
such as the derived category of coherent sheaves,
and discuss their relation to the categories of D-branes.
The mathematical description of branes in sigma models with nonzero
superpotential
will be used as an important tool in Section~\ref{subsec:PfEq}
and also quoted in Section~\ref{subsec:Gflow} as a remark,
both in the discussion of gauge theory lift of sigma model branes.

In Appendix~\ref{app:Weyl} and \ref{app:WCF},
we quote some facts on Weyl (anti-)invariant polynomial functions
on the Lie algebra of the maximal torus of a compact Lie group
as well as on representations of a compact disconnected Lie group.
These will be used in the description and
computation of the hemisphere partition function.

In Appendix~\ref{app:numerical}, we present an outline of
the numerical analysis of convergent contours for the hemisphere integrals.
We are interested in the region of $\usigma\in \ttt_{\C}$ where
the growth function $\AN_\rmV(\usigma)$ is positive.
By the scaling property of $\AN_\rmV$,
it is enough to look at a large 3-sphere $S^3_R$
inside $\ttt_{\C}=\C^2$ with which a generic contour intersects on a circle.
We use \texttt{RegionPlot} of Mathematica to plot
the region $\AN_\rmV(\usigma)>0$ on the two-dimensional slices of $S^3_R$
by the level hyperplanes of a particular coordinate.
As a preparation, we sketch the geography of the slices,
looking at the intersections with the walls and some of the contours
such as $\bgamma_0$ and $\bgamma_{\pm}$.
In Appendix~\ref{subapp:curved}, we study the transport along the curved paths.
We try to find a family on contours (circles in $S^3_R$) along such a path,
interpolating $\bgamma_+$ and $\bgamma_-$ in the two phases, which stay in
the region $\AN_\rmV(\usigma)>0$ but away from the walls.
Among four subtle representations $\rmV$, we find that
two admit such families but the other two do not.
The plot of the region $\AN_\rmV>0$ near the singular points
motivates the analytical study given in Section~\ref{subsec:catastroph}
and also supports the result obtained there.
In Appendix~\ref{subapp:genpaths}, we study the region $\AN_\rmV>0$
as we go away from the singular points.
We find that the no-entry zones extend indefinitely and 
check the result in Section~\ref{subsec:noentry} that their boundaries
correspond to the singularity of $\AN_\rmV$ on the walls.
We also observe and explain that some of such singularities
do not correspond to no-entry zones.
In Appendix~\ref{subapp:Weylinv}, we examine the Weyl invariance of
convergent contours for the windows of curved paths.
We confirm that there is no family of Weyl invariant
admissible contours for one member of the grade restricted subset
along any path in the windows.

In Appendix~\ref{app:periods},
we recall the Picard-Fuchs operators and
the series expansions of the solutions
in the two large volume limits of the R\o dland and Hosono-Takagi models.
This will be used in Sections~\ref{subsec:GHS}, \ref{subsec:PHS}
and \ref{subsec:HT}.

In Appendix~\ref{app:Lbconstruction},
we present the details of the construction of
the matrix factorizations $\Lb^h=\Lb^h(0)$ which are empty
in the Grassmannian phase,
by binding together the matrix factorization
completions of the twisted Lascoux complexes,
$(3-h)$ pairs of them.
The key process is to find the homomorphisms
between the matrix factorizations with which we bind them together.
Their existence and the fact that the edge components are
the special representations at one of the singular points will
be relevant in the brane replacement process.
Moreover, the explicit form of the arrows is needed in the study of their
low energy behaviour in the Pfaffian phase.

In Appendix~\ref{app:GRdetails}, we fill in some details of
the grade restriction process in the Grassmannian phase, that is,
we find a homomorphism from a given complex to another, $\Complex \to N'$.

In Appendix~\ref{app:redPf}, we study the low energy behaviour of
$\Lb^1$ and $\Lb^0$ in the Pfaffian phase, complementing the study
of $\Lb^2(-1)$ in Section~\ref{subsub:D6LE}.
After computing the R-charges of components in the Pfaffian phase,
we perform the tachyon condensation of invertible parts of
the arrows, and obtain a result that is consistent with the claim
that they indeed flow to D6-branes of the expected ranks.

In Appendix~\ref{app:chPf}, we compute the Chern characters of
some branes at an intermediate energy scale in the Pfaffian phase.
This is to provide additional confirmation of a claim
in Section~\ref{subsec:PfEq}
on the lift of branes from the Calabi-Yau $Y$ to gauge theory.

In Appendix~\ref{subsec:locfree}, we provide, with proof,
a fact on complexes of vector bundles used in several places in the paper.
In Appendix~\ref{subsec:ExtRGr} and \ref{subsec:ExtHT},
we compute the Ext groups of certain pairs of
vector bundles on the Calabi-Yau manifolds that appear in
the R\o dland model and the Hosono-Takagi model.
The result shows that the vector bundles that appear in the monodromies are
spherical objects, ensuring that the corresponding twists are autoequivalences,
and also that certain pairs of vector bundles
are orthogonal to each other, ensuring that the corresponding spherical twists
commute with each other.

In Appendix~\ref{app:someinv},
we compute or constrain Chern classes of vector bundles
that appear in the R\o dland model and the Hosono-Takagi model.
The results are mainly used for the computation
of the spherical twists at the level of D-brane charges,
which is relevant in the topological confirmation of
the identification of the twists as the
monodromies around the singular points,
done in Sections~\ref{subsec:monodromysing}, \ref{subsec:HT}
and \ref{subsec:dualHT}.

The present research has been carried out over more than ten years, and
some parts of it were presented in talks and reports, including
\cite{HoriStrings,BeijingLecture,HoriKinosaki,HoriSCGP,KnappSCGP,HoriMainz}.
We understand that some of the results at intermediate stages of this research
are used in the following papers.
Our discovery of the grade restricted subsets for the four types of windows
in the R\o dland model
and the resulting monodromies with a base point in the Grassmannian phase
was the source of \cite{DonovanPfGr}, as partially mentioned above.
An argument for gauge theory lift of D-branes in a weakly coupled phase,
given in an earlier draft of the present paper, was used in  \cite{LinRomo}.
The contour $\bgamma_0$ in (\ref{bgamma0}) for the straight paths
was used in \cite{GRS,GLZ} to find a part of
the grade restriction rule in related models.
Subsequent development of the present research addresses some of the
incompleteness contained in these papers.
For example, monodromies with a base point in the Pfaffian phase,
which was left open in \cite{DonovanPfGr}, are determined
in two ways: in Section~\ref{subsec:monodromysing} based on the construction
(Section~\ref{subsec:GRprocess} and Appendix~\ref{app:Lbconstruction})
and understanding (Section~\ref{subsub:D6LE}) of the branes $\Lb^h(i)$,
and in Section~\ref{subsec:dualRodland} using the dual description.
In Section~\ref{subsub:groupoid},
Section~\ref{subsub:MFcompletion} and Section~\ref{sec:RodlandGRR},
we point out and resolve some issues of \cite{DonovanPfGr}, \cite{LinRomo} and
\cite{GRS,GLZ}, respectively.

\subsection*{Remarks On Notation And Terminology}

For the part where models with $U(2)$ gauge group are discussed,
we use the following notation for representations of $U(2)$
or of its complexification $\GL_2(\C)$:
\beqa
&\fund=\mbox{the fundamental representation}\cong\C^2,\nn\\
&S^{\elll}\fund=\Sym^{\elll}\fund,\qquad
\det\fund=\textw^{\!2\,}\fund,\nn\\
&R(i)=R\otimes (\det\fund)^{\otimes i}~\mbox{for any representation
$R$ and $i\in\Z$}.\nn
\eeqa
An irreducible representation of $U(2)$ or $\GL_2(\C)$
is either one of
$$
S^{\elll}\fund(i),\quad\mbox{for $\elll\in \Z_{\geq 0}$ and $i\in\Z$.}
$$
It is understood that $S^0\fund=\Sym^0\fund$ is the one-dimensional
trivial representation $\C$, and we denote
$S^0\fund(i)=\det\fund^{\otimes i}$ simply by $\C(i)$.
Note that $\fund^*:=\Hom(\fund,\C)$  is isomorphic to $\fund(-1)$.

The imaginary unit is denoted by a roman $\im$.
Elements of the Lie algebra $\mathfrak{g}$
of a compact Lie group is regarded as purely imaginary, and elements of
$\im\mathfrak{g}$ are regarded as real.

By the shortage of letters and also to use notations not very far from
the conventional ones, we sometimes need to use the same symbol
or similar symbols for different purposes. Here are some examples:
\begin{itemize}
\item Italic $V$ is used for
  (i) the matter representation of a general GLSM,
  (ii) the flavor space for a subset of the matter fields
  in the R\o dland model and related models, and
  (iii) the vector superfield in 2d (2,2) supersymmetric gauge theories.
  Roman $\rmV$ is used for
  an irreducible representation of a group.
  The letter $\bV$ in upright calligraphy font is used for
  the Chan-Paton representation of a gauge theory brane.
\item Roman $\fund$ is used
  for the two-dimensional vector representation of $U(2)$ as described above,
  whereas a similar but different font $\sfund$ (typewriter) is used
  for the same doublet but regarded as the representation of $SU(2)$.
\item The scalar component of a 2d (2,2) supersymmetric vector multiplet
  is denoted by $\sigma={\rm Re}(\sigma)+\im{\rm Im}(\sigma)$ in
  usual Greek font. When its Cartan part appears as the integration variable
  of the hemisphere partition function, we use
  $\usigma=\Rs+\im\Is$ in upright Greek font in general.
  We use a slightly different version of upright Greek font $\cs=\rs+\im\is$ 
  for the case of gauge group $SU(2)$
\end{itemize}

\section{The Problem And Preparations}
\label{sec:Problem}

\subsection{The Primary Example}\label{subsec:model}

The primary focus of the present paper is
the gauged linear sigma model introduced in \cite{HoTo} which we call
the R\o dland model, after the name of a person who found a relevant
relation \cite{Rodland} between a pair of Calabi-Yau manifolds.
It has gauge group $U(2)$, seven fields
$x_1,\ldots,x_7$ in the fundamental representation $\fund$ and seven fields
$p^1,\ldots, p^7$ in the representation $\det^{-1}\fund=\C(-1)$,
the superpotential
\beq
W=\sum_{i,j,k=1}^7A^{ij}_kp^k[x_ix_j],
\label{theW}
\eeq
and the twisted superpotential $\wt{W}(\usigma)=-t\cdot{\rm tr}(\usigma)$.
Following \cite{Weyl} as in \cite{Hduality}, we denoted
\beq
    [x_ix_j]:=\epsilon_{ab}x^a_ix^b_j=x_i^1x_j^2-x_i^2x_j^1
\eeq
for the symplectic pairing of $x_i$ and $x_j$ that transforms as
$\det\fund=\C(1)$ under $U(2)$. The coefficients $A^{ij}_k$ can be taken
antisymmetric in the $i$-$j$ exchange. We may view
\beq
    A^{ij}(p):=\sum_{k=1}^7A^{ij}_kp^k
\eeq
as entries of a $7\times 7$ antisymmetric matrix $A(p)$.
We assume that $A^{ij}_k$ are generic so that the rank of
the matrix $A(p)$ cannot drop to $2$ for non-zero $p$.

The model has axial $U(1)$ R-symmetry since the gauge group 
lies in the special unitary group of the matter representation
(Calabi-Yau condition).
It also has vector $U(1)$ R-symmetry since the superpotential is 
quasi-homogeneous --- there is an R-charge assignment
$R_i$ for each chiral multiplet $\phi_i$ such that
$W(\lambda^R\phi)=\lambda^2 W(\phi)$: we may set
\beq
R_p=2-2\epsilon,\quad R_x=\epsilon.
\label{Rassignment}
\eeq
Moreover, there is charge integrality ---
the action of $\e^{\pi \im R}$ coincides with the action of
a gauge group element, $\e^{\pi i \epsilon}{\bf 1}\in U(2)$.
The theory is expected to flow to a family of superconformal field theories of
$\widehat{c}=3$, with the exactly marginal 
chiral parameters $A^{ij}_k$ and the exactly marginal twisted chiral parameter
$t=\zeta-\im\theta$, the FI-theta parameter.

The classical vacua are solutions to the D-term equation
\beq
\sum_{i=1}^7x^a_i(x^b_i)^*-\sum_{k=1}^7|p^k|^2\delta^a_b=\zeta\delta^a_b,
\eeq
the F-term equations
\beq
\sum_{j=1}^7A^{ij}(p)x^a_j=0,\quad
A_k(x):=\sum_{i,j=1}^7A^{ij}_k[x_ix_j]=0,
\eeq
and
\beq
    [\sigma,\bsigma]=0,\quad \sigma x_i=\bsigma x_i=0,\quad
    {\rm tr}(\sigma)p_k=0.
\eeq
The model has two phases.
$\zeta\gg 0$ is a usual geometric phase where the gauge group is completely
broken on the classical vacua, and the transversal modes are all massive.
The low energy theory is described by the non-linear sigma model
whose target space is the vacuum manifold.
The manifold is the complete intersection $X$
of seven hypersurfaces
$A_1(x)=\cdots =A_7(x)=0$
in the Grassmannian $G(2,7)$.
It is a smooth Calabi-Yau threefold with Hodge numbers
$h^{1,1}=1$ and $h^{2,1}=50$. This phase is often referred to as
the {\it Grassmannian phase}.
A Picard group generator $H$ and the Chern classes of $X$ obey
\beq
H^3=42,\quad c_2\cdot H=84,\quad c_3=-98.
\label{topGra}
\eeq
On the other hand, $\zeta\ll 0$ is a strongly coupled phase.
The D-term and F-term equations require $p\ne 0$
and $x=0$ and hence the $SU(2)$ subgroup is unbroken.
We have $SU(2)$ supersymmetric QCD (SQCD) with seven flavors
with varying mass matrix $A^{ij}(p)$.
After understanding the strongly coupled $SU(2)$ dynamics,
we find that the low energy theory in the end is again
a non-linear sigma model.
The target space is the Pfaffian variety
$Y$ defined as the locus of $p\in \CP^6$ at which the rank of $A(p)$
drops to $4$. It is also a smooth Calabi-Yau threefold with Hodge numbers
$h^{1,1}=1$ and $h^{2,1}=50$.
This phase is often referred to as the {\it Pfaffian phase}.
A Picard group generator $H$ and the Chern classes of $Y$ obey
\beq 
H^3=14,\quad c_2\cdot H=56,\quad c_3=-98.
\label{topPfa}
\eeq
$X$ and $Y$ have the same Hodge diamond
but are not birationally equivalent to each
other. 

\newcommand{\bfa}{{\bf a}}

The theory is singular when there is a noncompact Coulomb branch.
As a maximal torus $T$ of $U(2)$ we choose the diagonal subgroup,
so that the scalar component of the vector multiplet for $T$
is parametrized as
\beq
\sigma=\left(\begin{array}{cc}
  \sigma_1&0\\
  0&\sigma_2
\end{array}\right).
\label{sigmapara}
\eeq
The effective twisted superpotential at large $\sigma_1$, $\sigma_2$
and $\sigma_1\pm \sigma_2$ is
\beqa
\lefteqn{\wt{W}_{\!\it eff}(\sigma)=-t(\sigma_1+\sigma_2)
  +\pi \im (\sigma_1-\sigma_2)}\nn\\
&&
-7\sigma_1\left(\log\sigma_1-1\right)
-7\sigma_2\left(\log\sigma_2-1\right)
-7(-\sigma_1-\sigma_2)\left(\log(-\sigma_1-\sigma_2)-1\right).
\label{Weff}
\eeqa
The vacuum equation \cite{Wittenphases,Coleman}
\beq
\partial_{\sigma_1}\wt{W}_{\!{\it eff}}
\equiv\partial_{\sigma_2}\wt{W}_{\!{\it eff}}
\equiv 0\quad\mbox{mod $2\pi\im \Z$},
\label{vaceq0}
\eeq
reads
\beq
\e^t=(\sigma_1+\sigma_2)^7/\sigma_1^7=(\sigma_1+\sigma_2)^7/\sigma_2^7.
\eeq
A solution exists only when $\e^t=(\omega+1)^7$ for $\omega^7=1$,
and it is  $\sigma_1=\omega\sigma_2$.
Note that the Weyl reflection maps $\omega\to \omega^{-1}$
and the fixed point $\omega=1$ with enhanced $SU(2)$ symmetry
is excluded \cite{HoTo,longflow}.
Thus, there are three singular points
\beq
\e^t=(\omega_\bfa+1)^7=:\alpha_\bfa,\qquad
\omega_\bfa:=\e^{2\pi \im \bfa/7},\qquad
\bfa=1,2,3,
\label{defalphas}
\eeq
at which there is a one-dimensional noncompact Coulomb branch,
\beq
\sigma_1=\omega_\bfa^{\pm 1}\sigma_2.
\label{CoulombBranch}
\eeq
The numerical values of $\alpha_\bfa
=\omega_\bfa^{7\over 2}
\left(\omega_\bfa^{1\over 2}+\omega_\bfa^{-{1\over 2}}\right)^7
=\e^{\pi \im \bfa}\left(2\cos\left({\pi \bfa\over 7}\right)\right)^7$
are
\beq
\alpha_1\simeq -\e^{4.122},\quad
\alpha_2\simeq\e^{1.545},\quad
\alpha_3\simeq -\e^{-5.667}.
\eeq

\subsection{D-Branes Of Type B In Gauged Linear Sigma Models}
\label{subsec:Bbranes}

Our main interest is the low energy behaviour of B-branes
in gauged linear sigma models.
Let us describe how they are defined and discuss the issue of regularity
in a general model.

Let us consider a gauged linear sigma model with gauge group $G$, matter
representation $V$, superpotential $W$, and linear twisted superpotential
$\wt{W}(\sigma)=-t(\sigma)$, $t=\zeta-\im\theta$.
Our convention for the Lagrangian and supersymmetry
is put down in Appendix~\ref{app:SUSY}.
We assume the Calabi-Yau condition, $\mathfrak{g}\subset \mathfrak{sl}(V)$,
and homogeneity of the superpotential,
$W(\lambda^R\phi)=\lambda^2W(\phi)$ for all $\lambda\in \C^{\times}$,
for a diagonalizable element $R\in {\rm End}(V)$ commuting with $G$ such that
$\e^{\pi \im R}$ is the same as the transformation of a gauge group element
$J\in G$.
This is so that the theory has the axial and vector $U(1)$ R-symmetries
with charge integrality.
The theory is expected to flow to a non-trivial
superconformal field theory.
One important condition is regularity \cite{Wittenphases}
--- absence of noncompact Coulomb or mixed branches ---
so that the superconformal field theory has a discrete spectrum.

When formulated on a surface with a boundary,
an appropriate boundary interaction must be placed and
a boundary condition on the bulk fields must be specified.
To be specific, let us consider the left half of the Minkowski spacetime,
$\Sigma=\{(x^0,x^1)| x^1\leq 0\}$, with the boundary
$\partial\Sigma=\{(x^0,0)\}$ parametrized by the time $t=x^0$.
The kinetic terms and the FI-theta term have canonical boundary counterterms
so that the totals have B-type supersymmetry.
See Appendix~\ref{app:SUSY} for details.
In particular, the one for the FI-theta term is
\beq
\int_{\partial\Sigma}\dd t\,{1\over 2\pi}
\bigl(\,-\zeta({\rm Im}(\sigma))+\theta({\rm Re}(\sigma))\,\bigr).
\label{FIthct}
\eeq
The most important part of the boundary interaction is
the Chan-Paton factor, which is also necessary when the
superpotential $W$ is non-trivial.
It is specified by a $G$-equivariant matrix factorization of $W$ on $V$,
that is, a $\Z_2$-graded unitary representation $\rho:G\to U^{\rm ev}(\bV)$
and a $G$-equivariant holomorphic map
$Q:V\to {\rm End}^{\rm od}_{}(\bV)$ such that
\beq
Q(\phi)^2=W(\phi){\rm id}_{\bV}.
\eeq
Let us spell out what is meant by this:
$\bV=\bV^{\rm ev}_{}\oplus \bV^{\rm od}_{}$
is a finite dimensional $\Z_2$-graded hermitian vector space,
$\rho$ is a homomorphism from $G$ to the group of even unitary transformations
of $\bV$ and $G$-equivariance of $Q$ means
\beq
\rho(g)^{-1}Q(g\phi)\rho(g)=Q(\phi)
\label{Qgi}
\eeq
for all $g\in G$.
Note that $\rho$ extends to a complex representation of $G_{\C}$,
denoted by the same letter $\rho:G_{\C}\to \GL^{\rm ev}(\bV)$,
and that $G$-equivariance and holomorphy of $Q$ imply its $G_{\C}$-equivariance,
that is, (\ref{Qgi}) holds for all $g\in G_{\C}$.
The Chan-Paton factor is given by
\beq
    {\rm Pexp}\left(\,-\im \int_{\partial\Sigma} \dd t \,{\mathcal A}_t\,\right)
    \label{CPfactor}
\eeq
with
\beq
    {\mathcal A}_t=\rho(v_t-{\rm Re}(\sigma))+{1\over 2}\{Q,Q^{\dag}\}
    -{1\over 2}\psi^i\partial_iQ+{1\over 2}\bpsi^{\bi}\partial_{\bi}Q^{\dag},
    \label{At}
\eeq
where $v_t$ is the gauge potential on the boundary and
$\psi$ is the boundary value of a combination of the matter fermion components.
We consider B-branes that preserve the $U(1)$ vector R-symmetry
and charge integrality: there is a diagonalizable element
${\bf r}_*\in {\rm End}^{\rm ev}(\bV)$ commuting with $\rho(G)$
such that
\beqa
&&\lambda^{{\bf r}_*}Q(\lambda^R\phi)\lambda^{-{\bf r}_*}
=\lambda Q(\phi)\quad \mbox{for all $\lambda\in \C^{\times}$},
\label{QRs}\\
&&\e^{\pi \im {\bf r}_*}\rho(J)=\pm 1\quad \mbox{on $\bV^{\rm\, ev/od}$.}
\label{CI}
\eeqa
We shall refer to $\brane=(\bV,\rho,Q,{\bf r}_*)$, or
$\brane=(\bV,Q)$ for short, as a \emph{brane datum}
and denote the set of brane data by $D_G(V,W)$.

The brane data can also be described in terms of modules over the ring
$\C[\phi]$ of polynomials of the matter fields $\phi$.\footnote{To be precise,
this applies when $Q(\phi)$ is a polynomial function of $\phi$.
This is necessarily the case for an arbitrary brane datum
in a theory in which it is possible to assign positive R-charges
to all components of $\phi$, e.g. the R\o dland model
(choose $0<\epsilon<1$ in (\ref{Rassignment})), so that
the entries of $Q(\phi)$ are necessarily polynomials of $\phi$
as they have definite R-charges.}
For a brane datum $\brane=(\bV,Q)$, we consider the
free $\C[\phi]$-module
\beq
M=\bV\otimes_{\C} \C[\phi].
\label{MvsV}
\eeq
It is equipped with an action of $G_{\C}$,
$g:v\otimes P(\phi)\mapsto \rho(g)v\otimes P(g^{-1}\phi)$,
R-symmetry
$R_\lambda:v\otimes P(\phi)\mapsto \lambda^{{\bf r}_*}v\otimes P(\lambda^R\phi)$,
and a $\Z_2$-grading that agrees with $R_{\e^{\pi\im}}J$.
$Q(\phi)$ is an odd endomorphism of this $\Z_2$-graded
$\C[\phi]$-module $M$ which commutes with gauge group,
has R-charge $1$,
and squares to multiplication by $W(\phi)$.
We may regard a brane datum $\brane$ also as the pair $(M,Q)$
of such structures.
For the theory with vanishing superpotential,
  $W(\phi)=0$, with trivial
R-charge assignment on the fields and $J=1$,
$M$ is a $\Z$-graded $\C[\phi]$-module with a $G_{\C}$ action,
and $(M,Q)$ can be regarded as a $G_{\C}$-equivariant complex of
$\C[\phi]$-modules. We shall call such a complex $(M,Q)$,
with $M$ of the form (\ref{MvsV}),
a {\it complex of representations over $\C[\phi]$ of $G_{\C}$}.

The theta parameter enters into the path integral measure as
\beq
\exp\left(\,\im\int_{\Sigma}\dd^2 x\,{1\over 2\pi}\theta(v_{01})
+\im\int_{\partial\Sigma}\dd t\,\,{1\over 2\pi}\theta({\rm Re}(\sigma))\,\right)
\eeq
via the bulk theta term and the boundary counter term (\ref{FIthct}).
Let $\chi:G\to U(1)$ be a character of $G$ that determines
a one-dimensional representation $\C(\chi)$.
Its infinitesimal form, denoted by the same letter $\chi$,
is an adjoint invariant $\R$-linear form of $\im \mathfrak{g}$.
Under $\theta\to\theta+2\pi\chi$,
the theta term shifts by a term that can be written as a boundary term.
Together with the shift of ${1\over 2\pi}\theta({\rm Re}(\sigma))$
in the boundary counter term, the effect is the shift
$\rho\to \rho+\chi$ in (\ref{At}), that is,
tensoring the Chan-Paton representation $(\bV,\rho)$ with $\C(\chi)$.
This means
\beq
\theta\,\to\, \theta+2\pi \chi~\,\mbox{with $\brane$ fixed}
~~\Longleftrightarrow~~
\brane\,\to\, \brane\otimes\C(\chi)~\,\mbox{with $\theta$ fixed}.
\label{thetashift}
\eeq

The simplest choice for the boundary condition on the bulk fields is
a supersymmetric completion of the Neumann boundary condition
both on the gauge potential and the matter scalar field,
that is,
\beqa
&&v_1|={\rm Im}(\sigma)|=\partial_1({\rm Re}(\sigma)-v_0)|
=(\lambda_++\lambda_-)|=0,\label{vctBC}
\\
&&\partial_1\phi|=(\psi_+-\psi_-)|=\partial_1(\psi_++\psi_-)|=F|=0,
\label{chiBC}
\eeqa
where $O|$ stands for the boundary values of $O$.
In particular, the boundary value of the scalar component $\sigma$
of the vector multiplet is in the real locus,
\beq
\sigma|\in \im\mathfrak{g}\subset \mathfrak{g}_{\C}.
\label{NeumannBC}
\eeq
This condition may be deformed so that the boundary value of $\sigma$
is in the adjoint orbit of a Weyl invariant
Lagrangian submanifold $\bgamma$ of the Cartan subalgebra $\ttt_{\C}$,
\beq
\sigma|\in G\bgamma\subset \mathfrak{g}_{\C}.
\label{BCbgamma}
\eeq
See Appendix~\ref{app:SUSY} for details.

Each pair $(\brane,\bgamma)$ determines a D-brane in the classical system.
The question is whether it behaves nicely in the quantum theory,
in particular, whether the regularity of the bulk theory is not spoiled
in the boundary sector.
As long as regularity is respected, we expect that the details of $\bgamma$
are irrelevant for the low energy behaviour of the D-brane,
and the datum $\brane$ alone determines
the D-brane in the infra-red superconformal field theory.
The question is which brane data respect regularity
and determine D-branes in the quantum theory.
When the FI-parameter $\zeta$ is deep inside a weakly coupled phase
where the gauge symmetry is broken to finite subgroups by solutions to the
D-term equation,
we expect that any $\brane$ is fine since the vector multiplet does
not remain as a degree of freedom below the Higgs scale $e\sqrt{|\zeta|}$.
However, the condition on $\brane$ will be non-trivial when $\zeta$ is
in a strongly coupled phase or near a phase boundary where a part of
the vector multiplet remains as the low energy degrees of freedom.
A sign of possible irregularity can be seen in the boundary interactions
(\ref{FIthct}) and (\ref{At}).
They include a linear potential in the vector multiplet scalar
\beq
V_{\it cl}(\sigma)
={1\over 2\pi}\Bigl[\,\zeta({\rm Im}(\sigma))
  -(\theta+2\pi\rho)({\rm Re}(\sigma))\,\Bigr]
\label{Vclass}
\eeq
which can be unbounded below on $G\bgamma$.

This problem is important in identifying the category of B-branes.
A na\"ive analysis ignoring the regularity problem
would yield the following result. See Apeendix~\ref{app:CatGT}.
The space of morphisms for a pair of branes
$\brane_1=(\bV_1,Q_1)$ and $\brane_2=(\bV_2,Q_2)$
is the cohomology of the complex
\beq
\left(\,\Omega^{0,*}(V,\Hom(\bV_1,\bV_2))^G\,,\,\bartial_{Q_1.Q_2}\,\right)
\label{UVcomplex}
\eeq
with $\bartial_{Q_1,Q_2}O=\bartial O+Q_2O-(-1)^{|O|}OQ_1$
where $|O|$ is the total R-charge of $O$.
This is a correct answer if $G$ is a finite group.
If $G$ is continuous, this is correct only
in the strict ultraviolet limit where the gauge coupling constant is
sent to zero $e\to 0$ and the vector multiplet is frozen,
so that the regularity issue is totally absent.
We shall regard $D_G(V,W)$ as the set of brane data
equipped with this structure of category.
When the gauge interaction is turned on, $e\ne 0$, the structure of the
category must be modified.
When $\zeta$ is deep inside a weakly coupled phase so that the vector multiplet
is frozen at energies below the Higgs scale $e\sqrt{|\zeta|}$,
the modification is simply to take into account the Higgs mechanism:
We keep all the objects of $D_G(V,W)$ but change the morphism space
by replacing $V$ in (\ref{UVcomplex})
by the subspace $V^{\rm ss}(\zeta)\subset V$ consisting of $\phi$'s
whose $G_{\C}$ orbits go through the solutions to the D-term equation.
The question is how to modify the ultraviolet category
$D_G(V,W)$ when $\zeta$ is close to the phase boundary where the
regularity issue becomes non-trivial.

In \cite{HHP}, for theories with Abelian gauge groups,
this problem was discussed as a guide to find the rule
of D-brane transport along paths between different phases.
Note that $G=T$ and $G\bgamma=\bgamma$ when $G$ is Abelian.
The quantum correction $\sDelta V(\sigma)$ to the classical boundary potential
(\ref{Vclass}) was computed by integrating out the matter fields
at large values of $|\sigma|$.
It yields a bounded contribution
that might solve the possible instability problem.
It also has a singularity  at the half-hyperplanes,
${\rm Re}Q_i(\sigma)=0$ and ${\rm Im}Q_i(\sigma)>0$,
where $Q_i$ ranges over the matter charges.
The singularity is due to the presence of a normalizable zero mode of the
matter which is supported near the boundary.
The following condition for the brane $(\brane,\bgamma)$
to respect regularity was proposed in \cite{HHP}: 
the effective boundary potential
$V_{\it eff}(\sigma)=V_{\it cl}(\sigma)+\sDelta V(\sigma)$
is bounded from below on $\bgamma$
and $\bgamma$ does not meet the singular half-hyperplanes
except at the region where $|\sigma|$ is very small.
When the FI parameter is deep inside a phase, which is always weakly coupled
when the gauge group is Abelian,
there is a deformation $\bgamma$ of the real locus $i\ttt\subset \ttt_{\C}$
which satisfies the condition for any $\brane$.
However, as the FI parameter is deformed and goes across a phase boundary, 
the chosen $\bgamma$ may need to be rotated, and, depending on
the gauge charges $(\bV,\rho)$ of $\brane$,
it may be unavoidable that $\bgamma$ goes through the singular half-hyperplanes.
Only when all the gauge charges in $\brane$ lie in a certain set,
 $\bgamma$ can be continuously deformed all the way from one phase to the other
without hitting the singular half-hyperplanes.
This condition on the gauge charges is called the {\it grade restriction rule}.
The set of allowed charges and hence the rule depend on the window ${\bf w}$
in the FI-theta parameter space through which the path of continuation
 goes.

For a linear sigma model with $U(1)$ gauge group, with phases
$+$ and $-$,
this yields the following picture:
\beq
\begin{tikzcd}
  &D_{G}(V,W)\arrow[d,phantom,"\mbox{\large $\cup$}"]\arrow[ddl,bend right,"\pi_-"']\arrow[ddr,bend left,"\pi_+"]&\\[-0.3cm]
  &{\mathcal T}_{\bf w}\arrow[dl,"\cong"']\arrow[dr,"\cong"]\\
  D(X_-)&&D(X_+)
\end{tikzcd}
\label{HHPdiagram}
\eeq
$D(X_{\alpha})$ is the set of D-branes of type B in the low energy theory
$X_{\alpha}$ in phase $\alpha$. The downward arrow $\pi_{\alpha}$ stands
for the renormalization group flow. As is always the case in RG flows,
infinitely many inequivalent branes in $D_G(V,W)$ can flow to
branes which are equivalent to each other in $D(X_{\alpha})$.
${\mathcal T}_{\bf w}$ is the set of brane data
satisfying the grade restriction rule for the window ${\bf w}$.
Only those inside this subset can be safely transported from
one phase to the other along a path through the window ${\bf w}$.
It turns out that, in each phase,
the arrow $\pi_{\alpha}$ restricted to ${\mathcal T}_{\bf w}$
is essentially bijective.
That is, it yields a bijection between the chiral equivalence classes
of branes in ${\mathcal T}_{\bf w}$ and in $D(X_{\alpha})$.
To find how a brane is transported along the path, 
say from phase $+$ to phase $-$, 
we first lift the brane in $X_+$ to ${\mathcal T}_{\bf w}$,
and then take the low energy limit in phase $-$.
This picture holds as a diagram of categories and functors.
$D_G(V,W)$ is the ultra-violet category described above and
${\mathcal T}_{\bf w}$ is its full subcategory,
while $D(X_{\pm})$ is the category of D-branes
in the low energy theory $X_{\pm}$.
The functors $\pi_{\pm}:{\mathcal T}_{\bf w}\to D(X_{\pm})$
are equivalences of categories, and hence
determine an equivalence $D(X_+)\cong D(X_-)$.
If we change the window ${\bf w}$, the rule of brane transport changes and
in particular the equivalence $D(X_+)\to D(X_-)$ also changes.
Along a loop in the FI-theta parameter space, going from one phase
to another through a window and coming back through another,
we obtain an autoequivalence of the category of the starting phase.
For a system with higher rank Abelian gauge group,
we have almost the same picture at a phase boundary between two neighboring
phases. The difference is that
the restricted maps $\pi_{\pm}:{\mathcal T}_{\bf w}\to D(X_{\pm})$ 
are not one to one. However, they have a common kernel and hence
we have a rule of brane transport in exactly the same way.

\noindent
{\bf Remark.} (This remark applies equally well to 
theories with non-Abelian gauge groups, as long as weakly coupled phases
are concerned.)
The low energy theory $X_{\alpha}$ may be the sigma model with superpotential
$(E_{\alpha},W_{\rm ind})$ where $E_{\alpha}$ is the quotient by $G$
of the space of solutions to the D-term equation
and $W_{\rm ind}$ is the function induced from $W$.
Note that $E_{\alpha}$ can also be regarded as the quotient by
$G_{\C}$ of $V_{\alpha}=V^{\rm ss}(\zeta)$ for $\zeta$ in the phase $\alpha$.
The category  of branes in this theory, $D(E_{\alpha},W_{\rm ind})$,
may be identified as the category $D_G(V_{\alpha},W)$
of $G$-equivariant matrix factorizations of $W$ on $V_{\alpha}$
whose morphism space is the cohomology of a complex as in (\ref{UVcomplex})
in which $V$ is replaced by $V_{\alpha}$.
In view of the above identification of the category of
GLSM D-branes at this energy scale, the flow
$\pi_{\alpha}:D_G(V,W)\to D(E_{\alpha},W_{\rm ind})$ can be identified as
the functor $i_{\alpha}^*:D_G(V,W)\to D_G(V_{\alpha},W)$
associated to the inclusion $i_{\alpha}:V_{\alpha}\hookrightarrow V$.
If $W_{\rm ind}$ has massive modes, $X_{\alpha}$ can be the theory
obtained by integrating them out, in which case the flow
$\pi_{\alpha}:D_G(V,W)\to D(X_{\alpha})$
factors through $D(E_{\alpha},W_{\rm ind})$.
The second flow $D(E_{\alpha},W_{\rm ind})\to D(X_{\alpha})$
is an equivalence of categories known as
the Kn\"orrer periodicity \cite{Knorrer}.

The above result was obtained through an attempt to physically understand
the proof by Orlov \cite{OrlovCY/LG} on
the Calabi-Yau/Landau-Ginzburg correspondence for the category of B-branes.
Examples with $W=0$ include some cases of the derived McKay correspondence
which had been proven earlier (and more generally) in \cite{BKR}.
The case of $W=0$ in general corresponds to the derived equivalences
of toric varieties that appear in different phases of an Abelian GLSM,
and had been proven in \cite{Kawamata01,VDB02,Kawamata03}.
In particular, the proofs in \cite{VDB02,Kawamata03} goes along the same line
as in the mathematical part of \cite{HHP}.
A mathematical formulation of a part of the above picutre was given by Segal
in \cite{Segal} in which the categories $D_G(V,W)$, $D(X_{\alpha})$ etc are
replaced by their mathematical counterparts
$\Db_{G_{\C}}(V,W)$, $\Db(X_{\alpha})$ etc.
In particular, it was first explicitly stated there that
the equivalence
${\mathcal T}_{\bf w}\stackrel{\cong}{\to}\Db(X_{\alpha})$
is associated to the restriction of
the functor $i_{\alpha}^*:\Db_{G_{\C}}(V,W)\to \Db_{G_{\C}}(V_{\alpha},W)$
to ${\mathcal T}_{\bf w}$.
The equivalence proof was extended in \cite{DHL,BFK}
to the case where $G$ is a general compact Lie group which can be non-Abelian,
$V$ is a class of algebraic varieties which is not necessarily a vector space,
and the two phases $X_+$ and $X_-$ are both weakly coupled.
The case of a Grassmannian flop,
a non-Abelian generalization of the Atiyah flop,
is worked out in \cite{DS,Donovan}, with the derivation of autoequivalences
for changes of ${\mathcal T}_{\bf w}$.
More recently, there have been works which provides
equivalences among categories that appear in phases of
non-Abelian gauged linear sigma models \cite{OtRe,KuPe,KeSe,DHKR,KRS},
in addition to \cite{ADS,Rennemo} that will be described below.
However, in these mathematical works,
only the FI parameter is under consideration whereas
\emph{the theta parameter},
let alone the singular points in the FI-theta parameter space, is not.
Consequently,
the ${\mathcal T}_{\bf w}$'s are just subcategories of $\Db_{G_{\C}}(V,W)$
which have no relation at all to the windows.
Nevertheless they are called {\it window subcategories}.
(However, possible relation to windows in the FI-theta parameter space
is behind some of the constructions,
as mentioned in \cite{Segal,DHKR,KRS} for example.)
A notable exception is the work by Halpern-Leistner and Sam \cite{DHLSam}
in which the equivalences are studied for the case when $V$ is a
``quasi-symmetric'' (see Section~\ref{subsec:SU2GRR}) representation
of $G_{\C}$.
A space that looks like the FI-\emph{theta}
parameter space is constructed 
combinatorially and is used to describe the subcategories called
\emph{magic windows}. 
We shall make a comment on this work in Section~\ref{subsec:SU2GRR}.

\newcommand{\TADS}{{\mathcal T}_{\,\texttt{KU}}}
\newcommand{\RADS}{\texttt{KU}}
 
We would like to know if
the same or a similar picture as in (\ref{HHPdiagram}) exists
in theories with non-Abelian gauge groups.
We are particularly interested in systems like the R\o dland model
where one of the phases is strongly coupled.
In the sense that a continuous unbroken gauge group exists,
a strongly coupled phase is like a phase boundary in Abelian theories,
and it is possible that some kind of grade restriction rule is in action
even deep inside the phase. If that is the case, and if
the remaining aspects are the same as in the Abelian case,
we would have the following picture:
\beq
\begin{tikzcd}
  &D_{G}(V,W)\arrow[d,phantom,"\mbox{\large $\cup$}"]\arrow[dddr,bend left,"\pi_X"]&\\[-0.3cm]
  &{\mathcal S}\arrow[ddl,bend right,"\pi_Y"']\arrow[d,phantom,"\mbox{\large $\cup$}"]&\\[-0.3cm]
  &{\mathcal T}_{\bf w}\arrow[dl,"\cong"']\arrow[dr,"\cong"]\\
  D(Y)&&D(X)
\end{tikzcd}
\label{maindiagram}
\eeq
$D(X)$ and $D(Y)$ are the sets of branes in the
low energy non-linear sigma models with the target spaces $X$ and $Y$.
${\mathcal S}$ is the set of brane data satisfying the grade restriction rule
in the strongly coupled phase, while ${\mathcal T}_{\bf w}$ is
the set of brane data satisfying the grade restriction rule for
a window ${\bf w}$ at the phase boundary.
We would like to know: Does the above picture exist at all?
What is ${\mathcal S}$? What is ${\mathcal T}_{\bf w}$ for each window
${\bf w}$?
And: Are the maps $\pi_X:{\mathcal T}_{\bf w}\to D(X)$
and $\pi_Y:{\mathcal T}_{\bf w}\to D(Y)$ essentially bijective?
If the answers are yes and if we know ${\mathcal S}$ and
the ${\mathcal T}_{\bf w}$'s, then we will have the rule of
D-brane transport along paths through the windows as well as monodromies
along loops around the singular points.
We would also like the picture to hold as a diagram of categories and functors
just as in (\ref{HHPdiagram}). Here again, $D_G(V,W)$ is the UV category
of the gauged linear sigma model
and ${\mathcal T}_{\bf w}\subset {\mathcal S}$ are its full subcategories,
while $D(X)$ and $D(Y)$ are the categories of low energy sigma models,
which are equivalent to the bounded derived categories
of coherent sheaves $\Db(X)$ and $\Db(Y)$ on $X$ and $Y$.
Then, this would yield equivalences of categories $\Db(X)\cong \Db(Y)$
as well as autoequivalences of $\Db(X)$ and of $\Db(Y)$ that
represent the monodromies.

\newcommand{\BBr}{\mathcal{BB}r}

In fact, the derived equivalence of $X$ and $Y$ in the R\o dland model
were proved by Addington, Donovan and Segal in \cite{ADS}
by showing equivalences $\Db(X)\cong \TADS$ and $\Db(Y)\cong \TADS$
for a full subcategory $\TADS$ of $\Db_{G_{\C}}(V,W)$.
Here, $\TADS$ consists of brane data
whose Chan-Paton vector spaces are direct sums of copies of members
of the following set of representations of the gauge group $G=U(2)$,
\beq
\RADS=\{\,\Sym^h\fund(i)\,|\,0\leq h\leq 2,~0\leq i\leq 6\,\}.
\label{ADSset}
\eeq
It is possible that $\TADS$ matches with ${\mathcal T}_{\bf w}$ for one of
the windows ${\bf w}$ and the equivalence of \cite{ADS} may fit
into the picture (\ref{maindiagram}).
The present work is very much inspired by their work. 
The proof was extended to Hosono-Takagi model and its cousins in \cite{Rennemo}.

We take the picture (\ref{maindiagram}) as our guide to explore
the low energy behaviour of the branes.
To examine this and to find the details,
instead of repeating the analysis of \cite{HHP},
we shall use the hemisphere partition function \cite{SuTe,HoOk,HR}.
Indeed, its use was proposed in \cite{HR}
as a means to find the grade restriction rules for D-brane transport
along paths between different phases.
Also, in the strongly coupled phase where the low energy behaviour of branes
is difficult to study directly,
the information of the Ramond-Ramond charge
of the branes can be read off from the hemisphere partition function.

\subsection{The Hemisphere Partition Function}
\label{subsec:HS}

The hemisphere partition function is defined for a two-dimensional
$(2,2)$ supersymmetric field theory with a half BPS boundary condition
\cite{SuTe,HoOk,HR}.
Let us consider a gauged linear sigma model $(G,V,W,R)$
obeying the Calabi-Yau condition
and a D-brane $(\mathfrak{B},\bgamma)=(\bV,\rho, Q,{\bf r}_*;\bgamma)$
of the model.
We take the weight decomposition of the matter representation
\beq
V=\bigoplus_i\C(Q_i,R_i)
\eeq
with respect to a maximal torus $T$ of $G$ and the vector $U(1)$
R-symmetry group. We assume that all the R-charges are in the range
\beq
0<R_i<2.
\label{Rrange}
\eeq
Then, the partition function is expressed as an integral
over the contour $\bgamma$ that determines the boundary condition on the
vector multiplet,
\beq
Z_{D^2}(\brane)\,=\,
\int\limits_{\,\,\,\,\,\sgamma}
\dd^{\ell}\usigma
\prod_{\alpha>0}\alpha(\usigma)
\sinh(\pi\alpha(\usigma))
\prod_{i}\Gamma\left(\im Q_i(\usigma)+{R_i\over 2}\right)
\,\e^{\im t(\usigma)}\,f_{\mathfrak{B}}(\usigma),
\label{Zhs}
\eeq
with
\beq
f_{\mathfrak{B}}(\usigma)\,:=\,
{\rm tr}_\bV^{}\left[\e^{\pi \im {\bf r}_*}\!\rho(\e^{2\pi\usigma})\right],
\label{deffB}
\eeq
which we shall call the {\it brane factor}.
The measure $\dd^{\ell}\usigma$ is the flat holomorphic volume form
on $\ttt_{\C}$ ($\ell$ is the rank of $G$),
the first product is over positive roots $\alpha$
and the second product is over the chiral multiplets.
They are the one-loop determinants of the vector and the chiral multiplets,
respectively. The factor $\e^{\im t(\usigma)}$ comes from
the classical twisted superpotential $\widetilde{W}(\usigma)=-t(\usigma)$.
Note that the combination $\e^{\im t(\usigma)}f_{\brane}(\usigma)$
respects the relation (\ref{thetashift}).
To be precise, there is an overall renormalization factor in
(\ref{Zhs}) which we set equal to $1$ in the present paper.\footnote{It is
$(r\Lambda)^{\wh{c}/2}$ where $r$ is the radius of the hemisphere,
$\Lambda$ is a renormalization scale and
$\widehat{c}=\sum_i(1-R_i)-\dim G$.
The integration variable $\usigma$ comes from $r$ times the zero mode of the Cartan part of the scalar component $\sigma$ of the vector multiplet.
If not Calabi-Yau, $G\not\subset SL(V)$,
the FI parameter is also renormalized as $t_R=t-{\rm tr}_V\log(r\Lambda)$.}
If we take the irreducible decomposition
of the Chan-Paton representation $\bV$,
\beq
\bV=\bigoplus_j\rmV_j[r_j],
\label{decM}
\eeq
where $r_j$ is the R-charge of the irreducible component $\rmV_j$,
the brane factor and the partition function are decomposed accordingly,
\beq
f_{\mathfrak{B}}(\usigma)
=\sum_j\e^{\pi \im r_j}{\rm tr}^{}_{\rmV_j}\e^{2\pi\usigma},\qquad
Z_{D^2}(\brane)=\sum_j\e^{\pi\im r_j}Z_{\sgamma}(\rmV_j).
\eeq

The contribution of an irreducible representation $\rmV$,
\beq
Z_{\sgamma}(\rmV)=\int\limits_{\,\,\,\,\,\sgamma}
\dd^{\ell}\usigma
\prod_{\alpha>0}\alpha(\usigma)
\sinh(\pi\alpha(\usigma))
\prod_{i}\Gamma\left(\im Q_i(\usigma)+{R_i\over 2}\right)
\,\e^{\im t(\usigma)}\,{\rm tr}_{\rmV}\e^{2\pi\usigma},
\label{ZVdef}
\eeq
which we shall call the {\it  hemisphere integral of $\rmV$ on $\bgamma$},
will play an important r\^ole.
The expression can be simplified.
When $G$ is connected, we have the Weyl character formula
\beq
{\rm tr}^{}_{\rmV}\e^{2\pi\usigma}
={\displaystyle\sum_{w\in W}(-1)^{\ell(w)}\e^{2\pi w(\lambda+\rho)
(\usigma)}\over
\displaystyle \prod_{\alpha>0}\left(\e^{\pi\alpha(\usigma)}
-\e^{-\pi\alpha(\usigma)}\right)},
\label{Weylcharacter}
\eeq
where $W$ is the Weyl group of $G$,
$\lambda$ is the highest weight of $\rmV$
and $\rho$ is half the sum of positive roots of $G$.
We notice that the denominator is nothing
but the factor $\prod_{\alpha>0}\sinh(\pi\alpha(\usigma))$
from the one-loop determinant of the vector
multiplet in (\ref{ZVdef}), up to a multiple of $2$.
$t(\usigma)$ is Weyl invariant,  and so is
the factor from the one-loop determinant of the chiral multiplet.
The latter is true because $R$ commutes with $G$ and hence
the Weyl group action induces permutations on the set $(Q_i,R_i)$.
The measure $\dd^{\ell}\usigma\,\prod_{\alpha>0}\alpha(\usigma)$
is Weyl invariant. See Appendix~\ref{app:Weyl}.
The contour $\bgamma$ is Weyl invariant but
its orientation is Weyl odd since it is a deformation of the real locus
$\im\ttt$.
Combining all of this, we find that the hemisphere integral can be written as
\beq
Z_{\sgamma}(\rmV)\,=\,
{\rm Const}
\int\limits_{\,\,\,\,\,\sgamma}
\dd^{\ell}\usigma
\prod_{\alpha>0}\alpha(\usigma)
\prod_{i}\Gamma\left(\im Q_i(\usigma)+{R_i\over 2}\right)
\,\e^{\im t(\usigma)+2\pi (\lambda+\rho)(\usigma)},
\label{ZV}
\eeq
where ${\rm Const}=|W|/2^{|\Delta_+|}$. Here $|W|$ is the order of the Weyl group
and $|\Delta_+|$ is the number of positive roots.
When $G$ is not connected, each irreducible representation $\rmV$
decomposes into a sum of irreducible representations of the identity
component $G_0$ of $G$, and these components are Weyl images of a single
irreducible representation, say $\rmV_{0}$.
In particular, (\ref{ZVdef}) can also be written
as (\ref{ZV}), where $\lambda$ is the highest weight of the
representation $\rmV_{0}$ of $G_0$ and
Const depends on the number of Weyl images of $\rmV_0$ in $\rmV$.
See Appendix~\ref{app:WCF} for details.

The integrand of (\ref{Zhs}) or (\ref{ZV}) has poles at infinitely many
hyperplanes
\beq
~~~~~~~~~~~~~~~~~~
Q_i(\usigma)~=~\im\left(n_i+{R_i\over 2}\right),\qquad n_i=0,1,2,3,\ldots\, .
\label{poles}
\eeq
The asymptotic behaviour of the integrand 
can be found using Stirling's formula.
In the region of $\usigma$ where
$|Q_i(\usigma)|\gg 1$ and $Q_i(\usigma)\not\in\im \R_{>0}$ for all $i$,
we find
\beq
\mbox{integrand}\mbox{(\ref{ZV})}~\sim~
P(\usigma)\cdot\exp\Bigl(-\im\wt{W}_{\!{\it eff},\lambda+\rho}(\usigma)\Bigr),
\label{ZVint}
\eeq
where $P(\usigma)$ is a product of powers of linear functions of
$\usigma$ and
\beqa
\wt{W}_{\!{\it eff}, \wtq}(\usigma)
=-t(\usigma)+2\pi\im \wtq(\usigma)
-\sum_iQ_i(\usigma)\left(\log \im Q_i(\usigma)-1\right),
\eeqa
in which the imaginary part of the logarithms take values in the interval
$(-\pi,\pi)$.
Note that $\wt{W}_{\!{\it eff}, \lambda+\rho}$ for a weight $\lambda$ of $T$
is a particular representative of the effective twisted superpotential
$\wt{W}_{\!{\it eff}}$ which is defined only modulo $2\pi\im$ times a weight
function of $T$.
The absolute value of (\ref{ZVint}) behaves as
\beq
\Bigl|\,\mbox{integrand}\mbox{(\ref{ZV})}\,\Bigr|~\sim~
|P(\usigma)|\cdot\exp\Bigl(-\AN_{\lambda+\rho}(\usigma)\Bigr),
\label{behav}
\eeq
with
\beqa
\AN_{\wtq}(\usigma)\!&=&\!
{\rm Re}\left(\im\wt{W}_{\!{\it eff},\wtq}\right)
\,=\, -\,{\rm Im}\left(\wt{W}_{\!{\it eff},\wtq}(\usigma)\right)\nn\\[0.2cm]
\!&=&\!
\zeta(\Is)-(\theta+2\pi \wtq)(\Rs)\nn\\[0.1cm]
&&+\sum_i\Bigl\{\,
Q_i(\Is)(\log|Q_i(\usigma)|-1)
+Q_i(\Rs){\rm Arg}(\im Q_i(\usigma))
\,\Bigr\},~~~~~~
\label{Aq}
\eeqa
with
\beq
\Rs={\rm Re}(\usigma),\quad
\Is={\rm Im}(\usigma).
\label{newn}
\eeq
The Arg's in (\ref{Aq}) are defined to take values in the interval
$(-\pi,\pi)$. In (\ref{behav}), $\wtq=\lambda+\rho$
may be replaced by any of its Weyl images as long as $\gamma$ is
Weyl invariant.
We shall refer to (\ref{Aq}) as the {\it growth function}
as its growth at the asymptotic region is important for the
absolute convergence of the integral (\ref{ZV}).
(This is the growth function in the Introduction
where it is denoted by $\AN_{\rmV}$ for simplicity.)
Likewise, 
the term of the integrand of (\ref{Zhs}) corresponding to
the $T$-weight $q$ part $\e^{2\pi q(\usigma)}$ of the brane factor
$f_{\brane}(\usigma)$
behaves in the same limit as
\beq
\Bigl|\,\mbox{integrand}_q\mbox{(\ref{Zhs})}\,\Bigr|~\sim~
|P(\usigma)|\cdot\exp\Bigl(-\AO_q(\usigma)\Bigr),
\label{behavO}
\eeq
with
\beqa
\AO_q(\usigma)\!&=&\!
\zeta(\Is)-(\theta+2\pi q)(\Rs)-\sum_{\alpha>0}\pi|\alpha(\Rs)|\nn\\[0.1cm]
&&+\sum_i\Bigl\{\,
Q_i(\Is)(\log|Q_i(\usigma)|-1)
+Q_i(\Rs){\rm Arg}(\im Q_i(\usigma))
\,\Bigr\}.~~~~~~
\label{AOq}
\eeqa
This may be interpreted as the effective boundary potential
for the weight $q$ part of the brane,
times the circumference $2\pi$ of the boundary of the hemisphere.
Indeed, the part $\zeta(\Is)-(\theta+2\pi q)(\Rs)$ 
is nothing but $2\pi V_{\it cl}(\usigma)$ in (\ref{Vclass}).
Also, when the gauge group is Abelian,
the rest matches with the quantum correction
$2\pi\sDelta V(\usigma)$ computed in \cite{HHP}.

As the condition for the brane $(\brane,\bgamma)$ to behave nicely
in the quantum theory, we require that the integral (\ref{Zhs})
is absolutely convergent. For this it would be sufficient
if the function $\AO_q$ for all weights $q$ in $(\bV,\rho)$ were to blow up
at the ends of the contour $\bgamma$.
This generalizes the idea of \cite{HHP} that the effective boundary potential
is bounded from below on $\bgamma$.
Also, the poles (\ref{poles}) are on
${\rm Re}Q_i(\usigma)=0$ and ${\rm Im}Q_i(\usigma)>0$,
which are the singular half-hyperplanes of the effective potential
of \cite{HHP}.
This is not a coincidence. The pole (\ref{poles}) is due to
the presence of a matter zero mode on the hemisphere, and
that exists only when there is a normalizable zero mode near the boundary.
In view of this, the following proposal was made in \cite{HR}
concerning the correct choice of $\bgamma$ for $\brane$
in a theory with a general gauge group $G$:

\fbox{
\begin{minipage}{14.35cm}

\medskip
{\it The Lagrangian submanifold $\gamma$ is a
deformation of the real locus $\im \ttt$, avoiding the poles
(\ref{poles}), so that for any Chan-Paton weight $q$ of the brane
$\brane$, $\AO_q(\usigma)$ grows to infinity
in every asymptotic
direction of $\bgamma$.}

\medskip
\end{minipage}
}

\vspace{0.2cm}
\noindent
When there is no $\bgamma$ satisfying this condition,
$\brane$ is not acceptable as a part of the datum for a brane.
This is the proposal made in \cite{HR}
for the grade restriction rule.
For Calabi-Yau models with Abelian gauge groups,
this reproduces the grade restriction rule of \cite{HHP}
for the reason just stated.
The rule was also examined in non-Calabi-Yau models
and simple models with non-Abelian gauge groups,
and reasonable results were obtained in \cite{HR}.
(See also \cite{CLFM} for a further study.)
In the present paper, we shall examine 
this proposal in the R\o dland model and see
if the picture (\ref{maindiagram}) emerges.

An equivalent but a simpler test can be done using the function $\AN_{\wtq}$
in (\ref{Aq}): the latter part can be replaced by

\fbox{
\begin{minipage}{14.35cm}

\medskip
    {\it \ldots, so that for each irreducible Chan-Paton component of $\brane$,
      $\AN_{\lambda+\rho}(\usigma)$ for the highest weight $\lambda$
      grows to infinity in every asymptotic direction of $\bgamma$.}

\medskip
\end{minipage}
}

\vspace{0.2cm}
\noindent
That is, for an irreducible representation $\rmV$ of $G$,
  growth of $\AO_q$ for all weights $q$ of $\rmV$
  is equivalent to growth of $\AN_{\lambda+\rho}$
  for the highest weight $\lambda$ of $\rmV$.
To see this, we first note that for any $q\in i\ttt^*$
\beq
\AN_{q+\rho}(\usigma)-\AO_q(\usigma)
=\pi\sum_{\alpha>0}(|\alpha(\Rs)|-\alpha(\Rs))\geq 0
\label{ineq1}
\eeq
where equality holds if and only if $\Rs$ is in the closure
$\overline{C}_+$ of a positive Weyl chamber $C_+\subset i\ttt$.
Thus, growth of $\AO_q$ for all weights $q$ of $\rmV$ implies
growth of $\AN_{\lambda+\rho}$ for the highest weight $\lambda$.
To see the converse, let $q$ be an arbitrary weight of $\rmV$
and let
$\usigma$ be an arbitrary element of $\ttt_{\C}$.
Let us take a Weyl group element $w$ such that
$w^{-1}\Rs$ is in $\overline{C}_+$.
Then,
\beqa
\AO_q(\usigma)&=&\AO_{wq}(w^{-1}\usigma)
=\AN_{wq+\rho}(w^{-1}\usigma)
\nn\\
&=&\AN_{\lambda+\rho}(w^{-1}\usigma)+2\pi(\lambda-wq)(w^{-1}\Rs)\geq
\AN_{\lambda+\rho}(w^{-1}\usigma),
\label{ineq2}
\eeqa
where $w^{-1}\Rs\in \overline{C}_+$ is used in the second equality
and in the last inequality in which the fact
that $\lambda-wq$ is a non-negative span of simple roots is also used.
Thus, growth of $\AN_{\lambda+\rho}$ for the highest weight $\lambda$
implies growth of $\AO_q$ for all weights $q$ of $\rmV$,
provided the contour $\bgamma$ is Weyl invariant.

As we will see, it turns out that
this proposal for the grade restriction rule is too strong:
The picture (\ref{maindiagram}) does not hold for a general window
${\bf w}$. A weaker condition is needed.
In the present paper, we will propose a new one
so that the picture (\ref{maindiagram}) holds for all possible windows
${\bf w}$.
Expression (\ref{ZV}) for the hemisphere integral and the
use of the function $\AN_{\lambda+\rho}$, rather than $\AO_q$,
will be essential in the formulation.

It was observed in \cite{HR} and later proved in \cite{BaPl}
that the hemisphere partition function
computes the central charge of the D-brane. See also
\cite{HR2} for a supplementary remark on the proof
as well as on other results on D-brane central charges.
The central charge of a B-brane ${\mathcal B}$ in the non-linear sigma model
with target $M$ behaves in the large volume limit as
\beq
Z_M({\mathcal B})\,=\,\int_M
\widehat{\Gamma}_M\e^{{1\over 2\pi}(B+\im\omega)}\,
{\rm ch}({\mathcal B})\,+\,\cdots,
\label{Zgeometric}
\eeq
where $\widehat{\Gamma}_M$ is the Gamma class of $M$ (the characteristic class
associated with the function $\Gamma(1-{x\over 2\pi\im})$), 
$\omega-\im B$ is the complexified K\"ahler form,\footnote{The
  conventions for the sigma model are as in Appendix~\ref{app:NLSM}.
In particular the B-field class $[B]$ is regarded as an element of
$\rmH^2(M,\R/2\pi\Z)$.}
and $+\cdots$ consists of terms which decay as
$\e^{-\int_{\beta}(\omega-\im B)}$ for some effective homology classes $\beta$.
The formula (\ref{Zgeometric}) came from
mathematical studies of the central charge
\cite{Hosono,Iritani09,Katzarkov,Iritani11}.
(Further studies, including the relation to the hemisphere partition function,
can be found in \cite{KnRoSc,AlLi}.)
Moreover, the central charge obeys differential equations with respect to the
complexified K\"ahler parameters which correspond to
the Picard-Fuchs equations on the mirror side.
For a Calabi-Yau threefold $M$ of Picard number one, with a divisor
$H$ and a line $\ell$ such that $H\cdot\ell=1$, just as for
$M=X$ or $Y$ in the R\o dland model,
the D0 brane at a point, the D2 brane wrapped on $\ell$,
the D4 brane wrapped on $H$ and the D6 brane filling $M$, or their
Chern characters,
$[{\rm pt}]$, $\ell+[{\rm pt}]$,
$H-{1\over 2}H^2+{1\over 6}H^3$ and $1$, to be more precise,
form an integral basis of the lattice of Chern character images
of all B-branes.
They have the following central charges
\beqa
Z_M({\mathcal O}_{\rm pt})&=&\Pi_0,\label{Zpt}\\
Z_M({\mathcal O}_{\ell})&=&\Pi_1+\Pi_0,\label{ZOell}\\
Z_M({\mathcal O}_H)&=&{H^3\over 2}\Pi_2-{H^3\over 2}\Pi_1
+\left({c_2H\over 24}+{H^3\over 6}\right)\Pi_0,\label{ZOH}\\
Z_M({\mathcal O}_M)&=&{H^3\over 6}\Pi_3+{c_2H\over 24}\Pi_1
+\im{\zeta(3)c_3\over (2\pi)^3}\Pi_0,\label{ZOX}
\eeqa
where $\Pi_j$ are the periods, i.e. solutions to the Picard Fuchs equations,
which behave as
\beq
\Pi_j\,=\,\left({\log z\over 2\pi\im}\right)^j\,+\,O(z),
\label{Pij}
\eeq
for $[\omega-\im B]=-\log z H+O(z)$.
Computing the hemisphere partition function and comparing
the results with the above form, we can obtain some idea of
the geometric image of the brane.
This is what we shall carry out in the two phases of the R\o dland model
and in  the Hosono-Takagi model.
This is particularly useful in the strongly coupled phase
where it is not straightforward to read off the geometric interpretation
of a given linear sigma model brane.
Also, we will be interested in some subsets of $D_G(V,W)$ and examine whether
they are large enough to produce all possible branes in the low energy
theories. A diagnostic tool at the level of Ramond-Ramond charges
is provided by the hemisphere partition function.

\subsection{The Grade Restriction Rule Of $SU(2)$ Supersymmetric QCD}
\label{subsec:SU2GRR}

Let us consider the $SU(2)$ gauge theory with $N$ doublets
$x_1,\ldots, x_N$, with or without superpotential.
We only consider the case where $N$ is odd so that the theory is regular
--- the classical Coulomb branch is lifted by quantum corrections \cite{HoTo}.
As the gauge group is simple, the model has
no FI-theta parameter.

We choose the diagonal subgroup as a maximal torus of $SU(2)$, so that
the scalar component of the vector multiplet is parametrized as
(\ref{sigmapara}) with $\sigma_1+\sigma_2=0$.
We shall write $\cs=\rs+\im\is$ for $\usigma_1=-\usigma_2$.
For the spin $h/2$ representation, that is,
the $h$-th symmetric power of the fundamental doublet,
the shifted highest weight is
\beq
(\lambda_h+\rho)(\usigma)=(h+1)\cs
\eeq
and the growth function (\ref{Aq}) is given by
\beqa
\AN_{\lambda_h+\rho}(\usigma)&=&-2\pi(h+1)\rs\nn\\
&&
+N\left\{\is(\log|\cs|-1)+\rs\Arg(\im\cs)\right\}
+N\left\{-\is(\log|-\cs|-1)-\rs\Arg(-\im\cs)\right\}
\nn\\
&=&\pi(N|\rs|-2(h+1)\rs)
\eeqa
where we used
\beq
\Arg(\im\cs)-\Arg(-\im\cs)=\pi\,{\rm sign}(\rs).
\label{Argid}
\eeq
The growth condition is $\AN_{\lambda_h+\rho}\to \infty$ as $|\rs|\to\infty$.
It is satisfied if and only if $h+1<N/2$. That is,
\beq
h=0,1,\ldots, {N-3\over 2}.
\label{SU2QCDGRR}
\eeq
This is the grade restriction rule of the model.

For $N=1$, no representation is allowed. This may appear strange but
that is not the case: The $N=1$ model exhibits supersymmetry breaking
\cite{HoTo} and there are no supersymmetric D-branes to consider.
For $N=3$, only the trivial representation, the singlet, is allowed.
For $N=5$, only the singlet and the doublet are allowed.
For $N=7$, only the singlet, the doublet and the triplet are allowed.
These will be relevant in the study of D-branes in the Pfaffian phase
of the R\o dland model.

This generalizes to any theory
with matter in a {\it quasi-symmetric representation}, that is, when
the sum of weights that lie in each
line through the origin in $\im\ttt^*$ vanishes.
The logarithms and Args disappear from 
the growth function and the analysis of convergence is particularly simple.
Indeed, using (\ref{Argid}) we find
\beq
\AN_{\wtq}(\usigma)
=(\zeta-\zeta_V)(\Is)-(\theta+2\pi\wtq)(\Rs)+{\pi\over 2}\sum_i|Q_i(\Rs)|,
\eeq
where $\zeta_V$ is a Weyl invariant element of $\im\ttt^*$ determined by
$\zeta_V(\Is)=-\sum_iQ_i(\Is)\log|Q_i(\Is)|$.
This yields the following condition on the Chan-Paton representations:
the highest weight $\lambda$ satisfies
\beq
\left(\lambda+\rho+{\theta\over 2\pi}\right)(\Rs)~<~{1\over 4}\sum_i|Q_i(\Rs)|
\label{QSGRR}
\eeq
for all non-zero $\Rs\in \im\ttt$.
This is equivalent, by the argument with (\ref{ineq1})-(\ref{ineq2}),
to the condition that all the weights $q$ satisfy
$\left(q+{\theta\over 2\pi}\right)(\Rs)<{1\over 4}\sum_i|Q_i(\Rs)|-{1\over 4}\sum_{\alpha}|\alpha(\Rs)|$ for all non-zero $\Rs\in \im\ttt$.
The condition is well posed ---
there are no weights on the edge of the inequalities ---
as long as the theory is regular \cite{HoriNCCRtalk}.

When the theory has \emph{no FI parameter}, 
$\mathfrak{g}^{*G}=\{0\}$, as in the $SU(2)$ SQCD,
this is the same as the condition that determines the
\emph{twisted\,\footnote{Even if there is no FI parameter,
  there can be a non-trivial discrete theta parameter
  which affects the regularity condition. In such a case, the
  Chan-Paton representation is a \emph{projective} representation compatible
  with the theta parameter \cite{HoriNCCRtalk},
  to which the term \emph{twisted} corresponds.}\!
non-commutative crepant resolution}
by \v{S}penko and Van den Bergh \cite{SpVdB}
of the quotient singularity $V/G_{\C}$.
See \cite{HoriNCCRtalk} for details.
In Section~\ref{sec:other},
we shall spell out and utilise this condition for
other supersymmetric QCDs:
(\ref{USp4GRR}) in $USp(4)$ with seven fundamentals,
(\ref{O2GRR}) in $O(2)$ with five fundamentals, and
(\ref{SO4GRR}) in $SO(4)$ with five fundamentals.

In a theory \emph{with an FI parameter}, $\mathfrak{g}^{*G}\ne \{0\}$,
in which each phase is weakly coupled,
this condition is the same as the one
that determines the magic window category of \cite{DHLSam},
provided the space constructed there is identified with
the FI-theta parameter space (up to the shift of $\zeta$ by $\zeta_V$).
The identification of the space of \cite{DHLSam}
with the FI-theta parameter space
is discussed in Abelian theories in \cite{Kite} using a known form
of the latter
and in a theory with a $U(2)$ gauge group in \cite{Varshney}
using a construction of the space of Bridgeland stability conditions.
The precise relation in the general case is explained in \cite{HoriNCCRtalk}.

\section{The Grade Restriction Rule Of The R\o dland Model}
\label{sec:RodlandGRR}

In this section, we attempt to determine the grade restriction rule
in the R\o dland model using
the hemisphere partition function
\beq
Z_{D^2}(\brane)=\int\limits_{\,\,\,\,\sgamma}
\dd\usigma_1\,\dd\usigma_2\,
(\usigma_1-\usigma_2)\sinh(\pi(\usigma_1-\usigma_2))
\,\Gamma_{\rm R\o dland}(\usigma)\,\e^{\im t(\usigma)}\,f_{\brane}(\usigma)
\label{HSRod}
\eeq
where
\beq
\Gamma_{\rm R\o dland}(\usigma)=
\Gamma\bigl(-\im(\usigma_1+\usigma_2)+1-\epsilon\bigr)^7\,
\Gamma\left(\im\usigma_1+\mbox{${\epsilon\over 2}$}\right)^7\,
\Gamma\left(\im\usigma_2+\mbox{${\epsilon\over 2}$}\right)^7.
\label{GammaRod}
\eeq
We take the R-charge parameter $\epsilon$ of (\ref{Rassignment})
to be in the range $0<\epsilon <1$ to ensure (\ref{Rrange}).
We would like to find which irreducible representations $\rmV$ of $U(2)$
can enter into the Chan-Paton vector space of $\brane$.
There must be a contour $\bgamma$ on which the hemisphere integral
\beqa
Z_{\sgamma}(\rmV)&=&\int\limits_{\,\,\,\,\sgamma}
\dd\usigma_1\,\dd\usigma_2\,
(\usigma_1-\usigma_2)\sinh(\pi(\usigma_1-\usigma_2))
\,\Gamma_{\rm R\o dland}(\usigma)\,\e^{\im t(\usigma)}\,
        {\rm tr}_{\rmV}\e^{2\pi\usigma}
        \label{ZVRoddef}\\
&=&\int\limits_{\,\,\,\,\sgamma}
\dd\usigma_1\,\dd\usigma_2\,
(\usigma_1-\usigma_2)\,\Gamma_{\rm R\o dland}(\usigma)\,\e^{\im t(\usigma)}\,
\e^{2\pi(\lambda+\rho)(\usigma)}
\label{ZVRod}
\eeqa
is convergent. $\lambda$ in the last expression is the highest weight of $\rmV$.
The second equality holds as long as
$\bgamma$ is a Weyl invariant deformation of the real locus $\im\ttt$,
as shown in (\ref{ZVdef})-(\ref{ZV}).
The asymptotic behaviour of the integrand of (\ref{ZVRod}) is determined by
$\wt{W}_{\!{\it eff},\lambda+\rho}$ and $\AN_{\lambda+\rho}$ 
as in (\ref{ZVint}) and (\ref{behav}) where
\beqa
\wt{W}_{\!{\it eff},\wtq}(\usigma)
&=&-t(\usigma_1+\usigma_2)+2\pi\im (\wtq^1\usigma_1+\wtq^2\usigma_2)\nn\\
&&-7\usigma_1\log\im\usigma_1
-7\usigma_2\log\im\usigma_2
+7(\usigma_1+\usigma_2)\log(-\im(\usigma_1+\usigma_2)),
\label{W}
\eeqa
and 
\beqa
\AN_{\wtq}(\usigma)&=&{\rm Re}\left(\im \wt{W}_{\!{\it eff},\wtq}\right)
=-{\rm Im}\left(\wt{W}_{\!{\it eff},\wtq}(\usigma)\right)\nn\\
&=&\zeta(\Is_1+\Is_2)-\theta(\Rs_1+\Rs_2)-2\pi (\wtq^1\Rs_1+\wtq^2\Rs_2)
\nn\\
&&+7\Rs_1\Arg(\im\usigma_1)+7\Rs_2\Arg(\im \usigma_2)
-7(\Rs_1+\Rs_2)\Arg(-\im(\usigma_1+\usigma_2))\nn\\
&&+7\Is_1\log|\usigma_1|+7\Is_2\log|\usigma_2|
-7(\Is_1+\Is_2)\log|\usigma_1+\usigma_2|,
\label{A}
\eeqa
with
\beq
\usigma_a=\Rs_a+\im\Is_a,\quad a=1,2.
\eeq
In these expressions, ${\rm Im} \log=\Arg$ is assumed to take values in
the open interval $(-\pi,\pi)$.
For absolute convergence of the integral,
we need $\AN_{\lambda+\rho}$ to blow up at infinity of $\bgamma$.
The shifted highest weight $\lambda+\rho$ of 
\beq
S^{\elll}\fund(i)=\Sym^{\elll}\fund\otimes (\det\fund)^{\otimes i},
\eeq
is
\beq
\wtq_{\elll,i}=\left(i+\elll+{1\over 2},
i-{1\over 2}\right).
\label{wtqli}
\eeq
$\bgamma$ must also be a Lagrangian submanifold of $\ttt_{\C}$
obtained by deforming the real locus $\im\ttt$ without meeting the poles
of $\Gamma_{\rm R\o dland}(\usigma)$.
The latter is assured if $\bgamma$ avoids the walls
\beqa
W_1&=&\{\,\usigma_1\in\im\R_{>0}\,\}\,=\,\{\,\Rs_1=0,\,\Is_1>0\,\},\label{W1}\\
W_2&=&\{\,\usigma_2\in\im\R_{>0}\,\}\,=\,\{\,\Rs_2=0,\,\Is_2>0\,\},\label{W2}\\
W_p&=&\{\,-\usigma_1-\usigma_2\in\im\R_{>0}\,\}\,=\,
\{\,\Rs_1+\Rs_2=0, \,\Is_1+\Is_2<0\,\},\label{Wp}
\eeqa
on which the poles lie.
When we regard $\bgamma$ as a graph of a function $\Rs\mapsto \Is(\Rs)$,
the condition to avoid these walls is
\beqa
\Rs_1=0&\Longrightarrow&\Is_1\leq 0,\label{Wedge1}\\
\Rs_2=0&\Longrightarrow&\Is_2 \leq 0,\label{Wedge2}\\
\Rs_1+\Rs_2=0&\Longrightarrow&\Is_1+\Is_2\geq 0.\label{Wedgep}
\eeqa

We will examine the behaviour of the growth function $\AN_{\wtq}(\usigma)$
when $\usigma_1$, $\usigma_2$ and $-\usigma_1-\usigma_2$ have large
absolute values and are away from $\im\R_{>0}$.
In the analysis, it is often convenient to decompose it
into two parts
\beqa
\AN^{\Rs}_{\wtq}(\usigma)\!\!&:=&\!\!
-\theta(\Rs_1+\Rs_2)-2\pi(\wtq^1\Rs_1+\wtq^2\Rs_2)\nn\\
&&+7\Rs_1\Arg(\im\usigma_1)+7\Rs_2\Arg(\im \usigma_2)
-7(\Rs_1+\Rs_2)\Arg(-\im(\usigma_1+\usigma_2)),\nn\\
\AN^{\Is}_{\wtq}(\usigma)\!\!&:=&\!\!\zeta(\Is_1+\Is_2)
+7\Is_1\log|\usigma_1|+7\Is_2\log|\usigma_2|
-7(\Is_1+\Is_2)\log|\usigma_1+\usigma_2|~~~~\nn
\eeqa
and look at them separately.

Before starting, let us discuss the symmetries of the growth function.
Obvious but important is the scaling property
\beq
\AN_{\wtq}(\lambda\usigma)\,=\,\lambda\AN_{\wtq}(\usigma)\quad
\mbox{for $\lambda>0$.}
\label{scaleinv}
\eeq
Another is the invariance under the Weyl reflection,
$\usigma_1\leftrightarrow \usigma_2$
and $\wtq^1\leftrightarrow \wtq^2$. Less obvious is the invariance
under $\theta\to -\theta$, $\wtq\to -\wtq$, $(\Rs_a,\Is_a)\to (-\Rs_a,\Is_a)$.
This is because $\Arg(\im x-y)$ flips its sign under $(x,y)\to (-x,y)$.
Combining this with the Weyl symmetry and the equivalence (\ref{thetashift}),
we find that $\AN_{\wtq_{h,i}}$ is invariant under
\beq
\theta\to -\theta+2\pi n,~~
i\to -n-h-i,~~
(\Rs_1,\Is_1,\Rs_2,\Is_2)\to (-\Rs_2,\Is_2,-\Rs_1,\Is_1).
\label{reflectionsym}
\eeq
for each integer $n$.
We shall call this the {\it reflection symmetry} at $n\pi$ as it involves the
refection of $\theta$ at $n\pi$.

\subsection{Weakly Coupled Phase --- No Grade Restriction Rule}
\label{subsec:weak}

Let us first discuss the condition
deep inside the weakly coupled phase $\zeta\gg 0$.
In view of the term $\zeta(\Is_1+\Is_2)$ in (\ref{A}), we would like 
to have $\Is_1+\Is_2\gg 0$ at infinity of $\bgamma$.
Paying attention to the wall-avoiding conditions
(\ref{Wedge1}) and (\ref{Wedge2}),
we can take a contour $\bgamma_+^{\alpha}$ defined by
\beq
\Is_1=|\Rs_1|^{\alpha},\quad
\Is_2=|\Rs_2|^{\alpha}
\label{gammaplus}
\eeq
for some positive $\alpha$.
This is indeed a Weyl invariant and Lagrangian submanifold of $\ttt_{\C}$
which satisfies also the remaining wall-avoiding condition (\ref{Wedgep}).
In fact, this contour with $\alpha>1$ works for an arbitrary representation.

Indeed the growth function $\AN_{\wtq}(\usigma)$ blows up at the infinity of
$\gamma_+^{\alpha}$ for any $\wtq$.
To see this, let us look at the
behaviour of $\AN_{\wtq}|_{\sgamma_+^{\alpha}}$
along the ray $(\Rs_1,\Rs_2)=(\ulambda \Rs_1^0,\ulambda \Rs_2^0)$
as $\ulambda\to +\infty$
for a fixed $(\Rs_1^0,\Rs_2^0)$ on the circle
$|\Rs_1^0|^{\alpha}+|\Rs_2^0|^{\alpha}=1$.
In this direction, the part $\AN^{\Rs}_{\wtq}|_{\sgamma_+^{\alpha}}$
is essentially linear in $\ulambda$ since the range of $\Arg$ is in the
interval $(-\pi,\pi)$. On the other hand,
\beqa
\AN^{\Is}_{\wtq}|_{\sgamma_+^{\alpha}}
&=&\Is_1\left(\zeta+7\log{|\usigma_1|\over |\usigma_1+\usigma_2|}\right)
+\Is_2\left(\zeta+7\log{|\usigma_2|\over |\usigma_1+\usigma_2|}\right)
\nn\\
&\geq&
\ulambda^{\alpha}\Bigl(\,
  \zeta+7|\Rs_1^0|^{\alpha}\log |\Rs_1^0|^{\alpha}
  +7|\Rs_2^0|^{\alpha}\log |\Rs_2^0|^{\alpha}+O(1/\ulambda^{\alpha-1})\,\Bigr),
  \label{AyG}
\eeqa
where we used 
\beq
{|\usigma_a|\over |\usigma_1+\usigma_2|}
\geq{|\Is_a|\over |\Is_1+\Is_2|}(1+O(1/\ulambda^{\alpha-1}))
=|\Rs_a^0|^{\alpha}(1+O(1/\ulambda^{\alpha-1})).
\eeq
The function $x\log x+(1-x)\log(1-x)$
for $0\leq x\leq 1$
has a minimum $-\log 2$ at $x=1/2$.
Therefore, if $\zeta>7\log 2$,
the part (\ref{AyG}) grows to positive infinity as
$\ulambda^{\alpha}$ and is dominant for $\alpha>1$
in $\AN_{\wtq}|_{\sgamma_+^{\alpha}}$.
This means that there is a positive constant $C$ such that 
\beq
\AN_{\wtq}|_{\sgamma_+^{\alpha}}\geq C\|\Rs\|^{\alpha}
\eeq
at large enough $\|\Rs\|=\sqrt{|\Rs_1|^2+|\Rs_2|^2}$.
The function $\AN_{\wtq}$ blows up at infinity of $\bgamma_+^{\alpha}$
in this way, provided $\zeta>7\log 2$ and $\alpha>1$.

Thus, any representation can be included
in the Chan-Paton vector space.
For all representations, the hemisphere integral
is absolutely convergent on the contour
$\gamma_+^{\alpha}$ defined in (\ref{gammaplus}) with $\alpha>1$.
By continuous change of $\alpha$, we may take the contour
to be $\bgamma_+:=\bgamma_+^2$.
As expected, there is no grade restriction rule
in the weakly coupled phase.\footnote{A proof of this expectation
  that appies in some generality was recently given in \cite{SuetakeThesis}.}

\subsection{Strongly Coupled Phase --- $SU(2)$ Grade Restriction Rule}
\label{subsec:SU2GRRRodland}

Let us next examine the condition deep inside the
strongly coupled phase $\zeta\ll 0$.
In view of the term
$\zeta(\Is_1+\Is_2)$ in (\ref{A}), we would like 
to have $\Is_1+\Is_2\ll 0$ at infinity of $\bgamma$, but that is
in conflict with the wall-avoiding condition (\ref{Wedgep}).
We may, however, take $\bgamma$ such that
$\Is_1+\Is_2$ goes to negative infinity in almost all directions
except along the line $\Rs_1+\Rs_2=0$.
For example, we can take $\bgamma_-^{\alpha}$ for some positive
$\alpha$ defined by
\beq
\Is_1=-|\Rs_1+\Rs_2|^{\alpha},\qquad
\Is_2=-|\Rs_1+\Rs_2|^{\alpha},
\label{gammaminus}
\eeq
which is a Weyl invariant Lagrangian submanifold of $\ttt_{\C}$
satisfying also the other
wall-avoiding conditions (\ref{Wedge1}) and (\ref{Wedge2}).
This contour includes a line
$l=\{\Is_1=\Is_2=\Rs_1+\Rs_2=0\}$
that extends to infinity,
along which
\beqa
\AN_{\wtq}|_l&=&-2\pi (\wtq^1-\wtq^2)\Rs_1
+7\Rs_1(\Arg(\im \Rs_1)-\Arg(-\im\Rs_1))\nn\\
&=&-2\pi (\wtq^1-\wtq^2)\Rs_1+7\pi|\Rs_1|,
\label{essASU2}
\eeqa
where we used
$\Arg(\pm\im\Rs_1)=\pm{\pi\over 2}{\rm sgn}(\Rs_1)$.
This grows to positive infinity as $\Rs_1\to\pm\infty$
provided $|\wtq^1-\wtq^2|<7/2$.
For the shifted highest weight $\wtq_{\elll,i}$ of $S^{\elll}\fund(i)$ given in
(\ref{wtqli}), $\wtq_{\elll,i}^1-\wtq_{\elll,i}^2=\elll+1$ and hence
the condition is $\elll<5/2$, that is,
\beq
\elll=0,1,2.
\eeq
In fact, the contour $\bgamma_-^{\alpha}$ with $\alpha>1$
works for all representations within this range.

Let us first examine the behaviour of
$\AN_{\wtq}|_{\sgamma_-^{\alpha}}$ at infinity in the directions of
$(\Rs_1,\Rs_2)$ other than the line $l$, that is, along the ray
 $(\Rs_1,\Rs_2)=(\ulambda \Rs_1^0,\ulambda \Rs_2^0)$
as $\ulambda\to +\infty$
for a fixed $(\Rs_1^0,\Rs_2^0)$ such that $\Rs^0:=\Rs_1^0+\Rs_2^0\ne 0$.
The part $\AN^{\Rs}_{\wtq}|_{\sgamma_-^{\alpha}}$ is essentially linear in
$\ulambda$ as the range of $\Arg$ is bounded. On the other hand,
\beq
\AN^{\Is}_{\wtq}|_{\sgamma_-^{\alpha}}
\,=\,
\ulambda^{\alpha}|\Rs^0|^{\alpha}\left(
-2\zeta+7\log{|\usigma_1+\usigma_2|^2\over |\usigma_1\usigma_2|}\right)
\,=\,
\ulambda^{\alpha}|\Rs^0|^{\alpha}\left(-2\zeta+7\log 4
+O(1/\ulambda^{\alpha-1})\right),
\eeq
where we used
\beq
{|\usigma_1+\usigma_2|^2\over |\usigma_1\usigma_2|}
={(\Is_1+\Is_2)^2\over |\Is_1\Is_2|}(1+O(1/\ulambda^{\alpha-1}))
=4(1+O(1/\ulambda^{\alpha-1})).
\eeq    
Thus, if $-2\zeta+7\log 4>0$, the part
$\AN^{\Is}_{\wtq}|_{\sgamma_-^{\alpha}}$ grows to positive infinity
as $\ulambda^{\alpha}$ and is dominant for $\alpha>1$
in $\AN_{\wtq}|_{\sgamma_-^{\alpha}}$.
In this way, we find that $\AN_{\wtq}|_{\sgamma_-^{\alpha}}$  for any $\wtq$
blows up in all directions other than $l$
as long as $\zeta<7\log 2$.

This leaves us to consider the region where $|\Rs_1+\Rs_2|$ is
small compared to $\varepsilon|\Rs_1|$ and $\varepsilon|\Rs_2|$
for some $\varepsilon\ll 1$.
In such a regime, $\Rs_1\simeq-\Rs_2$ with an error of order
$\varepsilon|\Rs_1|$, but
\beq
    {|\Rs_1+\Rs_2|^{\alpha}\over \Rs_1}=:1/x
\label{defx}
\eeq
can take an arbitrary value since the upper bound 
$\varepsilon^{\alpha}|\Rs_1|^{\alpha-1}$ can be large. 
Note that
\beq
{|\usigma_1+\usigma_2|^2\over |\usigma_1\usigma_2|}
\geq{(2|\Rs_1+\Rs_2|^{\alpha})^2\over (\Rs_1)^2+|\Rs_1+\Rs_2|^{2\alpha}}
(1+O(\varepsilon))
={4\over x^2+1}(1+O(\varepsilon)).
\eeq 
Using this, we find
\beqa
\AN_{\wtq_{h,i}}|_{\sgamma_-^{\alpha}}&\geq&
\Rs_1\Bigl(7\Arg(1+\im x)-7\Arg(1-\im x)-2\pi(h+1)+O(\varepsilon)\Bigr)\nn\\
&&+|\Rs_1+\Rs_2|^{\alpha}\left(-2\zeta+7\log\left({4\over x^2+1}\right)
+O(\varepsilon)\right)
\nn\\
&=&2|\Rs_1+\Rs_2|^{\alpha}\Bigl(-\zeta+f_h(x)+O(\varepsilon)\Bigr),
\eeqa
with
\beq
f_h(x)={7\over 2}\log\left({4\over x^2+1}\right)
+{x\over 2}\Bigl(7\Arg(1+\im x)-7\Arg(1-\im x)
-2\pi(h+1)\Bigr).
\label{deffh}
\eeq
For $-\infty <x<+\infty$, this function
is bounded from below with a minimum at positive $x$ of order 1.
With the substitution $x=\tan\uvarphi$, it simplifies as
$f_h(\tan\uvarphi)=7\log(2\cos\uvarphi)+\tan\uvarphi(7\uvarphi-\pi(h+1))$
and is minimized at $\varphi={\pi\over 7}(h+1)$, with the value
\beq
f_{h\,{\rm min}}=7\log\left(2\cos{\pi(h+1)\over 7}\right).
\label{fhmin}
\eeq
It is less than the bound $7\log 2$ from the other regime. Therefore,
as long as $\zeta<f_{h\,{\rm min}}$,
$\AN_{\wtq_{h,i}}$ is positive on $\bgamma_-^{\alpha}$.
Note that this $f_{h\,{\rm min}}$ is the value 
of the FI parameter at the singular point $\alpha_{\bfa}$
given in (\ref{defalphas}) with $\bfa=h+1$ ($h=0,1,2$).
This fact will be useful in further analysis of convergent contours.

Thus, in the strongly coupled phase, the hemisphere integral
on the contour $\bgamma_-^{\alpha}$ defined in (\ref{gammaminus}) with
$\alpha>1$ is absolutely convergent if and only if the representation is
from
\beq
\C(i),\quad \fund(i),\quad \Symt(i)\quad\mbox{for all $i\in\Z$}.
\label{GRRstrong}
\eeq
By continuous change of $\alpha$, we may take the contour
to be $\bgamma_-:=\bgamma_-^2$.
(\ref{GRRstrong}) is the set of representations that can be included in the
Chan-Paton vector space of branes.
This can be regarded as the grade restriction rule in the strongly
coupled phase.

The set ${\mathcal S}$ of D-branes satisfying this rule is the one we would
like to have in the diagram (\ref{maindiagram}). The set of representations
in (\ref{GRRstrong}) includes the set $\RADS$ (\ref{ADSset})
corresponding to the subcategory $\TADS$ of $D_G(V,W)$ which is shown in
\cite{ADS} to be equivalent to the derived category of coherent sheaves $D(Y)$
on the Pfaffian Calabi-Yau variety. This is reassuring:
This suggests that the set ${\mathcal S}$ is large enough to produce all
branes in the low energy sigma model along the renormalization group flow
$\pi_Y:{\mathcal S}\to D(Y)$.

The essential part in the derivation of the rule (\ref{GRRstrong})
is in the discussion on the contour inside the subspace
$\usigma_1+\usigma_2=0$ of $\ttt_{\C}$, which is the Cartan subalgebra
of the $SU(2)$ subgroup of $U(2)$. In fact, the rule reflects
the grade restriction rule of 
the $SU(2)$ gauge theory with
seven copies of the fundamental doublets $\C^2$ discussed in
Section~\ref{subsec:SU2GRR}:
only $\C$, $\C^2$ and $\Sym^2\C^2$
are acceptable as Chan-Paton representations.
This is consistent with the physics of the phase $\zeta\ll 0$:
at an intermediate energy scale,
the theory reduces to the $SU(2)$ gauge theory with seven fundamentals
which is fibred over $\CP^6$ spanned by the $p$-fields.

\subsection{D-Brane Transport: The Setup}
\label{subsec:setup}

We now embark on the problem to determine the grade restriction rule
for D-brane transport along paths between the two phases.
First, let us fix our terminology and describe some basics facts.

Let us choose a path ${\mathcal P}$ in the FI-theta parameter space
between the two phases that avoids the singular points.
A representation $\rmV$ of the gauge group $U(2)$ is said to be
{\it grade restricted with respect to the path ${\mathcal P}$}
if there is a family of contours $\bgamma$ along ${\mathcal P}$
which interpolates between
the contour $\bgamma_+^{\alpha_+}$ at $\zeta\gg 0$
and the contour $\bgamma_-^{\alpha_-}$ at $\zeta\ll 0$, with $\alpha_{\pm}>1$,
on which the hemisphere integral of $\rmV$ is absolutely convergent all
the way along ${\mathcal P}$.
We recall that $\bgamma$ must be a Weyl invariant Lagrangian submanifold
obtained by a deformation of the real locus $\im\ttt$ without hitting
the poles of $\Gamma_{\rm R\o dland}(\usigma)$ given in (\ref{GammaRod}).
We expect that the condition is invariant under continuous
deformation of ${\mathcal P}$ and hence depends only on the homotopy class
of ${\mathcal P}$. We shall call a homotopy class of such paths
a {\it window} between the two phases. Thus, we also use the terminology
{\it grade restricted with respect to a window}.

The problem is to determine the set $\rmGRR_{\bf w}$
of grade restricted representations
for each window ${\bf w}$ between the two phases.
The representations must be taken from the set
(\ref{GRRstrong}) obeying the $SU(2)$ grade restriction rule in the strongly
coupled phase.
We denote by ${\mathcal T}_{\bf w}$ the set of D-brane data
$\brane$ whose Chan-Paton vector space is a direct sum of
members in the set $\rmGRR_{\bf w}$.
This set must be large enough
so that under the renormalization group flow
it produces all possible branes in the low energy theory
at any point on the path.

For a window ${\bf w}$ and an integer $i$,
we denote by ${\bf w}(i)$ the window obtained from
${\bf w}$ by the shift $\theta\to \theta +2\pi i$.
Since $\e^{\im t(\usigma)}{\rm tr}_{\rmV}\e^{2\pi\usigma}$
that appears in the hemisphere integral (\ref{ZVRoddef}) is invariant under 
$V\to V(i)=V\otimes \det^i\fund$ and $\theta\to \theta-2\pi i$,
if $\rmV$ is grade restricted with respect to a window
${\bf w}$, then, $\rmV(i)$ is grade restricted with respect to the window
${\bf w}(-i)$.
Therefore,
$\rmGRR_{{\bf w}(-i)}=\rmGRR_{\bf w}(i)$.
Of course, these are reflections of the fundamental
equivalence (\ref{thetashift}).

\begin{figure}[thb]
\psfrag{A}{\mbox{\small ${\mathcal P}_{\!\bf A}$}}
\psfrag{B}{\mbox{\small ${\mathcal P}_{\bf B}$}}
\psfrag{C}{\mbox{\small ${\mathcal P}_{\bf C}$}}
\psfrag{D}{\mbox{\small ${\mathcal P}_{\bf D}$}}
\psfrag{a1}{$t_1$}
\psfrag{a2}{$t_2$}
\psfrag{a3}{$t_3$}
\psfrag{a1'}{$t_1'$}
\psfrag{a2'}{$t_2'$}
\psfrag{a3'}{$t_3'$}
\psfrag{z}{$\zeta$}
\psfrag{th}{$\theta$}
\psfrag{-8pi}{$-8\pi$}
\psfrag{-6pi}{$-6\pi$}
\centerline{\includegraphics[page=1]{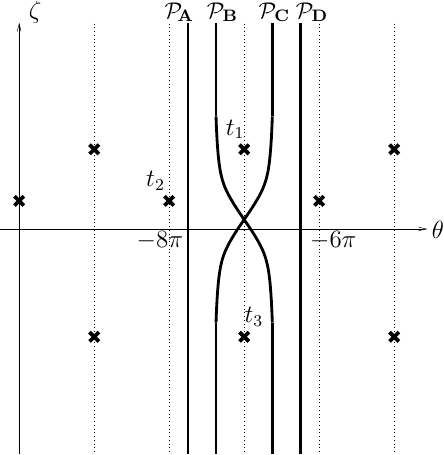}}
\caption{Four windows between the phases}
\label{fig:RodpathsX}
\end{figure}
We recall that there are three singular points
$\e^t=\alpha_1$, $\alpha_2$ and $\alpha_3$ given in (\ref{defalphas}).
Their inverse images under the exponential map
$t\mapsto \e^t$ form three arrays of points.
We give names to specific representatives as
\beqa
t_1&=&7\pi \im +7\log\left(2\cos\left({\pi\over 7}\right)\right),
\label{t1}\\
t_2&=&8\pi\im+7\log\left(2\cos\left({2\pi\over 7}\right)\right),
\label{t2}\\
t_3&=&7\pi\im+7\log\left(2\cos\left({3\pi\over 7}\right)\right).
\label{t3}
\eeqa
Other points are shifts of these by $2\pi\im\Z$.
As the typical windows between the phases, we consider
the four classes of paths, ${\mathcal P}_{\!\bf A}$,
${\mathcal P}_{\bf B}$, ${\mathcal P}_{\bf C}$ and ${\mathcal P}_{\bf D}$
as depicted in Fig. \ref{fig:RodpathsX}, and their $2\pi$
horizontal shifts.
Two of them, ${\mathcal P}_{\!\bf A}$ and ${\mathcal P}_{\bf D}$,
can be taken to be paths with fixed values of
$\theta$. We shall call them the straight paths.
Along two others, ${\mathcal P}_{\bf B}$ and ${\mathcal P}_{\bf C}$,
the values of $\theta$ must vary. We call them the curved paths.
These are ``fundamental'' in the sense that any path
between the two phases can be obtained by a combination of these and
their shifts by $\theta\to\theta+2\pi i$ with $i\in\Z$.
They are not independent.
A downward path in the window ${\bf D}$ followed by an upward path in
${\bf B}$ which is then followed by
a downward path in ${\bf A}$ can be continuously deformed to
a downward path in the window ${\bf C}$.
Such a homotopy equivalence relation of paths must be reflected in the
D-brane transport which will eventually result from
the grade restriction rule.

\subsection{Straight Paths}\label{subsec:straight}

As a warm-up, let us examine the convergence condition of
the hemisphere integral along the real locus $\bgamma=\im\ttt$.
Since $\Arg(\im \Rs_a)=\pm{\pi\over 2}{\rm sgn}(\Rs_a)$ for real
$\Rs_a$, the growth function on the real locus is
\beq
\AN_{\wtq}|_{\Is=0}=-(\theta+2\pi \wtq^1)\Rs_1-(\theta+2\pi\wtq^2)\Rs_2
+{7\pi\over 2}|\Rs_1|+{7\pi\over 2}|\Rs_2|
+{7\pi\over 2}|\Rs_1+\Rs_2|.
\label{ANreal}
\eeq
In the regions (i), (ii), (iii), (iv)
of the $\Rs$-plane as shown in Fig.~\ref{fig:sixreg},
\begin{figure}[htb]
\psfrag{Rs1}{\footnotesize $\Rs_1$}
\psfrag{Rs2}{\footnotesize $\Rs_2$}
\psfrag{i}{i}
\psfrag{ii}{ii}
\psfrag{iii}{iii}
\psfrag{iv}{iv}
\psfrag{i'}{i'}
\psfrag{ii'}{ii'}
\psfrag{iii'}{iii'}
\psfrag{iv'}{iv'}
\centerline{\includegraphics[page=2]{main-pics.pdf}}
\caption{Regions of the $\Rs$ plane}
\label{fig:sixreg}
\end{figure}
it is given by
\beqa
\mbox{(i):}&&|\Rs_1|(-\theta-2\pi\wtq^1+7\pi)
+|\Rs_2|(-\theta-2\pi\wtq^2+7\pi),\nn\\
\mbox{(ii):}&&|\Rs_1+\Rs_2|(-\theta-2\pi\wtq^1+7\pi)
+|\Rs_2|(-2\pi(\wtq^1-\wtq^2)+7\pi),\nn\\
\mbox{(iii):}&&|\Rs_1+\Rs_2|(\theta+2\pi\wtq^2+7\pi)
+|\Rs_1|(-2\pi(\wtq^1-\wtq^2)+7\pi),\nn\\
\mbox{(iv):}&&|\Rs_1|(\theta+2\pi\wtq^1+7\pi)
+|\Rs_2|(\theta+2\pi\wtq^2+7\pi),\nn
\eeqa
while the expressions on the other regions are obtained by Weyl reflection
$1\leftrightarrow 2$. We see that the growth of $\AN_{\wtq}|_{\Is=0}$
is equivalent to
\beq
2\pi |\wtq^1-\wtq^2|<7\pi,\quad
|\theta+2\pi\wtq^1|\,\,\&\,\,
|\theta+2\pi\wtq^2|<7\pi,
\label{condreal}
\eeq
which reads for the shifted highest weight $\wtq_{\elll,i}$ of
$S^{\elll}\fund(i)$ as
\beq
\elll=0,1,2,\quad -3<i+{\theta\over 2\pi}<3-\elll.
\eeq
The set of these representations
does not change as long as $\theta$ does not cross
$2\pi\Z$.
For example, in the region $-8\pi<\theta<-6\pi$, the set is
\beq
\begin{array}{cccccc}
\C(1)&\C(2)&\C(3)&\C(4)&\C(5)&\C(6)\\
\fund(1)&\fund(2)&\fund(3)&\fund(4)&\fund(5)&\\
\symt(1)&\symt(2)&\symt(3)&\symt(4)&&
\end{array}
\label{toosmall}
\eeq
This is too small to reproduce all branes in the low energy sigma models, both
in the Grassmannian phase $\zeta\gg 0$ and in the Pfaffian phase $\zeta\ll 0$,
since it is only a proper subset of the set $\RADS$ (\ref{ADSset})
corresponding to the subcategory $\TADS\subset D_G(V,W)$
which is equivalent both to $D(X)$ and $D(Y)$ \cite{ADS}.
The condition $-3<i+{\theta\over 2\pi}<3-\elll$ is too strong,
and that comes from the conditions $|\theta+2\pi\wtq^a|<7\pi$, $a=1,2$.
We would like to relax them by taking a better $\bgamma$.

So, let us turn on the imaginary part $\Is$. Consider
the graph of a non-zero function $\Rs\mapsto \Is(\Rs)$.
We maintain $\Is_1(\Rs)+\Is_2(\Rs)=0$ to simplify the analysis. Then,
the growth function is independent of $\zeta$, 
\beqa
\AN_{\wtq}|_{\Is=\Is(\Rs)}&=&-(\theta+2\pi \wtq^1)\Rs_1-(\theta+2\pi\wtq^2)\Rs_2
+7\Rs_1\Arg(\im\usigma_1)
+7\Rs_2\Arg(\im\usigma_2)
+{7\pi\over 2}|\Rs_1+\Rs_2|\nn\\
&&+7\Is_1\log|\usigma_1|+7\Is_2\log|\usigma_2|.
\label{ANgraph}
\eeqa
Compared to the real locus case (\ref{ANreal}),
instead of ${7\pi\over 2}|\Rs_a|$ we have $7\Rs_a\Arg(\im\usigma_a)$
that can be as close to $7\pi|\Rs_a|$ as possible.
Thus, there is a chance that the condition can be relaxed.
However, we do not want the second line to do any harm.
For this, let us consider $\Is(\Rs)$ so that $|\Is_1|=|\Is_2|$
is much bigger than $|\Rs_1|$ and $|\Rs_2|$ almost everywhere.
In such a region, 
\beqa
\Is_1\log|\usigma_1|+\Is_2\log|\usigma_2|
&=&\Is_1\log|\Is_1(1-\im \Rs_1/\Is_1)|
+\Is_2\log|\Is_2(1-\im \Rs_2/\Is_2)|\nn\\
&=&\Is_1\log|1-\im \Rs_1/\Is_1|
+\Is_2\log|1-\im \Rs_2/\Is_2|\nn\\
&\approx&\pm{1\over 2}|\Rs_1|^2/|\Is_1|\pm{1\over 2}|\Rs_2|^2/|\Is_2|
\label{lastline}
\eeqa
where we used $\Is_1\log|\Is_1|+\Is_2\log|\Is_2|=0$ in the second equality.
Its absolute value is much smaller than ${1\over 2}|\Rs_1|+{1\over 2}|\Rs_2|$,
and it is negligible compared to the first line of (\ref{ANgraph}),
as we will see in a moment.
In view of the wall-avoiding condition (\ref{Wedge1})-(\ref{Wedge2}),
let us take
\beqa
\Is_1\gg 0\gg \Is_2&&\mbox{in region (i), (ii), (iii'), (iv')}\label{sgnP}\\
\Is_1\ll 0\ll \Is_2&&\mbox{in region (i'), (ii'), (iii), (iv),}\label{sgnN}
\eeqa
with a sharp transition along the diagonal and the offdiagonal lines.
Away from the transition lines,
the Arg functions take the following values
$$
\begin{array}{c|cc}
  &\Arg(\im \usigma_1)&\Arg(\im\usigma_2)\\
  \hline
  \mbox{(i)}&\pi-\epsilon_1&\epsilon_2\\
  \mbox{(ii)}&\pi-\epsilon_1&-\epsilon_2\\
  \mbox{(iii)}&\epsilon_1&-\pi+\epsilon_2\\
  \mbox{(iv)}&-\epsilon_1&-\pi+\epsilon_2
\end{array}
$$
where $\epsilon_a$ are small positive number of the order of
$|\Rs_a|/|\Is_a|$.
Hence the first line of (\ref{ANgraph}) is given by
\beqa
\mbox{(i):}&&|\Rs_1-\Rs_2|\left(-\theta-2\pi\wtq^1+7{3\pi\over 2}
-7\epsilon_1\right)
+|\Rs_2|(-2\theta-2\pi(\wtq^1+\wtq^2)+14\pi-7\sDelta\epsilon),\nn\\
\mbox{(ii):}&&|\Rs_1+\Rs_2|\left(-\theta-2\pi\wtq^1+7{3\pi\over 2}
-7\epsilon_1\right)
+|\Rs_2|(-2\pi(\wtq^1-\wtq^2)+7\pi-7\sDelta\epsilon),\nn\\
\mbox{(iii):}&&|\Rs_1+\Rs_2|\left(\theta+2\pi\wtq^2+7{3\pi\over 2}
-7\epsilon_2\right)
+|\Rs_1|(-2\pi(\wtq^1-\wtq^2)+7\pi+7\sDelta\epsilon),\nn\\
\mbox{(iv):}&&|\Rs_1-\Rs_2|\left(\theta+2\pi\wtq^1+7{3\pi\over 2}
-7\epsilon_2\right)
+|\Rs_2|(2\theta+2\pi(\wtq^1+\wtq^2)+14\pi+7\sDelta\epsilon),\nn
\eeqa
with $\sDelta\epsilon=\epsilon_1-\epsilon_2$,
whereas the expressions in the other regions are found  by Weyl reflection
$1\leftrightarrow 2$.
Note that $\epsilon_1$ and $\epsilon_2$ can be made as small as possible
by taking $|\Rs|/|\Is|\to 0$.
Compared to this, as long as it is non-zero,
the second line of (\ref{ANgraph}), which is
(\ref{lastline}), is negligible.
The growth of (\ref{ANgraph}) requires
\beq
2\pi|\wtq^1-\wtq^2|<7\pi,\quad |\theta+\pi(\wtq^1+\wtq^2)|<7\pi,\quad
|\theta+2\pi\wtq^1|\,\,\&\,\,|\theta+2\pi\wtq^2|
<7{3\pi\over 2}.
\label{Cond111}
\eeq
It remains to consider the transition lines $|\Rs_1|=|\Rs_2|$.
The second line of (\ref{ANgraph}) vanishes as $|\usigma_1|=|\usigma_2|$
and $\Is_1=-\Is_2$,
and we have
$\Arg(\im\usigma_1)\pm\Arg(\im\usigma_2)=\pi\cdot{\rm sgn}(\Rs_1)$
for $\Rs_1=\pm\Rs_2$.
This yields
\beq
\AN_{\wtq}|_{\rm transition \,line}=\left\{\begin{array}{ll}
-(2\theta+2\pi(\wtq^1+\wtq^2))\Rs_1+14\pi|\Rs_1|&\Rs_2=\Rs_1\\
-2\pi(\wtq^1-\wtq^2)\Rs_1+7\pi|\Rs_1|&\Rs_2=-\Rs_1.
\end{array}\right.
\eeq
Its growth is guaranteed by the first two inequalities of of (\ref{Cond111}).
Thus, the full condition remains to be (\ref{Cond111}).
The condition $|\theta+2\pi\wtq^a|<7\pi$ is indeed relaxed to
$|\theta+2\pi\wtq^a|<7{3\pi\over 2}-7\epsilon$.
For the shifted highest weight $\wtq_{\elll,i}$ of $S^{\elll}\fund(i)$,
the condition reads as
\beq
\elll=0,1,2,\qquad
-{7\over 2}<i+{\elll\over 2}+{\theta\over 2\pi}<{7\over 2}.
\label{Cond11}
\eeq
The set of these representations does not change
as long as $\theta$ does not cross $\pi\Z$.
For example, the set for $-8\pi<\theta<-7\pi$ is 
\beq
\begin{array}{cccccccc}
&&\C(1)&\cdots&\C(5)&\C(6)&\C(7)\\
&\fund&\fund(1)&\cdots&\fund(5)&\fund(6)&\\
  ~~~~~~~~~~&\symt&\symt(1)&\cdots&\symt(5)&\symt(6)&
\end{array}
\label{oGRR1}
\eeq
and the set for $-7\pi<\theta<-6\pi$ is 
\beq
\begin{array}{cccccccc}
&\C&\C(1)&\cdots&\C(5)&\C(6)&\\
&\fund&\fund(1)&\cdots&\fund(5)&\fund(6)&\\
\symt(-1)&\symt&\symt(1)&\cdots&\symt(5)&&
\end{array}
\label{oGRR4}
\eeq
They do not agree with the set $\RADS$ (\ref{ADSset}) corresponding
to $\TADS$ of \cite{ADS}, but are no longer its proper subsets.
Also, the number of representations, twenty-one, matches with the cardinality
of $\RADS$.
Thus, they have a potential to be large enough to produce all branes in
the low energy theories in the two phases, and
they may even be of the right size to be the grade restricted subsets.

As a concrete example of a contour of the above type, that is,
satisfying $\Is_1+\Is_2=0$, (\ref{sgnP}) and (\ref{sgnN}),
we can take $\bgamma_0$ defined by
\beq
\Is_1=|\Rs_1|^2-|\Rs_2|^2,\quad
\Is_2=-|\Rs_1|^2+|\Rs_2|^2.
\label{bgamma0}
\eeq
Indeed, in the outskirts of the $\Rs$-plane,
$\Is_1=-\Is_2$ has much bigger absolute value than $|\Rs_1|$ and $|\Rs_2|$
with the sign as in (\ref{sgnP}) and (\ref{sgnN}),
except along the diagonal and offdiagonal lines
where it makes a sharp transition.
In particular, $\bgamma_0$ asymptotes to
the union of the transition planes, $\Rs_1=\pm \Rs_2$ and $\Is_1=-\Is_2$,
that is,
\beq
\overline{\bgamma}_0\,:=\,
\{\usigma_1=\overline{\usigma_2}\}\cup
\{\usigma_1=-\usigma_2\}.
\eeq
The contour $\bgamma_0$ is Weyl invariant and
is designed to obey the wall-avoiding conditions
(\ref{Wedge1})-(\ref{Wedgep}).
However, this is very far from Lagrangian.
It looks difficult to find a Lagrangain submanifold of the above type.
For example, one of the transition planes is
$\{\usigma_1=-\usigma_2\}$ which is a complex submanifold,
very oppsosite to being Lagrangian with respect to any
compatible symplectic structure. It is not even totally real.
For now and also in what follows,
we shall ignore the Lagrangian condition.
We will come back to this problem in Section~\ref{subsec:revision}.

The condition (\ref{Cond11}) 
for the integral $Z_{\sgamma_0}(S^{\elll}\fund(i))$ to converge
depends only on $\theta$ and not on $\zeta$.
Thus, the constant contour $\bgamma_0$ and
the set of representations $\sym^h\fund(i)$ satisfying (\ref{Cond11})
may be used for D-brane transport along the straight paths
between $\zeta\gg 0$ and $\zeta\ll 0$
with a fixed $\theta\not\in\pi\Z$.

For completion, we need to deform this contour $\bgamma_0$ to
$\bgamma_+^{\alpha_+}$ in the weakly coupled phase $\zeta\gg 0$
and to $\bgamma_-^{\alpha_-}$ in the strongly coupled phase $\zeta\ll 0$,
for some $\alpha_{\pm}>1$, so that $\AN_{\wtq}|_{\sgamma}$ grows
at infinity for $\wtq=\wtq_{\elll, i}$ in the range (\ref{Cond11})
for any $\bgamma$ in the deformation.
For $\alpha_{\pm}>2$, this can be done as follows.
In the phase $\pm\zeta\gg 0$, we consider a family of contours
$\bgamma_{\pm,t}$ parametrized by $t\in [0,1]$ defined by
\beqa
\bgamma_{+,t}&=&\left\{\begin{array}{c}
\Is_1=(1-t)(|\Rs_1|^2-|\Rs_2|^2)+t|\Rs_1|^{\alpha_+},\\
\Is_2=(1-t)(|\Rs_2|^2-|\Rs_1|^2)+t|\Rs_2|^{\alpha_+}.
\end{array}\right\},
\\
\bgamma_{-,t}&=&\left\{\begin{array}{c}
\Is_1=(1-t)(|\Rs_1|^2-|\Rs_2|^2)-t|\Rs_1+\Rs_2|^{\alpha_-},\\
\Is_2=(1-t)(|\Rs_2|^2-|\Rs_1|^2)-t|\Rs_1+\Rs_2|^{\alpha_-}.
\end{array}\right\}.
\eeqa
Among all the required properties, the only non-trivial one is the growth
condition. As $\bgamma_{\pm,0}=\bgamma_0$
for which the growth is already known for $\wtq=\wtq_{\elll,i}$
in the range (\ref{Cond11})
we only have to show it for $t>0$.
Let us first discuss the phase $\zeta\gg 0$.
$\AN^{\Is}_{\wtq}|_{\sgamma_{+,t}}$ dominates $\AN^{\Rs}_{\wtq}|_{\sgamma_{+,t}}$
and is given by
\beqa
\AN^{\Is}_{\wtq}|_{\sgamma_{+,t}}
&=&|\Rs_1|^{\alpha_+}t(\zeta-7\log|(\usigma_1+\usigma_2)/\usigma_1|)
+|\Rs_2|^{\alpha_+}t(\zeta-7\log|(\usigma_1+\usigma_2)/\usigma_2|)
\nn\\
&&+7(1-t)(|\Rs_1|^2-|\Rs_2|^2)\log|\usigma_1/\usigma_2|.
\eeqa
The second line is non-negative for any $(\Rs_1,\Rs_2)$.
At large $\|\Rs\|$, using $\alpha_+>2$,
we find $\log|(\usigma_1+\usigma_2)/\usigma_a|\leq C$ for some $C$.
Thus, this is bounded from below by
$t(|\Rs_1|^{\alpha_+}+|\Rs_2|^{\alpha_+})(\zeta-7C)$.
Therefore, as long as $\zeta>7C$, the required growth condition is
satisfied.
Next consider the phase $\zeta\ll 0$.
As in the case of $\bgamma_-$, we perform the analysis separately in
the two regions that cover the large $\|\Rs\|$ domain:
(i) $(\Rs_1,\Rs_2)=(\lambda\Rs_1^0,\lambda\Rs_2^0)$ as $\lambda\to\infty$
for a fixed $(\Rs_1^0,\Rs_2^0)$ subject to $\Rs_1^0+\Rs_2^0\ne 0$,
and (ii) $|\Rs_1+\Rs_2|<\varepsilon |\Rs_1|,\varepsilon|\Rs_2|$
for $\varepsilon\ll 1$.
In region (i), we find $\AN_{\wtq}|_{\sgamma_{-,t}}> 0$ for any $\wtq$
and $\zeta<7\log 2$
in exactly the same way as in the case of $\bgamma_-$.
In region (ii), we find
\beqa
\AN_{\wtq_{h,i}}|_{\sgamma_{-,t}}
&\geq&
2t|\Rs_1+\Rs_2|^{\alpha_-}\Bigl(-\zeta+f_h(x/t)+O(\varepsilon)\Bigr)
\nn\\
&&+7(1-t)(|\Rs_1|^2-|\Rs_2|^2)\log|\usigma_1/\usigma_2|,
\eeqa
for $x$ and $f_h(x)$ as defined in (\ref{defx}) and (\ref{deffh}).
This is positive as long as $\zeta$ is lower than
$f_{h\,{\rm min}}$ defined in (\ref{fhmin}).
Thus, the growth condition is satisfied in
the $\zeta\ll 0$ phase as well.
Thus, $\bgamma_0$ can be connected to $\bgamma_{\pm}^{\alpha_{\pm}}$ with
$\alpha_{\pm}>2$ in the phase $\pm\zeta\gg 0$.
Once that is done, $\alpha_{\pm}$ can be changed to any value
greater than $1$, say $2$.

To summarize, along a straight path between the two phases with a fixed
$\theta\not\in \pi\Z$, there is a family of contours
connecting $\bgamma_+$ in the weakly coupled phase $\zeta\gg 0$
and $\bgamma_-$ in the strongly coupled phase $\zeta\ll 0$
on which the hemisphere integral of the representations
$S^{\elll}\fund(i)$ in the range (\ref{Cond11}) is absolutely convergent
all the way along the path.
These are members of the grade restricted subset
for the window of the straight path.
For example, the sets (\ref{oGRR1}) and (\ref{oGRR4}) are included in
the grade restricted subsets $\rmGRR_{\bf A}$ and $\rmGRR_{\bf D}$
for the windows ${\bf A}$ and ${\bf D}$
represented by the paths ${\mathcal P}_{\!\bf A}$ and
${\mathcal P}_{\bf D}$ in Fig.~\ref{fig:RodpathsX}.
Regarding their cardinality, twenty-one, they may even agree with
$\rmGRR_{\bf A}$ and $\rmGRR_{\bf D}$.
However, we cannot claim so at this moment
since we have only considered a particular class of contours
and there remains a possibility that
other representations are admissible for more general contours.\footnote{This
  is the problem of \cite{GRS,GLZ} mentioned in the introduction.}
We will reach a decisive conclusion later, in Section~\ref{subsec:noentry}.

\subsection{Curved Paths ---  Numerical Analysis}
\label{subsec:numerical}

Next, let us consider the windows ${\bf B}$ and ${\bf C}$
that require the representing paths to be curved.
The paths may be taken to fit inside the strip $-8\pi<\theta<-6\pi$
and therefore the hemisphere integrals of representations in
the overlap of GRR${}_{\bf A}$ and GRR${}_{\bf D}$ are convergent
on the contour $\bgamma_0$. Thus, the grade restricted sets
for both ${\bf B}$ and ${\bf C}$ should include this overlap,
\beq
    {\rm GRR}_{\bf A}\cap {\rm GRR}_{\bf D}\,\subset\,
    {\rm GRR}_{\bf B},\,\,{\rm GRR}_{\bf C}.
\eeq
The question is which representations are added for completion.
Likely candidates are those in either of GRR${}_{\bf A}$ and GRR${}_{\bf D}$
but not in the overlap. If (\ref{oGRR1}) and (\ref{oGRR4}) indeed agree with
them, there are four such representations,
\beq
\C(0), \,\,\C(7),\,\,\Symt(-1),\,\,\Symt(6).
\eeq
We examine these candidates numerically to see whether some of them
are grade restricted with respect to ${\bf B}$ or ${\bf C}$.

Using the scaling property (\ref{scaleinv}) of the growth function,
we may project the problem onto the sphere in $\ttt_{\C}\cong \C^2$,
say, that of a radius $R$
\beq
S^3_R=\Bigl\{\,(\usigma_1,\usigma_2)\in\C^2\,\Bigl|\,
|\usigma_1|^2+|\usigma_2|^2=R^2 \,\Bigr\},
\label{S3def}
\eeq
that is, to find a circle in the region on $S^3_R$ where
$\AN_{\wtq}$ is positive. The cone over such a circle can be taken as a
contour $\bgamma$ on which the hemisphere integral of the representation
with the shifted highest weight $\wtq$ is convergent.
The circle must be a Weyl invariant deformation of the real locus
$C_{\rm real}=S^3_R\cap \R^2$ inside the complement of
the walls $W_1$, $W_2$ and $W_p$.
Inside $S^3_R$, these walls form a chain of two-dimensional discs which
intersect pairwise along a segment. In particular,
the pair $(S^3_R,S^3_R\cap (W_1\cup W_2\cup W_p))$
is homotopy equivalent to the pair of $S^3$ and an unknot $S^1\subset S^3$,
and therefore
\beq
\pi_1(S^3_R\setminus S^3_R\cap (W_1\cup W_2\cup W_p))
\cong \Z.
\label{pione}
\eeq
The real locus $C_{\rm real}$ represents a generator of this group,
and so do the circles
$C_{\alpha}=S^3_R\cap \bgamma_{\alpha}$ for $\alpha=\pm, 0$. 
Thus, the problem is to find a family of Weyl invariant
circles generating the fundamental group (\ref{pione})
that stays in the admissible region $\AN_{\wtq}>0$ 
all along a path representing the window in question,
connecting $C_+$ at $\zeta\gg 0$ and $C_-$ at $\zeta\ll 0$.
We examined this by plotting the admissible region in Mathematica.
An outline of the numerical analysis is presented in
Appendix~\ref{subapp:curved}.
The result is as follows.

\noindent
\underline{Window {\bf C}}:\\
For $\C(0)$ and $\Symt(6)$, there are continuous families
of generating circles connecting $C_+$ at $\zeta\gg 0$
and $C_-$ at $\zeta\ll 0$ that are admissible along a path of window ${\bf C}$.
For $\C(0)$, we can take the circles to be Weyl invariant
all the way.
For $\Symt(6)$, Weyl invariance seems difficult to achieve
as we will discuss in more detail in Section~\ref{subsec:Weyl}.
For $\C(7)$ and $\Symt(-1)$, there does not exist
such a continuous family, Weyl invariant or not.
If a family starts from $C_+$ at $\zeta\gg 0$,
as $t$ approaches the region to the upper-right of $t_1$
({\it resp}. upper-left of $t_3$) from above,
the circle is forced to either go out of the admissible region
for $\C(7)$ ({\it resp}. $\Symt(-1)$) or hit the wall $W_1$ ({\it resp}. $W_2$).
This is depicted in Figs.~\ref{fig:07contours1}, \ref{fig:coasts} and \ref{fig:W1}
({\it resp}. Figs.~\ref{fig:2m1contours}, \ref{fig:2m1coasts} and \ref{fig:W2})
in Appendix~\ref{subapp:curved}.
If a family starts from $C_-$ at $\zeta\ll 0$,
as $t$ approaches the right of $t_1$
({\it resp}. left of $t_3$) from below,
the circle must 
either go outside the admissible region for $\C(7)$
({\it resp}. $\Symt(-1)$) or hit the wall $W_p$.
This is depicted in Figs.~\ref{fig:0702}, \ref{fig:z1pp} and \ref{fig:07Wp}
({\it resp}. Fig.~\ref{fig:2m1pen}, \ref{fig:2m1EL} and \ref{fig:2m1Wp})
in Appendix~\ref{subapp:curved}.

\noindent
\underline{Window {\bf B}}:\\
By the reflection symmetry (\ref{reflectionsym}) at $-7\pi$,
the same as above holds with the replacement
$\C(0)\leftrightarrow \C(7)$, $\Symt(6)\leftrightarrow \Symt(-1)$,
$W_1\leftrightarrow W_2$,
(upper-)right of $t_1$ $\leftrightarrow$ (upper-)left of $t_1$
and (upper-)left of $t_3$ $\leftrightarrow$ (upper-)right of $t_3$.

We must note that this method based on Mathematica plots
works only in low rank systems.
The rank is two in the current model and that seems to be the maximum.
Rank two corresponds to real dimension four which was reduced to
three by the scaling property (\ref{scaleinv}) of the growth function.
For a gauge group of higher rank,
the dimension would be five or higher and it is hard to
imagine a method to plot the admissible region in such a
high dimensional space.
Instead, we found an alternative.
After looking at the numerical results,
we managed to extract the essence, and developed it into an analytical method.
And that may be applicable
to more general models including those with higher ranks.
Indeed, the method is successfully applied to other models, including
those with rank three, as presented in Section~\ref{sec:other}.

\subsection{Behaviour Near The Singular Points --- Special Representations}
\label{subsec:catastroph}

\newcommand{\CB}{{\rm CB}}
\newcommand{\Cc}{\usigma_{\rm c}}

Let us discuss the analytical method and its results.
The key is the behaviour of the growth function
$\AN_{\wtq}={\rm Re}\left(\im \wt{W}_{\!{\it eff},\wtq}\right)$
near the singular points, $t_\bfa+2\pi\im \Z$, $\bfa=1,2,3$.
($t_1$, $t_2$ and $t_3$ are given in (\ref{t1}), (\ref{t2}) and (\ref{t3}).)

Exactly at the singular point $t\equiv t_\bfa$ mod $2\pi\im \Z$,
there is a one-dimensional Coulomb branch (\ref{CoulombBranch})
which solves the vacuum equation (\ref{vaceq0}). The vacuum equation can
also be written as
\beq
\dd\wt{W}_{\!{\it eff}}\equiv 0\quad\mbox{mod $2\pi\im \rmP$}
\label{vaceq}
\eeq
where $\rmP$ is the weight lattice of $T$.
Recall that $\wt{W}_{\!{\it eff},\wtq}$ for $\wtq\in\rmP+\rho$
is a specific representative of $\wt{W}_{\!{\it eff}}$.
The ``vanishing'' (\ref{vaceq}) means that,
at each $t=t_*\in t_\bfa+2\pi\im \Z$, there is a special element
$\wtq_*\in \rmP+\rho$ such that the first derivative vanishes
identically on the Coulomb branch,
\beq
\dd\wt{W}_{\!{\it eff},\wtq_*}
=0,\label{vaceqnose}
\eeq
not just modulo $2\pi\im \rmP$,
which implies
\beq
\wt{W}_{\!{\it eff},\wtq_*}=0,
\eeq
by the Calabi-Yau property
$\wt{W}_{\!{\it eff},\wtq_*}=\sum_{a=1}^2
\usigma_a\partial_{\usigma_a}\wt{W}_{\!{\it eff},\wtq_*}$.

\begin{figure}[thb]
  \psfrag{w1}{$W_1$}
  \psfrag{w2}{$W_2$}
  \psfrag{wp}{$W_p$}
  \psfrag{a1}{${2\pi\over 7}$}
  \psfrag{a2}{${4\pi\over 7}$}
  \psfrag{a3}{${6\pi\over 7}$}
  \psfrag{sp}{$\Cc$}
  \psfrag{I1}{${\rm I}_1$}
  \psfrag{II1}{${\rm II}_1$}
  \psfrag{III1}{${\rm III}_1$}
  \psfrag{I2}{${\rm I}_2$}
  \psfrag{II2}{${\rm II}_2$}
  \psfrag{III2}{${\rm III}_2$}
  \psfrag{I3}{${\rm I}_3$}
  \psfrag{II3}{${\rm II}_3$}
  \psfrag{III3}{${\rm III}_3$}
\centerline{\includegraphics[page=3]{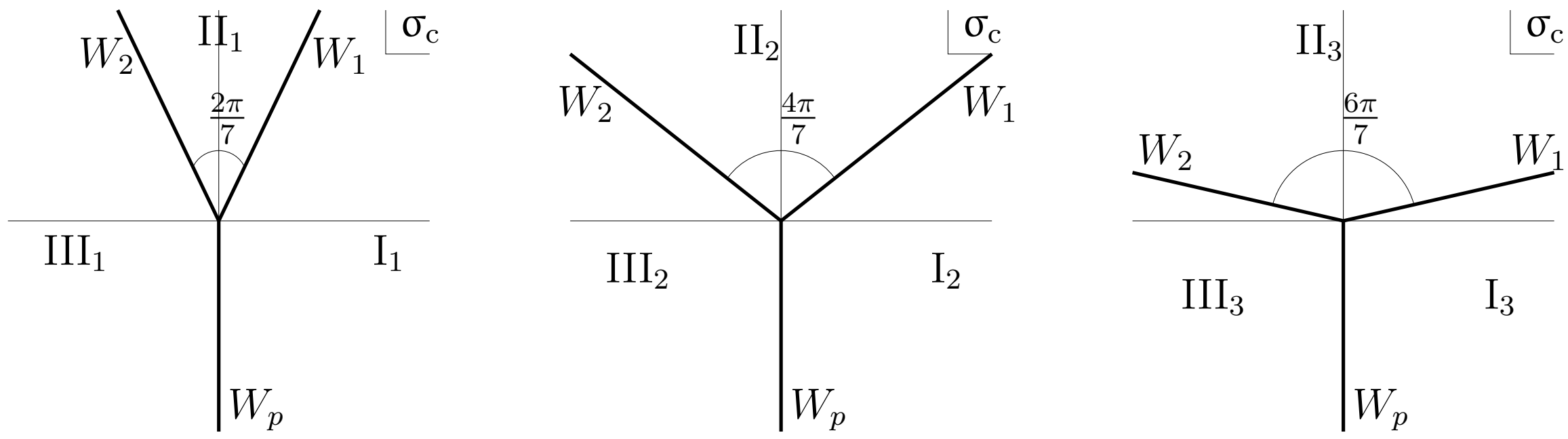}}
\caption{Chambers of the Coulomb branch:
  $\CB_1^+$, $\CB_2^+$, $\CB_3^+$ (left, middle, right)}
\label{fig:CB}
\end{figure}

To be precise, the Coulomb branch
$\CB_\bfa^{\pm}=\left\{\usigma_1=\omega_\bfa^{\pm 1}\usigma_2\right\}$
is separated by the walls
$W_1,W_2$ and $W_p$ into chambers and the special element
$\wtq_*$ depends on the chamber.
We first consider the branch $\CB_\bfa^+$ which we
parametrize by
\beq
\usigma_1=\omega_\bfa^{{1\over 2}}\Cc,
\quad
\usigma_2=\omega_\bfa^{-{1\over 2}}\Cc.
\label{CBplus}
\eeq
The walls $W_1$, $W_2$, $W_p$ intersect with $\CB_\bfa^+$
along the rays 
$\im\omega_\bfa^{-{1\over 2}}\R_{>0}$,
$\im\omega_\bfa^{{1\over 2}}\R_{>0}$,
$-\im\R_{>0}$ in the $\Cc$ plane and separate it into three chambers
${\rm I}_\bfa$, ${\rm II}_\bfa$
and ${\rm III}_\bfa$ as depicted in Fig.~\ref{fig:CB}.
The element $\wtq_*$ can be found by regarding (\ref{vaceqnose})
as an equation to determine it, with $\usigma\in \CB_\bfa^+$:
\beqa
2\pi\im \wtq_*&=&t_*+\sum_iQ_i\log(\im Q_i(\usigma))\nn\\
&=&t_*(1,1)
+7(1,0)\log(\im \omega_{\bfa}^{{1\over 2}}\Cc)
+7(0,1)\log(\im\omega_{\bfa}^{-{1\over 2}}\Cc)\nn\\
&&+7(-1,-1)
\log(-\im(\omega_{\bfa}^{1\over 2}+\omega_\bfa^{-{1\over 2}})\Cc)
\nn\\[0.2cm]
&=&-\im\theta_*(1,1)+7(1,0)\im\Arg(\im \omega_{\bfa}^{{1\over 2}}\Cc)
+7(0,1)\im\Arg(\im\omega_{\bfa}^{-{1\over 2}}\Cc)\nn\\
&&+7(-1,-1)
\im\Arg(-\im\Cc)
\label{findwtq}
\eeqa
where we used $\zeta_*=\zeta_\bfa
=7\log\left(\omega_{\bfa}^{1\over 2}+\omega_\bfa^{-{1\over 2}}\right)$.
To evaluate the Arg functions, let us denote the phase of $\Cc$
by $\varphi$ which is assumed to take values in the intervals
$\left(-{\pi\over 2},{\pi\over 2}-{\pi\bfa\over 7}\right)$,
$\left({\pi\over 2}-{\pi\bfa\over 7},{\pi\over 2}+{\pi\bfa\over 7}\right)$,
$\left({\pi\over 2}+{\pi\bfa\over 7},{3\pi\over 2}\right)$
in the chambers ${\rm I}_\bfa$, ${\rm II}_\bfa$,
${\rm III}_\bfa$. Then,
\beq
\begin{array}{c|ccc|c}
  &\Arg(\im \omega_{\bfa}^{1\over 2}\Cc)&
  \Arg(\im\omega_{\bfa}^{-{1\over 2}}\Cc)&\Arg(-\im\Cc)
  &{1\over 2\pi}\sum_iQ_i\Arg(\im Q_i(\usigma))\\
  \hline
      {\rm I}_\bfa&{\pi\over2}+{\pi\bfa\over 7}+\varphi
      &{\pi\over 2}-{\pi\bfa\over 7}+\varphi&-{\pi\over 2}+\varphi
      &({7+\bfa\over 2},{7-\bfa\over 2})\\
      {\rm II}_\bfa&{}_{''}-2\pi
      &{}_{''}
      &{}_{''}
      &({-7+\bfa\over 2},{7-\bfa\over 2})\\
      {\rm III}_\bfa&{}_{''}
      &{}_{''}-2\pi&{}_{''}
      &({-7+\bfa\over 2},{-7-\bfa\over 2})\\
\end{array}
\label{tableArg}
\eeq
where $\,{}_{''}\,$ stands for ``the same expression as above''.
The sum ${1\over 2\pi}\sum_iQ_i\Arg(\im Q_i(\usigma))$
must be equated with $\wtq_*+{\theta_*\over 2\pi}(1,1)$.
For example, at the particular points
$t_*=t_\bfa$ where $\theta_*=-7\pi,-8\pi,-7\pi$
for $\bfa=1,2,3$, we find
\beqa
t_*=t_1&&\begin{array}{c|ccc}
  &{\rm I}_1&{\rm II}_1&{\rm III}_1\\
  \hline
  \wtq_*&\left(7+{1\over 2},7-{1\over 2}\right)&
  \left({1\over 2},7-{1\over 2}\right)&
  \left({1\over 2},-{1\over 2}\right)
\end{array}
\label{wtq1Rod}
\\
t_*=t_2&&\begin{array}{c|ccc}
  &{\rm I}_2&{\rm II}_2&{\rm III}_2\\
  \hline
  \wtq_*&\left(8+{1\over 2},7-{1\over 2}\right)&
  \left(1+{1\over 2},7-{1\over 2}\right)&
  \left(1+{1\over 2},-{1\over 2}\right)
\end{array}
\label{wtq2Rod}
\\
t_*=t_3&&\begin{array}{c|ccc}
  &{\rm I}_3&{\rm II}_3&{\rm III}_3\\
  \hline
  \wtq_*&\left(7+{3\over 2},7-{3\over 2}\right)&
  \left({3\over 2},7-{3\over 2}\right)&
  \left({3\over 2},-{3\over 2}\right)
\end{array}
\label{wtq3Rod}
\eeqa
We observe that $\wtq_*$ in the chambers ${\rm I}_\bfa$ and ${\rm III}_\bfa$
are the shifted highest weights $\wtq_{\elll,i}$ (\ref{wtqli})
of the $SU(2)$ grade restricted representations (\ref{GRRstrong}):
\beqa
t_*=t_1:&&\wtq_*=\left\{\begin{array}{ll}
\wtq_{0,7}&\mbox{in ${\rm I}_1$}\\
\wtq_{0,0}&\mbox{in ${\rm III}_1$}
\end{array}\right.\label{wtqstR1}\\
t_*=t_2:&&\wtq_*=\left\{\begin{array}{ll}
\wtq_{1,7}&\mbox{in ${\rm I}_2$}\\
\wtq_{1,0}&\mbox{in ${\rm III}_2$}
\end{array}\right.\label{wtqstR2}\\
t_*=t_3:&&\wtq_*=\left\{\begin{array}{ll}
\wtq_{2,6}&\mbox{in ${\rm I}_3$}\\
\wtq_{2,-1}&\mbox{in ${\rm III}_3$}
\end{array}\right.\label{wtqstR3}
\eeqa
We next consider the other branch $\CB_\bfa^-$.
Its chamber decomposition is obtained from the one for
$\CB_\bfa^+$ by Weyl transformation
and the special shifted weight $\wtq_*$ in each chamber is the
Weyl image of the one for the corrersponding chamber in $\CB_\bfa^+$.
Indeed, if $\usigma$ is in a chamber of $\CB_\bfa^+$ and
$w$ is a Weyl group element,
\beq
t_*+\sum_iQ_i\log(\im Q_i(w\usigma))
=w\left(t_*+\sum_{i'}Q_{i'}\log(\im Q_{i'}(\usigma))\right)
\eeq
where we used the fact that $w^{-1}$ induces a permutation
of the matter weights $\{Q_i\}$.
Looking at the results (\ref{wtq1Rod})-(\ref{wtq3Rod}),
we see that $\wtq_*$ in the Weyl image of ${\rm II}_\bfa\subset \CB_\bfa^+$
are shifted highest weights $\wtq_{\elll, i}$ with
$(\elll,i)=(5,1), (4,2), (3,2)$ for $\bfa=1,2,3$.
None of these representations are $SU(2)$ grade restricted.

\begin{figure}[htb]
  \psfrag{x}{$x$}
  \psfrag{y}{$y$}
\centerline{\includegraphics[page=4]{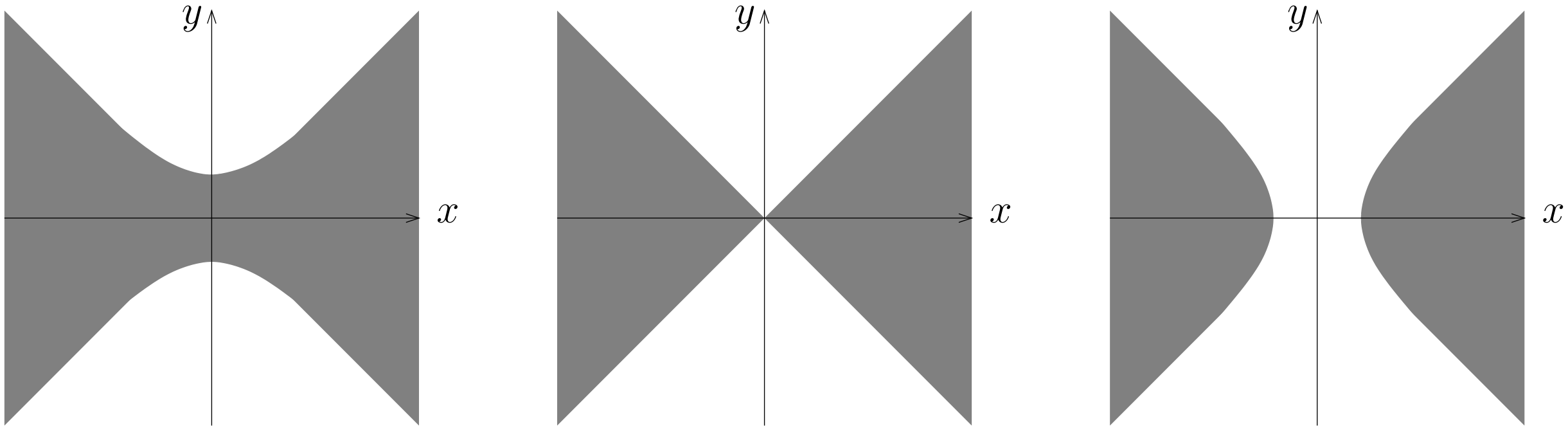}}
\caption{Admissible regions}
\label{fig:shaded}
\end{figure}

Let us take a closer look at the behaviour of $\wt{W}_{\!{\it eff},\wtq_*}$
and $\AN_{\wtq_*}$ at and near $t=t_*$.
Some of the second derivatives of $\wt{W}_{\!{\it eff},\wtq_*}$
$$
\partial_{\usigma_1}^2\wt{W}_{\!{\it eff},\wtq_*}
=-{7\over\usigma_1}+{7\over \usigma_1+\usigma_2},\quad
\partial_{\usigma_2}^2\wt{W}_{\!{\it eff},\wtq_*}
=-{7\over\usigma_2}+{7\over \usigma_1+\usigma_2},\quad
\partial_{\usigma_1}\partial_{\usigma_2}\wt{W}_{\!{\it eff},\wtq_*}
={7\over \usigma_1+\usigma_2}
$$
are non-zero on the Coulomb branch at $t_*$. In particular,
\beq
(\partial_{\usigma_1}-\partial_{\usigma_2})^2\,\wt{W}_{\!{\it eff},\wtq_*}
=-{7\over\usigma_1}-{7\over\usigma_2}
=-{7\cdot 2\cos{\pi\bfa\over 7}
  \over \Cc}\ne 0\quad\mbox{on $\CB_\bfa^+$.}
\label{Hessian}
\eeq
Along with this, the fact that
$\wt{W}_{\!{\it eff},\wtq_*}=\dd\wt{W}_{\!{\it eff},\wtq_*}=0$
at $t=t_*$ in a chamber of the
Coulomb branch means that the chamber has a neighborhood 
with a transversal coordinate $z$ on which
\beq
\wt{W}_{\!{\it eff},\wtq_*}|_{t=t_*}=-\im z^2.\label{Wzsq}
\eeq
Indeed, if we put
\beq
\usigma_1=\omega_\bfa^{1\over 2}\Cc+\tilde{z},\quad
\usigma_2=\omega_\bfa^{-{1\over 2}}\Cc-\tilde{z},
\label{ztilde}
\eeq
we find from (\ref{Hessian}) that
\beq
\wt{W}_{\!{\it eff},\wtq_*}|_{t=t_*}
=-{7\cos{\pi\bfa\over 7}
  \over \Cc}\tilde{z}^2+O(\tilde{z}^3)
\eeq
and we obtain (\ref{Wzsq}) under a change of coordinate
$\tilde{z}\to z$ of the form
\beq
\tilde{z}=\e^{\pi\im\over 4}\sqrt{\Cc\over 7\cos{\pi\bfa\over 7}}\,z
+O(z^2).
\label{relzz}
\eeq
In particular, the growth function at $t=t_*$ is given by
\beq
\AN_{\wtq_*}|_{t=t_*}={\rm Re}(z^2)=x^2-y^2
\eeq
for $z=x+\im y$ and the region where it is positive is as in
Fig.~\ref{fig:shaded}-middle.
When $t$ is moved off from $t_*$,
\beqa
\wt{W}_{\!{\it eff},\wtq_*}|_{t=t_*+\sDelta t}
&=&-\im z^2-\sDelta t(\usigma_1+\usigma_2)
=-\im z^2-2\cos\mbox{$\left({\pi\bfa\over 7}\right)$}\sDelta t\,\Cc,\\
\AN_{\wtq_*}|_{t=t_*+\sDelta t}
&=&{\rm Re}\Bigl(\,z^2
-\underbrace{2\cos\mbox{$\left({\pi\bfa\over 7}\right)$}
  \im\sDelta t\,\Cc}_{\displaystyle ~~~\,\,:=s}\,\Bigr)
=x^2-y^2-{\rm Re}(s),
\eeqa
and the region where the latter is positive changes.
We see that the topology of the region depends very much on the sign of
${\rm Re}(s)$:
It remains the same as for $t=t_*$ if ${\rm Re}(s)=0$
(Fig.~\ref{fig:shaded}-middle),
it is smooth and connected if ${\rm Re}(s)<0$
(Fig.~\ref{fig:shaded}-left),
while it is disconnected if ${\rm Re}(s)>0$
(Fig.~\ref{fig:shaded}-right).

Let $\rmV_*$ be the representation with the shifted highest weight $\wtq_*$
in ${\rm I}_\bfa$ ({\it resp}. in ${\rm III}_\bfa$)
as in (\ref{wtqstR1}), (\ref{wtqstR2}), (\ref{wtqstR3}).
Note that $t_*$ is on the right ({\it resp}. left) edge of
the region for the straight paths along which the representation $\rmV_*$
is grade restricted:
The hemisphere integral of $\rmV_*$ along
the contour $\bgamma_0$ is barely convergent
along the straight path that goes through $t_*$,
safely convergent if the path is moved to the left ({\it resp}. right)
of $t_*$,
and is divergent if it is moved to the right ({\it resp}. left) of $t_*$.
We would like to know what happens to the integral
$Z_{\sgamma}(\rmV_*)$ for $\bgamma=\bgamma_0$ or its deformation
when the straight path through $t_*$ is deformed away from $t_*$
only locally around $t_*$.

\begin{figure}[htb]
  \psfrag{x}{$x$}
  \psfrag{y}{$y$}
\centerline{\includegraphics[page=5]{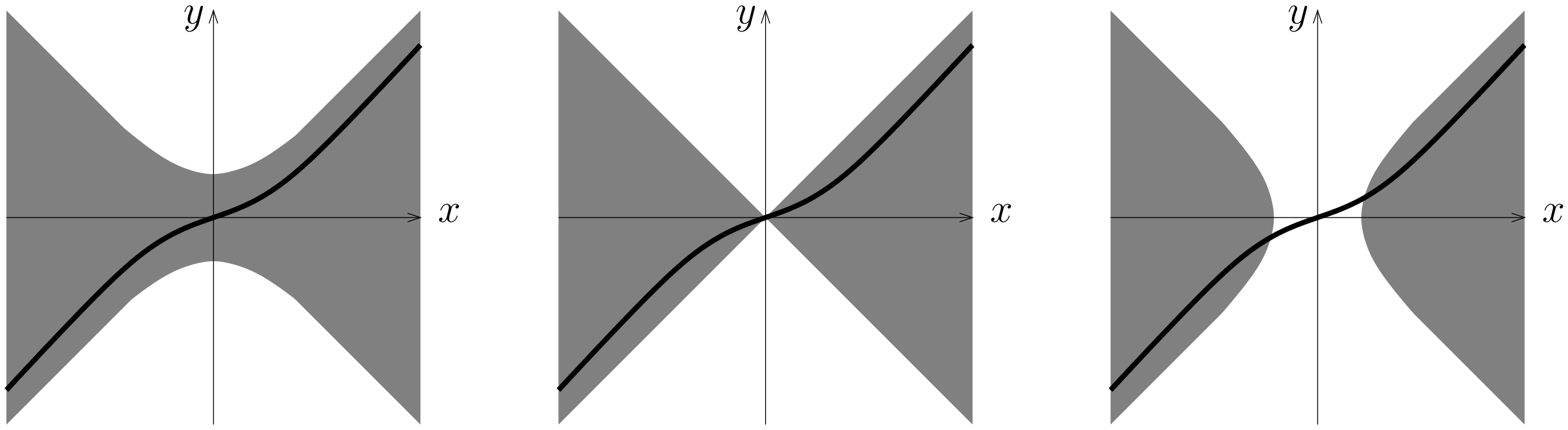}}
\caption{Contours and admissible regions}
\label{fig:shadecontour}
\end{figure}

The fact that $Z_{\sgamma_0}(V_*)$ is barely convergent on the straight path
through $t_*$ means that the infinity $\overline{\bgamma}_0$
of the contour $\bgamma_0$ is in the region $\AN_{\wtq_*}\geq 0$ at $t=t_*$.
In particular, it must go through the shaded region in
Fig.~\ref{fig:shaded}-middle.
To confirm this, let us describe the location of $\overline{\bgamma}_0$
in the coordinate system $(\Cc,z)$.
Recall that $\overline{\bgamma}_0$ is the union of
$\{\usigma_1=\overline{\usigma_2}\}$ and $\{\usigma_1+\usigma_2=0\}$.
We do not have to worry about the latter component
since the representation $\rmV_*$ is $SU(2)$ grade restricted.
In the coordinate system $(\Cc,\tilde{z})$ as in (\ref{ztilde}),
the equation $\usigma_1=\overline{\usigma_2}$ is equivalent to
$\Cc\in\R$ and $\tilde{z}\in\im \R$.
Thus, it intersects with the Coulomb branch $\CB_\bfa^+$ in the chambers
${\rm I}_\bfa$ and ${\rm III}_\bfa$.
In view of the relation (\ref{relzz}), for $\Cc\in {\rm I}_\bfa$
({\it resp}. $\Cc\in {\rm III}_\bfa$)
it approaches the
$45^\circ$ ({\it resp}. $-45^\circ$) line in the $z$-plane as $z\to 0$,
which is on the edge of the shaded region in Fig.~\ref{fig:shaded}-middle.
Thus, exactly at $t=t_*$, the contour
$\overline{\bgamma}_0$ is indeed in the region $\AN_{\wtq_*}|_{t=t_*}\geq 0$.
It is on the edge of the region but can be deformed inside.
When $t$ is moved away from $t_*$ by $\sDelta t$,
the contour $\overline{\bgamma}_0$ or its deformation
may or may not be able to stay in the admissible region.
See Fig.~\ref{fig:shadecontour}.
In particular, when ${\rm Re}(s)$ is positive,
it is impossible for any deformation of
$\overline{\bgamma}_0$ to stay inside the admissible region
(Fig.~\ref{fig:shadecontour}-right).
Since $\Arg(s)=\Arg(\im\sDelta t)+\Arg(\Cc)$
and $\Arg(\Cc)$ for $\Cc$ in ${\rm I}_\bfa$ ({\it resp}. ${\rm III}_\bfa$)
ranges over the open interval
$(-{\pi\over 2},{\pi\over 2}-{\pi\bfa\over 7})$ 
({\it resp}. $({\pi\over 2}+{\pi\bfa\over 7},{3\pi\over 2})$),
${\rm Re}(s)$ is positive
for any $\Cc$ inside the chamber ${\rm I}_\bfa$ ({\it resp}. ${\rm III}_\bfa$)
as long as
\beq
R_\bfa:~0\leq \Arg(\im\sDelta t)\leq {\pi\bfa \over 7}\quad
\left({\it resp}.~
L_\bfa:~\pi-{\pi\bfa\over 7}\leq\Arg(\im\sDelta t)\leq\pi\right).
\label{RandL}
\eeq
Thus, when $t$ is in this sector,
it is impossible for any deformation of $\bgamma_0$ to stay
inside the admissible region $\AN_{\wtq_*}>0$.
In particular, when $t$ is moved along a small loop around $t_\bfa$,
it has to go through both $R_\bfa$ and $L_\bfa$. Then,
the two representations, one with shifted highest weight
$\wtq_*$ in I${}_\bfa$ and another with $\wtq_*$ in III${}_\bfa$,
undergo a catastrophe --- no deformation of $\bgamma_0$ can stay inside
the admissible regions of those representations.

\begin{figure}[thb]
  \psfrag{P}{${\mathcal P}$}
\psfrag{t1}{$t_1$}
\psfrag{t2}{$t_2$}
\psfrag{t3}{$t_3$}
\psfrag{z}{$\zeta$}
\psfrag{th}{$\theta$}
\psfrag{-8pi}{$-8\pi$}
\psfrag{-6pi}{$-6\pi$}
\centerline{\includegraphics[page=6]{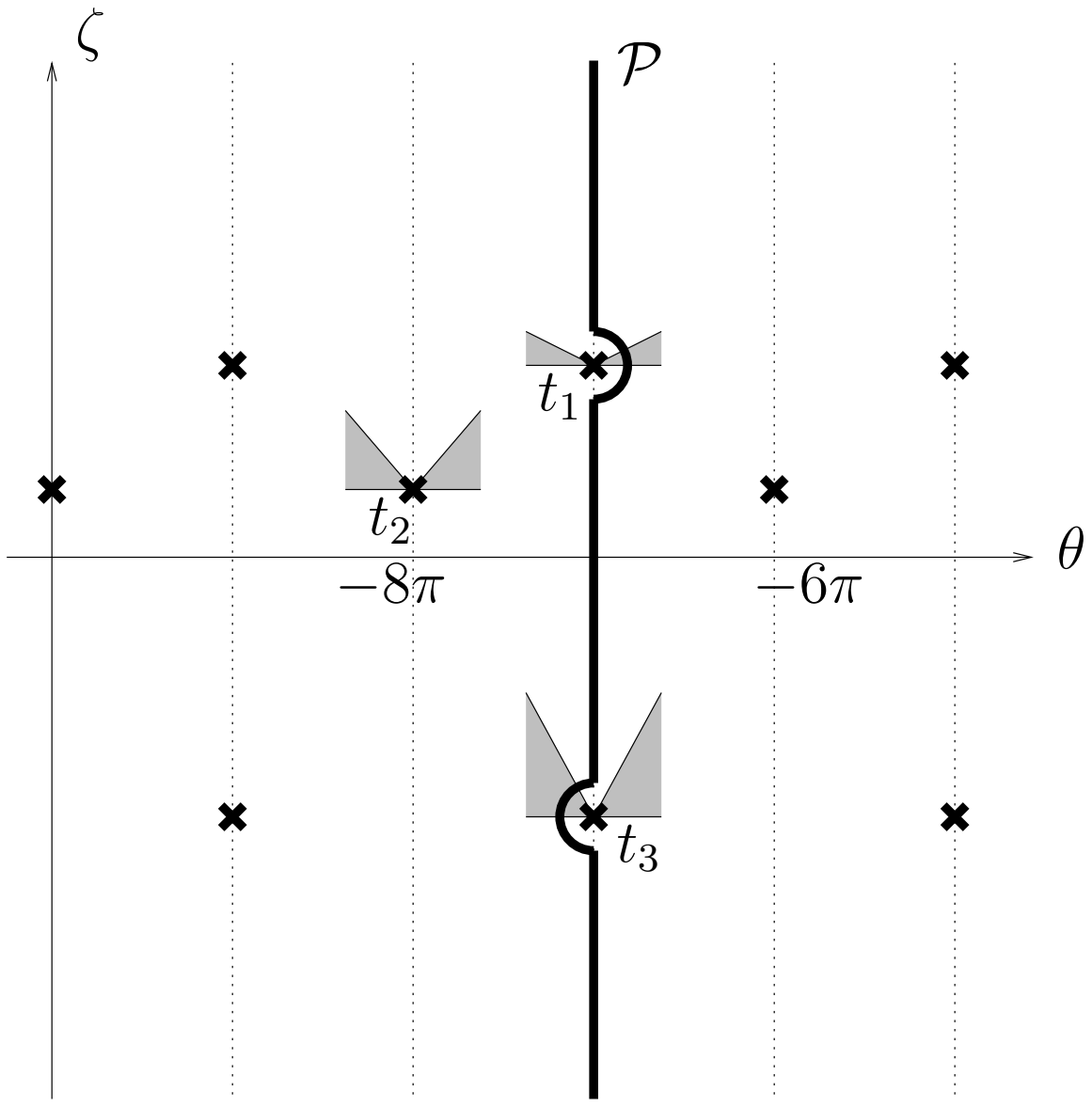}}
\caption{A path ${\mathcal P}$ representing the window ${\bf C}$.
The shaded regions to the right and the left
of $t_\bfa$ are the sectors $R_\bfa$ and $L_\bfa$ of (\ref{RandL}).
A particular representation is not allowed in each sector:
$\C(7)$ in $R_1$, $\C$ in $L_1$,
$\fund(7)$ in $R_2$, $\fund$ in $L_2$,
$\Sym^2\fund(6)$ in $R_3$, and
$\Sym^2\fund(-1)$ in $L_3$.} 
\label{fig:thepathP}
\end{figure}

Let us consider a path ${\mathcal P}$
which goes mostly along the straight line at $\theta=-7\pi$
but circumvents $t_1$ and $t_3$ by going to the right of $t_1$
and to the left of $t_3$,
see Fig.~\ref{fig:thepathP}.
It represents the window ${\bf C}$.
Such a path must go through the sectors $R_1$ and $L_3$,
and hence any deformation of $\bgamma_0$ cannot stay in
the admissible region for $\wtq_{0,7}$ and for $\wtq_{2,-1}$.
Thus, $\C(7)$ and $\Symt(-1)$ cannot be grade restricted
with respect to the window ${\bf C}$.
On the other hand, the path neither goes through sector
$L_1$ nor through sector $R_3$. Thus, the representations
$\C(0)$ and $\Symt(6)$ are not excluded from the grade restricted set
GRR${}_{\bf C}$.

Likewise, let us consider a path ${\mathcal P}'$ obtained by
reflecting ${\mathcal P}$ across the $\theta=-7\pi$ line,
which represents the window ${\bf B}$.
It must go through the sectors $L_1$ and $R_3$,
and hence no deformation of $\bgamma_0$ can stay in
the admissible region for $\wtq_{0,0}$ and for $\wtq_{2,6}$.
Thus, $\C(0)$ and $\Symt(6)$ cannot be grade restricted
with respect to the window ${\bf B}$.
On the other hand, the path does not go through the sectors
$R_1$ nor $L_3$. Thus, the representations
$\C(7)$ and $\Symt(-1)$ are not excluded from the grade restricted set
GRR${}_{\bf B}$.

We may also consider a path 
which goes mostly along the straight line at $\theta=-8\pi$
but circumvents $t_2$ by going to
its right ({\it resp}. left).
It represents the window ${\bf A}$ ({\it resp}. ${\bf D}(-1)$).
Such a path must go through the sector $R_2$ ({\it resp}. $L_2$),
and hence no deformation of $\bgamma_0$ can stay in
the admissible region for $\wtq_{1,7}$ ({\it resp}. $\wtq_{1,0}$).
Thus, $\fund(7)$ ({\it resp}. $\fund$) cannot be grade restricted
with respect to the window ${\bf A}$ ({\it resp}. ${\bf D}(-1)$).
On the other hand, the path does not go through the sector
$L_2$ ({\it resp}. $R_2$). Thus, the representation
$\fund$ ({\it resp}. $\fund(7)$) is not excluded from the grade restricted
set GRR${}_{\bf A}$ ({\it resp}. GRR${}_{{\bf D}(-1)}=\rmGRR_{\bf D}(1)$).
This is consistent with the result in Section~\ref{subsec:straight}
that $\rmGRR_{\bf A}$ and $\rmGRR_{\bf D}$ include
(\ref{oGRR1}) and (\ref{oGRR4}) respectively
and the possibility that ``include'' can be replaced by ``agree with''.
Indeed, $\fund$ is in (\ref{oGRR1}) but not in
(\ref{oGRR4})$\otimes \C(1)$
while $\fund(7)$ is in (\ref{oGRR4})$\otimes \C(1)$ but not in (\ref{oGRR1}).

Let us now turn our attention to the representations that are not excluded
along the curved paths.
For transport along the path ${\mathcal P}$ of
Fig.~\ref{fig:thepathP} representing the window {\bf C}, they are
$\C$ and $\Sym^2\fund(6)$.
We would like to know if there is a family of admissible contours.
We look for such a family within
deformations of $\bgamma_0$ or its limit $\overline{\bgamma}_0$.

First, let us consider $\Sym^2\fund(6)$.
Recall that the $\theta=-7\pi$ line is the right
boundary of the strip where $\bgamma_0$ is admissible for this
representation and hence $\AN_{\wtq_{2,6}}$ there is non-negative
on $\overline{\bgamma}_0$.
The smooth part of the region $\AN_{\wtq_{2,6}}\geq 0$ deforms continuously
as $t$ deformes. Thus, the part of $\overline{\bgamma}_0$ in the smooth part
of $\AN_{\wtq_{2,6}}\geq 0$ on the line $\theta=-7\pi$ can be deformed
so that it is in the region $\AN_{\wtq_{2,6}}>0$ on the path ${\mathcal P}$.
What is non-trivial is the part of $\overline{\bgamma}_0$ that goes through
the singular locus of the region $\AN_{\wtq_{2,6}}\geq 0$ on the line.
This is where $\overline{\bgamma}_0$ goes through the chamber ${\rm I}_3$
of the Coulomb branch $\CB_3^+$ at $t_3$.
And this is so everywhere on the line, not
just near the singular point $t_3$.
We use the coordinate system $(\Cc,z)$ on a neighborhood of the chamber
${\rm I}_3$, defined by (\ref{ztilde}) and (\ref{relzz}),
in which $\overline{\bgamma}_0$ is at $\Cc\in\R_{>0}$ and
$z\in \e^{\pi\im/4}\R$.
There, we have
\beqa
\AN_{\wtq_{2.6}}&=&
\AN_{\wtq_{2,6}}|_{t=t_3+\im\sDelta \theta}+(\zeta-\zeta_3)(\Is_1+\Is_2)\nn\\
&=&{\rm Re}\left[\,z^2-2\cos\left({3\pi\over 7}\right)\sDelta \theta\,\Cc
  \right]
+(\zeta-\zeta_3){\rm Im}\left[2\cos\left({3\pi\over 7}\right)\Cc\right]\nn\\
&=&x^2-y^2-2\cos\left({3\pi\over 7}\right)
\underbrace{\bigl[\,\sDelta\theta\,{\rm Re}(\Cc)
    +(\zeta_3-\zeta){\rm Im}(\Cc)\bigr]}_{\displaystyle ~~~\,\,=:c},
\label{deform26}
\eeqa
where $\theta=-7\pi+\sDelta\theta$.
We deform $\overline{\bgamma}_0$ by turning on ${\rm Im}(\Cc)$ if necessary
so that $c$ is {\it negative} all along the path ${\mathcal P}$.
For the part of ${\mathcal P}$ above ({\it resp}. below)
the semicircle around $t_3$,
we turn on a positive ({\it resp}. negative) ${\rm Im}(\Cc)$.
On the semicircle around $t_3$, we can achieve $c<0$
since $\sDelta\theta$ is negative (and ${\rm Re}(\Cc)$ is positive)
when $(\zeta_3-\zeta){\rm Im}(\Cc)$ vanishes.
Then, at any point of the path ${\mathcal P}$,
we are in the situation of Fig.~\ref{fig:shaded}-left,
and we can find a deformation that stays in the admissible region
as in Fig.~\ref{fig:shadecontour}-left.
The deformation is especially important along the semicircle around $t_1$
where $\sDelta\theta {\rm Re}(\Cc)$ is positive ---
without turning on ${\rm Im}(\Cc)$, $c$ would be positive and 
the picture would be the opposite, Fig.~\ref{fig:shaded}-right and
Fig.~\ref{fig:shadecontour}-right.
In this way, we find a family of
deformations of $\overline{\bgamma}_0$ which are admissible for
the representation $\Sym^2\fund(6)$ all along the path ${\mathcal P}$.

Next, let us consider the representation $\C$.
What is non-trivial is the part of $\overline{\bgamma}_0$
that goes through the chamber ${\rm III}_1$
of the Coulomb branch $\CB_1^+$ at $t_1$.
Using the coordinate system $(\Cc,z)$ on a neighborhood of that chamber,
in which $\overline{\bgamma}_0$ is at $\Cc\in\R_{<0}$ and
$z\in \e^{-\pi\im/4}\R$, we have
\beqa
\AN_{\wtq_{0.0}}&=&
\AN_{\wtq_{0,0}}|_{t=t_1+\im\sDelta \theta}+(\zeta-\zeta_1)(\Is_1+\Is_2)\nn\\
&=&{\rm Re}\left[\,z^2-2\cos\left({\pi\over 7}\right)\sDelta \theta\,\Cc
  \right]
+(\zeta-\zeta_1){\rm Im}\left[2\cos\left({\pi\over 7}\right)\Cc\right]\nn\\
&=&x^2-y^2-2\cos\left({\pi\over 7}\right)
\underbrace{\bigl[\,\sDelta\theta\,{\rm Re}(\Cc)
    +(\zeta_1-\zeta){\rm Im}(\Cc)\bigr]}_{\displaystyle ~~~\,\,=:c}.
\label{deform00}
\eeqa
We deform $\overline{\bgamma}_0$ by turning on ${\rm Im}(\Cc)$ if necessary
so that $c$ is {\it negative} all along the path ${\mathcal P}$.
Then, at any point of the path ${\mathcal P}$,
we are in the situation of Fig.~\ref{fig:shaded}-left,
and we can find a deformation that stays in the admissible region
as in the $-45^{\circ}$ version of Fig.~\ref{fig:shadecontour}-left.
The deformation is especially important along the semicircle around $t_3$
--- without it,
the picture would be the opposite.
For the other parts of $\overline{\bgamma}_0$,
we do not have to deform it at all or deform it towards the interior of
the admissible region,
just as in the case of $\Sym^2\fund(6)$.
In this way, we find a family of
deformations of $\overline{\bgamma}_0$ which are admissible for
the representation $\C$ all along the path ${\mathcal P}$.

Thus, for both $\Sym^2\fund(6)$ and $\C$, we managed to find families of
admissible contours along the path ${\mathcal P}$ representing
the window {\bf C}.
The same can be done for $\Sym^2\fund(-1)$ and $\C(7)$
along the path ${\mathcal P}'$ representing the window {\bf B}.
However, we have not payed attention to Weyl invariance of the contours.
The coordinates $(\Cc,z)$ which we have used are defined near
particular Weyl representatives of the Coulomb branches,
$\CB_3^+$ and $\CB_1^+$, and are quite inconvenient
to examine Weyl invariance.
We shall return to this problem in Section~\ref{subsec:Weyl}.

We shall call an $SU(2)$ grade restricted irreducible representation $\rmV_*$
 \emph{special} at a singular point $t_*$ when
its shifted highest weight $\wtq_*$
is special in the sense of (\ref{vaceqnose})
on a chamber of the Coulomb branch at $t_*$.
The special representations at $t_1$, $t_2$ and $t_3$ of
(\ref{t1}), (\ref{t2}) and (\ref{t3}) are the pairs $\{\C(0),\C(7)\}$,
$\{\fund(0),\fund(7)\}$ and $\{\Sym^2\fund(-1),\Sym^2\fund(6)\}$, respectively.
If $\rmV_*$ is special at $t_*$, then $\rmV_*(j)$ is special at
$t_*+2\pi\im j$.

\subsection{No-Entry Zones}
\label{subsec:noentry}

From the local analysis near the singular points,
we have seen that there are sectors (\ref{RandL}) in which
particular representations are forbidden.
See also Fig.~\ref{fig:thepathP}.
A natural question is whether and how they extend
away from the singular points.
With the help of numerical analysis described
in Appendix~\ref{app:numerical}, we shall see that
they extend to areas to which the representations cannot enter
from above and from below.

Let us consider the representation $\C(7)$.
It has the forbidden sector $R_1$ at $t_1$ in which
any deformation of $\bgamma_0$ is not admissible.
On the other hand, as sketched in Section~\ref{subsec:numerical},
as $t$ approaches this sector from above ({\it resp}. below),
any deformation of $\bgamma_+$ ({\it resp}. $\bgamma_-$),
whether Weyl invariant or not,
is forced to either go out of the admissible region
or hit the wall $W_1$ ({\it resp}. $W_p$).
These are consistent with each other as the upper ({\it resp}. lower)
boundary of the chamber ${\rm I}_1$ of the Coulomb branch,
which is responsible for the upper bound ${\rm Arg}(\im\sDelta t)\leq \pi/7$
({\it resp}. lower bound ${\rm Arg}(\im\sDelta t)\geq 0$)
of the sector $R_1$, is the wall $W_1$ ({\it resp}. $W_p$).
The numerical analysis shows that this no-entry from above and from below
due to blocking of the contour
by the walls $W_1$ and $W_p$
continues to hold away from the singular point $t_1$, i.e., along
the straight path at all $\theta>-7\pi$.
In fact, the exact boundary of the no-entry zone can also be determined.
For $\theta$ not very far from $-7\pi$, the upper boundary
is where the positive region $\AN_{\wtq_{0,7}}\geq 0$ on the wall
$W_1$ is singular.
The singularity is caused by the squeezing of the region
which chases the contour to hit the wall.
The growth function on the wall $W_1=\{\Rs_1=0,\,\,\Is_1>0\}$
(for a general $\wtq$) is
\beqa
\AN_{\wtq}|_{W_1}&=&\zeta(\Is_1+\Is_2)-\theta\Rs_2-2\pi\wtq^2\Rs_2\nn\\
&&+7\Rs_2\Arg(\im\usigma_2)-7\Rs_2\Arg(-\im(\usigma_1+\usigma_2))\nn\\
&&+7\Is_1\log|\Is_1|+7\Is_2\log|\usigma_2|
-7(\Is_1+\Is_2)\log|\usigma_1+\usigma_2|.
\eeqa
Singularity of the region $\AN_{\wtq}|_{W_1}\geq 0$
is equivalent to the singularity of the hypersurface $\AN_{\wtq}|_{W_1}=0$.
By the scaling property (\ref{scaleinv}), it is the common zero
of the first derivetives of $\AN_{\wtq}|_{W_1}$,
\beqa
\partial_{\Rs_2}\AN_{\wtq}|_{W_1}&=&
-\theta-2\pi\wtq^2+7\Arg(\im\usigma_2)-7\Arg(-\im(\usigma_1+\usigma_2)),\\
\partial_{\Is_1}\AN_{\wtq}|_{W_1}&=&
\zeta+7\log|\Is_1|-7\log|\usigma_1+\usigma_2|,\\
\partial_{\Is_2}\AN_{\wtq}|_{W_1}&=&
\zeta+7\log|\usigma_2|-7\log|\usigma_1+\usigma_2|.
\eeqa
Again by the scaling property, we may set $\Is_1=1$
which simplifies the equations to
\beq
\theta+2\pi\wtq^2=7\Arg(\im\usigma_2)-7\Arg(1-\im\usigma_2),\quad
|\usigma_2|=1,\quad
\zeta=7\log|1-\im\usigma_2|.
\eeq
The first two equations imply $\im\usigma_2=-\e^{2\im(\theta+2\pi\wtq^2)/7}$
and then the third reads
\beq
\zeta=7\log\left|2\cos\left({\theta+2\pi\wtq^2\over 7}\right)\right|
=:\zeta(\theta;\wtq^2).
\label{threshold1}
\eeq
For the case $\wtq=\wtq_{0,7}$,
the value of $\zeta$ at $\theta=-7\pi$ is $7\log(2\cos(\pi/7))=\zeta_1$
and hence the curve (\ref{threshold1}) indeed goes through the point $t_1$.
Moreover, the slope of the curve at that point,
${\dd\over\dd\theta}\zeta|_{\theta=-7\pi}=\tan(\pi/7)$,
is nothing but the slope of the upper bound of the sector $R_1$.
$\zeta(\theta;\wtq^2_{0,7})$ increases as $\theta$ moves away from $-7\pi$
and reaches $7\log 2$ at $\theta=-6\pi$ with slope zero.
$\theta=-6\pi$ is special in that the singular point
of $\AN_{\wtq_{0,7}}|_{W_1}\geq 0$ is $(\usigma_1,\usigma_2)=(\im,\im)$ and
is also on the wall $W_2$ where the function $\AN_{\wtq_{0,7}}|_{W_1}$
is not smooth.
Recall from Section~\ref{subsec:weak} that $\zeta> 7\log 2$ is the
sufficient condition for the contour $\bgamma_+$ to be admissible
for any representation.
In fact, the numerical analysis shows that
$\zeta=7\log 2$ with $\theta\geq -6\pi$
is the lower bound for any deformation of $\bgamma_+$ for the
representation $\C(7)$:
As $t$ moves down along the straight path at $\theta\geq -6\pi$,
there exists an admissible deformation of $\bgamma_+$ as long as $\zeta$
is above $7\log 2$, but, as it reaches $7\log 2$,
any deformation of $\bgamma_+$ must 
either go out of the admissible region or hit the wall $W_1$.
Thus, the upper boundary of the no-entry zone is
$\zeta=\zeta(\theta;\wtq_{0,7}^2)$ for $-7\pi\leq \theta\leq -6\pi$
and $\zeta=7\log 2$ for $\theta\geq -6\pi$.
On the other hand, the lower boundary of the no-entry zone is
where the positive region $\AN_{\wtq_{0,7}}\geq 0$ on the wall
$W_p$ has a singularity, which is caused by the squeezing of the region.
The growth function on the wall $W_p=\{\Rs_1+\Rs_2=0,\,\,\Is_1+\Is_2<0\}$ is
\beqa
\AN_{\wtq}|_{W_p}&=&
\zeta(\Is_1+\Is_2)-2\pi(\wtq^1-\wtq^2)\Rs
\nn\\
&&+7\Rs\Arg(\im\usigma_1)-7\Rs\Arg(\im\usigma_2)\nn\\
&&+7\Is_1\log|\usigma_1|+7\Is_2\log|\usigma_2|-7(\Is_1+\Is_2)\log|\Is_1+\Is_2|,
\eeqa
where $\Rs:=\Rs_1=-\Rs_2$.
The singularity of the region $\AN_{\wtq}|_{W_p}\geq 0$ or equivalently
of the hypersurface $\AN_{\wtq_{0,7}}|_{W_p}=0$ is the common zero of
\beqa
\partial_{\Rs}\AN_{\wtq}|_{W_p}&=&
-2\pi(\wtq^1-\wtq^2)+7\Arg(\im\usigma_1)-7\Arg(\im\usigma_2),\\
\partial_{\Is_1}\AN_{\wtq}|_{W_p}&=&
\zeta+7\log|\usigma_1|-7\log|\Is_1+\Is_2|,\\
\partial_{\Is_2}\AN_{\wtq}|_{W_p}&=&
\zeta+7\log|\usigma_2|-7\log|\Is_1+\Is_2|.
\eeqa
We may set $\Is_1+\Is_2=-1$ which simplifies the equations as
\beq
2\pi(\wtq^1-\wtq^2)
=7\Arg\left({{1\over 2}+\im\Rs\over {1\over 2}-\im\Rs}\right),\quad
\zeta=-7\log\left|{1\over 2}+\im \Rs\right|.
\eeq
Eliminating $\Rs$, we find 
\beq
\zeta=7\log\left|2\cos\left({\pi(\wtq^1-\wtq^2)\over 7}\right)\right|.
\eeq
For $\wtq=\wtq_{0,7}$, it is $\zeta=\zeta_1$.
Recall from Section~\ref{subsec:SU2GRRRodland} that $\zeta=\zeta_1$ is the
upper bound for $\bgamma_-$ to be admissible for any $\C(i)$.
In fact, the numerical analysis shows that
$\zeta=\zeta_1$ with $\theta\geq -7\pi$
is also the upper bound for any deformation of $\bgamma_-$ for $\C(7)$:
as $t$ moves up along the stright path at $\theta\geq -7\pi$,
there exists an admissible deformation of $\bgamma_-$ as long as $\zeta$
is below $\zeta_1$, but, as it reaches $\zeta_1$,
any deformation of $\bgamma_-$ is forced to either
go outside the admissible region or hit the wall $W_p$.
Thus, the no-entry zone is exactly
\beq
\zeta_1\leq \zeta\leq
\left\{
\begin{array}{ll}
\zeta(\theta;\wtq^2_{0,7})&-7\pi\leq\theta\leq -6\pi\\
7\log 2& \theta\geq -6\pi,
\end{array}\right.
\eeq
which is depicted in Fig.~\ref{fig:NoEntry1}.
\begin{figure}[htb]
  \psfrag{t1}{$t_1$}
\psfrag{t2}{$t_2$}
\psfrag{t3}{$t_3$}
\psfrag{z}{$\zeta$}
\psfrag{th}{$\theta$}
\psfrag{-7pi}{$-7\pi$}
\psfrag{-6pi}{$-6\pi$}
 \centerline{\includegraphics[page=7]{main-pics.pdf}}
\caption{}
\label{fig:NoEntry1}
\end{figure}
By the reflection symmetry, there is also a no-entry zone for $\C(7)$ that
extends to the left of the singular point $(\zeta,\theta)=(\zeta_1,-21\pi)$,
\beq
\zeta_1\leq \zeta\leq
\left\{\begin{array}{ll}
7\log 2&\theta\leq -22\pi\\
\zeta(\theta;\wtq^1_{0,7})&-22\pi\leq\theta\leq -21\pi.\\
\end{array}\right.
\eeq
The upper ({\it resp}. lower)
bound is where any deformation of $\bgamma_+$ ({\it resp}. $\bgamma_-$)
must 
either go out of the admissible region or
hit the wall $W_2$ ({\it resp}. $W_p$).
Its picture is obtained from the one in Fig.~\ref{fig:NoEntry1} by
reflection symmetry.

This analysis generalizes straightforwardly to other $SU(2)$ grade restricted
representations. Let us take $\Sym^h\fund(i)$ with
$h\in\{0,1,2\}$ and $i\in\Z$, which is a special representation
at the singular points $(\zeta,\theta)=(\zeta_{h+1},\theta_{h,i}^{\pm})$
where
\beq
\theta^{\pm}_{h,i}:=(\pm 7-h-2i)\pi.
\label{thetapmhi}
\eeq
Recall from (\ref{Cond11}) that it can be transported
via the contour $\bgamma_0$ along the straight path
in the interval
\beq
\theta_{h,i}^-<\theta<\theta_{h,i}^+
\eeq
To the right of $(\zeta_{h+1},\theta_{h,i}^+)$ and to the left of 
$(\zeta_{h+1},\theta_{h,i}^-)$ extend the no-entry zones
\beqa
{\mathscr R}_{h,i}:&&
\zeta_{h+1}\leq \zeta\leq
\left\{\begin{array}{ll}
\zeta(\theta;i-{1\over 2})&\,{\rm for}~\,
\theta_{h,i}^+\leq\theta\leq \theta_{h,i}^++(h+1)\pi\\
7\log 2&\,{\rm for}~\, \theta\geq \theta_{h,i}^++(h+1)\pi
\end{array}\right.\label{Rhi}\\
{\mathscr L}_{h,i}:&&
\zeta_{h+1}\leq\zeta\leq
\left\{\begin{array}{ll}
7\log 2&\,{\rm for}~\,\theta\leq \theta_{h,i}^--(h+1)\pi\\
\zeta(\theta;h+i+{1\over 2})&\,{\rm for}~\,
\theta_{h,i}^--(h+1)\pi\leq\theta\leq\theta_{h,i}^-\\
\end{array}\right.\label{Lhi}
\eeqa
As $t$ reaches
the upper boundaries of ${\mathscr R}_{h,i}$ and ${\mathscr L}_{h,i}$ from above,
any deformation of $\bgamma_+$ must 
either go outside the admissible region or
hit the walls $W_1$ and $W_2$, respectively.
As $t$ reaches the lower boundary of
${\mathscr R}_{h,i}$ and ${\mathscr L}_{h,i}$ from below,
any deformation of $\bgamma_-$ must 
either go outside the admissible region or hit the wall $W_p$.
This holds irrespective of whether the deformations of
$\bgamma_+$ and $\bgamma_-$ are Weyl invariant.
The  forbidden sectors $R_1$, $L_1$, $R_2$, $L_2$, $R_3$ and $L_3$
of Fig.~\ref{fig:thepathP} extend to the no-entry zones
${\mathscr R}_{0,7}$, ${\mathscr L}_{0,0}$,
${\mathscr R}_{1,7}$, ${\mathscr L}_{1,0}$,
${\mathscr R}_{2,6}$, and ${\mathscr L}_{2,-1}$, respectively.
${\mathscr R}_{0,7}$ is depicted in Fig.~\ref{fig:NoEntry1} as noted already.
${\mathscr R}_{1,7}$ and ${\mathscr R}_{2,6}$ are in
Fig.~\ref{fig:NoEntry23}-left and right.
The pictures of the others are obtained from these by the reflection symmetry.

\begin{figure}[htb]
  \psfrag{t1}{$t_1$}
\psfrag{t2}{$t_2$}
\psfrag{t3}{$t_3$}
\psfrag{z}{$\zeta$}
\psfrag{th}{$\theta$}
\psfrag{-8pi}{$-8\pi$}
\psfrag{-6pi}{$-6\pi$}
\psfrag{-4pi}{$-4\pi$}
\psfrag{-7pi}{$-7\pi$}
\centerline{\includegraphics[page=8]{main-pics.pdf}~~~~~~~~
\includegraphics[page=9]{main-pics.pdf}}
\caption{}
\label{fig:NoEntry23}
\end{figure}

\noindent
{\bf Remark.} The upper boundaries of ${\mathscr R}_{2,i}$,
${\mathscr R}_{1,i}$ and ${\mathscr R}_{0,i}$ partially overlap ---
they are all on the same curve $\zeta=\zeta(\theta;i-{1\over 2})$
or on the line $\zeta=7\log 2$.
For the curved part,
this is because $\AN_{\wtq_{h,i}}|_{W_1}$ depends on $i$ but not on $h$.
This means that $\AN_{\wtq_{h,i}}|_{W_1}\geq 0$ is singular also
on the part $\theta<\theta_{h,i}^+$ of the curve,
but we know that the representation $\Sym^h\fund(i)$ can be safely transported
across the curve.
Similarly, 
$\AN_{\wtq_{h,i}}|_{W_2}$ depends only on the combination $h+i$ and
the upper boundaries of ${\mathscr R}_{2,i-2}$,
${\mathscr R}_{1,i-1}$ and ${\mathscr R}_{0,i}$ are on a common curve.
In particular, $\AN_{\wtq_{h,i}}|_{W_2}\geq 0$ is singular on the curve also
for $\theta>\theta_{h,i}^-$, even though transport of
$\Sym^h\fund(i)$ across the curve is perfectly fine.
Likewise,
$\AN_{\wtq_{h,i}}|_{W_p}$ is independent of $\theta$ and
$\AN_{\wtq_{h,i}}|_{W_p}\geq 0$ is singular on the line $\zeta=\zeta_{h+1}$ also
for $\theta_{h,i}^-<\theta<\theta_{h,i}^+$, even though
the transport of $\Sym^h\fund(i)$ across the line is perfectly fine.
These observations show that the singularity of the region
$\AN_{\wtq_{h,i}}\geq 0$ on a wall does not necessarily mean obstruction
of the transport of $\Sym^h\fund(i)$
due to blocking of
the contour by the wall.
This is because the event near the wall responsible for the singularity
happens away from the contours.
Some details on such a ``harmless singularity'' processes are described in
Appendix~\ref{subapp:genpaths}.

This result can be used to narrow down possible
representations that can be included in the grade restricted subsets.
Any path in window {\bf w} $\in \{{\bf A}, {\bf B}, {\bf C}, {\bf D}\}$
must go through the no-entry zones of
the representation in the row {\bf w} in the table below.
\beq
\begin{array}{c||c|c|c}
  {\bf w}&\C(i)&\fund(i)&\Sym^2\fund(i)\\
  \hline
  {\bf A}&i\leq 0,~~ i\geq 8&i\leq -1,~~i\geq 7&i\leq -1,~~i\geq 7\\
  {\bf B}&i\leq 0,~~ i\geq 8&i\leq -1,~~i\geq 7&i\leq -2,~~i\geq 6\\
  {\bf C}&i\leq -1,~~ i\geq 7&i\leq -1,~~i\geq 7&i\leq -1,~~i\geq 7\\
  {\bf D}&i\leq -1,~~ i\geq 7&i\leq -1,~~i\geq 7&i\leq -2,~~i\geq 6
\end{array}
\eeq
These representations are excluded from the grade restricted set.
For window {\bf A}, the remaining representations precisely match with
those in (\ref{oGRR1}). And we already know that they are included
in $\rmGRR_{\bf A}$ by interpolation via the contour $\bgamma_0$.
This proves that (\ref{oGRR1}) precisely matches with $\rmGRR_{\bf A}$.
Similarly, for window {\bf D}, the remaining representations precisely
match with those in (\ref{oGRR4}) which we already know to be in
$\rmGRR_{\bf D}$ again via $\bgamma_0$.
This proves $\rmGRR_{\bf D}=$ (\ref{oGRR4}).
For window {\bf C}, the remaining representations are those in the overlap of
(\ref{oGRR1}) and (\ref{oGRR4}) plus $\C$ and $\Sym^2\fund(6)$.
For window {\bf B}, the remaining representations are those in the same overlap
plus $\C(7)$ and $\Sym^2\fund(-1)$.

\subsection{Interlude: Abelian Theories}
\label{subsec:Abelian}

Let us take a moment of digression to discuss our observations,
the special representations at singular points and the no-entry zones,
in models with rank one Abelian gauge groups.
To be specific, we consider the quintic model,
with gauge group $U(1)$, matter fields $x_1,\ldots, x_5,p$
of charges $1,\ldots, 1,-5$, and superpotential $W=pf(x_1,\ldots, x_5)$
where $f$ is a quintic polynomial.
Extension to a general model is straightforward.

The effective twisted superpotential is
$\wt{W}_{\!\it eff}=-(t-5\log(-5))\usigma$
and the Coulomb branch exists when $\e^t=(-5)^5$, that is,
$\zeta=5\log 5$ and $\theta\equiv \pi$ mod $2\pi\Z$.
The growth function reads
\beqa
\AN_q&=&\zeta\Is-\theta\Rs-2\pi q\Rs+5\Rs\Arg(\im\usigma)-5\Rs\Arg(-5\im\usigma)
+5\Is\log|\usigma|-5\Is\log|5\usigma|
\nn\\
&=&(\zeta-5\log 5)\Is-(\theta+2\pi q)\Rs+5\pi|\Rs|,\label{ANquintic}
\eeqa
and the walls are
\beq
W_x=\{\,\Rs=0,\,\,\Is>0\,\},\quad
W_p=\{\,\Rs=0,\,\,\Is<0\,\}.
\eeq
In the geometric/Landau-Ginzburg phase $\pm\zeta\gg 0$,
the hemisphere integral is convergent for any representation
if we take the contour $\bgamma_{\pm}=\{\Is=\pm\Rs^2\}$.
The expression (\ref{ANquintic}) is simple enough and can be used
directly to find the grade restriction rule.
Instead, let us proceed in a way that highlights our observations.

The Coulomb branch at a singular point $t_*=5\log 5-\im \theta_*$
with $\theta_*\in \pi+2\pi\Z$,
which is the entire $\usigma$-plane, is divided by the walls
into two chambers, I and II,
\begin{figure}[htb]
  \psfrag{wx}{$W_x$}
  \psfrag{wp}{$W_p$}
  \psfrag{I}{I}
  \psfrag{II}{II}
  \psfrag{sp}{$\usigma$}
\centerline{\includegraphics[page=10]{main-pics.pdf}}
\caption{}
\label{fig:CBab}
\end{figure}
as shown in Fig.~\ref{fig:CBab}.
The special representation at $t_*$ is $\C(q_*)$ with
\beqa
2\pi \im q_*&=&t_*+5\log(\im \usigma)-5\log(-5\im\usigma)\nn\\
&=&-\im\theta_*+5\im\Arg(\im\usigma)-5\im \Arg(-\im\usigma).
\eeqa
If we parametrize $\usigma=|\usigma|\e^{\im\varphi}$ with
$-{\pi\over 2}<\varphi<{3\pi\over 2}$, we find
\beq
\begin{array}{c|cc|c}
  &\Arg(\im \usigma)&
  \Arg(-\im\usigma)
  &q_*
  \\
  \hline
      {\rm I}&{\pi\over2}+\varphi
      &-{\pi\over 2}+\varphi
      &{5\over 2}-{\theta_*\over 2\pi}\\
      {\rm II}&{}_{''}-2\pi
      &{}_{''}
      &-{5\over 2}-{\theta_*\over 2\pi}
\end{array}
\label{tableArgab}
\eeq
For example, at the singular point with $\theta_*=-5\pi$,
the special representations are $\C(5)$ on chamber I and $\C(0)$ on chamber II.
In a neighborhood of this point,
$\zeta=5\log 5+\sDelta\zeta$ and $\theta=-5\pi+\sDelta\theta$,
we have
\beq
\AN_5|^{}_{\rm I}=\sDelta\zeta\,\Is-\sDelta\theta\,|\Rs|,\quad
\AN_0|^{}_{\rm II}=\sDelta\zeta\,\Is+\sDelta\theta\,|\Rs|.
\label{AN50}
\eeq
We see that $\AN_5$ ({\it resp}. $\AN_0$) is negative everywhere on
chamber I ({\it resp}. chamber II) for
$\sDelta\zeta=0$ and $\sDelta\theta>0$ ({\it resp}. $\sDelta\theta<0$).
Thus $\C(5)$ ({\it resp}. $\C(0)$) is forbidden in the horizontal direction
to the right ({\it resp}. left) of this singular point.
In fact, the formula (\ref{AN50}) is exact and
holds for an arbitrary $(\zeta,\theta)$,
not just for small $(\sDelta\zeta,\sDelta\theta)$. Therefore
$\C(5)$ ({\it resp}. $\C(0)$) is forbidden anywhere on the
line $\zeta=5\log 5$ to the right ({\it resp}. left) of the point.
Let us see what happens to $\C(5)$ if $(\zeta,\theta)$ crosses
the line at a finite distance on the right of the point.
\begin{figure}[htb]
  \psfrag{Wx}{\tiny $W_x$}
  \psfrag{Wp}{\tiny $W_p$}
  \psfrag{!}{$!$}
\centerline{\includegraphics[page=11]{main-pics.pdf}}
\caption{}
\label{fig:Vabove}
\end{figure}
\begin{figure}[htb]
  \psfrag{Wx}{\tiny $W_x$}
  \psfrag{Wp}{\tiny $W_p$}
  \psfrag{!}{$!$}
\centerline{\includegraphics[page=12]{main-pics.pdf}}
\caption{}
\label{fig:Vbelow}
\end{figure}
Fig.~\ref{fig:Vabove} and Fig.~\ref{fig:Vbelow}
show the $\usigma$-planes as $\zeta$ crosses the line from above and
from below at $\sDelta\theta=5\pi$ (i.e. $\theta=0$),
in which the region $\AN_5\geq 0$ is shaded.
$\sDelta\zeta$ is $100$, $20$, $5$, $0$, $-5$ from left to right
in Fig.~\ref{fig:Vabove}
and $-100$, $-20$, $-5$, $0$, $5$ in Fig.~\ref{fig:Vbelow}.
Admissible contours are also depicted starting from 
a deformation of the contours $\bgamma_+$ and $\bgamma_-$
at $\sDelta\zeta=100$ and $-100$.
We see that the contour is forced to hit the wall
$W_x$ as $\zeta$ approaches the line from above
and to hit $W_p$ for an approach from below.
Thus, the whole line to the right of the singular point
can be regarded as a no-entry zone for $\C(5)$.
As it has no interior point, it might better be called a ``no-crossing line''.
Similarly, the line to its left
is a no-crossing line for $\C(0)$.
In general, $\C(q)$ has two no-crossing lines,
one to the right of $\theta=(5-2q)\pi$
and one to the left of $\theta=(-5-2q)\pi$, both at $\zeta=5\log 5$.

This can be used to narrow down the set of representations that can be
transported along paths between the two phases.
Let us consider a path through the window
$-7\pi<\theta<-5\pi$ at $\zeta=5\log 5$.
The no-crossing line of $\C(q)$ includes this window if $q\leq 0$ and $q\geq 6$
and hence $\C(q)$ for such $q$'s cannot go through.
This leaves us with the five representations
$\C(1)$, $\C(2)$, $\C(3)$, $\C(4)$, $\C(5)$.
All of these can go through the window via the real contour,
and hence consitute the grade restricted subset for this window.

\subsection{Issue Of Weyl Invariance}
\label{subsec:Weyl}

Returning to the end of Section~\ref{subsec:catastroph},
we examine Weyl invariance of the deformations of 
$\overline{\bgamma}_0$ along the path ${\mathcal P}$
for the representations $\Sym^2\fund(6)$ and $\C$.
Note that the part
$\{\usigma_1+\usigma_2=0\}$ of $\overline{\bgamma}_0$ is admissible
as the representations are $SU(2)$ grade restricted.
Therefore, the focus will be deformations of the other part
$\{\usigma_1=\overline{\usigma_2}\}=:\overline{\bgamma}_0^{(0)}$.
We shall denote $\usigma$ on this part as $\usigma^{(0)}$
and parametrize it as
\beq
\usigma_1^{(0)}=\Rs+\im\Is,\quad
\usigma_2^{(0)}=\Rs-\im\Is.
\eeq
The growth function on this locus
$\AN_{\wtq}^{(0)}:=\AN_{\wtq}|_{\overline{\sgamma}_0^{(0)}}$ is given by
\beqa
\AN_{\wtq}^{(0)}
  &=&-2\theta\Rs-2\pi(\wtq^1+\wtq^2)\Rs
+7\Rs\Arg(\im\Rs-\Is)+7\Rs\Arg(\im\Rs+\Is)
-7\cdot 2\Rs\Arg(-2\im\Rs)\nn\\
&=&-2\theta\Rs-2\pi(\wtq^1+\wtq^2)\Rs+7\cdot 2\pi|\Rs|\nn\\
&=&
-2\sDelta\theta\,\Rs+7\cdot 2\pi(|\Rs|\mp \Rs)
\quad\mbox{for}~~~
\wtq=\left\{\begin{array}{c}
\wtq_{2,6}\\
\wtq_{0,0}
  \end{array}\right.
\label{gfA}
\eeqa
where we used $\Arg(\im\Rs-\Is)+\Arg(\im\Rs+\Is)
=\pi\cdot {\rm sgn}(\Rs)$
and $\Arg(-2\im\Rs)=-{\pi\over 2}{\rm sgn}(\Rs)$.
We also note that the power factor $P(\usigma)$
in (\ref{ZVint}) that appears from the use of Stirling's formula 
has the leading behaviour
\beq
|P(\usigma)|\simeq {\rm constant}\cdot
{|\Rs|^{{7\over 2}-7\epsilon}\over |\Rs+\im\Is|^{7-7\epsilon}}.
\label{power00}
\eeq
This would give a convergence factor in the direction
where $\AN_{\wtq}$ vanishes.
In particular, 
for the part of ${\mathcal P}$ on the $\theta=-7\pi$ line,
where $\sDelta\theta=0$,
there is no need to deform $\overline{\bgamma}_0$
even though we did so in Section~\ref{subsec:catastroph} just to be safe.
However, the factor $P(\usigma)$ is of no help
in the direction where $\AN_{\wtq}$ goes negative.

Let us first consider $\Symt(6)$.
The integral is convergent on $\overline{\bgamma}_0$ along ${\mathcal P}$
except for the small semicircle to the right of $t_1$,
thanks to the last remark.
Thus, the focus will be for $t=t_1+\sDelta t$ on this semicircle.
As $\sDelta\theta>0$ there,
$\AN^{(0)}_{\wtq_{2,6}}$ is positive if $\Rs<0$ but becomes negative if $\Rs>0$.
Thus, the problem is present only for the part $\Rs>0$
of $\overline{\bgamma}_0^{(0)}$.
For the deformation
\beq
\usigma_1=\usigma_1^{(0)}+\varepsilon_1,\quad
\usigma_2=\usigma_2^{(0)}+\varepsilon_2,
\label{defsz}
\eeq
we have
\beq
\AN_{\wtq_{2,6}}|_{\usigma^{(0)}+\varepsilon}
=\AN_{\wtq_{2,6}}^{(0)}
+{\rm Re}\left(\sum_{a=1}^2\varepsilon_a\im\partial_{\usigma_a}
\wt{W}_{\!{\it eff},\wtq_{2,6}}^{(0)}\right)+O(\varepsilon^2)
\eeq
If we write
\beq
\usigma_1^{(0)}=\xi\e^{\im \psi},\quad
\usigma_2^{(0)}=\xi\e^{-\im\psi};\quad\,\,\xi>0,~
-{\pi\over 2}<\psi<{\pi\over 2},
\label{parapos}
\eeq
we have for $t=t_1+\sDelta t$
\beqa
\partial_{\usigma_1}\wt{W}_{\!{\it eff},\wtq_{2,6}}^{(0)}
&=&-t+2\pi\im\wtq_{2,6}^1-7\log(\im\usigma_1^{(0)})
+7\log(-\im(\usigma_1^{(0)}+\usigma_2^{(0)}))\nn\\
&=&-7\log\left({\cos{\pi\over 7}\over\cos\psi}\right)
-\sDelta\zeta+\im(3\pi-7\psi+\sDelta\theta),\\
\partial_{\usigma_2}\wt{W}_{\!{\it eff},\wtq_{2,6}}^{(0)}
&=&-t+2\pi\im\wtq_{2,6}^2-7\log(\im\usigma_2^{(0)})
+7\log(-\im(\usigma_1^{(0)}+\usigma_2^{(0)}))\nn\\
&=&-7\log\left({\cos{\pi\over 7}\over\cos\psi}\right)
-\sDelta\zeta+\im(-3\pi+7\psi+\sDelta\theta),
\eeqa
and hence
\beqa
\AN_{\wtq_{2,6}}|_{\usigma^{(0)}+\varepsilon}
&=&-2\sDelta\theta\,\Rs
+{\rm Im}(\varepsilon_1+\varepsilon_2)
\cdot 7\log\left({\cos{\pi\over 7}\over\cos\psi}\right)
+{\rm Re}(\varepsilon_1-\varepsilon_2)(7\psi-3\pi)\nn\\
&&+{\rm Im}(\varepsilon_1+\varepsilon_2)\sDelta\zeta
-{\rm Re}(\varepsilon_1+\varepsilon_2)\sDelta \theta
+O(|\varepsilon|^2).\label{Aondef1}
\eeqa
In the region where $|\varepsilon|$ is of the same order
as the radius $|\sDelta t|$ of the semicircle, we need the first line of
(\ref{Aondef1}) to be positive.
An example is given by
\beq
\varepsilon_1=(-1+\im)\sDelta\theta\,\Rs,\quad
\varepsilon_2=(1+\im)\sDelta\theta\,\Rs.
\label{sol1}
\eeq
Indeed, for this $(\varepsilon_1,\varepsilon_2)$, the first line of
(\ref{Aondef1}) is $2\sDelta\theta\,\Rs$ times
$7\log\left({\cos{\pi\over 7}\over\cos\psi}\right)+3\pi-7\psi-1$
which is positive for $\psi\in (-{\pi\over 2},{\pi\over 2})$.
However, this deformation is not Weyl invariant.
In fact, it is impossible to find a Weyl invariant deformation.
Note that the Weyl reflection
$\usigma_1^{(0)}\leftrightarrow \usigma_2^{(0)}$ flips the sign of
$\psi$ and hence Weyl invariance of the deformation requires that
$\varepsilon_1$ and $\varepsilon_2$ are exchanged
as $\psi\to-\psi$.
Let us look at the coefficients of ${\rm Im}(\varepsilon_1+\varepsilon_2)$
and ${\rm Re}(\varepsilon_1-\varepsilon_2)$
in the first line of (\ref{Aondef1}).
The former, $7\log(\cos{\pi\over 7}/\cos\psi)$,
vanishes at $\psi=\pm{\pi\over 7}$
while the latter, $(7\psi-3\pi)$,
is negative at either of $\psi=\pm{\pi\over 7}$. Therefore,
in order for (\ref{Aondef1}) to be positive,
${\rm Re}(\varepsilon_1-\varepsilon_2)$ must be negative for both
$\psi={\pi\over 7}$ and $-{\pi\over 7}$,
but this is in conflict with the Weyl invariance.
We conclude that any admissible deformation of
$\overline{\bgamma}_0$ cannot be Weyl invariant,
at least when there is a regime where the deviation
$|\varepsilon|$ from $\overline{\bgamma}_0$ is of the same order
as $|\sDelta t|$.

Let us next consider $\C(0)$. The integral on $\overline{\bgamma}_0$
is convergent along ${\mathcal P}$ except for the small semicircle to the
left of $t_3$, on which we shall focus.
As $\sDelta\theta<0$ there,
$\AN_{\wtq_{0,0}}$ is positive if $\Rs>0$ but becomes negative if $\Rs<0$.
Thus, the problem is present only for the part $\Rs<0$. If we write
\beq
\usigma_1^{(0)}=-\xi\e^{\im \psi},\quad
\usigma_2^{(0)}=-\xi\e^{-\im\psi};\quad\,\,\xi>0,~
-{\pi\over 2}<\psi<{\pi\over 2},
\label{paraneg}
\eeq
we have for $t=t_3+\sDelta t$
\beqa
\partial_{\usigma_1}\wt{W}_{\!{\it eff},\wtq_{0,0}}^{(0)}
&=&-t+2\pi\im\wtq_{0,0}^1-7\log(\im\usigma_1^{(0)})
+7\log(-\im(\usigma_1^{(0)}+\usigma_2^{(0)}))\nn\\
&=&-7\log\left({\cos{3\pi\over 7}\over\cos\psi}\right)
-\sDelta\zeta+\im(\pi-7\psi+\sDelta\theta),\\
\partial_{\usigma_2}\wt{W}_{\!{\it eff},\wtq_{0,0}}^{(0)}
&=&-t+2\pi\im\wtq_{0,0}^2-7\log(\im\usigma_2^{(0)})
+7\log(-\im(\usigma_1^{(0)}+\usigma_2^{(0)}))\nn\\
&=&-7\log\left({\cos{3\pi\over 7}\over\cos\psi}\right)
-\sDelta\zeta+\im(-\pi+7\psi+\sDelta\theta),
\eeqa
and hence
\beqa
\AN_{\wtq_{0,0}}|_{\usigma^{(0)}+\varepsilon}
&=&-2\sDelta\theta\,\Rs
+{\rm Im}(\varepsilon_1+\varepsilon_2)
\cdot 7\log\left({\cos{3\pi\over 7}\over\cos\psi}\right)
+{\rm Re}(\varepsilon_1-\varepsilon_2)(7\psi-\pi)\nn\\
&&+{\rm Im}(\varepsilon_1+\varepsilon_2)\sDelta\zeta
-{\rm Re}(\varepsilon_1+\varepsilon_2)\sDelta \theta
+O(|\varepsilon|^2).\label{Aondef2}
\eeqa
An example of $(\varepsilon_1,\varepsilon_2)$ where this is positive is
\beq
\varepsilon_1=\sDelta\theta\,\Rs(\psi+\im (7|\psi|-3\pi)),\quad
\varepsilon_2=\sDelta\theta\,\Rs(-\psi+\im (7|\psi|-3\pi)).
\label{sol2}
\eeq
This is Weyl invariant:
$\varepsilon_1$ and $\varepsilon_2$ are exchanged as $\psi\to-\psi$.
The proof of Weyl non-invariance does not apply here:
At the two points $\psi=\pm{3\pi\over 7}$
at which $7\log(\cos{3\pi\over 7}/\cos\psi)$ vanishes,
$(7\psi-\pi)$ has the opposite signs.
Thus ${\rm Re}(\varepsilon_1-\varepsilon_2)$ must have opposite signs for
$\psi={3\pi\over 7}$ and $-{3\pi\over 7}$, which is consistent
with Weyl invariance.

The above discussion is limited to deformations of
$\overline{\bgamma}_0$ along the path ${\mathcal P}$
with a region where the deviation from $\overline{\bgamma}_0$
is of the same order as $|\sDelta t|$.
We would like to examine the Weyl invariance for more general contours,
and also along more general paths in the window {\bf C}.

For the representation $\C(0)$, we know from the numerical analysis that
there is a family of Weyl invariant admissible contours.
The key is the fact that there is
a region of the FI-theta parameter space where both
$\bgamma_0$ and $\bgamma_-$ are valid:
the former and the latter are valid respectively for $\theta>-7\pi$ and 
$\zeta<\zeta_1$ and the two regions overlap.
Any path in the window {\bf C} goes through the ovelap and we
may try to connect the two contours for that part of the path.
It is easy to observe in the numerical
analysis that one can do the interpolation while
preserving the Weyl invariance. With a little more effort, one may
also be able to do so analytically.

For the representation $\Sym^2\fund(6)$,
the above computation does not exclude the possibility of 
Weyl invariant and convergent contours 
which are very different from $\overline{\bgamma}_0$.
We would like to know if that is possible.
Let us first make a general remark.
Suppose a contour $\bgamma$ is a graph of a function
$\Rs\mapsto \Is(\Rs)$. Then, its Weyl transform is the graph of
$\Rs\mapsto \Is'(\Rs')$ where the prime $'$ stands for the exchange of the
two coordinates, and the Weyl invariance of
$\bgamma$ means that $\Is'(\Rs')=\Is(\Rs)$. In particular,
the function sends the diagonal $\Rs_1=\Rs_2$ to the diagonal $\Is_1=\Is_2$.
On the diagonal $\usigma_1=\usigma_2=:\usigma$, the growth function simplies as
\beqa
\AN_{\wtq_{h,i}}|_{\rm diagonal}
&=&2(\zeta-7\log 2)\Is+2\left\{7\pi|\Rs|-(\theta+\pi(h+2i))\Rs\right\}\nn\\
&=&2(\zeta-7\log 2)\Is+2\times\left\{\begin{array}{ll}
(\theta^+_{h,i}-\theta)|\Rs|&\Rs>0\\
(\theta-\theta^-_{h,i})|\Rs|&\Rs<0.
\end{array}\right.
\label{ANdiag}
\eeqa
where $\theta_{h,i}^{\pm}$ are given by (\ref{thetapmhi}).
This must be non-negative
if $\bgamma$ is admissible for $\Sym^h\fund(i)$.
The hemisphere integral has a factor $(\usigma_1-\usigma_2)$
which vanishes on the diagonal.
Also, the measure $\dd\usigma_1\wedge\dd\usigma_2$ also vanishes
on the diagonal.
However, if $\AN_{\wtq_{h,i}}$ is negative
on a digonal locus of $\bgamma$, it remains negative
in its neighborhood, putting the convergence at risk.
In particular, if $\zeta=7\log 2$, $\theta$ must be in the interval
$\theta^-_{h,i}\leq \theta\leq \theta^+_{h,i}$.
Thus, if there is a family of Weyl invariant admissible
contours along a path between the two phases, the path must
cross the $\zeta=7\log 2$ line only through this interval.
For the representation $\Sym^2\fund(6)$,
the interval is $-21\pi\leq \theta\leq -7\pi$.
For example,
the path ${\mathcal P}_{\bf C}$ as drawn in Fig.~\ref{fig:RodpathsX}
does not qualify.
(It qualifies for the representation $\C$
for which the interval is $-7\pi\leq \theta\leq 7\pi$.)
Of course, one may bend the path so that it goes through the interval,
avoiding the half-line $\{\zeta=7\log 2,\,\theta>-7\pi\}$,
just as the path ${\mathcal P}$ in Fig.~\ref{fig:thepathP} barely does.
However, any path in the window {\bf C} must go through the open segment
$\{\zeta=\zeta_1,\,-7\pi<\theta<-5\pi\}=:S_1$ stretched between
$t_1$ and its $2\pi$ shift to the right.
As $\zeta_1<7\log 2$,
$\Is$ must be negative on the diagonal locus with $\Rs>0$
in order for (\ref{ANdiag}) to be non-negative on $S_1$.
Using this as a key constraint, one can show
by numerical analysis that there is no Weyl invariant and admissible
contour anywehere on $S_1$,
as described in Appendix~\ref{subapp:Weylinv}.
This proves that for any path in the window {\bf C},
there cannot exist a family of Weyl invariant contours
on which $\AN_{\wtq_{2,6}}$
is non-genative all the way on the path.

By the reflection symmetry (\ref{reflectionsym}) at $-7\pi$,
what is discussed above applies equally well to 
 the representations $\Sym^2\fund(-1)$ and $\C(7)$
 concerning transport through the window {\bf B}:
 A family of admissible contours can be Weyl invariant for $\C(7)$
 but not for $\Sym^2\fund(-1)$.

\noindent
{\bf Remark.}
The Weyl invariance condition has no effect on
the exclusion of the representations
on the forbidden sectors or the no-entry zones, since they are
excluded already without requiring Weyl invariance.

\subsection{Revision Of The Grade Restriction Criterion}
\label{subsec:revision}

Let us discuss which representations are grade restricted with respect to
the window ${\bf C}$.
The same can be done for the window ${\bf B}$.
We already know that those in the overlap of
$\rmGRR_{\bf A}=$ (\ref{oGRR1}) and $\rmGRR_{\bf D}=$ (\ref{oGRR4})
must be included and that
the only additional possibilities are $\C(0)$ and $\Sym^2\fund(6)$:
any path in {\bf C} must go through the no-entry zones of other
representations.
$\C(0)$ has a continuous family of contours obeying all
the required properties. Thus, it must be included.
What about $\Symt(6)$?
There is a family of convergent contours but Weyl invariance cannot be achieved.
Thus, we must exclude it from the set according to our own criterion.
However, if we did so, the set GRR${}_{\bf C}$ would be a proper subset of
the set $\RADS$ (\ref{ADSset}) corresponding to $\TADS$ of \cite{ADS}
and the picture (\ref{maindiagram}) would fail for the window ${\bf C}$.
If we include it instead, GRR${}_{\bf C}$ is exactly the same as
the set $\RADS$ and the picture (\ref{maindiagram}) may hold.
This suggests that we need to include $\Symt(6)$.

This in turn means that we need to change our criterion for representations
to be included into the grade restricted subset:
we need to drop the Weyl invariance condition on the contour $\bgamma$.
Other than the current problem, there is no instance where
Weyl invariance is the reason for exclusion from the grade restricted sets.
Thus, dropping the Weyl invariance condition would not create a new problem.
At this point, we recall that we have also ignored the other condition
on the contour $\bgamma$, namely that it must be a Lagrangian submanifold of
$\ttt_{\C}$.
We have not been able to find a family of Lagrangian submanifolds
for grade restricted representations with respect to the windows
${\bf A}$ and ${\bf D}$, and we completely ignored the Lagrangian condition
in the discussion afterwards.
This suggests that we may also drop the Lagrangian condition in the criterion.

Let us take the new criterion where we drop both the Weyl invariance and
Lagrangian conditions on the contour $\bgamma$.
Then, both $\C$ and $\Sym^2\fund(6)$ are included in $\rmGRR_{\bf C}$
and both $\C(7)$ and $\Sym^2\fund(-1)$ are included in $\rmGRR_{\bf B}$.
Together with $\rmGRR_{\bf A}$ and $\rmGRR_{\bf D}$
which had already been established, let us write down
the grade restricted subsets of the four windows in one place:
\beqa
\!\!\!\!\!\!\!\!\!\!\!\!
{\rm GRR}_{\bf A}&&
\begin{array}{cccccccc}
&&\C(1)&\cdots&\C(5)&\C(6)&\C(7)\\
&\fund&\fund(1)&\cdots&\fund(5)&\fund(6)&\\
  ~~~~~~~~~~&\symt&\symt(1)&\cdots&\symt(5)&\symt(6)&
\end{array}
\label{GRRA}\\[0.2cm]
\!\!\!\!\!\!\!\!\!\!\!\!
{\rm GRR}_{\bf B}&&
\begin{array}{ccccccc}
&&\C(1)&\cdots&\C(5)&\C(6)&\C(7)\\
&\fund&\fund(1)&\cdots&\fund(5)&\fund(6)&\\
\symt(-1)&\symt&\symt(1)&\cdots&\symt(5)&&
\end{array}
\label{GRRB}\\[0.2cm]
\!\!\!\!\!\!\!\!\!\!\!\!
{\rm GRR}_{\bf C}&&
\begin{array}{ccccccc}
&\C&\C(1)&\cdots&\C(5)&\C(6)&\\
&\fund&\fund(1)&\cdots&\fund(5)&\fund(6)&\\
~~~~~~~~~~&\symt&\symt(1)&\cdots&\symt(5)&\symt(6)&
\end{array}
\label{GRRC}\\[0.2cm]
\!\!\!\!\!\!\!\!\!\!\!\!
{\rm GRR}_{\bf D}&&
\begin{array}{cccccccc}
&\C&\C(1)&\cdots&\C(5)&\C(6)&\\
&\fund&\fund(1)&\cdots&\fund(5)&\fund(6)&\\
\symt(-1)&\symt&\symt(1)&\cdots&\symt(5)&&
\end{array}
\label{GRRD}
\eeqa
The sets for the windows with $\theta$ shifted by $2\pi\Z$
are obtained from these via
\beq
    {\rm GRR}_{{\bf w}(-i)}={\rm GRR}_{\bf w}(i).
    \label{shiftGRR}
\eeq
The rule is compatible with
the reflection symmetry (\ref{reflectionsym}) at $-7\pi$ which exchanges
the windows as ${\bf A}\leftrightarrow {\bf D}$ and
${\bf B}\leftrightarrow {\bf C}$
and the representations as
$\Sym^h\fund(i)\leftrightarrow \Sym^h\fund(7-h-i)$.

Dropping the two conditions makes perfect sense in one respect:
analytic continuation of the hemisphere integral
along the paths between the phases.
At both ends of the paths, $\zeta\gg 0$ and $\zeta\ll 0$,
the two expressions of the integral, (\ref{ZVRoddef}) and (\ref{ZVRod}),
give the same results, since $\bgamma_+$ and $\bgamma_-$ are Weyl invariant.
The results in $\zeta\gg 0$ and in $\zeta\ll 0$
are analytic continuations of each other
as long as there is a family of contours $\bgamma$, interpolating
between $\bgamma_+$ and $\bgamma_-$,
on which the second integral (\ref{ZVRod}) is convergent all the way.
For that, $\bgamma$'s over the middle of the paths
do not have to be Weyl invariant or Lagrangian.

However, the contour $\bgamma$ was originally regarded as the datum to specify
the boundary condition of the vector multiplet scalar as in (\ref{BCbgamma}).
Then, it does need to be Weyl invariant and Lagrangian.
The need to drop the conditions suggests that
we should abandon the idea that the contour $\bgamma$
coincides with the datum to specify the boundary condition.
Then, what would be the boundary condition of the vector multiplet?
The only natural choice is the original condition: the
supersymmetric completion of the Neumann boundary condition
for the vector potential, which requires the scalar component to
take boundary values in the real locus (\ref{NeumannBC}).

Then, we are forced to revise the proposal as follows:
Consider the boundary interaction specified by a brane datum $\brane=(\bV,Q)$
and the supersymmetric completion of the Neumann boundary condition 
both on the matter scalar fields and the vector potential.
This defines a well-behaved D-brane in the quantum theory when,
for each irreducible component of $\bV$, 
the real locus $\im\ttt$ can be deformed,
without hitting the poles of $\Gamma_{\rm R\o dland}(\usigma)$,
to a contour $\bgamma\subset \ttt_{\C}$ on which the hemisphere integral
of the second form (\ref{ZVRod}) is convergent.
Here we do not require $\bgamma$ to be Weyl invariant or Lagrangian.

The new proposal frees us from several problems in the original
proposal of \cite{HHP} and \cite{HR}
in which $\bgamma$ is regarded as the datum to specify
the boundary condition.
The first is about the boundary condition on the full vector multiplet.
In both \cite{HHP} and \cite{HR}, it is assumed that
the boundary condition (\ref{BCbgamma}) associated to $\bgamma$
on the scalar component
can be completed to a boundary condition on the full vector mutiplet.
However, as discussed in Appendix~\ref{app:SUSY},
we do not know how to complete it to a full boundary condition,
especially the condition on the gauge potential.
In the new proposal, there is no need to look for such a condition.
We take the supersymmetric completetion of the Neumann boundary condition,
(\ref{vctBC}) and (\ref{chiBC}), where everything is spelled out explicitly.
Second, the convergent contour could be
different for different irreducible representations.
For example, for the window {\bf C}, the families of convergent contours
for $\C$ and for $\Sym^2\fund(6)$ found in Section~\ref{subsec:Weyl}
are different from each other
and also from $\bgamma_0$ that works for the other 19 representations
in $\rmGRR_{\bf C}$.
In the original proposal, the contour $\bgamma$ determines the boundary
condition and hence must be common to
all the irreducible components of the Chan-Paton representation
$(\bV,\rho)$ of $\brane$.
This would pose the problem of finding a universal
contour that works for all the members of the grade restricted set.
In the new proposal, we do not have to worry about such a problem.
The third is about the Lagrangian condition:
With respect to which symplectic structure of $\ttt_{\C}$
should $\bgamma$ be Lagrangian?
Of course the gauge coupling constant defines a symplectic structure
but that is modified by quantum corrections.
Should the classical gauge coupling be used or the effective
one after the matter fields are integrated out?
In the new proposal, the ultra-violet boundary condition is
associated to the real locus $\im\ttt$
which is a Lagrangian submanifold of $\ttt_{\C}$
with respect to the ultra-violet symplectic structure for any choice
of the gauge coupling constant.
All these problems disappear in the new proposal.

In the following three sections,
we shall examine whether the picture (\ref{maindiagram}) holds
with the grade restriction rule (\ref{GRRA})-(\ref{GRRD}) and
whether reasonable monodromies around singular points come out.
We shall take the limit $\epsilon\searrow 0$ of the R-charge parameter,
so that the R-charges of the $p$'s and the $x$'s are
\beq
R_p=2\quad\mbox{and}\quad R_x=0.
\label{Grframe}
\eeq
The bound (\ref{Rrange}) is violated but that is saved by
a small shift of contours \cite{HR}: $\bgamma$ is a deformation of
the shifted real locus
\beq
\im\ttt -\im\delta (1,1),\quad 0<\delta\ll 1.
\label{shiftcontour}
\eeq
The expression for the hemisphere partition function is
\beqa
Z_{D^2}(\mathfrak{B})
&=&\int\limits_{\,\,\,\,\,\mbox{$\gamma$}}
\dd\usigma_1\,\dd\usigma_2\,
(\usigma_1-\usigma_2)\sinh(\pi(\usigma_1-\usigma_2))
\times\nn\\
&&~~~~~~\Gamma(-\im(\usigma_1+\usigma_2)+1)^7\,
\Gamma(\im\usigma_1)^7\,
\Gamma(\im\usigma_2)^7\,
\e^{\im t(\usigma_1+\usigma_2)}
f_{\mathfrak{B}}(\usigma_1,\usigma_2).~~~~~~~~
\label{Zthemodel}
\eeqa
The integrand has poles at
\beqa
\usigma_1&=&0,\, \im,\, 2{}_{\,}\im,\ldots,\label{pole1}\\
\usigma_2&=&0,\,\im,\, 2{}_{\,}\im,\ldots,\label{pole2}\\
\usigma_1+\usigma_2&=&-\im, -2{}_{\,}\im, -3{}_{\,}\im,\ldots.\label{poleP}
\eeqa

\section{Grassmannian Phase}
\label{sec:Grassmannian}

In this section, we describe the boundary renormalization group flow
in the Grassmannian phase $\zeta\gg 0$. Since the phase is weakly coupled,
the low energy limit of a gauge theory brane is straightforward to find.
All the necessary tools have been developed in \cite{HHP}
for Abelian theories and can be extended to non-Abelian theories
with little effort \cite{HR}.
We will compute the hemisphere partition function
and check the results against expression (\ref{Zgeometric})
for the central charge of the expected low energy brane
in the non-linear sigma model with target $X$.

The main goal is to describe the process to replace a given brane by
another one without changing its low energy behaviour.
This is done by binding a brane that becomes empty in the infra-red limit.
Before doing so, we will make a rather long digression to
introduce necessary mathematical tools, which will also
give us a technique to lift low energy branes in $X$ to
high energy branes in gauge theory.

\subsection{The Flow $\pi_X:D_G(V,W)\to D(X)$}
\label{subsec:Gflow}

If the Higgsing scale $e\sqrt{\zeta}$ is much bigger than the energy scale
of the superpotential,
\beq
W=\sum_{k=1}^7p^kA_k(x),
\label{theWGr}
\eeq
the renormalization group flow can be separated
into two steps. The theory at an intermediate energy scale
has the total space $\fb$ of a vector bundle
over the Grassmannian $\Gr=G(2,7)$ as the target space,
where $p=(p^k)$ and $x=(x^a_i)$ parametrize the fibers and the base
respectively, 
and the superpotential $W_{\rm ind}$ induced from the original one
(\ref{theWGr}). The superpotential $W_{\rm ind}$ is a Morse-Bott function
with the critical locus
$p^k=A_k(x)=0$ which is nothing but the Calabi-Yau manifold $X$.
Integrating out the pairs of massive modes $(p^k,A_k(x))$,
we end up with the sigma model with target $X$.

Let us introduce some notation.
A representation $\rmV$ of the gauge group $G=U(2)$ determines
a holomorphic vector bundle over the space $\mathfrak{X}$ involved, such as
$\mathfrak{X}=\fb, \Gr, X$.
We denote it by ${\mathcal O}_{\mathfrak{X}}(\rmV)$.
For the representation $\C(i)=(\det\fund)^{\otimes i}$, we denote
${\mathcal O}_{\mathfrak{X}}(\C(i))$ simply by ${\mathcal O}_{\mathfrak{X}}(i)$.
For a vector bundle ${\mathcal V}$ on $\mathfrak{X}$,
we write ${\mathcal V}(i):={\mathcal V}\otimes {\mathcal O}_{\mathfrak{X}}(i)$.
In this notation, $\fb$ is the total space
of ${\mathcal O}_{\Gr}(-1)^{\oplus 7}$.
Note that ${\mathcal O}_{\Gr}(\fund^*)$ is the tautological subbundle
$S$ on $\Gr$ as $(x^a_i)$ defines an embedding $\fund^*\hookrightarrow \C^7$,
and $\det S={\mathcal O}_{\Gr}(-1)$.
We denote by $D(\,\fb,W_{\rm ind})$ the set of B-branes in the
theory at the intermediate energy scale.

We now describe the flow of a gauge theory brane  $\brane=(\bV,Q)$.
Under the R-charge assignment (\ref{Grframe}), we may set $J=1_G$
and charge integrality (\ref{CI}) requires that $\e^{\pi\im {\bf r}_*}$
is the $\Z_2$-grading of $\bV$ itself. 
That is, ${\bf r}_*$ determines a $\Z$-grading of $\bV$
that reduces modulo 2 to the original $\Z_2$-grading.
In particular,
$\bV=(\bV,\rho,{\bf r}_*)$ is a $\Z$-graded representation of the gauge group.
We shall also denote by ${\mathcal O}_{\mathfrak{X}}(\bV)$
the associated $\Z$-graded vector bundle on $\mathfrak{X}$.
The flow is separated into two steps.
The first step corresponds to the Higgsing
\beq
D_G(V,W)\longrightarrow D(\,\fb,W_{\rm ind}),
\label{Higgsingflow}
\eeq
and is straightforward to describe.
The brane $\brane=(\bV,Q)$ simply reduces to
the brane supporting the $\Z$-graded vector bundle
${\mathcal O}_{\fb}(\bV)$ on $\fb$
with the boundary interaction determined by $Q_{\rm ind}$
induced from $Q$. It is a matrix factorization 
of $W_{\rm ind}$ which we denote by
${\mathcal O}_{\fb}(\brane)=({\mathcal O}_{\fb}(\bV),Q_{\rm ind})$.
The second step of the flow corresponds to integrating out
the pairs of modes with superpotential masses
\beq
D(\,\fb,W_{\rm ind})\longrightarrow D(X).
\label{Knorrerflow}
\eeq
To describe this,
we shall use the results obtained in \cite{HHP}, Sections 3.4 and 10.3.1,
by fiberwise application of the Kn\"orrer periodicity \cite{Knorrer}.
As a guide, let us introduce a diagram of spaces
\beq
\begin{tikzcd}
  &\fb\arrow[dr,"q^{}_{\Gr}"]\\
  \fb|_X\arrow[dr,"q^{}_X"]\arrow[ur, hook,"\hat{i}^{}_X"]&&\Gr\\
  &X\arrow[ur, hook, "i^{}_X"]
\end{tikzcd}
\label{Knorrer}
\eeq
$\fb|_X$ is the restriction of $\fb$ on $X$,
$q_{\Gr}$ and $q_X$ are projections, $i_X$ is the inclusion of $X$ into $\Gr$,
and $\hat{i}_X$ is its lift to total spaces of the vector bundles.
Since $W_{\rm ind}$ vanishes on $\fb|_X$,
$Q_{\rm ind}$ squares to zero there. That is,
the restriction ${\hat{i}}^{\,*}_X{\mathcal O}_{\fb}(\brane)
=({\mathcal O}_{\fb|_X}(\bV),{\hat{i}}^{\,*}_XQ_{\rm ind})$ is
a $\Z_2$-graded complex of vector bundles on $\fb|_X$,
with an action of $U(1)$ R-symmetry that reduces modulo 2 to the $\Z_2$-grading.
We may regard it as a complex of vector bundles on $X$ in which the $p^k$'s
are regarded as extra free variables.
To understand this, it is helpful to take the sheaf theoretical view point.
For an open subset $U\subset X$, the sections of
${\mathcal O}_{\fb|_X}(\bV)$ over $q_X^{-1}(U)\subset \fb|_X$
form a (free) module
over the ring of functions on $q_X^{-1}(U)$.
When regarded as a module over the ring
of functions on $U$,
it is free by construction of ${\mathcal O}_{\fb|_X}(\bV)$
but is of infinite rank because infinitely many powers of $p^k$'s are all
independent.
However, as the $p^k$'s carry R-charge 2 and the $x^a_i$'s zero (\ref{Grframe}),
we can decompose it into a sum of finite rank components of definite R-charges,
although there are infinitely many R-charges with non-zero components.
This corresponds to the $\Z$-graded vector bundle
\beq
{\mathcal O}_X(\bV)
\otimes {\rm Sym}\left({\mathcal O}_X(1)^{\oplus 7}[-2]\right)
\label{MX}
\eeq
on $X$.
${\hat{i}}^{\,*}_XQ_{\rm ind}$ then induces a differential of degree $1$.
In this way, we have a $\Z$-graded complex of vector bundles on $X$.
The sheaf theoretic notation for this would be
$q_{X*}{\hat{i}}^{\,*}_X{\mathcal O}_{\fb}(\brane)$.
We simply denote it by $\pi_X(\brane)$ for brevity.
The complex is unbounded to the right but is exact beyond a
certain degree. Thus, it has a truncation to a bounded complex of
coherent sheaves. (In fact, a bounded complex of vector bundles.
See Appendix~\ref{subsec:locfree}.)
As argued in \cite{HHP},
this is the low energy limit of the gauge theory brane $\brane$.
To be precise, it should be regarded
as the brane in the sigma model with the B-field parameter
\beq
\theta_B=\theta+7\pi.
\label{Bvstheta}
\eeq
The parameter $\theta_B$ is uniquely defined by the property that 
the complexified K\"ahler class depends holomorphically on
$\zeta-\im \theta_B$ with
$[\omega-\im B]=(\zeta-\im \theta_B+O(\e^{-\zeta+\im\theta_B}))H$ at large
$\zeta$.
Another fine point is that the choice of overall R-charge given in (\ref{MX})
is a matter of convention and we have made one choice
which we shall use throughout. One obvious but important property
of $\pi_X$ is
\beq
\pi_X(\brane(i))\,\,=\,\,
\pi_X(\brane)(i).
\label{shiftGr}
\eeq
Note that only the left part of the diagram (\ref{Knorrer})
is used in this description. The right part will be used momentarily
when we describe an uplift from $D(X)$ to $D(\,\fb,W_{\rm ind})$
and to $D_G(V,W)$.

\subsection*{Examples}

Let us provide some examples of branes in the gauge theory
and describe their low energy limits in the Grassmannian phase.

\noindent
\underline{The $p$-Koszul brane $\Kos$}

We introduce the following brane which will play an important r\^ole.
The matrix factorization is
\beq
Q=\sum_{k=1}^7p^k\eta_k+\sum_{k=1}^7A_k(x)\bareta^k
\label{QKp}
\eeq
realized on the irreducible representation $\bV$
of the Clifford algebra 
\beq
\{\eta_k,\bareta^l\}=\delta_k^l,~\{\eta_k,\eta_l\}=\{\bareta^k,\bareta^l\}=0,
\quad 1\leq k,l\leq 7.
\label{Cliff7}
\eeq
For gauge invariance (\ref{Qgi}) and vector $U(1)$ R-symmetry (\ref{QRs}),
the $\eta_k$'s and the $\bareta^k$'s
are given the gauge and R-charges $(\det^{-1}\fund,-1)$ and $(\det\fund,+1)$
respectively.
A hermitian inner product on $\bV$ with hermiticity
$\eta_i^{\dag}=\bareta^i$ preserves this symmetry.
We assign the trivial gauge and R-charges to the
Clifford vacuum $|0\rangle\in \bV$ annihilated by the $\eta_k$'s.
This matrix factorization may be described by the diagram
\beq
\begin{tikzcd}
  \C\arrow[r,yshift=0.5ex, "A(x)"] &
  \arrow[l,yshift=-0.5ex, "p"]\C(1)^{\oplus 7}\arrow[r,yshift=0.5ex, "A(x)"]
  & \arrow[l,yshift=-0.5ex, "p"] \C(2)^{\oplus {7\choose 2}}
  \arrow[r,yshift=0.5ex, "A(x)"]
  & \arrow[l,yshift=-0.5ex, "p"]
  ~~\cdots~~\arrow[r,yshift=0.5ex, "A(x)"]
  &\arrow[l,yshift=-0.5ex, "p"] \C(6)^{\oplus 7}
  \arrow[r,yshift=0.5ex, "A(x)"]
  &\arrow[l,yshift=-0.5ex, "p"] \C(7)\\[-0.7cm]
  {}_0&{}_1&{}_2&&{}_6&{}_7
\end{tikzcd}
\label{Kp}
\eeq
The $i$-th term from the left is spanned by $i$-th powers of
the $\bareta^k$'s on $|0\rangle$.
The left arrow $p$ and the right arrow $A(x)$ are
respectively the first and the second terms of (\ref{QKp}).
This is a matrix factorization completion of the Koszul complex for
the $p$-variables. For this reason, we shall call it the
{\it $p$-Koszul brane} and denote it by $\Kos$.

\noindent
{\bf Remark on the notation and terminology}:
The diagram (\ref{Kp}) would make technical sense
as a diagram of gauge equivariant modules of the polynomial ring $\C[p,x]$
of the variables $p=(p^k)$ and $x=(x^a_i)$,
if each component
$\C(i)^{\oplus {7\choose i}}$ were tensored with $\C[p,x]$.
Indeed, this is what was done in \cite{HHP}.
In the present paper, to save the space, we simply omit $\otimes \C[p,x]$
as long as there is no danger of confusion.
We shall also refer to (\ref{Kp}) as {\it a diagram of
representations over $\C[p,x]$ of the gauge group $G$ (or of $G_{\C}$)}.
The numbers below the components are their R-charges. We shall write them
when we would like to emphasize them, as in the present case.
Sometimes they will be written as subscripts.
We omit them when the R-charge is clear from the context, or when
we decide not to specify it.

Let us describe the low energy limit $\pi_X(\Kos)$ of the $p$-Koszul brane.
For this purpose, we note that $\Kos$ is the tensor product
$\otimes_{k=1}^7\Kos_k$
where $\Kos_k$ is the matrix factorization 
\beq
\begin{tikzcd}
 \C\arrow[r,yshift=0.5ex, "A_k(x)"] &
  \arrow[l,yshift=-0.5ex, "p_k"]\C(1)\\[-0.7cm]
  {}_0&{}_1
\end{tikzcd}
\eeq
of $p^kA_k(x)$.
Then,
\beq
\pi_X(\Kos)=\pi_X\left(\bigotimes_{k=1}^7\Kos_k\right)
=\bigotimes_{k=1}^7\pi_X^{(k)}(\Kos_k),
\eeq
in which $\pi_X^{(k)}$ is the same operation as $\pi_X$ but where
$\fb$ in (\ref{Knorrer}) is replaced by the total space
of a single ${\mathcal O}_{\Gr}(-1)$ over $\Gr$ corresponding to
$p^k$. What is $\pi_X^{(k)}(\Kos_k)$?
Note that
$(p^k)^s\C$ and $(p^k)^s\C(1)$ descend to ${\mathcal O}_X(s)$ at R-charge $2s$
and ${\mathcal O}_X(s+1)$ at R-charge $2s+1$, respectively.
The right arrow $A_k(x)$ descends to zero on $X$ while the left arrow
$$
\begin{array}{ccccccc}
  (p^k)^{s+1}\C&\!\!\stackrel{\,\,p^k\!\!}{\longleftarrow}\!\!&(p^k)^s\C(1)
&\mbox{descends to}&
 {\mathcal O}_X(s+1)&\!\!\stackrel{\,{\rm id}\!}{\longleftarrow}\!\!&
  {\mathcal O}_X(s+1).\\
 {}_{2(s+1)}&&{}_{2s+1}&& {}_{2(s+1)}&&{}_{2s+1}
\end{array}
$$
Aligning the components in the order of increasing R-charges,
we find that $\pi_X^{(k)}(\Kos_k)$ is
\beq
\begin{array}{ccccccccccc}
  {\mathcal O}_X&\!\!\!\stackrel{0}{\longrightarrow}\!\!\!&
  {\mathcal O}_X(1)&\!\!\!\stackrel{{\rm id}}{\longrightarrow}\!\!\!&
  {\mathcal O}_X(1)&\!\!\!\stackrel{0}{\longrightarrow}\!\!\!&
  {\mathcal O}_X(2)&\!\!\!\stackrel{{\rm id}}{\longrightarrow}\!\!\!&
  {\mathcal O}_X(2)&\!\!\!\stackrel{0}{\longrightarrow}&\cdots\\
  {}_0&&{}_1&&{}_2&&{}_3&&{}_4&
\end{array}
\eeq
and hence $\pi_X(\Kos)$ is
\beq
\left(\begin{array}{ccccccccccc}
  {\mathcal O}_X&\!\!\!\stackrel{0}{\longrightarrow}\!\!\!&
  {\mathcal O}_X(1)&\!\!\!\stackrel{{\rm id}}{\longrightarrow}\!\!\!&
  {\mathcal O}_X(1)&\!\!\!\stackrel{0}{\longrightarrow}\!\!\!&
  {\mathcal O}_X(2)&\!\!\!\stackrel{{\rm id}}{\longrightarrow}\!\!\!&
  {\mathcal O}_X(2)&\!\!\!\stackrel{0}{\longrightarrow}&\cdots
  \end{array}
\right)^{\otimes 7}.
\eeq
Note that
${\mathcal O}_X(i)\stackrel{{\rm id}}{\longrightarrow}{\mathcal O}_X(i)$
are all exact. Since the tensor product of an exact complex with any
complex is exact, it is the direct sum of
${\mathcal O}_X^{\otimes 7}\cong {\mathcal O}_X$ at degree zero and
(infinitely many) exact complexes.
As the exact complexes are empty in the infra-red limit, we find
\beq
\pi_X(\Kos)~\cong~{\mathcal O}_X.
\label{KostoO}
\eeq
That is, $\Kos$ flows to a single D6-brane wrapped
on $X$ and supporting a trivial line bundle.

We could have guessed this result by looking at the boundary potential,
\beq
\{Q,Q^{\dag}\}=\sum_{k=1}^7|p^k|^2+\sum_{k=1}^7|A_k(x)|^2.
\label{QQdag1}
\eeq
It is zero exactly at $X\subset \fb$.

\noindent
\underline{Variants}

There are several variants of the $p$-Koszul brane $\Kos$.
An obvious one is obtained by taking the tensor product of
$\Kos$ with any representation $\rmV$ of the gauge group.
Its low energy limit is isomorphic to
the vector bundle associated with $\rmV$,
\beq
\pi_X(\Kos\otimes \rmV)\cong {\mathcal O}_X(\rmV).
\eeq
In particular,
$\pi_X(\Kos(i))\cong {\mathcal O}_X(i)$ and 
$\pi_X(\Kos\otimes \fund^*)\cong i_X^*S$ which we denote by $S_X$.
More generally,
\beq
\pi_X(\Kos\otimes \Sym^h \fund^*(i))\cong \Sym^h S_X(i).
\label{KosSymfate}
\eeq

Instead of a representation $\rmV$, we may consider
a complex $\Complex=(M_{\Complex},d_{\Complex})$ of representations over $\C[x]$
of the gauge group.
That is, a $\Z$-graded representation
$(\bV_{\Complex},\rho_{\Complex},{\bf r}_{\Complex *})$
of $G=U(2)$ and a polynomial function $x\mapsto d_{\Complex}(x)$
with values in degree $1$ endomorphisms of $\bV_{\Complex}$
satisfying $d_{\Complex}(x)^2=0$ and the gauge invariance
condition (\ref{Qgi}).
It determines a complex of vector bundles on $X$
which we denote by ${\mathcal O}_X(\Complex)$.
The tensor product $\Kos\otimes\Complex$ defines a gauge theory brane,
and its low energy limit is isomorphic to that complex
\beq
\pi_X(\Kos\otimes\Complex)\cong {\mathcal O}_X(\Complex).
\label{complexcase}
\eeq

As an example, consider the complex
\beq
\begin{tikzcd}
  \C(-1)\arrow[r, "g(x)"]&\C\\[-0.7cm]
  {}_{-1}&{}_0
\end{tikzcd}
\eeq
where $g(x)$ is a linear combination of $[x_ix_j]$.
Let $\Kos(g)$ be the tensor product of $\Kos$ with this complex.
The low energy limit is isomorphic to
the associated complex on $X$,
\beq
\begin{tikzcd}
  {\mathcal O}_X(-1)\arrow[r, "g"]&{\mathcal O}_X\\[-0.7cm]
  {}_{-1}&{}_0
\end{tikzcd}
\eeq
If $g$ and $A_1,\ldots, A_7$ are in general position in $\Gr=G(2,7)$,
it is further isomorphic to the skyscraper sheaf
supported on the divisor $D_g=\{g(x)=0\}\subset X$,
\beq
\pi_X(\Kos(g))\cong {\mathcal O}_{D_g}.
\eeq
That is, the D4-brane wrapped on $D_g$.

This can be generalized to a set $g_1(x),\ldots, g_n(x)$
of linear combinations of $[x_ix_j]$.
We have the Koszul complex for $g=(g_1,\ldots, g_n)$
\beq
\begin{tikzcd}
  \C(-n)\!\!\!\arrow[r, "g", shorten=2mm]
  &\!\!\!\C(-n+1)^{\oplus n}\!\!\!\arrow[r, "g", shorten=2mm]&
  \!\!\!\C(-n+2)^{\oplus {n\choose 2}}\!\!\!\arrow[r, "g", shorten=2mm]
  &\cdots\arrow[r, "g", shorten=2mm]
&\!\!\!\C(-1)^{\oplus r}\!\!\!\arrow[r, "g", shorten=2mm]
  &\!\!\!\C
  \end{tikzcd}
\label{gKoszul}
\eeq
where the rightmost component $\C$ is at degree zero.
Let $\Kos(g_1,\cdots,g_n)$ be the tensor product of $\Kos$ with this complex.
Its low energy limit is isomorphic to
the complex of vector bundles on $X$ associated with
(\ref{gKoszul}). If $g_1,\ldots, g_n$ and $A_1,\ldots, A_7$
are in general position in $\Gr$, the result is further isomorphic to
the skyscraper sheaf supported on the intersection
of the divisors $D_{g_{\alpha}}=\{g_{\alpha}(x)=0\}\subset X$,
\beq
\pi_X(\Kos(g_1,\ldots, g_n))\cong
   {\mathcal O}_{D_{g_1}\cap\cdots\cap D_{g_n}}.
\eeq
That is, the D$(6-2n)$-brane wrapped on $D_{g_1}\cap\cdots\cap D_{g_n}$.
This may also be realized by the matrix factorization
\beq
Q=\sum_{k=1}^7p^k\eta_k+\sum_{k=1}^7A_k(x)\bareta^k+\sum_{\alpha=1}^n
g_{\alpha}(x)\xi^{\alpha}
\eeq
realized on the representation of
the Clifford algebra (\ref{Cliff7}) and
\beq
\{\xi^{\alpha},\barxi_{\beta}\}=\delta^{\alpha}_{\beta},\quad
\{\xi^{\alpha},\xi^{\beta}\}=\{\barxi_{\alpha},\barxi_{\beta}\}=0,
\eeq
with an invariant inner product such that $\eta_i^{\dag}=\bareta^i$ and
$\xi^{\alpha\dag}=\barxi_{\alpha}$.
Again, the result could have been guessed by looking at the boundary potential
\beq
\{Q,Q^{\dag}\}=\sum_{k=1}^7|p^k|^2+\sum_{k=1}^7|A_k(x)|^2
+\sum_{\alpha=1}^n|g_{\alpha}(x)|^2.
\label{QQdag2}
\eeq

\noindent
\underline{A Further Generalization}

The computation leading to $\pi_X(\Kos)\cong {\mathcal O}_X$
can be applied to a more general setup.
Let $I$ be a subset of $\{1,\ldots, 7\}$ and
let
\beq
\Kos_I=\otimes_{k\in I}\Kos_k
\label{KosI}
\eeq
be the Koszul brane
for the components in $I$, which is a matrix factorization of
$\sum_{k\in I}p^kA_k(x)$.
Let $\brane_{I^c}$ be a matrix factorization of $\sum_{k\not\in I}p^kA_k(x)$,
depending only on the variables $(p^k)_{k\not\in I}$ and $(x^a_i)$,
with the usual required properties such as gauge and R-symmetry.
Then, $\Kos_I\otimes \brane_{I^c}$ is a matrix factorization of
the superpotential (\ref{theWGr}) and defines a brane in the gauge theory.
Proceeding in the same way as with $\Kos=\otimes_{k=1}^7\Kos_k$, we find
\beq
\pi_X(\Kos_I\otimes \brane_{I^c})
=\pi_X^{(I^c)}(\brane_{I^c}),
\label{appli}
\eeq
where $\pi_X^{(I^c)}$ is an operation as $\pi_X$ in which $\fb$ in
(\ref{Knorrer}) is replaced by the total space of the sum of
${\mathcal O}_{\Gr}(-1)$ corresponding to $p^k$ for $k\not\in I$.
In the case where $I=\{1,\ldots, 7\}$, it reduces to 
(\ref{complexcase}).

\subsection*{An Uplift From $D(X)$ To $D_G(V,W)$}

It is useful to know how to lift a low energy brane
$\mathcal{E}\in D(X)$ to a brane in the gauge theory.
It can be obtained by steps provided in \cite{HHP}, Section 10.3.2,
for Abelian theories
which are adapted to the submanifold $X$ of the Grassmannian $\Gr=G(2,7)$.
We employ the right part of the diagram (\ref{Knorrer}) in this
construction.

Let ${\mathcal E}$ be a complex of coherent sheaves on $X$.
We first push it forward to a complex $i_{X*}{\mathcal E}$ on $\Gr$.
We next take its resolution 
by homogeneous vector bundles on $\Gr$.
That is, a complex ${\mathcal O}_{\Gr}(\Complex)$
which is isomorphic in $D(\Gr)$ to $i_{X*}{\mathcal E}$ where
$\Complex=(M_{\Complex},d_{\Complex})$ is a complex of representations
over $\C[x]$ of the gauge group.
The third step is to extend the differential
$d_{\Complex}(x)$ to a matrix factorization $Q(p,x)$ of
$\sum_{k=1}^7p^kA_k(x)$.
Existence of a resolution of $i_{X*}{\mathcal E}$ together
with its matrix factorization completion, along with
some related facts, will be explained in Sections~\ref{subsub:IT} and
\ref{subsub:MFcompletion}.
Let us denote such a resolution
of $i_{X*}{\mathcal E}$ by $\widehat{i_{X*}{\mathcal E}}$
and its matrix factorization completion by
$\widehat{i_{X*}{\mathcal E}}^{\rm mf}$.

As argued in \cite{HHP},
this $\widehat{i_{X*}{\mathcal E}}^{\rm mf}$ is a lift of
${\mathcal E}\in D(X)$ to the gauge theory, provided appropriate shifts both in
gauge and R charges are made.
If we want to keep the relationship (\ref{Bvstheta})
between the gauge theory
theta parameter and the low energy B-field, and if we want to
match with the convention (\ref{MX})
of the overall R-charge of the low energy limit, 
a lift is
\beq
    {\mathcal E}
    ~~\rightsquigarrow~~
    \widehat{i_{X*}{\mathcal E}}^{\rm mf\!\!}(7)[-7]=:\brane_{\mathcal E}.
\label{alift}
\eeq

\noindent
\underline{Example}

Consider the structure sheaf ${\mathcal O}_X$ of the Calabi-Yau $X$.
Its pushforward
$i_{X*}{\mathcal O}_X$ to the Grassmannian $\Gr$ has a Koszul resolution
$\widehat{i_{X*}{\mathcal O}_X}$
\beq
\begin{tikzcd}
  {\mathcal O}_{\Gr}(-7)\!\!\arrow[r,shorten=0.5mm, "A(x)"] &
    \!\!{\mathcal O}_{\Gr}(-6)^{\oplus 7}\!\!\arrow[r,shorten=2mm,"A(x)"]&   
    \!\!\cdots\!\!\arrow[r,shorten=2mm,"A(x)"]&
    \!\!\!{\mathcal O}_{\Gr}(-2)^{\oplus {7\choose 2}}\!\!
    \arrow[r,shorten=0.5mm,"A(x)"]&
    \!\!{\mathcal O}_{\Gr}(-1)^{\oplus 7}\!\!
    \arrow[r,shorten=0.5mm,"A(x)"]&
    \!\!{\mathcal O}_{\Gr}\\[-0.8cm]
  {}_{-7}&{}_{-6}&&\!\!{}_{-2}&\!\!{}_{-1}&{}_0
\end{tikzcd}
\label{KosOX}
\eeq
The corresponding complex $\Complex$ of representations over $\C[x]$ of the
gauge group is obtained by replacing the component
at degree $-j$ by $\C(-j)^{\oplus {7\choose j}}$, and its
matrix factorization completion $\widehat{i_{X*}{\mathcal O}_X}^{\!\!\rm mf}$
is given by adding the left arrows depending on $p$,
\beq
\begin{tikzcd}
  \C(-7)\arrow[r,yshift=0.5ex, "A(x)"] &
  \arrow[l,yshift=-0.5ex, "p"]\C(-6)^{\oplus 7}
  \!\!\arrow[r,shorten=1.5mm,yshift=0.5ex, "A(x)"]
  & \!\!\arrow[l,shorten=1.5mm,yshift=-0.5ex,"p"]\,
  \cdots\arrow[r,shorten=1.5mm,yshift=0.5ex,"A(x)"]
  & \!\!\arrow[l,shorten=1.5mm,yshift=-0.5ex, "p"] \C(-5)^{\oplus {7\choose 2}}
  \arrow[r,yshift=0.5ex,"A(x)"]
  &\arrow[l,yshift=-0.5ex,"p"] \C(-1)^{\oplus 7}
  \arrow[r,yshift=0.5ex, "A(x)"]
  &\arrow[l,yshift=-0.5ex, "p"] \C\\[-0.8cm]
  {}_{-7}&{}_{-6}&&{}_{-2}&{}_{-1}&{}_0
\end{tikzcd}
\label{KosOXlift}
\eeq
Its shift
$\widehat{i_{X*}{\mathcal O}_X}^{\!\!\rm mf\!\!}(7)[-7]$ is nothing but (\ref{Kp}), that is,
the $p$-Koszul brane $\Kos$.
Thus, a lift is
\beq
    {\mathcal O}_X~~\rightsquigarrow~~\Kos.
    \label{OXlift}
\eeq
Indeed, we had already seen in (\ref{KostoO})
that $\Kos$ flows to ${\mathcal O}_X$.

\medskip
\noindent
{\bf Remarks.} {\bf 1.} The relationship (\ref{Bvstheta}) between the gauge
theory theta parameter and the B-field of the low energy sigma model,
its modulo $2\pi$ version to be precise, was first
found in \cite{MorrisonPlesser} in the context of closed topological A-model
of Abelian gauged linear sigma models.\\
{\bf 2.} The second step of the flow, (\ref{Knorrerflow}),
is expected to hold as an equivalence of categories.
Indeed, fiberwise Kn\"orrer periodicity as categorical equivalence
is constructed in \cite{Orlov05,Isik,Shipman,Preygel}
for the categories $\Db(\fb,W_{\rm ind})$ and $\Db(Y)$
which are equivalent to $D(\fb,W_{\rm ind})$ and $D(Y)$.
As described in Appendix~\ref{app:Catmath},
$\Db(Y)$ is the derived category of coherent sheaves on $Y$
while $\Db(\fb,W_{\rm ind})$ is the full subcategory of
the derived category of quasicoherent curved dg-sheaves on 
$(\fb,W_{\rm ind})$ consisting of objects isomorphic to matrix factorizations.
The equivalence is given by a functor in the inverse direction
$\kappa:\Db(Y)\to \Db(\fb,W_{\rm ind})$
which can be described using the maps in the diagram (\ref{Knorrer}) as
\beq
\kappa({\mathcal E})=\Bigl(\,\hat{i}^{}_{X*}q_X^*{\mathcal E},\,0\,\Bigr).
\eeq
As $\hat{i}^{}_{X*}q_X^*{\mathcal E}$ is supported inside $\fb|_X$
on which $W_{\rm ind}$ vanishes, $\kappa({\mathcal E})$ is indeed
a curved dg-sheaf on $(\fb,W_{\rm ind})$. As $\hat{i}_X$ and $i_X$ are
closed embeddings, $\hat{i}^{}_{X*}q_X^*{\mathcal E}$ is isomorphic in
$\Db(\fb)$ to $q_{\Gr}^*i^{}_{X*}{\mathcal E}$.
As described above (and as will be explained in
Section~\ref{subsub:IT} and \ref{subsub:MFcompletion}),
it has a resolution $q_{\Gr}^*\widehat{i_{X*}{\mathcal E}}$
which can be completed to a matrix factorization of $W_{\rm ind}$.
Thus, we find that $\kappa({\mathcal E})$ is indeed in $\Db(\fb,W_{\rm ind})$
and its lift to gauge theory is the same as
$\widehat{i_{X*}^{}{\mathcal E}}^{\rm mf}$ described above.

\subsection*{The Charge Of The Low Energy Brane}

We propose a formula for the Chern character of
the low energy limit $\pi_X(\brane)$ of a gauge theory brane
$\brane=(\bV,Q)$:
\beq
(1-\e^H)^7\,\ch\bigl(\pi_X(\brane)\bigr)
\,=\,\ch\bigl({\mathcal O}_{\Gr}(\bV)\bigr)
\quad\mbox{in ${\rm H}^*(\Gr,\Q)$},
\label{chformula}
\eeq
where $H\in {\rm H}^2(\Gr,\Z)$ is the Picard group generator of $\Gr$.
$(1-\e^H)^7$ is the Poincar\'e dual
\beq
    [X]_{{}_{\rm PD}}=H^7\in {\rm H}^{14}(\Gr,\Z)
    \label{PDX}
\eeq
of the class of $X\subset \Gr$ times an invertible element of
${\rm H}^*(\Gr,\Q)$.
Therefore, even though
$\ch(\pi_X(\brane))$ is an element of ${\rm H}^*(X,\Q)$
the left-hand side of (\ref{chformula})
makes sense as an element of ${\rm H}^*(\Gr,\Q)$.
Equivalently, 
$\ch(\pi_X(\brane))$
is a polynomial with rational coefficients
of the generator $H$ of ${\rm H}^2(X,\Z)$
and can be regarded as an element of ${\rm H}^*(\Gr,\Q)$
by regarding $H$ as the generator of ${\rm H}^2(\Gr,\Z)$.
The formula (\ref{chformula}) says that
$\ch({\mathcal O}_{\Gr}(\bV))$ is a multiple of $[X]_{{}_{\rm PD}}$ and
uniquely determines $\ch(\pi_X(\brane))\in {\rm H}^*(X,\Q)$.

A formal argument for (\ref{chformula}) is as follows.
The underlying graded vector bundle for $\pi_X(\brane)$ is (\ref{MX})
before finite truncation and let us formally take its Chern character
\beqa
\ch(\pi_X(\brane))&=&\ch\left({\mathcal O}_X(\bV)
\otimes {\rm Sym}\left({\mathcal O}_X(1)^{\oplus 7}[-2]\right)\right)\nn\\
&=&
\sum_{n_1\geq 0}\cdots\sum_{n_7\geq 0}
\ch\bigl({\mathcal O}_X(\bV)(n_1+\cdots+n_7)\bigr).
\eeqa
When multiplied to $(1-\e^H)^7=[X]_{{}_{\rm PD}}\cdot$(an invetible element),
the ${\mathcal O}_X$ on the right-hand side
can be replaced by ${\mathcal O}_{\Gr}$, and we have
\beqa
(1-\e^H)^7\,\ch\bigl(\pi_X(\brane)\bigr)&=&
(1-\e^H)^7\sum_{n_1\geq 0}\cdots\sum_{n_7\geq 0}
\ch\bigl({\mathcal O}_{\Gr}(\bV)(n_1+\cdots+n_7)\bigr)\nn\\
&=&
\left(\sum_{n_1\geq 0}-\sum_{n_1\geq 1}\right)\cdots
\left(\sum_{n_7\geq 0}-\sum_{n_7\geq 1}\right)
\ch\bigl({\mathcal O}_{\Gr}(\bV)(n_1+\cdots+n_7)\bigr)\nn\\[0.2cm]
&=&\ch({\mathcal O}_{\Gr}(\bV)).
\eeqa
This is only a formal manipulation and
an appropriate regularization has to be made.
Instead of attempting to do so, we shall provide two pieces of evidence
to support the proposal. One, given right below, is the proof
in the case when the brane $\brane$
is the gauge theory lift $\brane_{\mathcal E}$ (\ref{alift})
of an object ${\mathcal E}$ of $D(X)$.
Another is that the hemisphere partition function of a general brane $\brane$
is, in the Grassmannian phase,
proportional to the central charge (\ref{Zgeometric})
for $\pi_X(\brane)$ provided (\ref{chformula}) holds.
This will be seen in the next subsection.

The Chern characters of
${\mathcal E}\in D(X)$ and its pushforward
$i_{X*}{\mathcal E}\in D(\Gr)$ are related by
the Grothendieck-Riemann-Roch formula,
\beq
i_{X*}\bigl(\ch({\mathcal E})\,{\rm td}(X)\bigr)
\,=\,\ch(i_{X*}{\mathcal E})\,{\rm td}(\Gr).
\label{GrRiRo}
\eeq
The cohomology pushforward $i_{X*}$ on the left-hand side,
defined via Poincar\'e duality and homology pushforward, satisfies
\beq
i_{X*}(\eta)\,\omega\,=\,i_{X*}(\eta\, i_X^*\omega)
\quad\mbox{for $\eta\in {\rm H}^*(X)$ and $\omega\in {\rm H}^*(\Gr)$.}
\label{pullpull}
\eeq
Thus, (\ref{GrRiRo}) can also be written as 
\beq
i_{X*}\left(\ch({\mathcal E}){{\rm td}(X)\over i_X^*{\rm td}(\Gr)}\right)
\,=\,\,\ch(i_{X*}{\mathcal E}).
\label{GrRiRo2}
\eeq
Using the short exact sequence $0\to T_X\to T_{\Gr}|_X\to N_{X/\Gr}\to 0$,
we find
\beq
{{\rm td}(X)\over i_X^*{\rm td}(\Gr)}={1\over {\rm td}(N_{X/\Gr})}
={(1-\e^{-H})^7\over H^7}.
\eeq
Inside the big parenthisis of the left-hand side of
(\ref{GrRiRo2}), including $\ch({\mathcal E})$,
can be regarded as the pullback $i_X^*$ of
an element of ${\rm H}^*(\Gr,\Q)$,
by regarding $H$ as the pullback of $H\in {\rm H}^2(\Gr,\Z)$.
Using (\ref{pullpull}) again, now with $\eta=1$, and using the fact that
$i_{X*}(1)$ is the Poincar\'e dual of the homology class $i_{X*}[X]$,
which is $H^7$ (\ref{PDX}),
the left-hand side of (\ref{GrRiRo2}) can be expressed as
$(1-\e^{-H})^7\,\ch({\mathcal E})$.
On the other hand, if $\bV_{\mathcal E}$ is the
Chan-Paton vector space of a gauge theory lift (\ref{alift})
of ${\mathcal E}$,
\beq
\ch\bigl({\mathcal O}_{\Gr}(\bV_{\mathcal E})\bigr)
\,=\,\ch\left(\widehat{i_{X*}{\mathcal E}}(7)[-7]\right)
\,=\,\ch\left(\widehat{i_{X*}{\mathcal E}}\right)(-e^H)^7
\,\,=\ch(i_{X*}{\mathcal E})(-e^H)^7.
\eeq
Multiplying both sides of (\ref{GrRiRo2}) by $(-\e^H)^7$, we obtain
$(1-e^H)^7\,\ch({\mathcal E})=
\ch\bigl({\mathcal O}_{\Gr}(\bV_{\mathcal E})\bigr)$,
which is nothing but (\ref{chformula}) for the case $\brane=\brane_{\mathcal E}$.

\subsection{Computation Of The Hemisphere Partition Function}
\label{subsec:GHS}

Let us compute the hemisphere partition function (\ref{Zthemodel})
in the Grassmannian phase. As the contour, we can take $\bgamma_+$
with a small shift (\ref{shiftcontour}), that is,
\beq
\Is_1=(\Rs_1)^2-\delta,\quad
\Is_2=(\Rs_2)^2-\delta,\quad
0<\delta\ll 1.
\eeq
By deforming the contour, the integral (\ref{Zthemodel})
picks up the poles at (\ref{pole1}) and (\ref{pole2}),
\beq
Z_{D^2}(\mathfrak{B})\,=\,
\sum_{n_1,n_2=0}^{\infty}\oint\limits_{|x_1|
=|x_2|=\varepsilon}i^2\dd x_{1\,}\dd x_2\,
\Bigl(\cdots\Bigr)\Bigr|_{\usigma_1=\im (n_1+x_1),\,\,
\usigma_2=\im (n_2+x_2)}.
\label{forZLV}
\eeq

The integral (\ref{forZLV}) has been processed in \cite{HR}
as follows.\footnote{There is a minor point to clarify
  on the procedure in \cite{HR}. See Appendix~\ref{app:Weyl}.}
The contour $|x_1|=|x_2|=\varepsilon$ is the one
to which the Jeffrey-Kirwan localization formula can be applied,
so that the residue integral can be written as an integral over $\Gr=G(2,7)$.
It can further be written as an integral over
$X$ as
\beq
Z_{D^2}(\mathfrak{B})\,=\,
(2\pi)^{13}\sum_{n=0}^{\infty}\e^{-nt'}\int_{X}
\widehat{\Gamma}_X(n)\e^{{\im\over 2\pi}t'H}
{\rm ch}(\pi_X(\mathfrak{B})),
\label{LVZ}
\eeq
provided that the Chern character of $\pi_X(\brane)$ is given by
(\ref{chformula}). To arive at this, we used 
\beq
\ch\bigl({\mathcal O}_{\Gr}(\bV)\bigr)
\,=\,
    {\rm tr}_{\bV}^{}\left(\e^{\pi\im {\bf r}_*}\left(\begin{array}{cc}
      \e^{H_1}&0\\
      0&\e^{H_2}
    \end{array}\right)\right)
\,=\,    
f_{\brane}\left({H_1\over 2\pi},{H_2\over 2\pi}\right)
\label{chvsbf}
\eeq
where $H_1$ and $H_2$ are the Chern roots of the dual $S^*$ of
the tautological subbundle on $\Gr$.
Note that $H_1+H_2=H$ as $\det S^*={\mathcal O}_{\Gr}(1)$.
On the right-hand side of (\ref{LVZ}),
$t'=t-7\pi\im $ and
\beqa
\!\!\!\widehat{\Gamma}_X(n)
\!\!&:=&\!\!\widehat{\rm A}_X\cdot\Biggl[\,\sum_{n_1,n_2\geq 0\atop
  n_1+n_2=n}
\left({\Gamma\left(1+n_1+n_2+x_1+x_2\right)\over
\Gamma\left(1+n_1+x_1\right)
\Gamma\left(1+n_2+x_2\right)}\right)^7
\times
\nn\\[-0.2cm]
&&~~~~~~~~~~\Gamma(1+n_1-n_2+x_1-x_2)
\Gamma(1+n_2-n_1+x_2-x_1)
\Biggr]_{x_1\to {H_1\over 2\pi\im},\,x_2\to {H_2\over 2\pi\im}}.
\nn\\
&&\label{whGamma12}
\eeqa
In the large volume limit $\zeta\to\infty$, where
$t'H$ approaches the  complexified K\"ahler class
$[\omega-\im B]$,
the $n=0$ term is dominant in the sum (\ref{LVZ}).
Since $\widehat{\Gamma}_X(0)$ is nothing but the Gamma class of $X$,
the result (\ref{LVZ}) indeed takes the form
(\ref{Zgeometric}) up to the numerical factor $(2\pi)^{13}$.
This suggests
\beq
Z_{D^2}(\brane)\,=\,(2\pi)^{13}Z_X(\pi_X(\brane)).
\label{HSvsCCGr}
\eeq
The integral (\ref{forZLV}) can also be performed more directly.
Since the integrand has seventh order pole for each $x_1$ and $x_2$,
it can be processed as follows
\beqa
Z_{D^2}(\mathfrak{B})&=&
\sum_{n_1,n_2=0}^{\infty}\e^{-(n_1+n_2)t}
\!\!\!\!\!
\oint\limits_{|x_0|=|x_1|=\varepsilon}\!\!
i^2{\dd x_{1\,}\dd x_2\over x_1^7x_2^7}\,F_{n_1,n_2}(x_1,x_2;t)
\nn\\
&=&\sum_{n_1,n_2=0}^{\infty}
\e^{-(n_1+n_2)t}{(2\pi)^2\over 6!^2}
\left({\partial^{12}
F_{n_1,n_2}(x_1,x_2;t)\over
\partial x_1^6\partial x_2^6}\right)_{x_1=x_2=0}\nn\\[0.2cm]
&=:&Z_{\rm Gr}\{f_{\brane}\}
\label{ZexpGra}
\eeqa
with
\beqa
F_{n_1,n_2}(x_1,x_2;t)\!\!&:=&\!\!(n_1-n_2+x_1-x_2)\sin(\pi 
(x_1-x_2))(-1)^{n_1-n_2+1}\times
\nn\\
&&
\Gamma(1+n_1+n_2+x_1+x_2)^7
x_1^7\Gamma(-n_1-x_1)^7 x_2^7\Gamma(-n_2-x_2)^7
\times\nn\\
&&
\e^{-t (x_1+x_2)}\,f_{\mathfrak{B}}(\im x_1,\im x_2).
\label{Fn1n2}
\eeqa
Let us compute this for the examples of branes given above.
The brane factor for the $p$-Koszul brane is
\beq
f_{\Kos}(\usigma_1,\usigma_2)=
\sum_{j=0}^7(-1)^j{7\choose j}\e^{2\pi j(\usigma_1+\usigma_2)}
=(1-\e^{2\pi (\usigma_1+\usigma_2)})^7.
\eeq
The brane factors for its variants are
$f_{\Kos\otimes\Complex}=f_{\Kos}f_{\Complex}$.
In particular, for the $g$-Koszul complex (\ref{gKoszul}),
\beq
f_{\Complex}(\usigma_1,\usigma_2)
=\sum_{j=0}^n(-1)^j{n\choose j}\e^{-2\pi j(\usigma_1+\usigma_2)}
=(1-\e^{-2\pi (\usigma_1+\usigma_2)})^n,
\eeq
and hence
\beq
f_{\Kos(g_1,\ldots, g_n)}(\usigma_1,\usigma_2)
=(1-\e^{2\pi (\usigma_1+\usigma_2)})^7(1-\e^{-2\pi(\usigma_1+\usigma_2)})^n.
\label{bfKosn}
\eeq
The first few terms of (\ref{ZexpGra}) can easily be computed,
and the result is compared with the expansion of the mirror periods
as recorded in Appendix~\ref{app:periods}, Eqns~(\ref{periodRodGra}).
If we set $\log z=-t'=-(t-7\pi\im)$,
 we find
\beqa
Z_{\rm Gr}\{f_{\Kos}\}&=&(2\pi)^{13}\left(7\Pi_3+{7\over 2}\Pi_1
-\im {98\zeta(3)\over (2\pi)^3}\Pi_0\right),\label{d6ex1}\\
Z_{\rm Gr}\{f_{\Kos(g)}\}&=&(2\pi)^{13}\left(21\,\Pi_2-21\,
\Pi_1+{21\over 2}\Pi_0\right),\\
Z_{\rm Gr}\{f_{\Kos(g_1,g_2)}\}
&=&(2\pi)^{13}\cdot 42\cdot\left(\Pi_1-\Pi_0\right),\label{d2ex1}\\
Z_{\rm Gr}\{f_{\Kos(g_1,g_2,g_3)}\}&=&(2\pi)^{13}\cdot 42\cdot \Pi_0.
\label{d0ex1}
\eeqa
to the order which we computed.
Recall that
$\pi_X(\Kos(g_1,\ldots, g_n))$ is the D$(6-2n)$-brane
wrapped on $D_{g_1}\cap \cdots \cap D_{g_n}$, whose Chern character is
${\rm ch}({\mathcal O}_H)^n=(1-\e^{-H})^n$.
The result (\ref{d6ex1})-(\ref{d0ex1})
is consistent with what is expected, (\ref{HSvsCCGr})-(\ref{ZexpGra})
and (\ref{Zgeometric})
with (\ref{Zpt})-(\ref{ZOX}), in view of (\ref{topGra}) and
$H^2=42\ell$.

We note that the intersections $D_{g_1}\cap D_{g_2}$ and 
$D_{g_1}\cap D_{g_2}\cap D_{g_3}$
do not produce objects with minimal charges.
The branes whose charges are as in (\ref{Zpt}) and (\ref{ZOell}) cannot be 
constructed by restricting branes from the ambient geometry to the Calabi-Yau. 
To construct such objects, we have to find a suitable locus in the complex 
structure moduli space, i.e~we make a specific choice for the coefficients 
$A_k^{ij}$ such that all genericity requirements are still satisfied. Such a 
choice is far from unique, but a suitable possibility is given
by\footnote{This choice is inspired by the mirror of the Pfaffian
  Calabi-Yau \cite{boehmthesis}.} 
\begin{equation}
  \label{mirrora}
A(p)=\left(
\begin{array}{ccccccc}
 0 & \psi p^2 & p^3 & 0 & 0 & -p^6 & -\psi p^7 \\
 -\psi p^2 & 0 & \psi p^4 & p^5 & 0 & 0 & -p^1 \\
 -p^3 & -\psi p^4 & 0 & \psi p^6 & p^7 & 0 & 0 \\
 0 & -p^5 & -\psi p^6 & 0 & \psi p^1 & p^2 & 0 \\
 0 & 0 & -p^7 & -\psi p^1 & 0 & \psi p^3 & p^4 \\
 p^6 & 0 & 0 & -p^2 & -\psi p^3 & 0 & \psi p^5 \\
 \psi p^7 & p^1 & 0 & 0 & -p^4 & -\psi p^5 & 0 \\
\end{array}
\right),
\end{equation}
which is a restriction to a subfamily depending only on a single complex 
structure parameter $\psi$. As one can easily check, for non-zero $\psi$
and $p$ 
the rank is never less than $4$. Note also that at $\psi=0$ the genericity 
condition is violated i.e., there exist loci where $p\neq 0$ and the rank of 
$A(p)$ becomes $2$. Therefore, we must keep 
$\psi$ non-zero. By using (\ref{mirrora}) in the GLSM 
superpotential, we can produce matrix factorizations that lead to brane factors 
of the form  
\begin{equation}
  \label{fabdef}
c(1-e^{-2\pi (\usigma_1+\usigma_2)})^a(1-e^{2\pi\usigma_1})^b(1-e^{2\pi\usigma_2})^b, 
\end{equation}
where $c$ is a factor depending on the choice of $\bV$.
Brane factors of this type lead to the following branes in the Grassmannian phase:
\begin{eqnarray}
D6:&\quad& a=7-2i\quad b=i\qquad i=0,1,2,3\label{fd6}\\
D4:&\quad& a=7+1-2i\quad b=i\qquad i=0,1,2,3,4\label{fd4}\\
D2:&\quad& a=7+2-2i\quad b=i\qquad i=0,1,2,3,4\label{fd2}\\
D0:&\quad& a=7+3-2i\quad b=i\qquad i=0,1,2,3,4,5\label{fd0}.
\end{eqnarray}
In this case we refer to a brane as a D$p$-brane if the hemisphere partition
function only contains periods with labels $\leq\frac{p}{2}$.
We refrain from giving explicit examples of matrix factorizations
as the expressions are long and not very enlightening.
To produce a single factor $(1-e^{-2\pi(\usigma_1+\usigma_2)})$
we have to factor out a $p$-field. In other words,
our matrix factorization has to contain a term
\begin{equation}
\label{mfcomp1}
p\bar{\eta}_1+f(x^1x^2)\eta^1,
\end{equation}
where $\eta^1,\bar{\eta}_1$ are elements of the Clifford basis and $f(x^1x^2)$ 
denotes a polynomial in products $x_i^1x_j^2$. To produce a factor 
$(1-e^{2\pi\usigma_1})(1-e^{2\pi\usigma_2})$ our matrix factorization must contain 
a term of the form
\begin{equation}
\label{mfcomp2}
x_m^1\bar{\eta}_1+f_1(px^2)\eta^1+x_m^2\bar{\eta}_{2}+f_2(px^1)\eta^2,
\end{equation}
where $x_m^1,x_m^2$ have fixed $m$ and $f_1(px^2),f_2(px^1)$ are polynomials in 
$x_j^1,x_j^2$ for some $j\neq m$. Then (\ref{fabdef}) can be intuitively 
understood: it corresponds to a subvariety of codimension $2b+a$ inside 
$G(2,7)$. It is not hard to see that, making use of (\ref{mirrora}), one can 
realize brane factors (\ref{fd6}), (\ref{fd4}) of D6-branes and D4-branes with 
$i=0$ as well as brane factors (\ref{fd2}), (\ref{fd0}) for D2-branes and 
D0-branes for all values of $i$. One can show for instance that matrix 
factorizations with (\ref{fd2}) with $i=4$ and (\ref{fd0})
with $i=5$ correspond 
to branes with minimal charge, i.e.~the coefficient $42$ in (\ref{d2ex1}), 
(\ref{d0ex1}) is replaced by a $1$.

\subsection{Some Mathematical Background}

We would like to pause to explain some mathematical background
that will fascilitate the subsequent discussions
such as grade restriction processes.
The contents include basic facts on homological algebra,
the Borel-Weil-Bott theorem and invariant theory.
We shall also describe the procedure to lift branes in the sigma model
with target $X$ to the gauge theory.
At the end, we quote some results which are useful to understand
the equivalence ${\mathcal T}_{\bf w}\cong D(X)$.

\subsubsection{Homological Algebra}
\label{subsub:HA}

\subsubsection*{I. Some Brane Gymnastics}

Useful techniques in brane gymnastics are binding branes by open strings
and brane-antibrane annihilation via tachyon condensation.
In what follows, ``a brane $(E,Q_E)$'' is a graded obeject $E$ in an
additive category with a differential $Q_E$ of degree one.
We have in mind
a complex of vector bundles on the target space in a non-linear sigma model,
a matrix factorization of the superpotential in a Landau-Ginzburg model,
and a brane datum in a gauged linear sigma model,
in which the grading is by $\Z$ or $\Z_2$ depending on whether or not
there is a $U(1)$ R-symmtry with charge integrality.

Let $(E,Q_E)$ and $(F,Q_F)$ be branes
and let $f:(E,Q_E)\to (F,Q_F)$ be a homomorphism, that is, a degree zero
map $f:E\to F$ obeying $Q_F\circ f=f\circ Q_E$.
The {\it cone} of $f$ binds the two branes together into one,
\beq
    {\rm Cone}(f)=
    \left(E[1]\oplus F,
    \left(\begin{array}{cc}
      -Q_E&0\\
      f&Q_F
    \end{array}\right)\right).
    \label{cone1}
\eeq
where $(E[1],-Q_E)$ is the degree shift of $(E,Q_E)$ by $+1$.\footnote{Here,
  we take the standard convention which flips the sign of
$Q_E$. However, there is a different convention (\ref{newdiff})
which we shall use at times in what follows.}

In addition,
let $g$ be a homomorphism from $(F,Q_F)$
to another brane $(G,Q_G)$.
Can we extend it to a homomorphism
from ${\rm Cone}(f)$ to $(G,Q_G)$
so that its cone can be taken to bind $(E,Q_E)$, $(F,Q_F)$ and $(G,Q_G)$
together?
Alternatively, can we extend $f:(E,Q_E)\to (F,Q_F)$
to a homomorphism
from $(E,Q_E)$ to ${\rm Cone}(g)$ for the same purpose?
It is easy to see that the condition for either to be possible is that
the composition $g\circ f$ is null homotopic, that is,
there is a degree $-1$ map $h:E\to G$ such that
$g\circ f=Q_G\circ h+h\circ Q_E$.
In such a case, there are homomorphisms
\beqa
\hat{g}&:=&(h,g):{\rm Cone}(f)\longrightarrow (G,Q_G),\\
\check{f}&:=&\left(\begin{array}{c}f\\-h
\end{array}\right):(E,Q_E)[1]\longrightarrow {\rm Cone}(g),
\eeqa
and the two cones are isomorphic
${\rm Cone}(\hat{g})\cong {\rm Cone}(\check{f})$.
Indeed, the shifts of these cones by $-1$ are both isomorphic to
\beq
\left(E[1]\oplus F\oplus G[-1],
\left(\begin{array}{ccc}
  -Q_E&0&0\\
  f&Q_F&0\\
  h&g&-Q_G
\end{array}\right)\right).
\label{cone2}
\eeq
This can be interpreted as the associativity of the cone operation,
which may be described as
\beq
((E\to F)\to G)\,\,\cong\,\, (E\to (F\to G)).
\label{associativity}
\eeq

For another homomorphism in a different direction,
$g':(E,Q_E)\to (G',Q_{G'})$ or $f'':(G'',Q_{G''})\to (F,Q_F)$,
the successive cone is possible without any condition.
There are two orders for doing it
and the successive cones in both orders are the same as the
cone of the sum of the homomorphisms,
$f\oplus g'$ or $f\oplus f''$.

Let us present a brane ${\mathcal B}=(E,Q_E)$ as
\beq
{\mathcal B}:~
\begin{tikzcd}
  E^{\rm od} \arrow[r,yshift=0.5ex, "f"]
  & \arrow[l,,yshift=-0.5ex,"g"] E^{\rm ev}
\end{tikzcd}
\label{branestart}
\eeq
where $E^{\rm ev}$ and $E^{\rm od}$ are the direct sums of
even and odd degree components of $E$.
Suppose a part of the arrows is invertible, that is,
there is a decomposition $E=E_1\oplus E_2$ with 
$f_{ij}:E_j^{\rm od}\to E_i^{\rm ev}$
and $g_{ij}:E_j^{\rm ev}\to E_i^{\rm od}$, and $f_{22}$ (say) is an isomorphism.
Then, the brane splits into the direct sum
\beq
{\mathcal B}\,~\cong\,~{\mathcal B}_1\oplus {\mathcal B}_2,
\label{splits}
\eeq
where
\beq
{\mathcal B}_1:~
\begin{tikzcd}
  E_1^{\rm od}~~~~~~
  \arrow[r,shorten=-0.8cm,yshift=0.5ex, "f^{}_{11}-f^{}_{12}f_{22}^{-1}f^{}_{21}"]
  & \arrow[l,shorten=-0.8cm,yshift=-0.5ex, "g^{}_{11}"]
  ~~~~~~ E_1^{\rm ev}
\end{tikzcd}
\label{braneend1}
\eeq
and
\beq
{\mathcal B}_2:~
\begin{tikzcd}
  E_2^{\rm od}~~~~~~
  \arrow[r,shorten=-0.8cm,yshift=0.5ex, "f^{}_{22}"]
  & \arrow[l,shorten=-0.8cm,yshift=-0.5ex, "g^{}_{22}+\Delta g^{}_{22}"]
  ~~~~~~ E_2^{\rm ev}
\end{tikzcd}
\label{braneend2}
\eeq
in which $\Delta g^{}_{22}=g^{}_{21}f^{}_{12}f_{22}^{-1}
=f_{22}^{-1}f^{}_{21}g^{}_{12}$.
Indeed, there is an isomorphism
$\varphi:{\mathcal B}\to{\mathcal B}_1\oplus {\mathcal B}_2$
given by
\beq
\varphi^{\rm ev}=\left(\begin{array}{cc}
  {\rm id}_{E^{\rm ev}_1}&-f^{}_{12}f_{22}^{-1}\\
    0&{\rm id}_{E^{\rm ev}_2}
\end{array}\right)
\quad\mbox{and}\quad
\varphi^{\rm od}=\left(\begin{array}{cc}
  {\rm id}_{E^{\rm od}_1}&0\\
    f_{22}^{-1}f^{}_{21}&{\rm id}_{E^{\rm od}_2}
\end{array}\right).
\eeq

In the physical setup, a brane $(E,Q_E)$ is empty in the
infra-red limit when the boundary potential $\{Q_E,Q_E^{\dag}\}$
is positive definite.
A brane does not change its low energy behaviour
if an empty brane is bound to it. For example,
if the source ({\it resp}. the target) of
$f:(E,Q_E)\to (F,Q_F)$ is empty in the infra-red limit,
its cone (\ref{cone1}) flows to the same brane as the target
({\it resp}. the source shifted by $+1$).
Also, when a part of the arrows 
is invertible and the brane splits
${\mathcal B}\cong{\mathcal B}_1\oplus{\mathcal B}_2$ as in (\ref{splits}),
the component ${\mathcal B}_2$ with the invertible arrow $f_{22}$
is empty in the infra-red limit
as $f_{22}^{\dag}f_{22}^{}>0$, $f_{22}^{}f_{22}^{\dag}>0$ and
hence $\{Q_2,Q_2^{\dag}\}>0$.
The brane $E_2^{\rm ev}$ and the antibrane $E_2^{\rm od}$ annihilate each other
  by tachyon condensation of $f_{22}$.
In particular, the brane ${\mathcal B}$ flows to the same brane
as ${\mathcal B}_1$.
The net effect of the brane-antibrane annihilation by tachyon condensation
is the removal of the isomorphic pair $(E_2^{\rm ev},E_2^{\rm od})$
and the addition of a new arrow $-f^{}_{12}f_{22}^{-1}f^{}_{21}$.
The latter is called the {\it zigzag map} \cite{HHP}
as it is (the negative of)
the composition of the forward arrow $f^{}_{21}$,
the inverse arrow $f_{22}^{-1}$,
and the forward arrow $f^{}_{12}$, through the annihilated pair.

What is said above has some mathematical counterpart,
in which ``empty brane'' and ``flows to the same branes as''
are replaced by a ``zero object'' and ``is isomorphic to'' respectively.
In the derived category over an Abelian category,
a complex is a zero object if and only if it is exact.
If the source or the target of a homomorphism is a zero object,
the cone is isomorphic to the target or the shifted source.
Also, when an object splits by an invertible arrow
as in (\ref{splits}),
the component ${\mathcal B}_2$ with the invertible arrow
is homotopy equivalent to $0$: 
\beq
\mbox{It has}~~
Q_2=\left(\begin{array}{cc}0&f^{}_{22}\\
  g^{}_{22}+\Delta g^{}_{22}&0\end{array}\right)
  \quad\mbox{and}\quad
  \left\{Q_2,\left(\begin{array}{cc}
    0&0\\
    f_{22}^{-1}&0\end{array}\right)\right\}={\rm id}_{E_2}.
\eeq
Therefore,    
${\mathcal B}$ is isomorphic to ${\mathcal B}_1$ in the homopoty category.

\subsubsection*{II. The Extension Theorem}

Let $\phi$ be a set of variables on which a group $G$ acts.
$G$ acts on the polynomial ring $\C[\phi]$ by
$g:P(\phi)\mapsto g(P)(\phi):=P(g^{-1}\phi)$.
A $G$-equivariant $\C[\phi]$-module
(or $(G,\C[\phi])$-module for short)
is a $\C[\phi]$-module $M$ with a $G$-action
satisfying $g(Pm)=g(P)g(m)$.
An example is $\rmV\otimes^{}_{\C}\C[\phi]$ for a complex representation
$\rmV$ of $G$ where the $G$ action is $g:v\otimes P\mapsto gv\otimes g(P)$.
A homomorphism of
$(G,\C[\phi])$-modules is a homomorphism
of $\C[\phi]$-modules that is $G$-equivariant.
In this language,
a complex of representations over $\C[\phi]$ of $G$
is a complex of $(G,\C[\phi])$-modules
whose terms are of the form $\rmV\otimes^{}_{\C}\C[\phi]$.
We describe some properties of such complexes.

Let $(M,d_M)$ and $(N,d_N)$ be complexes of representations over $\C[\phi]$
of $G$ where the latter terminates at degree $0$.
Suppose there is a homomorphism $f^0:M^0\to N^0$ of $(G,\C[\phi])$-modules.
Then, we consider a problem to extend
this to a cochain map from $(M,d_M)$ to $(N,d_N)$:
\beq
\begin{tikzcd}
\cdots \arrow[r]& M^{-2}\arrow[r]\arrow[d,dashed]
  &  M^{-1} \arrow[r]\arrow[d, dashed] & M^0\arrow[r]\arrow[d,"f^0"]
  &M^1\arrow[r]&\cdots\\
\cdots\arrow[r]& N^{-2} \arrow[r]&N^{-1}\arrow[r]&N^0&&
\end{tikzcd}
\eeq
We shall say that a cochain complex is {\it acyclic} when it is exact except
possibly at the last non-zero term.\footnote{Warning:
  this may not be a standard
  terminology.}\\[0.2cm]
{\bf Theorem.}
This problem has a solution
when $G$ is a compact group, $(N,d_N)$ is acyclic,
and $f^0{\rm Im}\,d_M^{-1}\subset {\rm Im}\,d_N^{-1}$.
\\[0.1cm]
{\bf Proof.} 
By the last condition, $f^0$ induces a homomorphism
from the cokernel of $d_M^{-1}$ to the cokernel of $d_N^{-1}$, and hence
the problem can be truncated to
\beq
\begin{tikzcd}
\cdots \arrow[r]& M^{-2}\arrow[r]\arrow[d,dashed]
  &  M^{-1} \arrow[r]\arrow[d, dashed] & M^0\arrow[r]\arrow[d,"f^0"]
  &{\rm Cok}\,d_M^{-1}\arrow[d,"f^0"]\\
\cdots\arrow[r]& N^{-2} \arrow[r]&N^{-1}\arrow[r]&N^0\arrow[r]&
{\rm Cok}\,d_N^{-1}
\end{tikzcd}
\eeq
As $(N,d_N)$ is acyclic, the bottom complex is exact.
Therefore, as $M^{-1},M^{-2},\ldots$ are free $\C[\phi]$-modules,
one can find homomorphisms
$\tilde{f}^i:M^i\to N^i$ of $\C[\phi]$-modules ($i=-1,-2,\ldots$)
which provide an extension of $f^0$
to a cochain map of complexes of $\C[\phi]$-modules.
These $\tilde{f}^i$ may not be $G$-equivariant but the averages
\beq
f^i~:=~\int_Gg\circ \tilde{f}^i\circ g^{-1}\,\,\dd g
\label{averaging}
\eeq
are. By the $G$-equivariance of $d_M^i, f^0$ and $d_N^i$,
the squares remain to commute also with the $f^i$'s.
Thus, these provide an extension of $f^0$
to a cochain map of complexes of $(G,\C[\phi])$-modules.
{\bf Q.E.D.}
\\
{\bf Remark.} When $G$ is a compact Lie group, $G$-equivariance implies
equivariance with respect to its complexification,
which is a complex reductive group. 
What we have shown is that, when $G$ is one of such groups,
$\rmV\otimes^{}_{\C}\!\C[\phi]$
is a projective object in the abelian category of $(G,\C[\phi])$-modules.

For a $(G,\C[\phi])$-module $M$, its
$\C[\phi]$-dual $M^*:=\Hom_{\C[\phi]}(M,\C[\phi])$ is also a
$(G,\C[\phi])$-module:
the transform of $\omega\in M^*$ by $g\in G$
is $(g\omega)(m)=g(\omega(g^{-1}m))$.
For a complex representation $\rmV$ of $G$,
there is a canonical isomorphism of $(G,\C[\phi])$-modules,
\beq
\left(\rmV\otimes^{}_{\C}\!\C[\phi]\right)^*
~\cong~ \rmV^*\otimes^{}_{\C}\!\C[\phi],
\eeq
where $\rmV^*$ is the $\C$-dual representation of $\rmV$.
Therefore, if $(M,d_M)$ is a complex of representations over $\C[\phi]$ of $G$,
so is its transpose $(M,d_M)^T:=(M^*,d_M^T)$.
Let $(M,d_M)$ and $(N,d_N)$ be complexes of representations over $\C[\phi]$ of
$G$ where the latter begins at degree $0$. Suppose
there is a homomorphism
$f^0:N^0\to M^0$ of $(G,\C[\phi])$-modules.
Then, we consider the problem to extend this to a
cochain map from $(N,d_N)$ to $(M,d_M)$:
\beq
\begin{tikzcd}
  &&N^0\arrow[r]\arrow[d,"f^0"]&N^1\arrow[r]\arrow[d,dashed]
  &N^2\arrow[r]\arrow[d,dashed]&\cdots\\
  \cdots\arrow[r]&M^{-1}\arrow[r]&M^0\arrow[r]&M^1\arrow[r]&M^2\arrow[r]&\cdots
\end{tikzcd}
\eeq
{\bf Corollary.} This problem has a solution when
$G$ is a compact group, the transpose $(N,d_N)^T$ is acyclic, and
${\rm Ker}\,d_N^1\subset (f^0)^{-1}{\rm Ker}\,d_M^1$.

\subsubsection*{III. Homotopy vs Derived Categories}

Let ${\mathcal A}$ be an abelian category.
We denote by ${\rm K}^b({\mathcal A})$ and $\Db({\mathcal A})$
the homotopy category and the derived category of
bounded complexes over ${\mathcal A}$.
The Ext groups for a pair of objects of ${\mathcal A}$ are defined by
$\Ext^p_{\mathcal A}(X,Y)=\Hom_{\Db({\mathcal A})}(X[0],Y[p])$,
in which
$X[0]$ and $Y[p]$ are the one-term complexes, $X$ at degree
$0$ and $Y$ at degree $-p$ respectively.
Then, we have
\\[0.1cm]
{\bf Lemma.} Let $(E,d_E)$ and $(F,d_F)$ be bounded complexes
over ${\mathcal A}$. If $\Ext^p_{\mathcal A}(E^i,F^j)=0$
for $p>0$ and all $i,j$, then the canonical homomorphism
\beq
\Hom_{{\rm K}^b({\mathcal A})}^{}((E,d_E),(F,d_F))
~\longrightarrow~
\Hom^{}_{\Db({\mathcal A})}((E,d_E),(F,d_F))
\eeq
is an isomorphism.\\[0.2cm]
This is stated in \cite{KapranovGr}, Lemma 0.1,
\cite{KapranovHomog}, Lemma 1.6
and \cite{GM}, IV.2.2c).
We shall use this on many occasions, for the case
where ${\mathcal A}={\it Coh}(M)$, the category of coherent sheaves
on an algebraic variety $M$.
In that case, $\Ext^p_{{\it Coh}(M)}(E,F)$ will be denoted by
$\Ext^p_M(E,F)$. Note that it is non-zero only for $0\leq p\leq \dim M$.
We shall call $\Ext^p_M(E,F)$ for $p>0$ {\it higher Ext groups}.

\subsubsection{Borel-Weil-Bott Theorem}
\label{subsub:BWB}

We next compute the Ext groups for pairs of homogeneous vector bundles
on $\Gr=G(2,7)$ using the Borel-Weil-Bott theorem.

Let us first describe the theorem.
(See for example, \cite{Kostant}, Theorem 6.4. See also
\cite{KapranovGr,KapranovHomog}.)
For a vector space $P$ of dimension $n$
and for an $n$-vector $\lambda=(\lambda_1,\ldots,\lambda_n)$
with a non-increasing sequence of integers
$\lambda_1\geq\lambda_2\geq\cdots\geq\lambda_n$ as its entries, we denote by
$\Sigma^{\lambda}P$ the irreducible representation of $\GL(P)$
with highest weight $\lambda$. It is realized by the Young symmetrization
of a tensor power of $P$ corresponding to the partition
$\lambda_1-\lambda_n,\ldots, \lambda_{n-1}-\lambda_n$
tensored with $\det^{\lambda_n} P$. See for example \cite{Weyl}.
The operation $\Sigma^{\lambda}$ can hence be applied not just to vector spaces
but also to vector bundles.
For a complex vector space $V$ of dimension $N$,
let $G(k,V)$ be the Grassmann manifold of $k$-dimensional subspaces in $V$.
It is a homogeneous space of $\GL(V)$.
The Borel-Weil-Bott theorem describes the
cohomology groups of the sheaves of sections of homogeneous vector bundles
as representations of $\GL(V)$.
We denote by $S$ the tautological subbundle of rank $k$ on $G(k,V)$.
Let $\rho=(N,N-1,\ldots, 1)$ be half the sum of positive roots of $\GL(V)$
shifted by a permutation invariant vector.
For a $k$-vector $\alpha$, we denote by
$(\alpha,0)$ the $N$-vector $(\alpha_1,\ldots,\alpha_k,0,\ldots, 0)$.
Then, we have
\beq
\rmH^p(G(k,V),\Sigma^{\alpha}S^*)~\cong~
\left\{\begin{array}{cl}
\Sigma^{\sigma((\alpha,0)+\rho)-\rho}V^*&\mbox{Case $\ast$ and $p=\ell(\sigma)$}
  \\
0&\mbox{otherwise}
\end{array}
\right.
\label{BWB}
\eeq
where Case $\ast$ is when the entries of $(\alpha,0)+\rho$ are distinct
so that there is a unique permutation $\sigma\in\mathfrak{S}_N$
such that $\sigma((\alpha,0)+\rho)$ is strictly decreasing,
and $\ell(\sigma)$ is the minimum number of exchanges to present $\sigma$.
In particular $\rmH^p(G(k,V),\Sigma^{\alpha}S^*)=0$ for any $p>0$
if and only if 
$(\alpha,0)+\rho$ has coincident entries or is strictly decreasing.
For example, in the case $k=1$ where $G(k,V)$
is the projective space $\PP(V)=\CP^{N-1}$ and
$\Sigma^{\alpha}S^*={\mathcal O}^{}_{\CP^{N-1}}(\alpha)$ for an integer $\alpha$,
Case $\ast$ is when $N+\alpha,N-1,\ldots, 1$ are distinct, and we have
\beq
\rmH^p(\CP^{N-1},{\mathcal O}^{}_{\CP^{N-1}}(\alpha))
~\cong~
\left\{\begin{array}{cl}
\Sym^{\alpha}V^*&\alpha\geq0,~\,p=0,\\
\det V\otimes \Sym^{|N+\alpha|}V&\alpha\leq -N,~\,p=N-1,\\
  0&\mbox{otherwise.}
\end{array}\right.
\label{BWBProj}
\eeq

Let us apply this theorem to compute the Ext groups of
pairs of $\Sym^h S^*(i)$'s over $\Gr=G(2,7)$.
Recall that $S^*={\mathcal O}_{\Gr}(\fund)$.
Note that $\Sym^h\fund(i)$ has highest weight $x_1^h(x_1x_2)^i=x_1^{h+i}x_2^i$
so that
\beq
\Sym^h\fund(i)=\Sigma^{h+i,i}\fund.
\label{SymvsY}
\eeq
The $SU(2)$ Clebsch-Gordan rule
combined with $U(1)$ charge conservation,
or the Littlewood-Richardson rule, reads
(see for example \cite{LL} or \cite{FultonHarris})
\beqa
\Hom(\Sym^{h_1}\fund(i_1),\Sym^{h_2}\fund(i_2))
&=&\Sym^{h_1}\fund(-i_1-h_1)\otimes\Sym^{h_2}\fund(i_2)\nn\\
&=&\bigoplus_{j=0}^{{\rm min}\{h_1,h_2\}}
\Sym^{h_1+h_2-2j}\fund(i_2-i_1-h_1+j)\nn\\
&=&\bigoplus_{j=0}^{{\rm min}\{h_1,h_2\}}
    \Sigma^{h_2+i_2-i_1-j,i_2-i_1-h_1+j}\fund,
\eeqa
and this means
\beq
\Ext^p_{\Gr}(\Sym^{h_1}S^*(i_1),\Sym^{h_2}S^*(i_2))
=\bigoplus_{j=0}^{{\rm min}\{h_1,h_2\}}
\rmH^p(\Gr,\Sigma^{h_2+i_2-i_1-j,i_2-i_1-h_1+j}S^*).
\label{ExtComp}
\eeq
By Borel-Weil-Bott,
the summand depends on the relations among the entries of
\beq
    {\bf x}:=
    (\underbrace{7+h_2+i_2-i_1-j}_{\displaystyle \,\,\,\,\,\,\,\,\,\,=:a},
    \underbrace{6-h_1+i_2-i_1+j}_{\displaystyle \,\,\,\,\,\,\,\,\,\,=:b},
    5,4,3,2,1).
    \label{vectorx}
\eeq
Note that $a$ is always larger than $b$.
In particular, higher Ext groups vanish if, for all $j$, $b\geq 1$ (so that
$b$ coincides with $1,2,\ldots, 5$ or $b\geq 6$ in which case
${\bf x}$ is strictly decreasing)
or $a$ coincides with $5,4,3,2,1$.

\newcommand{\SetR}{{\rm R}}

For a set $\SetR$ of representations of $U(2)$ and an integer $m$, we define\\
{\bf Property EV${}_m$}:
For vector bundles $E$ and $E'$ in
${\mathcal O}_{\Gr}(\SetR)$,
\beq
\Ext^p_{\Gr}(E,E'(m))
=0\quad \mbox{for $p>0$}.
\label{ClaimExt}
\eeq
Then, we have\\[0.1cm]
{\bf Theorem.}
The set $\rmGRR_{\bf w}$ has Property EV${}_m$ for $m\geq 0$.
\\[0.1cm]
  {\bf Proof.}
As $\Ext_{\Gr}^p(E,E'(m))$ is invariant under $(E,E')\to (E(i),E'(i))$,
it is enough to show this for the windows {\bf A}, {\bf B}, {\bf C},
{\bf D}. Let us start with the window {\bf C}.
The set $\rmGRR_{\bf C}$ consists of $\Sym^h\fund(i)$
for $h=0,1,2$ and $i=0,1,\ldots, 6$.
We put $(h_1,i_1)=(h,i)$ and $(h_2,i_2)=(h',i'+m)$ for which
$a=7+h'+i'-i+m-j$ and $b=6-h+i'-i+m+j$.
Note that $i'-i+m\geq -6$.
If $i'-i+m\geq -3$, then $b\geq 3-h+j\geq 1$ since
$h\leq 2$ and $j\geq 0$.
If $-6\leq i'-i+m\leq -4$, then $1+h'-j\leq a\leq 3+h'-j$.
The lower bound $1+h'-j$ is at least $1$ as $j\leq h'$ and
the upper bound $3+h'-j$ is at most $5$ as $h'\leq 2$ and $j\geq 0$.
Thus, $a$ must coincides with $5,4,3,2$ or $1$.
This proves the vanishing of the higher Ext groups, (\ref{ClaimExt}).
Let us next consider the window {\bf B}.
The set $\rmGRR_{\bf B}$ consists of
$\Sym^h\fund(i-h)$ for $h=0,1,2$ and $i=1,2,\ldots,7$.
We put $(h_1,i_1)=(h,i-h)$ and $(h_2,i_2)=(h',i'-h'+m)$ for which
$a=7+h+i'-i+m-j$ and $b=6-h'+i'-i+m+j$.
Compared to $\rmGRR_{\bf C}$, the only difference is $h\leftrightarrow h'$
and the simultaneous shift of $i,i'$ by $1$.
Thus, $b\geq 1$ or $1\leq a\leq 5$ holds for each $j$
and the higher Ext groups all vanish, (\ref{ClaimExt}).
Next, consider the window {\bf A}.
The set $\rmGRR_{\bf A}$ differs
from $\rmGRR_{\bf C}$ in that it contains $\C(7)$ instead of $\C$, and
from $\rmGRR_{\bf B}$ in that it contains $\Sym^2\fund(6)$
instead of $\Sym^2\fund(-1)$. Therefore only the pair of
$\C(7)$ and $\Sym^2\fund (6)$ has to be examined.
The two are members of $\rmGRR_{\bf C}(1)$
and hence (\ref{ClaimExt}) holds.
Let us finally consider the window {\bf D}.
The set $\rmGRR_{\bf D}$ differs
from $\rmGRR_{\bf B}$ in that it contains $\C$ instead of $\C(7)$, and
from $\rmGRR_{\bf C}$ in that it contains $\Sym^2\fund(-1)$
instead of $\Sym^2\fund(6)$. Therefore only the pair of
$\C$ and $\Sym^2\fund (-1)$ has to be examined.
The two are again members of $\rmGRR_{\bf C}(-1)$
and hence (\ref{ClaimExt}) holds.
{\bf Q.E.D.}\\[0.1cm]
{\bf Corollary.} 
$\rmGRR_{\bf w}\cup \rmGRR_{\bf w}(1)$ has Property EV${}_m$ for $m\geq 1$.

We also have the following results for $\Hom^{}_{\Gr}=\Ext^0_{\Gr}$:
\beqa
&&\begin{array}{ll}
  \Hom^{}_{\Gr}(\Sym^{h_1}S^*(i_1),\Sym^{h_2}S^*(i_2))\ne 0\\
  \mbox{~~~~~~~~~~~~~~~~~~~~
    if and only if $i_1\leq i_2$ and $h_1+i_1\leq h_2+i_2$,}
  \end{array}
\label{Hom11}\\
&&\Hom^{}_{\Gr}(\Sym^hS^*(i),\Sym^hS^*(i))=\C.
\label{End11}
\eeqa
{\bf Proof.}
(\ref{ExtComp}) for $p=0$ is non-vanishing if and only if
{\bf x} is strictly decreasing for some $j$,
that is, $b>5$ for both $j=h_1$ and $j=h_2$.
In the former case, $b=6+i_2-i_1$ and the condition is
$i_2\geq i_1$, while in the latter case, $b=6-h_1+i_2-i_1+h_2$
and the condition is $h_2+i_2\geq h_1+i_1$.
For $(h_1,i_1)=(h_2,i_2)=(h,i)$, we have
$a=7+h-j$ and $b=6-h+j$. For $j=h$,
${\bf x}=\rho$ and the cohomology $\rmH^0$ is $\Sigma^0V^*=\C$.
For $j<h$, {\bf x} is not strictly decreasing and the
cohomology $\rmH^0$ vanishes. {\bf Q.E.D.}
    
\noindent
{\bf Terminology.}
Continuing the last step of the proof of (\ref{End11}), {\bf x}
has distinct but nondescreasing entries for $j\leq h-6$, which is possible
only when $h\geq 6$. In other words, $E=\Sym^hS^*(i)$
with $h\leq 5$ satisfies
$\Ext^p_{\Gr}(E,E)=\delta_{p,0}\C$.
Such an object $E$ of $D(\Gr)$ is said to be {\it exceptional}.
By Property EV${}_0$ and (\ref{Hom11}),
the set ${\mathcal O}_{\Gr}(\rmGRR_{\bf w})$ is
said to form a {\it strong exceptional collection} of $D(\Gr)$.

\subsubsection{Invariant Theory}
\label{subsub:IT}

Let us describe some more details about
homomorphisms between homogeneous vector bundles on $\Gr$.
We regard $\Gr=G(2,7)$ as the quotient of the space
$S(2,7)$ of $2\times 7$ matrices $x=(x^a_i)$ of maximal rank
by the free $\GL_2(\C)$ action, $g:x\mapsto gx$.
For a finite dimensional representation $\rmV$ of $\GL_2(\C)$,
$\rho:\GL_2(\C)\to \GL(\rmV)$,
the homogeneous vector bundle ${\mathcal O}_{\Gr}(\rmV)$
is realized as the quotient of $S(2,7)\times \rmV$ by the
$\GL_2(\C)$ action, $g:(x,v)\mapsto (gx,\rho(g)v)$.
Then a holomorphic section of ${\mathcal O}_{\Gr}(\rmV)$
is realized as a
holomorphic function $f$ on $S(2,7)$ with values in $\rmV$
which is $\GL_2(\C)$-equivariant,
$f(gx)=\rho(g)f(x)$.
Note that $S(2,7)$ is an open subset of the space $\Hom(\C^7,\C^2)$
of all $2\times 7$ matrices and the complement has codimension $7+1-2=6\geq 2$.
By Hartogs's theorem, $f$ extends to a holomorphic function
on the affine space $\Hom(\C^7,\C^2)$.
If $\rmV$ is irreducible,
$\rho(z{\bf 1}_2)=z^{n_{\rmV}}{\rm id}_{\rmV}$ for some integer $n_{\rmV}$
and $f$ satisfies $f(zx)=z^{n_{\rmV}}f(x)$. This means that
the function $f(x)$ is a homogenous polynomial of the $x^a_i$'s
of degree $n_{\rmV}$.
(It must be zero if $n_{\rmV}<0$.) 
For a general $\rmV$, as it is completely reducible,
$f(x)$ is also a polynomial of the $x^a_i$'s though it may not be homogeneous.
To summarize, holomorphic sections of ${\mathcal O}_{\Gr}(\rmV)$
are in one to one correspondence with
$\GL_2(\C)$-equivariant polynomial functions of
$x=(x^a_i)$ with values in $\rmV$.

Therefore, homomorphisms
${\mathcal O}_{\Gr}(\rmV_1)\to {\mathcal O}_{\Gr}(\rmV_2)$
are in one to one correspondence with homomorphisms 
$\rmV_1\otimes\C[x]\to\rmV_2\otimes\C[x]$
of $(\GL_2(\C),\C[x])$-modules.
In particular, complexes of homogeneous vector bundles on $\Gr$
may be identified with complexes of
representations over $\C[x]$ of $\GL_2(\C)$.
However, we should note that the notions of exactness differ
between the two. The matrix $(x^a_i)$
that determines the differentials for the former is of maximal rank
while such a condition is absenet in the latter.
Therefore, an exact complex of homogeneous vector bundles on $\Gr$
may not be exact as a complex of $\C[x]$-modules.

The homomorphisms can be described more explicitly.
To be specific, let us consider
\beq
\varphi:\Sym^{h_1}\fund(i_1)\otimes\C[x]\to
\Sym^{h_2}\fund(i_2)\otimes\C[x],
\label{hommm}
\eeq
corresponding to the case of (\ref{Hom11}).
Note that $z{\bf 1}_2$ acts on $\Hom(\rmV_1,\rmV_2)$ for this pair 
  by multiplication by $z^n$ with $n=h_2+2i_2-(h_1+2i_1)$,
and $n\geq 0$ is guaranteed if $i_2\geq i_1$ and $h_2+i_2\geq h_1+i_1$,
consistently with (\ref{Hom11}).
A basis $\{e_a\}$ of $\fund$ determines a basis $\{e_{a_1\cdots a_h}\!(i)\}$
of $\Sym^h\fund(i)$. Let us represent (\ref{hommm}) with respect to the bases
of this form,
\beq
\varphi(e_{b_1\cdots b_{h_1}\!}\!(i_1))
=e_{a_1\cdots a_{h_2}\!}\!(i_2)\,
\varphi^{a_1\cdots a_{h_2}}_{\,\,\,\,b_1\cdots b_{h_1}\!}\!(x).
\eeq
The matrix elements are polynomials in the $x^a_i$'s, homogeneous of degree
$h_2+2i_2-(h_1+2i_1)$, and
satisfy
\beq
\varphi^{a_1\cdots a_{h_2}}_{\,\,\,\,b_1\cdots b_{h_1}\!}\!(gx)
=g^{a_1}_{\,\,c_1}\cdots g^{a_{h_2}}_{\,\,c_{h_2}}\,
\varphi^{c_1\cdots c_{h_2}}_{\,\,\,\,d_1\cdots d_{h_1}\!}\!(x)\,
(g^{-1})^{d_1}_{\,\,b_1}\cdots
(g^{-1})^{d_{h_1}}_{\,\,b_{h_1}}\times(\det{g})^{i_2-i_1}.
\eeq
Possible forms of such $\varphi^{a_1a_2\cdots}_{\,\,\,\,b_1b_2\cdots}(x)$
can be determined by the theory of
algebraic invariants \cite{Weyl}.
As a guide, let us introduce contravariant and covariant variables
$\xi_a$ and $\eta^a$ with respect to $\GL_2(\C)$, and put
\beq
F(\xi,x,\eta):=\xi_{a_1}\cdots\xi_{a_{h_2}}
\varphi^{a_1\cdots a_{h_2}}_{\,\,\,\,b_1\cdots b_{h_1}\!}\!(x)\,
\eta^{b_1}\cdots \eta^{b_{h_1}}.
\eeq
It satisfies $F(\xi g^{-1}, gx,g\eta)=(\det g)^{i_2-i_1}F(\xi,x,\eta)$.
In particular, it is $\SL_2(\C)$ invariant.
By Theorem 2.6.A of \cite{Weyl}, it can be written as a polynomial of
$\xi x_i=\xi_a x^a_i$, $\xi\eta=\xi_a\eta^a$ ,
$[\eta x_j]=\epsilon_{ab}\eta^ax^b_j$
and $[x_ix_j]=\epsilon_{ab}x^a_ix^b_j$.
From the condition for $g\not\in\SL_2(\C)$, we see that the total
number of powers of $[\eta x_i]$ and $[x_ix_j]$ in each term must be
$i_2-i_1$. This shows that
\beq
\boxed{\begin{array}{l}
\mbox{$\varphi^{a_1\cdots a_{h_2}}_{\,\,\,\,b_1\cdots b_{h_1}}\!(x)$
is a sum of products of $x^{a_s}_i$, $\delta^{a_s}_{b_t}$,
$x_{b_tj}$ and $[x_ix_j]$,}\\
\mbox{~~which is symmetric and multilinear
  in $a_1,\ldots, a_{h_2}$ and $b_1,\ldots,b_{h_1}$,}\\
\mbox{~~where the
  total number of powers of $x_{b_tj}$ and $[x_ix_j]$ in each term
  is $i_2-i_1$.}
\end{array}}
\label{formofmael}
\eeq
In this description, we used
\beq
x_{ai}:=\epsilon_{ab}x^b_i
\label{xlowa}
\eeq
so that $[\eta x_j]=\eta^a x_{aj}$.
Note that there is a relation 
\beq
x^a_ix_{bj}-x^a_jx_{bi}=\delta^a_b[x_ix_j].
\label{fundrel}
\eeq
Thus, the expression for $\varphi^{a_1\cdots a_{h_2}}_{\,\,\,\,b_1\cdots b_{h_1}}\!(x)$
as in (\ref{formofmael}) is not unique.

\subsubsection{Matrix Factorization Completion}
\label{subsub:MFcompletion}

\newcommand{\Kap}{\texttt{KA}}

Let us introduce a set of irreducible representations of the gauge group:
\beq
\Kap~~~~
\begin{array}{cccccc}
  \C&\C(1)&\C(2)&\C(3)&\C(4)&\C(5)\\
  \fund&\fund(1)&\fund(2)&\fund(3)&\fund(4)&\\
  \sym^2\fund&\sym^2\fund(1)&\sym^2\fund(2)&\sym^2\fund(3)&&\\
  \sym^3\fund&\sym^3\fund(1)&\sym^3\fund(2)&&&\\
  \sym^4\fund&\sym^4\fund(1)&&&&\\
  \sym^5\fund&&&&&
\end{array}
\label{KapranovEC}
\eeq
It has Property EV${}_m$ for $m\geq 0$, as can be easily seen by applying
Borel-Weil-Bott (\ref{BWB}) to (\ref{ExtComp}).
This is indeed what was observed by Kapranov
in \cite{KapranovGr,KapranovHomog} as a part of a fundamental result:
the category $D(\Gr)$ is equivalent to
the homotopy category of bounded complexes of
direct sums of vector bundles in ${\mathcal O}_{\Gr}(\Kap)$.

This result can be used to complete the open steps of
Section~\ref{subsec:Gflow} in
the construction of a gauge theory lift of a sigma model brane.\footnote{As
  we shall see, higher Ext vanishing between certain pairs of sheaves will
  be the key. This was missed in \cite{LinRomo}.}
We started with ${\mathcal E}\in D(X)$
and pushed it forward to $i_{X*}{\mathcal E}\in D(\Gr)$.
By the fundamental result of \cite{KapranovGr,KapranovHomog}, 
$i_{X*}{\mathcal E}$ is isomorphisc in $D(\Gr)$ to
${\mathcal O}_{\Gr}(\Complex)$ where
$\Complex=(\bV_{\Complex}, d_{\Complex})$
(or $(M_{\Complex},d_{\Complex})$ with
$M_{\Complex}=\bV_{\Complex}\otimes^{}_{\C}\C[x]$)
is a bounded complex of representations
over $\C[x]$ of the gauge group
whose terms are direct sums of members of $\Kap$.
We shall refer to
such an ${\mathcal O}_{\Gr}(\Complex)$ (and also to $\Complex$)
a $\Kap$-resolution of $i_{X*}{\mathcal E}$.
We would like to extend $d_{\Complex}$ to a matrix factorization $Q$ of
$\sum_{k=1}^7p^kA_k(x)$.

Let us first consider the case where $\mathcal{E}$ is a single term complex,
that is, a coherent sheaf on $X$.
A treatment of this case is basically given in \cite{HHP}.
Note that $(\bV_{\Complex},d_{\Complex})=(\bV_{\Complex}, \rho_{\Complex *}, {\bf r}_{\Complex *},d_{\Complex})$
already satisfies gauge invariance (\ref{Qgi})
and R-symmetry (\ref{QRs}) with charge integrality (\ref{CI}).
We make the ansatz
\begin{eqnarray}
  Q(p,x)=d_{\Complex}(x)+\underbrace{\sum_{k=1}^7p^kQ_k^{(1)}(x)}_{Q^{(1)}}
  +\underbrace{\sum_{k_1,k_2=1}^7
  p^{k_1}p^{k_2}Q_{k_1k_2}^{(2)}(x)}_{Q^{(2)}}+\cdots
\end{eqnarray}
$Q^{(n)}_{k_1\ldots k_n}(x)$ is a polynomial function of $x=(x^a_i)$
with values in $\mathrm{End}^{\mathrm{od}}(\bV_{\Complex})$.
Gauge invariance (\ref{Qgi}) and R-symmetry (\ref{QRs}) of $Q$
require
\beqa
&&\rho_{\Complex}(g)^{-1}Q^{(n)}_{k_1\ldots k_n}(gx)\rho_{\Complex}(g)
  =(\det g)^nQ^{(n)}_{k_1\ldots k_n}(x),\label{giLift}\\
&&\lambda^{{\bf r}_{\Complex *}}Q^{(n)}_{k_1\ldots k_n}(x)
  \lambda^{-{\bf r}_{\Complex *}}=\lambda^{1-2n}Q^{(n)}_{k_1\ldots k_n}(x).
  \label{RsLift}
\eeqa
The condition for the matrix factorization splits to
\beqa
\{d_{\Complex},Q^{(1)}_k\}&=&A_k\cdot {\rm id}_{M_{\Complex}},\label{mfdeg1}\\
\{d_{\Complex},Q^{(n)}\}&=&-\sum_{n_1+n_2=n}Q^{(n_1)}Q^{(n_2)},\qquad n\geq 2.
\label{mfdegn}
\eeqa
We shall find the $Q^{(n)}_{k_1\ldots k_n}$'s recursively starting with the
$Q^{(1)}_k$'s. Multiplication by $A_k(x)$ defines a cochain map
$\Complex \to \Complex(1)$.
Since $i_{X*}{\mathcal E}$ is supported at $A_1(x)=\cdots =A_7(x)=0$,
this corresponds to the zero map $i_{X*}{\mathcal E}\to i_{X*}{\mathcal E}(1)$.
Since $\Kap$ has Property EV${}_1$,
by the Lemma in Section~\ref{subsub:HA}-III,
this means that this cochain map has to be null homotopic.
That is, there is a map
$Q^{(1)}_k:\Complex\to \Complex(1)$ of degree $-1$ such that
$\{d_{\Complex},Q^{(1)}_k\}$ equals this multiplication map. 
We have found $Q^{(1)}_k(x)$ satisfying (\ref{giLift}) and
(\ref{RsLift}) for $n=1$ and (\ref{mfdeg1}).
The rest are found by induction. Suppose we found
$Q^{(2)},\ldots, Q^{(m-1)}$ that satisfy the three conditions
for all $n$ below $m$.
Let us denote the right-hand side of (\ref{mfdegn})
for $n=m$ by
\beq
{\rm RHS}^{(m)}=\sum_{k_1,\ldots, k_m=1}^7p^{k_1}\cdots p^{k_m}
{\rm RHS}^{(m)}_{k_1\cdots k_m}.
\eeq
By the induction assumption, each coefficient
defines a map ${\rm RHS}^{(m)}_{k_1\ldots k_m}:\Complex\to \Complex(m)$
of degree $2-2m$. Also,
\beqa
[d_{\Complex}, {\rm RHS}^{(m)}]
&=&-\sum_{n+p=m}[d_{\Complex},Q^{(n)}Q^{(p)}]\nn\\
&=&
-\sum_{n+p=m}\Big(\underbrace{\{d_{\Complex},Q^{(n)}\}
}_{-\sum_{k+l=n}Q^{(k)}Q^{(l)}}\!\!\!\!\!\!Q^{(p)}
-Q^{(n)}\!\!\!\!\!\!\!\underbrace{\{d_{\Complex},Q^{(p)}\}
}_{-\sum_{q+r=p}Q^{(q)}Q^{(r)}}\Bigr)
\nn\\
&=&\sum_{k+l+p=m}Q^{(k)}Q^{(l)}Q^{(p)}-\sum_{n+q+r=m}Q^{(n)}Q^{(q)}Q^{(r)}~=~0.
\eeqa
Each coefficient ${\rm RHS}^{(m)}_{k_1\ldots k_m}$
of course commutes with $d_{\Complex}$.
This means that it is a cochain map and defines
an element of ${\rm Ext}^{2-2m}_{\Gr}(i_{X*}{\mathcal E},i_{X*}{\mathcal E}(m))$.
However, this group is zero for $m\geq 2$ since $2-2m<0$.
(This is where the assumption that ${\mathcal E}$ is a single term
complex is used.)
Again using Property EV${}_m$ of $\Kap$ and
the Lemma in Section~\ref{subsub:HA}-III,
we find that the cochain map ${\rm RHS}^{(m)}_{k_1\ldots k_m}$
has to be null homotopic.
That is, there is a map
$Q^{(m)}_{k_1\ldots k_m}:\Complex\to \Complex(m)$ of degree $1-2m$
such that
$\{d_{\Complex}, Q^{(m)}_{k_1\ldots k_m}\}={\rm RHS}^{(m)}_{k_1\ldots k_m}$.
We have found
$Q^{(m)}=\sum_{k_1\ldots k_m}p^{k_1}\cdots p^{k_m}Q^{(m)}_{k_1\ldots k_m}$
satisfying all the three conditions.
This process terminates at some point --- RHS${}^{(m)}$
vanishes at large enough $m$.
This is because RHS${}^{(m)}_{k_1\cdots k_m}$ has degree $2-2m$
and has nowhere to go if $m$ is too large, as
the complex $\Complex$ is bounded.

Let us next consider the problem to lift
a homomorphism $\varphi:{\mathcal E}_1\to {\mathcal E}_2$
of coherent sheaves on $X$ to a homomorphisms
$f:(M_{\Complex_1},Q_1)\to (M_{\Complex_2},Q_2)$ of the corresponding
matrix factorizations.
Let $f_{\Complex}:\Complex_1\to \Complex_2$ be a homomorphism
that corresponds to
$i_{X*}\varphi:i_{X_*}{\mathcal E}_1\to i_{X*}{\mathcal E}_2$
under the equivalence of \cite{KapranovGr,KapranovHomog}.
(We sall call $f_{\Complex}$ a $\Kap$-resolution of $i_{X*}\varphi$.)
We try to find a completion of $f_{\Complex}$ of the form
\beq
f(p,x)=f_{\Complex}(x)+\sum_{k=1}^7p^kf^{(1)}(x)
+\sum_{k_1,k_2=1}^7p^{k_1}p^{k_2}f^{(2)}_{k_1k_2}(x)+\cdots
\label{homcompletion}
\eeq
satisfying $Q_2f=fQ_1$, 
where $f^{(m)}_{k_1\cdots k_m}(x)$ are required to satisfy the gauge invariance
and R-symmetry conditions, which can be stated in a form similar to
(\ref{giLift}) and (\ref{RsLift}).
The $f^{(m)}_{k_1\cdots k_m}(x)$'s
can be found recursively in almost the same way
as in the case of finding the $Q^{(m)}_{k_1\ldots k_m}$'s,
using Property EV${}_{m\geq 1}$ of $\Kap$.
The only difference is that the corresponding RHS${}^{(m)}$
represents an element of
${\rm Ext}^{-2m}_{\Gr}(i_{X*}{\mathcal E}_1,i_{X*}{\mathcal E}_2(m))$
which vanishes for all $m\geq 1$: The recursive argument works uniformly from
the first step $m=1$ in the present case.

This can be applied immediately to find a lift of a two term complex,
${\mathcal E}_1\stackrel{\varphi}{\longrightarrow}{\mathcal E}_2$.
The cone of $f_{\Complex}:\Complex_1\to\Complex_2$ 
provides a $\Kap$-resolution of the pushed forward complex,
and its matrix factorization completion
is given by the cone of $f:(M_{\Complex_1},Q_1)\to (M_{\Complex_2},Q_2)$.
Let us next discuss the case of a three term complex
${\mathcal E}_1\stackrel{\varphi_{12}}{\longrightarrow}{\mathcal E}_2
\stackrel{\varphi_{23}}{\longrightarrow}{\mathcal E}_3$.
Let $f_{\Complex_{ij}}:\Complex_i\to\Complex_j$
be a $\Kap$-resolution of
$i_{X*}\varphi_{ij}:i_{X*}{\mathcal E}_i\to i_{X*}{\mathcal E}_j$,
and let $f_{ij}:(M_{\Complex_i},Q_i)\to (M_{\Complex_j},Q_j)$
be its completion to a homomorphism of matrix factorizations.
Since the composition of $\varphi_{12}$ and $\varphi_{23}$ vanishes,
so does the composition of $i_{X*}\varphi_{12}$ and $i_{X*}\varphi_{23}$,
and hence, by Property EV${}_{m\geq 1}$ of $\Kap$,
the composition of
$f_{\Complex_{12}}$ and $f_{\Complex_{23}}$ is null-homotopic.
By a now obvious method, the homotopy can be completed to a homotopy
of the composition of the homomorphisms $f_{12}$ and $f_{23}$
of matrix factorizations.
Then, we can define the successive cone of
$f_{12}$ and $f_{23}$ using the homotopy,
as described in Section \ref{subsub:HA}-I.
This provides the wanted lift.
The construction for a complex of arbitrary length is a straightforward
generalization.

\noindent
{\bf Remarks.} {\bf (1)} What has been said holds equally for any shift
$\Kap(i)$ of $\Kap$.\\
{\bf (2)}
The above is only one of many different ways to lift ${\mathcal E}\in D(X)$
to a matrix factorization.
For example, the Koszul resolution (\ref{KosOX}) of $i_{X*}{\mathcal O}_X$
is made of the representations
$\C(-7),\ldots,\C(-1),\C$ which are obviously not in $\Kap$ nor its shifts
$\Kap(i)$. Existence of its matrix factorization completion
is guaranteed also by the above argument
since this set has Property EV${}_m$ for $m\geq 1$,
as one can see from the fact that it is in
$\rmGRR_{\bf w}\cup\rmGRR_{\bf w}(1)$ for some ${\bf w}$.
As in this example, Property EV${}_m$ with $m\geq 1$ is key to
guarantee a completion. However, this is a sufficient condition but not
a necessary one
--- we will later encounter examples of
complexes violating the condition but with
matrix factorization completions.

\subsubsection{On The Equivalence
  $\pi_X:{\mathcal T}_{\bf w}\stackrel{\cong}{\longrightarrow} D(X)$}

For a set $\SetR$ of representations of $U(2)$,
let ${\mathcal T}_{\SetR}\subset D_G(V,W)$ be the set of brane data
whose Chan-Paton vector spaces are direct sums of members of $\SetR$.
Note that ${\mathcal T}_{\rmGRR_{\bf w}}={\mathcal T}_{\bf w}$.
If $\SetR$ has Property EV${}_{m}$ for $m\geq 0$,
the Higgsing flow $D_G(V,W)\to D(\,\fb,W_{\rm ind})$
restricted to ${\mathcal T}_{\SetR}$,
regarded as a functor, is fully faithful \cite{ADS}.
As the mass flow $D(\,\fb,W_{\rm ind})\to D(X)$ is an equivalence
by Kn\"orrer periodicity,
the entire flow $\pi_X:{\mathcal T}_{\rm R}\to D(X)$
is also fully faithful.
This means that the flow $\pi_X:{\mathcal T}_{\SetR}\to D(X)$ is
essentially injective.
For example, this is so for $\SetR=\Kap$ and
$\rmGRR_{\bf w}$ for any ${\bf w}$.

Moreover, in the case of ${\mathcal T}_{\,\Kap}$ the flow is
also essentially surjective.
Indeed, we have just seen that any object of $D(X)$ has a lift in
${\mathcal T}_{\,\Kap}$.
The main property of $\Kap$ that makes it possible is that
${\mathcal O}_{\Gr}(\Kap)$ {\it generates} the category $D(\Gr)$
via homological shifts, direct sums, and cones.
In particular the flow
$\pi_X:{\mathcal T}_{\,\Kap}\to D(X)$ is an equivalence of categories.

What about ${\mathcal T}_{\bf w}$?
In fact, the grade restricted set for the window ${\bf C}$ (\ref{GRRC})
coincides with $\RADS$, and it was shown by Kuznetsov in \cite{KuznetsovEC}
that ${\mathcal O}_{\Gr}(\RADS)$ generates $D(\Gr)$.
Therefore $\pi_X:{\mathcal T}_{\bf C}\to D(X)$ is an equivalence,
which is nothing but the result of \cite{ADS}.
We claim that the same holds also for other windows,
${\bf A}$, ${\bf B}$, ${\bf D}$. (This was also shown in \cite{DonovanPfGr}
after the input of the present work.)
In the rest of this section, we shall 
prove generation of $D(\Gr)$ for all the windows (including ${\bf C}$)
starting with $\Kap$.
We shall do so manually by constructing and employing empty branes
in the Grassmannian phase building on a known class of exact sequences
of vector bundles on the Grassmann manifold $\Gr$.
These empty branes will also play important r\^oles later in this paper,
especially in the study of low enegy behaviour of branes
in the Pfaffian phase.


\noindent
{\bf Terminology.}
A set of objects of $D(\Gr)$ is said to be {\it full} when it generates
the category $D(\Gr)$ in the sense described above. Thus,
what we are going to show is that ${\mathcal O}_{\Gr}(\rmGRR_{\bf w})$
form {\it full and strong exceptional collections} of $D(\Gr)$
as $\Kap$ does.

\subsection{Empty Branes}
\label{subsec:Gempty}

\begin{figure}[thb]
\psfrag{0m}{\footnotesize $\C$(-1)}
\psfrag{00}{\footnotesize $\C(0)$}
\psfrag{01}{\footnotesize $\C(1)$}
\psfrag{02}{\footnotesize $\C(2)$}
\psfrag{03}{\footnotesize $\C(3)$}
\psfrag{04}{\footnotesize $\C(4)$}
\psfrag{1m}{\footnotesize $\fund^*(0$)}
\psfrag{10}{\footnotesize $\fund^*(1)$}
\psfrag{11}{\footnotesize $\fund^*(2)$}
\psfrag{12}{\footnotesize $\fund^*(3)$}
\psfrag{13}{\footnotesize $\fund^*(4)$}
\psfrag{2mm}{\footnotesize $S^2\fund^*$(0)}
\psfrag{2m}{\footnotesize $S^2\fund^*(1)$}
\psfrag{20}{\footnotesize \!\!\!$S^2\fund^*(2)$}
\psfrag{21}{\footnotesize \!\!\!$S^2\fund^*(3)$}
\psfrag{22}{\footnotesize \!\!\!$S^2\fund^*(4)$}
\psfrag{23}{\footnotesize \!\!\!$S^2\fund^*(5)$}
\psfrag{3mm}{\footnotesize $S^3\fund^*(1)$}
\psfrag{3m}{\footnotesize $S^3\fund^*(2)$}
\psfrag{30}{\footnotesize \!\!\!$S^3\fund^*(3)$}
\psfrag{31}{\footnotesize \!\!\!$S^3\fund^*(4)$}
\psfrag{32}{\footnotesize \!\!\!$S^3\fund^*(5)$}
\centerline{\includegraphics[page=13]{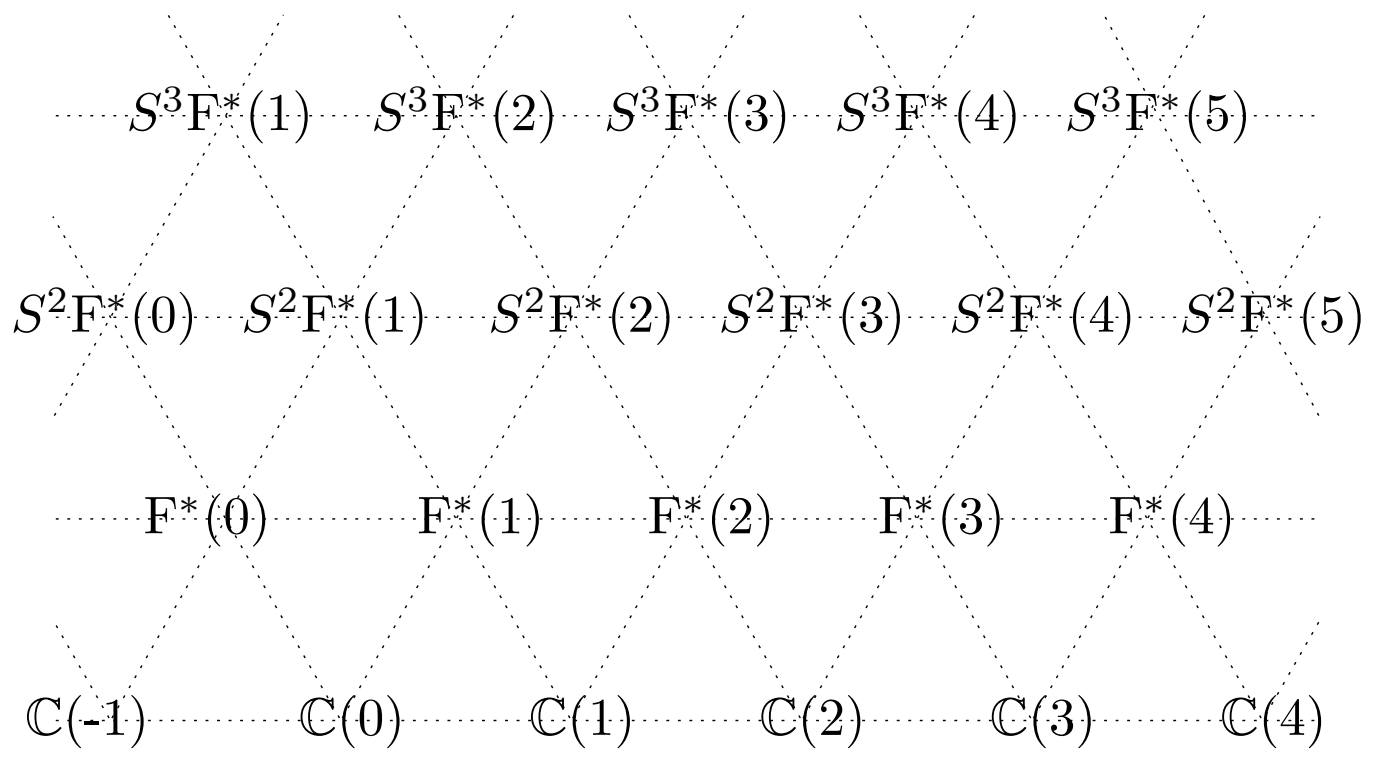}}
\caption{Representations of $U(2)$ on a triangular grid.}
\label{fig:grid}
\end{figure}

Let us introduce empty branes in the Grassmannian phase,
that is, branes in the R\o dland model that can be ignored
in the low energy theory at $\zeta\gg 0$
by complete brane-antibrane annihilation.
For this purpose, we introduce a system to present representations of
the gauge group $U(2)$ on a triangular grid as in Fig.~\ref{fig:grid}.
The representations of symmetric power $h$ of $\fund^*$ are placed
at the sites of height $h$ in the grid, while the tensor product with
$\det\fund=\C(1)$ corresponds to a one step shift to the right.

\subsubsection{Twisted Lascoux Complexes}

Donovan and Segal \cite{DS,Donovan} introduced certain
$\GL_k(\C)\times\GL_N(\C)$-equivariant complexes of free
$\C[\C^k\otimes\C^N]$-modules, called the twisted Lascoux complexes,
based on earlier works on resolutions of determinantal varieties
\cite{Lascoux,Weyman}.
We describe them for the case $k=2$ relevant for us. 

As before, $\fund$ is the two-dimensional complex vector space
on which the gauge group $U(2)$ is represented.
Let $V$ be an external $N$ dimensional complex vector space.
A twisted Lascoux complex is a complex of 
$\GL(\fund)\times \GL(V)$ representations over the polynomial ring
$\C[x]$ of coordinates $x=(x^a_i)$ of $\fund\otimes V$.
It has $N$ or $N+1$ terms, each of which is of the form
\beq
\Sym^h\fund^*(i)\otimes \textw^{\!j\,}V^*
\label{Lascouxcomponents}
\eeq
for some $(h,i,j)$.
It starts with a term of $j=N$ and ends with one of $j=0$.\footnote{There are
  exceptions. See the next footnote.}
It has at most three parts: 
left $\to$ middle $\to$ right. 
The left part is a sequence of arrows of the type
$(h,i,j)\to (h-1,i,j-1)$, 
the middle part is a single arrow of the type $(0,i,j)\to (0,i+1,j-2)$, 
and the right part is a sequence of arrows of the type
$(h,i,j)\to (h+1,i+1,j-1)$.
In particular, it has $N$ terms when it involves the middle part,
while it has $(N+1)$ terms if it does not, i.e., when it consists only of
the left part or the right part.\footnote{There are exceptions:
  (a) left-part-only starting with $(h,j)=(N-1,N)$ has only $N$ terms and
  ends with $(h,j)=(0,1)$, and
  (b) right-part-only ending with $(h,j)=(N-1,0)$ has only
  $N$ terms and starts with $(h,j)=(0,N-1)$.
  These may be regarded as non-exceptional by adding $\{0\}$ as an
  extra term: (a) add extra end with $(h,j)=(-1,0)$, and
  (b) add extra start with $(h,j)=(-1,N)$, where we regard
  $\Sym^{-1}\fund^*=\{0\}$.}

\begin{figure}[htb]
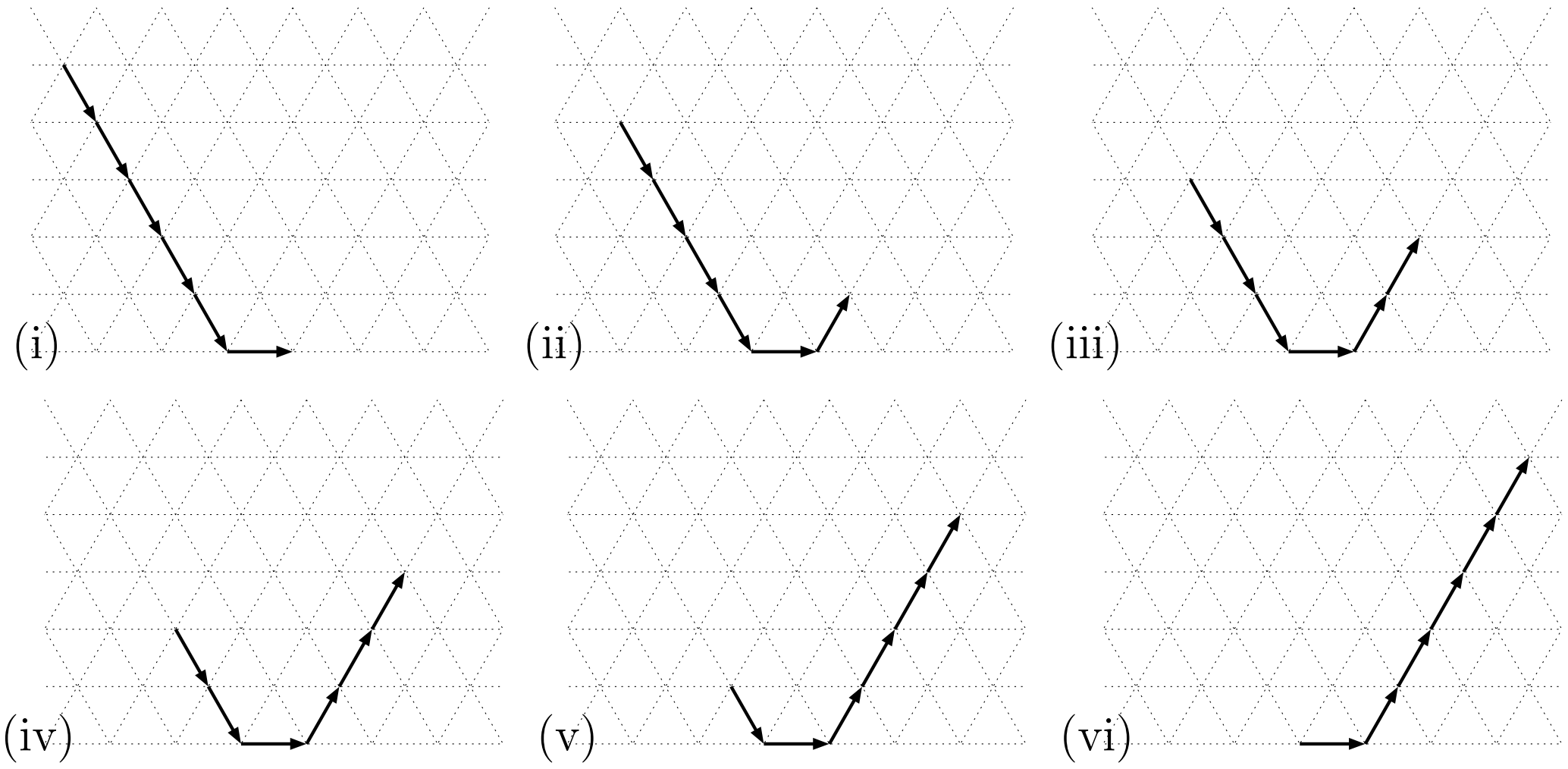

\centerline{(i)\!\!\!\!\includegraphics[page=14]{main-pics.pdf}~~
(ii)\!\!\!\!\includegraphics[page=15]{main-pics.pdf}~~
(iii)\!\!\!\!\includegraphics[page=16]{main-pics.pdf}}

\medskip
\centerline{(iv)\!\!\!\!\includegraphics[page=17]{main-pics.pdf}~~
(v)\!\!\!\!\includegraphics[page=18]{main-pics.pdf}~~
(vi)\!\!\!\!\includegraphics[page=19]{main-pics.pdf}}
\caption{Some twisted Lascoux complexes for $(k,N)=(2,7)$}
\label{fig:Lc}
\end{figure}
It can be presented in a triangular grid as in Fig.~\ref{fig:grid}.
The left part is a sequence of unit-length down-right
arrows, the middle part is a unit-length horizontal right arrow at the bottom,
and the right part is a sequence of unit-length up-right arrows.
In Fig.~\ref{fig:Lc}, we present
those with middle part for the case $(k,N)=(2,7)$.
For example, (iv) in Fig.~\ref{fig:Lc} with $i=0$ at the tail
is a complex of the form
\beqa
&\begin{tikzcd}
  \Sym^2\fund^*\otimes \textw^{\!7\,} V^* \arrow[r, shorten=2mm]&
  \fund^*\otimes \textw^{\!6\,} V^* \arrow[r, shorten=2mm]&
  \C\otimes \textw^{\!5\,}V^*\arrow[r, shorten=2mm]&
  \C(1)\otimes \textw^{\!3\,}V^*
\end{tikzcd}\nn\\
&\begin{tikzcd}
 ~~ \arrow[r, shorten=2mm]&
  \fund^*(2)\otimes\textw^{\!2\,}V^*
  \arrow[r, shorten=2mm]&
  \Sym^2\fund^*(3)\otimes V^*\arrow[r, shorten=2mm]&
  \Sym^3\fund^*(4)
 \end{tikzcd}
\label{exLc}
\eeqa

\newcommand{\bigast}{(\cdots)}

To describe the arrows, it is convenient to introduce a
Weyl-Clifford algebra.
Let $\{e_a\}_{a=1,2}$ and $\{\eta^i\}_{i=1}^N$
be bases of $\fund$ and $V$ respectively, and
$\{e^a\}_{a=1,2}\subset \fund^*$ and $\{\bareta_i\}_{i=1}^N\subset V^*$
be their duals.
We let them have the (anti)commutation relations
\beqa
&&[e_a,e^b]=\delta^b_a, \quad
[e_a,e_b]=[e^a,e^b]=0,\label{Heisenberg}\\
&&\{\eta^i,\bareta_j\}=\delta^i_j,\quad
\{\eta^i,\eta^j\}=\{\bareta_i,\bareta_j\}=0.
\label{Clifford1}
\eeqa
Then, $\Sym\, \fund^*(i)\otimes \textw_{} V^*$ for each $i\in \Z$
can be regarded as a representation of this algebra
with a ``vacuum'' vector $|0\rangle_i$ annihilated
by the $e_a$'s and $\eta^i$'s which is $\GL(V)$ invariant but transforms
as $\C(i)$ under $\GL(\fund)$.
The arrows $(\searrow)$, $(\to)$, $(\nearrow)$ of the left, middle,
right parts, respectively, are
\beqa
(\searrow)&=&x_i^ae_a\eta^i\label{Lse}\\
(\to)&=&\half[x_ix_j]\eta^i\eta^j\label{Le}\\
(\nearrow)&=&x_{ai}e^a\eta^i\label{Lne}
\eeqa
where $x_{ai}:=\epsilon_{ab}x^b_i$ as in (\ref{xlowa}).
This description is somewhat informal.
The precise meaning can be found by noting that
the arrows $(\searrow)$ preserve
the $i$-label of (\ref{Lascouxcomponents}) while
the arrows $(\to)$ and $(\nearrow)$ increase the $i$-label by $+1$.
That is,
\beqa
(\searrow):&&\!\!\!\!\!\!\!
\bigast |0\rangle_i\,\longmapsto\, x_i^ae_a\eta^i\,\bigast |0\rangle_i,\\
(\to):&&\!\!\!\!\!\!\!
\bigast |0\rangle_i\,\longmapsto\,
\half[x_ix_j]\eta^i\eta^j\,\bigast |0\rangle_{i+1},\\
(\nearrow):&&\!\!\!\!\!\!\!
\bigast |0\rangle_i\,\longmapsto\,
x_{ai}e^a\eta^i\,\bigast |0\rangle_{i+1},
\eeqa
in which $\bigast$ are polynomials of the $x^a_i$'s, $e^a$'s and $\bareta_i$'s
of the specified degrees in the latter two.
In the present paper, in order to avoid notational complexity,
we shall use this kind of informal description of the arrows
in complexes and/or matrix factorizations.
It is easy to see that the arrows commute with $\GL(\fund)\times \GL(V)$.
In particular, if the complex is presented as the pair $(\bV,Q_0)$
where $\bV$ is the underlying Chan-Paton vector space and
$Q_0$ is the collection of the arrows, it satisfies the gauge invariance
condition as in (\ref{Qgi}).
As it is a complex, the R-charges of the components are assigned so that
each arrow increases them by $+1$.
The R-symmetry condition (\ref{QRs}) is realized by
the following R-charge assignment:
the $\eta^i$'s and $\bareta_i$'s have R-charges $+1$ and $-1$ respectively
while the (single) ``vacuum'' vector $|0\rangle_i$ for the components of
the left part has one higher R-charge compared to the ``vacuum'' vectors
$|0\rangle_i$'s for the components of the right part.
The statistics of these vacuum vectors are opposite accordingly,
which ensures that $Q_0$ is fermionic even though (\ref{Le})
appears to be bosonic.
One can also see that $Q_0$ squares to zero, $Q_0^2=0$.
At the $h>0$ positions, we use commutativity of the $e_a$'s or of the $e^a$'s
and anticommutativity of the $\eta^i$'s.
At the $h=0$ positions, we use the fact that the antisymmetrization of
$x^a_ix^b_jx^c_k$ in $i,j,k$ vanishes since $a,b,c$ can take only two values.

The representation of the Weyl-Clifford algebra is unitary:
there is a Hermitian inner product on $\Sym\, \fund^*(i)\otimes \textw_{} V^*$
such that $(e_a)^{\dag}=e^a$ and $(\eta^i)^{\dag}=\bareta_i$.
And the inner product is induced on the Chan-Paton vector space
$\bV\subset \bigoplus_i\Sym\, \fund^*(i)\otimes \textw_{} V^*$.
One can check that $\{Q_0,Q_0^{\dag}\}$ is positive definite
if the $2\times n$ matrix $(x^a_{\,i})$ has maximal rank.
This means that that complex is exact for a fixed $x=(x^a_i)$
of maximal rank. In particular, the associated
complex of vector bundles on $G(2,V)$ is exact.

More importantly, when regarded as a complex of
$\C[x]$-modules, the twisted Lascoux complex is acyclic.
For example, the complex (\ref{exLc})
is exact except at the last term $\Sym^3\fund^*(4)$.
This is because the complex is a free resolution of a certain
$\C[x]$-module which is supported at the locus where 
$(x^a_i)$ does not have maximal rank (\cite{DS}, Theorem A.7).
By the extension theorem in Section~\ref{subsub:HA}-II,
this fact allows us to use the twisted Lascoux complexes
(or their descendants which we shall describe)
in the grade restriction process in which
a given complex or a brane is replaced by a ``better'' representative.

\subsubsection{Matrix Factorization Completions}

Let $(\bV,Q_0)$ be one of the twisted Lascoux complexes with $N=7$.
It has a matrix factorization completion:
We can add arrows $\Delta Q$ so that $Q=Q_0+\Delta Q$
is a matrix factorization of $W=A^{ij}(p)[x_ix_j]$.
The additional arrow $\Delta Q$
consists of a left part, a middle part, and a right part, 
\begin{figure}[htb]
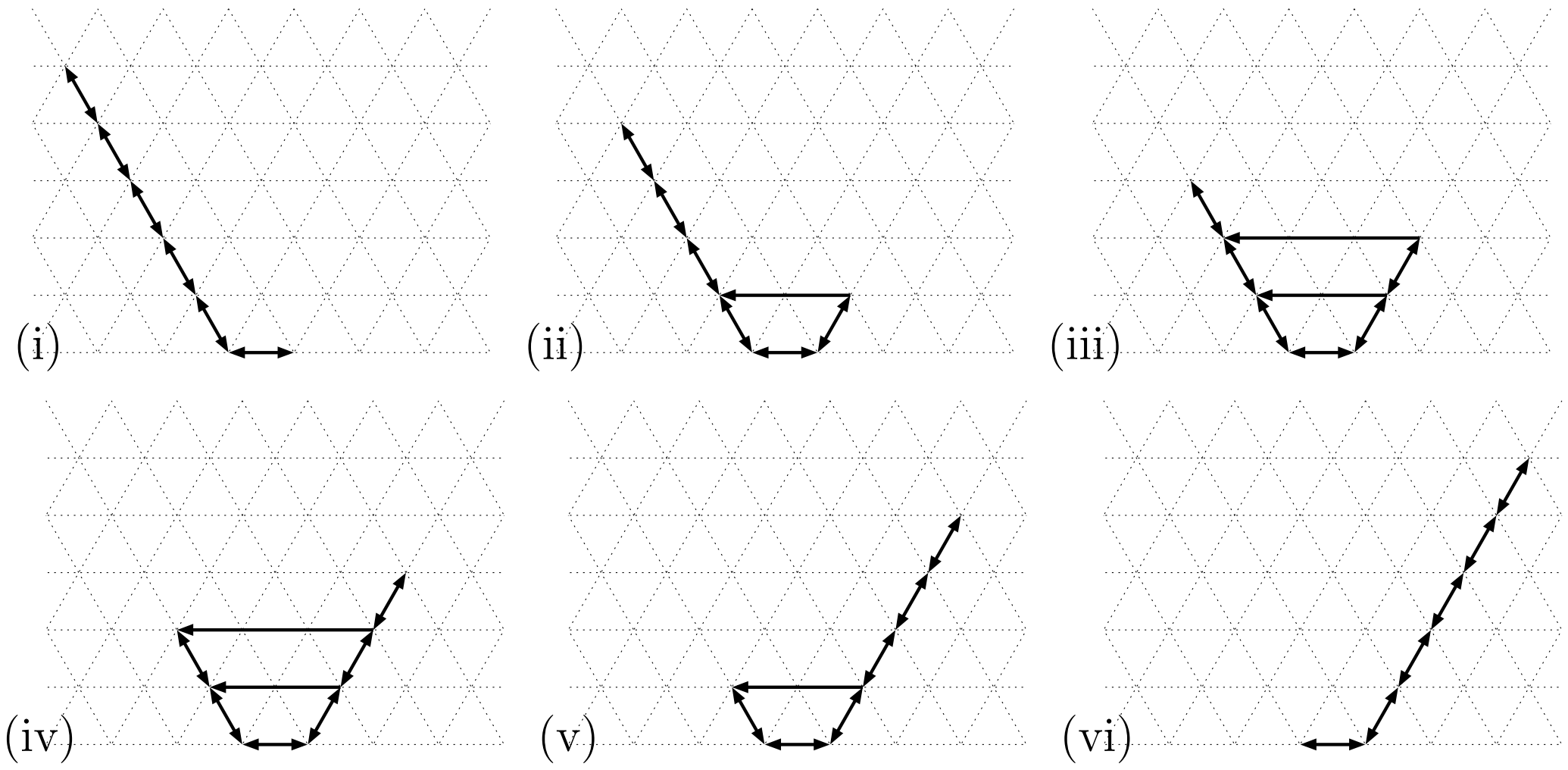

\centerline{(i)\!\!\!\!\includegraphics[page=20]{main-pics.pdf}~~
(ii)\!\!\!\!\includegraphics[page=21]{main-pics.pdf}~~
(iii)\!\!\!\!\includegraphics[page=22]{main-pics.pdf}}

\medskip
\centerline{(iv)\!\!\!\!\includegraphics[page=23]{main-pics.pdf}~~
(v)\!\!\!\!\includegraphics[page=24]{main-pics.pdf}~~
(vi)\!\!\!\!\includegraphics[page=25]{main-pics.pdf}}
\caption{The matrix factorization completions of Fig.~\ref{fig:Lc} complexes}
\label{fig:Lmf}
\end{figure}
where the left (right) part is a sequence of arrows in
the opposite direction to the left (right) part of $Q_0$, 
and the middle part consists
of horizontal arrows from the right part to the left part
(the arrow is present only when its
head and tail existed in the original Lascoux complex).
See Fig.~\ref{fig:Lmf} for the matrix factorization completions of
the examples of Fig.~\ref{fig:Lc}.
Each arrow can again be described by the Weyl-Clifford algebra.
The expressions for the arrows $(\nwarrow)$, $(\leftarrow)$, $(\swarrow)$
in the left, middle and right parts, respectively,
depend on the height $h$ and are given by
\beqa
(h\nwarrow h-1)&=&-{2\over h+1}A^{ij}(p)x_{ia}e^a\bareta_j
+{1\over h(h+1)}A^{ij}(p)x_{ka}e^a\bareta_i\bareta_j\eta^k
\label{Lnw}\\
(h\leftarrow h)&=&
     {(-1)^{h+1}\over (h+1)(h!)^2}\widehat{A}(p)^{h+1}
     \label{Lw}\\
(h-1\swarrow h)&=&{2\over h+1}A^{ij}(p)x_i^ae_a\bareta_j
+{1\over h(h+1)}A^{ij}(p)x_k^ae_a\eta^k\bareta_i\bareta_j
\label{Lsw}
\eeqa
where $\widehat{A}(p)=A^{ij}(p)\bareta_i\bareta_j$.
It is easy to see that the gauge invariance
and R-symmetry conditions,  (\ref{Qgi}) and (\ref{QRs}), are satisfied.
It is also fairly straightforward to see that 
$Q=Q_0+\Delta Q$ squares to $W=A^{ij}(p)[x_ix_j]$.
Thus, we obtain a brane in the gauged linear sigma model
which we refer to as a
{\it Lascoux matrix factorization} or
{\it Lascoux brane} in short.

One can also see that $\{Q,Q^{\dag}\}$ is still positive definite
on the terms involved when $(x^a_i)$ is of maximal rank. Thus,
the Lascoux branes are indeed empty in the Grassmannian phase.
Compatibly with this, the hemisphere partition functions of these branes are
all vanishing 
\beq
Z_{D^2}({\rm Lascoux~ brane})=0\quad \mbox{at $\zeta\gg 0$.}
\label{Lbvanishing}
\eeq

\medskip
\noindent
{\bf Remark.}
The complex ${\mathcal C}={\mathcal O}_{\Gr}(\bV,Q_0)$
of vector bundles on $\Gr$ is exact and hence
$\Hom_{D(\Gr)}({\mathcal C},{\mathcal C}(m)[2-2m])=0$ for any $m$.
Therefore, provided that the set of terms of $\bV$ has
Property EV${}_m$ for $m\geq 1$, existence of a matrix factorization is
guaranteed by the argument of Section~\ref{subsub:MFcompletion}.
That is indeed the case for those whose terms have
height at most $h=6$, (e.g. all of Fig.~\ref{fig:Lc}).
However, it is not the case for those with height $h\geq 7$ term(s),
as $\Ext^5_{\Gr}(E,E(1))\ne 0$ for $E=\Sym^hS^*(i)$ for $h\geq 7$,
which can be seen easily from Section~\ref{subsub:BWB}.
The above construction shows that 
a matrix factorization completion may exist even if the argumant fails.

\subsubsection{The Second Version}
\label{subsub:otherversion}

It is useful to introduce another version of Lascoux complexes and
matrix factorizations. This is obtained by taking the tensor product with
$\textw^{\!N}V$ and using
$\textw^jV^*\otimes \textw^{\!N} V\cong \textw^{\!N-j}V$.
Then, the terms are of the form
\beq
\Sym^h\fund^*(i)\otimes \textw^{\!j\,}V
\label{otherLascouxcomponents}
\eeq
and a Lascoux complex starts from $j=0$ and ends with $j=N$.
For example, the tensor product of (\ref{exLc}) with $\textw^{\!7}V$ is
\beqa
&\begin{tikzcd}
  \Sym^2\fund^*\arrow[r, shorten=2mm]&
  \fund^*\otimes V \arrow[r, shorten=2mm]&
  \C\otimes \textw^{\!2\,}V\arrow[r, shorten=2mm]&
  \C(1)\otimes \textw^{\!4\,}V
\end{tikzcd}\nn\\
&\begin{tikzcd}
 ~~ \arrow[r, shorten=2mm]&
  \fund^*(2)\otimes\textw^{\!5\,}V
  \arrow[r, shorten=2mm]&
  \Sym^2\fund^*(3)\otimes \textw^{\!6\,}V\arrow[r, shorten=2mm]&
  \Sym^3\fund^*(4)\otimes \textw^{\!7\,}V.
 \end{tikzcd}
\label{otherexLc}
\eeqa
For this version, the differentials are described by
(\ref{Heisenberg}) and
\beq
\{\eta_i,\bareta^j\}=\delta^j_i,\quad
\{\eta_i,\eta_j\}=\{\bareta^i,\bareta^j\}=0,
\label{Clifford2}
\eeq
where $\{\bareta^i\}\subset V$ 
and $\{\eta_i\}\subset V^*$
play the r\^ole of the creation and annihilation operators, respectively.
The arrows are the same as
(\ref{Lse}), (\ref{Le}), (\ref{Lne}), (\ref{Lnw}), (\ref{Lw})
and (\ref{Lsw}) provided we make the replacement
\beq
(\eta^i,\bareta_i)~\longrightarrow~(\bareta^i,\eta_i).
\eeq
This version of Lascoux complexes and matrix factorizations of the type
that appear in Fig.~\ref{fig:Lc} and Fig.~\ref{fig:Lmf}
will be denoted by
\beq
\mbox{(i)', (ii)', (iii)', (iv)', (v)', (vi)'.}
\label{otherLcLb}
\eeq
Of course, this version is also acyclic
as a complex of $\C[x]$-modules and also determines
an exact complex of vector bundles on $\Gr$.

It is useful to note that the transpose of a Lascoux complex or
matrix factorization is isomorphic to a Lascoux complex or matrix factorization.
For example, the transpose of those of type (i), (ii), (iii), (iv), (v), (vi)
are isomorphic to those of type (vi)', (v)', (iv)', (iii)', (ii)', (i)',
respectively.
This fact will be especially important in the grade restriction process
in view of the Corollary in Section~\ref{subsub:HA}-II.

\subsection{Grade Restriction Process}
\label{subsec:GRprocess}

Let us describe the grade restriction process
which replaces a given brane by a grade restricted representative,
or the analogous process for complexes of vector bundles on
$\Gr=G(2,7)$.
This will complete the proof that
the renormalization group flow in the Grassmannian phase is
essentially bijective when restricted to the subset ${\mathcal T}_{\bf w}$
consisting of the grade restricted branes.
Also, this allows us to see how one can switch representatives,
from one in ${\mathcal T}_{\bf w}$ to another in ${\mathcal T}_{{\bf w}'}$

For this purpose, it is usefull to
plot the grade restricted subsets GRR${}_{\bf w}$
in the triangular grid.
\begin{figure}[thb]
\psfrag{0m}{\footnotesize $\C$(-1)}
\psfrag{00}{\footnotesize $\C(0)$}
\psfrag{01}{\footnotesize $\C(1)$}
\psfrag{02}{\footnotesize $\C(2)$}
\psfrag{03}{\footnotesize $\C(3)$}
\psfrag{04}{\footnotesize $\C(4)$}
\psfrag{1m}{\footnotesize $\fund$(-1)}
\psfrag{10}{\footnotesize $\fund(0)$}
\psfrag{11}{\footnotesize $\fund(1)$}
\psfrag{12}{\footnotesize $\fund(2)$}
\psfrag{13}{\footnotesize $\fund(3)$}
\psfrag{2mm}{\footnotesize $S^2\fund$(-2)}
\psfrag{2m}{\footnotesize $S^2\fund$(-1)}
\psfrag{20}{\footnotesize $S^2\fund(0)$}
\psfrag{21}{\footnotesize $S^2\fund(1)$}
\psfrag{22}{\footnotesize $S^2\fund(2)$}
\psfrag{23}{\footnotesize $S^2\fund(3)$}
\psfrag{3mm}{\footnotesize $S^3\fund$(-2)}
\psfrag{3m}{\footnotesize $S^3\fund$(-1)}
\psfrag{30}{\footnotesize $S^3\fund(0)$}
\psfrag{31}{\footnotesize $S^3\fund(1)$}
\psfrag{32}{\footnotesize $S^3\fund(2)$}
\centerline{\includegraphics[page=26]{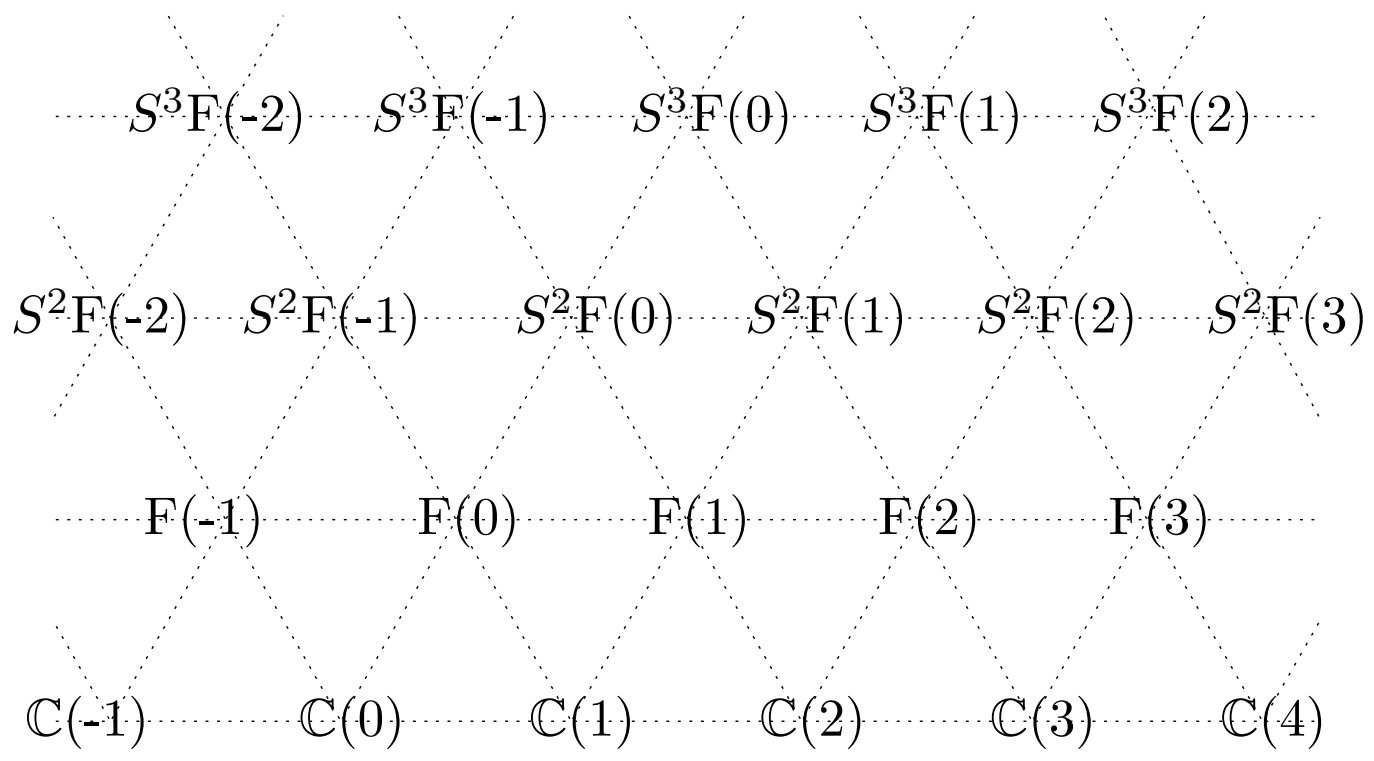}}
\caption{The representations in terms of the symmetric powers of $\fund$.}
\label{fig:grid2}
\end{figure}
Using the relation $\Sym^h\fund^*\cong\Sym^h\fund(-h)$,
the plot of Fig.~\ref{fig:grid} can equivalently be described as
Fig.~\ref{fig:grid2}.
In view of this, we see that
the grade restricted subsets 
are as plotted in
Fig.~\ref{fig:GRRplot}.
\begin{figure}[htb]
\psfrag{A}{${\bf A}$}
\psfrag{B}{${\bf B}$}
\psfrag{C}{${\bf C}$}
\psfrag{D}{${\bf D}$}
  \centerline{\includegraphics[page=27]{main-pics.pdf}}
  \caption{Plot of the grade restricted subsets GRR${}_{\bf w}$
    for ${\bf w}={\bf A},
    {\bf B}, {\bf C}, {\bf D}$.}
\label{fig:GRRplot}
\end{figure}
In each grid, the lower-left and the lower-right edges are
the sites for $\C(-1)$ and $\C(8)$, respectively.

We first describe the grade restriction process for complexes of vector bundles
on $\Gr$ which shows that $D(\Gr)$ is generated by
${\mathcal O}_{\Gr}(\rmGRR_{\bf w})$.
Our starting point is Kapranov's result that $D(\Gr)$ is generated by
${\mathcal O}_{\Gr}(\Kap)$ or any of its shifts ${\mathcal O}_{\Gr}(\Kap)(i)$.
If we aim at the grade restriction to ${\bf w}={\bf A},
{\bf B}, {\bf C}, {\bf D}$, it is most convenient to start with
$\Kap(1)$.
Members of this set form a pyramid in the $U(2)$ grid
as shown in Fig.~\ref{fig:Kapplot} in which
the lower edges of the grid are $\C(-1)$ and $\C(8)$ just as in
Fig.~\ref{fig:GRRplot}.
\begin{figure}[htb]
  \centerline{\includegraphics[page=28]{main-pics.pdf}}
  \caption{Plot of the set $\Kap(1)$.}
  \label{fig:Kapplot}
\end{figure}
The basic idea is to cut it down to one of
the four sets in Fig.~\ref{fig:GRRplot},
via gluing twisted Lascoux complexes.
Let us take an arbitrary object of $D(\Gr)$.
It is isomorphic in $D(\Gr)$ to ${\mathcal O}_{\Gr}(\Complex)$
where $\Complex$ is a bounded complex of representations over $\C[x]$
of $\GL(\fund)$ whose terms are direct sums of members of $\Kap(1)$.
Suppose a term of $\Complex$ includes $\Sym^5\fund(1)$ which is
on the top of the pyramid. We consider binding $\Complex$ and
the twisted Lascoux complex of the type (i) or (vi) in Fig.~\ref{fig:Lc}
which has $\Sym^5\fund(1)$ as the first or the last term.
Let us choose the former, the complex $C_{\rm i}$ of type (i). 
We can find a homomorphism from
$C_{\rm i}$ to $\Complex$ that extends the identity map
from $\Sym^5\fund(1)$ in $C_{\rm i}$
to $\Sym^5\fund(1)$ in $\Complex$.
(We will explain the details momentarily.)
Let us take the cone of this homomorphism and
perform brane-antibrane annihilation,
removing the $\Sym^5\fund(1)$ pair and introducing the zigzag map.
We do the same for all the remaining
$\Sym^5\fund(1)$ components, one after another,
and let $\Complex'$ be the resulting complex.
\begin{figure}[htb]
  \centerline{\includegraphics[page=29]{main-pics.pdf}}
  \caption{Grade restricting $\Kap(1)$.}
  \label{fig:KapGR}
\end{figure}
Its terms are direct sums of members of a new set of representations
obtained from the pyramid by trading
the summit for a new member at the foot ---
the white box marked 1'' in Fig.~\ref{fig:KapGR}.
Note that ${\mathcal O}_{\Gr}(\Complex')$ is isomorphic to
${\mathcal O}_{\Gr}(\Complex)$ in $D(\Gr)$
because binding an exact complex
${\mathcal O}_{\Gr}(C_{\rm i})$ preserves the isomorphism class
and brane-antibrane annihilation is a homotopy
equivalence.
As the second step, we remove one of the two
height 4 members of $\Kap(1)$, $\Sym^4\fund(2)$,
shown in Fig.~\ref{fig:KapGR} as the black box with mark 2,
using the twisted Lascoux complex $C_{\rm v}$ of the type (v)
which has $\Sym^4\fund(2)$ as the last term.
We can find a homomorphism from $\Complex'$ to $C_{\rm v}$
extending the identity map between $\Sym^4\fund(2)$ components of the two.
(Again, see below for an explanation.)
Repeating the same procedure as in the first step,
$\Complex'$ is replaced by a complex $\Complex''$ of direct sums of members of
another set of representations
obtained from the last one by trading
the height 4 member for a new member of height 1
--- the white box marked 2 in Fig.~\ref{fig:KapGR}.
The $n$-th step does the same
($n=3,4,5$): the black box $n$ is replaced by the
white box $n$,
 using the twisted Lascoux complexes of the type
(ii), (iv), (iii) respectively.
The final step is to remove the remaining height 3 member of $\Kap(1)$,
shown in Fig.~\ref{fig:KapGR} as the black box with mark 6.
Here, as in the first step, a choice must be made: either we bind the twisted
Lascoux complex of type (iii) or of type (iv). Let us choose
type (iii). Then, the black box 6 is replaced by the white box 6''.
In the end, we obtain a complex of direct sums of
members of a set of representations,
consisting of height $\leq 2$ members of
the pyramid and the white boxes in Fig.~\ref{fig:KapGR}
marked 1'', 2, 3, 4, 5 and 6''.
We see that the final set coincides with $\rmGRR_{\bf A}$.
We have seen that an arbitrary object of
$D(\Gr)$ is represented by
a complex of direct sums of members of
${\mathcal O}_{\Gr}(\rmGRR_{\bf A})$.
If we make different choices in the first and the last steps, we obtain
different representatives.
The result is summarized in the table below.
\beq
\begin{array}{c|c|c|c}
  \mbox{1st step}&\mbox{6th step}&\mbox{added members}&\mbox{the final set}\\
  \hline
    \mbox{(i)}&\mbox{(iii)}&\mbox{1'', 2, 3, 4, 5, 6''}&\rmGRR_{\bf A}\\
    \mbox{(i)}&\mbox{(iv)}&\mbox{1'', 2, 3, 4, 5, 6'}&\rmGRR_{\bf B}\\
    \mbox{(vi)}&\mbox{(iii)}&\mbox{1', 2, 3, 4, 5, 6''}&\rmGRR_{\bf C}\\
    \mbox{(vi)}&\mbox{(iv)}&\mbox{1', 2, 3, 4, 5, 6'}&\rmGRR_{\bf D}  
\end{array}
\label{table1}
\eeq
Similarly, an arbitrary morphism of $D(\Gr)$
can be represented by a homomorphism between complexes of direct sums of
members of ${\mathcal O}_{\Gr}(\rmGRR_{\bf w})$.
The idea is to repeat the grade restriction process
together with the homomorphism.
Let $g:\Complex_1\to\Complex_2$ be the homomorphism
of $\Kap(1)$-resolutions corresponding to the original morphism. 
Let us consider the initial step of binding twisted Lascoux complexes of
type (i).
Let us bind a twisted Lascoux complex $C_{\rm i}$ to
$\Complex_1$ by a homomorphism $f_1:C_{\rm i}\to \Complex_1$
extending the identity of a $\Sym^5\fund(1)$ pair,
and consider the problem of extending $g$ to
a homomorphism from the bound complex ${\rm Cone}(f_1)$ to $\Complex_2$.
This is the problem we encountered in Section~\ref{subsub:HA}-I, and we found
that this is possible provided $g f_1$ is null homotopic.
In fact, it is the case for an appropriate choice of $f_1$.
To see this, we first find a homomorphism $F$
from $C_{\rm i}$ to ${\rm Cone}(g)[-1]$
extending the identity of the $\Sym^5\fund(1)$ pair.
(Again, see below for an explanation.)
Let us denote the components of $F$ to $\Complex_1$ and $\Complex_2[-1]$
by $f_1$ and $-h$. As $F$ is a cochain map,
\beq
0=\left(\begin{array}{cc}
  d_1&0\\
  -g&-d_2\end{array}\right)
  \left(\begin{array}{c}
    f_1\\
    -h\end{array}\right)
   -\left(\begin{array}{c}
    f_1\\
    -h\end{array}\right)d_{\rm\,i}
    =\left(\begin{array}{c}
      d_1 f_1-f_1 d_{\rm\, i}\\
      -gf_1+d_2h+hd_{\rm \, i}
    \end{array}\right),
\eeq
we see that $f_1$ is a homomorphism $C_{\rm i}\to \Complex_1$ 
extending the identity map of the $\Sym^5\fund(1)$ pair
and $h$ is the homotopy of $g f_1$.
Thus, $\hat{g}=(h,g):{\rm Cone}(f_1)\to \Complex_2$ is the wanted extension.
Now we can perform the brane-antibrane annihilation of
the $\Sym^5\fund(1)$ pair.
Repeating this for all the remaining $\Sym^5\fund(1)$ components
in $\Complex_1$,
we have a homomorphism $\tilde{f}:\Complex_1'\to \Complex_2$.
It is straightforward to extend $\tilde{f}$ to a homomorphism from
$\Complex_1'$ to the result of binding $\Complex_2$ and a twisted
Lascoux complex ---
we just have to set to $0$ the map to the twisted Lascoux complex.
Repeating this,
we have a homomorphism $f':\Complex_1'\to \Complex_2'$
between the complexes after the first step.
It is now clear how to proceed. After completing the six steps,
with the choice (iii) in the last step,
we obtain a homomorphism between complexes of direct sums of
members of ${\mathcal O}_{\Gr}(\rmGRR_{\bf A})$.
For other choices in the first and the last steps,
we obtain a homomorphism between
complexes of direct sums of members of
different sets as shown in (\ref{table1}).
This completes the proof that ${\mathcal O}_{\Gr}(\rmGRR_{\bf w})$
generates $D(\Gr)$ for ${\bf w}={\bf A}, {\bf B}, {\bf C}, {\bf D}$.
If we start with $\Kap(i+1)$, we find the same conclusion
for $\rmGRR_{\bf w}(i)=\rmGRR_{{\bf w}(-i)}$.

Let us explain how to find a homomorphism
between a given complex and a twisted Lascoux complex,
extending the identity map between common components of the two.
This is the situation where 
the Theorem or the Corollary in Section~\ref{subsub:HA}-II can be applied,
where the given complex and the twisted Lascoux complex
play the r\^oles of $(M,d_M)$ and $(N,d_N)$, respectively,
possibly with a shift of degree.
Indeed, twisted Lascoux complexes and their transposes are acyclic
as complexes of $\C[x]$-modules.
The only thing to check is the condition 
$f^0{\rm Im} d_M^{-1}\subset {\rm Im} d_N^{-1}$ (for Theorem) or
${\rm Ker} d_N^1\subset (f^0)^{-1}{\rm Ker} d_M^1$ (for Corollary).
It is enough to check the former as the latter is related to it by transpose.
In this situation,
$N^0=\Sym^h\fund^*(i)\otimes\C[x]$ for some $(h,i)$,
$M^0$ has the same as a direct summand,
and $f^0:M^0\to N^0$ is the projection to that summand.
We may suppose that each $d_M$ has at least one power of $x$
by removing a constant part of $d_M$ by brane-antibrane annihilation.
The map $f^0d_M^{-1}:M^{-1}\to N^0$
splits to the sum of maps of the form
\beq
\varphi:\Sym^{h'}\fund^*(i')\otimes \C[x]
\to \Sym^h\fund^*(i)\otimes\C[x],
\label{varphi}
\eeq
for some set of $(h',i')$.
Since we are cutting $\Kap(1)$ from the top and downward,
$h'$ is less than or equal to $h$. Note also that we are considering $h\geq 3$.
We would like to show that the image of each such $\varphi$ is
included in the image of the last arrow $d_N^{-1}$
of the twisted Lascoux complex,
which is given in (\ref{Lne}) as $(\nearrow)=x_{ia}e^a\eta^i$.
The matrix elements 
$\varphi^{\,\,\,b_1\cdots b_{h'}}_{a_1\cdots a_h}(x)$ of (\ref{varphi})
are sums of products of $x_{ia_s}$, $x^{b_t}_j$,
$\delta_{a_s}^{b_t}$ and $[x_ix_j]$,
just as in (\ref{formofmael}).
If $h'<h$, each term has at least one factor of the form
$x_{ia_s}$ and hence the image of $\varphi$ is included in
the image of $(\nearrow)$.
If $h'=h$, the matrix elements may also include a term of the form
$\delta^{b_1\cdots b_h}_{a_1\cdots a_h}[x_{i_1}x_{j_1}]\cdots[x_{i_q}x_{j_q}]$
for $q=i-i'$.
(Note that $q>0$ by the assumption that $d_M$
has at least one power of $x$.)
The image of this part of $\varphi$ is also in the image of $(\nearrow)$
thanks to the relation (\ref{fundrel}).
Indeed,
\beq
    [x_ix_j]e^{bc\cdots}=[x_ix_j]\delta^b_ae^{ac\cdots}
    =(x^b_ix_{ja}-x^b_jx_{ia})e^{ac\cdots}
    =x_{ja}e^a(x^b_ie^{c\cdots})-x_{ia}e^a(x^b_je^{c\cdots}).
\eeq
This completes the proof that the condition $f^0{\rm Im}d_M^{-1}\subset
{\rm Im}d_N^{-1}$ is satisfied.
Thus, the extension of $f^0$ to a cochain map
$(M,d_M)\to (N,d_N)$ can be found.

Given the above result, the argument of Section~\ref{subsub:MFcompletion}
for matrix factorization completion can be applied for $\rmGRR_{\bf w}$
in place of $\Kap$. In particular, any object of
$D(X)$ has a lift in ${\mathcal T}_{\bf w}$.
Combined with the injectivity result that follows from
Property EV${}_{m\geq 0}$ of $\rmGRR_{\bf w}$, we find that the flow
$\pi_X:{\mathcal T}_{\bf w}\to D(X)$ is essentially bijective.

The above procedure can be applied also to a complex
whose representations
do not fit into a pyramid as in Fig.~\ref{fig:Kapplot}
but are ``scattered around'' in the grid.
The positions of the representations can be lowered by
binding with the twisted Lascoux complexes and performing
brane-antibrane annihilation.
It is clear that this process can bring all
the representations to fit into the rows of height 0, 1 and 2.
However, the occupied region may be wide in the horizontal directions
and may not fit into a set of the form $\rmGRR_{\bf w}$.
We would like to know how to move the representations in the horizontal
direction while maintaining the heights to stay in $\{0, 1, 2\}$, so that
all are moved into a target set $\rmGRR_{\bf w}$.
Such a method would also allow us to switch
the grade restricted representatives
to the target set $\rmGRR_{\bf w}$ from another one $\rmGRR_{\bf w'}$.
Of course, we could use the result of Kapranov \cite{KapranovGr,KapranovHomog}
to go back to the $\Kap(i)$-resolution for an appropriate $i$
and then restart from there to get to the target $\rmGRR_{\bf w}$.
However, we would like to find a direct and explicit
way without such a detour.
We may also want to do the same for matrix factorizations.

For this purpose, we would like to have
complexes made only of height 0, 1, 2 representations
which are exact on the locus where $x$ has maximal rank.
In fact, we will construct three types of such height 2
complexes by binding together twisted Lascoux complexes.
We will also find their matrix factorization completions.

The simplest one is obtained by binding just two Lascoux complexes,
one of the type (iv) in Fig.~\ref{fig:Lc} and another
of the type (iii)' in the sense of (\ref{otherLcLb}). 
Let us describe the construction of this one in some detail.
With appropriate shifts in the horizontal direction,
the head of a type (iv) complex $C_{\rm iv}$ and
the tail of a type (iii)' complex $C_{\rm iii}$
can be brought into
a common site, say, $\Sym^3\fund^*$.
Then, there is a homomorphism from $C_{\rm iii}$ to $C_{\rm iv}$
which is the identity of $\Sym^3\fund^*$ on the common site and zero elsewhere.
Let us take the cone of this homomorphism, a part of which is as follows:
\beq
\begin{tikzcd}
  &&\Sym^3\fund^*\ar[draw=none]{d}[sloped,auto=false]{\bigoplus}
  \arrow[r,shorten=1mm]\arrow[dr,"\!{\rm id}"]
  &\Sym^2\fund^*\otimes V\ar[draw=none]{d}[sloped,auto=false]{\bigoplus}
  \!\!\arrow[r,shorten=1mm]&\!\!\cdots
  &\!\!\!\!C_{\rm iii}\\
  C_{\rm iv}\!\!\!\!&\cdots\!\!\arrow[r,shorten=1mm]&\!\!
  \Sym^2\fund^*(-1)\otimes V^*\arrow[r]&\Sym^3\fund^*&&
\end{tikzcd}
\eeq
It still has height 3 components, but they can be eliminated by
brane-antibrane annihilation via tachyon condensation of
an invertible map between them, the identity of $\Sym^3\fund^*$.
As described in Section~\ref{subsub:HA}-I, this annihilates the $\Sym^3\fund^*$
pair and introduces the zigzag map that goes through the pair.
\begin{figure}[htb]
\centerline{\includegraphics[page=30]{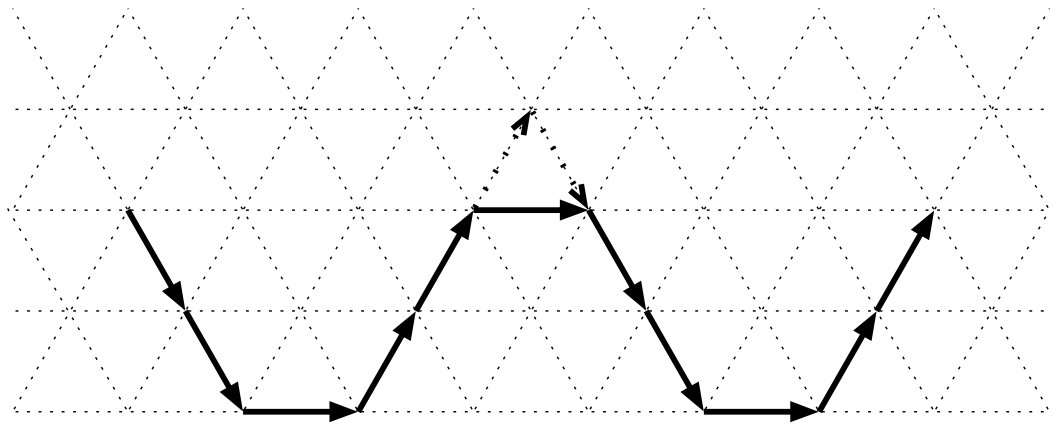}}
\caption{Two Lascoux complexes bound together.}
\label{fig:D6c}
\end{figure}
The result is a W-shaped complex that fits into
the rows of height 0, 1, 2, as shown in Fig.~\ref{fig:D6c}.
The zig-zag map, shown as the rightward arrow 
that connects the two complexes,
is given by
\beq
-x_i^ax_{jb}e_ae^b\bareta^i\eta^j.
\label{zigzag1}
\eeq
This description is again informal: To be precise,
between the $\bareta^i$ and the $\eta^j$
there is a map that sends $(\cdots)|0\rangle_{-1}$
to $(\cdots)|\bar{0}\rangle_0$ where
$|0\rangle_{-1}$ and $|\bar{0}\rangle_0$ are the ``vacuum'' vectors for
the algebras (\ref{Heisenberg})-(\ref{Clifford1})
and (\ref{Heisenberg})-(\ref{Clifford2}), respectively, and
$(\cdots)$ is a polynomial of the $x^a_i$'s and $e^a$'s,
quadratic in the latter.
In the present construction, the complex starts from
$\Sym^2\fund^*(-4)\otimes \textw^{\!7}V^*$ 
and ends at $\Sym^2\fund^*(3)\otimes \textw^{\!7} V$.
If we forget $\GL(V)$ and make identification
$\textw^{\!7}V^*=\textw^{\!7}V=\C$, which we shall do often below, it starts
from $\Sym^2\fund^*(-4)$ and ends at $\Sym^2\fund^*(3)$.
We have thus constructed a complex made only of height 0, 1, 2 representations.
It is exact on the locus where $x$ has maximal rank
as the two constituent complexes are.

The complex can also be completed into a matrix factorization of the
superpotential.
We keep all possible arrows of the matrix factorization completions
of $C_{\rm iii}$ and $C_{\rm iv}$, but that would not be sufficient
since $Q^2=W$ fails at 
the head and the tail of the zigzag map (\ref{zigzag1}).
\begin{figure}[htb]
\centerline{\includegraphics[page=31]{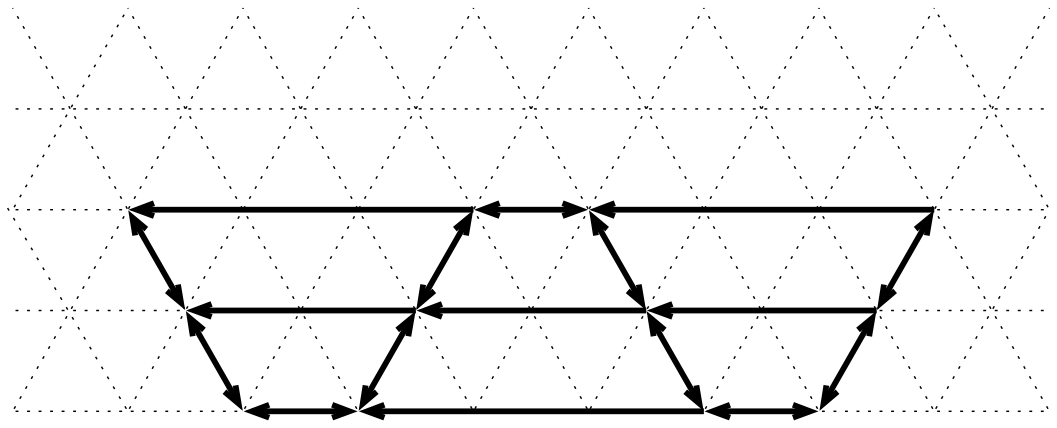}}
\caption{The matrix factorization completion of Fig.~\ref{fig:D6c}}
\label{fig:D6mf}
\end{figure}
It turns out that we need to add three extra horizontal arrows
from the $C_{\rm iii}$ part on the right
to the $C_{\rm iv}$ part on the left, one for each height
in the triangle grid as shown in Fig.~\ref{fig:D6mf}.
The expressions of the additional arrows are
\beqa
\!\!\!h=2:&&{2\over 3}A^{ij}(p)\bareta_i\eta_j,
\label{D6mf2}\\
\!\!\!h=1:&&{1\over 18}\bar{A}(p)A(p)
-{1\over 3}A^{ij}(p)A^{kl}(p)\bareta_i\bareta_k\eta_j\eta_l,
\label{D6mf1}\\
\!\!\!h=0:&&
{1\over 6}\bar{A}(p)A^{ij}(p)\bareta_i\eta_jA(p)
-{2\over 9}A^{ij}(p)A^{kl}(p)A^{mn}(p)
\bareta_{i}\bareta_{k}\bareta_{m}
\eta_{j}\eta_{l}\eta_{n},~~~~
\label{D6mf0}
\eeqa
in which $\bar{A}(p)=A^{ij}(p)\bareta_i\bareta_j$ and
$A(p)=A^{ij}(p)\eta_i\eta_j$.
We denote by $\Lc^2(i)$ and $\Lb^2(i)$
the complex and its matrix factorization completion
such that the left-most component is $\Symt(i)$ and has R-charge zero
(the right-most component is $\Symt(i+7)$ and has R-charge $11$).

\begin{figure}[htb]
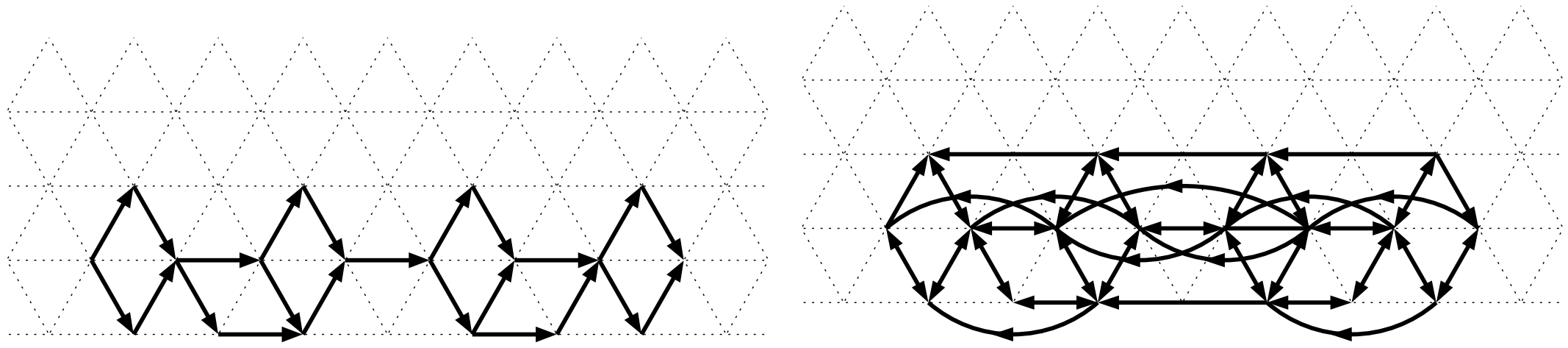

  \centerline{\includegraphics[page=32]{main-pics.pdf}~~
  \includegraphics[page=33]{main-pics.pdf}}
  \caption{Four types of Lascoux complexes/branes
    bound together.}
\label{fig:Lemp1}
\end{figure}

The next simplest height two complex
is obtained by binding four types of Lascoux complexes.
We first bind together two Lascoux complexes with ends at height 1,
that is, type (v) and type (ii)'.
We can glue them together as in the case above:
Take a cone of a homomorphism consisting of an isomorphism
from the tail of (ii)' to the head of (v) at height 4,
and then do the brane-antibrane annihilation by that map.
The result still has two components of height 3,
with multiplicities $V$ and $V^*$.
To remove them, we bind to it the Lascoux complexes with height 3 ends,
that is, (iv) and (iii)', tensored with $V$ and $V^*$ respectively.
After gluing and additional brane-antibrane annihilation,
we find a height 2 complex of the form shown in Fig.~\ref{fig:Lemp1} (left).
It can be completed to a matrix factorization of the form
as in Fig.~\ref{fig:Lemp1} (right).
The construction is described in Appendix~\ref{app:Lbconstruction}.
We denote by $\Lc^1(i)$ and $\Lb^1(i)$
the complex and its matrix factorization completion
such that the left-most component is $\fund(i)$ and has R-charge zero
(the right-most component is $\fund(i+7)$ and has R-charge $11$).

\begin{figure}[htb]
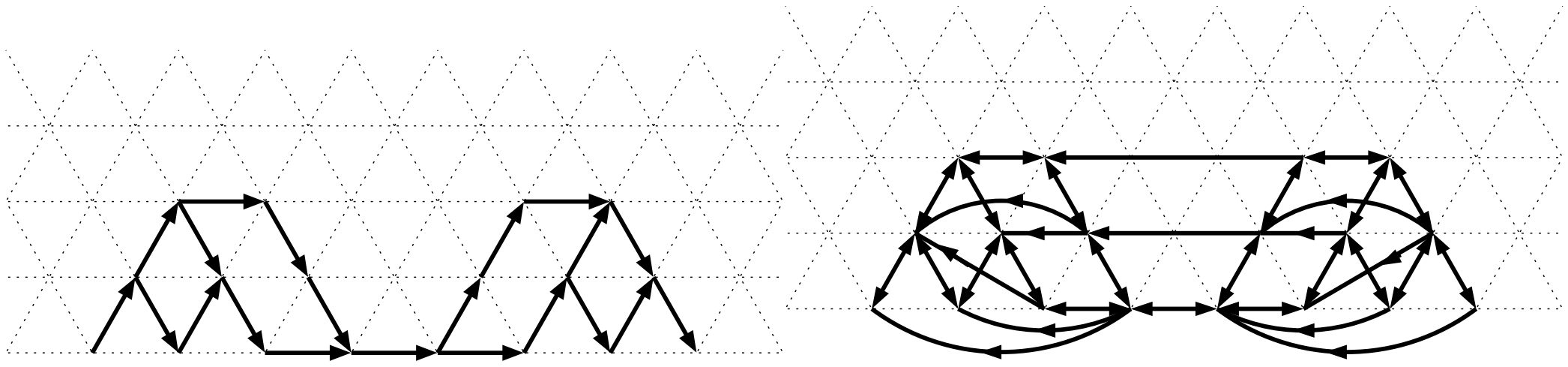

  \centerline{\includegraphics[page=34]{main-pics.pdf}
  \includegraphics[page=35]{main-pics.pdf}}
\caption{Six types of Lascoux complexes/branes bound together.}
\label{fig:Lemp0}
\end{figure}

The most complicated complex
consists of six types of Lascoux complexes.
We first bind together two Lascoux complexes
of height 0 ends, that is, type (vi) and type (i)'.
We can glue them together but the result
still has two components of height 4,
with multiplicities $V$ and $V^*$.
To remove them, we bind to it the Lascoux complexes with height 4 ends,
that is, (v) and (ii)', tensored with $V$ and $V^*$ respectively.
After gluing and additional brane-antibrane annihilation,
we find a complex which still has two components of height 3,
with multiplicities $\textw^{\!2}V$ and $\textw^{\!2}V^*$.
To remove them, we bind to it the Lascoux complexes with height 3 ends,
that is, (iv) and (iii)',
tensored with $\textw^{\!2}V$ and $\textw^{\!2}V^*$ respectively.
After gluing and additional brane-antibrane annihilation,
we find a height 2 complex
of the form shown in Fig.~\ref{fig:Lemp0} (left).
It can be completed to a matrix factorization of the form
as in Fig.~\ref{fig:Lemp0} (right).
See Appendix~\ref{app:Lbconstruction}.
We denote by $\Lc^0(i)$ and
$\Lb^0(i)$ the complex and its matrix factorization completion
such that the left-most component is $\C(i)$ and has R-charge zero
(the right-most component is $\C(i+7)$ and has R-charge $11$).

\noindent
{\bf Remark 1.} That the twisted Lascoux complexes can be bound in
the way described above is guaranteed since
the conditions for 
the Theorem and the Corollary of Section~\ref{subsub:HA}-II hold.
Existence of matrix factorization completions of the resulting complexes
is also guaranteed since the sets of representations involved (after
brane-antibrane annihilation) have Property EV${}_m$ for $m\geq 1$.
We neverthless constructed the complexes
and matrix factorizations explicitly, as these will play important r\^oles
later and
the explicit form of the arrows will be needed on some occasions.

\noindent
{\bf Remark 2.}
The above construction of the complex $\Lc^1$ may be regarded as the result of
a sequence of cones
$((C_{\rm iv}\otimes V
\leftarrow (C_{\rm v}\leftarrow C_{\rm ii}))\leftarrow
C_{\rm iii}\otimes V^*)$,
with brane-antibrane annihilation at various stages.
By the associativity (\ref{associativity}) of the cone operation,
this can also be constructed by successive cones in a different order,
such as
\beq
(((C_{\rm iv}\!\otimes\! V
\leftarrow C_{\rm v})\leftarrow C_{\rm ii})\leftarrow
C_{\rm iii}\!\otimes\! V^*)
~~\mbox{or}~~
(C_{\rm iv}\!\otimes\! V\leftarrow (C_{\rm v}\leftarrow (C_{\rm ii}\leftarrow
C_{\rm iii}\!\otimes\! V^*))).
\label{anotherC}
\eeq
Likewise, the complex $\Lc^0$ can be constructed by parenthesizing
\beq
C_{\rm iv}\!\otimes\!\textw^{\!2}V
\leftarrow C_{\rm v}\!\otimes\! V\leftarrow C_{\rm vi}
\leftarrow C_{\rm i}\leftarrow C_{\rm ii}\!\otimes\!V^*\leftarrow
C_{\rm iii}\!\otimes\! \textw^{\!2}V^*
\label{anotherCC}
\eeq
in an arbitrary way. Exactly the same can be said for the branes
$\Lb^1$ and $\Lb^0$ as well.

\noindent
{\bf Remark 3.}
The transpose of the above complexes and matrix factorizations
are isomorphic to themselves up to a shift,
\beq
(\Lc^h(i))^T\,\,\cong\,\,\Lc^h(-i-h-7)[11],\quad
(\Lb^h(i))^T\,\,\cong\,\,\Lb^h(-i-h-7)[11].
\label{dualLcLb}
\eeq

As complexes of $(\GL_2(\C),\C[x])$-modules,
the $\Lc^h(i)$ are quasi-isomorphic to complexes of the modules
which are resolved by the constituent twisted Lascoux complexes (tensored
with multiplicities $\textw^qV$ or $\textw^qV^*$).
In particular, they are not acyclic since the cohomologies at a
central degree are non-vanishing.
However, as the modules resolved by twisted Lascoux complexes are
supported on the non-maximal rank locus of
$(x^a_i)$, the complexes $\Lc^h(i)$ define exact complexes of vector bundles
on the Grassmannian $\Gr$.
As a consequence, the branes $\Lb^h(i)$ are empty in the Grassmannian phase.
This is compatible with the vanishing of the hemisphere partition function
\beq
Z_{D^2}(\Lb^h(i))=0,\quad \mbox{at $\zeta\gg 0$,}
\eeq
which follows from (\ref{Lbvanishing}).
We would like to use $\Lc^h(i)$ and $\Lb^h(i)$ 
to replace a given height two complex or brane
by another one with the same low energy behaviour in the Grassmannian phase,
in which the Chan-Paton representations are moved in the horizontal direction
while maintaining the heights to stay in $\{0,1,2\}$.

Let us see how to use $\Lc^h(i)$ 
to replace complexes.
The process is not as straightforward as using the twisted Lascoux complexes,
since $\Lc^h(i)$ are not acyclic as $\C[x]$-modules.
However, with a little more effort, 
they can still be used in the replacement process as we now explain.
In what follows, we shall describe the process of moving the representations
towards the left.
The move in the opposite direction, towards the right, is obtained as
its transpose thanks to the relation (\ref{dualLcLb}).

Let us first descibe how to use $\Lc^2$ and its shifts.
Let $\Complex$ be a height two complex including
a component $\Sym^2\fund(7)$ which we would like to replace by
representations that are located on its left.
We may assume that $\Complex$ is of reduced form
--- all of its arrows are at least linear in $x$ ---
by brane-anti-brane annihilation if necessary.
A twisted Lascoux complex $C_{\rm iii}$ of type (iii)
with $\Sym^2\fund(7)$ as its last term
can be bound to it by a homomorphism $\Complex\to C_{\rm iii}$
that extends the identity map of $\Sym^2\fund(7)$.
Existence of such an extension is guaranteed by the acyclicity of
$C_{\rm iii}$ and the height argument similar to the one around (\ref{varphi})
in which $h=2$ and $h'\leq 2$ as $\Complex$ has height two.
However, a height three component $\Sym^3\fund(3)$ at the tail
of $C_{\rm iii}$ ``sticks out'' of the compound.
In order to take care of it,
let us bind the twisted Lascoux complex $C_{\rm iv}$ of type (iv)
with $\Sym^3\fund(3)$ as its last term,
by a homomorphism from the compound to $C_{\rm iv}$ that extends the identity
map of $\Sym^3\fund(3)$.
Existence of such an extension is guaranteed by the acyclicity of
$C_{\rm iv}$ and the height argument similar to the one around (\ref{varphi})
in which $h=3$ and $h'\leq 2$ as
the other part of the compound has height two.
After eliminating the $\Sym^3\fund(3)$ pair by brane-antibrane annihilation,
we obtain a new height two complex.
The net effect in the end is to bind $\Lc^2$ to $\Complex$
by a homomorphism $\Complex\to \Lc^2$
that extends the identity map of $\Sym^2\fund(7)$.
The replacement process is completed by
eliminating the $\Sym^2\fund(7)$ pair by brane-anitibrane annihilation.

Let us next describe how to use $\Lc^1$ and its shifts.
In this case, an important r\^ole is played by
the composite $N=(C_{\rm ii}\leftarrow C_{\rm iii}\!\otimes\! V^*)$
that appears in the construction (\ref{anotherC}) of $\Lc^1$.
Let $\Complex$ be a height two complex of reduced form, including
a component $\fund(7)$ which we would like to replace by
representations that are located on its left.
The composite $N$ with $\fund(7)$ as its last term
can be bound to it by a homomorphism $\Complex\to N$
that extends the identity map of $\fund(7)$.
Existence of such an extension is not as easy to see as in the case above,
because $N$ may not be acyclic and also the height argument does not apply.
Nevertheless, with a much more  \label{page:x}
involved argument as presented in Appendix~\ref{app:GRdetails},
one can show that the wanted extension exists.
The height four and three components of $C_{\rm ii}$
and $C_{\rm iii}\!\otimes\! V^*$ ``stick out'' of the compound, but
they can be eliminated by gluing $C_{\rm v}$ and $C_{\rm iv}\!\otimes\!V$
successively.
Existence of homomorphisms required for the gluing can be shown
using the acyclicity of $C_{\rm v}$ and $C_{\rm iv}\!\otimes\!V$
as well as the height argument.
The net effect in the end is to bind $\Lc^1$ to $\Complex$
by a homomorphism $\Complex\to \Lc^1$ that extends the identity map
of $\fund(7)$.
The replacement process is completed
by eliminating the $\fund(7)$ pair by brane-antibrane annihilation.

Finally, let us describe how to use $\Lc^0$ or its shifts.
In this case, the important player is the composite
$N'=((C_{\rm i}\leftarrow C_{\rm ii}\!\otimes\!V^*)
\leftarrow C_{\rm iii}\!\otimes\!\textw^{\!2}V^*)$
that appears in one way of constructing $\Lc^0$, see (\ref{anotherCC}).
The process is similar to the previous cases.
Given a height two complex $\Complex$ of reduced form with a component
$\C(7)$, we first bind $N'$ to it by a homomorphism $\Complex\to N'$
extending the identity map of $\C(7)$. Then we glue
$C_{\rm vi}$, $C_{\rm v}\otimes V$ and $C_{\rm iv}\otimes\textw^{\!2}V$
successively to eliminate the stuck-out components of height 5, 4, 3.
The net effect is to bind $\Lc^0$ to $\Complex$ by a homomorphism
$\Complex\to\Lc^0$ extending the identity map of $\C(7)$.
The most difficult part, finding a homomorphism $\Complex\to N'$
extending the identity map of $\C(7)$, \label{page:y}
is explained in Appendix~\ref{app:GRdetails}.

Using the $\Lc^h(i)$'s we can now freely move height two complexes
in the horizontal direction
within the same isomorphism classes of $D(\Gr)$. 
In view of the shape of $\Lc^h(i)$ shown in Fig.~\ref{fig:D6c},
\ref{fig:Lemp1} and \ref{fig:Lemp0}, 
it is clear that this process can make all representations fit into
any of the finite sets GRR${}_{\bf w}$
in Fig.~\ref{fig:GRRplot} or their shifts.

We would like to extend this to
replacement of  matrix factorizations, i.e. use
$\Lb^h(i)$'s to move height two branes in the horizontal direction
without changing their low energy behaviour in the Grassmannian phase.
Let $\brane$ be a height two matrix factorization including a
component $\Sym^h\fund(7+i)$
which we would like to replace by representations that are located on its left.
As in the case of complexes, we would like to find a homomorphism
$\brane\to \Lb^h(i)$ that extends the identity map of $\Sym^h\fund(7+i)$.
Let $\Complex$ be the complex obtained from $\brane$ by setting $p=0$.
As we have seen above,
there is a homomorphism $\varphi:\Complex\to\Lc^h(i)$
that extends the identity map of $\Sym^h\fund(7+i)$.
We would like to extend this to
a homomorphism $\widehat{\varphi}:\brane\to\Lb^h(i)$ of
matrix factorizations.
This is a problem encountered in (\ref{homcompletion}), and 
a solution can be found under two conditions
on the complexes
\beq
K={\mathcal O}_{\Gr}(\Complex)\quad\mbox{and}\quad
L={\mathcal O}_{\Gr}(\Lc^h(i)).
\eeq
The first is that $\Hom_{D(\Gr)}(K,L(m)[-2m])$ vanishes for all $m\geq 1$,
which always holds since ${\mathcal O}_{\Gr}(\Lc^h(i))$ is exact.
The second is that 
\beq
\Ext^p_{\Gr}(K^q,L^r(m))=0,\quad p>0, ~~m\geq 1,~\forall(q,r).
\label{secondC}
\eeq
This is the case when $\Sym^h\fund(7+i)$ sits
at the right-most position of $\brane$ in the triangular grid.
More generally, this holds when all the representations of $\brane$
fit into the union of $\rmGRR_{\bf w}$ and its shifts to the left
in the triangular grid, where ${\bf w}$ is a window such that
$\Sym^h\fund(7+i)$ is the right-most member at height $h$ of $\rmGRR_{\bf w}$.
(For example, $\fund(6)$ is the right-most member at height 1
of $\rmGRR_{\bf w}$ for ${\bf w}={\bf A}$, ${\bf B}$, ${\bf C}$ and ${\bf D}$.)
In such a case, $\Lb^h(i)$ fits into $\rmGRR_{\bf w}(-1)\cup \rmGRR_{\bf w}$
and hence (\ref{secondC}) holds as $\rmGRR_{\bf w}$ has Propetry EV${}_m$
for $m\geq 0$ (Theorem in Section~\ref{subsub:BWB}).
Once an extension $\widehat{\varphi}$ is found,
we can use it to bind $\Lb^h(i)$ to $\brane$
and then annihilate the $\Sym^h\fund(7+i)$ pair.
This completes a process of brane replacement.
Repeating this process successively, we can move the entries of
a given brane $\brane$ towards left as much as we want.
Motion towards right can also be done in the similar way.
Suppose $\brane$ has a component $\Sym^{h'}\fund(i')$
which is the left-most member at height $h'$ of $\rmGRR_{{\bf w}'}$
and $\brane$ fits into the union of $\rmGRR_{{\bf w}'}$ and its shifts
to the right in the triangular grid.
(For example, suppose
$\Sym^{h'}\fund(i')$ sits at the left-most position of $\brane$.)
Then, we can find a homomorphism from
$\Lb^{h'}(i')$ to $\brane$ that extends the identity map
from $\Sym^{h'}\fund(i')$ to itself.
Taking its cone and annihilating the $\Sym^{h'}\fund(i')$ pair,
a process of brane replacement is completed.
Repeating this, we can move the entries of
$\brane$ towards right as much as we want.

Using this process, we can freely move height two branes
in the horizontal direction
within the same isomorphism classes of $D(X)$. 
In view of the shape of the $\Lb^h(i)$'s shown in Fig.~\ref{fig:D6mf},
\ref{fig:Lemp1} and \ref{fig:Lemp0}, 
it is clear that this process can make all the representations fit into
any of the finite sets
GRR${}_{\bf w}$ in Fig.~\ref{fig:GRRplot} or their shifts.
This will be relevant when we transport a given brane in the Grassmannian phase
to the Pfaffian phase through the window ${\bf w}$,
which we shall discuss in Section~\ref{sec:transport}.

This process can also be used to switch the grade restricted representatives,
from one subset to another.
To be specific, let us consider the switch from
${\mathcal T}_{\bf A}$ to ${\mathcal T}_{\bf B}$.
The difference between  $\rmGRR_{\bf A}$ and $\rmGRR_{\bf B}$
is that $\Sym^2\fund(6)$ is in the former but not in the latter
while $\Sym^2\fund(-1)$ is in the latter but not in the former.
We need to trade $\Sym^2\fund(6)$ for $\Sym^2\fund(-1)$
while keeping all other representations in
$\rmGRR_{\bf A}\cap \rmGRR_{\bf B}$.
A brane that does the job is $\Lb^2(-1)$
which has $\Sym^2\fund(-1)$ and $\Sym^2\fund(6)$ as the left-most and
the right-most components as well as members of
$\rmGRR_{\bf A}\cap \rmGRR_{\bf B}$ in between.
Suppose a brane $\brane$ is in ${\mathcal T}_{\bf A}$.
Then, a $\Sym^2\fund(6)$ component in $\brane$, if there is, is in the position
which admits a homomorphism from $\brane$ to $\Lb^2(-1)$
that extends the identity map from the $\Sym^2\fund(6)$ in $\brane$
to the $\Sym^2\fund(6)$ in $\Lb^2(-1)$.
Applying this to all the $\Sym^2\fund(6)$ components
in $\brane$, we obtain another representative $\brane'$
that has no $\Sym^2\fund(6)$ component but has
$\Sym^2\fund(-1)$ components instead, as well as components in
$\rmGRR_{\bf A}\cap \rmGRR_{\bf B}$.
That is, another brane $\brane'$ in ${\mathcal T}_{\bf B}$
that has the same infra-red limit as $\brane$ in ${\mathcal T}_{\bf A}$
in the Grassmannian phase.
This procedure will be relevant when we discuss monodromy along loops
with a base point in the Pfaffian phase
(Section~\ref{subsec:monodromy}).

\section{Pfaffian Phase}
\label{sec:Pfaffian}

In this section, we study the low energy behaviour of
gauge theory branes in the Pfaffian phase $\zeta\ll 0$.
This phase is strongly coupled due to the unbroken $SU(2)$ gauge symmetry.
Understanding of $SU(2)$ dynamics was essential to find the bulk flow to
the sigma model with Pfaffian Calabi-Yau target \cite{HoTo},
and that needs to be extended to the boundary sector.

After making preliminary remarks,
we first present a ``blindfolded'' study of branes in which only the
hemisphere partition function is used.
We shall find brane factors that produce the central
charges of D0, D2, D4 and D6-branes in the Pfaffian Calabi-Yau $Y$.
We then try to find the matrix factorizations that realize some of
these factors.
Next, we move on to the analysis of the boundary conditions in
the $SU(2)$ supersymmetric QCDs that flow to the free theory of baryons.
We identify a brane $\brane^{SU(2)}_{3,7}$ in such a QCD with seven fundamental
doublets that flows to the Neumann boundary condition of the free baryons.
We apply this result to the Pfaffian phase of the R\o dland model.
We will prove that the candidates for D0 and D6 branes
reduce to branes that include a $\brane^{SU(2)}_{3,7}$ factor
in the local behaviour,
and indeed flow to D0 and D6-branes on $Y$, respectively.
We next discuss the brane replacement process.
Information of higher Ext groups shows that
the set of representations in a brane can be brought to fit into
any one of the grade restricted subsets and that
nothing else can be such a terminal subset.
Finally,
we argue that the flow ${\mathcal T}_{\bf w}\to D(Y)$ is an equivalence.
The most important step is to find
a lift of branes from $Y$ to the gauge theory.
We show that the functor constructed in \cite{ADS} is indeed such a lift
(up to a shift), using the understanding of the boundary flow in $SU(2)$ QCD.

\subsection{Preliminary Remarks On The Flow
$D_G(V,W)\supset{\mathcal S}\stackrel{\pi_Y}{\longrightarrow} D(Y)$}
\label{subsec:Pflow}

\subsubsection{Empty Branes}

The $p$-Koszul brane $\Kos$
introduced in Section~\ref{subsec:Gflow}
has an $SU(2)$ invariant Chan-Paton representation 
and hence defines a brane in the Pfaffian phase.
As it has the matrix factorization $Q$ given in
(\ref{QKp}), the boundary potential is
\beq
\{Q,Q^{\dag}\}=\sum_{k=1}^7|p^k|^2+\sum_{k=1}^7|A_k(x)|^2.
\eeq
It is always positive in the Pfaffian phase where $(p^1,\ldots,p^7)$
is non-vanishing by the D-term equation.
Then, the complete brane-antibrane annihilation
via tachyon condenseation is at work,
and the brane $\Kos$ is empty at low energies.
If we denote an empty brane as $0$ we can say
\beq
\pi_Y(\Kos)\cong 0.
\label{Kosempty}
\eeq
The same applies to the tensor product of $\Kos$
and a representation $\rmV$ of the gauge group
or a complex $\Complex$ of representations over $\C[p,x]$,
\beq
\pi_Y(\Kos\otimes \rmV)\cong\pi_Y(\Kos\otimes\Complex)\cong 0.
\label{Kosvarempty}
\eeq
Of course, the representation $\rmV$ and those in $\Complex$
must be $SU(2)$ grade restricted in the Pfaffian phase.

\subsubsection{Weakly Coupled Sector}

\newcommand{\wC}{\mathscr{C}}

In the Pfaffian phase,
even though the $SU(2)$ subgroup of the $U(2)$ gauge group is strongly coupled,
the $U(1)$ quotient group as in
\beq
1\to SU(2)\longrightarrow U(2)\stackrel{\det\,}{\longrightarrow} U(1)\to 1
\eeq
is weakly coupled as it is broken by the non-zero values
of $p=(p^1,\ldots,p^7)$ of charge $-1$.
In particular, classical reasoning can be
employed to read off the information on the low energy behaviour of
that part of the theory. If the theory consisted only of this
$U(1)$ gauge group and the fields $p=(p^k)$,
the low energy theory would be the non-linear sigma model
with target $\CP^6$, and 
the charge $i$ representation placed on the Chan-Paton factor
would reduce to the line bundle ${\mathcal O}_{\CP^6}(-i)$ on $\CP^6$.
More generally,
a complex $\wC$ of representations over $\C[p]$ of $U(1)$
would reduce to the corresponding complex of vector bundles on
$\CP^6$, which we denote by ${\mathcal O}_{\CP^6}(\wC)$.

Now let us consider branes in our gauge theory.
The charge $i$ representation of $U(1)$ can be regarded as a representation
of the full gauge group $U(2)$ via the homomorphism $\det:U(2)\to U(1)$
--- it is nothing but $\det^i\fund=\C(i)$.
The complex $\wC$ can likewise be regarded
as a complex of $U(2)$ representations over $\C[p]$.
While they alone do not determine branes in the full theory,
they can be used to construct new branes from a given one.
If $\brane$ is a matrix factorization of the superpotential $W$, so are
the tensor products $\brane\otimes \C(i)=\brane(i)$
and $\brane\otimes \wC$.
Since the $\C(i)$'s are $SU(2)$ singlets,
if $\brane$ is $SU(2)$ grade restricted, they are also $SU(2)$
grade restricted and hence define branes in the Pfaffian phase.
Employing the classical argument in the weakly coupled sector,
we expect that the low energy limits
of $\brane(i)$ and $\brane\otimes \wC$
are the low energy limits of $\brane$
tensored with ${\mathcal O}_Y(-i)$ and ${\mathcal O}_Y(\wC)$.
For a vector bundle ${\mathcal V}$ (or a complex thereof) on $Y$,
we shall write ${\mathcal V}(i):={\mathcal V}\otimes {\mathcal O}_Y(i)$.

There is one subtlety though. In the above discussion, we have assumed
that the $p$ fields have vanishing R-charges. That is, the
$\epsilon\nearrow 1$ limit in (\ref{Rassignment}):
\beq
R^{\rm Pf}_p=0\quad\mbox{and}\quad R^{\rm Pf}_x=1.
\label{Pfframe}
\eeq
However, we have decided to take the other limit $\epsilon\searrow 0$,
(\ref{Grframe}), and are planning to discuss
the transport of branes from one phase to another
using that choice throughout.
In particular, $\pi_Y:{\mathcal S}\to D(Y)$ is a map of gauge theory
branes in the frame (\ref{Grframe}) to branes in the low energy sigma model.
Thus, we need to move our ``R-charge frame'' back to that choice.
This involves a
change of brane R-charges.
As discussed in Section 3.4 of \cite{HR}, if we shift the R-charge
of the matter fields by $R\to R+{\mit\Delta}$ by a central element
${\mit\Delta}$ of the Lie algebra of the gauge group,
the brane R-charge is shifted by
${\bf r}_*\to {\bf r}_*-\rho({\mit\Delta})$.
In the present case, the change from the Grassmannian frame (\ref{Grframe})
to the Pfaffian frame (\ref{Pfframe}) is done by $\Delta={\rm diag}(1,1)$,
and hence the brane R-charge changes as
\beq
    {\bf r}_*\,\longrightarrow\,{\bf r}_*^{\rm Pf}
    ={\bf r}_*-\rho\left(\begin{array}{cc}
      1&0\\
      0&1
    \end{array}\right).
\label{RPf}
\eeq
Let us denote by $\brane\to\brane^{\rm Pf}$ the corresponding change of
the brane datum.
For the representation $\C(i)$, the shift is by $-2i$, and hence
$(\brane(i))^{\rm Pf}=\brane^{\rm Pf}(i)[2i]$.
Therefore, what we have seen above implies
\beq
\pi_Y(\brane(i))=\pi_Y(\brane)(-i)[2i],
\label{shiftPf}
\eeq
and
\beq
\pi_Y(\brane\otimes\wC)=
\pi_Y(\brane)\otimes {\mathcal O}_Y(\wC^{\rm Pf}).
\label{Wtensor}
\eeq

\subsection{Computation Of The Hemisphere Partition Function}
\label{subsec:PHS}

Let us compute the partition function (\ref{Zthemodel}) in the Pfaffian phase.
As the contour, we can take $\bgamma_-$ with a small shift
(\ref{shiftcontour}). If we change the integration variables as
$\usigma_0=\usigma_1+\usigma_2$ and $\usigma_{12}=(\usigma_1-\usigma_2)/2$,
it reads
\beq
\Is_0=-2(\Rs_0)^2-2\delta,\quad \Is_{12}=0,\qquad 0<\delta\ll 0.
\eeq
The poles (\ref{pole1}), (\ref{pole2}), (\ref{poleP}) are at
\beqa
&\usigma_0=-\im (n_p+1),\label{polePagain}\\
&\usigma_{12}+{\usigma_0\over 2}=\im n_1,\quad
-\usigma_{12}+{\usigma_0\over 2}=\im n_2.
\eeqa
We first perform the $\usigma_0$ integration.
By deforming the contour, it picks up the poles at
$\usigma_0=-\im (n_p+1)$.
For each $n_p$, we have the $\usigma_{12}$
integration along the real line and the integrand has poles
on the positive and negative imaginary axis,
$\usigma_{12}={\im\over 2}(n_p+1)+\im n_1$ and
$-{\im\over 2}(n_p+1)-\im n_2$.
When the $SU(2)$ grade restriction rule is satisfied,
this integral is absolutely convergent. In particular, we may
include a factor $\e^{\im \xi \usigma_{12}}$ for a real positive $\xi$
in the integrand
and take the limit $\xi\searrow 0$ after the integration, without changing
the result. This factor allows us to deform the contour toward
the positive imaginary direction. 
Then, the integration picks up the poles on the positive
imaginary axis. Thus, we find
\beq
Z_{D^2}(\mathfrak{B})\,=\,
\lim_{\xi\searrow 0}
\sum_{n_p,n_1=0}^{\infty}
\oint\limits_{{\im\over 2}(n_p+1)+\im n_1}\dd\usigma_{12}\,
\e^{\im\xi\usigma_{12}}
\oint\limits_{-\im (n_p+1)}\dd\usigma_0\,\,\Bigl(\cdots\Bigr),
\label{sumZS}
\eeq
where the ellipsis is
\beq
2\usigma_{12}\sinh(2\pi\usigma_{12})
\Gamma(-\im \usigma_0+1)^7
\Gamma(\im\usigma_1)^7
\Gamma(\im\usigma_2)^7\e^{\im t\usigma_0}
f_\brane(\usigma_1,\usigma_2)
\Bigr|_{\usigma_1={1\over 2}\usigma_0+\usigma_{12}\atop
  \usigma_2={1\over 2}\usigma_0-\usigma_{12}}.
\label{ellipsis}
\eeq
For each term of the sum,
if we write $\usigma_0=-\im (n_p+1)+z_0$ and
$\usigma_{12}={\im\over 2}(n_p+1)+\im n_1+z_{12}$, the contour 
is $|z_0|=\varepsilon_0$ and
$|z_{12}|=\varepsilon_{12}$ with $\varepsilon_0\ll \varepsilon_{12}$
while the pole loci are $z_0=0$ and $z_{12}+{z_0\over 2}=0$.
Rewriting $x_0=-\im z_0$ and $x_1=\im\left(z_{12}+{z_0\over 2}\right)$,
the contour is homotopic to $|x_0|=|x_1|=\varepsilon$, and
(\ref{sumZS}) is computed as
\beqa
Z_{D^2}(\mathfrak{B})&=&
\lim_{\xi\searrow 0}\sum_{n_p,n_1=0}^{\infty}
\e^{t(n_p+1)-\xi\, n_1}\!\!\!\!\!
\oint\limits_{|x_0|=|x_1|=\varepsilon}\!\!
\im^2{\dd x_{0\,}\dd x_1\over x_0^7x_1^7}\,F_{n_p,n_1}(x_0,x_1;t)
\nn\\
&=&\lim_{\xi\searrow 0}\sum_{n_p,n_1=0}^{\infty}
\e^{t(n_p+1)-\xi\, n_1}{(2\pi)^2\over 6!^2}
\left({\partial^{12}
F_{n_p,n_1}(x_0,x_1;t)\over
\partial x_0^{6}\partial x_1^{6}}\right)_{x_0=x_1=0}\nn\\[0.2cm]
&=:&Z_{\rm Pf}\{f_{\BBB}\},
\label{Zstrongexp}
\eeqa
with
\beqa
F_{n_p,n_1}(x_0,x_1;t)\!\!&:=&\!\!(1+n_p+2n_1-x_0-2x_1)\sin(\pi 
(x_0+2x_1))(-1)^{n_p+1}\times
\nn\\
&&
x_0^7\Gamma(-n_p+x_0)^7 x_1^7\Gamma(-n_1+x_1)^7
\Gamma(1+n_p+n_1-x_0-x_1)^7\times\nn\\
&&
\e^{-t x_0}\,f_{\mathfrak{B}}(-\im x_1,\im(x_0+x_1)).
\label{defF}
\eeqa
It takes the form of an expansion in powers of $\e^t$, but to evaluate
each order, the infinite sum over $n_1$ must be carried out.
For this, we use the Mathematica packages {\tt Sigma.m} \cite{RISC203}
and {\tt HarmonicSums.m} \cite{harmonic} which
can perform symbolic manipulations of sums.
Despite these powerful computational tools, the expressions in
(\ref{Zstrongexp}) become increasingly complicated for higher orders
in $\e^t$ and one quickly reaches technical limits. 
In the R\o dland model, we managed to
perform relevant computations up to the second order (i.e.
$n_p=2$).

Let us compute the hemisphere partition function of the $p$-Koszul
brane $\Kos$. Its brane factor is
\beq
f_{\Kos}(\usigma)
  =(1-\e^{2\pi(\usigma_1+\usigma_2)})^7=(1-\e^{2\pi\usigma_0})^7.
\eeq
This cancels the seventh order pole of $\Gamma(-\im\usigma_0+1)^7$
in (\ref{ellipsis}) at (\ref{polePagain}) and the $\usigma_0$ integration
vanishes. Thus, we find
\beq
Z_{D^2}(\Kos)\,=\,0.
\label{ZKos}
\eeq
This is consistent with the result  (\ref{Kosempty}) that the
$p$-Koszul brane is empty in the infra-red limit in the Pfaffian phase.
The same applies to its variants as their brane factors are divisible by
$(1-\e^{2\pi\usigma_0})^7$,
\beq
Z_{D^2}(\Kos\otimes \rmV)\,=\,Z_{D^2}(\Kos\otimes\Complex)\,=\,0,
\label{ZKosvar}
\eeq
which is consistent with (\ref{Kosvarempty}).

At this point, we only have a limited information
such as empty branes and the r\^ole of the weakly coupled sector.
In this situation, we first do some experiments.
That is, we compute $Z_{\rm Pf}\{f\}$ defined in
(\ref{Zstrongexp})-(\ref{defF}) for some trial functions
$f(\usigma_1,\usigma_2)$,
instead of the functions $f_{\BBB}(\usigma_1,\usigma_2)$
that correspond to actual branes, and see for which functions
$f$ it produces the expected forms
(\ref{Zpt})-(\ref{ZOX}) of the central charges
in the large volume limit of $Y$.

A random choice of $f$ may not be a brane factor of an actual brane,
in which case $Z_{\rm Pf}\{f\}$ does not solve the Picard-Fuchs equations.
It is a non-trivial task even to find a function such that
$Z_{\rm Pf}\{f\}$ can be written as a linear combination of periods.
It turns out to be useful to consider the functions
\beq
f_{a,b}(\usigma_1,\usigma_2):=
(1-\e^{2\pi(\usigma_1+\usigma_2)})^a(1-\e^{2\pi\usigma_1})^b
(1-\e^{2\pi\usigma_2})^b.
\label{deffab}
\eeq
Note that $Z_{\rm Pf}\{f_{a,b}\}=0$ if $a\geq 7$ since the factor
$(1-\e^{2\pi (\usigma_1+\usigma_2)})^7$ would cancel the seventh order pole
of $\Gamma(-\im\usigma_0+1)^7$
and the $\usigma_0$ integration would vanish, as in (\ref{ZKos})
or (\ref{ZKosvar}).
Note also that we can only consider $f_{a,b}$ with $b=0,1,2$
since those with $b\geq 3$ would correspond to representations
violating the $SU(2)$ grade restriction rule.

After some trial and error, we find that $Z_{\rm Pf}\{ -\}$ of
certain linear combinations of these functions
can be matched with the periods $\Pi_0,\ldots, \Pi_3$
given in Appendix~\ref{app:periods}, Eqns
(\ref{per-struct1})-(\ref{per-struct4}),
(\ref{periodRodPfa}):
Under the identification $\log z=t-7\pi\im$,
\beqa
Z_{\rm Pf}\{f_{6,2}\}&=&-(2\pi)^{13}\e^t\,\Pi_0,\label{Z62}\\
Z_{\rm Pf}\{f_{5,2}\}&=&-(2\pi)^{13}\e^t\,\Pi_1,\label{Z52}\\
Z_{\rm Pf}\{f_{\rm D4?}\}&=&-(2\pi)^{13}\e^t\left(\Pi_2-\Pi_1+{2\over 3}\Pi_0\right),
\label{ZD4Q}\\
Z_{\rm Pf}\{f_{\rm D6?}\}&=&-(2\pi)^{13}\e^t\left({1\over 3}\Pi_3
+{1\over 3}\Pi_1-\im {14\zeta(3)\over (2\pi)^3}\Pi_0\right),
\label{ZD6Q}
\eeqa
with
\beqa
f_{\rm D4?}&:=&~~~~~~~~~~~~~~~~\,-2f_{6,1}+2f_{4,2}-2f_{5,2}+f_{6,2},\\
f_{\rm D6?}&:=&-2f_{5,1}+2f_{3,2}+f_{6,1}-2f_{4,2}+f_{5,2}.
\eeqa
to the order $z^2$ (i.e. $n_p=2$) which we computed.
In comparison with the central charge formula (\ref{Zpt})-(\ref{ZOX})
with the aid of (\ref{topPfa}),
we see that
(\ref{Z62}) through (\ref{ZD6Q}) coincides with
$-Z_Y({\mathcal O}_{\rm pt})$,
$-Z_Y({\mathcal O}_{\ell})+Z({\mathcal O}_{\rm pt})$,
$-Z_Y({\mathcal O}_H)/7$ and
$-Z_Y({\mathcal O}_Y)/7$,
times a common factor $(2\pi)^{13}\e^t$.

This result implies that the gauge theory branes that flow to
${\mathcal O}_{\rm pt}$,  ${\mathcal O}_{\ell}$, ${\mathcal O}_H$,
${\mathcal O}_Y$ in the low energy sigma model have brane factors
\beq
f_{\rm D0}=-f_{6,2},\quad
f_{\rm D2}=-f_{5,2}-f_{6,2},\quad
f_{\rm D4}=-7f_{\rm D4?},\quad
f_{\rm D6}=-7f_{\rm D6?}
\eeq
respectively,
modulo addition of functions annihilated by $Z_{\rm Pf}\{-\}$,
provided the hemisphere partition function is related to the central charge by
\beq
Z_{D^2}(\brane)\,\stackrel{!}{=}\,(2\pi)^{13}\e^t \cdot Z_Y(\pi_Y(\brane)).
\label{HSvsCCPf}
\eeq
In the Pfaffian phase, the natural choice of the R-charge is (\ref{Pfframe})
which is obtained from our choice (\ref{Grframe}) by the gauge
shift by ${\mit\Delta}={\rm diag}(1,1)$.
As explained in Section 3.4 of \cite{HR}, the outcome of this shift is
the shift of the classical action by $\Delta S={1\over 2}t({\mit\Delta})=t$
which corresponds to the overall multiplication of the partition function
by $\e^{-t}$.
That would eliminate the overall factor $\e^t$ in (\ref{HSvsCCPf}).

We can perform a consistency check on the above claim.
In the low energy sigma model, intersecting a brane
with a hyperplane $D_h=\{h(p)=\sum_{k=1}^7h_kp^k=0\}$
is realized by taking the tensor product with
${\mathcal O}_{D_h}$, which is isomorphic to the complex
\beq
\begin{tikzcd}
  \displaystyle\mathop{{\mathcal O}(-1)}_{{}_{-1}}\arrow[r, "h(p)"]&
         \displaystyle\mathop{{\mathcal O}}_{{}^{}_0}.
\end{tikzcd}
\eeq
Via (\ref{Wtensor}), this lifts to
the tensor product of a gauge theory brane $\brane$ with the complex
\beq
\begin{tikzcd}
  \displaystyle\mathop{\C(1)}_{{}_1}\arrow[r, "h(p)"]&
  \displaystyle\mathop{\C}_{{}_0}.
\end{tikzcd}
\label{wCh}
\eeq
On the brane factor, this has the effect
\beq
f_{\brane}(\usigma)\,\longrightarrow\,
f_{\brane}(\usigma)(1-\e^{2\pi(\usigma_1+\usigma_2)}).
\eeq
Note that this makes the modification $f_{a,b}\to f_{a+1,b}$ on
the functions (\ref{deffab}).
Starting with $f_{\rm D6}$, interating this operation produces
\beq
f_{\rm D6}\,\longrightarrow\,
f_{\rm D4}\,\longrightarrow\,
14(f_{\rm D2}-2f_{\rm D0})\,\longrightarrow\,
14f_{\rm D0},
\eeq
modulo $f_{a,b}$ with $a\geq 7$. When $((2\pi)^{13}\e^{t})^{-1}Z_{\rm Pf}\{-\}$
is applied, this sequence becomes
\beq
Z_Y({\mathcal O}_{Y})\,\longrightarrow\,
Z_Y({\mathcal O}_H)\,\longrightarrow\,
14 \bigl(Z_Y({\mathcal O}_{\ell})-2Z_Y({\mathcal O}_{\rm pt})\bigr)
  \longrightarrow\,
  14 Z_Y({\mathcal O}_{\rm pt}).
\eeq
This is indeed the sequence of central charges
for the intersection with hyperplanes
\beq
{\mathcal O}_Y\,\longrightarrow\,
{\mathcal O}_{D_{h_1}}\,\longrightarrow\,
{\mathcal O}_{D_{h_1}\cap D_{h_2}}\,\longrightarrow\,
{\mathcal O}_{D_{h_1}\cap D_{h_2}\cap D_{h_3}},
\eeq
as can be seen by
\beq
{\rm ch}({\mathcal O}_{D_{h_1}\cap \cdots\cap D_{h_n}})
={\rm ch}({\mathcal O}_H)^n=\left(1-\e^{-H}\right)^n,
\eeq
with the aid of $H^2=14\ell$, $H^3=14[{\rm pt}]$,
${\rm ch}({\mathcal O}_{\ell})=\ell+[{\rm pt}]$
and ${\rm ch}({\mathcal O}_{\rm pt})=[{\rm pt}]$.

\subsection{Candidates Of Gauge Theory Branes For D0 And D6-Branes}

We identify some branes in the gauge theory that have
brane factors corresponding to D0-branes and D6-branes.

\subsubsection{D0-Branes}
\label{subsub:D0}

The brane factor $f_{\rm D0}=-f_{6,2}$
corresponding to a D0-brane is $(-1)$ times the character of
the exterior algebra generated by six copies of $\det\fund$
and two copies of $\fund$.
We can find an example of branes with such
a Chan-Paton representation, as described blow.

Let us choose an arbitrary point $p_*$ in the Pfaffian Calabi-Yau
$Y\subset \CP^6$.
By linear reparametrizations of the $p$'s and $x$'s,
we may assume that $p_*=(0,\ldots,0,1)$ and that $A(p_*)=A_7$ has
$A^{45}(p_*)=-A^{54}(p_*)=A^{67}(p_*)=-A^{76}(p_*)={1\over 2}$ as its only
non-zero entries.
Then, we have a matrix factorization of $W$,
\beq
Q_{p_*}=\sum_{k=1}^6\left(p^k\eta_k+A_k(x)\bareta^k\right)
+\sum_{a=1,2}\left(
p^7x_4^a\xi_a^4+x_5^a\xi_a^5
+p^7x_6^a\xi_a^6+x_7^a\xi_a^7\right),
\label{Qpstar}
\eeq
realized on the irreducible representation $\bV_{p_*}$ of
the Clifford algebra
\beqa
&\{\eta_k,\bareta^l\}=\delta_k^l,\quad 1\leq k,l\leq 6,\nn\\
&\{\xi_a^{\mu},\xi_b^{\nu}\}=2A(p_*)^{\mu\nu}\epsilon_{ab},
\quad 4\leq \mu,\nu\leq 7,~~\,a,b=1,2,\\
&\mbox{other anticommutators =0}.\nn
\eeqa
For gauge invariance (\ref{Qgi}) and vector $U(1)$ R-symmetry (\ref{QRs}),
the Clifford generators have gauge and R-charges as follows:
\beqa
&\eta_1,\ldots,\eta_6:\,(\det^{-1}\fund,-1),~\,\,
\bareta^1,\ldots,\bareta^6:\,(\det\fund, +1),\label{D0charge1}\\
&(\xi_a^4),\,(\xi^6_a):\,(\fund(-1),-1),~\,\,
(\xi_a^5),\,(\xi_a^7)\, :\,(\fund,+1).\label{D0charge2}
\eeqa
The space $\bV_{p_*}$ is built on the Clifford vacuum $|0\rangle$,
annihilated by the $\eta_k$'s, $\xi^4_a$'s and $\xi^6_a$'s,
by multiplying powers of the $\bareta^k$'s, $\xi_a^5$'s and $\xi_a^7$'s.
Thus, it is indeed the exterior algebra generated by six copies of 
$\det\fund$ and two copies of $F$ as we wanted.
We assume that the state $|0\rangle$ is gauge invariant and has an odd
R-charge,
\beq
|0\rangle:\,(\C,i_0),\quad i_0\in 2\Z+1,
\label{D0charge3}
\eeq
so that the brane factor is $-f_{6,2}$.
Note that $\bV_{p_*}$ has a gauge invariant and $U(1)$ R-symmetric
hermitian inner product with hermiticity
$\eta_k^{\dag}=\bareta^k$ and $(\xi_a^{4,6})^{\dag}=\epsilon^{ab}\xi_b^{5,7}$.
($\epsilon^{ab}$ is the inverse of
$\epsilon_{ab}$; $\epsilon^{ab}\epsilon_{bc}=\delta^a_c$.)
We denote this brane by $\brane_{p^*}^{(i_0)}=(\bV_{p_*}^{(i_0)},Q_{p_*})$.

Recall that the classical D-term and F-term equations forbid $p$ to vanish
and require $x$ to vanish. On this locus, the boundary potential is
\beq
\{Q_{p_*},Q_{p_*}^{\dag}\}\,=\,\sum_{i=1}^6|p^i|^2.
\label{bpPf}
\eeq
Therefore, at the classical level, the brane $\brane_{p_*}^{(i_0)}$
is supported at low energies
at the locus  $p^1=\cdots=p^6=0$ and $p^7\ne 0$, that is, at the point $p_*$.
This and $Z_{D^2}(\brane_{p_*}^{(i_0)})\propto Z_Y({\mathcal O}_{p_*})$
indicate that the low energy limit of $\brane_{p_*}^{(i_0)}$
is the D0-brane at $p_*\in Y$. Later we shall provide a proof of
this identification.

\subsubsection{D6-Branes}
\label{subsub:D6}

Let us next compute the brane factors of
the branes $\mathfrak{L}^h(i)$
introduced in Section~\ref{subsec:GRprocess} as
$SU(2)$ grade restricted branes which are empty in the Grassmannian phase.
As we will see, they are the brane factors corresponding to D6-branes
of specific charges.

Let $C_{\rm i}$,\ldots,$C_{\rm vi}$ be the twisted Lacoux complexes
of the type (i),\ldots,(vi) in Fig.~\ref{fig:Lc} such that the tail
components are $\Sym^h\fund(0)$ for $h=5,\ldots,0$ at an even
degree.\footnote{In the present computation, we can ignore the action of
  $\GL(V)$ so that there is no disctinction between Lascoux complexes
  of the types (i),\ldots,(vi) and of the types (i)',\ldots,(vi)'
  in (\ref{otherLcLb}).}
They have characters
\beqa
f_{\rm i}&=&f_{S^5\fund}-7f_{S^4\fund(1)}+21 f_{S^3\fund(2)}
-35 f_{S^2\fund(3)}+35 f_{\fund(4)}-21 f_{\C(5)}+f_{\C(6)},\\
f_{\rm ii}&=&f_{S^4\fund}-7f_{S^3\fund(1)}+21f_{S^2\fund(2)}
-35 f_{\fund(3)}+35 f_{\C(4)}-7 f_{\C(5)}+f_{\fund(5)},\\
f_{\rm iii}&=&f_{S^3\fund}-7f_{S^2\fund(1)}+21 f_{\fund(2)}
-35 f_{\C(3)}+21 f_{\C(4)}-7f_{\fund(4)}+f_{S^2\fund(4)},\\
f_{\rm iv}&=&f_{S^2\fund}-7f_{\fund(1)}+21f_{\C(2)}-35 f_{\C(3)}
+21 f_{\fund(3)}-7f_{S^2\fund(3)}+f_{S^3\fund(3)},\\
f_{\rm v}&=&f_{\fund}-7f_{\C(1)}+35f_{\C(2)}-35f_{\fund(2)}
+21f_{S^2\fund(2)}-7f_{S^3\fund(2)}+f_{S^4\fund(2)},\\
f_{\rm vi}&=&f_{\C}-21f_{\C(1)}+35 f_{\fund(1)}-35f_{S^2\fund(1)}
+21 f_{S^3\fund(1)}-7f_{S^4\fund(1)}+f_{S^5\fund(1)}.
\eeqa
The construction
described in Section~\ref{subsec:GRprocess} yields
the following expressions for
$\mathfrak{LC}^2,\mathfrak{LC}^1,\mathfrak{LC}^0$
as $\Z_2$-graded representations of the gauge group,
\beqa
[\mathfrak{LC}^2]&=&[C_{\rm iv}]-[C_{\rm iii}(3)],\\
{[}\mathfrak{LC}^1{]}&=&[C_{\rm v}]-[C_{\rm ii}(2)]
-7[C_{\rm iv}]+7[C_{\rm iii}(2)],\\
{[}\mathfrak{LC}^0{]}&=&[C_{\rm vi}]-[C_{\rm i}(1)]-7[C_{\rm v}]
+7[C_{\rm ii}(1)]
+21 [C_{\rm iv}]-21 [C_{\rm iii}(1)].
\eeqa
Their characters are the brane factors of
$\mathfrak{L}^2,\mathfrak{L}^1,\mathfrak{L}^0$.
For computation, it is convenient to
introduce $u=\e^{2\pi(\usigma_1+\usigma_2)}$ and note
$f_{\rmV(i)}=f_{\rmV}u^i$. We also simplify the notation as
$f_{S^h\fund}=:f_h$. Then,
\beqa
f_{\mathfrak{L}^2}&=&f_{\rm iv}-f_{\rm iii}u^3\nn\\
&=&(1-u)^3\Bigl(7u^2(3u^2+4u+3)f_0-7u(u+1)^3 f_1\nn\\
&&+(u^4+3u^3+6u^2+3u+1)f_2\Bigr),\label{fL2}\\
f_{\mathfrak{L}^1}&=&f_{\rm v}-f_{\rm ii}u^2-7f_{\rm iv}+7f_{\rm iii}u^2\nn\\
&=&-7u(u+1)(u^2+18u+1)f_0
-7(u+1)^3f_2\nn\\
&&+(1-u)^3(u^4+52u^3+118u^2+52u+1)f_1\label{fL1}\\
f_{\mathfrak{L}^0}&=&f_{\rm vi}-f_{\rm i}u-7f_{\rm v}+7f_{\rm ii}u+21f_{\rm iv}
-21f_{\rm iii}u
\nn\\
&=&(1-u)^3\Bigl(7(3u^2+4u+3)f_2
-7(u^3+19u^2+19u+1)f_1\nn\\
&&+(u^4+31u^3+286u^2+31u+1)f_0\Bigr).\label{fL0}
\eeqa
We indeed see that $f_{S^h\fund(i)}=f_hu^i$ with $h=5,4,3$ all cancel,
 and we can insert these expressions into $Z_{\rm Pf}\{~\}$.
Note that $f_0=1$, $f_1=u_1+u_2$ and $f_2=u_1^2+u_1u_2+u_2^2$
for $u_a=\e^{2\pi\usigma_a}$.
Writing $f_{0,b}=(1-u_1)^b(1-u_2)^b$ for $b=1,2$ in
terms of $f_1, f_2$ and $u$, and inverting the relations, we find
\beqa
f_1&=&1+u-f_{0,1},\\
f_2&=&1+u+u^2-2(1+u)f_{0,1}+f_{0,2}.
\eeqa
Inserting these into (\ref{fL2}), (\ref{fL1}), (\ref{fL0})
and expanding in powers of $x:=1-u$, we find
\beqa
f_{\mathfrak{L}^2}&=&7\left(\,f_{0,1}(-2+3x)x^5+f_{0,2}x^3(2-4x+3x^2-x^3)\,\right)
+O(x^7)\nn\\
&\equiv&7\bigl(\, -2f_{5,1}+3f_{6,1}+2f_{3,2}-4f_{4,2}+3f_{5,2}-f_{6,2}\,\bigr)
\nn\\
&=&-f_{\rm D6}+f_{\rm D4},\\[0.1cm]
f_{\mathfrak{L}^1}&=&7\left(\,8f_{0,1}(1-x)x^5+f_{0,2}(-2+x)^3x^3\,\right)
+O(x^7),\nn\\
&\equiv&7\bigl(\,8f_{5,1}-8f_{6,1}-8f_{3,2}+12f_{4,2}-6f_{5,2}+f_{6,2}\,\bigr)
\nn\\
&=&4 f_{\rm D6}-2f_{\rm D4}-14f_{\rm D2}+21f_{\rm D0},\\[0.1cm]
f_{\mathfrak{L}^0}&=&7\left(\,5f_{0,1}(-2+x)x^5+f_{0,2}x^3(10-10x+3x^2)\,\right)
+O(x^7)\nn\\
&\equiv&7\bigl(\,-10f_{5,1}+5f_{6,1}+10f_{3,2}-10f_{4,2}+3f_{5,2}\,\bigr)\nn\\
&=&-5f_{\rm D6}+14f_{\rm D2}-14f_{\rm D0}.
\eeqa
In the above expressions,
$\equiv$ stands for equality modulo terms divisible by $x^7$ that vanish
when inserted into $Z_{\rm Pf}\{~\}$. We also used $f_{0,b}x^a=f_{a,b}$.

The result implies that the D-brane charges of
the low energy limits of the $\mathfrak{L}^h$'s are\footnote{In
  the present paper, we use $[{\mathcal F}]$ as the shorthand notation for
  the Chern character ${\rm ch}({\mathcal F})$ for a complex ${\mathcal F}$
  of coherent sheaves on a K\"ahler manifold $M$.
  As in Type II superstring theory \cite{WittenK},
  the lattice of D-brane charges in a non-linear sigma model
  with target space $M$ may be identified as the topological K-theory of $M$.
  For $X$ and $Y$ in the R\o dland model, the topological K-theory has
  no torsion (see the footnote on page \pageref{page:foot}).
  Thus, $[{\mathcal F}]={\rm ch}({\mathcal F})$
  captures the full information of the D-brane charge of ${\mathcal F}$
  for $M=X$ or $Y$ in this model.}
\beqa
[\pi_Y(\Lb^2)]\!&=&\!-[{\mathcal O}_Y]
+[{\mathcal O}_H]\,=\,-[{\mathcal O}_Y(-1)],
\label{chL2}\\
{[}\pi_Y(\Lb^1){]}\!&=&\!
4[{\mathcal O}_Y]-2[{\mathcal O}_H]
-14[{\mathcal O}_{\ell}]+21[{\mathcal O}_{\rm pt}],
\label{chL1}\\
{[}\pi_Y(\Lb^0){]}\!&=&\!
-5[{\mathcal O}_Y]+14[{\mathcal O}_{\ell}]
-14[{\mathcal O}_{\rm pt}].
\label{chL0}
\eeqa
We see that $\pi_Y(\mathfrak{L}^2)$,
$\pi_Y(\mathfrak{L}^1)$ and $\pi_Y(\mathfrak{L}^0)$
have the same charge as a single anti-D6-brane,
a D6-brane supporting a rank 4 vector bundle
and an anti-D6-brane supporting a rank 5 vector bundle.
In view of (\ref{shiftPf}), we find that (\ref{chL2}) implies
$[\pi_Y(\Lb^2(-1))]=-[{\mathcal O}_Y]$.
We shall prove below that they indeed flow to the (anti-)D6-branes
of the stated ranks at low energies.

\subsection{Exact Results On $SU(2)$ Supersymmetric QCD With Boundary}
\label{subsec:SU2}

\newcommand{\tti}{\texttt{i}}
\newcommand{\ttj}{\texttt{j}}
\newcommand{\ttk}{\texttt{k}}
\newcommand{\ttl}{\texttt{l}}
\newcommand{\triv}{\brane_{\bf 1}}

The main result of \cite{HoTo} relevant for the 
Pfaffian phase of the R\o dland model is that the $SU(2)$ gauge theory
with three doublets $x_1,x_2,x_3$ and vanishing superpotential
$W=0$ flows in the infra-red limit
to the free theory of baryons $b_{12}=[x_1x_2]$,
$b_{23}=[x_2x_3]$ and $b_{31}=[x_3x_1]$,
that is, the sigma model with target $\C^3$.
We would like to extend this to the boundary sector:
Given a D-brane in the gauge theory, where does it
flow to in the free theory at low energies?
Recall from Section~\ref{subsec:SU2GRR} that only the $SU(2)$ singlet
is allowed as the Chan-Paton representation.
Let $\triv$ be the
purely Neumann boundary condition with the $SU(2)$ singlet on the boundary
with R-charge $0$.
The problem is to identify its infra-red limit.
Since there is no condition on the boundary values of the $x_\tti$'s,
it is expected that there will be no condition on the boundary values
of the baryons $[x_\tti x_\ttj]$.
Thus, we propose that the infra-red limit is the Neumann boundary
condition on the fields $b_{12}, b_{23}, b_{31}$, that is, the D6-brane on
$\C^3$.

Let us test this proposal using the hemisphere partition function.
Since the brane is non-compact, we need some regularization in order
to make the partition function finite.
For this purpose,
we turn on the twisted mass $\wt{m}_\tti$ and assign R-charge
$R_\tti$ for each doublet $x_\tti$.
This should correspond in the infra-red limit
to the twisted mass $\wt{m}_\tti+\wt{m}_\ttj$ and R-charge
$R_\tti+R_\ttj$ for the baryon $b_{\tti\ttj}$.
The hemisphere partition functions in the ultra-violet and
the infra-red theories are respectively,
\beqa
Z_{D^2}^{{\rm UV}}(\triv)
&=&\int_{\sgamma}\dd\cs \cdot 2\cs\sinh(2\pi\cs)
\prod_{\tti=1}^3\Gamma(\im(\cs+\wt{m}'_\tti))
\Gamma(\im(-\cs+\wt{m}'_\tti)),\\
Z_{D^2}^{{\rm IR}}({\rm D6})
&=&\Gamma(\im(\wt{m}'_1+\wt{m}'_2))\Gamma(\im(\wt{m}'_2+\wt{m}'_3))
\Gamma(\im(\wt{m}'_3+\wt{m}'_1)),
\eeqa
where $\wt{m}'_\tti:=\wt{m}_\tti-\im {R_\tti\over 2}$ \cite{HR}.
The contour $\bgamma\subset \C$ is the real line if each $\wt{m}'_\tti$
is given a small negative imaginary part, and obtained from that by analytic
continuation for a general configuration of $\wt{m}'_\tti$'s \cite{HR}.
The two partition functions must agree under the proposal.
Indeed, one can show that $Z_{D^2}^{{\rm UV}}(\triv)$ has the same
singularity structure as $Z_{D^2}^{{\rm IR}}({\rm D6})$:
it has a simple pole at $i(\wt{m}'_\tti+\wt{m}'_\ttj)=0,-1,-2,\ldots$.
If we compute the residues at some of them, we find that
they match up to a coefficient $(2\pi)^2$. 
This suggests that
\beq
Z_{D^2}^{{\rm UV}}(\triv)
=(2\pi)^2\cdot Z_{D^2}^{{\rm IR}}({\rm D6}).
\label{RGflow1}
\eeq
In fact, this relation can be proven --- it follows from an
identity found in \cite{Gustafson}: Take the $a_4\to\infty$ limit in
Eqn (9.4) with $n=1$ of that paper.
This gives strong support for the proposal that
the Neumann boundary condition with an $SU(2)$ singlet on the boundary flows
under the renormalization group flow to the Neumann boundary condition
on the fields $b_{12},b_{23},b_{31}$.

Once this is established, it is a trivial matter to find the ultra-violet 
lifts of other infra-red branes.
Any brane in $\C^3$ can be realized as a complex
of trivial vector bundles on $\C^3$. All the maps in the complex
are holomorphic functions of the baryons, which are $SU(2)$ invariants,
and hence its lift to the gauge theory consists only of the
trivial representation of $SU(2)$. Thus, the lift obeys the grade
restriction rule.

\newcommand{\mm}{n}

Let us add to $x_1,x_2,x_3$ an even number of
doublets $x_4,\ldots, x_N$ ($N=3+2\mm$)
with superpotential mass terms
\beq
W=\sum_{\mu,\nu=4}^NA_*^{\mu\nu}[x_\mu x_\nu]
\label{massm}
\eeq
where $A_*=(A_*^{\mu\nu})$ is an antisymmetric and
non-degenerate matrix of size $2\mm$.
We may assign R-charge $1$ to $x_4,\ldots, x_N$.
The doublets $x_4,\ldots, x_N$ decouple at energies below the smallest
eigenvalues of $A_*$ and the theory flows
in the infra-red limit again to the free theory of
the three baryons $b_{\tti\ttj}=[x_{\tti} x_{\ttj}]$.
The grade restriction rule found in Section~\ref{subsec:SU2GRR} says that
only $\Sym^h\C^2$ with $h=0,1,\ldots, {N-3\over 2}=\mm$ are allowed
as the Chan-Paton representations. 
The simplest choice for a matrix factorization of $W$ would be
$\brane^{SU(2)}_{3,N}$ given by
\beq
Q=\sum_{\mu=4}^N\sum_{a=1}^2\,x_{\mu}^a\xi^{\mu}_a
\label{QSU2}
\eeq
on the unique irreducible representation $\bV$ of the Clifford algebra
\beq
\{\xi_a^{\mu},\xi_b^{\nu}\}=2A_*^{\mu\nu}\epsilon_{ab},\quad
4\leq \mu,\nu\leq N,~\,a,b=1,2.
\label{xiSU2}
\eeq
$(\xi^{\mu}_a)_{a=1,2}$ for each $\mu$
form an $SU(2)$ doublet and have R-charge zero.
With a suitable basis change, we may assume
that $A_*$ is of the block-diagonal form
\beq
A_*={1\over 2}\left(\begin{array}{ccccc}
  0&1&&&\\-1&0&&&\\
  &&\ddots&&\\
  &&&0&1\\
  &&&-1&0
\end{array}\right).
\label{Astar}
\eeq
Then, $\bV$ may be generated on the algebra
by a ``vacuum vector'' $|0\rangle$ annihilated by
$\xi_a^{2j+2}$ for $a=1,2$ and $j=1,\ldots,\mm$.
The vector $|0\rangle$ is $SU(2)$ invariant and may be assigned R-charge zero.
Then the entire space $\bV$ has R-charge zero.
In order to view $\bV$ as a representation of $SU(2)$,
note that the creation operators may be decomponsed into 
$\mm$ pairs $\{\xi^{2j+3}_1,\xi_2^{2j+3}\}$ ($j=1,\ldots, \mm$).
Note that $1$ and $\xi_1^{2j+3}\xi_2^{2j+3}$ are
$SU(2)$ singlets whereas $\xi_1^{2j+3}$ and $\xi_2^{2j+3}$
form an $SU(2)$ doublet $\sfund$.\footnote{We distinguish the notations
  for the $U(2)$ and $SU(2)$ doublets: $\fund$ for the former
  and $\sfund$ for the latter.}
Thus, as a representation of
$SU(2)$, $\bV$ is isomorphic to
\beq
\bV\cong (\C\oplus \sfund\oplus \C)^{\otimes \mm}.
\label{MSU2}
\eeq
It is a direct sum of $\Sym^h\sfund$ for $h=0,\ldots, \mm$
with certain multiplicities.
As $\mm={N-3\over 2}$, the brane $\brane^{SU(2)}_{3,N}$ is grade restricted.
We propose that
\beq
\boxed{\begin{array}{l}
 \mbox{$\brane^{SU(2)}_{3,N}$ flows in the infra-red limit
   to the purely Neumann boundary}\\
 \mbox{condition on the fields
   $b_{12}, b_{23}, b_{31}$, that is, the D6-brane on $\C^3$.}
 \end{array}}
\label{mainprop}
\eeq

\newcommand{\scs}{\mbox{\footnotesize \textsigma}}

Let us test this proposal by computing the hemisphere partition function.
Again, we give twisted mass $\wt{m}_\tti$ and R-charge $R_\tti$ to $x_\tti$
for $\tti=1,2,3$.
The ultra-violet partition function is
\beqa
Z^{\rm UV}_{D^2}(\brane^{SU(2)}_{3,N})
&=&\int_{\sgamma}\dd\cs \cdot 2\cs\sinh(2\pi\cs)
\prod_{\tti=1}^3
\Gamma(\im(\cs+\wt{m}'_\tti))\Gamma(\im(-\cs+\wt{m}'_\tti))\nn\\
&&~~~~~~~~~~
\left(\Gamma(\im\cs+{\textstyle {1\over 2}})
\Gamma(-\im\cs+{\textstyle {1\over 2}})
\right)^{2\mm}f_{\brane^{SU(2)}_{3,N}}(\cs),
\label{ZUV}
\eeqa
where the brane factor can be read off from (\ref{MSU2}) as
\beq
f_{\brane^{SU(2)}_{3,N}}(\cs)=(1+\e^{2\pi\scs}+\e^{-2\pi\scs}+1)^{\mm}
=(\e^{\pi\scs}+\e^{-\pi\scs})^{2\mm}.
\label{bfSU2}
\eeq
In view of Euler's reflection formula
\beq
\Gamma(\im\cs+{\textstyle {1\over 2}})
\Gamma(-\im\cs+{\textstyle {1\over 2}})
={2\pi\over \e^{\pi\scs}+\e^{-\pi\scs}},
\label{Gid2}
\eeq
we see that the brane factor cancels against
the one loop determinant of the massive multiplets
 up to a constant
\beq
\left(\Gamma(\im\cs+{\textstyle {1\over 2}})
\Gamma(-\im\cs+{\textstyle {1\over 2}})
\right)^{2\mm}f_{\brane^{SU(2)}_{3,N}}(\cs)=(2\pi)^{2\mm}.
\eeq
Using (\ref{RGflow1}), we obtain
\beq
Z_{D^2}^{\rm UV}(\brane^{SU(2)}_{3,N})
=(2\pi)^{N-1}\cdot
Z_{D^2}^{{\rm IR}}({\rm D6}),
\eeq
which supports the proposal (\ref{mainprop}).

Note that the non-degenerate antisymmetric matrix $A_*$
determines a symplectic structure on the flavor space
$\cktt\cong \C^{2\mm}$ for the massive doublets $x_4,\ldots, x_N$.
Then, the system with superpotential (\ref{massm})
has a flavor symmetry $Sp(\cktt)$.
The symmetry is preserved by the matrix factorization $Q$ of (\ref{QSU2})
if we identify $\cktt^*$ as the flavor space for $\xi^4,\ldots,\xi^N$.
However, the symmetry is not obvious in the presentation (\ref{MSU2})
of the Chan-Paton vector space $\bV$.
In fact, we can describe the brane $\brane^{SU(2)}_{3,N}$ in a way that the
$Sp(\sfund)\times Sp(\cktt)$ symmetry is manifest:
\beqa
&&  \Sym^{\mm}\sfund^*~\longleftrightarrow~
  \Sym^{\mm-1}\sfund^*\otimes \cktt~\longleftrightarrow~
  \Sym^{\mm-2}\sfund^*\otimes \textw_0^{\!2}\cktt~\longleftrightarrow~
  \cdots
\nn\\
&&
 ~~~~~\cdots ~\longleftrightarrow~
  \Sym^2\sfund^*\otimes \textw_0^{\!\mm-2}\cktt~\longleftrightarrow~
  \sfund^*\otimes \textw_0^{\!\mm-1}\cktt~\longleftrightarrow~
  \textw_0^{\!\mm}\cktt
  \label{massmfleft}
  \eeqa
In this expression,
$\textw_0^j\cktt$ is the subspace of $\textw^j\cktt$ consisting of
elements $\omega$ which are
$A_*$-traceless,
$A_*^{\mu_1\mu_2}\omega_{\mu_1\mu_2\cdots \mu_j}=0$.
The arrows are
the ones for the left part of the Lascoux matrix factorization
(the second version, see Section~\ref{subsub:otherversion}) for the case
$(k,N)=(2,2\mm)$, in which we replace $A(p)$ by $A_*$,
composed with the inclusion and a projection for
$\textw_0^j\cktt\subset \textw^j\cktt$.\footnote{Note that
  $\sfund^*\cong\sfund$ and
  $\textw^{2\mm-j}\cktt^*\cong \textw^j\cktt$ as representations of
  $Sp(\sfund)$ and $Sp(\cktt)$ respectively,
  and the projection $\textw^j\cktt\twoheadrightarrow \textw_0^j\cktt$
  is the unique one with $Sp(\cktt)$ equivariance.}
Equivalently, it is also given by
\beqa
&&\textw_0^{\!\mm}\cktt^*  ~\longleftrightarrow~
  \sfund^*\otimes \textw_0^{\!\mm-1}\cktt^*~\longleftrightarrow~
  \Sym^2\sfund^*\otimes \textw_0^{\!\mm-2}\cktt^*~\longleftrightarrow~
  \cdots
\nn\\
&&
 ~~~~~\cdots ~\longleftrightarrow~
\Sym^{\mm-2}\sfund^*\otimes \textw_0^{\!2}\cktt^*
~\longleftrightarrow~
  \Sym^{\mm-1}\sfund^*\otimes \cktt^*~\longleftrightarrow~
  \Sym^{\mm}\sfund^*
  \label{massmfright}
\eeqa
where the arrows are the ones for the right part of the Lascoux matrix
factorization (the original version), composed with the inclusion and
a projection for $\textw^j_0\cktt^*\subset \textw^j\cktt^*$.
These descriptions for the case $N=7$ will be used later.

\subsection{The Proof Of The Claimed Flows}
\label{subsec:theproof}

\newcommand{\ttx}{x}
\newcommand{\nhU}{{\mathcal U}}

We would like to confirm that $\brane_{p_*}^{(i_0)}$ and the $\Lb^h$'s
indeed flow to the D0 and D6-branes in the Pfaffian Calabi-Yau threefold $Y$.
Let us first recall how the bulk theory with $\zeta\ll 0$
flows to the sigma model with target $Y$ \cite{HoTo}.
Let $e_1$ and $e_2$ be the gauge coupling constants for the $U(1)$
and the $SU(2)$ factors of the gauge group, respectively.
The $U(1)$ Higgsing scale is $e_1\sqrt{-\zeta}$ and
the $SU(2)$ gauge interaction is strong at energies below  $e_2$.
If $x$ and $p$ have the canonical kinetic terms,
the superpotential coefficients $A^{ij}_k$ have the dimension of a mass.
Let $E_A$ be their typical energy scale.
We shall work in the parametric regime with the hierachy of
energy scales
\beq
E_A\,\,\ll \,\,e_2\,\,\ll\,\, E_A\sqrt{-\zeta}\,\,\ll\,\,  e_1\sqrt{-\zeta}.
\label{Pfhierarchy}
\eeq
At energies below $e_1\sqrt{-\zeta}$, the gauge group is broken to
$SU(2)$ by the non-zero values of $p$.
In this situation, it is natural to change the ``R-charge frame'' so that
the $p^k$'s have R-charge 0 and the $x_i$'s have R-charge 1.
We have $SU(2)$ gauge theories fibered over $\CP^6$,
with seven fundamentals $x_i$ with mass matrix $A^{ij}(p)$.
Away from the Pfaffian locus $Y\subset \CP^6$,
this matrix has rank six and leaves only a single massless fundamental.
As the $SU(2)$ SQCD with a single fundamental has no zero energy state
\cite{HoTo}, the theory at the energies below $e_2$ is localized near $Y$
where three fundamentals remain light.

Let us choose an arbitrary point $p_*$ in $Y$ and look at the behaviour of
the theory around this point.
As in Section~\ref{subsub:D0}, we may perform $\C$-linear reparametrizations of
the $p^k$'s and $x_i$'s so that
$p_*=(0,\ldots,0,1)$ and that $A(p_*)=A_7$ has
$A^{45}(p_*)=-A^{54}(p_*)=A^{67}(p_*)=-A^{76}(p_*)={1\over 2}$ as its only
non-zero entries.
If we write $A(p)$ according to the decomposition $7=3+4$ as
\beq
A(p)=\left(\begin{array}{cc}
  a(p)&b(p)\\
  -b(p)^T&c(p)
\end{array}\right),
\eeq
then $a(p_*)=0$, $b(p_*)=0$ and $c(p_*)=A_*$
given in (\ref{Astar}) with $\mm=2$.
$c(p)$ remains invertible
in a neighborhood $\nhU$ of $p_*$ in $\CP^6$.
There is a $p$-dependent $4\times 4$ regular matrix $\gamma(p)$
such that $\gamma(p)^Tc(p)\gamma(p)=A_*$
and $\gamma(p_*)={\bf 1}_4$.
Then, for
\beq
M(p)=\left(\begin{array}{cc}
  {\bf 1}_3&0\\
  c(p)^{-1}b(p)^T&\gamma(p)
  \end{array}\right),
\eeq
we have
\beq
M(p)^TA(p)M(p)\,=\,\left(\begin{array}{cc}
  \widetilde{a}(p)&0\\
  0&A_*
\end{array}\right),
\label{AinU}
\eeq
with $\widetilde{a}(p)=a(p)+b(p)c(p)^{-1}b(p)^T$.
The Pfaffian $Y\cap\,\nhU$ is the zero locus of $\widetilde{a}(p)$.
Put
\beq
p^{\tti\ttj}:=2\widetilde{a}^{\tti\ttj}(p), \quad1\leq\tti,\ttj\leq 3.
\eeq
Then, $p^{12}, p^{23},p^{31}$ and three of $p^1,\ldots, p^6$,
form coordinates of $\CP^6$ around $p_*$.
For example, with a $\C$-linear reparametrization of $p^1,\ldots, p^6$,
we may assume
$A^{12}_k={1\over 2}\delta_k^4$,
$A^{23}_k={1\over 2}\delta_k^5$,
$A^{31}_k={1\over 2}\delta_k^6$, so that
\beq
p^{12}=p^4+\cdots,~~\,
p^{23}=p^5+\cdots,~~\,
p^{31}=p^6+\cdots,
\label{p12etc}
\eeq
where the ellipsis are quadratic or higher in $p^1,\ldots,p^6$.
Then, $p^{12}$, $p^{23}$, $p^{31}$, $p^1$, $p^2$, $p^3$ form coordinates on
$\nhU$.
The Pfaffian $Y\cap\, \nhU$ is at
$p^{12}=p^{23}=p^{31}=0$ and is parametrized by $p^1,p^2,p^3$.

Let us make the change of variables
\beq
x^a_i\,\longrightarrow\,\sum_jM(p)_i^{\,\,j}\ttx^a_j.
\label{replacex}
\eeq
Then, the superpotential is expressed as
\beq
W={1\over 2}\sum_{\tti,\ttj}p^{\tti\ttj}[\ttx_{\tti}\ttx_{\ttj}]
+\sum_{\mu,\nu}A_*^{\mu\nu}[\ttx_{\mu}\ttx_{\nu}].
\label{Wpstar}
\eeq
The second term gives mass to the four doublets $\ttx_4,\ldots, x_7$.
As the mass matrix $A^{\mu\nu}(p_*)$ is of order $E_A\sqrt{-\zeta}$
before the change of variables, they decouple at energies below this scale.
At much lower energies, $E\ll e_2$ to be precise, the three doublets
$\ttx_1,\ttx_2,\ttx_3$ are glued together by the $SU(2)$ gauge interaction
to form the baryons
$b_{\tti\ttj}=[\ttx_\tti \ttx_\ttj]$, 
which are coupled to the $p$-fields via
the superpotential
\beq
W_{\rm int}={1\over 2}\sum_{\tti,\ttj}p^{\tti\ttj}b_{\tti\ttj}.
\label{Wint}
\eeq
This superpotential gives mass of order $E_A$ to
the fields $p^{12},p^{23},p^{31},b_{12},b_{23},b_{31}$.
Thus, at energies below $E_A,$ we are left with
the theory of the three fields
$p^1$, $p^2$, $p^3$ only, which is the sigma model with target
$Y\cap\,\nhU$.

Let us now consider examples of D-branes.
As a matrix factorization of (\ref{Wpstar}), let us consider 
\beq
Q=\sum_{\tti<\ttj}\left(p^{\tti\ttj}\xi_{\tti\ttj}
+[\ttx_{\tti}\ttx_{\ttj}]\barxi^{\tti\ttj}\right)
+\sum_{a,\mu}\ttx^a_{\mu}\xi^{\mu}_a,
\eeq
realized on the irreducible representation $\bV$
of the Clifford algebra (\ref{xiSU2}) adjoined with
\beqa
&\{\xi_{\tti\ttj},\barxi^{\ttk\ttl}\}=
\delta_{\tti}^{\ttk}\delta_{\ttj}^{\ttl}-\delta_{\tti}^{\ttl}\delta_{\ttj}^{\ttk},
\quad
\{\xi_{\tti\ttj},\xi_{\ttk\ttl}\}=\{\barxi^{\tti\ttj},\barxi^{\ttk\ttl}\}=0,
\label{xiij}\\
&
~~\xi_{\tti\ttj}=-\xi_{\ttj\tti},~~
\barxi^{\tti\ttj}=-\barxi^{\ttj\tti},\quad
1\leq \tti,\ttj,\ttk,\ttl\leq 3.\nn
\eeqa
The R-charge assignment is
\beq
[\xi_{\tti\ttj}]=1,~\,
[\barxi^{\tti\ttj}]=-1,~\,
[\xi_a^{\mu}]=0,
\label{Retaxi}
\eeq
and zero on the states annihilated by the $\xi_{\tti\ttj}$'s.
The result of Section~\ref{subsec:SU2}, especially
(\ref{mainprop}), tells us the fate of this brane
at the energy scale $E_A\ll E\ll e_2$
where the theory is described by
$p^{1,2,3}$, $p^{\tti\ttj}$, $b_{\tti\ttj}$:
the Neumann boundary condition on all the variables
with the matrix factorization
\beq
\sum_{\tti<\ttj}\left(p^{\tti\ttj}\xi_{\tti\ttj}
+b_{\tti\ttj}\barxi^{\tti\ttj}\right)
\eeq
 of (\ref{Wint})
represented on the irreducible representation of (\ref{xiij}).
At lower energy scales $E\ll E_A$
where $p^{\tti\ttj}$ and $b_{\tti\ttj}$ are integrated out,
this brane reduces
to the Neumann boundary condition on the fields $p^{1,2,3}$.
That is, the D6-brane on the Pfaffian $Y\cap\,\nhU$.
Let us denote this brane $(\bV,Q)$ by $\brane^{\rm D6}_{\nhU}$.

As another matrix factorization of (\ref{Wpstar}), consider
\beq
Q
=\sum_{k=1}^3p^k\eta_k+\sum_{\tti<\ttj}\left(p^{\tti\ttj}\xi_{\tti\ttj}
+[\ttx_{\tti}\ttx_{\ttj}]\barxi^{\tti\ttj}\right)
+\sum_{a,\mu}\ttx^a_{\mu}\xi^{\mu}_a,
\label{QD0U}
\eeq
realized on the irreducible representation $\bV$
of the Clifford algebra
\beq
\{\eta_k,\bareta^l\}=\delta_k^l,\quad k,l=1,2,3,
\eeq
adjoined with (\ref{xiij}) and (\ref{xiSU2}).
The R-charge assignment is $[\eta_k]=1$, $[\bareta^k]=-1$,
(\ref{Retaxi}), and zero on the states annihilated by
$\eta_k$'s and $\xi_{\tti\ttj}$'s.
This is the tensor product of $\brane^{\rm D6}_{\nhU}$
and the complex
\beq
\C\,\stackrel{p}{\longrightarrow}\,\C^3\,
\stackrel{p}{\longrightarrow}\,\C^3\,
\stackrel{p}{\longrightarrow}\,\C
\eeq
where the arrow $p$ is the first term of (\ref{QD0U})
and the rightmost $\C$ has R-charge $0$.
In the low energy sigma model with target $Y\cap\,\nhU$,
it reduces to the complex
\beq
{\mathcal O}\,\stackrel{p}{\longrightarrow}\,{\mathcal O}^3\,
\stackrel{p}{\longrightarrow}\,{\mathcal O}^3\,
\stackrel{p}{\longrightarrow}\,{\mathcal O}
\eeq
which is isomorphic to the skyscraper sheaf at $p^1=p^2=p^3=0$,
that is, the D0-brane at $p_*$.
Let us denote this brane $(\bV,Q)$ by
$\brane^{\rm D0}_{p_*,\,\nhU}$.

One important subtlety is that the overall R-charge of the branes in the low
energy sigma model is not canonically determined,
and we need to fix it once and for all.
Let us do it now by declaring that $\brane^{\rm D6}_{\nhU}$
flows to the structure sheaf of $Y\cap\,\nhU$ with a specific
R-charge. Then, $\brane^{\rm D0}_{p_*,\,\nhU}$ flows to
the skyscraper sheaf at $p_*$ with the same R-charge.
It turns out that this R-charge must be an odd integer for
compatibility with the relation (\ref{HSvsCCPf})
between the hemisphere partition function and the central charge.
Let us choose it to be $+1$, so that the flow is
\beq
\brane^{\rm D6}_{\nhU}\stackrel{\rm flow}{\longrightarrow}
      {\mathcal O}_{Y\cap\,\nhU}[-1],\qquad
\brane^{\rm D0}_{p_*,\,\nhU}\stackrel{\rm flow}{\longrightarrow}
      {\mathcal O}_{p_*}[-1].
      \label{choicePf}
\eeq

In what follows, we study the behaviour of the branes $\brane_{p_*}^{(i_0)}$
and $\Lb^h$ 
in the part of the field space
corresponding to the neighborhood $\nhU\subset\CP^6$ of $p_*$,
and compare them with $\brane^{\rm D0}_{p_*,\,\nhU}$
and $\brane^{\rm D6}_{\nhU}$.
For this purpose
we need to switch the R-charge frame from our choice (\ref{Grframe})
to the Pfaffian frame (\ref{Pfframe}) under which the $p$-field is neutral.
We recall that this involves the change of the brane R-charge as in (\ref{RPf}).

\subsubsection{D0-Branes}
\label{subsec:D0proof}

Recall that the boundary potential of $\brane_{p_*}^{(i_0)}$,
given in (\ref{bpPf}),
is strictly positive away from $p_*$ and hence the low energy
dynamics is localized near $p_*$. 
Thus, the analysis in our neighborhood $\nhU$ of $p_*$ will
tell us everything about the low energy behaviour of $\brane_{p_*}^{(i_0)}$.
In the Pfaffian frame (\ref{RPf}), the Clifford generators
of $\brane_{p_*}^{(i_0)}$ have R-charges
\beqa
&[\eta_1]=\cdots=[\eta_6]=-1+2=1,\quad [\bareta^1]=\cdots=[\bareta^6]=1-2=-1,\\
&[\xi^4_a]=[\xi^6_a]=-1+1=0,\quad [\xi^5_a]=[\xi^7_a]=1-1=0,
\eeqa
while the Clifford vacuum $|0\rangle$ keeps the same R-charge $i_0$
as in (\ref{D0charge3}).
Let us express the matrix factorization $Q_{p_*}$ in (\ref{Qpstar})
in terms of $p^{1,2,3}$, $p^{\tti\ttj}$'s and
the new $\ttx^a_i$'s after the change (\ref{replacex}).
Inverting (\ref{p12etc}), we have
\beq
p^4=p^{12}+\cdots,~~\,
p^5=p^{23}+\cdots,~~\,
p^6=p^{31}+\cdots,
\eeq
where the ellipsis are quadratic or higher in
$p^{1,2,3}$ and the $p^{\tti\ttj}$'s.
In view of this and $A^{12}_k={1\over 2}\delta_k^4$ etc, we find 
\beqa
&A_k(x)=\cdots,~~\,k=1,2,3,\\
&A_4(x)=[\ttx_1\ttx_2]+\cdots,~~\,
A_5(x)=[\ttx_2\ttx_3]+\cdots,~~\,
A_6(x)=[\ttx_3\ttx_1]+\cdots,
\eeqa
where the ellipsis consists of
terms which are at least linear in $\ttx_{\mu}$ or quadratic
in $p$ and/or $[\ttx_\tti\ttx_\ttj]$.
Let us rename
$\eta_4=\xi_{12}$, $\eta_5=\xi_{23}$, $\eta_6=\xi_{31}$,
$\bareta^4=\barxi^{12}$, $\bareta^5=\barxi^{23}$, $\bareta^6=\barxi^{31}$.
Then,
we find that $\bV_{p_*}^{(i_0)}$ and
$\bV^{\rm D0}_{p_*,\,\nhU}[-i_0]$ are isomorphic and
\beq
Q_{p_*}|_{\nhU}=Q^{\rm D0}_{p_*,\,\nhU}+\cdots,
\eeq
where the ellipsis $\cdots$ are of higher order compared to
$Q^{\rm D0}_{p_*,\,\nhU}$ and are expected to be negligible at
low energies. Thus, we find that $\brane_{p_*}^{(i_0)}|_{\nhU}$ and
$\brane^{\rm D0}_{p_*,\,\nhU}[-i_0]$ have the same infra-red limit.
This means, via (\ref{choicePf}), that $\brane_{p_*}^{(i_0)}$ flows
to ${\mathcal O}_{p_*}[-1-i_0]$.
Recall that the hemisphere partition function of $\brane_{p_*}^{(i_0)}$
for odd $i_0$ matches with the central charge of ${\mathcal O}_{p_*}$
under the relation (\ref{HSvsCCPf}).
This supports the need of an odd shift as in (\ref{choicePf}).
The brane that flows exactly to ${\mathcal O}_{p_*}$ is
$\brane_{p_*}^{(i_0)}$ with $i_0=-1$ which
we shall denote simply by $\brane_{p_*}$ in what follows.

\subsubsection{D6-Branes}
\label{subsub:D6LE}

Let us first discuss the fate of the brane $\Lb^2(-1)$
whose hemisphere partition function is the same as the central charge of
the D6-brane ${\mathcal O}_Y$ with an odd shift.
It has $S^2\fund(-1)\otimes\textw^{\!7}V^*
=S^2\fund^*(1)\otimes\textw^{\!7}V^*$ with R-charge 0 as its leftmost
component, and hence is of the from
\beq
\begin{array}{ccccccc}
  S^2\fund^*(1)\otimes\textw^{\!7}V^*{}_0&\longleftrightarrow&
  S^2\fund^*(4)\otimes V^*{}_5&\longleftrightarrow&
  S^2\fund^*(5)\otimes V{}_6&\longleftrightarrow&
  S^2\fund^*(8)\otimes\textw^{\!7}V{}_{11}\\
  \updownarrow&&\updownarrow&&\updownarrow&&\updownarrow\\
  \fund^*(1)\otimes\textw^{\!6}V^*{}_1&\longleftrightarrow&
  \fund^*(3)\otimes\textw^{\!2}V^*{}_4&\longleftrightarrow&
  \fund^*(5)\otimes \textw^{\!2}V{}_7&\longleftrightarrow&
  \fund^*(7)\otimes \textw^{\!6}V{}_{10}\\
  \updownarrow&&\updownarrow&&\updownarrow&&\updownarrow\\  
  \C(1)\otimes\textw^{\!5}V^*{}_2&\longleftrightarrow&
  \C(2)\otimes\textw^{\!3}V^*{}_3&\longleftrightarrow&
  \C(5)\otimes\textw^{\!3}V{}_8&\longleftrightarrow&
  \C(6)\otimes\textw^{\!5}V{}_9
\end{array}
\label{Lb2m}
\eeq
In the picture  of the triangular grid (Figure~\ref{fig:D6mf}),
the first and the thrid vertical lines of (\ref{Lb2m})
have a slope of 120${}^{\circ}$ while
the second the fourth have 60${}^{\circ}$.
The subscript of each component shows the R-charge in the original frame
(\ref{Grframe}).
Under (\ref{RPf}),
the R-charge of the component $S^h\fund^*(i)\otimes \cdots$ changes
as $r\to r+h-2i$. Thus, the R-charges of (\ref{Lb2m}) in the Pfaffian
frame (\ref{Pfframe}) are
\beq
\begin{array}{ccccccc}
  0+2-2=0&&5+2-8=-1&&6+2-10=-2&&11+2-16=-3\\
  1+1-2=0&&4+1-6=-1&&7+1-10=-2&&10+1-14=-3\\
  2+0-2=0&&3+0-4=-1&&8+0-10=-2&&9+0-12=-3.
\end{array}
\eeq
As the gauge group is broken to $SU(2)$, we only need to
keep the information of the components as representations of $SU(2)$.
Then, we are considering the matrix factorization
\beq
\begin{array}{ccccccc}
  S^2\sfund^*\otimes\textw^{\!7}V^*&\longleftrightarrow&
  S^2\sfund^*\otimes V^*&\longleftrightarrow&
  S^2\sfund^*\otimes V&\longleftrightarrow&
  S^2\sfund^*\otimes\textw^{\!7}V\\
  \updownarrow&&\updownarrow&&\updownarrow&&\updownarrow\\
  \sfund^*\otimes\textw^{\!6}V^*&\longleftrightarrow&
  \sfund^*\otimes\textw^{\!2}V^*&\longleftrightarrow&
  \sfund^*\otimes \textw^{\!2}V&\longleftrightarrow&
  \sfund^*\otimes \textw^{\!6}V\\
  \updownarrow&&\updownarrow&&\updownarrow&&\updownarrow\\  
  \textw^{\!5}V^*&\longleftrightarrow&
  \textw^{\!3}V^*&\longleftrightarrow&
  \textw^{\!3}V&\longleftrightarrow&
  \textw^{\!5}V\\[0.2cm]
  \hline
  \mbox{\footnotesize{0}}&&\mbox{\footnotesize{$-1$}}&&
  \mbox{\footnotesize{$-2$}}&&\mbox{\footnotesize{$-3$}}
\end{array}
\label{PfL0}
\eeq
We make the change of variables (\ref{replacex}) along with
\beqa
&&\eta^i\,\longrightarrow\,(M(p)^{-1})_j^{\,\,\,i}\,\eta^j,\qquad
\bareta_i\,\longrightarrow\,M(p)_i^{\,\,j}\,\bareta_j,\label{etacv}\\
&&\bareta^i\,\longrightarrow\,(M(p)^{-1})_j^{\,\,\,i}\,\bareta^j,\qquad
\eta_i\,\longrightarrow\,M(p)_i^{\,\,j}\,\eta_j.\label{baretacv}
\eeqa
Then the arrows maintain their forms as given in
Sections~\ref{subsec:Gempty} and \ref{subsec:GRprocess},
except that
$A(p)$ is replaced by $A'(p)=M(p)^TA(p)M(p)$.

Note that $A'(p)$ has the $3+4$ block-diagonal form (\ref{AinU}), whose lower
$4\times 4$ block $A_*$ is constant and invertible.
This means that the leftward horizontal arrows
are partially invertible.
Let $d_*$ be the collection of arrows in which we keep only $A_*$ in
(\ref{AinU}), discarding $\widetilde{a}(p)$ and $x$.
In effect, it is the same as setting $(p,x)=(p_*,0)$ in
the matrix factorization before the change of variables,
\beq
d_*=Q(p_*,0).
\label{dstar}
\eeq
It is a map of R-charge +1 that squares to zero, $d_*\circ d_*=0$,
and hence defines a complex.
In what follows, we shall use a shorthand notation
$\bareta_{ijk\cdots}=\bareta_i\bareta_j\bareta_k\cdots$.
The map $d_*$ from R-charge $-1$ 
to $0$ is a term in (\ref{Lw}) and is
proportional to $\widehat{A}_*^{\,h+1}$ at height $h$
where $\widehat{A}_*=\bareta_{45}+\bareta_{67}$.
It vanishes at height 2 as $\widehat{A}_*^{\,3}=0$.
At height 1, it is ${1\over 2}\widehat{A}_*^{\,2}=\bareta_{4567}$
and is of rank $2\times 3$, determining an isomorphism between
subspaces,
\beq
\sfund^*\otimes \langle\,\bareta_{\tti\ttj4567}\,\rangle
\,\stackrel{\simeq}{\longleftarrow}\,
\sfund^*\otimes\langle\,\bareta_{\tti\ttj}\,\rangle\,:\,d_*
\eeq
At height 0, it is $-\widehat{A}_*=-(\bareta_{45}+\bareta_{67})$
and is of rank 16,
\beq
\left\langle\,
  \bareta_{\tti\ttj\mu\nu\rho},\,
  \bareta_{12345}+\bareta_{12367},\,\bareta_{\tti4567}\,\right\rangle
\,\stackrel{\simeq}{\longleftarrow}\,
\langle\,
\bareta_{\tti\ttj\mu},\,\bareta_{123},\,\bareta_{\tti 45}+\bareta_{\tti 67}
\,\rangle
\,:\,d_*
\eeq
Similarly, the map $d_*$ from  R-charge $-3$ to $-2$
is a term in (\ref{Lw}) in the version of
Section~\ref{subsub:otherversion} and is proportional to
$\widehat{A}_*^{\,h+1}$ at height $h$
where $\widehat{A}_*=\eta_{45}+\eta_{67}$.
It is zero  at height 2,
is a rank $2\times 3$ map ${1\over 2}\widehat{A}_*^2=\eta_{4567}$ at height 1,
\beq
\sfund^*\otimes \langle\,\bareta^{\tti\ttj}\,\rangle
\,\stackrel{\simeq}{\longleftarrow}\,
\sfund^*\otimes \langle\,\bareta^{\tti\ttj 4567}\,\rangle
\,:\,d_*
\eeq
and is a rank 16 map $-\widehat{A}_*=-(\eta_{45}+\eta_{67})$ at height 0,
\beq
\langle\,\bareta^{\tti\ttj\mu},\,\bareta^{123},\,
\bareta^{\tti 45}+\bareta^{\tti 67}\,\rangle
\,\stackrel{\simeq}{\longleftarrow}\,
\left\langle\,
\bareta^{\tti\ttj\mu\nu\rho},\,
  \bareta^{12345}+\bareta^{12367},\,\bareta^{\tti4567}\,\right\rangle
\,:\,d_*
\eeq
The maps $d_*$ from R-charge $-2$ to $-1$
are the $A_*$-parts of (\ref{D6mf2}), (\ref{D6mf1}) and (\ref{D6mf0}).
At height 2, it is a rank $3\times 4$ map,
\beqa
&d_*={1\over 3}\Bigl(\,\bareta_4\eta_5-\bareta_5\eta_4
+\bareta_6\eta_7-\bareta_7\eta_6\,\Bigr)\,:\nn\\[0.1cm]
&S^2\sfund^*\otimes \langle\,\bareta_{\mu}\,\rangle
\,\stackrel{\simeq}{\longleftarrow}\,
S^2\sfund^*\otimes \langle\,\bareta^{\mu}\,\rangle
\eeqa
At height 1, it is a rank $2\times 6$ map,
\beqa
&d_*={1\over 6}\left(\,
\left({2\over 3}\bareta_{45}-{1\over 3}\bareta_{67}\right)\eta_{54}
+\left(-{1\over 3}\bareta_{45}+{2\over 3}\bareta_{67}\right)\eta_{76}
+\bareta_{46}\eta_{75}-\bareta_{56}\eta_{74}
-\bareta_{47}\eta_{65}+\bareta_{57}\eta_{64}\,\right)\,:
\nn\\[0.1cm]
&\sfund^*\otimes\langle\,\bareta_{\mu\nu}\,\rangle
\,\stackrel{\simeq}{\longleftarrow}\,
\sfund^*\otimes\langle\,\bareta^{\mu\nu}\,\rangle
\eeqa
At height 0, it is a rank 4 map,
\beqa
&d_*={1\over 12}\Bigl(\,\bareta_{456}\eta_{754}
-\bareta_{457}\eta_{654}+\bareta_{467}\eta_{765}
-\bareta_{567}\eta_{764}\,\Bigr)\,:\nn\\[0.1cm]
&\langle\,\bareta_{\mu\nu\rho}\,\rangle
\,\stackrel{\simeq}{\longleftarrow}\,
\langle\,\bareta^{\mu\nu\rho}\,\rangle
\eeqa

Tachyon condensation of these invertible parts of $d_*$ annihilates
the isomorphic pairs and induces zigzag maps.
This leaves us with the matrix factorization in the reduced form
\beq
\begin{tikzcd}
  \! S^2\sfund^*\!\otimes\!\textw^{\!7}V^*\!
  \arrow[r,yshift=0.5ex, "{[}x_\tti x_\ttj{]}\eta^{\tti\ttj 4567}"]
  &[4em]
    \arrow[l,yshift=-0.5ex, "-{1\over 4}p^{\tti\ttj}\bareta_{\tti\ttj 4567}"]
  \! S^2\sfund^*\!\otimes\! V^*_{\rm red}\!
    \arrow[r,yshift=0.5ex, "-3{[}x_\tti x_\ttj{]}\bareta^\tti\eta^\ttj"]
    &[4.5em]
    \arrow[l,yshift=-0.5ex, "{1\over 3}p^{\tti\ttj}\bareta_\tti\eta_\ttj"]
  \! S^2\sfund^*\!\otimes\! V_{\rm red} \!
    \arrow[r,yshift=0.5ex, "{[}x_\tti x_\ttj{]}\bareta^{\tti\ttj 4567}"]
    &[4em]
   \arrow[l,yshift=-0.5ex, "-{1\over 4}p^{\tti\ttj}\eta_{\tti\ttj 4567}"]
  \! S^2\sfund^*\!\otimes\!\textw^{\!7}V \!
   \\[-0.9cm]
   \Big\updownarrow &\Big\updownarrow &\Big\updownarrow &\Big\updownarrow
   \\[-0.9cm]
  \! \sfund^*\!\otimes\!\textw^{\!6}_{\rm red}V^* \!
  \arrow[r,yshift=0.5ex, "{1\over 2}{[}x_\tti x_\ttj{]}
    \eta^{\tti\ttj}(\eta^{45}+\eta^{67})"]
  &[4em]
  \arrow[l,yshift=-0.5ex, "{1\over 2}p^{\tti\ttj}
    \bareta_{\tti\ttj}(\bareta_{45}+\bareta_{67})"]
  \! \sfund^*\!\otimes\!\textw^{\!2}_{\rm red}V^* \!
  \arrow[r,yshift=0.5ex, "3{[}x_\tti x_\ttj{]}A_*^{-1}\!\!\!{}_{\mu\nu}
    \bareta^{\tti\mu}\eta^{\ttj\nu}"]
    &[4.5em]
  \arrow[l,yshift=-0.5ex, "-{1\over 3}p^{\tti\ttj}A_*^{\mu\nu}
    \bareta_{\tti\mu}\eta_{\ttj\nu}"]
 \!  \sfund^*\!\otimes\!\textw^{\!2}_{\rm red}V \!
  \arrow[r,yshift=0.5ex, "{1\over 2}{[}x_\tti x_\ttj{]}
    \bareta^{\tti\ttj}(\bareta^{45}+\bareta^{67})"]
    &[4em]
  \arrow[l,yshift=-0.5ex, "{1\over 2}p^{\tti\ttj}
    \eta_{\tti\ttj}(\eta_{45}+\eta_{67})"]
 \!  \sfund^*\!\otimes\!\textw^{\!6}_{\rm red}V \!
   \\[-0.9cm]
   \Big\updownarrow &\Big\updownarrow &\Big\updownarrow &\Big\updownarrow
   \\[-0.9cm]
 \!  \textw^{\!5}_{\rm red}V^* \!
  \arrow[r,yshift=0.5ex, "{1\over 2}{[}x_\tti x_\ttj{]}\eta^{\tti\ttj}"]
  &[4em]
  \arrow[l,yshift=-0.5ex, "-{1\over 2}p^{\tti\ttj}\bareta_{\tti\ttj}"]
 \! \textw^{\!3}_{\rm red}V^* \!
  \arrow[r,yshift=0.5ex, "{3\over 4}{[}x_\tti x_\ttj{]}
    A_*^{-1}\!\!\!{}_{\mu\nu}A_*^{-1}\!\!\!{}_{\rho\lambda}
    \bareta^{\tti\mu\rho}\eta^{\ttj\nu\lambda}"]
    &[4.5em]
  \arrow[l,yshift=-0.5ex, "-{1\over 3}p^{\tti\ttj}A_*^{\mu\nu}A_*^{\rho\lambda}
    \bareta_{\tti\mu\rho}\eta_{\ttj\nu\lambda}"]
 \!  \textw^{\!3}_{\rm red}V \!
  \arrow[r,yshift=0.5ex, "{1\over 2}{[}x_\tti x_\ttj{]}
    \bareta^{\tti\ttj}"]
    &[4em]
  \arrow[l,yshift=-0.5ex, "-{1\over 2}p^{\tti\ttj}
    \eta_{\tti\ttj}"]
  \! \textw^{\!5}_{\rm red}V \!\\[-0.5cm]
   {}_0&{}_{-1}&{}_{-2}&{}_{-3}
\end{tikzcd}
\label{redform}
\eeq
where 
\beq
\begin{array}{cccc}
  \textw^{\!7}V^*=\langle\,\bareta_{1234567}\,\rangle,
  &V^*_{\rm red}=\langle\,\bareta_\tti\,\rangle,
  &V_{\rm red}=\langle\,\bareta^\tti\,\rangle,
  &\textw^{\!7}V=\langle\,\bareta^{1234567}\,\rangle,\\
\textw_{\rm red}^{\!6}V^*=\langle\, \bareta_{123\mu\nu\rho}\,\rangle,
&\textw_{\rm red}^{\!2}V^*=\langle\,\bareta_{\tti\mu}\,\rangle,
&\textw_{\rm red}^{\!2}V=\langle\,\bareta^{\tti\mu}\,\rangle,
&\textw_{\rm red}^{\!6}V=\langle\,\bareta^{123\mu\nu\rho}\,\rangle,\\
\textw_{\rm red}^{\!5}V^*=\langle\,\bareta_{123\mu\nu}|\ast\rangle,
&\textw_{\rm red}^{\!3}V^*=\langle\,\bareta_{\tti\mu\nu}|\ast\rangle,
&\textw_{\rm red}^{\!3}V=\langle\,\bareta^{\tti\mu\nu}|\ast\rangle,
&\textw_{\rm red}^{\!5}V=\langle\,\bareta^{123\mu\nu}|\ast\rangle,\,
\end{array}
\label{redFF}
\eeq
with
\beqa
&&\langle\,\bareta_{\cdots\mu\nu}|\ast\rangle
:=\langle\,\bareta_{\cdots 45}-\bareta_{\cdots 67},~\bareta_{\cdots 46},~
\bareta_{\cdots 47},~\bareta_{\cdots 56},~\bareta_{\cdots 57}\,\rangle,\nn\\
&&\langle\,\bareta^{\cdots\mu\nu}|\ast\rangle
:=\langle\,\bareta^{\cdots 45}-\bareta^{\cdots 67},~\bareta^{\cdots 46},~
\bareta^{\cdots 47},~\bareta^{\cdots 56},~\bareta^{\cdots 57}\,\rangle,\nn
\eeqa
and the vertical arrows are the ones induced from the
Lascoux matrix factorizations.

\newcommand{\Ttt}{\texttt{T}}

Let us denote by $\Ttt$ and $\cktt$
the dimension 3 and 4 subspaces of $V$ spanned
by the $\bareta^{\tti}$'s and $\bareta^{\mu}$'s, respectively.
The dual spaces $\Ttt^*$ and $\cktt^*$ are spanned by the $\bareta_\tti$'s
and $\bareta_{\mu}$'s.
Let $\textw_0^{\!2}\cktt\subset\textw^{\!2\,}\cktt$ and
$\textw_0^{\!2}\cktt^*\subset\textw^{\!2\,}\cktt^*$
be the subspaces of $A_*$ and $A_*^{-1}$ traceless elements.
Then, the 
reduced flavor factors (\ref{redFF}) can be identified as
\beq
\begin{array}{cccc}
  \textw^{\!3\,}\Ttt^*\otimes \textw^{\!4\,}\cktt^*,
  &\Ttt^*,&\Ttt,
  &\textw^{\!3\,}\Ttt\otimes\textw^{\!4\,} \cktt,\\
  \textw^{\!3\,}\Ttt^*\!\otimes\! \textw^{\!3\,}\cktt^*,&
  \Ttt^*\!\otimes \cktt^*,&
  \Ttt\otimes \cktt,&
  \textw^{\!3\,}\Ttt\otimes \textw^{\!3\,}\cktt,\\
  \textw^{\!3\,}\Ttt^*\!\otimes \textw_0^{\!2}\cktt^*,&
  \Ttt^*\!\otimes \textw_0^{\!2}\cktt^*,&
  \Ttt\otimes \textw_0^{\!2}\cktt,&
  \textw^{\!3\,}\Ttt\otimes \textw_0^{\!2}\cktt.
\end{array}
\eeq
Note that the vertical arrows of (\ref{redform}) act trivially on the
$\textw^{\!p\,}\Ttt$ and $\textw^{\!p\,}\Ttt^*$ factors.
If we strip off these factors,
the four vertical lines are isomorphic to one another,
\beq
\begin{tikzcd}
  S^2\sfund^*\!\otimes\!\textw^{\!4\,}\cktt^*
  \arrow[r,yshift=0.5ex, "2\eta^{4567}"]
  &[2em]
    \arrow[l,yshift=-0.5ex, "{1\over 2}\bareta_{4567}"]
  S^2\sfund^*
    \arrow[r,yshift=0.5ex, "-3"]
    &[4.5em]
    \arrow[l,yshift=-0.5ex, "-{1\over 3}"]
  S^2\sfund^*
    \arrow[r,yshift=0.5ex, "2\bareta^{4567}"]
    &[2em]
   \arrow[l,yshift=-0.5ex, "{1\over 2}\eta_{4567}"]
  S^2\sfund^*\!\otimes\!\textw^{\!4\,}\cktt 
   \\[-0.9cm]
   \Big\updownarrow &\Big\updownarrow &\Big\updownarrow &\Big\updownarrow
   \\[-0.9cm]
  \sfund^*\!\otimes\!\textw^{\!3\,}\cktt^* 
  \arrow[r,yshift=0.5ex, "-(\eta^{45}+\eta^{67})"]
  &[2em]
  \arrow[l,yshift=-0.5ex, "\bareta_{45}+\bareta_{67}"]
  \sfund^*\!\otimes \cktt^* 
  \arrow[r,yshift=0.5ex, "3A_*^{-1}\!\!\!{}_{\mu\nu}
    \bareta^{\mu}\eta^{\nu}"]
    &[4.5em]
  \arrow[l,yshift=-0.5ex, "{1\over 3}A_*^{\mu\nu}
    \bareta_{\mu}\eta_{\nu}"]
  \sfund^*\!\otimes \cktt 
  \arrow[r,yshift=0.5ex, "-(\bareta^{45}+\bareta^{67})"]
    &[2em]
  \arrow[l,yshift=-0.5ex, "\eta_{45}+\eta_{67}"]
  \sfund^*\!\otimes\!\textw^{\!3\,}\cktt 
   \\[-0.9cm]
   \Big\updownarrow &\Big\updownarrow &\Big\updownarrow &\Big\updownarrow
   \\[-0.9cm]
  \textw^{\!2}_0\cktt^* 
  \arrow[r,yshift=0.5ex, "{\rm id}"]
  &[2em]
  \arrow[l,yshift=-0.5ex, "{\rm id}"]
  \textw^{\!2}_0\cktt^* 
  \arrow[r,yshift=0.5ex, "{3\over 4}
    A_*^{-1}\!\!\!{}_{\mu\nu}A_*^{-1}\!\!\!{}_{\rho\lambda}
    \bareta^{\mu\rho}\eta^{\nu\lambda}"]
    &[4.5em]
  \arrow[l,yshift=-0.5ex, "{1\over 3}A_*^{\mu\nu}A_*^{\rho\lambda}
    \bareta_{\mu\rho}\eta_{\nu\lambda}"]
  \textw^{\!2}_0\cktt 
  \arrow[r,yshift=0.5ex, "{\rm id}"]
    &[2em]
  \arrow[l,yshift=-0.5ex, "{\rm id}"]
   \textw^{\!2}_0\cktt
\end{tikzcd}
\label{iso4v}
\eeq
Unlike (\ref{PfL0}) and (\ref{redform}) which show matrix factorizations,
this is a commutative diagram --- each of the six squares commute.
Also, each pair of the horizonal arrows is inverse to each other.
Thus, (\ref{iso4v}) shows three isomorphisms between the four vertical
lines.
We see that (\ref{redform}) is isomorphic to the tensor product of
\beq
\begin{tikzcd}
  \textw^{\!3\,}\Ttt^*
  \arrow[r,yshift=0.5ex, "{1\over 2}{[}x_\tti x_\ttj{]}\eta^{\tti\ttj}"]
  &[5em]
    \arrow[l,yshift=-0.5ex, "-{1\over 2}p^{\tti\ttj}\bareta_{\tti\ttj}"]
    \Ttt^*\arrow[r,yshift=0.5ex, "{[}x_\tti x_\ttj{]}\bareta^\tti\eta^\ttj"]
    &[5em]
    \arrow[l,yshift=-0.5ex, "-p^{\tti\ttj}\bareta_\tti\eta_\ttj"]
    \Ttt\arrow[r,yshift=0.5ex, "{1\over 2}{[}x_\tti x_\ttj{]}\bareta^{\tti\ttj}"]
    &[5em]
   \arrow[l,yshift=-0.5ex, "-{1\over 2}p^{\tti\ttj}\eta_{\tti\ttj}"]
   \textw^{\!3\,}\Ttt\\[-0.5cm]
   {}_0&{}_{-1}&{}_{-2}&{}_{-3}
\end{tikzcd}
\label{mf11}
\eeq
and any of the four vertical lines in (\ref{iso4v}).

The diagram (\ref{mf11}) defines a matrix factorization $\brane_{\Ttt}$
of $W={1\over 2}p^{\tti\ttj}[x_{\tti}x_{\ttj}]$.
If we define
$\barxi^{\tti\ttj}$ and $\xi_{\tti\ttj}$ by the right and the left arrows of
\beq
\begin{tikzcd}
  0
  \arrow[r,yshift=0.5ex, "0"]&\arrow[l,yshift=-0.5ex, "0"]
  \textw^{\!3\,}\Ttt^*
  \arrow[r,yshift=0.5ex, "\eta^{\tti\ttj}"]
  &[1em]
    \arrow[l,yshift=-0.5ex, "-\bareta_{\tti\ttj}"]
    \Ttt^*\arrow[r,yshift=0.5ex, "\bareta^\tti\eta^\ttj-\bareta^\ttj\eta^\tti"]
    &[5em]
    \arrow[l,yshift=-0.5ex, "-(\bareta_\tti\eta_\ttj-\bareta_\ttj\eta_\tti)"]
    \Ttt\arrow[r,yshift=0.5ex, "\bareta^{\tti\ttj}"]
    &[1em]
   \arrow[l,yshift=-0.5ex, "-\eta_{\tti\ttj}"]
   \textw^{\!3\,}\Ttt
   \arrow[r,yshift=0.5ex, "0"]&\arrow[l,yshift=-0.5ex, "0"]
   0
\end{tikzcd}
\label{defxi}
\eeq
they satisfy the commutation relation (\ref{xiij}), and $\brane_{\Ttt}$
can be regarded as the matrix factorization
$Q={1\over 2}p^{\tti\ttj}\xi_{\tti\ttj}+{1\over 2}[x_\tti x_\ttj]\barxi^{\tti\ttj}$
realized on the irreducible representation of (\ref{xiij}).

On the other hand, the vertical lines of (\ref{iso4v})
are matrix factorizations of $W=A_*^{\mu\nu}[x_{\mu}x_{\nu}]$.
In particular, the second and the third lines
are nothing but (\ref{massmfright})
and (\ref{massmfleft}) with $\mm=2$ respectively. That is,
the four vertical lines of (\ref{iso4v}) are isomorphic to $\brane^{SU(2)}_{3,7}$
which is originally described as the matrix factorization
$Q=x^a_{\mu}\xi^{\mu}_a$ realized on the irreducible representation of
(\ref{xiSU2}).

Therefore, the tensor product $\brane_{\Ttt}\otimes\brane^{SU(2)}_{3,7}$
can be regarded as
the matrix factorization
$Q={1\over 2}p^{\tti\ttj}\xi_{\tti\ttj}
+{1\over 2}[x_\tti x_\ttj]\barxi^{\tti\ttj}+x^a_{\mu}\xi_a^{\mu}$
of (\ref{Wpstar}) realized on the representation
(\ref{xiij}) and (\ref{xiSU2}). This is nothing but
$\brane^{\rm D6}_{\nhU}$ that reduces to
${\mathcal O}_{Y\cap\,\nhU}[-1]$ at low energies, (\ref{choicePf}).

To summarize, we find
\beq
\Lb^2(-1)|_{\nhU,{\rm reduced}}\,\,\cong\,\,\brane_{\Ttt}\otimes\brane^{SU(2)}_{3,7}
\,\,=\,\,\brane^{\rm D6}_{\nhU}.
\eeq
Since this holds in a neighborhood $\nhU$ of any point $p_*$, 
this shows that $\Lb^2(-1)$ flows to a single D6-brane on the Pfaffian
Calabi-Yau $Y$ with a shift by $-1$.
This local analysis does not pay attention to the global structure on $Y$
and hence does not tell which line bundle is supported by the D6-brane.
Here we use the information obtained from the hemisphere partition function,
(\ref{chL2}) or $[\pi_Y(\Lb^2(-1)]=-[{\mathcal O}_Y]$.
This means that the line bundle has the same Chern character
as the trivial bundle ${\mathcal O}_Y$.
Since $Y$ is simply connected, it must be ${\mathcal O}_Y$ itself.
This concludes that $\Lb^2(-1)$ flows to the D6-brane ${\mathcal O}_Y$
shifted by $-1$,
\beq
\pi_Y(\Lb^2(-1))\,\,\cong\,\,{\mathcal O}_Y[-1].
\label{L2fate}
\eeq

This result can be used to find the gauge theory branes that
flow to the branes in $Y$ that come from the ambient space $\CP^6$
via restriction.
For a complex $\wC$ of representations of $U(1)$ over $\C[p]$,
in view of the relation (\ref{Wtensor}), it yields the flow
\beq
\pi_Y:\Lb^2(-1)\otimes\wC[1]\longmapsto  {\mathcal O}_Y(\wC^{\rm Pf}).
\eeq
For example, for the complex $\wC_h$ given by (\ref{wCh}),
the brane $\Lb^2(-1)\otimes \wC_h[1]$ flows to the D4-brane
${\mathcal O}_{D_h}$. Likewise, $\Lb^2(-1)\otimes \bigotimes_{i=1}^s\wC_{h_i}[1]$
flows to the D($6-2s$)-brane ${\mathcal O}_{D_{h_1}\!\cap\cdots\cap D_{h_s}}$.

\newcommand{\ck}{{\mathcal R}}

The fate of the branes $\Lb^1$ and $\Lb^0$ in the Pfaffian phase
can be studied in the same way.
The analysis has been carried out to a certain extent and some details
are described in Appendix~\ref{app:redPf}.
For $\Lb^1$, the possible R-charges in the Pfaffian phase
(\ref{RPf}) are $-1,-2,-3,-4$ for the various components.
On the part of the field space where
$p$ takes values in $\nhU$, we can consider
$d_*$ as in (\ref{dstar}) which defines a four term complex at each height.
Brane-antibrane annihilation by the tachyon condensation of the invertible
parts of $d_*$ leaves us with a matrix factorization in the reduced from.
The underlying vector space is the cohomology of the complex with
differential $d_*$ and the arrows are those induced from the original ones 
and the new ones from the zigzag maps.
The dimension of the reduced flavor factors
at height 2 is $4,3\times 4,3\times 4,4$
for the degrees $-1,-2,-3,-4$
and those at heights 1 and 0 are 4 times and 5 times as many, respectively.
See (\ref{redFF1}).
At this moment, we do not yet know the arrows
since the zigzag maps have not been computed. However,
we may expect from the dimension that
the reduced matrix factorization
is that of $\brane_{\Ttt}[1]\otimes\brane^{SU(2)}_{3,7}$
tensored with a vector space of dimension four.
Furthermore, from the structure of parts of the $d_*$-cohomology,
this four dimensional vector space may be identified with a copy of
$\cktt$ or of $\cktt^*$ which is isomorphic to $\cktt$ via $A_*$.
This leads us to conjecture that the reduced matrix
factorization is isomorphic to
\beq
\Lb^1|^{}_{\nhU,\,{\rm reduced}}\,\,\cong\,\,
\brane_{\Ttt}[1]\otimes \brane^{SU(2)}_{3,7}\otimes \cktt
\,\,=\,\,\brane^{\rm D6}_{\nhU}[1]\otimes \cktt.
\label{fateL1loc}
\eeq
Since this is in a neighborhood of any point $p_*$,
in view of (\ref{choicePf}),
the claim implies that $\Lb^1$ flows to D6-branes
supporting a rank four vector bundle ${\mathcal E}_4$,
which is consistent with (\ref{chL1}).
Note that $\cktt$ can be identified with the quotient space
$V/{\rm Ker} A(p_*)$.
Let $\ck$ be the vector bundle on $Y$ whose fiber over
$p\in Y$ is
\beq
\ck|_p:=V/{\rm Ker} A(p).
\label{ckdef}
\eeq
It is natural to expect that the vector bundle
${\mathcal E}_4$ is isomorphic to $\ck$, possibly up to 
tensor product with a line bundle since we used $A_*:\cktt\cong\cktt^*$
in finding (\ref{fateL1loc}) but the scale of $A_*=A(p_*)|_{\cktt}$
was not canonical.
That is, ${\mathcal E}_4\cong\ck(i)$ for some $i$.
In fact we can fix $i$ by comparing the first Chern classes.
The information (\ref{chL1}) of
${\rm ch}(\pi_Y(\Lb^1))=-{\rm ch}({\mathcal E}_4)$ 
tells us that $c_1({\mathcal E}_4)=-2H$. On the other hand,
$c_1(\ck)$ can be determined as follows.
The matrix $A(p)$ determines a non-degenerate skew-symmetric pairing
$\ck\times \ck\to {\mathcal O}_Y(1)$, which means that
$\ck(-1)$ is the dual of $\ck$,
\beq
\ck(-1)\,\cong\,\ck^*.
\label{ckckdual}
\eeq
Since a vector bundle and its dual have opposite odd Chern classes, we find
$0=c_1(\ck)+c_1(\ck(-1))=2c_1(\ck)-4H$.
As ${\rm H}^2(Y,\Z)=\Z H$ has no torsion, we find
\beq
c_1(\ck)=2 H.
\label{c1R}
\eeq
Matching $c_1(\ck(i))=(2+4i)H$ with
$c_1({\mathcal E}_4)=-2H$, we find ${\mathcal E}_4=\ck(-1)$.
Thus, our expectation is
\beq
\pi_Y(\Lb^1)\,\,\cong\,\,\ck(-1)\,\cong\,\ck^*.
\label{L1fate}
\eeq
One can also check that the consequence from the other relation,
$c_3(\ck)+c_3(\ck(-1))=0$, is also satisfied by (\ref{chL1}),
giving further support for (\ref{L1fate}).

For $\Lb^0$, the possible R-charges in the Pfaffian phase
(\ref{RPf}) are $0,-1,-2,-3$ for the various components.
The dimension of the reduced flavor factors
at height 2 is $5,3\times 5,3\times 5,5$
and those at heights 1 and 0 are 4 times and 5 times as many, respectively.
See (\ref{redFF0}).
We expect from the dimension that
the reduced matrix factorization
is that of $\brane_{\Ttt}\otimes\brane^{SU(2)}_{3,7}$
tensored with a vector space of dimension five
which may be identified with a copy of
$\textw^{\!2}_0\cktt\cong \textw^{\!2}_0\cktt^*$, where $A_*$ is used
in this isomorphism again.
This leads us to conjecture that the reduced matrix
factorization is isomorphic to
\beq
\Lb^0|^{}_{\nhU,\,{\rm reduced}}\,\,\cong\,\,
\brane_{\Ttt}\otimes \brane^{SU(2)}_{3,7}\otimes \textw^{\!2}_0\cktt
\,\,=\,\,\brane^{\rm D6}_{\nhU}\otimes \textw^{\!2}_0\cktt.
\label{fateL0loc}
\eeq
As it is in a neighborhood of any point $p_*$, in view of (\ref{choicePf}),
the claim implies that $\Lb^0$ flows at low energies to
a rank five vector bundle ${\mathcal E}_5$ shifted by $-1$,
which is consistent with (\ref{chL0}).
Note that $A(p)$ defines a surjective
map $\textw^{\!2}\ck\to {\mathcal O}_Y(1)$
and we denote its kernel by $\textw^{\!2}_0\ck$.
The fiber of the latter at $p_*\in \nhU$ is isomorphic to
$\textw^{\!2}_0\cktt$.
We expect that ${\mathcal E}_5$ is isomorphic to $\textw^{\!2}_0\ck$,
possibly up to tensor product with a line bundle
as the scale of the isomorphism
$A_*:\textw^{\!2}_0\cktt\cong \textw^{\!2}_0\cktt^*$ is not canonical.
That is, ${\mathcal E}_5\cong \textw^{\!2}_0\ck(i)$ for some $i$.
This $i$ can be fixed by computation of the first Chern class.
The information (\ref{chL0}) of 
${\rm ch}(\pi_Y(\Lb^0))={\rm ch}({\mathcal E}_5)$
tells us that $c_1({\mathcal E}_5)=0$.
On the other hand, 
\beq
c_1(\textw^{\!2}_0\ck)\,=\,
\underbrace{c_1(\textw^{\!2}\ck)}_{3c_1(\ck)}-c_1({\mathcal O}_Y(1))
=3\cdot 2H-H=5H.
\eeq
Matching $c_1(\textw^{\!2}\ck(i))=(5+5i)H$ with
$c_1({\mathcal E}_5)=0$, we find ${\mathcal E}_5=\textw^{\!2}_0\ck(-1)$.
Thus, our expectation is
\beq
\pi_Y(\Lb^0)\,\,\cong\,\,\textw^{\!2}_0\ck(-1)[-1]\,\,\cong\,\,
\textw^{\!2}_0\ck^*(1)[-1].
\label{L0fate}
\eeq

\subsection{Grade Restriction Process}
\label{subsec:GRprocessPf}

We would like to bring an arbitrary brane in ${\mathcal S}\subset D_G(V,W)$
to a given grade restricted subset ${\mathcal T}_{\bf w}$
without changing the low energy behaviour in the Pfaffian phase,
by a brane replacement procedure using empty branes.
Recall that the $p$-Koszul brane $\Kos$, given in (\ref{Kp}),
as well as its tensor product with representations of $G$
are empty in the Pfaffian phase.
Let $\brane$ be a brane in ${\mathcal S}$
and let us glue to it an empty brane at a component $\rmV$ of $\brane$.
The brane to be glued is an appropriate R-charge shift of
$\Kos\otimes \rmV$ or $\Kos(-7)\otimes \rmV$
which has $\rmV$ at the left or the right edge.
This is possible if we can find a homomorphism between $\brane$ and the
empty brane that extends the identity map of $\rmV$
--- we take its cone and annihilate the $\rmV$ pair by tachyon condensation
of the identity map.
As in the similar procedure in the Grassmannian phase,
we try to find a homomorphism in two steps:
first find a homomorphism between the complexes obtained by setting $x=0$
and then extend it to a homomorphism between the matrix factorizations.
We may assume that $\brane$ is in the reduced form where all the arrows are
at least linear in the variables $(p,x)$ --- otherwise, we just perform
brane-anti-brane annihilation at the non-reduced arrows.
Also, just to be specific,
we assume that the component $\rmV$ is at degree $0$.
Then, the empty brane to be glued is either $\Kos\otimes\rmV$
or $\Kos(-7)[7]\otimes \rmV$.

Let us first consider gluing $\Kos\otimes \rmV$.
The first step is to find a homomorphism between the complexes
$\brane|_{x=0}$ and $\Kos|_{x=0}\otimes\rmV$
that extends the identity map of $\rmV$.
This is always possible and straightforward.
Recall that the Koszul complex $\Kos|_{x=0}$ is acyclic
in the sense of Section~\ref{subsub:HA}-{\rm II},
that is, it is exact except at the last term $\rmV$.
Since the arrows of $\brane|_{x=0}$ are at least linear in $p$ and
the image of the last arrow of $\Kos|_{x=0}$ includes all $p^k\rmV$'s,
by the extension theorem,
there is a homomorphism $f_{\mathscr{C}}:\brane|_{x=0}\to\Kos|_{x=0}\otimes \rmV$
that extends the identity map from the $\rmV$ in $\brane$ to the $\rmV$ 
in $\Kos\otimes \rmV$.

The second step is to find its extension to a homomorphism
$f:\brane\to \Kos\otimes \rmV$ of matrix factorizations.
This requires some homological conditions on pairs of components of
$\brane$ and those of $\Kos\otimes \rmV$.
We may proceed as in the analogous case (\ref{homcompletion})
in the Grassmannian phase.
We first write the candidate extension as
\beqa
f(p,x)&=&f_{\mathscr{C}}(p)
+\sum_{a,i}x^a_if^i_a(p)
+\sum_{a,b,i,j}x^a_ix^b_jf^{ij}_{ab}(p)
+\cdots\nn\\
&=&f^{(0)}+f^{(1)}+f^{(2)}+\cdots.
\eeqa
Likewise, we write the matrix factorizations of $\brane_1=\brane$
and $\brane_2=\Kos\otimes \rmV$ as
$Q_I=Q_I^{(0)}+Q_I^{(1)}+Q_I^{(2)}+\cdots$ for $I=1,2$
where $Q_I^{(m)}$ is of order $m$ in the $x_i^a$'s.
Note that $Q_1^{(0)}$ and $Q_2^{(0)}$ are the differentials of the
complexes $\brane|_{x=0}$ and $\Kos|_{x=0}\otimes \rmV$, respectively.
Then, the condition
$Q_2f=fQ_1$ splits to
\beq
Q_2^{(0)}f^{(n)}-f^{(n)}Q_1^{(0)}
=-\sum_{l+m=n,\,l\geq 1}\left(Q_2^{(l)}f^{(m)}-f^{(m)}Q_1^{(l)}\right),
\qquad n=0,1,2,\ldots .
\label{EqnforHom}
\eeq
The zeroth order equation is satisfied as $f^{(0)}$ is a homomorphism from
$\brane|_{x=0}$ to $\Kos|_{x=0}\otimes \rmV$.
If the equations of order 0,1,\ldots, $(n-1)$ are solved, then
the right-hand side ${\rm RHS}^{(n)}$ of the $n$-th order equation satisfies
$Q_2^{(0)}{\rm RHS}^{(n)}+{\rm RHS}^{(n)}Q_1^{(0)}=0$.
Note that it takes the form
\beq
    {\rm RHS}^{(n)}=\sum_{a_1,\ldots a_n\atop i_1,\ldots,i_n}x^{a_1}_{i_1}\cdots
    x^{a_n}_{i_n}{\rm RHS}^{i_1\ldots i_n}_{a_1\ldots a_n}(p),
\eeq
and $({\rm RHS}^{i_1\ldots i_n}_{a_1\ldots a_n})$
can be regarded as a homomorphism from $\brane|_{x=0}$
to $\Kos|_{x=0}\otimes\rmV\otimes \Sym^n\fund^{*\oplus 7}[1]$.
The question is whether it is null-homotopic, so that
one can find $f^{(n)}$ solving the $n$-th order equation.
At this point, we make use of the fact that the complex $\Kos|_{x=0}$
is exact on the subset $\{p\ne 0\}$ relevant in the Pfaffian phase.
We consider the category ${\it Coh}(\Pf)$ of
$G_{\C}=\GL_2(\C)$-equivariant coherent sheaves on $\{p\ne 0\}$.
(Here $\Pf$ is the quotient stack of $\{p\ne 0\}$ by $G_{\C}$.
See Appendix~\ref{app:Catmath}.)
For a complex $\mathscr{C}$ of representations of $G$ over $\C[p]$,
we denote by ${\mathcal O}_{\Pf}(\mathscr{C})$ the associated
complex over ${\it Coh}(\Pf)$.
By the exactness on $\{p\ne 0\}$,
$\Kos|_{x=0}$ defines a zero object in the derived category $D^b(\Pf)$
of ${\it Coh}(\Pf)$,
and so is its tensor product with any representation of $G$.
Therefore the homomorphism $({\rm RHS}^{i_1\ldots i_n}_{a_1\ldots a_n})$
vanishes as a morphism in $D^b(\Pf)$.
Hence by the Lemma in Section~\ref{subsub:HA}-{\rm III},
it is null-homotopic provided
the higher Ext groups of the pair
$({\mathcal O}_{\Pf}(\rmV_1),{\mathcal O}_{\Pf}(\rmV_2))$ vanish
for all components $\rmV_1$ and $\rmV_2$ of
$\brane$ and $\Kos\otimes\rmV\otimes \Sym^n\fund^{*\oplus 7}$, respectively.
Thus, an important task is to compute ${\rm Ext}^p_{{\it Coh}(\Pf)}$ of
such a pair, which we denote by ${\rm Ext}^p_{\Pf}(\rmV_1,\rmV_2)$.
An instruction is given inside the proof of Lemma 4.3 of \cite{ADS}.
Let
\beq
{\rm Hom}^{}_{\C}(\rmV_1,\rmV_2)
=\bigoplus_{h,i}\left(\Sym^h\fund(i)\right)^{\oplus N_{h,i}}
\eeq
be the decomposition into irreducible representations.
The component with $h\geq 1$ has no $SU(2)\subset G$ invariant and
does not contribute. For $\Sym^0\fund(i)=\C(i)$, $SU(2)$ acts trivially,
and the $G_{\C}$-equivariant sheaf ${\mathcal O}_{\Pf}(\C(i))$ descends
on the $G/SU(2)=U(1)$ quotient $\CP^6$ to ${\mathcal O}_{\CP^6}(-i)$.
Thus,
\beq
    {\rm Ext}^p_{\Pf}(\rmV_1,\rmV_2)=\bigoplus_{i}
        {\rm H}^p(\CP^6,{\mathcal O}_{\CP^6}(-i))^{\oplus N_{0,i}}.
\eeq
The condition for ${\rm H}^p(\CP^6,{\mathcal O}_{\CP^6}(-i))\ne 0$ for $p>0$
is $i\geq 7$, see (\ref{BWBProj}).
Applying the Clebsch-Gordan or Littlewood-Richardson rule,
we find for $h_1,h_2=0,1,2$ and $i_1,i_2\in \Z$ that
\beq
{\rm Ext}^p_{\Pf}(\Sym^{h_1}\fund(i_1),\Sym^{h_2}\fund(i_2)\otimes
\Sym^n\fund^{*\oplus 7})=0,\qquad \forall p>0,~\forall n\geq 1,
\label{Extrelevant1}
\eeq
if and only if
\beq
\begin{array}{l}
h_1<h_2\quad\mbox{and}\quad i_2-i_1\leq 6,\\
h_1=h_2~\,\mbox{or}~\,h_2+1\quad\mbox{and}\quad i_2-i_1\leq 7,\\
h_1=h_2+2\quad\mbox{and}\quad i_2-i_1\leq 8.
\end{array}
\label{condEV}
\eeq
For $\rmV=\Sym^h\fund(i)$, the brane $\Kos\otimes \rmV$ has the term
$\Sym^{h_2}\fund(i_2)$ with $(h_2,i_2)=(h,i+7)$, for which these conditions
on $(h_1,i_1)$ are
\beq
\begin{array}{c|c|c|c}
  &h_1=0&h_1=1&h_1=2\\
  \hline
  h=0&i_1\geq i&i_1\geq i&i_1\geq i-1\\
  h=1&i_1\geq i+1&i_1\geq i&i_1\geq i\\
  h=2&i_1\geq i+1&i_1\geq i+1&i_1\geq i
\end{array}
\label{homoC1}
\eeq
Other terms of $\Kos\otimes\rmV$ only give weaker conditions.
Therefore,
the equations (\ref{EqnforHom}) can be solved recursively if and only if
all the components $\Sym^{h_1}\fund(i_1)$ of $\brane$ satisfy this condition.
This has a geometrical meaning that
$\rmV$ is a left-most component of $\brane$ in the triangular grid.

Let us next consider gluing $\Kos(-7)[7]\otimes \rmV$
to a brane $\brane$ in ${\mathcal S}$.
The first step is to find a homomorphism of the complexes
$f_{\mathscr{C}}:\Kos(-7)[7]|_{x=0}\otimes\rmV\to \brane|_{x=0}$
that extends the identity map of $\rmV$.
This can be done using the transpose of the first step in the case above.
The second step is to extend it to a homomorphism of matrix
factorizations $f:\Kos(-7)[7]\otimes\rmV\to \brane$.
It can be searched in the same way as above
and can be found provided some conditions for vanishing of higher Ext groups
are satisfied.
For $\rmV=\Sym^h\fund(i)$, the brane $\Kos(-7)\otimes\rmV$ has the term
$\Sym^{h_1}\fund(i_1)$ with $(h_1,i_1)=(h,i-7)$, for which the conditions
(\ref{condEV}) on $(h_2,i_2)$ read
\beq
\begin{array}{c|c|c|c}
  &h_2=0&h_2=1&h_2=2\\
  \hline
  h=0&i_2\leq i&i_2\leq i-1&i_2\leq i-1\\
  h=1&i_2\leq i&i_2\leq i&i_2\leq i-1\\
  h=2&i_2\leq i+1&i_2\leq i&i_2\leq i
\end{array}
\label{homoC2}
\eeq
These are the strongest conditions coming from terms of $\Kos(-7)\otimes\rmV$
and the homomorphism $f$ can be found if and only if all the components
$\Sym^{h_2}\fund(i_2)$ of $\brane$ satisfy this condition.
Geometrically, this means that
$\rmV$ is a right-most component of $\brane$ in the triangular grid.

Applying these procedures successively,
we can replace $\brane$ by another brane in which the horizontal extent
of the Chan-Paton representations in the tringular grid is reduced.
Eventually, the set of representations can be made to fit into any
one of the four shapes depicted in Fig.~\ref{fig:GRRplot}.
This will be useful in the D-brane transport from the Pfaffian phase
to the Grassmannian phase, as will be done in Section~\ref{sec:transport}.
From the geometrical interpretation of the homological
conditions for the gluing, (\ref{homoC1}) and (\ref{homoC2}),
it is clear that no shape other than these four can be such a minimal subset.
Switching from one subset to another can also be done.
For example, the swicth from ${\mathcal T}_{\bf B}$ to ${\mathcal T}_{\bf A}$
is done by gluing $\Kos\otimes \Sym^2\fund(-1)$ at each
$\Sym^2\fund(-1)$ component of a given brane in ${\mathcal T}_{\bf B}$.
This is possible since $\Sym^2\fund(-1)$ sits on the left edge
of $\rmGRR_{\bf B}$ in the triangular grid.
This procedure will be relevant when we find the monodromy along loops
with a base point in the Grassmannian phase (Section~\ref{subsec:monodromy}).

\subsection{On The Equivalence
  $\pi_Y:{\mathcal T}_{\bf w}\stackrel{\cong}{\longrightarrow} D(Y)$}
\label{subsec:PfEq}

\newcommand{\NYst}{{\mathcal N}_Y\!\!{\vrule height1.9ex width0pt}^*}
\newcommand{\VPf}{V_{\rm Pf}}
\newcommand{\NPf}{{\mathcal N}_{\VPf}}
\newcommand{\YU}{Y{\scriptscriptstyle \cap\,}\nhU}
\newcommand{\UY}{\nhU_Y}

We would like to see if the flow $\pi_Y:{\mathcal T}_{\bf w}\to D(Y)$
is an equivalence for each window ${\bf w}$.
We can say that it is essentially surjective
provided $\pi_Y:{\mathcal S}\to D(Y)$ is,
since we have shown in the previous subsection that any brane
in ${\mathcal S}$ has a representative in ${\mathcal T}_{\bf w}$
with the same low energy limit.
Other than that, the task might look a little too tall for us
as we have just begun to understand the low energy behaviour
of gauge theory branes in the strongly coupled phase.
On the other hand, an equivalence of categories
$D(Y)\cong \TADS(={\mathcal T}_{\bf C})$ is constructed in \cite{ADS}.
Combining this result and the understanding of boundary flow in $SU(2)$ QCD,
and with the aid of general theory of derived categories
\cite{BK1991,BondalLarsenLunts,BOSOD,BOreconstruction,BridgelandEq},
we can in fact argue that $\pi_Y:{\mathcal T}_{\bf w}\to D(Y)$
is indeed an equivalence.

The theory at the energy scale $E_A\sqrt{-\zeta}\ll E\ll e_1\sqrt{-\zeta}$
is the $SU(2)$ gauge theory with seven fundamentals fibered over $\CP^6$
with a superpotential $W_{\rm ind}$ induced from $W$.
The data for B-branes in this theory are $SU(2)$-equivariant matrix
factorizations of $W_{\rm ind}$ on the total space, or equivalently,
$G$-equivariant matrix factorizations of $W$ on $V_-=\{(x,p)|p\ne 0\}$.
The set $D_G(V_-,W)$ of such brane data has a structure of category.
The morphism space is as in (\ref{Homgtphase}) of Appendix~\ref{app:CatGT}
in which $V^{\rm ss}(\zeta)$ is set to be $V_-$. (This applies here
since the $U(1)$ part of $G$ is completely Higgsed.)
The restriction functor $i_-^*:D_G(V,W)\to D_G(V_-,W)$
may be identified as the flow of gauge theory branes to this energy scale.
In particular, the flow $\pi_Y$ from ${\mathcal S}\subset D_G(V,W)$ to $D(Y)$
can be broken into two steps,
\beq
\pi_Y:{\mathcal S}\stackrel{i_-^*}{\longrightarrow}
i_-^*{\mathcal S}\stackrel{\pi_Y'}{\longrightarrow}
D(Y).
\eeq
Note that the Koszul branes $\Kos(i)$ are empty already at this energy scale
where $p\ne 0$ is enforced.
In particular, $i_-^*(\Kos(i))$ are zero objects of $D_G(V_-,W)$.
By the grade restriction process discussed in the previous subsection,
we find that $i_-^*{\mathcal S}$ is equivalent to $i_-^*{\mathcal T}_{\bf w}$
for any window ${\bf w}$.

One important result of \cite{ADS} is that the functor
$i_-^*:{\mathcal T}_{\bf w}\to D_G(V_-,W)$ is fully faithful.
This means that the first step of the flow
$i^*:{\mathcal S}\to i_-^*{\mathcal S}$ restricted 
${\mathcal T}_{\bf w}\subset {\mathcal S}$ is an equivalence.
This is shown in \cite{ADS} for ${\bf w}={\bf C}$ but can be generalized to
any ${\bf w}$ as follows.
Let ${\it Coh}(\,\fb_-)$ be the category of $G_{\C}$-equivariant
coherent sheaves on $V_-$.
(Here $\,\fb_-$ is the quotient stack of $V_-$ by $G_{\C}$.)
For a pair $(\rmV_1,\rmV_2)$ of representations of $G$,
we shall denote the Ext groups of the corresponding pair of
$G_{\C}$-equivariant sheaves on $V_-$ by
$\Ext^p_{\fb_-\!}(\rmV_1,\rmV_2)$.
The claimed fully faithfulness follows from the vanishing of these
Ext groups for all $p>0$ for the representations $\rmV_1$ and $\rmV_2$
in $\rmGRR_{\bf w}$.
Indeed, we have
\beq
\Ext_{\fb_-\!}^p(\rmV_1,\rmV_2)
=\Ext_{\Pf}^p(\rmV_1,\rmV_2\otimes \Sym(\fund^{*\oplus 7})),
\eeq
where $\Ext_{\Pf}^p$ is the Ext group for the category ${\it Coh}(\Pf)$
introduced in the previous section.
For $\rmV_a=\Sym^{h_a}\fund(i_a)$ with $h_a=0,1,2$ ($a=1,2$),
the vanishing of this for all
$p>0$ is the same as the vanishing (\ref{Extrelevant1}) 
in which $\forall n\geq 1$ is replaced by $\forall n\geq 0$.
That is, we add the condition for $n=0$ which is
$h_1=h_2$ and $i_2-i_1\leq 6$. Including this to (\ref{condEV}), we find
\beq
    {\rm Ext}^p_{\fb_-}(\Sym^{h_1}\fund(i_1),\Sym^{h_2}\fund(i_2))=0,
    \qquad \forall p>0,
\label{Extrelevant2}
\eeq
if and only if
\beq
\begin{array}{l}
h_1\leq h_2\quad\mbox{and}\quad i_2-i_1\leq 6,\\
h_1=h_2+1\quad\mbox{and}\quad i_2-i_1\leq 7,\\
h_1=h_2+2\quad\mbox{and}\quad i_2-i_1\leq 8.
\end{array}
\label{condEV2}
\eeq
It is easy to check that any pair 
in $\rmGRR_{\bf w}$ satisfies this condition. One can also find that
a maximal set of representations satisfying this condition
is $\rmGRR_{\bf w}$ for some ${\bf w}$.

\newcommand{\PsiADS}{\Psi_{\!\scriptscriptstyle\rm ADS}}

The most important result of \cite{ADS} 
is the construction of
a lift $\PsiADS:D(Y)\to D_G(V_-,W)$.
It can be described using a diagram similar to (\ref{Knorrer}),
\beq
\begin{tikzcd}
  &\fb_-\arrow[dr,"q^{}_{\CP^6}"]\\
  \fb_-|_Y\arrow[dr,"q^{}_Y"]\arrow[ur, hook,"\hat{i}^{}_Y"]&&\CP^6\\
  &Y\arrow[ur, hook, "i^{}_Y"]
\end{tikzcd}
\label{Pfdiagram}
\eeq
where $\fb_-|_Y$ is the quotient stack of
$V_-|_Y=\{(x,p)|{\rm rk}A(p)=4\}$ by $G_{\C}$.
Let us employ the mathematical description of the branes and their categories
as in Appendix~\ref{app:Catmath}. The lift is
a functor $\PsiADS:\Db(Y)\to \Db(\fb_-,W_{\rm ind})$,
where the target category consists of curved dg-sheaves
on $(\fb_-,W_{\rm ind})$ isomorphic to matrix factorizations.
The image of a sheaf ${\mathcal F}\in {\it Coh}(Y)$ is
\beq
\PsiADS({\mathcal F})=\left(\hat{i}_{Y*}({\mathcal O}_{\Gamma}\otimes q_Y^*{\mathcal F}),0\right)
\label{ADSlift}
\eeq
in which $\Gamma$ is the locus of points with $(x,p)\in V_-|_Y$
such that $x(\fund^*)/x(\fund^*)\cap {\rm Ker}A(p)$ has dimension at most $1$
when $x$ is regarded as a map $\fund^*\to V$.
Note that the superpotential $W_{\rm ind}$ vanishes on $\Gamma$
and hence $\PsiADS({\mathcal F})$ is a curved dg-sheaf on $(\fb_-,W_{\rm ind})$.
The sheaf ${\mathcal O}_{\Gamma}$ has a resolution by a complex of
locally free sheaves called the \emph{Eagon-Northcott complex},
\beq
\Sym^2S(-1)\otimes\textw^{\!4}(V/K)^*\to
S(-1)\otimes\textw^{\!3}(V/K)^*\to
{\mathcal O}(-1)\otimes \textw^{\!2}(V/K)^*\to
{\mathcal O}
\label{ENcomplex}
\eeq
where $\Sym^hS(i)$ is the vector bundle associated  to the representation
$\Sym^h\fund^*(i)$ of $G$, with ${\mathcal O}:=\Sym^0S$,
and $K\subset V$ is the rank three subbundle 
whose fiber at $(x,p)$ is ${\rm Ker}A(p)$.
The arrows are precisely the ones for the twisted Lascoux complex with
$(k,N)=(2,4)$.
Note that $V/K$ is the pullback by $q_Y$ of the vector bundle $\ck$ on $Y$
defined in (\ref{ckdef}).
Thus, $\hat{i}_{Y*}({\mathcal O}_{\Gamma}\otimes q_Y^*{\mathcal F})$
is quasi-isomorphic to a complex similar to
(\ref{ENcomplex}) where $\textw^{\!j}(V/K)^*$ for $j=0,2,3,4$
in the degree $0, -1,-2,-3$ terms are replaced by
$\hat{i}_{Y*}q_Y^*(\textw^{\!j}R^*\otimes{\mathcal F})
=q_{\CP^6}^*i_{Y*}(\textw^{\!j}R^*\otimes{\mathcal F})$.
Recall that any sheaf on $\CP^6$ is resolved by a complex of sums of
${\mathcal O}_{\CP^6}(i)$'s where $i$ runs over any set of
seven consecutive integers, and that $q_{\CP^6}^*{\mathcal O}_{\CP^6}(i)$
is associated to the representation $\C(-i)$ of $G$.
Therefore, we find that
$\hat{i}_{Y*}({\mathcal O}_{\Gamma}\otimes q_Y^*{\mathcal F})$
has a locally free resolution
whose terms are direct sums of vector bundles associated to
the representations $\Sym^h\fund(i)$ with $h=0,1,2$ and $i\in I_h$
where $I_h$ is a set of seven consecutive integers.
We may choose $I_h$ so that the set agrees with $\rmGRR_{\bf w}$ for some
${\bf w}$. Then, by the higher Ext vanishing (\ref{condEV2})
for pairs in $\rmGRR_{\bf w}$,
the proof of Lemma 3.6 in \cite{Segal} (or Lemma 4.10 of \cite{ADS})
shows that we can add arrows so that the complex is completed
to a matrix factorization of $W_{\rm ind}$ and that it is isomorphic to
$\PsiADS({\mathcal F})$ in the derived category of
quasicoherent curved dg-sheaves on $(\fb_-,W_{\rm ind})$.
Thus, $\PsiADS$ lands in $\Db(\fb_-,W_{\rm ind})$.
Since the representations involved are in $\rmGRR_{\bf w}$ for some ${\bf w}$,
the image $\PsiADS({\mathcal F})$ is isomorphic to an object of
$i_-^*{\mathcal T}_{\bf w}$ or equivalently of $i_-^*{\mathcal S}$.
Let $\BBr$ be the subcategory of $\Db(\fb_-,W_{\rm ind})$
consisting of objects isomorphic to objects of $i_-^*{\mathcal S}$.
Then, $\PsiADS$ can be regarded as a functor from $\Db(Y)$ to $\BBr$.
It was shown in \cite{ADS} that this is an equivalence of categories.
To summarize what we have learned so far, we have a chain of equivalences
\beq
{\mathcal T}_{\bf w}
\,\xrightarrow[\cong]{i_-^*}\,
i_-^*{\mathcal T}_{\bf w}
~\cong~
i_-^*{\mathcal S}
~\cong~
\BBr
\,\xleftarrow[\cong]{\PsiADS} \,
\Db(Y)
\label{congcongcong}
\eeq

\newcommand{\braneA}{\brane_{\!A}}
\newcommand{\braneUY}{\braneA|_{\fb_-|_{\UY}}}

We would like to know whether this functor $\PsiADS$ is acceptable as a lift
of the second step of the flow $\pi_Y':i_-^*{\mathcal S}\to D(Y)$.
\footnote{Here and in what follows,
  we shall be somewhat loose in the domain and the target of functors,
  via compostions with the equivalences such as
  $D(Y)\cong \Db(Y)$, $i_-^*{\mathcal S}\cong \BBr$ and
  $D_G(V_-,W)\cong \Db(\fb_-,W_{\rm ind})$.}
A key observation relevant for this question
is that the complex (\ref{ENcomplex}) can be related to our brane
$\brane^{SU(2)}_{3,7}$ which plays an important r\^ole in the flow.
We may consider a family of $\brane^{SU(2)}_{3,7}$ over $Y$ by letting
$\cktt$ move as the fiber (\ref{ckdef}) of the vector bundle $\ck$.
In particular,
one presentation of $\brane^{SU(2)}_{3,7}$, which appears as the
left-most vertical line in (\ref{iso4v}), can be promoted to
a $G$-equivariant matrix factorization of $W$ on $V_-|_Y$
\beq
  \Sym^2\fund^*\otimes q_Y^*\textw^{\!4}\ck^*~\longleftrightarrow~
  \fund^*\otimes q_Y^*\textw^{\!3}\ck^*~\longleftrightarrow~
  \C\otimes q_Y^*\textw_0^{\!2}\ck^*
  \label{glB37A}
\eeq
where we denote the projection $V_-|_Y\to Y$ again by $q_Y$.
This is isomorphic to the $(k,N)=(2,4)$ Lascoux matrix factorization
\beq
  \Sym^2\fund^*\otimes q_Y^*\textw^{\!4}\ck^*~\longleftrightarrow~
  \fund^*\otimes q_Y^*\textw^{\!3}\ck^*~\longleftrightarrow~
  \C\otimes q_Y^*\textw^{\!2}\ck^*~\longleftrightarrow~
  \C(1).
\label{glB37B}
\eeq
To see this, note that $\C(1)\cong q_Y^*{\mathcal O}_Y(-1)$
as $G_{\C}$-equivariant sheaves on $V_-|_Y$ and that
the leftward arrow from
$\C(1)$ in (\ref{glB37B}) is induced from
$A:{\mathcal O}_Y(-1)\to\textw^{\!2}\ck^*$ which is $A(p)$ at $p\in Y$.
This arrow is injective and its tachyon condensation
results in (\ref{glB37A}), where we used 
\beq
\textw^{\!2}_0\ck^*~\cong~
\textw^{\!2}\ck^*/{\rm Im}\,A.
\eeq
The matrix factorization (\ref{glB37A}) $\cong$ (\ref{glB37B}) can be regarded
as a globalization of $\brane^{SU(2)}_{3,7}$ over $Y$, and so is its tensor
product with $\C(i)$ which we denote by $\brane_{A,i}$.
We see that $\brane_{A,-1}$ is nothing but
a matrix factorization completion of the complex (\ref{ENcomplex}).

Let us discuss how we may find a lift of ${\mathcal F}\in {\it Coh}(Y)$
from our point of view.
We shall first work locally, considering the lift of ${\mathcal F}$ over
a neighborhood of each point of $Y$.
We take the neighborhood $\nhU$ in $\CP^6$ of a point $p_*\in Y$
introduced at the beginning of Section~\ref{subsec:theproof}.
It has coordinates $p^{1,2,3}, p^{12}, p^{23}, p^{31}$ and
$\UY=Y\cap \,\nhU$ is the locus with $p^{\tti\ttj}=0$.
We separate the lift into two steps, inserting the intermediate energy scale
$E_A\ll E\ll e_2$ at which the theory has the target space
$\fb_{{\rm int},\,\nhU}$ with coordinates
$p^{1,2,3}$, $p^{\tti\ttj}$ and $b_{\tti\ttj}$
and the superpotential
$W_{\rm int}=\sum_{\tti<\ttj} p^{\tti\ttj}b_{\tti\ttj}$ as in (\ref{Wint}).
B-branes in this theory are matrix factorizations of $W_{\rm int}$
on $\fb_{{\rm int},\,\nhU}$.
Since $\UY\subset \fb_{{\rm int},\,\nhU}$ is the non-degenerate
critical locus of $W_{\rm int}$, we are in the situation where
Kn\"orrer periodicity \cite{Knorrer} can be applied.
Therefore, the lift of ${\mathcal F}|_{\UY}$ in the first step
is its image under the Kn\"orrer map
$\kappa:D(\UY)\to D(\fb_{{\rm int},\,\nhU},W_{\rm int})$,
\beq
\kappa({\mathcal F}|_{\UY}).
\eeq
The theory in the upper energy scale is described by the variables
$p^{1,2,3}$, $p^{\tti\ttj}$, $x$ which take values in
$V_-|_{\nhU}=\{(x,p)|p\in \nhU\}$.
We use the parametrization of $x$ after the change of variables
(\ref{replacex}).
The superpotential (\ref{Wpstar}) can be written as
\beq
q_1^*W|_{V_-|_{\UY}}+q_2^*W_{\rm int},
\eeq
where $q_1$ is the projection to
$V_-|_{\UY}=\{(x,p)|p\in \UY\}$ which sets $p^{\tti\ttj}=0$
and $q_2$ is the projection to $\fb_{{\rm int},\,\nhU}$ which maps $(x,p)$
to $(p,b)$ with $b_{\tti\ttj}=[x_{\tti}x_{\ttj}]$.
Let us choose a matrix factorization $\brane_{\UY}$ of $W|_{V_{\UY}}$ on
$V_-|_{\UY}$
that restricts to $\brane^{SU(2)}_{3,7}$ at each $p\in \UY$.
By (\ref{mainprop}), it flows to the Neumann boundary condition on
the variables $p^{1,2,3}$ and $b_{\tti\ttj}$ at energies below $e_2$.
Thus, as a lift of $\kappa({\mathcal F}|_{\UY})$ in the second step,
we may take
\beq
q_1^*\brane_{\UY}\otimes q_2^*\kappa({\mathcal F}|_{\UY}).
\label{LiftPf}
\eeq
We would like to know if the results for different open subsets
are compatible with each other and glue together to give
a global lift of ${\mathcal F}$ on $Y$.
For this purpose, we may choose a globalization $\braneA=\brane_{A,i}$
of $\brane^{SU(2)}_{3,7}$ over $Y$
and take $\brane_{\UY}$ to be its local form
\beq
\brane_{\UY}=\braneA|_{V_-|_{\UY}}.
\label{braneAUY}
\eeq
However, it is still not obvious whether (\ref{LiftPf}) glue together
since $q_2:V_-|_{\nhU}\to \fb_{{\rm int},\nhU}$ depends very much on the
choice of the reparametrization (\ref{replacex}) which is hardly canonical.
At this stage, we again employ the mathematical description
of the branes and their categories, which turns out to be useful.
We regard (\ref{braneAUY}) as an object $\braneA|_{\fb_-|_{\UY}}$
of $\Db(\fb_-|_{\UY},W_{\rm ind})$.
We shall also employ the diagram
\beq
\begin{tikzcd}
  &\arrow[dl,bend right,"q_1^{}"']\fb_-|_{\nhU}
  \arrow[dr,"q_2^{}"']\arrow[ddrr,bend left,"q^{}_{\nhU}"]\\
  \fb_-|_{\UY} \!\!
  \arrow[ddrr,bend right,"q^{}_{\UY}"']
  \arrow[dr,"\bar{q}_2^{}"]
  \arrow[ur,hook,"\hat{i}_{\UY}^{}"']&&
  \fb_{{\rm int},\,\nhU}\arrow[dr,"p^{}_{\nhU}\!\!\!"']\\
  &\!\!\!\fb_{{\rm int},\,\nhU}|_{\UY}\!\!\!\!\!\!
    \arrow[dr,"p^{}_{\UY}"]\arrow[ur, hook, "j_{\UY}^{}"]&&
     \nhU\\
  &&\,\,\UY\!\!\arrow[ur,hook,"i_{\UY}^{}"]
\end{tikzcd}
\label{Pfdiagram2}
\eeq
We use the same letters $q_1$ and $q_2$ for the projections
as the ones before the $G_{\C}$-quotient.
The meaning of the other arrows should be obvious.
The diagram (\ref{Pfdiagram2}) is commutative except the part involving
$q_1$ which satisfies $q_1\circ \hat{i}_{\UY}={\rm id}_{\fb_-|_{\UY}}$.
The Kn\"orrer map as described in
\cite{Orlov05,Isik,Shipman,Preygel} is given  by
\beq
\kappa({\mathcal F}|_{\UY})
\,=\,\left(\,j_{\UY*}p_{\UY}^*({\mathcal F}|_{\UY}),0\right)
\,\cong\,\left(\,p_{\nhU}^*i_{\UY *}({\mathcal F}|_{\UY}),0\right)
\eeq
where we used the fact that $i_{\UY}$ and $j_{\UY}$ are closed embeddings
in the isomorphism.
Plugging this into (\ref{LiftPf}), we find
\beqa
q_1^*\braneUY\!\otimes q_2^*\kappa({\mathcal F}|_{\UY})
&\cong&q_1^*\braneUY\!\otimes\underbrace{q_2^*\,p_{\nhU}^*}_{q_{\nhU}^*}\!
i_{\UY*}({\mathcal F}|_{\UY})\nn\\
&\cong&q_1^*\braneUY\!\otimes
\hat{i}_{\UY*}q_{\UY}^*({\mathcal F}|_{\UY})\nn\\
&\cong&
\hat{i}_{\UY*}\left(\braneUY\!\otimes
q_{\UY}^*({\mathcal F}|_{\UY})\right).~~~~~~~
\label{translift}
\eeqa
The second isomorphism holds because $i_{\UY}$ and $\hat{i}_{\UY}$ are
closed embeddings,
and the last isomorphism is the projection formula combined with
$\hat{i}_{\UY}^*q_1^*={\rm id}_{\fb_-|_{\UY}}^*$.
We have arrived at an expression that bypasses
the non-canonical $q_2:\fb_-|_{\nhU}\to\fb_{{\rm int},\nhU}$.
In this form, we see that the results for different open subsets glue
together to 
\beq
\Psi_{\braneA}({\mathcal F})=
\hat{i}_{Y*}(\braneA\!\otimes q_Y^*({\mathcal F}))
\label{ourlift}
\eeq
defined globally on $\fb_-$. Indeed, its restriction to $\fb_-|_{\nhU}$
\beqa
i^{\fb_-*}_{\fb_-|_{\nhU}}\Psi_{\braneA}({\mathcal F})
&=&\hat{i}_{\UY*}i^{\fb_-|_Y*}_{\fb_-|_{\UY}}(\braneA\!\otimes q_Y^*({\mathcal F}))
\nn\\
&=&\hat{i}_{\UY*}\Bigl(\,\underbrace{i^{\fb_-|_Y*}_{\fb_-|_{\UY}}\braneA}_{\braneUY}\otimes \underbrace{i^{\fb_-|_Y*}_{\fb_-|_{\UY}}q_Y^*({\mathcal F})}_{q_{\UY}^*i^{Y*}_{\UY}({\mathcal F})}\,\Bigr)
\label{inclincl}
\eeqa
is nothing but the right-hand side of (\ref{translift}).
In this computation, we denoted the inclusion map $U\hookrightarrow U'$
by $i^{U'}_U$.
We may take $\Psi_{\braneA}$ defined by (\ref{ourlift}) as
a candidate for the lift.
Being a matrix factorization completion of the complex (\ref{ENcomplex}),
$\brane_{A,-1}$ is isomorphic to $({\mathcal O}_{\Gamma},0)$
in $\Db(\fb_-|_Y,W_{\rm ind})$, up to a possible shift of R-charge to which
we have not yet been attentive.
See for example \cite{Orlov11}, Lemma 3.1a or \cite{KapustinLi}.
Therefore, our candidate $\Psi_{\braneA}$ for $\braneA=\brane_{A,-1}$
agrees with $\PsiADS$ in (\ref{ADSlift}) (up to an R-charge shift).
This shows that the lift $\PsiADS$ of \cite{ADS} is reasonable
also from the physics point of view.
Note also that $\Psi_{\brane_{A,i}}:\Db(Y)\to \BBr$ for any $i$
is also an equivalence
as it is the composition of $\PsiADS$ and the twist functor
on $\Db(Y)$ by the line bundle ${\mathcal O}_Y(-1-i)$ (and a shift).

We would like to see whether one of the candidates,
$\Psi_{\brane_{A,i}}$ for some $i$ and for some R-charge assignment,
is indeed acceptable as a lift of
$\pi_Y':i_-^*{\mathcal S}\to D(Y)$.
The condition is,
for any brane $\brane\in {\mathcal S}$ that flows to ${\mathcal F}\in D(Y)$,
\beq
\Psi_{\brane_{A,i}}({\mathcal F})~\cong~ i_-^*\brane
\label{isolift}
\eeq
in $\Db(\fb_-,W_{\rm ind})$.
Let us first find the correct R-charge assignment 
by looking at the $\Psi_{\brane_A,i}$-image
of a D0-brane ${\mathcal O}_{p_*}$ at $p_*\in Y$.
As $\hat{i}_{Y*}(q_Y^*\textw^{\!j}\ck^*\otimes q_Y^*{\mathcal O}_{p_*})\cong\textw^{\!j}\cktt^*\otimes q_{\CP^6}^*{\mathcal O}_{p_*}$
where $\cktt$ is the fiber of $\ck$ at $p_*$, we find, for any $i$,
\beq
\Psi_{\brane_{A,i}}({\mathcal O}_{p_*})~=~
\Bigl(\,\Sym^2\fund^*\otimes \textw^{\!4}\cktt^*~\longleftrightarrow~
  \fund^*\otimes \textw^{\!3}\cktt^*~\longleftrightarrow~
  \C\otimes \textw_0^{\!2}\cktt^*\,\Bigr)\otimes q_{\CP^6}^*{\mathcal O}_{p_*}.
\label{PsiOp}
\eeq
On the other hand, a lift of ${\mathcal O}_{p_*}$ in $D_G(V_-,W)$
is given by $\brane^{\rm D0}_{p_*,\,\nhU}[1]$ (\ref{choicePf})
where the Clifford vacuum $|0\rangle$ has R-charge $-1$.
This is isomorphic in $\Db(\fb_-,W_{\rm ind})$ to (\ref{PsiOp}) provided
the three terms inside the parenthesis in (\ref{PsiOp}) all have R-charge $-1$.
This requires that the R-charges of the three terms of $\brane_{A,i}$
as in (\ref{glB37A}) must all be $-1$, or quivalently, the R-charges of
the four terms as in (\ref{glB37B}) must be $-1,-1,-1,-2$ from left to right.
More explicitly, we will find in Section~\ref{sub:D0}
(see {\bf Remark} on page~\pageref{page:Rmk}) that
\beq
i_-^*\brane_{p_*}~\cong~
\Bigl(\,\mathop{\Sym^2\fund^*}_{{}_{-1}}~\longleftrightarrow~
  \mathop{\fund^*\otimes \cktt}_{{}_{-1}}~\longleftrightarrow~
  \mathop{\C\otimes \textw_0^{\!2}\cktt}_{{}_{-1}}\,\Bigr)\otimes i_-^*\Kos_{I_*}
\label{braneOpst}
\eeq
where $i_-^*\Kos_{I_*}$ is a completion of
the Koszul resolution of $q_{\CP^6}^*{\mathcal O}_{p_*}$ to
a matrix factorization of $\sum_{k\in I_*}p^kA_k(x)$.
This is isomorphic in $\Db(\fb_-,W_{\rm ind})$ to (\ref{PsiOp})
with the above assignment of the R-charge,
confirming (\ref{isolift}) for ${\mathcal F}={\mathcal O}_{p_*}$
and $\brane=\brane_{p_*}$.

This means that $\pi_Y'\circ \Psi_{\brane_{A,i}}$ (for any $i$)
sends ${\mathcal O}_p$
to ${\mathcal O}_p$ itself for any point $p$ of $Y$.
Combined with general results on derived categories,
this has the following significance.\footnote{We thank Alexey Bondal for help
  (September 2026) to form the argument given in this paragraph.}  
We expect that $\pi_Y'$ and $\Psi_{\brane_{A,i}}$
come from pre-exact dg-functors \cite{BK1991}
between a dg-enhancement of $i_-^*{\mathcal S}\cong\BBr$ and
the standard dg-enhancement of $D(Y)\cong \Db(Y)$.
(See Appendix~\ref{app:Catmath}.)
By the result of \cite{BondalLarsenLunts}, the functor
$\pi_Y'\circ \Psi_{\brane_{A,i}}$ can be represented as a Fourier-Mukai
transform from $\Db(Y)$ to itself. Then, 
by the results of \cite{BOSOD,BOreconstruction,BridgelandEq},
it must be an equivalence
as it sends ${\mathcal O}_p$ to itself for any $p\in Y$.
As $\Psi_{\brane_{A,i}}$ is an equivalence, so is $\pi_Y'$.
This shows what we wanted to know:
\beq
\pi_Y:{\mathcal T}_{\bf w}\,\xrightarrow[\cong]{i_-^*}\,
i_-^*{\mathcal S}\,\xrightarrow[\cong]{\pi_Y'}D(Y)
\eeq
is an equivalence.
Moreover, applying \cite{BOSOD}, Proposition 1.5 to the diagonal embedding
$j:Y\hookrightarrow Y\times Y$, we find that
$\pi_Y'\circ \Psi_{\brane_{A,i}}$ is the twist functor by a line bundle on $Y$.
In particular, $\pi_Y'\circ \Psi_{\brane_{A,i}}$ for one special $i$
is the identity functor of $\Db(Y)$.
In other words, $\Psi_{\brane_{A,i}}$ for that $i$ is a lift of $\pi_Y'$.

\newcommand{\chdot}{\mbox{\large $\cdot$}}

The correct value of $i$ can be found by examining a topological part of
(\ref{isolift}) for a pair $({\mathcal F},\brane)$ where
${\mathcal F}$ is extended over $Y$, such as
$({\mathcal F},\brane)=({\mathcal O}_Y,\Lb^2(-1)[1])$.
To each object $(\mathscr{E},d_{\mathscr{E}})$ of $\Db(\fb_-,W_{\rm ind})$,
we assign the Chern character of the $G$-equivariant complex of sheaves
at $x=0$, which we shall denote by
$\lfloor (\mathscr{E},d_{\mathscr{E}})\rfloor$.
It takes values in
\beq
    {\rm Rep}(U(2))\!\!\!\mathop{\otimes}_{\,\,\,\,{\rm Rep}(U(1))}
    \!\!\!{\rm H}^*(\CP^6,\Q),
\eeq
in which the class $\rmV\chdot\,\omega$ of $\rmV\otimes \omega$
obeys
$\rmV(i)\chdot\,\omega=\rmV\chdot\,{\rm ch}({\mathcal O}_{\CP^6}(-i))\omega=\rmV\chdot\e^{-iH}\omega$.
What we examine is
\beq
\lfloor \Psi_{\brane_{A,i}}({\mathcal F})\rfloor
=\lfloor i_-^*\brane \rfloor.
\label{cheqlift}
\eeq
In view of the presentation of $\brane_{A,i}$ as in (\ref{glB37A}) and using
the fact that $\hat{i}_{Y*}q_Y^*{\mathcal E}=q_{\CP^6}^*i_{Y*}{\mathcal E}$
restricts to $i_{Y*}{\mathcal E}$ at $x=0$,
we find 
\beqa
\lfloor \Psi_{\brane_{A,i}}({\mathcal F})\rfloor
&=&
-\Sym^2\fund^*(i)\,\chdot\,{\rm ch}(i_{Y*}(\textw^{\!4}\ck^*\otimes{\mathcal F}))\nn\\
&&-\fund^*(i)\,\chdot\,{\rm ch}(i_{Y*}(\textw^{\!3}\ck^*\otimes{\mathcal F}))
-\C(i)\,\chdot\,{\rm ch}(i_{Y*}(\textw^{\!2}_0\ck^*\otimes{\mathcal F})).
\eeqa
The minus signs come from the odd R-charges of the three terms
of (\ref{glB37A}).
For the evaluation, we use the Grothendieck-Riemann-Roch formula
\beq
    {\rm ch}(i_{Y*}{\mathcal E})
    =i_{Y*}\left({\rm ch}({\mathcal E}){{\rm td}(Y)\over
      i_Y^*{\rm td}(\CP^6)}\right)
    =14H^3\,{\rm ch}({\mathcal E})\left(1-{7H\over2}+{27H^2\over 4}
    -{28 H^3\over 3}\right)
\eeq
and the formulae
\beq
    {\rm ch}(\ck^*)=4-2H+{H^3\over 6},\qquad
    {\rm ch}(\textw^{\!2}_0\ck^*(1))=5-H^2
\label{chckst}
\eeq
that follow from (\ref{chL1}) and (\ref{chL2}) combined with
(\ref{L1fate}) and (\ref{L0fate}), which in turn yield
\beqa
    {\rm ch}(\textw^{\!4}\ck^*)&=&\e^{-2H},\\
    {\rm ch}(\textw^{\!3}\ck^*)&=&{\rm ch}(\ck^*(-1))=\left(4-2H+{H^3\over 6}\right)\e^{-H},\label{sndid}\\
    {\rm ch}(\textw^{\!2}_0\ck^*)&=&(5-H^2)\e^{-H}.
\eeqa
In (\ref{sndid}), we used the fact that there is a perfect paring between
$\textw^{\!3}\ck^*$ and $\ck^*$ with values in
$\textw^{\!4}\ck^*\cong {\mathcal O}_Y(-2)$
and that $\ck\cong\ck^*(1)$ from (\ref{ckckdual}).
This implies
\beqa
\lfloor\Psi_{\brane_{A,i}}({\mathcal F})\rfloor
  &=&-\Sym^2\fund^*\,\chdot\e^{-(i+1)H} 14 H^3\left(1-{9H\over 2}+{43H^2\over 4}-18H^3\right){\rm ch}({\mathcal F})\nn\\
&&  -\fund^*\,\chdot \e^{-(i+1)H}14H^3\left(4-16H+34H^2-{152H^3\over 3}\right){\rm ch}({\mathcal F})\nn\\
&&  -\C\,\chdot\e^{-(i+1)H} 14H^3\left(5-{35H\over 2}+{131H^2\over 4}-{259 H^3\over 6}\right){\rm ch}({\mathcal F}).
\label{CHPF}
\eeqa
For example, since $\ch({\mathcal O}_{p_*})=H^3/14$, we find
\beq
\lfloor \Psi_{\brane_{A,i}}({\mathcal O}_{p_*})\rfloor
=-\Sym^2\fund^*\,\chdot\,H^6-\fund^*\,\chdot\,4H^6-\C\,\chdot\,5H^6\quad
\eeq
for any $i$. This is indeed what we obtain from the expression (\ref{PsiOp})
with the R-charge fixed above, or equivalently from (\ref{braneOpst}),
since
$\lfloor q_{\CP^6}{\mathcal O}_{p_*}\rfloor=\lfloor i_-^*\Kos_{I_*}\rfloor=(1-\e^{-H})^6=H^6$.
Now, let us examine (\ref{cheqlift}) for
${\mathcal F}={\mathcal O}_Y$ and $\brane=\Lb^2(-1)[1]$.
In view of the Pfaffian frame R-charge assignment (\ref{PfL0}),
we find
\beqa
\lfloor i_-^*\Lb^2(-1)[1]\rfloor
&=&-\Bigl\lfloor\Bigl(\,\sym^2\fund^*(1)\otimes \textw^{\!7}V^*-\sym^2\fund^*(4)\otimes V^*
+\sym^2\fund^*(5)\otimes V-\sym^2\fund^*(8)\otimes \textw^{\!7}V\nn\\
&&+\fund^*(1)\otimes\textw^{\!6}V^*-\fund^*(3)\otimes \textw^{\!2}V^*
+\fund^*(5)\otimes\textw^{\!2}V-\fund^*(7)\otimes \textw^{\!6}V\nn\\
&&+\C(1)\otimes\textw^{\!5}V^*-\C(2)\otimes\textw^{\!3}V^*
+\C(5)\otimes\textw^{\!3}V-\C(6)\otimes\textw^{\!5}V\,\Bigr)\Bigr\rfloor
\nn\\
&=&-\Sym^2\fund^*\,\chdot\left(\e^{-H}-7\e^{-4H}+7\e^{-5H}-\e^{-8H}\right)\nn\\
&&-\fund^*\,\chdot\left(7\e^{-H}-21\e^{-3H}+21\e^{-5H}-7\e^{-7H}\right)\nn\\
&&-\C\,\chdot\left(21\e^{-H}-35\e^{-2H}+35\e^{-5H}-21\e^{-6H}\right).
\eeqa
One can check that this precisely matches with (\ref{CHPF}) for $i=-1$ and
${\mathcal F}={\mathcal O}_Y$.
In the same way, we can also show that (\ref{cheqlift}) holds for
$({\mathcal F},\brane)=(\ck^*,\Lb^1)$ and $(\textw^{\!2}_0\ck^*(1),\Lb^0[1])$
when $i=-1$ (see Appendix~\ref{app:chPf}).
We conclude that $i=-1$ is the correct value.
Therefore, $\Psi_{\brane_{A,-1}}$ is a lift of $\pi_Y'$.

Let us compare the lift $\Psi_{\brane_{A,-1}}$ with $\PsiADS$.
We already know that they agree up to a possible R-charge shift.
If we assume that ${\mathcal O}_{\Gamma}$ has R-charge zero,
which is the value in the convention of \cite{ADS},
the R-charges of the four terms of the resolution (\ref{ENcomplex}) are
$1,1,1,0$ from left to right. This is $+2$ compared to
$-1,-1,-1,-2$ of the corresponding four terms of $\brane_{A,-1}$.
Thus, the precise relation is $\PsiADS=[-2]\circ\Psi_{\brane_{A,-1}}$.

The lift can be used to demonstrate the relation (\ref{HSvsCCPf})
between the hemisphere partition function and the Calabi-Yau central charge
in the Pfaffian phase, that is, 
\beq
Z_{D^2}(\brane)
\,=\,(2\pi)^{13}\e^t\cdot Z_Y({\mathcal F})\quad\, \mbox{for\, $\zeta \ll 0$},
\label{claimedid}
\eeq
for each $\brane\in{\mathcal S}$ that flows to ${\mathcal F}\in D(Y)$.
We already know that it holds for
$(\brane,{\mathcal F})=(\Lb^2(-1)[1],{\mathcal O}_Y)$, $(\Lb^1,\ck^*)$,
$(\Lb^0[1],\textw^{\!2}_0\ck^*(1))$ and $(\brane_{p_*},{\mathcal O}_{p_*})$
but not yet in general.
We first note that $Z_{D^2}(\brane)$ in this phase
is determined by the Chern character $\lfloor i_-^*\brane\rfloor$.
Suppose the Chan-Paton vector space of $\brane$ has
$\Sym^h\fund^*(i)$ with multiplicity $n_{h,i}^r$ at R-charge $r\in \Z$
in the Grassmannian frame (\ref{Grframe}).
Of course $n_{h,i}^r$ is non-zero only for $h=0,1,2$.
Then, the brane factor $f_{\brane}(\usigma)$ and the Chern character of
$i_-^*\brane$ are given by
\beqa
f_{\brane}(\usigma)&=&\sum_{h,i,r}n_{h,i}^r(-1)^r{\rm tr}_{\Sym^h\fund^*}(\e^{2\pi\usigma})\e^{2\pi i\usigma_0},\\
\lfloor i_-^*\brane\rfloor&=&
\sum_{h,i,r}n_{h,i}^r(-1)^{r+h}\,\Sym^h\fund^*\cdot \e^{-iH}.
\eeqa
The sign $(-1)^h$ in the latter comes from
the R-charge frame change (\ref{RPf}).
Suppose $\lfloor i_-^*\brane\rfloor$ vanishes.
Then, for each $h$,
$\sum_{i,r}n_{h,i}^r(-1)^r\e^{-iH}$ vanishes as an element of
${\rm H}^*(\CP^6,\Q)$. This means that
$\sum_{i,r}n_{h,i}^r(-1)^r\e^{-ix}$
is divisible by $x^7$ in $\Q[[x]]$ and hence by
$(1-\e^{-x})^7$ in $\Q[\e^{-x},\e^x]$.
That is, $f_{\brane}(\usigma)$ is divisible by
$(1-\e^{2\pi\usigma_0})^7$. Then, as we have seen in Section~\ref{subsec:PHS},
$Z_{D^2}(\brane)$ vanishes for $\zeta\ll 0$.
For a sequence of rational numbers $c_1,c_2,\ldots$ and
branes $\brane_1$, $\brane_2, \ldots$, the same argument shows
\beq
\sum_kc_k\lfloor i_-^*\brane_k\rfloor=0
~\,\mbox{implies}~\,
\sum_kc_kZ_{D^2}(\brane_k)=0
~\mbox{for\, $\zeta \ll 0$}.
\label{floorsZZ}
\eeq
Note that the Chern characters $1$, $4-2H+H^3/6$, $5-H^2$ and $H^3/14$
of ${\mathcal O}_Y$, $\ck^*$, $\textw^{\!2}_0\ck^*(1)$
and ${\mathcal O}_{p_*}$ form a basis of
the linear subspace of ${\rm H}^*(Y,\Q)$ spanned by
$\{1,H,H^2,H^3\}$.
Therefore, for an arbitrary brane ${\mathcal F}$ in $Y$,
there are rational numbers $a_1,\ldots, a_4$
such that
\beq
{\rm ch}({\mathcal F})=a_1\ch({\mathcal O}_Y)+a_2\ch(\ck^*)+a_3\ch(\textw^{\!2}_0\ck^*(1))+a_4\ch({\mathcal O}_{p_*}).
\label{chreln}
\eeq
Inserting this into (\ref{CHPF}) with $i=-1$,
the same formula holds for $\lfloor \Psi_{\brane_{A,-1}}(-)\rfloor$
in the place of ${\rm ch}(-)$.
As $\Psi_{\brane_{A,-1}}$ is a lift of $\pi_Y'$, we find
$\Psi_{\brane_{A,-1}}({\mathcal F}) \cong i_-^*\brane$
if $\brane\in {\mathcal S}$ flows to ${\mathcal F}\in D(Y)$.
Thus, we find
\beq
\lfloor i_-^*\brane\rfloor
=
  a_1\lfloor i_-^*\Lb^2(-1)[1]\rfloor+a_2\lfloor i_-^*\Lb^1\rfloor+a_3 \lfloor i_-^*\Lb^0[1]\rfloor
  +a_4 \lfloor i_-^*\brane_{p_*}\rfloor.
\eeq
Then, it follows from (\ref{floorsZZ}) that
\beqa
\lefteqn{Z_{D^2}(\brane))
  ~\stackrel{\downarrow}{=}~
  a_1Z_{D^2}(\Lb^2(-1)[1])+a_2Z_{D^2}(\Lb^1)+a_3 Z_{D^2}(\Lb^0[1])
+a_4 Z_{D^2}(\brane_{p_*})}\nn\\
&=&(2\pi)^{13}\e^t\cdot\Bigl(\,
a_1Z_Y({\mathcal O}_Y)+a_2Z_Y(\ck^*)+a_3Z_Y(\textw^{\!2}_0\ck^*(1))
+a_4Z_Y({\mathcal O}_{p_*})\,\Bigr)\nn\\
&\stackrel{(\ref{chreln})}{=}&
(2\pi)^{13}\e^t\cdot Z_Y({\mathcal F}).
\eeqa

\section{D-Brane Transport And Monodromy}
\label{sec:transport}

\newcommand{\PtoG}{{\mathscr T}^{\,{}^{\uparrow}}}
\newcommand{\GtoP}{{\mathscr T}^{\,{}^{\downarrow}}}

Let us recall the main diagram (\ref{maindiagram})
\beq
\begin{tikzcd}
  &D_{G}(V,W)\arrow[d,phantom,"\mbox{\large $\cup$}"]\arrow[dddr,bend left,"\pi_X"]&\\[-0.3cm]
  &{\mathcal S}\arrow[ddl,bend right,"\pi_Y"']\arrow[d,phantom,"\mbox{\large $\cup$}"]&\\[-0.3cm]
  &{\mathcal T}_{\bf w}\arrow[dl,"\cong"']\arrow[dr,"\cong"]\\
  D(Y)&&D(X)
\end{tikzcd}
\label{q:maindiagram}
\eeq
Let us take an arbitrary brane ${\mathcal E}$
in the Grassmannian Calabi-Yau $X$
and consider its transport along a path
through a window {\bf w} to the Pfaffian Calabi-Yau $Y$.
The first step is to find a lift of ${\mathcal E}$
to a matrix factorization which is grade restricted with respect
to the window {\bf w}, that is, to find a $\brane
\in {\mathcal T}_{\bf w}$
such that $\pi_X(\brane)={\mathcal E}$.
It can be safely transported along the path to the Pfaffian phase.
Then, we just have to look at its low energy
limit to find the result of the transport. That is,
${\mathcal E}$ is transported to $\pi_Y(\brane)$.
The same holds for the transport in the opposite direction,
from the Pfaffian phase to the Grassmannian phase.
We shall denote the transport in the two directions by
\beq
\begin{tikzcd}
 D(Y)~~~~
 \arrow[r,shorten=-3mm,yshift=-0.5ex,"\displaystyle\PtoG_{\bf w}"']&
 \arrow[l,shorten=-3mm,yshift=0.5ex,"\displaystyle \GtoP_{\bf w}"']
~~~~D(X)
\end{tikzcd}
\eeq
If $\brane$ is in ${\mathcal T}_{\bf w}$, we have
\beq
\pi_Y(\brane)=\GtoP_{\bf w}(\pi_X(\brane))\quad\mbox{and}\quad
\pi_X(\brane)=\PtoG_{\bf w}(\pi_Y(\brane)).
\label{iddd}
\eeq
In this section, we describe this in some detail.
First, we determine the transport at the level of D-brane charges
using the hemisphere partition function.
Next, we work out some examples of transport.
Our main target is the transport of D6-branes from $X$ to $Y$
and from $Y$ to $X$, as well as transport of D0-branes in $Y$ to $X$.
The latter can be used to identify the kernel of the Fourier-Mukai
transform that describes the transport through each window.
Finally, we determine the monodromies along loops that go around
the singular points by employing the grade restriction processes
in the phase opposite to the phase of the base point.
We also discuss the large volume monodromies, where we find an important
subtlety concerning the choice of R-charge frame.
We close this section by confirming
that the homotopy equivalence relations of paths and loops
are correctly reflected in the transport and monodromy.

Before starting, let us describe an identity that shows how
the transport with respect to a window {\bf w} is related to
its $2\pi$ shift ${\bf w}(1)$.
We recall from (\ref{shiftGRR}) that,
if a brane $\brane$ is in ${\mathcal T}_{{\bf w}(1)}$,
then its gauge shift $\brane(1)$ is in ${\mathcal T}_{\bf w}$.
Together with the responses (\ref{shiftGr}) and (\ref{shiftPf})
of the projections $\pi_X$ and $\pi_Y$ to the gauge shift,
this yields 
\beq
\begin{array}{ccccc}
  && \pi_Y(\brane(1))\!\!\!&\stackrel{(\ref{iddd})}{=}
  &\GtoP_{\bf w}(\pi_X(\brane(1)))\\
  &&\!\!\!\mbox{\scriptsize (\ref{shiftPf})}\Vert\,\,\,&
  &\,\,\,\Vert\mbox{\scriptsize (\ref{shiftGr})}\!\!\!\\
  \GtoP_{{\bf w}(1)}(\pi_X(\brane))(-1)[2]&\stackrel{(\ref{iddd})}{=}&
  \pi_Y(\brane)(-1)[2]\!\!\!\!\!\!&&\GtoP_{\bf w}(\pi_X(\brane)(1))
\end{array}
\eeq
As this holds for any $\brane$ in ${\mathcal T}_{{\bf w}(1)}$,
we find for any ${\mathcal E}\in D(X)$,
\beq
\GtoP_{{\bf w}(1)}({\mathcal E})
=\GtoP_{\bf w}({\mathcal E}(1))(1)[-2].
\label{shiftGtoP}
\eeq
Similarly, for any ${\mathcal F}\in D(Y)$,
\beq
\PtoG_{{\bf w}(1)}({\mathcal F})
=\PtoG_{\bf w}({\mathcal F}(-1))(-1)[2].
\label{shiftPtoG}
\eeq

\subsection{D-Brane Transport At The Level Of Charges}
\label{sub:charge}

We first find the transport of D-brane charges
using the hemisphere partition function.
We shall use the integral basis of the lattice of D-brane charges
\beq
{\rm B}_X=\bigl([{\mathcal O}_X],[{\mathcal O}_H],
[{\mathcal O}_{\ell}], [{\mathcal O}_{\rm pt}]\bigr),\quad
{\rm B}_Y=\bigl([{\mathcal O}_Y],[{\mathcal O}_H],
[{\mathcal O}_{\ell}], [{\mathcal O}_{\rm pt}]\bigr).
\label{basis}
\eeq
The basis elements can be expressed
as a polynomial of the second cohomology class $H$: For $X$,
$[{\mathcal O}_X]=1$, 
$[{\mathcal O}_H]=1-\e^{-H}$,
$[{\mathcal O}_{\ell}]=\ell+[{\rm pt}]$,
$[{\mathcal O}_{\rm pt}]=[{\rm pt}]$,
with $\ell=H^2/42$ and $[{\rm pt}]=H^3/42$, while for $Y$,
$[{\mathcal O}_Y]=1$ and the same expression for the other three
with $\ell=H^2/14$ and $[{\rm pt}]=H^3/14$.
We would like to find the charge transport rule of the form
\beq
\GtoP_{\bf w}:{\rm B}_X
\,\longmapsto \,{\rm B}_Y\cdot T_{\bf w}
\label{chargetransport}
\eeq
for some matrix $T_{\bf w}\in \GL_4(\Z)$.

Recall that the functional
$f\mapsto Z_{\rm Gr}\{f\}$ ({\it resp}. $f\mapsto Z_{\rm Pf}\{f\}$)
defined by (\ref{ZexpGra}) ({\it resp}. (\ref{Zstrongexp}))
computes the hemisphere partition function 
for a brane $\brane$ with brane factor $f=f_{\brane}$
in the Grassmannian phase ({\it resp}. the Pfaffian phase).
The first step is to find, for each $[{\mathcal E}]\in {\rm B}_X$,
an integral linear combination of the characters
$f_{\rmV}(\usigma)={\rm tr}_{\rmV}(\e^{2\pi\usigma})$
of the grade restricted representations $\rmV$
\beq
f^{\bf w}_{\mathcal E}=\sum_{\rmV\in \rmGRR_{\bf w}}n_{\mathcal E}^{\!\rmV\,}f_{\rmV}
\eeq
whose $Z_{\rm Gr}$-value matches with the central charge of ${\mathcal E}$
\beq
Z_{\rm Gr}\{f^{\bf w}_{\mathcal E}\}\,\stackrel{!}{=}\,
(2\pi)^{13}Z_X({\mathcal E}).
\label{ZGrvsZ}
\eeq
The next step is to evaluate
the $Z_{\rm Pf}$-values of the functions $f^{\bf w}_{\mathcal E}$.
The final step is to expand the result by the Pfaffian phase central charges
for $[{\mathcal F}]\in {\rm B}_Y$.
That is,
\beqa
\lefteqn{\bigl(Z_{\rm Pf}\{f^{\bf w}_{{\mathcal O}_X}\},
Z_{\rm Pf}\{f^{\bf w}_{{\mathcal O}_H}\},
Z_{\rm Pf}\{f^{\bf w}_{{\mathcal O}_{\ell}}\},
Z_{\rm Pf}\{f^{\bf w}_{{\mathcal O}_{\rm pt}}\}\bigr)}\nn\\
&&~~~~~~~=(2\pi)^{13}\e^t\cdot\bigl(Z_Y({\mathcal O}_Y),
Z_Y({\mathcal O}_H),
Z_Y({\mathcal O}_{\ell}),
Z_Y({\mathcal O}_{\rm pt})\bigr)\cdot T_{\bf w}.~~~~~~~
\label{evaluateTw}
\eeqa
The actual computation is done using Mathematica.
To find the $4\times 21$ integers $n_{\mathcal E}^{\!\rmV}$ in the first step,
we only have to match the leading terms of (\ref{ZGrvsZ})
in the $\e^{-t}$ expansion: they include a sufficient
number of independent equations over the integers.
Also, in the evaluation of $Z_{\rm Pf}\{f^{\bf w}_{\mathcal E}\}$,
it is enough to compute only the leading terms of the
$\e^t$ expansion in order to read off the coefficients with respect to
$Z_Y({\mathcal F})$ for $[{\mathcal F}]\in {\rm B}_Y$
as in (\ref{evaluateTw}).

In this way, we find the charge transport matrix $T_{\bf w}$.
We present the results for six windows:
\beqa
&&
T_{\bf A}=\left(\begin{array}{cccc}
5&-70&-3&-4\\
0&9&1&1\\
-14&224&11&14\\
14&-259&-15&-18
\end{array}\right),\quad
T_{\bf B}=\left(\begin{array}{cccc}
5&-42&0&-1\\
0&9&1&1\\
-14&224&11&14\\
14&-259&-15&-18
\end{array}\right),
\nn\\
&&
T_{\bf C}=\left(\begin{array}{cccc}
5&0&2&1\\
0&9&1&1\\
-14&28&-3&0\\
14&-63&-1&-4
\end{array}\right),\quad
T_{\bf D}=\left(\begin{array}{cccc}
5&28&5&4\\
0&9&1&1\\
-14&28&-3&0\\
14&-63&-1&-4
\end{array}\right),
\label{RtM}\\
&&T_{{\bf A}(-1)}=\left(\begin{array}{cccc}
75& -280& 1& -4\\
-84& 331& -1& 5\\
-112& 224& -3& 0\\
259& -651& 6& -4
\end{array}\right),\quad
T_{{\bf D}(-1)}=\left(\begin{array}{cccc}
-23&154&1&4\\
14&-103&-1&-3\\
84&-560&-3&-14\\
-133&917&6&24
\end{array}\right).
\nn
\eeqa

The tensor products with ${\mathcal O}_X(1)$ and ${\mathcal O}_Y(1)$
are multiplication by $\e^H$ at the charge level
and are represented with respect to the basis (\ref{basis})
by matrices
\beq
M_X
=\left(\begin{array}{cccc}
1&0&0&0\\
1&1&0&0\\
42&42&1&0\\
-42&-42&1&1
\end{array}\right)
\quad\mbox{and}\quad
M_Y
=\left(\begin{array}{cccc}
1&0&0&0\\
1&1&0&0\\
14&14&1&0\\
-14&-14&1&1
\end{array}\right).
\eeq
As the R-charge shifts by even integers are trivial at the level of charges,
in view of (\ref{shiftGtoP}),
the transport matrices for the shifted windows are related by
\beq
T_{{\bf w}(1)}={M_Y}\cdot T_{\bf w}\cdot {M_X}.
\label{YTXcharge}
\eeq
For ${\bf w} = {\bf A}(-1)$ and ${\bf D}(-1)$, this identity is indeed
satisfied by the above matrices.

\newcommand{\redCone}{{\rm C}}

\subsection{Transport Of D6-Branes}
\label{sub:D6}

Let us consider the transport of D6-branes, especially the trivial line bundle,
from one phase to another.
We first consider the transport from the Grassmannian phase to
the Pfaffian phase.
A lift of the trivial line bundle ${\mathcal O}_X$ is provided by the
$p$-Koszul brane $\Kos$, see (\ref{KostoO}) or (\ref{OXlift}).
The Chan-Paton representation of $\Kos$ consists of
$\C$, $\C(1)^{\oplus 7}$, \ldots, $\C(7)$, spanning an interval of
length 8 at the bottom of the $U(2)$ grid.
This is not grade restricted with respect to any of the windows, since
the grade restricted subsets have length 7 at each height.
Before the transport, $\Kos$ must be replaced by another representative
which is grade restricted with respect to some of the windows.
We must at least trade either $\C$ or $\C(7)$ for
something else so that the length is shortened to seven.
This can be done by binding $\Lb^0$ which is empty in the Grassmannian phase,
with $\C$ and $\C(7)$ at the edge of the bottom line.
See Fig.~\ref{fig:Lemp0} or \ref{fig:L0}.
Let us first consider binding at $\C$.
There exists a homomorphism $\varphi$ from
$\Lb^0$ to $\Kos$ that extends the identity map from
$\C$ of the former to $\C$ of the latter, both at degree zero.
The binding is to take the cone of $\varphi$ and then remove the
$\C$ pair by brane-antibrane annihilation via tachyon condentation of
the identity map. We denote the resulting matrix factorization by
$\redCone(\varphi)$.
Its Chan-Paton representation consists of
the irreducible representations as shown in Fig~\ref{fig:L0transport}-left.
\begin{figure}[htb]
  \centerline{\includegraphics[page=36]{main-pics.pdf}}
  \caption{Representations in $\redCone(\varphi)$
  and $\redCone(\psi[-1])$.}
    \label{fig:L0transport}
\end{figure}
Comparing this with Fig.~\ref{fig:GRRplot},
we see that it is grade restricted with respect to the windows {\bf A}
and {\bf B}.
Thus,  $\redCone(\varphi)$ can be safely transported through
the window {\bf A} or {\bf B} to the Pfaffian phase.
After the transport, we reinstate the $\C$ pair
and bring $\redCone(\varphi)$ back to the cone of
$\varphi:\Lb^0\to \Kos$ before the brane-antibrane annihilation.
In the Pfaffian phase it is $\Kos$ that is empty.
Thus, the cone flows to the same brane as $\Lb^0[1]$ in the infra-red limit.
Therefore, the resulting brane is
$\pi_Y(\Lb^0[1])$ which is isomorphic to
$\textw^{\!2}_0\ck^*(1)$ by (\ref{L0fate}).
Let us next consider binding at $\C(7)$.
There exists a homomorphism $\psi$ from $\Kos$ to a shift of $\Lb^0$ that
extends the identity map from $\C(7)$ of the fomer to $\C(7)$ of the latter.
Since $\C(7)$ is at degree 7 in the former and at degree $11$ in the latter,
the shift of the latter is by four,
$\psi:\Kos\to \Lb^0[4]$.
The binding is to take the cone of $\psi[-1]$
and then remove the $\C(7)$ pair by brane-antibrane annihilation.
We made the shift $\psi\to \psi[-1]$ in order to
keep the $\Kos$ part unshifted in the cone.
We denote the resulting matrix factorization by
$\redCone(\psi[-1])$.
Its Chan-Paton representation consists of
the irreducible representations as shown in Fig~\ref{fig:L0transport}-right.
Comparison with Fig.~\ref{fig:GRRplot} tells us
that it is grade restricted with respect to the windows {\bf C} and {\bf D}.
Thus,  $\redCone(\psi[-1])$ can be safely transported through
the window {\bf C} or {\bf D} to the Pfaffian phase.
After the transport, we can bring $\redCone(\psi[-1])$
back to the cone of $\psi[-1]:\Kos[-1]\to \Lb^0[3]$
before the brane-antibrane annihilation of the $\C(7)$ pair.
As $\Kos$ is empty in this phase,
the cone flows to the same brane as $\Lb^0[3]$ in the infra-red limit.
Therefore, the resulting brane is $\pi_Y(\Lb^0[3])$
which is isomorphic to $\textw^{\!2}_0\ck^*(1)[2]$ by (\ref{L0fate}).
To summarize, we found the transport of ${\mathcal O}_X$ through
the four windows,
\beq
\begin{tikzcd}
 \textw^{\!2}_0\ck^*_Y(1)~~~~
 \arrow[r,leftrightarrow,shorten=-3mm,"\mbox{\small {\bf A},\,\,{\bf B}}"]&
 ~~~~{\mathcal O}_X\\
 \textw^{\!2}_0\ck^*_Y(1)[2]~~~~
 \arrow[r,leftrightarrow,shorten=-3mm,"\mbox{\small {\bf C},\,\,{\bf D}}"]&
~~~~{\mathcal O}_X
\end{tikzcd}
\label{OXtr}
\eeq

Let us next consider the transport from the Pfaffian phase to
the Grassmannian phase.
A lift of the structure sheaf ${\mathcal O}_Y$ is provided by
$\Lb^2(-1)[1]$, see (\ref{L2fate}).
The Chan-Paton representation of $\Lb^2(-1)[1]$ has $\Sym^2\fund(-1)$
and $\Sym^2\fund(6)$ at the upper left and upper right corners
in the grid, see Fig.~\ref{fig:D6mf} or \ref{fig:L2},
and therefore is not grade restricted.
We must trade either of the two representations for something else.
This can be done by binding the brane $\Kos\otimes \Sym^2\fund(-1)$
which is empty in the Pfaffian phase, with the two representations
at the edge.
Let us first consider binding at $\Sym^2\fund(-1)$.
There is a homomorphism $\varphi'$
from $\Lb^2(-1)[1]$ to $\Kos\otimes \Sym^2\fund(-1)[1]$
that extends the identity map from $\Sym^2\fund(-1)$ of the former to
$\Sym^2\fund(-1)$ of the latter, both at degree $-1$.
The binding is to take the cone of $\varphi'[-1]$
and then remove the $\Sym^2\fund(-1)$ pair by brane-antibrane annihilation.
Let $\redCone(\varphi'[-1])$ be the resulting matrix factorization.
Its Chan-Paton representation consists of those in
Fig.~\ref{fig:L2transport}-left.
\begin{figure}[htb]
  \centerline{\includegraphics[page=37]{main-pics.pdf}}
  \caption{Representations in $\redCone(\varphi'[-1])$
  and $\redCone(\psi')$.}
    \label{fig:L2transport}
\end{figure}
Comparison with Fig.~\ref{fig:GRRplot} tells us
that it is grade restricted with respect to the windows
{\bf A} and {\bf C}. Thus, $\redCone(\varphi'[-1])$ can be safely
transported through {\bf A} and {\bf C} to the Grassmannian phase.
After the transport, we can bring $\redCone(\varphi'[-1])$
back to the cone of
$\varphi'[-1]:\Lb^2(-1)\to \Kos\otimes\Sym^2\fund(-1)$
before the brane-antibrane annihilation of the $\Sym^2\fund(-1)$ pair.
As $\Lb^2(-1)$ is empty in the Grassmannian phase, the cone
flows to the same brane as $\Kos\otimes\Sym^2\fund(-1)$
in the infra-red limit. Therefore, the resulting brane is
$\pi_X(\Kos\otimes \Sym^2\fund(-1))$ which is isomporphic to
$\Sym^2 S^*_X(-1)$ by (\ref{KosSymfate}).
Let us next consider binding at $\Sym^2\fund(6)$.
There exists a homomorphism $\psi'$
from a shift of $\Kos\otimes \Sym^2\fund(-1)$
to $\Lb^2(-1)[1]$ that extends the identity map from
$\Sym^2\fund(6)$ of the former to
$\Sym^2\fund(6)$ of the latter.
Since $\Sym^2\fund(6)$ is at degree 7 in the former and
at degree 10 in the latter, the shift of the former is by $-3$,
$\psi':\Kos\otimes\Sym^2\fund(-1)[-3]\to \Lb^2(-1)[1]$.
We take the cone of $\psi'$ and then remove the $\Sym^2\fund(6)$ pair
by brane-antibrane annihilation. We denote the resulting matrix factorization
by $\redCone(\psi')$.
Its Chan-Paton representation consists of the representations as shown in
Fig.~\ref{fig:L2transport}-right.
Comparison with Fig.~\ref{fig:GRRplot} tells us
that it is grade restricted with respect to
the windows {\bf B} and {\bf D}.
Thus, $\redCone(\psi')$ can be safely transported through {\bf B}
and {\bf D} to the Grassmannian phase.
After the transport, we can bring $\redCone(\psi')$
back to the cone of
$\psi':\Kos\otimes\Sym^2\fund(-1)[-3]\to \Lb^2(-1)[1]$
before the brane-antibrane
annihilation of the $\Sym^2\fund(6)$ pair.
As $\Lb^2(-1)$ is empty in the Grassmannian phase, 
the cone flows to the same brane as
$\Kos\otimes\Sym^2\fund(-1)[-2]$ in the infra-red limit.
Therefore, the resulting brane is
$\pi_X(\Kos\otimes \Sym^2\fund(-1)[-2])$ which is isomorphic to
$\Sym^2S^*_X(-1)[-2]$.
To summarize, we found the transport of ${\mathcal O}_Y$ through the four
windows,
\beq
\begin{tikzcd}
 {\mathcal O}_Y~~~~
 \arrow[r,leftrightarrow,shorten=-3mm,"\mbox{\small {\bf A},\,\,{\bf C}}"]&
 ~~~~\Sym^2S_X^*(-1)\\
  {\mathcal O}_Y~~~~
 \arrow[r,leftrightarrow,shorten=-3mm,"\mbox{\small {\bf B},\,\,{\bf D}}"]&
~~~~\Sym^2S_X^*(-1)[-2]
\end{tikzcd}
\label{OYtr}
\eeq

We may also consider the transport of $\ck^*\in D(Y)$ which we
know to lift to the matrix factorization $\Lb^1$, (\ref{L1fate}).
The Chan-Paton representation of $\Lb^1$ has $\fund$ and $\fund(7)$
at the left and the right edges, see Fig.~\ref{fig:Lemp1} or \ref{fig:L1},
and therefore is not grade
restricted with respect to any of the windows.
Trading either of $\fund$ and $\fund(7)$ can be done by binding
the brane $\Kos\otimes\fund$.
Let us first consider  binding at $\fund$.
There is a homomorphism $\varphi''$ from $\Lb^1$ to $\Kos\otimes \fund$
that extends the identity map from the $\fund$ of the former
to the $\fund$ of the latter, both at degree zero.
Take the cone of $\varphi''[-1]$ and then remove the
$\fund$ pair by brane-antibrane annihilation.
Let $\redCone(\varphi''[-1])$ be the resulting matrix factorization.
Its Chan-Paton representation consists of those in
Fig.~\ref{fig:L1transport}-left.
\begin{figure}[htb]
  \centerline{\includegraphics[page=38]{main-pics.pdf}}
  \caption{Representations in $\redCone(\varphi''[-1])$
  and $\redCone(\psi'')$.}
    \label{fig:L1transport}
\end{figure}
Comparison with Fig.~\ref{fig:GRRplot} tells us
that it is grade restricted with respect to the window ${\bf D}(-1)$.
Thus, it can be safely transported through ${\bf D}(-1)$ to the Grassmannian
phase.
After the transport, we can replace
$\redCone(\varphi''[-1])$ by the cone of
$\varphi''[-1]:\Lb^1[-1]\to \Kos\otimes \fund[-1]$
before the brane-antibrane annihilation of the $\fund$ pair.
As $\Lb^1$ is empty in the Grassmannian phase, the cone
flows to the same brane as $\Kos\otimes \fund[-1]$ in the infra-red limit.
Therefore, the resulting brane is
$\pi_X(\Kos\otimes \fund[-1])$ which is isomorphic to $S_X^*[-1]$
by (\ref{KosSymfate}).
Let us next consider binding at $\fund(7)$.
There is a homomorphism $\psi''$ from a shift of $\Kos\otimes \fund$ to
$\Lb^1$ that extends $\fund(6)$ of the former to the $\fund(6)$
of the latter.
Since $\fund(6)$ is at degree 7 in the former and at degree 11 in the latter,
the shift of the former is by $-4$, $\psi'':\Kos\otimes \fund[-4]\to \Lb^1$.
Take its cone, remove the $\fund(6)$ pair by brane-antibrane
annihilation, and denote the result by $\redCone(\psi'')$.
Its Chan-Paton representation consists of those in
Fig.~\ref{fig:L1transport}-right.
Comparison with Fig.~\ref{fig:GRRplot} tells us
that it is grade restricted with respect to the window ${\bf A}$.
Thus, it can be safely transported through ${\bf A}$ to the Grassmannian
phase.
After the transport, we can replace
$\redCone(\psi'')$ by the cone of
$\psi'':\Kos\otimes \fund[-4]\to \Lb^1$ before the brane-antibrane
annihilation of the $\fund(6)$ pair.
As $\Lb^1$ is empty in the Grassmannian phase, the cone
flows to the same brane as $\Kos\otimes \fund[-3]$ in the infra-red limit.
Therefore, the resulting brane is
$\pi_X(\Kos\otimes \fund[-3])$ which is isomorphic to $S_X^*[-3]$.
To summarize, we found the transport of $\ck^*_Y$ through two windows,
\beq
\begin{tikzcd}
 \ck^*_Y~~~~
 \arrow[r,leftrightarrow,shorten=-3mm,"\mbox{\small {\bf A}}"]&
 ~~~~ S_X^*[-3]\\
 \ck^*_Y~~~~
 \arrow[r,leftrightarrow,shorten=-3mm,"\mbox{\small {\bf D}({\bf --}1)}"]&
~~~~ S_X^*[-1]
\end{tikzcd}
\label{D6XY}
\eeq
To find the transport of $\ck^*_Y[1]$  through other windows,
{\bf B}, {\bf C}, {\bf D} additional bindings with empty branes are
needed.

Let us check these results against the transport of D-brane charges
found in Section~\ref{sub:charge}. We note that the charges of
$S_X^*$ and $\Sym^2S_X^*(-1)$ are given by
\beqa
[S_X^*]&=&2[{\mathcal O}_X]+[{\mathcal O}_H]+28[{\mathcal O}_{\ell}]
-35[{\mathcal O}_{\rm pt}],
\label{chSv}\\
{[}\Sym^2S_X^*(-1){]}&=&3[{\mathcal O}_X]-14[{\mathcal O}_{\ell}]
+14[{\mathcal O}_{\rm pt}].
\label{chSym2Sv}
\eeqa
See Appendix~\ref{app:classicalGr}. On the other hand,
those of $\ck_Y^*$ and $\textw^2_0\ck_Y^*(1)$ can be read off from
(\ref{chL1}) and (\ref{chL0}) via
the identification $\pi_Y(\Lb^1)\cong\ck_Y^*$
and $\pi_Y(\Lb^0)\cong \textw^2_0\ck_Y^*(1)[-1]$ as
\beqa
[\ck_Y^*]&=&
4[{\mathcal O}_Y]-2[{\mathcal O}_H]
-14[{\mathcal O}_{\ell}]+21[{\mathcal O}_{\rm pt}],
\label{chck}\\
{[}\textw^{\!2}_0\ck_Y^*(1){]}&=&
5[{\mathcal O}_Y]-14[{\mathcal O}_{\ell}]
+14[{\mathcal O}_{\rm pt}].
\label{chw2ck}
\eeqa
One can confirm that the charge versions of
(\ref{OXtr}), (\ref{OYtr}) and (\ref{D6XY}) indeed hold:
\beqa
\left(\begin{array}{c}5\\0\\-14\\14\end{array}\right)
  &=&T_{\bf w}\left(\begin{array}{c}1\\0\\0\\0\end{array}\right)\qquad
    {\bf w}={\bf A}, {\bf B}, {\bf C}, {\bf D},\\
\left(\begin{array}{c}1\\0\\0\\0\end{array}\right)
  &=&T_{\bf w}\left(\begin{array}{c}3\\0\\-14\\14\end{array}\right)\qquad
    {\bf w}={\bf A}, {\bf B}, {\bf C}, {\bf D},\\
\left(\begin{array}{c}4\\-2\\-14\\21\end{array}\right)  
  &=&T_{\bf w}(-1)\left(\begin{array}{c}2\\1\\28\\-35\end{array}\right)\qquad
    {\bf w}={\bf A}, {\bf D}(-1).
\eeqa

\subsection{Transport Of D0-Branes In $Y$ To $X$}
\label{sub:D0}

Let us transport the D0-branes in the Pfaffian phase to the Grassmannian phase.
The gauge theory brane $\brane_{p_*}$ corresponding to
the D0-brane at a point $p_*\in Y$
is $\brane_{p_*}^{(i_0)}$ constructed in Section~\ref{subsub:D0} with $i_0=-1$.
See Section~\ref{subsec:D0proof}.
We use a parametrization of $p^k$ and $x^a_i$ such that
$p_*=(0,\ldots, 0,1)$ and $A(p_*)$ has the lower $4\times 4$ block
of the form (\ref{Astar}) as the non-zero entries.
We regard $\brane_{p_*}$ as the tensor product of
two matrix factorizations,
\beqa
Q_{1,\ldots, 6}&=&
\sum_{k=1}^6\left(\,p^k\eta_k+A_k(x)\bareta^k\,\right),
\label{QIst}\\
Q_7&=&
\sum_{a=1,2}\left(\,p^7x^a_4\xi_a^4+x^a_5\xi_a^5
+p^7x^a_6\xi_a^6+x^a_7\xi_a^7\,\right),
\eeqa
the former of $\sum_{k=1}^6p^kA_k(x)$ and the latter of $p^7A_7(x)$.
The former is nothing but the Koszul brane $\Kos_{I_*}$
for the components in $I_*:=\{1,\ldots, 6\}$, see (\ref{KosI}).
The latter is $\brane^{SU(2)}_{3,7}$ modified by multiplying
$p^7$ at certain positions, and is denoted by $\brane_7$.
Thus, we have
\beq
\brane_{p_*}=\Kos_{I_*}\otimes \brane_7.
\label{D0tensor}
\eeq
By the charge assignment (\ref{D0charge1}), (\ref{D0charge2}) and
(\ref{D0charge3}) of the Clifford generators and the Clifford vacuum
$|0\rangle$, with $i_0=-1$, the Chan-Paton vector space
is the tensor product of the following two,
\beqa
\bV_{I_*}&=&\mathop{\C}_{{}^{}_0}\oplus \mathop{\C(1)^{\oplus 6}}_{{}_1}\oplus
\mathop{\C(2)^{\oplus{6\choose 2}}}_{{}_2}
\oplus\cdots\oplus\mathop{\C(5)^{\oplus 6}}_{{}_5}\oplus
\mathop{\C(6)}_{{}_6},\\
\bV_7&=&\Bigl(\,\mathop{\C}_{{}^{}_0}\oplus \mathop{\fund}_{{}^{}_1}\oplus
\mathop{\C(1)}_{{}_2}\,\Bigr)^{\otimes 2}[1]
\nn\\
&=&\mathop{\C}_{{}^{}_{-1}}\oplus \mathop{\fund^{\oplus 2}_{}}_{{}_0}\oplus
\mathop{\C(1)^{\oplus 3}}_{{}_1}\oplus
\mathop{\Sym^2\fund}_{{}^{}_1}\oplus \mathop{\fund(1)^{\oplus 2}_{}}_{{}_2}
\oplus \mathop{\C(2)}_{{}_3}.
\eeqa
We plot the components of $\bV_{I_*}\otimes \bV_7$ in the $U(2)$ grid
in Fig.~\ref{fig:D0plot}.
\begin{figure}[htb]
  \centerline{\includegraphics[page=39]{main-pics.pdf}}
  \caption{The representations in $\brane_{p_*}$}
    \label{fig:D0plot}
\end{figure}
We see that it is not grade restricted with respect to any of
the four windows {\bf A}, {\bf B}, {\bf C}, {\bf D} or their shifts.
Thus, we need to change the representative of the brane before crossing
the phase boundary.
As a preparation, let us describe $\brane_7$ in the following way:
\beq
\begin{tikzcd}
  &&\arrow[dl,shorten=-0.0cm,yshift=-0.5ex,"\!\!\!(\nwarrow)_c"]
  \,\displaystyle\mathop{\C}_{\scriptscriptstyle -1}
  \\[-0.3cm]
  &\arrow[dl,shorten=-0.1cm,yshift=-0.2ex,"\!\!\!(\nwarrow)_a"]
  \displaystyle\mathop{\fund^*(1)^{\oplus 2}}_{\scriptscriptstyle 0}
  \arrow[ur,shorten=-0.0cm,yshift=0.5ex,"p^7(\searrow)_c\!\!\!"]
  \arrow[dr,shorten=-0.2cm,yshift=0.8ex,"\!\!\!(\searrow)_d"]&\\[-0.2cm]
  \displaystyle\mathop{\Sym^2\fund^*(2)}_{\scriptscriptstyle 1}
  \arrow[ur,shorten=-0.1cm,yshift=0.8ex,"p^7(\searrow)_a\!\!\!"]
  \arrow[dr,shorten=-0.25cm,xshift=-1.5ex,yshift=0.5ex,"\!\!\!(\searrow)_b"]&&
  \arrow[ul,shorten=-0.2cm,yshift=-0.2ex,"p^7(\nwarrow)_d\!\!\!\!\!"]
  \arrow[dl,shorten=-0.2cm,xshift=1.5ex,yshift=-0.5ex,"\!\!\!(\nwarrow)_e"]
  ~~\displaystyle\mathop{\C(1)^{\oplus 3}}_{\scriptscriptstyle 1}\\[-0.2cm]
   &\arrow[ul,shorten=-0.25cm,xshift=-1.5ex,yshift=-0.5ex,"p^7(\nwarrow)_b\!\!\!\!\!"]
   \displaystyle\mathop{\fund^*(2)^{\oplus 2}}_{\scriptscriptstyle 2}
   \arrow[ur,shorten=-0.2cm,xshift=1.5ex,yshift=0.5ex,"p^7(\searrow)_e\!\!\!"]
   \arrow[dr,shorten=-0.25cm,yshift=0.5ex,"\!\!\!(\searrow)_f"]
    &\\[-0.3cm]
   &&\arrow[ul,shorten=-0.25cm,yshift=-0.5ex,"p^7(\nwarrow)_f\!\!\!\!\!"]
   ~\displaystyle\mathop{\C(2)}_{\scriptscriptstyle 3}
\end{tikzcd}
\label{brane7}
\eeq
It is presented as a splitting of the $SU(2)$ brane (\ref{massmfleft})
\beq
\brane^{SU(2)}_{3,7}:~
\begin{tikzcd}
  \Sym^2\sfund^*\arrow[r,yshift=0.5ex,"(\searrow)"]&
  \arrow[l,yshift=-0.5ex,"(\nwarrow)"]\sfund^*\otimes \cktt
  \arrow[r,yshift=0.5ex,"(\searrow)"]&
  \arrow[l,yshift=-0.5ex,"(\nwarrow)"]\C\otimes\textw^{\!2}_0\cktt
\end{tikzcd}
\eeq
with respect to the charge under the
$U(1)=U(2)/SU(2)$ and the position of $p^7$.
Recall that $(\searrow)$ and $(\nwarrow)$ originate from the arrows
(\ref{Lse}) and (\ref{Lnw}) in the Lascoux matrix factorizations.
(See the description below (\ref{massmfleft}).)
We can employ the Koszul brane for the 7${}^{\rm th}$ component
\beq
\Kos_7:~~
\begin{tikzcd}
  \displaystyle\mathop{\C}_{{}^{}_0}
  \arrow[r,yshift=0.5ex,"A_7(x)"]&\arrow[l,yshift=-0.5ex,"p^7"]
  \displaystyle\mathop{\C(1)}_{{}_1}
\end{tikzcd}
\eeq
or its shifts and tensor product with other representations to convert
a component $\rmV$ in $\brane_7$ to $\rmV(\pm 1)$
and to swap the position of $p^7$ in the attached arrows at the same time.
For example, there is a homomorphism from $\brane_7$ to $\Kos_7[1]$
which maps $\C{}_{{}_{-1}}$ to
$\C{}_{{}_{-1}}$ by the identity map
and $\fund^*(1)^{\oplus 2}{}_{{}^{}_0}$ to
$\C(1){}_{{}^{}_0}$ by $(\searrow)_c$.
Taking the cone and annihilating the
$\C{}_{{}_{-1}}$-pair by tachyon condentation of the identity map,
we obtain a different matrix factorization $\brane_7'$ where $\C{}_{{}_{-1}}$
is replaced by $\C(1){}_{{}^{}_1}$ and the position of $p^7$
is swapped from the right arrow $(\searrow)_c$ to the left arrow
$(\nwarrow)_c$.
This process $\brane_7\mapsto\brane_7'$ does not change the low energy image
in the Pfaffian phase provided it is done inside (\ref{D0tensor}),
i.e. under the tensor product with $\Kos_{I_*}$.
This is because the tensor product of $\Kos_{I_*}$ and $\Kos_7$
is nothing but the full $p$-Koszul brane,
\beq
\Kos_{I_*}\otimes\Kos_7=\Kos,
\label{Kostens7}
\eeq
which is empty
in the Pfaffian phase (\ref{Kosempty}).
We can also make an overall shift of the $U(1)$ charge by $\pm 1$
without changing the positions of $p^7$
but with a shift of the R-charge by $\mp 2$.
To show this, suppose we have a matrix factorization
of $p^7A_7(x)$ which is at most linear in $p^7$,
\beq
\brane:~
\begin{tikzcd}
  M^{\rm ev}~~\arrow[r,shorten=-3mm,yshift=0.5ex,"f_1+p^7f_2"]&
  \arrow[l,shorten=-3mm,yshift=-0.5ex,"p^7g_1+g_2"]~~M^{\rm od}
\end{tikzcd}
\eeq
with
\beqa
&g_1f_1+g_2f_2=A_7(x){\rm id}_{M^{\rm ev}},\quad
f_1g_1+f_2g_2=A_7(x){\rm id}_{M^{\rm od}},\nn\\
&f_ig_j=g_jf_i=0,\quad i\ne j.\nn
\eeqa
Consider another brane
\beq
\brane':~
\begin{tikzcd}
  M^{\rm ev}(-1)[1]~~\arrow[r,shorten=-3mm,yshift=0.5ex,"-f_1"]
  \arrow[d,shorten=-1mm,xshift=0.5ex,"A_7(x)"]&
  \arrow[l,shorten=-3mm,yshift=-0.5ex,"-g_2"]
  \arrow[d,shorten=-1mm,xshift=0.5ex,"A_7(x)"]~~M^{\rm od}(-1)[1]\\[0.3cm]
  M^{\rm ev}~~\arrow[u,shorten=-1mm,xshift=-0.5ex,"p^7"]
  \arrow[r,shorten=-3mm,yshift=0.5ex,"f_1"]&
  \arrow[l,shorten=-3mm,yshift=-0.5ex,"g_2"]
  \arrow[u,shorten=-1mm,xshift=-0.5ex,"p^7"]~~M^{\rm od}
\end{tikzcd}
\eeq
Note that this is $\Kos_7$ tensored with
the complex obtained from $\brane(-1)[1]$ by setting $p^7=0$,
and hence $\Kos_{I_*}\otimes\brane'$
is empty in the Pfaffian phase, by
(\ref{Kosempty})-(\ref{Kosvarempty}) and  (\ref{Kostens7}).
There is a homomorphism from $\brane'$ to $\brane$ consisting of
the identity $M^{\rm ev/od}\to M^{\rm ev/od}$,
$f_2:M^{\rm ev}(-1)[1]\to M^{\rm od}$ and
$g_1:M^{\rm od}(-1)[1]\to M^{\rm ev}$.
Taking the cone and annihilating the $M^{\rm ev}$ and
$M^{\rm od}$ pairs by the tachyon condensation of the identity maps,
we obtain
\beq
\begin{tikzcd}
  M^{\rm ev}(-1)[2]~~\arrow[r,shorten=-3mm,yshift=0.5ex,"f_1+p^7f_2"]&
  \arrow[l,shorten=-3mm,yshift=-0.5ex,"p^7g_1+g_2"]~~M^{\rm od}(-1)[2]
\end{tikzcd}
\eeq
which is nothing but $\brane(-1)[2]$.
Since $\Kos_{I_*}\otimes \brane'$ is empty in the Pfaffian phase,
we can make a replacement
\beq
\Kos_{I_*}\otimes \brane
~\,\longmapsto ~\,
\Kos_{I_*}\otimes \brane(-1)[2]
\eeq
without changing the low energy limit.
That is, they flow to the same brane in the Pfaffian Calabi-Yau.
At this point, we recall the information (\ref{shiftPf})
from the weakly coupled sector.
Combining the two, we find
\beq
\pi_Y(\Kos_{I_*}\otimes \brane)\cong
\pi_Y(\Kos_{I_*}\otimes \brane(-1)[2])
\cong\pi_Y(\Kos_{I_*}\otimes \brane)(1).
\label{Pfrelations}
\eeq
This is indeed the property of a brane supported at a point ---
it is invariant under tensor products with line bundles,
${\mathcal O}_{p_*}={\mathcal O}_{p_*}(1)$.

In this way, the brane $\brane_{p_*}=\Kos_{I_*}\otimes \brane_7$
can be replaced by other representatives that can go through the windows.
The one that can go through {\bf w} is
$\Kos_{I_*}\otimes \brane_7^{\bf w}$ where
\beqa
\brane_7^{\bf A}:&&
\begin{tikzcd}
  \displaystyle\mathop{\Sym^2\fund^*(2)}_{{}_1}
  \arrow[r,yshift=0.5ex,"p^7(\searrow)"]&
  \arrow[l,yshift=-0.5ex,"(\nwarrow)"]
 \displaystyle \mathop{\fund^*(1)\otimes \cktt}_{{}_0}
  \arrow[r,yshift=0.5ex,"(\searrow)"]&
  \arrow[l,yshift=-0.5ex,"p^7(\nwarrow)"]
  \displaystyle\mathop{\C(1)\otimes\textw^{\!2}_0\cktt}_{{}_1}
\end{tikzcd}
\label{braneD0A}\\
\brane_7^{\bf B}:&&
\begin{tikzcd}
  \displaystyle\mathop{\Sym^2\fund^*(1)}_{{}_{-1}}
  \arrow[r,yshift=0.5ex,"(\searrow)"]&
  \arrow[l,yshift=-0.5ex,"p^7(\nwarrow)"]
 \displaystyle \mathop{\fund^*(1)\otimes \cktt}_{{}_0}
  \arrow[r,yshift=0.5ex,"(\searrow)"]&
  \arrow[l,yshift=-0.5ex,"p^7(\nwarrow)"]
  \displaystyle\mathop{\C(1)\otimes\textw^{\!2}_0\cktt}_{{}_1}
\end{tikzcd}
\label{braneD0B}\\
\brane_7^{\bf C}:&&
\begin{tikzcd}
  \displaystyle\mathop{\Sym^2\fund^*(2)}_{{}_1}
  \arrow[r,yshift=0.5ex,"p^7(\searrow)"]&
  \arrow[l,yshift=-0.5ex,"(\nwarrow)"]
 \displaystyle \mathop{\fund^*(1)\otimes \cktt}_{{}_0}
  \arrow[r,yshift=0.5ex,"p^7(\searrow)"]&
  \arrow[l,yshift=-0.5ex,"(\nwarrow)"]
  \displaystyle\mathop{\C\otimes\textw^{\!2}_0\cktt}_{{}_{-1}}
\end{tikzcd}
\label{braneD0C}\\
\brane_7^{\bf D}:&&
\begin{tikzcd}
  \displaystyle\mathop{\Sym^2\fund^*(1)}_{{}_{-1}}
  \arrow[r,yshift=0.5ex,"(\searrow)"]&
  \arrow[l,yshift=-0.5ex,"p^7(\nwarrow)"]
 \displaystyle \mathop{\fund^*(1)\otimes \cktt}_{{}_0}
  \arrow[r,yshift=0.5ex,"p^7(\searrow)"]&
  \arrow[l,yshift=-0.5ex,"(\nwarrow)"]
  \displaystyle\mathop{\C\otimes\textw^{\!2}_0\cktt}_{{}_{-1}}
\end{tikzcd}
\label{braneD0D}
\eeqa
and $\brane_7^{{\bf w}(i)}=\brane_7^{\bf w}(-i)[2i]$.

\noindent
{\bf Remark.}\label{page:Rmk}
As in most other places, the R-charges above are in
the Grassmannian frame (\ref{Grframe}).
R-charges in the Pfaffian frame are obtained from (\ref{RPf}).
For example, for
$\Kos_{I_*}\otimes \brane_7^{{\bf B}(1)}$, it is the tensor product of
\beq
\bigl(\Kos_{I_*}\bigr)^{\rm Pf}:
~~\begin{tikzcd}
  \displaystyle\mathop{\C}_{{}_0}\arrow[r,yshift=0.5ex, "A_{I_*}(x)"] &
  \arrow[l,yshift=-0.5ex, "p^{I_*}"]
  \displaystyle\mathop{\C(1)^{\oplus 6}}_{{}_{-1}}
  \arrow[r,yshift=0.5ex, "A_{I_*}(x)"]
  & \arrow[l,yshift=-0.5ex, "p^{I_*}"]
  ~~\cdots~~\arrow[r,yshift=0.5ex, "A_{I_*}(x)"]
  &\arrow[l,yshift=-0.5ex, "p^{I_*}"]
  \displaystyle\mathop{\C(5)^{\oplus 6}}_{{}_{-5}}
  \arrow[r,yshift=0.5ex, "A_{I_*}(x)"]
  &\arrow[l,yshift=-0.5ex, "p^{I_*}"]
  \displaystyle\mathop{\C(6)}_{{}_{-6}}
\end{tikzcd},
\eeq
where $p^{I_*}$ and $A_{I_*}(x)$ are the two terms of (\ref{QIst}), and
\beq
\bigl(\brane_7^{{\bf B}(1)}\bigr)^{\rm Pf}:~~
\begin{tikzcd}
  \displaystyle\mathop{\Sym^2\fund^*}_{{}_{-1}}
  \arrow[r,yshift=0.5ex,"(\searrow)"]&
  \arrow[l,yshift=-0.5ex,"p^7(\nwarrow)"]
 \displaystyle \mathop{\fund^*\otimes \cktt}_{{}_{-1}}
  \arrow[r,yshift=0.5ex,"(\searrow)"]&
  \arrow[l,yshift=-0.5ex,"p^7(\nwarrow)"]
  \displaystyle\mathop{\C\otimes\textw^{\!2}_0\cktt}_{{}_{-1}}
\end{tikzcd}
\label{B7Bp}
\eeq
This is used in (\ref{braneOpst}).

After entering the Grassmannian phase, we just look at their low
energy limits, applying the map $\pi_X:D_G(V,W)\to D(X)$
discussed in Section~\ref{subsec:Gflow}. This gives
the transport of the D0-brane at $p_*$,
\beq
\PtoG_{\bf w}({\mathcal O}_{p_*})
=\pi_X(\Kos_{I_*}\otimes \brane_7^{\bf w}).
\eeq
To evaluate this, we can use the formula (\ref{appli}),
$\pi_X(\Kos_{I_*}\otimes \brane)=\pi^{(7)}_X(\brane)$,
for a matrix factorization $\brane$ of $p^7A_7(x)$.
For $\brane$ as in
(\ref{braneD0A})-(\ref{braneD0D}),
$\pi^{(7)}_X(\brane)$ is a truncation from the left
of the infinitely long complex
\beq
\cdots
\stackrel{g_{-1}}\longrightarrow\,
S_X\!\otimes \!\cktt\stackrel{f_0}{\longrightarrow}
\!\begin{array}{c}
  {\mathcal O}_X\!\otimes\!\textw^{\!2}_0\cktt\\
  \oplus\\
  \Sym^2 S_X(1)
\end{array}
\!\stackrel{g_0}{\longrightarrow}\,
S_X(1)\!\otimes\! \cktt
\stackrel{f_1}{\longrightarrow}
\!\!\begin{array}{c}
  {\mathcal O}_X(1)\!\otimes\!\textw^{\!2}_0\cktt\\
  \oplus\\
  \Sym^2 S_X(2)
\end{array}
\!\stackrel{g_1}{\longrightarrow}
\cdots
\label{longcomplex}
\eeq
where the differentials are
\beq
f_i=\left(\begin{array}{c}(\searrow)\\
  (\nwarrow)\end{array}\right)
\quad\mbox{and}\quad g_i=\Bigl(\,(\nwarrow),(\searrow)\,\Bigr).
\eeq
The complex (\ref{longcomplex}) is exact. To see this, note that
the differential is basically that of
$\brane^{SU(2)}_{3,7}$, that is, $Q=\sum x^a_{\mu}\xi_a^{\mu}$ in (\ref{QSU2}).
As $\{Q,Q^{\dagger}\}=\sum |x^a_{\mu}|^2\cdot{\rm id}$,
it fails to be exact only on the locus $(x^a_{\mu})=0$, i.e.
$G(2,3)\subset G(2,7)$ which misses $X$.
Thus, the truncation is indeed exact at positions far enough to the right.
Cutting off an exact part, we obtain a finite length complex ---
in fact of one or two terms only:
\beqa
\PtoG_{\bf A}({\mathcal O}_{p_*})&=&
     {\rm Ker}(f_1)
\label{TAD0}
\\[0.2cm]
\PtoG_{\bf B}({\mathcal O}_{p_*})&=&
\mathop{\Sym^2S_X(1)}_{{}_{-1}}
\stackrel{(\searrow)}{\longrightarrow}\mathop{{\rm Ker} (f_1)}_{{}_0}
\label{TBD0}
\\[0.2cm]
\PtoG_{\bf C}({\mathcal O}_{p_*})&=&
\mathop{{\mathcal O}_X\otimes \textw^{\!2}_0\cktt}_{{}_{-1}}
\stackrel{(\nwarrow)}{\longrightarrow}\mathop{{\rm Ker}(f_1)}_{{}_0}
\label{TCD0}
\\[0.2cm]
\PtoG_{\bf D}({\mathcal O}_{p_*})&=&
   {\rm Ker}(g_0)[1]
\label{TDD0}
\eeqa
An alternative description of these branes exists.
By the exactness of (\ref{longcomplex}),
a truncation from the left is isomorphic to its complement
shifted to the right by one, which is unbounded to the left
but is exact at positions far enough to the left.
Cutting off an exact part, we have
\beqa
\PtoG_{\bf A}({\mathcal O}_{p_*})&=&
   {\rm Coker}(f_0)
\label{TAD0a}
\\[0.2cm]
\PtoG_{\bf B}({\mathcal O}_{p_*})&=&
\mathop{{\rm Coker}(g_{-1})}_{{}_{-1}}
  \stackrel{(\searrow)}{\longrightarrow}
  \mathop{{\mathcal O}_X\otimes \textw^{\!2}_0\cktt}_{{}_0}
\label{TBD0a}
\\[0.2cm]
\PtoG_{\bf C}({\mathcal O}_{p_*})&=&
\mathop{{\rm Coker}(g_{-1})}_{{}_{-1}}
\stackrel{(\nwarrow)}{\longrightarrow}\mathop{\Sym^2S_X(1)}_{{}_0}
\label{TCD0a}
\\[0.2cm]
\PtoG_{\bf D}({\mathcal O}_{p_*})&=&
   {\rm Coker}(g_{-1})[1]
\label{TDD0a}
\eeqa
At this point, we note that the kernels and cokernels
of the differentials of (\ref{longcomplex}), $f_i$ and $g_i$,
are all vector bundles. See Appendix~\ref{subsec:locfree}.
Thus the above are all (complexes of) vector bundles.

There is a duality relation among these, determined by
an isomorphism $\cktt\cong \cktt^*$ provided by
the symplectic form $A(p_*)$. It comes from
the self-duality of (\ref{longcomplex}) ---
its transpose is isomorphic to itself,
with $f_i^T\simeq g_{-i}$ and $g_i^T\simeq f_{-i}$.
This is natural from the origin (\ref{QSU2}) of the differential, 
and can be proven by combing the duality
among the Lascoux complexes mentioned at the end of
Section~\ref{subsub:otherversion} and the isomorphisms
(\ref{iso4v}). This implies
\beqa
&&{\rm Coker}(f_0)\cong {\rm Ker}(g_0)^*,\quad
    {\rm Coker}(g_{-1})\cong {\rm Ker}(f_1)^*\\
    &&\mathop{{\rm Coker}(g_{-1})}_{{}_0}\stackrel{(\nwarrow)}{\longrightarrow}
      \mathop{\Sym^2S_X(1)}_{{}_1}~\,\cong~\,
      \Bigl(\mathop{\Sym^2S_X(1)}_{{}_{-1}}
      \stackrel{(\searrow)}{\longrightarrow}\mathop{{\rm Ker}(f_1)}_{{}_0}
      \Bigr)^T
\eeqa
These yield relations among the four,
\beq
\PtoG_{\bf w}({\mathcal O}_{p_*})
\cong \mathfrak{D}(\PtoG_{\bf w'}({\mathcal O}_{p_*}))[1],
\label{D0duality}
\eeq
with ${\bf A}'={\bf D}$, ${\bf B}'={\bf C}$, ${\bf C}'={\bf B}$
and ${\bf D}'={\bf A}$,
where
$\mathfrak{D}({\mathcal E})$ for ${\mathcal E}\in D(X)$
is the transpose or the dual of
a complex of vector bundles that realizes ${\mathcal E}$.

Let us compute the charges of the resulting branes
$\PtoG_{\bf w}({\mathcal O}_{p_*})$, applying the Chern character
formula (\ref{chformula}).
For a power series  $f(\usigma_1,\usigma_2)$ of two variables, we shall denote
\beq
f|_{\Gr}:=f\left({H_1\over 2\pi},{H_2\over 2\pi}\right)
\eeq
where $H_1,H_2$ are the Chern roots of the dual tautological
subbundle $S^*$ on $\Gr$.
Then, the identity (\ref{chvsbf}) reads
$\ch({\mathcal O}_{\Gr}(\bV))=f_{\brane}|_{\Gr}$.
As $f_{\Kos_{I_*}}|_{\Gr}=(1-\e^H)^6$, the formula (\ref{chformula})
for $\brane=\Kos_{I_*}\otimes \brane^{\bf w}_7$ reads
\beq
(1-\e^H)^7\,\ch\bigl(\PtoG_{\bf w}({\mathcal O}_{p_*})\bigr)
\,=\,(1-\e^H)^6\,f_{\brane^{\bf w}_7}|_{\Gr}.
\label{chforD0}
\eeq
We find from (\ref{braneD0A})-(\ref{braneD0D})
\beqa
f_{\brane^{\bf A}_7}&=&-f_{\Sym^2\fund}+4f_{\fund}-5f_{\C(1)}
\,=\,4(1-u)-\delta^2,\\
f_{\brane^{\bf B}_7}&=&-f_{\Sym^2\fund(-1)}+4f_{\fund}-5f_{\C(1)}
\,=\,(1-u)u^{-1}(5u-4-4\delta)-u^{-1}\delta^2,\\
f_{\brane^{\bf C}_7}&=&-f_{\Sym^2\fund}+4f_{\fund(1)}-5f_{\C}
\,=\,-(1-u)-\delta^2,\\
f_{\brane^{\bf D}_7}&=&-f_{\Sym^2\fund(-1)}+4f_{\fund(1)}-5f_{\C}
\,=\,-4(1-u)u^{-1}(1+\delta)-u^{-1}\delta^2,
\eeqa
where 
$f_{\C}=1$, $f_{\fund}=u_1+u_2$, $f_{\Sym^2\fund}=u_1^2+u_1u_2+u_2^2$
and $f_{\rmV(i)}=(u_1u_2)^if_{\rmV}$ with $u_a:=\e^{2\pi\usigma_a}$
in the middle expressions,
and the results are written in terms of
$u:=u_1u_1$ and $\delta:=u_1+u_2-2$ on the right-hand sides.
In terms of $H_1+H_2=H$ and $H_1H_2=\sigma_{1,1}$ (a Schubert class), we have
\beq
u|_{\Gr}=\e^H,\quad
\delta|_{\Gr}=H+{1\over 2}H^2-\sigma_{1,1}+{1\over 6}H^3-{1\over 2}H\sigma_{1,1}
+\cdots
\eeq
where $\cdots$ are terms of degree 8 and higher.
We may use the ring relations in ${\rm H}^*(\Gr,\Q)$,
\beq
H^7\sigma_{1,1}=H^7\cdot {1\over 3}H^2,\qquad
H^6\sigma_{1,1}^2=H^7\cdot {5\over 42}H^3.
\label{cohrelGr}
\eeq
Then, we have
\beq
(1-\e^H)^6\,\delta^2|_{\Gr}=(1-e^H)^7\left(-H+{H^2\over 6}+{H^3\over 21}\right),
\eeq
and the right-hand side of (\ref{chforD0})
can be written as $(1-\e^H)^7$ times a polynomial in $H$.
This gives
\beqa
{[}\PtoG_{\bf A}({\mathcal O}_{p_*})]\!&=&\!
4+H-{H^2\over 6}-{H^3\over 21}\,=\,
4[{\mathcal O}_X]+[{\mathcal O}_H]+14[{\mathcal O}_{\ell}]-23[{\mathcal O}_{\rm pt}],\label{chAD0}\\
{[}\PtoG_{\bf B}({\mathcal O}_{p_*})]\!&=&\!
1+H+{H^2\over 6}-{H^3\over 21}\,=\,
[{\mathcal O}_X]+[{\mathcal O}_H]+28[{\mathcal O}_{\ell}]-37[{\mathcal O}_{\rm pt}],\\
{[}\PtoG_{\bf C}({\mathcal O}_{p_*})]\!&=&\!
-1+H-{H^2\over 6}-{H^3\over 21}\,=\,
-[{\mathcal O}_X]+[{\mathcal O}_H]+14[{\mathcal O}_{\ell}]-23[{\mathcal O}_{\rm pt}],\\
{[}\PtoG_{\bf D}({\mathcal O}_{p_*})]\!&=&\!
-4+H+{H^2\over 6}-{H^3\over 21}\,=\,
-4[{\mathcal O}_X]+[{\mathcal O}_H]+28[{\mathcal O}_{\ell}]-37[{\mathcal O}_{\rm pt}].
\eeqa
The result agrees with the charge transportation rule of
Section~\ref{sub:charge}:
the coefficients match the entries of $T_{\bf w}^{-1}(0,0,0,1)^T$ 
for ${\bf w}={\bf A}, {\bf B}, {\bf C}, {\bf D}$.

\medskip
\noindent
{\bf Comparison with Borisov-C\u ald\u araru}

The work \cite{BC} by Borisov and C\u ald\u araru
provides an equivalence $D(Y)\to D(X)$ by giving the images of the
skyscraper sheaves of points of $Y$.
So, it is appropriate at this point to compare our result with that work.
The image of ${\mathcal O}_p$ for $p\in Y$ under the equivalence of
\cite{BC} is the ideal sheaf ${\mathcal I}_{C_p}$ of a curve
$C_p\subset X$,
\beq
0\longrightarrow {\mathcal I}_{C_p}\longrightarrow {\mathcal O}_X
\longrightarrow {\mathcal O}_{C_p}\longrightarrow 0
\eeq
$C_p$ as a subvariety of $\Gr=G(2,V)$
is defined as the intersection of $X$ and the
$(3,0)$ Schubert cycle determined by $K_p:={\rm Ker}\,A(p)$,
\beq
\sigma_p:=\{T\subset V|\,\dim T=2,~T\cap K_p\ne \{0\}\,\}.
\eeq
Among the seven equations defining $X$, one of them,
$A_p(x):=\sum_{k=1}^7p^kA_k(x)=0$, is trivially satisfied on
$\sigma_p$. That is, $\sigma_p\subset H_p:=\{A_p=0\}$.
If the seven functions are rearranged as
$A_1^{(p)},\ldots, A^{(p)}_6,A_p$,  then $C_p$
is the intersection of $L^{(p)}:=\{A^{(p)}_1=\cdots=A^{(p)}_6=0\}$ and 
$\sigma_p$ with the expected codimension $6+3=9$.
And it is indeed a curve \cite{BC}.
Let us try to find a gauge theory lift of ${\mathcal I}_{C_p}$.
As the pushforward by a closed embedding is exact, we have
\beq
    i_{X*}{\mathcal I}_{C_p}\,\,\cong\,\,
    \left[\,\,\mathop{i_{X*}{\mathcal O}_X}_{{}_0}\longrightarrow
    \mathop{i_{X_*}{\mathcal O}_{C_p}}_{{}_1}\,\,\right]
    \,\,=\,\,
      \left[\,\,\mathop{{\mathcal O}_{L^{(p)}\cap H_p}}_{{}_0}\longrightarrow
      \mathop{{\mathcal O}_{L^{(p)}\cap \sigma_p}}_{{}_1}\,\,\right].
      \label{ICp1}
\eeq
The sheaves ${\mathcal O}_{L^{(p)}}$ and ${\mathcal O}_{H_p}$ on $\Gr$
have Koszul resolutions by the functions $A^{(p)}(x)$ and $A_p(x)$, while
${\mathcal O}_{\sigma_p}$ has a resolution by
the Eagon-Northcott complex
\beq
\begin{tikzcd}
  \Sym^2 S(-1)\otimes \textw^{\!4} K_p^{\perp}\arrow[r,"(\searrow)"]&
  S(-1)\otimes\textw^{\!3}K_p^{\perp}\arrow[r,"(\searrow)"]&
  {\mathcal O}_{\Gr}(-1)\textw^{\!2} K_p^{\perp}\arrow[r,"(\rightarrow)"]&
  {\mathcal O}_{\Gr}
\end{tikzcd}
\eeq
where $K_p^{\perp}$ is the subspace of $V^*$
orthogonal to $K_p$ and the arrows are those of the Lascoux complex.
Note that the homomorphism
${\mathcal O}_{H_p}\to {\mathcal O}_{\sigma_p}$
is extended to a homomorphism of the resolutions
\beq
\begin{tikzcd}
  &&{\mathcal O}_{\Gr}(-1)\arrow[d,"A(p)"]\arrow[r,"A_p(x)"]&
  {\mathcal O}_{\Gr}\arrow[d,"{\rm id}"]\\
  \Sym^2 S(-1)\otimes \textw^{\!4} K_p^{\perp}\arrow[r,"(\searrow)"]&
  S(-1)\otimes\textw^{\!3}K_p^{\perp}\arrow[r,"(\searrow)"]&
  {\mathcal O}_{\Gr}(-1)\otimes\textw^{\!2} K_p^{\perp}\arrow[r,"(\rightarrow)"]&
  {\mathcal O}_{\Gr}
\end{tikzcd}
\label{hom11}
\eeq
The complex $\mathop{{\mathcal O}_{H_p}}_{{}_0}\to
\mathop{{\mathcal O}_{\sigma_p}}_{{}_1}$ has a resolution by the cone of
(\ref{hom11}) shifted to the right by one.
Annihilating the ${\mathcal O}_{\Gr}$ and ${\mathcal O}_{\Gr}(-1)$ pairs
by tachyon condensation, we see that it is isomorphic to
\beq
\begin{tikzcd}
  \displaystyle\mathop{\Sym^2 S(-1)\otimes \textw^{\!4} K_p^{\perp}}_{{}_{-2}}
  \arrow[r,"(\searrow)"]&
  \displaystyle\mathop{S(-1)\otimes\textw^{\!3}K_p^{\perp}}_{{}_{-1}}
  \arrow[r,"(\searrow)"]&
  \displaystyle\mathop{{\mathcal O}_{\Gr}(-1)
    \otimes\textw^{\!2} K_p^{\perp}/\C A(p)}_{{}_0}
\end{tikzcd}
\label{cone11}
\eeq
Via (\ref{ICp1}),
$i_{X*}{\mathcal I}_{C_p}$ is isomorphic to
the tensor product of the Koszul resolution
of ${\mathcal O}_{L^{(p)}}$ and this complex (\ref{cone11}).
This provides a resolution of $i_{X*}{\mathcal I}_{C_p}$
by homogeneous vector bundles on $\Gr$.
We would like to find a matrix factorization completion
of the associated complex of representations over $\C[x]$ of the gauge group.
At this stage, we set
$p=p_*$ and use our parametrization of $p^k$ and $x^a_i$,
in which $A_{p_*}=A_7$ and we may take $A^{(p_*)}_k=A_k$
for $k=1,\ldots, 6$. Then,
a matrix factorization completion of the Koszul resolution
of ${\mathcal O}_{L^{(p_*)}}$ is given by $\Kos_{I_*}(-6)[6]$.
Concerning the complex (\ref{cone11}), we find a similarity with
$\brane_7^{\bf B}$ given in (\ref{braneD0B}).
Note that $K_{p_*}^{\perp}$ is identified with $\cktt^*$ and hence
the tensor product with $\textw^{\!4\,}\cktt\cong \C$ gives isomorphisms
\beq
\textw^{\!4}K_{p_*}^{\perp}\cong \C,\quad
\textw^{\!3}K_{p_*}^{\perp}\cong \cktt,\quad
\textw^{\!2}K_{p_*}^{\perp}/\C A(p_*)\cong \textw^{\!2}_0\cktt.
\eeq
We see that $\brane_7^{\bf B}(-2)[1]$ is nothing but
a matrix factorization completion of (\ref{cone11}).
Thus, a desired completion 
$\widehat{i_{X*}{\mathcal I}_{C_{p_*}}}^{\!\!\rm mf}$
is the tensor product of the two, which is
$\Kos_{I_*}\otimes \brane_7^{\bf B}(-8)[7]$.
By (\ref{alift}),
a gauge theory lift of ${\mathcal I}_{C_{p_*}}$ is given by
\beq
\brane_{i_{X*}{\mathcal I}_{C_{p_*}}}
=\Kos_{I_*}\otimes
\brane_7^{\bf B}(-1).
\label{wanted}
\eeq
The first isomorphism in (\ref{Pfrelations}) tells us that
$\Kos_{I_*}\otimes \brane_7^{\bf B}(-1)[2]$
is also a gauge theory lift of ${\mathcal O}_{p_*}$,
and it can go through the window ${\bf B}(1)$ by (\ref{shiftGRR}).
Therefore, we find
\beq
\PtoG_{{\bf B}(1)}({\mathcal O}_{p_*})
=\pi_X(\Kos_{I_*}\otimes \brane_7^{\bf B}(-1)[2])
={\mathcal I}_{C_{p_*}}[2]
\eeq
where we used (\ref{wanted}) in the last equality.
We find that the equivalence of \cite{BC}
is equal to the transport through the window ${\bf B}(1)$ followed by
the shift $[-2]$, as far as the images of skyscraper sheaves of
points of $Y$ are concerned.

The result (\ref{wanted}) also means
$\pi_X(\Kos_{I_*}\otimes \brane_7^{\bf B})
\cong{\mathcal I}_{C_{p_*}}(1)$, that is,
\beq
\PtoG_{\bf B}({\mathcal O}_{p_*})\,\cong\,
     {\mathcal I}_{C_{p_*}}(1).
\eeq
which means, by (\ref{D0duality}),
\beq
\PtoG_{\bf C}({\mathcal O}_{C_{p_*}})\,\cong\,
\mathfrak{D}( {\mathcal I}_{C_{p_*}}(1))[1].
\eeq
These provide a yet another description
of (\ref{TBD0}) and (\ref{TCD0}).

\subsection{The Integral Kernels}
\label{sub:kernel}

\newcommand{\bff}{\textit{\textbf{f}}}
\newcommand{\bfg}{\textit{\textbf{g}}}
\newcommand{\scrP}{{\mathscr P}}

As always \cite{OrlovEq},
an equivalence of categories, say $D(Y)\stackrel{\simeq}{\longrightarrow}D(X)$,
can be presented as a Fourier-Mukai transform
\beq
\Phi_{\scrP}({\mathcal F})=p_{X*}(\scrP\otimes p_Y^*({\mathcal F}))
\label{FM}
\eeq
by an object $\scrP\in D(X\times Y)$ called the integral kernel.
Using the information of Grassmannian images of Pfaffian points
(Section~\ref{sub:D0}) as well as the transport of D-brane
charges (Section~\ref{sub:charge}), we shall identify the integral kernel
$\scrP_{\bf w}$ corresponding to the equivalence
$\PtoG_{\bf w}:D(Y)\to D(X)$ for each window ${\bf w}$,
\beq
\PtoG_{\bf w}=\Phi_{\scrP_{\bf w}}.
\eeq

The image of the skyscraper sheaf of a point ${\mathcal O}_p\in D(Y)$
under the Fourier-Mukai transform (\ref{FM}) is the kernel $\scrP$
restricted to $X\times \{p\}$.
Thus, if the family of objects $\PtoG_{\bf w}({\mathcal O}_p)$
parametrized by $p\in Y$ ``glues'' to an object of $D(X\times Y)$,
that would be a candidate for the integral kernel $\scrP_{\bf w}$.
Of course, the ``gluing'' is always accompanied with
an ambiguity of taking the tensor product with a line bundle on $Y$.
In our problem, the family does glue: Look at the infinitely long exact
complex (\ref{longcomplex}) on $X$ associated with $p_*\in Y$.
Note that $\cktt$ can be identified with
$V/{\rm Ker}A(p_*)$ which glues to a vector bundle
$\ck_Y$ on $Y$, see (\ref{ckdef}).
Replacing $\cktt$ by this vector bundle and tensoring each component with
${\mathcal O}_Y(i)$ for an appropriate $i$, the complex
(\ref{longcomplex}) glues to an exact complex ${\mathscr C}$
on $X\times Y$ given by
\beq
\!\!\!\!\!\!\!\!\!\!\!\!
\cdots
\stackrel{\bfg_{-1}}\longrightarrow\,
S_X\boxtimes \ck_Y\stackrel{\bff_0}{\longrightarrow}
\!\!\!\!\begin{array}{c}
  {\mathcal O}_X\boxtimes\textw^{\!2}_0\ck_Y\\
  \oplus\\
  \Sym^2 S_X(1)\boxtimes {\mathcal O}_Y(1)
\end{array}
\!\!\!\!\stackrel{\bfg_0}{\longrightarrow}\,
S_X(1)\boxtimes \ck_Y(1)
\stackrel{\bff_1}{\longrightarrow}
\!\!\!\!\begin{array}{c}
  {\mathcal O}_X(1)\boxtimes\textw^{\!2}_0\ck_Y(1)\\
  \oplus\\
  \Sym^2 S_X(2)\boxtimes{\mathcal O}_Y(2)
\end{array}
\!\!\!\!\stackrel{\bfg_1}{\longrightarrow}
\cdots
\label{longcomplexXY}
\eeq
The arrows are
\beq
\bff_i=\left(\begin{array}{c}(\searrow)\\
  (\nwarrow)\end{array}\right)
\quad\mbox{and}\quad \bfg_i=\Bigl(\,(\nwarrow),(\searrow)\,\Bigr),
\eeq
where $(\searrow)$ and $(\nwarrow)$ are
(\ref{Lse}) and (\ref{Lnw}) in which $p^k$ are now
regarded as the coordinates of $Y$.
Then, the objects $\PtoG_{\bf w}({\mathcal O}_{p_*})$ for
${\bf w}={\bf A}, {\bf B}, {\bf C}, {\bf D}$ given in
(\ref{TAD0}), (\ref{TBD0}), (\ref{TCD0}), (\ref{TDD0}) glue to
the following objects of $D(X\times Y)$
\beqa
\scrP^0_{\bf A}&=&{\rm Ker}(\bff_1)
    \label{PA}
\\[0.2cm]
\scrP^0_{\bf B}&=&
\mathop{\Sym^2S_X(1)\boxtimes {\mathcal O}_Y(1)}_{{}_{-1}}
\stackrel{(\searrow)}{\longrightarrow}\mathop{{\rm Ker} (\bff_1)}_{{}_0}
\label{PB}
\\[0.2cm]
\scrP^0_{\bf C}&=&
\mathop{{\mathcal O}_X\boxtimes \textw^{\!2}_0\ck_Y}_{{}_{-1}}
\stackrel{(\nwarrow)}{\longrightarrow}\mathop{{\rm Ker}(\bff_1)}_{{}_0}
\label{PC}
\\[0.2cm]
\scrP^0_{\bf D}&=&{\rm Ker}(\bfg_0)[1]
\,\,\cong
\,\,\mathop{\begin{array}{c}
  {\mathcal O}_X\boxtimes\textw^{\!2}_0\ck_Y\\
  \oplus\\
  \Sym^2 S_X(1)\boxtimes {\mathcal O}_Y(1)
\end{array}}_{{}_{-1}}
\!\!\!\stackrel{\bfg_0}{\longrightarrow}\,\,
\mathop{{\rm Ker}(\bff_1)}_{{}_0}
\label{PD}
\eeqa
or $\scrP^j_{\bf w}:=\scrP^0_{\bf w}(0,j)$ for some $j\in\Z$, where
${\mathscr E}\mapsto {\mathscr E}(i,j)$ in $D(X\times Y)$
is the tensor product with
${\mathcal O}_X(i)\boxtimes{\mathcal O}_Y(j)$.
Note that ${\rm Ker}(\bff_1)$ and ${\rm Ker}(\bfg_0)$ are locally free
(see Appendix~\ref{subsec:locfree}).
These are the candidates for the integral kernel $\scrP_{\bf w}$.

In order to fix the ambiguity,
let us compute the effect of the Fourier-Mukai
transform by $\scrP^j_{\bf w}$ on the D-brane charge
and compare it with the effect of $\PtoG_{\bf w}$ obtained in
Section~\ref{sub:charge},
\beq
\ch(\Phi_{\scrP^j_{\bf w}}({\mathcal F}))\,\stackrel{!}{=}\,
\ch(\PtoG_{\bf w}({\mathcal F})).
\label{thematch}
\eeq
Only the cases
$[{\mathcal F}]=[{\mathcal O}_Y], [{\mathcal O}_H], [{\mathcal O}_{\ell}]$
need to be examined, as we know that it holds in
the case $[{\mathcal F}]=[{\mathcal O}_{\rm pt}]$ for any $j$.
In general, by the Grothendieck-Riemann-Roch formula, we have
\beq
\ch(\Phi_{\scrP}({\mathcal F}))
=\int_Y \ch(\scrP)\ch({\mathcal F}){\rm td}(Y).
\eeq
In the following computation,
we shall denote the hyperplane class $H$ on $X$ and $Y$ by $x$ and $y$
respectively. We know ${\rm td}(Y)=1+{1\over 12}c_2(Y)=1+{1\over 3}y^2$.
The problem is to compute the Chern character of $\scrP^j_{\bf w}$.
Note that
$\ch(\scrP^j_{\bf w})=\ch(\scrP^0_{\bf w})\e^{jy}$
and that (\ref{PA})-(\ref{PD}) yield
\beqa
\ch(\scrP^0_{\bf B})&=&-\ch(\Sym^2S_X(1)\boxtimes {\mathcal O}_Y(1))
+\ch(\scrP^0_{\bf A}),\label{BvsA}\\
\ch(\scrP^0_{\bf C})&=&-\ch({\mathcal O}_X\boxtimes \textw^{\!2}_0\ck_Y)
+\ch(\scrP^0_{\bf A}),\label{CvsA}\\
\ch(\scrP^0_{\bf D})&=&\ch(\scrP^0_{\bf B})+\ch(\scrP^0_{\bf C})
-\ch(\scrP^0_{\bf A})\label{ABCDrel}.
\eeqa
Thus, it is enough to compute $\ch(\scrP^0_{\bf A})$.
One should be able to compute it directly by the definition,
but the authors do not possess the techniques to determine it completely
at the moment.
However, we can partially determine it as follows.
Recall that $\scrP^0_{\bf A}$ is isomorphic to a semi-infinite complex
${\mathscr C}_{\geq {\bf A}}$ which is obtained from ${\mathscr C}$ in
(\ref{longcomplexXY})
by keeping only $S_X(1)\boxtimes \ck_Y(1)$ (placed at degree $0$)
and higher degree terms.
Note that ${\mathscr C}_{\geq {\bf A}}(1,1)[-2]$
appears as a subcomplex of ${\mathscr C}_{\geq{\bf A}}$,
with the complement given by the right two terms that appear in
(\ref{longcomplexXY}) (placed at degree 0 and 1).
Thus, we have
\footnote{To be precise,
  we should take finite length truncations of
  ${\mathscr C}_{\geq {\bf A}}$ in this computation.
  This can be done as follows.
  Let ${\mathscr C}_{\geq {\bf A}}^{\leq j}$
  be the truncation of ${\mathcal C}_{\geq {\bf A}}$
  where the degree $j$ term is replaced by
  ${\rm Ker} d^j$ and the terms of degree $j+1$ and higher are set
  to zero. Then we compare ${\mathscr C}_{\geq {\bf A}}^{\leq j}$
  and ${\mathscr C}_{\geq {\bf A}}^{\leq j-2}(1,1)[-2]$.}
\beqa
\lefteqn{(1-\e^{x+y})\ch(\scrP^0_{\bf A})
  \,=\,\ch({\mathscr C}_{\geq{\bf A}})
  -\ch({\mathscr C}_{\geq{\bf A}}(1,1))}\nn\\
&&=\,\ch(S_X(1)\boxtimes \ck_Y(1))
-\ch\left(\begin{array}{c}
  {\mathcal O}_X(1)\boxtimes\textw^{\!2}_0\ck_Y(1)\\
  \oplus\\
  \Sym^2 S_X(2)\boxtimes{\mathcal O}_Y(2)
\end{array}\right).
\eeqa
This determines $\ch(\scrP^0_{\bf A})$ up to four unknowns at order 3,4,5,6
in $x$ and $y$.
The knowledge (\ref{chAD0}) of
$\ch(\PtoG_{\bf A}({\mathcal O}_{p_*}))$ fixes
the unknown at order 3, leaving us with three unknowns.
In such a situation, we cheat a little:
we {\it use} the match (\ref{thematch})
for ${\bf w}={\bf A}$ and
${\mathcal F}={\mathcal O}_Y$
to {\it determine} the three unknowns and the right value of $j$
for $\scrP^j_{\bf A}$.
The result is $j=-1$ and
\beqa
\ch(\scrP^0_{\bf A})&=&4+x+5y-{x^2\over 6}+xy+{5y^2\over 2}
  -{x^3\over 21}-{2x^2y\over 7}+{2xy^2\over 7}+{23 y^3\over 42}\nn\\
&&
-{x^3y\over 21}-{17 x^2y^2\over 84}-{xy^3\over 21}
-{x^3y^2\over 84}-{x^2y^3\over 14}+{x^3y^3\over 252}.
\eeqa
Then, (\ref{thematch}) with $j=-1$ for ${\bf w}={\bf A}$
{\it holds} for ${\mathcal F}={\mathcal O}_H$ and ${\mathcal O}_{\ell}$
as well, which is a non-trivial consistency result.
Once $\ch(\scrP^0_{\bf A})$ is fixed, we know $\ch(\scrP^j_{\bf w})$ for
${\bf w}={\bf B},{\bf C}, {\bf D}$
by (\ref{BvsA})-(\ref{ABCDrel})
and can test if (\ref{thematch}) holds
for some $j$'s.
It turns out it holds for $j=-1$ for all the three windows as well.

In conclusion, the integral kernel for the transport $\PtoG_{\bf w}$ is
\beq
\scrP_{\bf w}=\scrP^0_{\bf w}(0,-1)\quad
\mbox{for ${\bf w}={\bf A}, {\bf B}, {\bf C}, {\bf D}$.}
\eeq
The kernel for the shifted windows can be obtained from this by
using
\beq
\scrP_{{\bf w}(i)}=\scrP_{\bf w}(-i,-i)[2i],
\eeq
that follows from  (\ref{shiftPtoG}).
$\scrP_{{\bf A}(i)}$ and $\scrP_{{\bf D}(i)}$
are vector bundles of rank $\pm 4$
while $\scrP_{{\bf B}(i)}$ and $\scrP_{{\bf C}(i)}$
are two term complexes of vector bundles
of effective rank $\pm 1$.

\newcommand{\bfC}{\textit{\textbf{C}}}

We would like to determine the relation of our equivalences to the one
$\PtoG_{\rm BC}$ by Borisov-C\u ald\u araru \cite{BC}.
Looking at the effect on the skyscraper sheaves at points,
we have already seen 
$\PtoG_{\rm BC}=[-2]\circ \PtoG_{{\bf B}(1)}\circ \otimes {\mathcal O}_Y(j)$
for some $j\in\Z$, and we would like to fix this $j$.
Their equivalence is the Fourier-Mukai transform by the ideal sheaf
${\mathcal I}_{\bfC}$ of a subvariety $\bfC$ of $X\times Y$ such that
$\bfC\cap (X\times\{p\})=C_p$.
One should be able to fix the relation by comparing $\ch({\mathcal I}_{\bfC})$
and $\ch(\scrP_{{\bf B}(1)})$, but
the authors do not have a technique to compute $\ch({\mathcal I}_{\bfC})$
directly at the moment. In this situation we proceed as follows.
We suppose that ${\mathcal I}_{\bfC}\in D(X\times Y)$ can be obtained by
na\"ively gluing the family of objects ${\mathcal I}_{C_p}$ of $D(X)$.
In order to compare it with the one of the form (\ref{PB}),
we lift it to a family of gauge theory branes and project the result to
a family of objects of $D(X)$.
In fact, the presentation in Section~\ref{sub:D0} is already given in a family
until the point where we specialize to $p=p_*$.
The key is the family of complexes (\ref{cone11}) on $\Gr$
and we would like to see if it glues to some complex on $\Gr\times Y$.
Note that $K_p^{\perp}=({\rm Ker} A(p))^{\perp}$
is the dual of $V/{\rm Ker}A(p)$ and hence is the fiber of $\ck_Y^*$ at
$p\in Y$.
Thus, $\textw^{\!4} K_p^{\perp}$ is the fiber of
\beq
\det \ck_Y^*=\det (\ck_Y(-1))=\det \ck_Y(-4)={\mathcal O}_Y(-2),
\label{detRst}
\eeq
where we used $\det \ck_Y={\mathcal O}_Y(2)$
that follows from $c_1(\ck_Y)=2H$ found in (\ref{c1R}).
$\textw^{\!3}K_p^{\perp}$ is the fiber of
$\textw^{\!3}\ck_Y^*$. It has a perfect pairing with $\ck_Y^*$ with values
in (\ref{detRst}) and hence is isomorphic to $\ck_Y(-2)$.
Finally, $\textw^{\!2}K_p^{\perp}/(A_p)$ has a perfect pairing with
$\textw^{\!2}_0(V/{\rm Ker}A(p))$ and hence is the fiber of
the dual of $\textw^{\!2}_0\ck_Y$. As $\textw^{\!2}_0\ck_Y$ has
a perfect pairing with itself by wedge product, 
with values in
$\det R_Y={\mathcal O}_Y(2)$, its dual is isomorphic to
$\textw^{\!2}_0\ck_Y(-2)$.
Thus, the family of complexes (\ref{cone11}) on $\Gr$ glues to a complex
\beq
\begin{tikzcd}
  \displaystyle\mathop{\Sym^2 S(-1)\boxtimes {\mathcal O}_Y(-2)}_{{}_{-2}}
  \arrow[r,"(\searrow)"]&
  \displaystyle\mathop{S(-1)\boxtimes \ck_Y(-2)}_{{}_{-1}}
  \arrow[r,"(\searrow)"]&
  \displaystyle\mathop{{\mathcal O}_{\Gr}(-1)
    \boxtimes\textw^{\!2}_0\ck_Y(-2)}_{{}_0}
\end{tikzcd}
\label{cone11GrY}
\eeq
on $\Gr\times Y$. Taking our procedure of
uplift to gauge theory and projection to $D(X)$, we obtain
an object of $D(X\times Y)$. It is the truncation of ${\mathscr C}(0,-2)$ with
$\Sym^2 S_X\boxtimes{\mathcal O}_Y(-2)$ as the first non-zero term at
degree $-1$. This is nothing but
$\scrP^0_{\bf B}(-1,-3)=\scrP_{{\bf B}(1)}(0,-1)[-2]$.
We are claiming that this is isomorphic to
${\mathcal I}_{\bfC}$ in $D(X\times Y)$.
Thus, we conclude that the equivalence of \cite{BC}
is related to ours by
\beq
\PtoG_{\rm BC}=[-2]\circ \PtoG_{{\bf B}(1)}\circ \otimes {\mathcal O}_Y(-1)
=\otimes{\mathcal O}_X(-1)\circ \PtoG_{\bf B}\circ \otimes {\mathcal O}_Y(-2)
\eeq

\subsection{Monodromy}
\label{subsec:monodromy}

\newcommand{\mdm}{{\mathscr M}}
\newcommand{\twist}{\mbox{\large $\utau$}}

\subsubsection{Monodromies Around Singular Points}
\label{subsec:monodromysing}

Let us next determine the monodromies
along loops around the singular points.
The method developed in \cite{HHP} will be used: we employ the
empty branes in the turn-around-phase to carry out the necessary
brane replacement.
We denote the monodromy along a loop ${\mathcal L}$ by
$\mdm_{\mathcal L}$.

\begin{figure}[thb]
\psfrag{t1}{\scriptsize $t_1$}
\psfrag{t2}{\scriptsize $t_2$}
\psfrag{t3}{\scriptsize $t_3$}
\psfrag{L1}{\small ${\mathcal L}_1$}
\psfrag{L2}{\small ${\mathcal L}_2$}
\psfrag{L3}{\small ${\mathcal L}_3$}
\psfrag{L3'}{\small $\widetilde{\mathcal L}_3$}
\psfrag{z}{\scriptsize $\zeta$}
\psfrag{th}{\scriptsize $\theta$}
\centerline{\includegraphics[page=40]{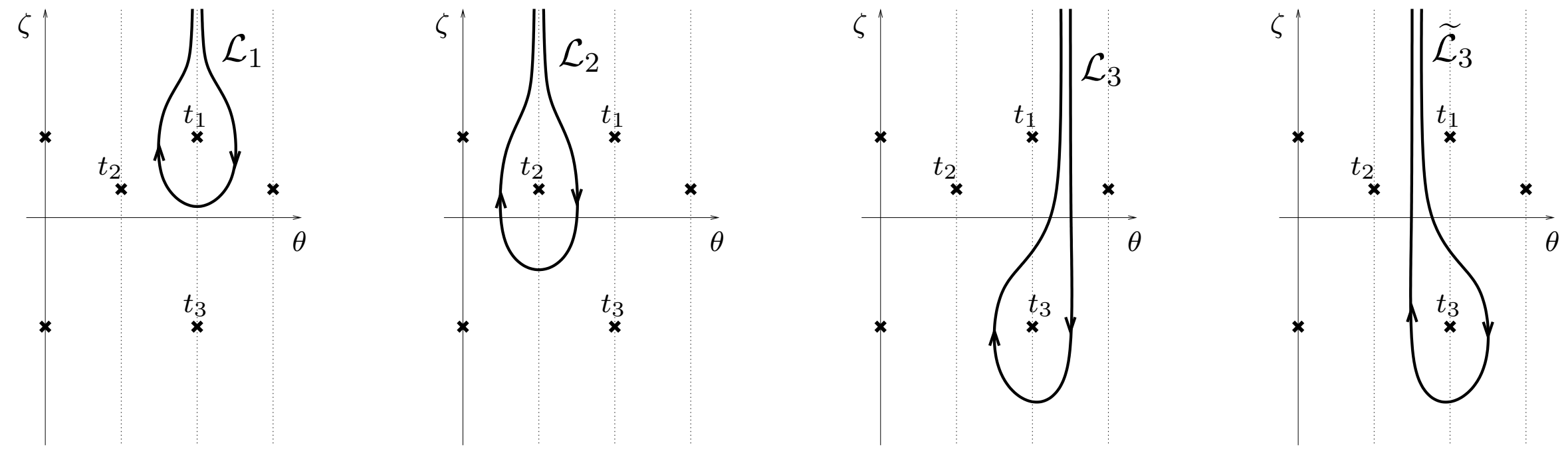}}
\caption{Loops around singular points with a base point in the Grassmannian
  phase.}
\label{fig:monodromy}
\end{figure}
We first consider loops with a base point in the Grassmannian phase.
A loop ${\mathcal L}_1$
around $t_1$ as in Fig.~\ref{fig:monodromy} can be regarded
as a downward path  through the window ${\bf C}$
followed by an upward path through the window ${\bf A}$.
Thus, the monodromy is the composition of the corresponding
D-brane transports, $\mdm_{{\mathcal L}_1}=\PtoG_{\bf A}\circ \GtoP_{\bf C}$.
Comparing GRR${}_{\bf A}$ and GRR${}_{\bf C}$ given by
(\ref{GRRA}) and (\ref{GRRC}), we see that the difference is
whether $\C$ or $\C(7)$ is in the set:
$\C$ is in GRR${}_{\bf C}$ but not in GRR${}_{\bf A}$
while $\C(7)$ is in GRR${}_{\bf A}$ but not in GRR${}_{\bf C}$.
Thus, if a brane in $X$ is represented by
a gauge theory brane $\brane$ in the subset ${\mathcal T}_{\bf C}$,
it can be safely
transported to the Pfaffian phase through the window ${\bf C}$.
However, if $\brane$ has a $\C$ component,
it cannot be safely transported back to the Grassmannian phase
through the window ${\bf A}$.
While in the Pfaffian phase, before coming back,
$\brane$ must be replaced by another brane $\brane'$ in ${\mathcal T}_{\bf A}$
which is isomorphic to $\brane$ at low energies in the Pfaffian phase.
The $\C$ components in $\brane$ must be traded for something else
that may include $\C(7)$ instead of $\C$.
This replacement
$\brane\in {\mathcal T}_{\bf C}\mapsto \brane'\in{\mathcal T}_{\bf A}$
can be done using the $p$-Koszul brane $\Kos$ (\ref{Kp}) which
is empty in the Pfaffian phase and has
$\C$ and $\C(7)$ as the edge components,
with the other components in the overlap GRR${}_{\bf A}\cap$ GRR${}_{\bf C}$.
Namely, we bind a copy of $\Kos$ at each $\C$-component of $\brane$
by a homomorphism between them: take its cone and annihilate the $\C$ pair
by tachyon condensation.
Since $\C$ appears at the head of the $p$-Koszul complex part of
$\Kos$, the homomorphism is from $\brane$ to (R-charge shifts of) the $\Kos$'s.
Once that is done, the resulting brane $\brane'$
can be safely transported back to the
Grassmannian phase through the window ${\bf A}$.
Since $\pi_X(\Kos)={\mathcal O}_X$,
the net effect is the Seidel-Thomas twist by ${\mathcal O}_X$, 
\beq
\twist_{{\mathcal O}_X}\,:\,\,{\mathcal E}\,\,\longmapsto\,\,
    {\rm Cone}\left(\,{\mathcal E}\,\longrightarrow\,
    \bigoplus_{j\in\Z}{\rm Hom}({\mathcal E},{\mathcal O}_X[j])
    \otimes {\mathcal O}_X[j]\,\right)[-1].
    \label{ST1}
\eeq    
Note that the loop ${\mathcal L}_1$ can also be regarded
as a downward path through the window ${\bf D}$
followed by an upward  path through the window ${\bf B}$.
Thus we should also be able to write the monodromy as
$\mdm_{{\mathcal L}_1}=\PtoG_{\bf B}\circ\GtoP_{\bf D}$.
The difference between GRR${}_{\bf B}$ and GRR${}_{\bf D}$ is the same as above:
$\C$ is in GRR${}_{\bf D}$ but not in GRR${}_{\bf B}$
while $\C(7)$ is in GRR${}_{\bf B}$ but not in GRR${}_{\bf D}$.
We again need to trade $\C$ for $\C(7)$ while in the Pfaffian phase,
which can be done using the $p$-Koszul brane $\Kos$.
This yields the same monodromy, the twist $\twist_{{\mathcal O}_X}$.
Thus, $\PtoG_{\bf B}\circ\GtoP_{\bf D}$
indeed agrees with $\PtoG_{\bf A}\circ \GtoP_{\bf C}$.

For the loop of the opposite orientation, we need to trade $\C(7)$ for
$\C$ while in the Pfaffian phase. Since $\C(7)$ appears at the tail of
the $p$-Koszul complex part of $\Kos$, the binding homomorphism is from
$\Kos$ to the given brane.
The net effect is another version of the twist\footnote{In \cite{SeiTho},
  $\twist_A$ and $\twist^{\,\rm ad}_A$
  are denoted by $T'_A$ and $T_A$ respectively.}
\beq
\twist^{\,\rm ad}_{{\mathcal O}_X}\,:\,\,
{\mathcal E}\,\,\longmapsto\,\,
 {\rm Cone}\left(\,
\bigoplus_{j\in\Z}{\rm Hom}({\mathcal O}_X[j],{\mathcal E})
    \otimes {\mathcal O}_X[j]
 \,\longrightarrow\,
     {\mathcal E} \,\right).
     \label{ST2}
\eeq

A loop ${\mathcal L}_2$ around $t_2$ as in Fig.~\ref{fig:monodromy}
can be regarded as a downward path through the window ${\bf A}$ followed by
an upward path through the window ${\bf D}(-1)$.
Thus, $\mdm_{{\mathcal L}_2}=\PtoG_{{\bf D}(-1)}\circ \GtoP_{\bf A}$.
The difference of the grade restriction rules is that
$\fund$ is in GRR${}_{\bf A}$ but not in GRR${}_{{\bf D}(-1)}$
while $\fund(7)$ is in GRR${}_{{\bf D}(-1)}$ but not in GRR${}_{\bf A}$.
We need to trade $\fund$ for $\fund(7)$ while in the Pfaffian phase.
This can be done using the brane $\Kos\otimes \fund$
which is empty in the Pfaffian phase and has
$\fund$ and $\fund(7)$ as the edge components,
with the other components in GRR${}_{\bf A}\cap$ GRR${}_{{\bf D}(-1)}$.
In the same way as in the case of the loop ${\mathcal L}_1$,
we find that the monodromy is the twist by
$\pi_X(\Kos\otimes\fund)\cong S^*_X$.

A loop ${\mathcal L}_3$ around $t_3$ as in Fig.~\ref{fig:monodromy}
can be regarded as a downward path through the window ${\bf D}$ followed by
an upward path through the window ${\bf C}$.
Thus, $\mdm_{{\mathcal L}_3}=\PtoG_{\bf C}\circ \GtoP_{\bf D}$.
The difference of the grade restriction rules is that
$\Symt(-1)$ is in GRR${}_{\bf D}$ but not in GRR${}_{\bf C}$
while $\Symt(6)$ is in GRR${}_{\bf C}$ but not in GRR${}_{\bf D}$.
We need to trade $\Symt(-1)$ for $\Symt(6)$ while in the Pfaffian phase.
This can be done using the brane $\Kos\otimes\Symt(-1)$
which is empty in the Pfaffian phase and has
$\Symt(-1)$ and $\Symt(6)$ as the edge components,
with the other components in GRR${}_{\bf C}\cap$ GRR${}_{\bf D}$.
The monodromy is the twist by 
$\pi_X(\Kos\otimes \Sym^2\fund(-1))\cong \Sym^2S^*_X(-1)$.

There is another loop that goes around $t_3$, shown as
$\widetilde{\mathcal L}_3$ in Fig.~\ref{fig:monodromy}. This
can be regarded as a downward path through the window ${\bf B}$ followed by
an upward path through the window ${\bf A}$.
Thus, $\mdm_{\widetilde{\mathcal L}_3}=\PtoG_{\bf A}\circ\GtoP_{\bf B}$.
The difference of the grade restriction rules is again that
$\Symt(-1)$ is in GRR${}_{\bf B}$ but not in GRR${}_{\bf A}$
while $\Symt(6)$ is in GRR${}_{\bf A}$ but not in GRR${}_{\bf B}$.
Since the components of $\Kos\otimes \Symt(-1)$
other than $\Symt(-1)$ and $\Symt(6)$ are also
in GRR${}_{\bf A}\cap$ GRR${}_{\bf B}$,
the monodromy must be the same as the one along ${\mathcal L}_3$, the twist
by $\Sym^2S^*_X(-1)$.
Note that ${\mathcal L}_3$ and $\widetilde{\mathcal L}_3$ are homotopically
inequivalent, and the coincidence of the corresponding monodromies
implies a non-trivial relation.
Note that ${\mathcal L}_1\circ {\mathcal L}_3$ is homotopic to
$\widetilde{\mathcal L}_3\circ {\mathcal L}_1$.
The coincidence thus
means that the monodromies of ${\mathcal L}_1$ and ${\mathcal L}_3$
must commute. That is, the twists by ${\mathcal O}_X$
and $\Sym^2S^*_X(-1)$ must commute,
\beq
\twist_{{\mathcal O}_X}\circ \twist_{\Sym^2S^*_X(-1)}
\cong     \twist_{\Sym^2S^*_X(-1)}\circ \twist_{{\mathcal O}_X}.
\label{commuteG}
\eeq
This would be the case provided
${\mathcal O}_X$ and $\Sym^2S^*_X(-1)$
are orthogonal in the sense that
${\rm Hom}_{D(X)}({\mathcal O}_X, \Sym^2S^*_X(-1)[i])=0$ for any $i$,
\cite{SeiTho}.
In the present case, the orthogonality indeed holds.
See Appendix~\ref{subsec:ExtRGr}.

To summarize
\beq
\mdm_{{\mathcal L}_1}=\twist_{{\mathcal O}_X},\quad
\mdm_{{\mathcal L}_2}=\twist_{S^*_X},\quad
\mdm_{{\mathcal L}_3}=\twist_{\Sym^2 S^*_X(-1)}.
\label{mdmbaseGr}
\eeq
Note that the monodromy must be an autoequivalence.
The twist $\twist_A$ in a $d$-dimensional Calabi-Yau category ${\mathcal C}$
is an autoequivalence if and only if $A\in {\mathcal C}$
is a spherical object \cite{SeiTho}, that is,
${\rm Hom}_{\mathcal C}(A,A[p])$ is $\C$ for $p=0,d$
and zero for other $p$.
The objects ${\mathcal O}_X$, $S^*_X$, $\Sym^2S^*_X(-1)$ are
indeed all spherical in $D(X)$. See Appendix ~\ref{subsec:ExtRGr}.

\begin{figure}[thb]
\psfrag{t1}{\scriptsize $t_1$}
\psfrag{t2}{\scriptsize $t_2$}
\psfrag{t3}{\scriptsize $t_3$}
\psfrag{L1}{\small ${\mathcal L}'_1$}
\psfrag{L2}{\small ${\mathcal L}'_2$}
\psfrag{L3}{\small ${\mathcal L}'_3$}
\psfrag{L1'}{\small $\widetilde{\mathcal L}'_1$}
\psfrag{z}{\scriptsize $\zeta$}
\psfrag{th}{\scriptsize $\theta$}
\centerline{\includegraphics[page=41]{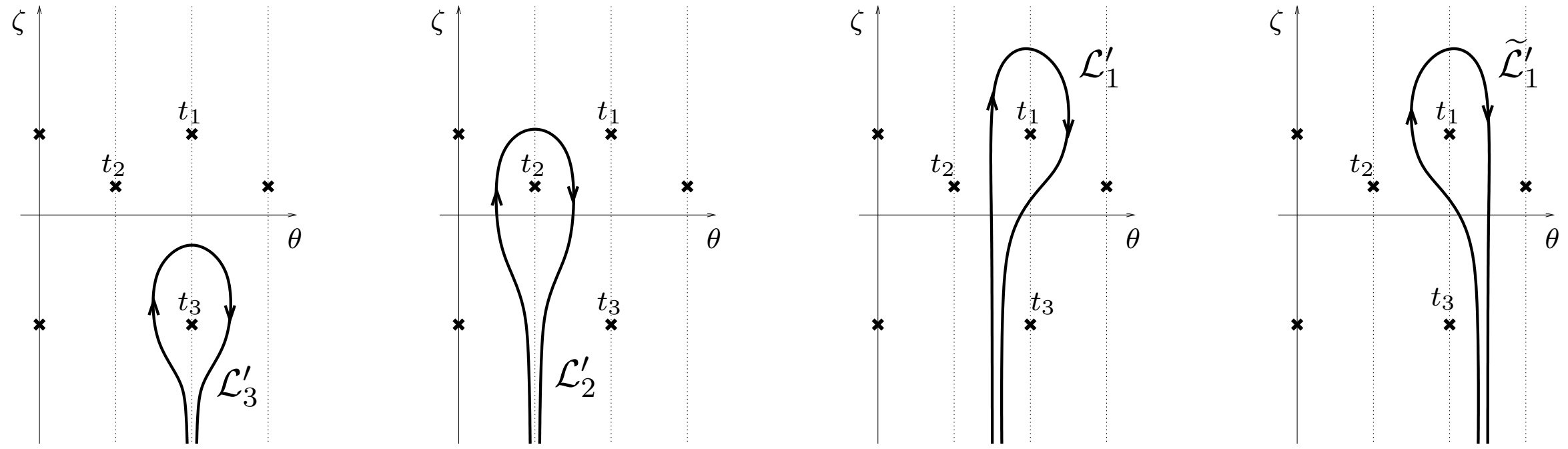}}
\caption{Loops around singular points with a base point in the Pfaffian phase.}
\label{fig:monodromy2}
\end{figure}
We next consider loops with a base point in the Pfaffian phase.
A loop ${\mathcal L}'_3$
around $t_3$ as in Fig.~\ref{fig:monodromy2} can be regarded
as an upward path through the window ${\bf A}$ followed by
a downward path through the window ${\bf B}$,
or equivalently
as an upward path through the window ${\bf C}$
followed by a downward path through the window ${\bf D}$.
Thus, $\mdm_{{\mathcal L}'_3}
=\GtoP_{\bf B}\circ\PtoG_{\bf A}=\GtoP_{\bf D}\circ\PtoG_{\bf C}$.
We need to trade $\Symt(6)$ for $\Symt(-1)$ while in the Grassmannian phase.
This can be done using the brane $\Lb^2(-1)$ obtained in
Section~\ref{subsec:GRprocess}
which is empty in the Grassmannian phase and has $\Symt(-1)$
and $\Symt(6)$ as the edge components, with the other components in
GRR${}_{\bf A}\cap$ GRR${}_{\bf B}$ and also in
GRR${}_{\bf C}\cap$ GRR${}_{\bf D}$.
See Fig.~\ref{fig:mono1}.
\begin{figure}[thb]
\psfrag{a}{\tiny ${\bf A}$}
\psfrag{b}{\tiny ${\bf B}$}
\psfrag{c}{\tiny ${\bf C}$}
\psfrag{d}{\tiny ${\bf D}$}
\psfrag{ab}{\tiny ${\bf A}$\!\!${\bf B}$}
\psfrag{cd}{\tiny ${\bf C}$\!\!${\bf D}$}
\centerline{\includegraphics[page=42]{main-pics.pdf}}
\caption{Grade restricted sets for the windows ${\bf A}$,
  ${\bf B}$, ${\bf C}$, ${\bf D}$ and the range of $\Lb^2(-1)$.}
\label{fig:mono1}
\end{figure}
Since $\Symt(6)$ appears as the head of the exact complex
$\Lc^2(-1)$, the binding is done by a homomorphism from
the given brane to the $\Lb^2(-1)$'s.
Thus, the monodromy is the twist by $\pi_Y(\Lb^2(-1))\cong {\mathcal O}_Y[-1]$.
As the twist $\twist_A$ is invariant under the shift $A\to A[1]$ by definition,
it is the same as the twist by ${\mathcal O}_Y$.

\begin{figure}[thb]
\psfrag{a}{\tiny ${\bf A}$}
\psfrag{d}{\tiny ${\bf D}'$}
\psfrag{ad}{\tiny ${\bf A}$\!\!${\bf D}'$}
\centerline{\includegraphics[page=43]{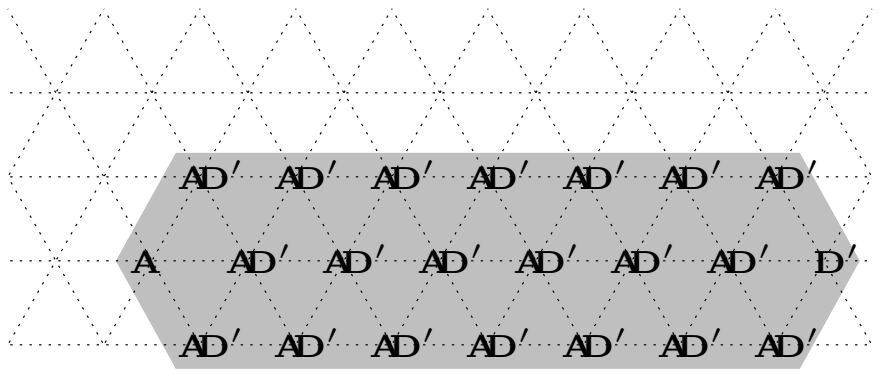}}
\caption{Grade restricted sets for the windows ${\bf A}$ and
  ${\bf D}(-1)={\bf D}'$ and the range of $\Lb^1$.}
\label{fig:mono2}
\end{figure}

A loop ${\mathcal L}'_2$
around $t_2$ as in Fig.~\ref{fig:monodromy2} can be regarded
as an upward path through the window ${\bf D}(-1)$
followed by a downward path through the window ${\bf A}$.
Thus, $\mdm_{{\mathcal L}'_2}=\GtoP_{\bf A}\circ \PtoG_{{\bf D}(-1)}$.
We need to trade $\fund(7)$ for $\fund$ while in the Grassmannian phase.
This can be done using the brane $\Lb^1$ 
which is empty in the Grassmannian phase and has $\fund$
and $\fund(7)$ as the edge components, with the other components in
GRR${}_{{\bf D}(-1)}\cap$ GRR${}_{\bf A}$.
See Fig.~\ref{fig:mono2}.
Thus, the monodromy is the twist by $\pi_Y(\Lb^1)\cong \ck^*_Y$.

A loop ${\mathcal L}'_1$
around $t_1$ as in Fig.~\ref{fig:monodromy2} can be regarded
as an upward path through the window ${\bf A}$
followed by a downward path through the window ${\bf C}$.
Thus, $\mdm_{{\mathcal L}'_1}=\GtoP_{\bf C}\circ \PtoG_{\bf A}$.
We need to trade $\C(7)$ for $\C$ while in the Grassmannian phase.
This can be done using the brane $\Lb^0$ 
which is empty in the Grassmannian phase and has $\C$
and $\C(7)$ as the edge components, with the other components in
GRR${}_{\bf A}\cap$ GRR${}_{\bf C}$. See Fig.~\ref{fig:mono3}-left.
\begin{figure}[thb]
\psfrag{a}{\tiny ${\bf A}$}
\psfrag{c}{\tiny ${\bf C}$}
\psfrag{b}{\tiny ${\bf B}$}
\psfrag{d}{\tiny ${\bf D}$}
\psfrag{ac}{\tiny ${\bf A}$\!\!${\bf C}$}
\psfrag{bd}{\tiny ${\bf B}$\!\!${\bf D}$}
\centerline{\includegraphics[page=44]{main-pics.pdf}}
\caption{Grade restricted sets for the windows ${\bf A}$, ${\bf B}$, ${\bf C}$,
  ${\bf D}$ and the range of $\Lb^0$.}
\label{fig:mono3}
\end{figure}
Thus, the monodromy is the twist by
$\pi_Y(\Lb^0)\cong \textw^{\!2}_0\ck^*_Y(1)[-1]$
or equivalently by $\textw^{\!2}_0\ck^*_Y(1)$.

There is another loop around $t_1$, shown as $\widetilde{\mathcal L}'_1$
in Fig.~\ref{fig:monodromy2}. This
can be regarded as an upward path through the window ${\bf B}$ followed by
a downward path through the window ${\bf D}$.
Thus, $\mdm_{\widetilde{\mathcal L}'_1}=\GtoP_{\bf D}\circ\PtoG_{\bf B}$.
We again need to trade $\C(7)$ for $\C$ while in the Grassmannian phase.
Since the components other than $\C$ and $\C(7)$
in $\Lb^0$ are also in GRR${}_{\bf B}\cap$ GRR${}_{\bf D}$
(see Fig.~\ref{fig:mono3}-right),
the monodromy is again the twist by
$\pi_Y(\Lb^0)\cong \textw^{\!2}_0\ck^*_Y(1)[-1]$.
Thus, we find that the monodromies along ${\mathcal L}'_1$
and $\widetilde{\mathcal L}'_1$ coincide.
This again implies a non-trivial relation.
Since ${\mathcal L}_3'\circ {\mathcal L}_1'
=\widetilde{\mathcal L}'_1\circ {\mathcal L}_3'$,
the coincidence of the monodromies of ${\mathcal L}'_1$ and
$\widetilde{\mathcal L}_1'$ means that
the monodromies of ${\mathcal L}_3'$ and ${\mathcal L}'_1$ commute, that is, 
\beq
    \twist_{{\mathcal O}_Y}\circ
    \twist_{\wedge^{2}_0\ck^*_Y(1)}
    \cong    \twist_{\wedge^{2}_0\ck^*_Y(1)}\circ
             \twist_{{\mathcal O}_Y}.
             \label{commuteP}
\eeq
A direct proof of this statement is left open.

To summarize
\beq
\mdm_{{\mathcal L}'_3}=\twist_{{\mathcal O}_Y},\quad
\mdm_{{\mathcal L}'_2}=\twist_{{\mathcal R}_Y^*},\quad
\mdm_{{\mathcal L}'_1}=\twist_{\wedge^2_0{\mathcal R}_Y^*(1)}.
\label{mdmbasePf}
\eeq
${\mathcal O}_Y$ is a spherical object of $D(Y)$ as $Y$
is a simply connected Calabi-Yau manifold. We leave the sphericalness of
$\ck^*_Y$ and $\textw^{\!2}_0\ck^*_Y(1)$ as an open problem.

\medskip
\noindent
{\bf Monodromy on D-brane charges}

Let us look at the effect of the twist operator on the D-brane charges.
By definition (\ref{ST1}),
\beqa
[\twist_{\mathcal F}({\mathcal E})]&=&[{\mathcal E}]
-\sum_{j\in \Z}\dim {\rm Hom}({\mathcal E},{\mathcal F}[j])
(-1)^j[{\mathcal F}]\nn\\
&=&
[{\mathcal E}]-\chi({\mathcal E},{\mathcal F})[{\mathcal F}],
\eeqa
In the last expression, $\chi$ is the Euler pairing 
\beq
\chi({\mathcal E},{\mathcal F})
:=\sum_{j\in\Z}(-1)^j\dim {\rm Hom}({\mathcal E}, {\mathcal F}[j]),
\eeq
which is antisymmetric on a Calabi-Yau threefold
$\chi({\mathcal E},{\mathcal F})=-\chi({\mathcal F},{\mathcal E})$
by Serre duality
${\rm Hom}({\mathcal E},{\mathcal F})\cong
{\rm Hom}({\mathcal F},{\mathcal E}[3])^*$.
We may compute the pairing using the Riemann-Roch formula
\beq
\chi({\mathcal E},{\mathcal F})
=\int_M{\rm td}(M)\overline{{\rm ch}({\mathcal E})}{\rm ch}({\mathcal F}),
\eeq
where $\overline{\omega}=(-1)^p\omega$ for $\omega\in {\rm H}^{p,p}(M)$.
In either of the two phases,
we describe the monodromy with respect to the integral basis (\ref{basis})
of the charge lattice.

\medskip
\noindent
\underline{Grassmannian base point}

The D-brane charges of $S_X^*$ and $\Sym^2S_X^*(-1)$ are
given in (\ref{chSv}) and (\ref{chSym2Sv}).
The Euler pairing is
\beq
\begin{array}{c|cccc}
-
&{\mathcal O}_X&{\mathcal O}_H&{\mathcal O}_{\ell}&{\mathcal O}_{\rm pt}\\
\hline
\chi({\mathcal O}_X,-)&0&14&1&1\\
\chi(S_X^*,-)&-7&56&1&2\\
  \chi(\Sym^2S_X^*(-1),-)&0&28&3&3
\end{array}
\label{EulerG}
\eeq
Thus, the twists on the charges are represented by
\beqa
\twist_{{\mathcal O}_X}:&&\!\!\!{\bf 1}_4
+\left(\begin{array}{c}1\\0\\0\\0\end{array}\right)(0,14,1,1)
  =\left(\begin{array}{cccc}
    1&14&1&1\\
    0&1&0&0\\
    0&0&1&0\\
    0&0&0&1
    \end{array}\right)
\label{GTO}\\
\twist_{S_X^*}:&&\!\!\!{\bf 1}_4
+\left(\begin{array}{c}2\\1\\28\\-35\end{array}\right)(-7,56,1,2)
  =\left(\begin{array}{cccc}
    -13&112&2&4\\
    -7&57&1&2\\
    -196&1568&29&56\\
    245&-1960&-35&-69
    \end{array}\right)
\label{GTS}
\\
\twist_{\Sym^2S_X^*(-1)}:&&\!\!\!{\bf 1}_4
+\left(\begin{array}{c}3\\0\\-14\\14\end{array}\right)(0,28,3,3)
  =\left(\begin{array}{cccc}
    1&84&9&9\\
    0&1&0&0\\
    0&-392&-41&-42\\
    0&392&42&43
    \end{array}\right)
\label{GTSS}
\eeqa      
One can check that these match respectively with the monodromy matrices
obtained by the compositions of the transport matrices in (\ref{RtM})
\beq
T_{\bf A}^{-1}T_{\bf C}=T_{\bf B}^{-1}T_{\bf D},
\quad
T_{{\bf D}(-1)}^{-1}T_{\bf A},
\quad
T_{\bf C}^{-1}T_{\bf D}=T_{\bf A}^{-1}T_{\bf B}.
\label{MMMeq}
\eeq
The last equality of (\ref{MMMeq}) is the agreement of
the monodromies along ${\mathcal L}_3$ and $\widetilde{\mathcal L}_3$
at the charge level
and is compatible with the commutativity (\ref{commuteG}).
This is equivalent to the vanishing of the Euler pairing between
${\mathcal O}_X$ and $\Sym^2S_X^*(-1)$ (see the lower-left corner of
the table (\ref{EulerG})) which of course also follows
from the orthogonality of the two.

\medskip
\noindent
\underline{Pfaffian base point}

The D-brane charges of $\ck_Y^*$ and $\textw^{\!2}_0\ck_Y^*(1)$
are given in (\ref{chck}) and (\ref{chw2ck}).
The Euler pairing is
\beq
\begin{array}{c|cccc}
-
&{\mathcal O}_Y&{\mathcal O}_H&{\mathcal O}_{\ell}&{\mathcal O}_{\rm pt}\\
\hline
\chi({\mathcal O}_Y,-)&0&7&1&1\\
\chi(\ck^*_Y,-)&7&14&6&4\\
  \chi(\textw^{\!2}_0\ck^*_Y(1),-)&0&21&5&5
\end{array}
\label{EulerP}
\eeq
Thus, the twists on the charges are represented by
\beqa
\twist_{{\mathcal O}_Y}:&&\!\!\!{\bf 1}_4
+\left(\begin{array}{c}1\\0\\0\\0\end{array}\right)(0,7,1,1)
  =\left(\begin{array}{cccc}
    1&7&1&1\\
    0&1&0&0\\
    0&0&1&0\\
    0&0&0&1
    \end{array}\right)
\label{PTL2}\\
\twist_{\ck^*_Y}:&&\!\!\!{\bf 1}_4
+\left(\begin{array}{c}4\\-2\\-14\\21\end{array}\right)(7,14,6,4)
  =\left(\begin{array}{cccc}
    29&56&24&16\\
    -14&-27&-12&-8\\
    -98&-196&-83&-56\\
    147&294&126&85
    \end{array}\right)~~~~
\label{PTL1}
\\
\twist_{\wedge^2_0\ck^*_Y(1)}:&&\!\!\!{\bf 1}_4
+\left(\begin{array}{c}5\\0\\-14\\14\end{array}\right)(0,21,5,5)
  =\left(\begin{array}{cccc}
    1&105&25&25\\
    0&1&0&0\\
    0&-294&-69&-70\\
    0&294&70&71
    \end{array}\right)
\label{PTL0}
\eeqa      
One can check that these match respectively with the monodromy matrices
\beq
T_{\bf B}T_{\bf A}^{-1}=T_{\bf D}T_{\bf C}^{-1},
\quad
T_{\bf A}T_{{\bf D}(-1)}^{-1}
\quad
T_{\bf C}T_{\bf A}^{-1}=T_{\bf D}T_{\bf B}^{-1}.
\label{MMMeqP}
\eeq
The last equality of (\ref{MMMeqP}) is the agreement of
the monodromies along ${\mathcal L}_1'$ and $\widetilde{\mathcal L}_1'$
at the charge level
and supports the commutativity claim (\ref{commuteP}).
This is due to the vanishing of the Euler pairing between
${\mathcal O}_Y$
and $\textw^{\!2}_0\ck^*_Y(1)$ (see the lower-left corner of
the table (\ref{EulerP})).

\subsubsection{Large Volume Monodromies}

We recall that the physics is invariant under the shift
$\theta\to \theta +2\pi$ and the \emph{moduli space} of theories
is parametrized by $\e^t\in S^2$,
\beq
\mathfrak{M}_{\rm t.c.}=S^2\setminus\{\alpha_1,\alpha_2,\alpha_3,0,\infty\}.
\eeq
In this space, we also have loops around
two special points, $\e^t=\infty$ and $0$ where $X$ and $Y$ respectively
have infinite volume.
Let us discuss the monodromies $\mdm_X$ and $\mdm_Y$ along loops
${\mathcal L}_X$ and ${\mathcal L}_Y$ in $\mathfrak{M}_{\rm t.c.}$
that encircle $\e^t=\infty$ and $0$ in the counter-clockwise direction.

The loop ${\mathcal L}_X$ can be the gradual \emph{increase} of $\theta$
by $2\pi$ at a fixed value of $\zeta\gg 0$. 
In view of (\ref{thetashift}) and (\ref{shiftGr}), we see that
the monodromy is
\beq
\mdm_X({\mathcal E})={\mathcal E}\otimes {\mathcal O}_X(1).
\label{mdmX}
\eeq
This is consistent with the fact that the gradual increase
$\theta\to \theta+2\pi$ corresponds to the gradual change of the B-field class
$[B]\to [B]+2\pi H$, with $H=c_1({\mathcal O}_Y(1))$,
and the identification (\ref{Bshift}) in the non-linear sigma model.

The loop ${\mathcal L}_Y$ can be the gradual \emph{decrease} of $\theta$
by $2\pi$ at a fixed value of $\zeta\ll 0$. 
In view of (\ref{thetashift}) and (\ref{shiftPf}), we see that
the monodromy is
\beq
\mdm_Y({\mathcal F})={\mathcal F}\otimes {\mathcal O}_Y(1)[-2].
\label{mdmY}
\eeq
As $\theta\to \theta-2\pi$ corresponds to $[B]\to [B]+2\pi H$,
with $H=c_1({\mathcal O}_Y(1))$,
the part $\otimes {\mathcal O}_Y(1)$ is consistent with (\ref{Bshift}).
However, we also have a shift $[-2]$ in homological degree which is somewhat
surprising.
Is it related to the fact that $\zeta\ll 0$ is a strongly coupled phase?

The key is the \emph{flow of grading}, an idea introduced by Douglas
in the discussion of stability conditions of D-branes
\cite{Douglas,Aspinwall-Douglas}.
Suppose that a B-brane ${\mathcal E}$ is \emph{stable}, that is,
it flows to an indecomposable superconformal boundary condition
in the infra-red limit with a certain property.\footnote{The property
  is ``preserving spacetime supersymmetry'' when
  the bulk superconformal field theory is used for string compactification,
  which is the case for the infra-red limit of a Calabi-Yau sigma model.
  More generally, it is stated using
  the boundary condition on the ``spectral flow operators''.}
Then, the choice of overall integral R-charge of ${\mathcal E}$
corresponds to a choice of branch for the Arg function in
\beq
\varphi({\mathcal E})={1\over \pi}\Arg Z({\mathcal E})
\eeq
where $Z({\mathcal E})$ is the central charge of ${\mathcal E}$.
The function $\varphi$ controls the R-charge of local operators inserted
between the superconformal boundary conditions for stable branes.
We recall from (\ref{HSvsCCGr}) and (\ref{HSvsCCPf})
that the hemisphere partition function in the R-charge frame
with $\epsilon\searrow 0$ (Grassmannian frame) behaves as
\beq
Z_{D^2}(\brane)=\left\{\begin{array}{cc}
(2\pi)^{13}\,Z_X(\pi_X(\brane))&\e^t\to \infty,\\
(2\pi)^{13}\,\e^t\cdot Z_Y(\pi_Y(\brane))&\e^t\to 0
\end{array}\right.
\label{ZD2vsZLV}
\eeq
where $Z_X$ and $Z_Y$ are the large volume central charges
for $X$ and $Y$,
with the properties (\ref{Zpt})-(\ref{ZOX}) and (\ref{Pij}).
A D0-brane ${\mathcal O}_{\rm pt}$ in a Calabi-Yau manifold $M$
is expected to be stable in the large volume regime,
and its large volume central charge $Z_M({\mathcal O}_{\rm pt})$
approaches $1$ with an exponentially small correction.
Therefore, ${\mathcal O}_x$ for $x\in X$ and ${\mathcal O}_y$ for $y\in Y$
are stable in the respective large volume regimes $\e^t\sim\infty$
and $\e^t\sim 0$ where $Z_X({\mathcal O}_x)=1+O(\e^{-t})$ 
and $Z_Y({\mathcal O}_y)=1+O(\e^t)$.
In particular, $\Arg Z_X({\mathcal O}_x)$
and $\Arg Z_Y({\mathcal O}_y)$ are single valued along
small loops around $\e^t=\infty$ and $\e^t=0$, respectively.
Thus, if we take $Z=Z_{D^2}$ in the Grassmannian frame,
$\varphi({\mathcal O}_x)$ is invariant as $\theta\to\theta+2\pi$
while $\varphi({\mathcal O}_y)$ shifts by $+2$ as $\theta\to\theta-2\pi$.
This supports our conclusion that there is no shift in $\mdm_X$
and there is a shift by $[-2]$ in $\mdm_Y$.

The shift $[-2]$ in $\mdm_Y$ has nothing to do with
the strongly coupledness of the phase.
To show this, let us work in the Pfaffain frame $\epsilon\nearrow 1$.
Then, the relations (\ref{shiftGr}) and (\ref{shiftPf})
are modified: the shift $[-2i]$ is added to the former and the shift $[2i]$
is removed from the latter.
Also, $\e^{-t}$ is multipled to the right-hand sides of (\ref{ZD2vsZLV})
so that $\varphi({\mathcal O}_x)$ shifts by $+2$ as $\theta\to\theta+2\pi$
while $\varphi({\mathcal O}_y)$ is invariant as $\theta\to\theta-2\pi$.
Thus, we have a shift $[-2]$ in $\mdm_X$ and no shift in $\mdm_Y$.
This exhibits the fact that the monodromy of the chiral sector depends on
the choice of R-charge frame so that the physics is independent of any choice.

\subsubsection{Groupoid Relations}
\label{subsub:groupoid}

Let us fix two points in the FI-theta parameter space,
$t_X$ in the Grassmannian phase and $t_Y$ in the Pfaffian phase.
The paths ${\mathcal P}_{\bf w}$ in the windows ${\bf w}$
can all be taken to connect these two points.
\begin{figure}[thb]
\psfrag{a1}{\scriptsize $\alpha_1$}
\psfrag{a2}{\scriptsize $\alpha_2$}
\psfrag{a3}{\scriptsize $\alpha_3$}
\psfrag{infty}{\scriptsize $\infty$}
\psfrag{0}{\scriptsize $0$}
\psfrag{bX}{\scriptsize $b_X$}
\psfrag{bY}{\scriptsize $b_Y$}
\psfrag{LX}{\small ${\mathcal L}_X$}
\psfrag{LY}{\small ${\mathcal L}_Y$}
\psfrag{A}{\small ${\mathcal P}_{\bf A}$}
\psfrag{B}{\small ${\mathcal P}_{\bf B}$}
\psfrag{C}{\small ${\mathcal P}_{\bf C}$}
\psfrag{D}{\small ${\mathcal P}_{\bf D}$}
\psfrag{D'}{\small ${\mathcal P}_{{\bf D}(-1)}$}
\centerline{\includegraphics[page=45]{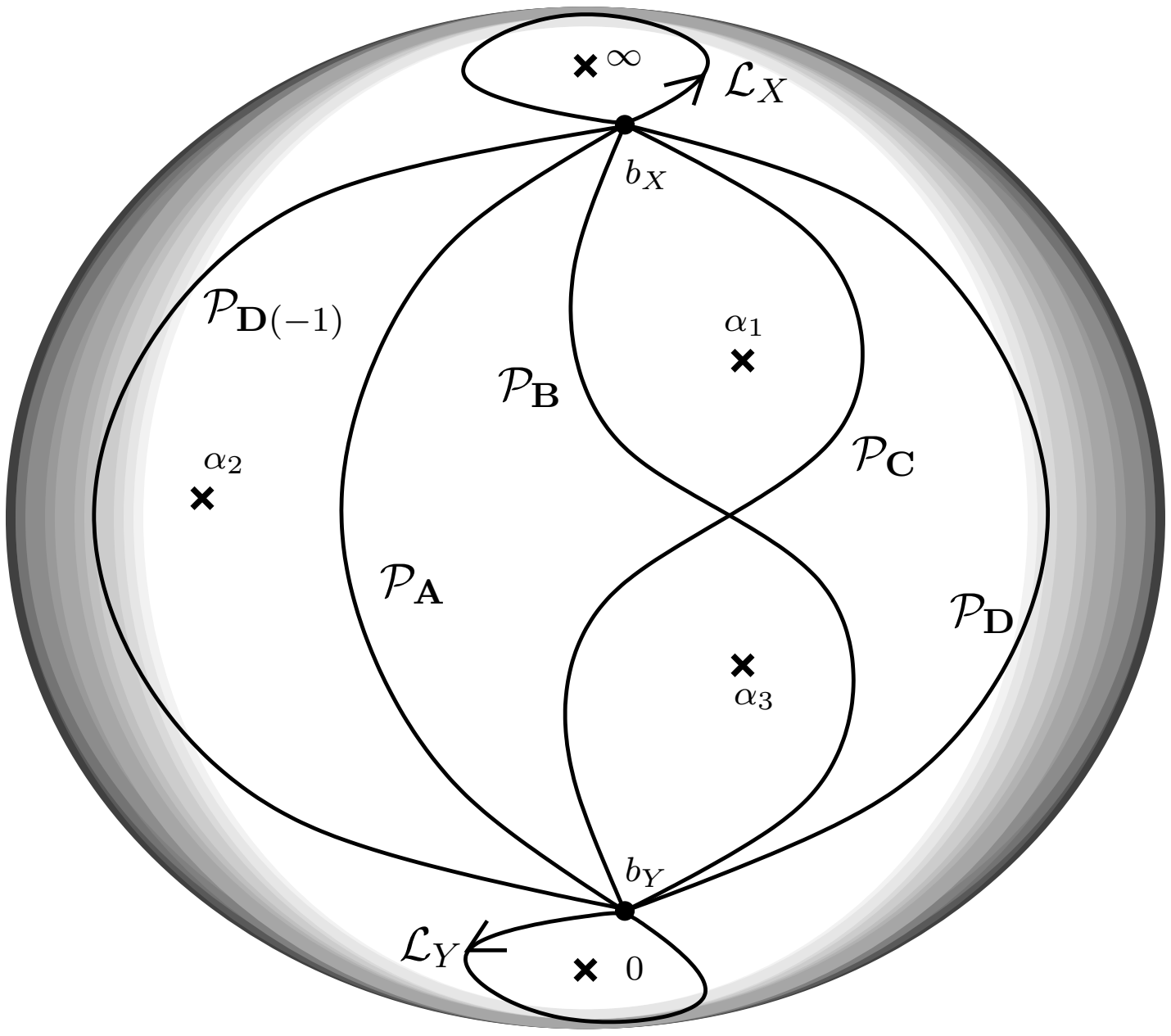}}
\caption{Paths and loops on the moduli space $\mathfrak{M}_{\rm t.c.}$.}
\label{fig:tcmoduli}
\end{figure}
 By the map $t\mapsto \e^t$,
these paths may be regarded as the paths in the moduli space
$\mathfrak{M}_{\rm t.c.}$ connecting $b_X=\e^{t_X}$ and $b_Y=\e^{t_Y}$.
The points $b_X$ and $b_Y$ may also be taken as the base points of 
the loops ${\mathcal L}_X$ and ${\mathcal L}_Y$, respectively.
Five paths and the two loops are shown in Fig.~\ref{fig:tcmoduli}.
(The picture reflects only the topology: Everything is depicted on the
front side of the sphere simply for illustration purpose.)
We shall write $\pathGtoP_{\bf w}$ and $\pathPtoG_{\bf w}$
for the paths with the downward and upward orientations.

Let us confirm that the transport and monodromy obey the relations
that are expected from the homotopy equivalence among the compositions
of paths and loops in the moduli space $\mathfrak{M}_{\rm t.c.}$. 
The composition
${\mathcal L}_Y\ast \pathGtoP_{\bf w}\ast {\mathcal L}_X$
can be continuously deformed to $\pathGtoP_{{\bf w}(1)}$
as a path from $b_X$ to $b_Y$ in the moduli space.
See Fig.~\ref{fig:tcmoduli} for the case ${\bf w}={\bf D}(-1)$:
The middle part of
${\mathcal L}_Y\ast \pathGtoP_{{\bf D}(-1)}\ast {\mathcal L}_X$
can be deformed towards the back side of the sphere and then moved further
until it comes back to the front side and coincides with $\pathGtoP_{\bf D}$.
Therefore, the corrersponding transport must agree,
\beq
\mdm_Y\circ \GtoP_{\bf w}\circ \mdm_X=\GtoP_{{\bf w}(1)}.
\label{YTX}
\eeq
In view of the formulae (\ref{mdmX}) and (\ref{mdmY})
for $\mdm_X$ and $\mdm_Y$, this is nothing but (\ref{shiftGtoP}).
It is important that we have the shift $[-2]$ in either
$\mdm_Y$ (the Grassmannian frame) or
in $\mdm_X$ (the Pfaffian frame) for this relation to hold.
The relation at the level of D-brane charges,
already noted in (\ref{YTXcharge}), is ignorant of this shift.
We also recall from Section~\ref{subsec:setup}
that $\pathGtoP_{\bf A}\ast\pathPtoG_{\bf B}\ast\pathGtoP_{\bf D}$
can be continuously deformed to $\pathGtoP_{\bf C}$
which is also evident in Fig.~\ref{fig:tcmoduli}.
This requires
\beq
\GtoP_{\bf A}\circ \PtoG_{\bf B}\circ \GtoP_{\bf D}
=\GtoP_{\bf C}.
\label{ABD=C}
\eeq
This immediately follows from the relation
$\PtoG_{\bf B}\circ \GtoP_{\bf D}=\PtoG_{\bf A}\circ \GtoP_{\bf C}$
found earlier --- see the paragraph below (\ref{ST1}).
This shows that the transport and monodromy provide an action of
the fundamental groupoid of $\mathfrak{M}_{\rm t.c.}$
with objects $\{b_X,b_Y\}$ on the chiral sector of B-branes,
and on the categories of B-branes in particular.

We also found a relation that does not follow from homotopy equivalence:
The monodromies along the loops ${\mathcal L}_1$ and ${\mathcal L}_3$
around $\alpha_1$ and $\alpha_3$ with base point $b_X$,
as in Fig.~\ref{fig:monodromy},
commute with each other,
\beq
\mdm_{{\mathcal L}_1}\circ \mdm_{{\mathcal L}_3}
=\mdm_{{\mathcal L}_3}\circ \mdm_{{\mathcal L}_1}.
\label{ML1ML3}
\eeq
The monodromies along the loops ${\mathcal L}_1'$ and ${\mathcal L}_3'$
with base point $b_Y$, as in Fig.~\ref{fig:monodromy2},
also commute but that follows from
(\ref{ML1ML3}) by the conjugation with $\GtoP_{\bf C}$.

\medskip
\noindent
{\bf Remark.}
The shift $[-2]$ in the relation (\ref{YTX}) was missed in \cite{DonovanPfGr}.
A similar relation was also written in \cite{HerbstWalcher} for Abelian GLSMs,
for example in Eqn (16). This equation is correct since the monodromy
${\mathcal M}_{\hat{\Phi}_i}^q$ is defined ``upstairs'' in GLSM.
If we describe it ``downstairs'' in the derived category, an appropriate
shift must be accompanied.
The fact that the monodromy relations at the charge level
do not faithfully reflect the relations at the level of category
was also noted earlier in \cite{Aspinwall-Douglas}: the fifth power
of the Gepner point monodromy in the quintic model is the shift
$[-2]$ which is invisible at the charge level.\footnote{Refs.~\cite{Aspinwall-Douglas,Aspinwall-lecture} consider the \emph{clockwise} 
  loops around the Gepner point for which the shift is $[+2]$.}
This is a non-trivial computation in the derived category of coherent
sheaves on the quintic threefold \cite{Aspinwall-Douglas,Aspinwall-lecture}
but is an immediate consequence of the flow from GLSM to
the Landau-Ginzburg orbifold (see Eqn (10.12) in \cite{HHP}).
(The same relation was discussed also in \cite{WalcherStability}
from the viewpoint of the Landau-Ginzburg orbifold.)
Here again, the R-charge frame plays an important r\^ole.
The hemisphere partition function in the Calabi-Yau frame behaves as \cite{HR}
\beq
Z_{D^2}(\brane)
=\left\{\begin{array}{cc}
-(2\pi)^5\,Z_{\rm CY}(\pi_{\rm CY}(\brane))&\e^t\to \infty,\\
{2\pi\over 5}\Gamma({1\over 5})^5\e^{t/5}\cdot Z_{\rm LG}(\pi_{\rm LG}(\brane))&\e^t\to 0
\end{array}\right.
\label{ZD2vsZLVquintic}
\eeq
where $\pi_{\rm CY}$ and $\pi_{\rm LG}$ are the flows to low energy branes
in the Calabi-Yau and the Landau-Ginzburg phases and
$Z_{\rm CY}$ and $Z_{\rm LG}$ are the central charges in the respective theories.
We see that $\varphi(\pi_{\rm LG}(\brane))$ for $Z=Z_{D^2}$ in this frame
shifts by $+2$ as $\e^t$ goes around 0 five times,
as long as $Z_{\rm LG}(\pi_{\rm LG}(\brane))\ne 0$ at $\e^t=0$,
which is the case when
$\pi_{\rm LG}(\brane)$ is nontrivial and stable at $\e^t=0$.

\section{Other Models}
\label{sec:other}

\newcommand{\ydt}{\Box\!\Box}
\newcommand{\ydoo}{\!\!\begin{array}{c}
  \Box\\[-0.34cm]
  \Box
  \end{array}\!\!}

\newcommand{\ydoop}{\!\!{\begin{array}{c}
  \Box\\[-0.34cm]
  \Box
  \end{array}}_{\!\!\!\!{}^{\scalebox{0.7}{+}}}\!\!}

\newcommand{\ydoom}{\!\!{\begin{array}{c}
  \Box\\[-0.34cm]
  \Box
\end{array}}_{\!\!\!\!{}^{\scalebox{0.7}{$-$}}}\!\!}

\newcommand{\ydoopm}{\!\!{\begin{array}{c}
  \Box\\[-0.34cm]
  \Box
\end{array}}_{\!\!\!\!{}^{\scalebox{0.7}{$\pm$}}}\!\!}

\newcommand{\mydoo}{\!\!{\small\begin{array}{c}
  \Box\\[-0.32cm]
  \Box
  \end{array}}\!\!}

\newcommand{\mydoop}{\!\!{\small\begin{array}{c}
  \Box\\[-0.32cm]
  \Box
  \end{array}}_{\!\!\!\!{}^{\scalebox{0.7}{+}}}\!\!}

\newcommand{\mydoom}{\!\!{\small\begin{array}{c}
  \Box\\[-0.32cm]
  \Box
  \end{array}}_{\!\!\!\!{}^{\scalebox{0.7}{$-$}}}\!\!}

\newcommand{\mydoopm}{\!\!{\small\begin{array}{c}
  \Box\\[-0.32cm]
  \Box
  \end{array}}_{\!\!\!\!{}^{\scalebox{0.7}{$\pm$}}}\!\!}

\newcommand{\sydoo}{\!\!\!{\tiny\begin{array}{c}
  \Box\\[-0.15cm]
  \Box
  \end{array}}\!\!\!}

\newcommand{\sydoop}{\!\!\!{\tiny\begin{array}{c}
  \Box\\[-0.15cm]
  \Box
  \end{array}}_{\!\!\!\!\scalebox{0.7}{+}}\!\!}

\newcommand{\sydoom}{\!\!\!{\tiny\begin{array}{c}
  \Box\\[-0.15cm]
  \Box
  \end{array}}_{\!\!\!\!\scalebox{0.7}{$-$}}\!\!}

\newcommand{\Yd}{\texttt{Y}}

In this section, we determine the grade restriction rule
in theories which are closely related to the R\o dland model:
the dual of the R\o dland model, the Hosono-Takagi model and its dual,
and $N$-generalizations of the R\o dland model.
The main purpose is to show that the method we developed
is powerful enough to be applicable to other models.
There is a lot to be done in each of the models, such as
constructing empty branes in the weakly coupled phase,
finding the low energy behaviour of branes in the strongly coupled phase,
presentation of the equivalence of categories as
Fourier-Mukai transforms, etc.
We limit ourselves to the determination of the grade restriction rule
and some simple consequences that can be drawn from that.
However, even with this limited discussion, we find a glimpse of
a number of interesting features such as secret strong coupling,
the monodromy around the singular point at which there are multiple
Coulomb branchs, and duality between branes.
More details will be presented in separate accounts.

In these models, the gauge group is of the form
\beq
G={U(1)\times H\over \{(\pm 1, \pm {\bf 1}_H)\}},
\label{Gtype}
\eeq
where $H$ is a symplectic or (special) orthogonal group.
Its subsets $\{[z,{\bf 1}_H];z\in U(1)\}$
and $\{[1,h];h\in H\}$ are subgroups isomorphic to $U(1)$ and $H$ respectively,
and they generate $G$.
An irreducible representation of $G$ is of the form
$\Yd(i)$ where $\Yd$ is a label of
an irreducible representation of $H$
and $i$ is an integer: $H\subset G$ acts as $\Yd$
and the $U(1)$ charge is $|\Yd|+2i$ where $|\Yd|$ is an integer associated to
$\Yd$.
We follow the convention of \cite{Weyl} for the label $\Yd$:
basically\footnote{An additional classifier other than a Young diagram
  is needed for some representations of (special) orthogonal groups.
See Sections~\ref{subsec:HT} and \ref{subsec:dualHT}.}
it is a Young diagram and $|\Yd|$
is the number of boxes.
The defining vector representation of $H$
is labelled by the diagram with one box, $\Yd=\Box$, with $|\Box|=1$.
Also, the trivial one-dimensional representation of $H$ is labeled by
the empty diagram, $\Yd=\emptyset$, with $|\emptyset|=0$.
We shall denote $\emptyset(i)$ by $\C(i)$.
We oftern denote $\Yd(0)$ simply by $\Yd$.
Then, $\Yd(i)\cong \Yd\otimes\C(i)$.
Likewise, we shall write $\rmV\otimes \C(i)=:\rmV(i)$
for a general (not necessarily irreducible) representation $\rmV$. 
The gauge group $U(2)$ of the R\o dland model is also of the type
(\ref{Gtype}) with $H=SU(2)=USp(2)$:
\beq
U(2)\cong {U(1)\times USp(2)\over \{(\pm 1,\pm{\bf 1}_2)\}}
\label{U2iso}
\eeq
In the diagramatic description,
$\Sym^h \fund (i)$ is $\Box\cdots \Box(i)$ (with $h$ boxes).
For $h=0$, it is $\det^{\otimes i}\fund$ and its notation
$\C(i)$ is already used earlier in this paper.

\subsection{Dual R\o dland Model}
\label{subsec:dualRodland}

\newcommand{\tx}{\tilde{x}}

The dual R\o dland model is obtained by applying the 2d Seiberg duality
\cite{Hduality} to the $USp(2)$ factor of the gauge group (\ref{U2iso})
of the R\o dland model.
The gauge group is
\beq
G^{\vee}={U(1)\times USp(4)\over \{(\pm 1,\pm {\bf 1}_4)\}}.
\eeq
The matter consists of seven fields $\tx^1,\ldots, \tx^7$
in $\Box(-1)$, seven fields $p^1,\ldots, p^7$ in $\C(-1)$,
and ${7\choose 2}=21$ fields $a_{ij}=-a_{ji}$ ($i,j=1,\ldots, 7$)
in $\C(1)$. The superpotential and the twisted superpotential are
\beqa
W^{\vee}&=&\sum_{i,j}a_{ij}[\tx^i\tx^j]+\sum_{i,j,k}A^{ij}_kp^ka_{ij},
\label{WdualRodland}\\
\wt{W}^{\vee}&=&-2t^{\vee}\sigma_{\mathfrak{u}(1)}.
\label{twistedWdualRodland}
\eeqa
The expression $[\tx^i\tx^j]$ in the superpotential
is the $USp(4)$ invariant skewsymmetric pairing.

The model has two phases. The $\zeta^{\vee}\ll 0$ phase is weakly coupled.
A solution to the vacuum equation breaks the gauge group completely,
and the vacuum manifold is the locus of $(p,\tx,a)$ with $a=0$,
$p\ne 0$ and $(p,\tx)$ solving the F-term equation
\beq
F^{ij}(p,\tx):=A^{ij}(p)+[\tx^i\tx^j]=0,
\label{FdualRod}
\eeq
modded out by the $G^{\vee}_{\C}$ action. This is isomorphic to
the Pfaffian Calabi-Yau manifold $Y$ of the R\o dland model.
Indeed, if $(p,\tx)$ solves (\ref{FdualRod}), then $A(p)$ is of rank 4,
and conversely, for each such $p$ there is an $\tx$ solving (\ref{FdualRod}) 
which is unique up to gauge. The $\zeta^{\vee}\gg 0$ phase is strongly coupled.
The D-term equation requires $a\ne 0$ which leaves
the $USp(4)$ subgroup unbroken.
After understanding the $USp(4)$ gauge dynamics, we find that
the theory reduces at low energies to
the sigma model on the rank 2 locus of $a$ satisfying the F-term equation
\beq
\sum_{i,j}A^{ij}_ka_{ij}=0.
\label{FdualRodstrong}
\eeq
Regarding $a_{ij}$ as the Pl\"ucker coordinate for the Grassmannian
$G(2,7)\subset \CP^{20}$,
the target is nothing but our Calabi-Yau manifold $X$.
As opposed to the original R\o dland model,
$X$ appears in the strongly coupled phase
and $Y$ appears in the weakly coupled phase in the dual model.

\newcommand{\ul}{\underline}

A maximal torus $T^{\vee}\subset G^{\vee}$ can be taken to be the set
of elements of the form
\beq
\left[\left(h_{\ul{0}},\left(\begin{array}{cccc}
    h_{\ul{1}}&&&\\
    &h_{\ul{2}}&&\\
    &&h_{\ul{1}}^{-1}&\\
    &&&h_{\ul{2}}^{-1}
  \end{array}\right)\right)\right],
\eeq
where $h_{\ul{a}}$ are on the unit circle $U(1)$ with identification
$(h_{\ul{0}},h_{\ul{1}},h_{\ul{2}})\sim (-h_{\ul{0}},-h_{\ul{1}},-h_{\ul{2}})$.
A single valued $U(1)^3$ parametrization is
$(h_0,h_1,h_2)=(h_{\ul{0}}^2,h_{\ul{0}}h_{\ul{1}},h_{\ul{0}}h_{\ul{2}})$.
The scalar components of the vector multiplet in the two parametrizations
are related by
\beq
\usigma_0=2\usigma_{\ul{0}},\quad
\usigma_1=\usigma_{\ul{0}}+\usigma_{\ul{1}},\quad
\usigma_2=\usigma_{\ul{0}}+\usigma_{\ul{2}}.
\label{sigmaulsigma}
\eeq
Note that $\usigma_{\mathfrak{u}(1)}$ in (\ref{twistedWdualRodland})
is $\usigma_{\ul{0}}$.
We shall use both coordinate systems of $\ttt_{\C}$: we shall refer to
the ones for $(h_{\ul{0}},h_{\ul{1}},h_{\ul{2}})$ and $(h_0,h_1,h_2)$
the {\it original frame} and 
the {\it single valued frame}, respectively.
The Weyl group is $(\Z_2)^3$ generated by the exchange
$\usigma_{\ul{1}}\leftrightarrow\usigma_{\ul{2}}$ and the sign flips
of $\usigma_{\ul{1}}$ and $\usigma_{\ul{2}}$, which are the exchange
$\usigma_1\leftrightarrow \usigma_2$ and $\usigma_s\to \usigma_0-\usigma_s$
for $s=1,2$.
Half the sum of positive roots $\rho$ is
\beq
h^{\rho}=h_{\ul{1}}^2h_{\ul{2}}=h_0^{-{3\over 2}}h_1^2h_2.
\eeq
It is equal to $h_0^{1\over 2}$ modulo a weight.
The weights of the matter fields are
$h_{\ul{0}}^{-2}=h_0^{-1}$ for the $p^k$,
$h_{\ul{0}}^{-1}(h_{\ul{1}},h_{\ul{2}},h_{\ul{1}}^{-1},h_{\ul{2}}^{-1})
=(h_0^{-1}h_1,h_0^{-1}h_2,h_1^{-1},h_2^{-1})$ for the $(\tx_1^i,\ldots, \tx_4^i)$
and $h_{\ul{0}}^2=h_0$ for the $a_{ij}$.
The effective twisted superpotential is thus
\beqa
\wt{W}^{\vee}_{\it eff}
&=&-t^{\vee}\usigma_0+\pi \im \usigma_0-7(-\usigma_0)(\log(-\usigma_0)-1)
-21\usigma_0(\log\usigma_0-1)\nn\\
&&-7\sum_{s=1}^2(\usigma_s-\usigma_0)(\log(\usigma_s-\usigma_0)-1)
-7\sum_{s=1}^2(-\usigma_s)(\log(-\usigma_s)-1).
\eeqa
The vacuum equation reads
\beq
\e^{t^{\vee}}={(\usigma_1-\usigma_0)^7(\usigma_2-\usigma_0)^7\over
  \usigma_0^{14}},\quad
  {(\usigma_s-\usigma_0)^7
    \over (-\usigma_s)^7}=1~~(s=1,2).
\eeq
A solution is $\usigma_0/\usigma_s-1=\omega_s$ with $\omega_s^7=1$ for $s=1,2$,
that is,
\beq
\usigma_1={\usigma_0\over \omega_1+1},\quad
\usigma_2={\usigma_0\over \omega_2+1},\qquad
\omega_1^7=\omega_2^7=1,
\label{dualRCB}
\eeq
and it exists only when $\e^{t^{\vee}}=1/((1+\omega_1)^7(1+\omega_2)^7)$.
The Weyl group generators act as
$(\omega_1,\omega_2)\to(\omega_2,\omega_1),(\omega_1^{-1},\omega_2), (\omega_1,\omega_2^{-1})$ and the fixed points are excluded as
they have enhanced $SU(2)$ symmetry \cite{HoTo,longflow}.
There are only three inequivalent cases:
$(\omega_1,\omega_2)=(\omega,\omega^2), (\omega,\omega^3), (\omega^2,\omega^3)$
with $\omega=\e^{2\pi\im/7}$.
As $(1+\omega)(1+\omega^2)(1+\omega^3)=\omega^3$, these correspond
respectively to
\beq
\e^{t^{\vee}}\,=\,(1+\omega^3)^7,~\,(1+\omega^2)^7,~\,(1+\omega)^7.
\eeq
These are nothing but the values of $\e^t$ of the singular points
of the original model labelled by $\bfa=3,2,1$ in (\ref{defalphas}).
This means that the FI-theta parameters of
the original and the dual models are related by
\beq
t=t^{\vee}\quad\mbox{modulo $2\pi\im$.}
\eeq

The label $\Yd$ of an irreducible representation of $USp(4)$ is
a Young diagram of height at most two \cite{Weyl}.
It is represented on
\beq
\rmV_\Yd=c_{\Yd}\left[(\C^4)^{\otimes |\Yd|}\right]_0
\eeq
where $\C^4$ is the defining representation of $USp(4)$,
$|\Yd|$ is the number of boxes in $\Yd$,
$c_{\Yd}$ is the Young symmetrizer for $\Yd$,
and $[-]_0$ is the subspace consisting of vectors that are traceless with
respect to the symplectic form on $\C^4$.
As described earlier, an irreducible representation of $G^{\vee}$
is labeled by $\Yd(i)$.
If the number of boxes of the first and the second rows of
$\Yd$ are $\Yd_1$ and $\Yd_2$, the highest weight of $\Yd(i)$ is
$h_{\ul{0}}^{2i+|\Yd|}h_{\ul{1}}^{\Yd_1}h_{\ul{2}}^{\Yd_2}=h_0^ih_1^{\Yd_1}h_2^{\Yd_2}$.
Thus the shifted highest weight $\wtq$ for this representation is
\beq
h^{\wtq}=h_{\ul{0}}^{2i+|\Yd|}h_{\ul{1}}^{\Yd_1+2}h_{\ul{2}}^{\Yd_2+1}
=h_0^{i-{3\over 2}}h_1^{\Yd_1+2}h_2^{\Yd_2+1}.
\label{wtqforYi}
\eeq
The growth function that controles
the convergence of the hemisphere integral (\ref{ZV}) is
\beqa
\AN_{\wtq}&=&2\zeta^{\vee}\Is_{\ul{0}}-2\theta^{\vee}\Rs_{\ul{0}}
-2\pi \langle\wtq,\Rs\rangle\nn\\
&&+7(-2\Rs_{\ul{0}})\Arg(-2\im\usigma_{\ul{0}})
+21(2\Rs_{\ul{0}})\Arg(2\im\usigma_{\ul{0}})\nn\\
&&+\sum_{s=1}^2\bigl(7(-\Rs_{\ul{0}}+\Rs_{\ul{s}})
\Arg(\im(-\usigma_{\ul{0}}+\usigma_{\ul{s}}))
+7(-\Rs_{\ul{0}}-\Rs_{\ul{s}})
\Arg(\im(-\usigma_{\ul{0}}-\usigma_{\ul{s}}))\bigr)\nn\\
&&+7(-2\Is_{\ul{0}})\log|2\usigma_{\ul{0}}|
+21(2\Is_{\ul{0}})\log|2\usigma_{\ul{0}}|\nn\\
&&+\sum_{s=1}^2\bigl(
7(-\Is_{\ul{0}}+\Is_{\ul{s}})\log|-\usigma_{\ul{0}}+\usigma_{\ul{s}}|
+7(-\Is_{\ul{0}}-\Is_{\ul{s}})\log|-\usigma_{\ul{0}}-\usigma_{\ul{s}}|
\bigr).
\label{ANdualR}
\eeqa
The integration contour $\bgamma\subset\ttt_{\C}$
is a deformation of the real locus
$\im\ttt$ without meeting the poles of the Gamma function factors,
which is assured if it avoids the walls
\beqa
W_p&=&\{\,-2\usigma_{\ul{0}}\in\im\R_{>0}\,\},\label{WpdualR}\\
W_a&=&\{\,2\usigma_{\ul{0}}\in \im\R_{>0}\,\},\\
W_1&=&\{\,-\usigma_{\ul{0}}+\usigma_{\ul{1}}\in\im \R_{>0}\,\},\\
W_2&=&\{\,-\usigma_{\ul{0}}+\usigma_{\ul{2}}\in\im \R_{>0}\,\},\\
W_3&=&\{\,-\usigma_{\ul{0}}-\usigma_{\ul{1}}\in\im \R_{>0}\,\},\\
W_4&=&\{\,-\usigma_{\ul{0}}-\usigma_{\ul{2}}\in\im \R_{>0}\,\}.
\label{W4dualR}
\eeqa
If we regard $\bgamma$ as a graph of a function $\Rs\mapsto\Is(\Rs)$,
the condition to avoid the walls is
\beqa
\Rs_{\ul{0}}=0&\Longrightarrow& \Is_{\ul{0}}=0,\label{Wedgepa}\\
\Rs_{\ul{0}}\pm \Rs_{\ul{1}}=0&\Longrightarrow&
\Is_{\ul{0}}\pm \Is_{\ul{1}}\geq 0,\label{Wedge13}\\
\Rs_{\ul{0}}\pm \Rs_{\ul{2}}=0&\Longrightarrow&
\Is_{\ul{0}}\pm \Is_{\ul{2}}\geq 0.\label{Wedge24}
\eeqa
In addition to the scaling property and Weyl symmetry, 
the growth function for the shifted highest weight (\ref{wtqforYi})
has reflection symmetry
at $\theta^{\vee}=\pi n$ for $n\in\Z$:
\beq
\theta^{\vee}\to -\theta^{\vee}+2\pi n,~~
\Yd(i)\to \Yd(-n-i-|\Yd|),~~
\left\{\begin{array}{c}
(\Rs_{\ul{0}},\Rs_{\ul{1}},\Rs_{\ul{2}})\to
(-\Rs_{\ul{0}},\Rs_{\ul{1}},\Rs_{\ul{2}}),\\
(\Is_{\ul{0}},\Is_{\ul{1}},\Is_{\ul{2}})\to
(\Is_{\ul{0}},-\Is_{\ul{1}},-\Is_{\ul{2}}).
\end{array}\right.
\label{dualRodrefsym}
\eeq
This exchanges the walls as $W_1\leftrightarrow W_3$
and $W_2\leftrightarrow W_4$ while keeping $W_p$ and $W_a$ fixed.

Let us discuss the implication of the condition (\ref{Wedgepa}).
Along the hyperplane $\Rs_{\ul{0}}=0$, we must take $\Is_{\ul{0}}=0$.
In this direction, the growth function reads
\beqa
\AN_{\wtq}&=&-2\pi\sum_{s=1}^2\wtq^{\ul{s}}\Rs_{\ul{s}}
+\sum_{s=1}^2\left(\Rs_{\ul{s}}\Arg(\im\usigma_{\ul{s}})
-\Rs_{\ul{s}}\Arg(-\im\usigma_{\ul{s}})\right)\nn\\
&=&2\pi\sum_{s=1}^2\left(-\wtq^{\ul{s}}\Rs_{\ul{s}}+{7\over 2}|\Rs_{\ul{s}}|\right)
\label{AqUSp4QCD}
\eeqa
where we used (\ref{Argid}) again.
Requiring this to grow at infinity, we find
the condition
\beq
|\wtq^{\ul{1}}|, ~|\wtq^{\ul{2}}|\,<\,{7\over 2}
\label{wtq12dualR}
\eeq
As $\wtq^{\ul{1}}=\Yd_1+2>\wtq^{\ul{2}}=\Yd_2+1$ by (\ref{wtqforYi}),
this is equivalent to $\Yd_1<3/2$ which is solved by $\Yd_1=0,1$.
That is, the width $\Yd_1$ of the Young diagram $\Yd$ is at most $1$.
Only the three series of representations
\beq
\C(i),\quad
\Box(i),\quad
\ydoo(i)
\label{USp4GRR}
\eeq
are allowed.

Requiring (\ref{AqUSp4QCD}) to grow at infinity is
nothing but the grade restriction rule for $USp(4)$ SQCD with seven
fundamentals.
We may refer to (\ref{USp4GRR}) as the $USp(4)$ grade restriction rule.
This requirement is understandable in the strongly coupled phase
$\zeta^{\vee}\gg 0$ where we have a $USp(4)$ SQCD with seven fundamentals
fibered over the space of $a$'s at an intermediate energy scale.
This is just like the $SU(2)$ grade restriction rule (\ref{GRRstrong})
is required in the strongly coupled phase of the original R\o dland model.
However, we are requiring (\ref{USp4GRR}) irrespective of the value of
$\zeta^{\vee}$. This especially sounds odd in the weakly coupled phase
$\zeta^{\vee}\ll 0$ where the gauge group is completely broken.
At this point, we recall that the F-term equation (\ref{FdualRod})
played the important r\^ole in the complete breaking of the gauge symmetry.
Configurations with $a\ne 0$, $p=0$, $\tx=0$ can solve the D-term equation
and leave the $USp(4)$ gauge symmetry unbroken.
At the Higgsing scale, if it is set higher than
the superpotential scale, we have the $USp(4)$ SQCD with seven fundamentals
fibered over the space of $p$'s.
In this sense, the theory at $\zeta^{\vee}\ll 0$ is secretly strongly
coupled --- it is weakly coupled only at low enough energy scales
where the effect of the superpotential is not negligible.
This may be responsible for the $USp(4)$ grade restriction rule
(\ref{USp4GRR}) in the $\zeta^{\vee}\ll 0$ phase.
This is in contrast to the $\zeta\gg 0$ phase of the original R\o dland model
where the D-term equation alone breaks the gauge symmetry completely.

As long as the $USp(4)$ grade restriction rule (\ref{USp4GRR})
is satisfied, convergent contours can be found in the two phases.
Let us consider the contours
\beqa
\bgamma_+&=&
\bigl\{\,\Is_{\ul{0}}=\Rs_{\ul{0}}^2,~\Is_{\ul{1}}=\Is_{\ul{2}}=0\,\bigr\},
\label{bgammapdual}\\
\bgamma_-&=&\bigl\{\,\Is_{\ul{0}}=-\Rs_{\ul{0}}^2,~
\Is_{\ul{1}}=-\Rs_{\ul{0}}\Rs_{\ul{1}},~
\Is_{\ul{2}}=-\Rs_{\ul{0}}\Rs_{\ul{2}}
\,\,\bigr\},
\label{bgammamdual}
\eeqa
which satisfy the wall avoiding conditions (\ref{Wedgepa})-(\ref{Wedge24}).
The function $\AN_{\wtq}$ for $\pm \zeta^{\vee}\gg 0$ grows at infinity of
$\bgamma_{\pm}$ if $\wtq$ satisfies (\ref{wtq12dualR}). That is, 
for any of the representations in (\ref{USp4GRR}),
the hemisphere integral on the contour $\bgamma_+$ ({\it resp}. $\bgamma_-$)
is absolutely convergent in the phase $\zeta^{\vee}\gg 0$
({\it resp}. $\zeta^{\vee}\ll 0$).

\begin{figure}[thb]
\psfrag{A}{\mbox{\small ${\mathcal P}_{\!\bf A}$}}
\psfrag{B}{\mbox{\small ${\mathcal P}_{\bf B}$}}
\psfrag{C}{\mbox{\small ${\mathcal P}_{\bf C}$}}
\psfrag{D}{\mbox{\small ${\mathcal P}_{\bf D}$}}
\psfrag{a1}{$t^{\vee}_1$}
\psfrag{a2}{$t^{\vee}_2$}
\psfrag{a3}{$t^{\vee}_3$}
\psfrag{z}{$\zeta^{\vee}$}
\psfrag{th}{$\theta^{\vee}$}
\psfrag{-8pi}{\!\!\!$-22\pi$\,\,\,}
\psfrag{-6pi}{\!$-20\pi$\,}
\centerline{\includegraphics[page=46]{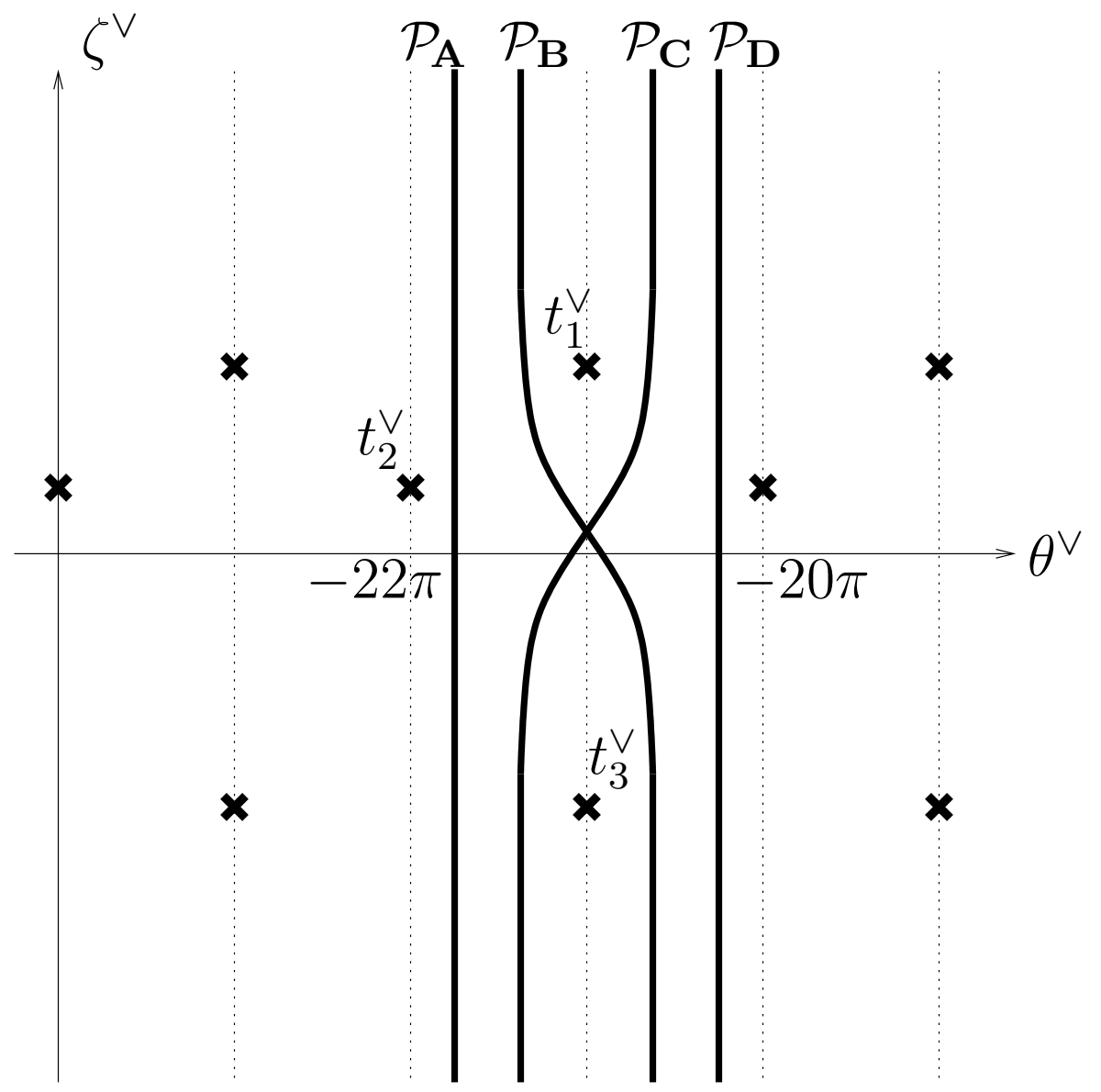}}
\caption{Four windows between the phases}
\label{fig:dualRodpathsX}
\end{figure}
Next, let us determine the grade restriction rule for paths between
the two phases. Let us consider the four classes of paths
${\mathcal P}_{\bf A},$
${\mathcal P}_{\bf B},$
${\mathcal P}_{\bf C},$
${\mathcal P}_{\bf D}$
that go near the three points
\beqa
t^{\vee}_1&=&21\pi \im +7\log\left(2\cos\left({\pi\over 7}\right)\right),
\label{dualt1}
\\
t^{\vee}_2&=&22\pi\im+7\log\left(2\cos\left({2\pi\over 7}\right)\right),
\label{dualt2}\\
t^{\vee}_3&=&21\pi\im+7\log\left(2\cos\left({3\pi\over 7}\right)\right),
\label{dualt3}
\eeqa
as depicted in Fig.~\ref{fig:dualRodpathsX}.
We take the following procedure to find the rule:
we first determine the special representations at the singular points,
make a guess for the rule, and then confirm it by careful analysis.

\begin{figure}[thb]
  \psfrag{w1}{$W_1$}
  \psfrag{w2}{$W_2$}
  \psfrag{w3}{$W_3$}
  \psfrag{w4}{$W_4$}
  \psfrag{wp}{$W_p$}
  \psfrag{wa}{$W_a$}
  \psfrag{sp}{$\usigma_0$}
  \psfrag{I}{${\rm I}$}
  \psfrag{II}{${\rm II}$}
  \psfrag{III}{${\rm III}$}
  \psfrag{IV}{${\rm IV}$}
  \psfrag{V}{${\rm V}$}
  \psfrag{VI}{${\rm VI}$}
\centerline{\includegraphics[page=47]{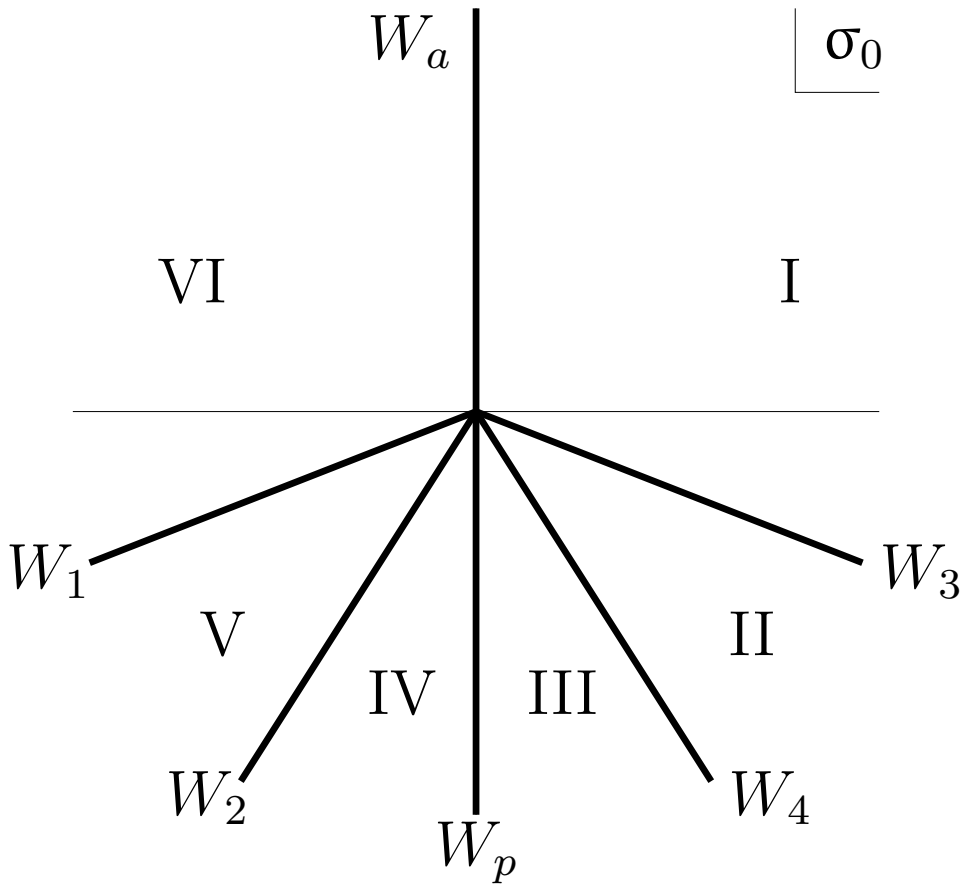}}
\caption{Chambers of the Coulomb branch.}
\label{fig:CBdualR}
\end{figure}

The special representation at the singular point $t^{\vee}=t^{\vee}_*$
has shifted highest weight $\wtq_*$ such that
$\dd \wt{W}_{\!{\it eff}, \wtq_*}=0$ in a chamber of the Coulomb branch
(\ref{dualRCB}). In the single valued frame, the equation reads
\beqa
2\pi\im\wtq_*&=&t^{\vee}_*+\sum_iQ_i\log(\im Q_i(\usigma))\nn\\
&=&(t^{\vee}_*,0,0)+7(-1,0,0)\log(-\im\usigma_0)
+21(1,0,0)\log(\im\usigma_0)\nn\\
&&+7(-1,1,0)\log(\im(\usigma_1-\usigma_0))
+7(-1,0,1)\log(\im(\usigma_2-\usigma_0))\nn\\
&&+7(0,-1,0)\log(-\im\usigma_1)
+7(0,0,-1)\log(-\im\usigma_2),
\label{1stwtq}
\eeqa
in which we insert (\ref{dualRCB}).
For $\omega_s=\e^{2\pi\im n_s/7}$, $s=1,2$, we have
\beq
\usigma_s={\e^{-{\pi\im n_s\over 7}}\usigma_0\over
  2\cos\left({\pi n_s\over 7}\right)},\quad
\usigma_s-\usigma_0=-{\e^{{\pi\im n_s\over 7}}\usigma_0\over
  2\cos\left({\pi n_s\over 7}\right)}.
\eeq
Recall that $t_*^{\vee}=t^{\vee}_3, t^{\vee}_2, t^{\vee}_1$ correspond
to
$(\omega_1,\omega_2)=(\omega,\omega^2), (\omega,\omega^3), (\omega^2,\omega^3)$
with $\omega:=\e^{2\pi \im/7}$ or their Weyl images.
It is easy to see that the real part on the right-hand side
of (\ref{1stwtq}) vanishes under this correspondence.
To find the imaginary part, we need to evaluate
$\Arg(\im Q_i(\usigma))$ which are assumed to take values in
$(-{\pi\over 2},{\pi\over 2})$.
They depend on the chamber of the Coulomb branch separated by the walls
(\ref{WpdualR})-(\ref{W4dualR}).
The chambers for the Weyl representative of the Coulomb branch
with $1\leq n_2<n_1\le 3$
are as depicted in Fig.~\ref{fig:CBdualR},
and $\Arg_i=\Arg(\im Q_i(\usigma))$
are given by (\ref{tableArgsDR})
\beq
\begin{array}{c|cccccc|c}
  &\Arg_p&\Arg_a&\Arg_1
  &\Arg_2&\Arg_3&\Arg_4
  &{1\over 2\pi}\sum_iQ_i\Arg_i\\
  \hline
              {\rm I}&\varphi&\varphi+\pi
              &\varphi+{\pi n_1\over 7}
              &\varphi+{\pi n_2\over 7}
              &\varphi-{\pi n_1\over 7}
              &\varphi-{\pi n_2\over 7}
              &({21-n_1-n_2\over 2},n_1,n_2)\\
              {\rm II}&{}_{''}&{}_{''}&{}_{''}&{}_{''}&{}_{''}+2\pi&{}_{''}
              &({21-n_1-n_2\over 2},n_1-7,n_2)\\
              {\rm III}&{}_{''}&{}_{''}&{}_{''}&{}_{''}&{}_{''}&{}_{''}+2\pi
              &({21-n_1-n_2\over 2},n_1-7,n_2-7)\\
              {\rm IV}&{}_{''}+2\pi&{}_{''}&{}_{''}&{}_{''}&{}_{''}&{}_{''}
              &({7-n_1-n_2\over 2},n_1-7,n_2-7)\\
              {\rm V}&{}_{''}&{}_{''}&{}_{''}&{}_{''}+2\pi&{}_{''}&{}_{''}
              &({-7-n_1-n_2\over 2},n_1-7,n_2)\\
              {\rm VI}&{}_{''}&{}_{''}&{}_{''}+2\pi&{}_{''}&{}_{''}&{}_{''}
              &({-21-n_1-n_2\over 2},n_1,n_2)
\end{array}
\label{tableArgsDR}
\eeq
where $\usigma_0=\im|\usigma_0|\e^{\im\varphi}$
with $-2\pi<\varphi<0$.
The sum ${1\over 2\pi}\sum_iQ_i\Arg_i$ must be equated with
$\wtq_*+({\theta^{\vee}_*\over 2\pi},0,0)$.
We notice that $\wtq_*$ in the chambers I and VI
are shifted highest weights of some $USp(4)$ grade restricted
representations of the gauge group,
$(i-{3\over 2},\Yd_1+2,\Yd_2+1)$ for $\Yd(i)$ as shown in (\ref{wtqforYi}):
\beq
\begin{array}{c|c|c|c}
  t_*^{\vee}&(n_1,n_2) &\mbox{$\wtq_*$:$\Yd(i)$ on I}
  &\mbox{$\wtq_*$:$\Yd(i)$ on VI}\\
  \hline
  t_3^{\vee}&(2,1)&(21-{3\over 2},2,1):\C(21)&(-{3\over 2},2,1):\C\\
  t_2^{\vee}&(3,1)&(21-{3\over 2},3,1):\Box(21)&(-{3\over 2},3,1):\Box\\
  t_1^{\vee}&(3,2)&(21-{5\over 2},3,2):\ydoo(20)&(-{5\over 2},3,2):\ydoo(-1)
\end{array}
\label{specialrepdualR}
\eeq
On the other hand, $\wtq_*$ in the other chambers or on the other Weyl
representatives are not the shifted highest weights of any of
the $USp(4)$ grade restricted representations.
Thus, we conclude that those in the table (\ref{specialrepdualR})
are the special representations at the three points.
The special representations at $t^{\vee}_*+2\pi\im j$
are obtained from those at
$t^{\vee}_*$ by $\rmV\mapsto \rmV(j)$.

We anticipate that the two representations (\ref{specialrepdualR})
at $t_*^{\vee}$ undergo a catastrophe along a small loop around $t_*^{\vee}$,
as described in Section~\ref{subsec:catastroph} for the
original R\o dland model.
A more precise expectation is:
the representation $\C(21)$ ({\it resp}. $\C$) cannot stay admissible
on the right ({\it resp}. left) of  $t_3^{\vee}$,
and similarly for
$\Box(21)$ ({\it resp}. $\Box$) on the right ({\it resp}. left) of $t_2^{\vee}$,
and $\mydoo(20)$ ({\it resp}. $\mydoo(-1)$) on the right ({\it resp}. left)
of $t_1^{\vee}$.
If $\rmV(21)$ and $\rmV$ are bad on the right and the left of $t_*^{\vee}$,
so are $\rmV(21+j)$ and $\rmV(j)$ on the right and the left of
$t_*^{\vee}+2\pi\im j$.
Then, the grade restriction rule with respect to the four windows would be
\beqa
\!\!\!\!\!\!\!\!\!\!\!\!
{\rm GRR}_{\bf A}&&
\begin{array}{cccccccc}
&&\C(1)&\cdots&\C(19)&\C(20)&\C(21)\\
&~\Box~&\Box(1)&\cdots&\Box(19)&\Box(20)&\\
  ~~~~~~~~~&~\ydoo~&\ydoo(1)&\cdots&\ydoo(19)&\ydoo(20)&
\end{array}
\label{dGRRA}\\[0.2cm]
\!\!\!\!\!\!\!\!\!\!\!\!
{\rm GRR}_{\bf B}&&
\begin{array}{ccccccc}
&~\C~&\C(1)&\cdots&\C(19)&\C(20)&\\
&~\Box~&\Box(1)&\cdots&\Box(19)&\Box(20)&\\
~~~~~~~~~&~\ydoo~&\ydoo(1)&\cdots&\ydoo(19)&\ydoo(20)&
\end{array}
\label{dGRRB}\\[0.2cm]
\!\!\!\!\!\!\!\!\!\!\!\!
{\rm GRR}_{\bf C}&&
\begin{array}{ccccccc}
&&\C(1)&\cdots&\C(19)&\C(20)&\C(21)\\
&~\Box~&\Box(1)&\cdots&\Box(19)&\Box(20)&\\
\,\,\ydoo(-1)\!\!&~\ydoo~&\ydoo(1)&\cdots&\ydoo(19)&&
\end{array}
\label{dGRRC}\\[0.2cm]
\!\!\!\!\!\!\!\!\!\!\!\!
{\rm GRR}_{\bf D}&&
\begin{array}{cccccccc}
&~\C~&\C(1)&\cdots&\C(19)&\C(20)&\\
&~\Box~&\Box(1)&\cdots&\Box(19)&\Box(20)&\\
\,\,\ydoo(-1)\!\!&~\ydoo~&\ydoo(1)&\cdots&\ydoo(19)&&
\end{array}
\label{dGRRD}
\eeqa
This is the target proposal which we would like to confirm.

\newcommand{\brq}{\breve{q}}

First, let us see which representations are allowed if we had chosen the real
contour $\bgamma=\im\ttt$.
The growth function on the real locus $\Is=0$ reads
\beqa
\AN_{\wtq}&=&-2\theta^{\vee}\Rs_{\ul{0}}
-2\pi\left(\wtq^{\ul{0}}\Rs_{\ul{0}}+\wtq^{\ul{1}}\Rs_{\ul{1}}+\wtq^{\ul{2}}\Rs_{\ul{2}}\right)
+28\pi |\Rs_{\ul{0}}|\nn\\
&&+{7\pi\over 2}\left(
|\Rs_{\ul{0}}+\Rs_{\ul{1}}|+
|\Rs_{\ul{0}}-\Rs_{\ul{1}}|+
|\Rs_{\ul{0}}+\Rs_{\ul{2}}|+
|\Rs_{\ul{0}}-\Rs_{\ul{2}}|\right).
\eeqa
Decomposing the $\Rs$-space into regions of various signs of
$\Rs_{\ul{0}}$, $\Rs_{\ul{0}}\pm\Rs_{\ul{s}}$
and working on each region, we find that the growth of
$\AN_{\wtq}$ at infinity requires
\beq
\epsilon_0\left(\wtq^{\ul{0}}+{\theta^{\vee}\over \pi}\right)
+\epsilon_1\wtq^{\ul{1}}+\epsilon_2\wtq^{\ul{2}}+21\geq 0,
\quad\epsilon_1\wtq^{\ul{1}}+{7\over 2}\geq 0,
\quad\epsilon_2\wtq^{\ul{2}}+{7\over 2}\geq 0,
\label{condrealdualR}
\eeq
for $\epsilon_0,\epsilon_1,\epsilon_2\in \{\pm 1\}$.
The latter two inequalities are nothing but (\ref{wtq12dualR}), which is
the $USp(4)$ grade rerstriction rule, and is satisfied by (\ref{USp4GRR}):
$\C(i)$, $\Box(i)$ and $\mydoo(i)$ for any $i\in\Z$.
The first one imposes a condition on $i$. If the theta parameter is in the
interval $-22\pi<\theta^{\vee}<-20\pi$, it is satisfied by
\beq
\begin{array}{cccccc}
  \C(2)&\C(3)&\cdots&\C(17)&\C(18)&\C(19)\\
  \Box(2)&\Box(3)&\cdots&\Box(17)&\Box(18)&\\
  \ydoo(2)&\ydoo(2)&\cdots&\ydoo(17)
\end{array}
\label{realGRRdualR}
\eeq
This is a common subset of of all four sets (\ref{dGRRA})-(\ref{dGRRD}).
In view of the fact that
the paths ${\mathcal P}_{\bf A}$, ${\mathcal P}_{\bf B}$,
${\mathcal P}_{\bf C}$ and ${\mathcal P}_{\bf D}$ all fit into this interval,
this is consistent with the proposal that (\ref{dGRRA})-(\ref{dGRRD})
are the grade restriction rules for the four windows.

Let us try to enlarge the set of allowed representations by turning on
the imaginary part $\Is$. Maintaining $\Is_{\ul{0}}=0$
as a simplifying assumption, we have
\beqa
\AN_{\wtq}&=&-2\theta^{\vee}\Rs_{\ul{0}}
-2\pi(\wtq^{\ul{0}}\Rs_{\ul{0}}+\wtq^{\ul{1}}\Rs_{\ul{1}}+\wtq^{\ul{2}}\Rs_{\ul{2}})
+28\pi |\Rs_{\ul{0}}|\nn\\
&&+\sum_{s=1}^2\bigl(7(-\Rs_{\ul{0}}+\Rs_{\ul{s}})
\Arg(\im(-\Rs_{\ul{0}}+\usigma_{\ul{s}}))
+7(-\Rs_{\ul{0}}-\Rs_{\ul{s}})
\Arg(\im(-\Rs_{\ul{0}}-\usigma_{\ul{s}}))\bigr)\nn\\
&&+\sum_{s=1}^2
7\Is_{\ul{s}}\log{|\usigma_{\ul{s}}-\Rs_{\ul{0}}|\over
|\usigma_{\ul{s}}+\Rs_{\ul{0}}|}.
\eeqa
If we take $|\Is_{\ul{s}}|\gg |\Rs_{\ul{s}}\pm\Rs_{\ul{0}}|$ at most places,
we can ignore the logarithm terms (the third line). After some trials,
we find that the largest number of representations is allowed if
$\Is_{\ul{s}}$ and $\Rs_{\ul{0}}\Rs_{\ul{s}}$ have the opposite signs
for $s=1,2$.
Then, the growth of $\AN_{\wtq}$ at infinity requires, in addition to
$|\wtq^{\ul{1}}|, |\wtq^{\ul{2}}|<7/2$,
\beq
\left|\,\wtq^{\ul{0}}+{\theta^{\vee}\over \pi}\,\right|<21.
\label{relaxeddualR}
\eeq
As an example of contour that satisfies the conditions, we may take
\beq
\bgamma_0=\Bigl\{\,\Is_{\ul{0}}=0,~
\Is_{\ul{1}}=-\Rs_{\ul{0}}\Rs_{\ul{1}}\|\Rs\|^2,~
\Is_{\ul{2}}=-\Rs_{\ul{0}}\Rs_{\ul{2}}\|\Rs\|^2\,\Bigr\},
\label{bgamma0dual}
\eeq
where $\|\Rs\|^2:=(\Rs_{\ul{0}})^2+(\Rs_{\ul{1}})^2+(\Rs_{\ul{2}})^2$.
It asymptotes to
\beq
\overline{\bgamma}_0=\underbrace{\left\{\begin{array}{l}
\Is_{\ul{0}}=\Rs_{\ul{1}}=\Rs_{\ul{2}}=0,\\
\Rs_{\ul{0}},\,\,\Is_{\ul{1}},\,\,\Is_{\ul{2}}\,\,\,\mbox{free}.
\end{array}
\right\}}_{\,\,\,\,\,\,\,\,\,\,\,\,\,\,\,\,\,\,\,\,=:\,\overline{\bgamma}_0^{(0)}}
\cup
\underbrace{\left\{\begin{array}{l}
\Rs_{\ul{0}}=\Is_{\ul{0}}=0,\\
(\Is_{\ul{1}},\Is_{\ul{2}})=(u\Rs_{\ul{1}},u\Rs_{\ul{2}}),\,\,u\in\R.
\end{array}
\right\}}_{\,\,\,\,\,\,\,\,\,\,\,\,\,\,\,\,\,\,\,\,=:\,\overline{\bgamma}_0^{(1)}},
\eeq  
on which
\beqa
\AN_{\wtq}|_{\overline{\sgamma}_0^{(0)}}&=&
-2\theta^{\vee}\Rs_{\ul{0}}-2\pi\wtq^{\ul{0}}\Rs_{\ul{0}}
  +42\pi|\Rs_{\ul{0}}|,\\
  \AN_{\wtq}|_{\overline{\sgamma}_0^{(1)}}&=&
      -2\pi\wtq^{\ul{1}}\Rs_{\ul{1}}-2\pi \wtq^{\ul{2}}\Rs_{\ul{2}}
      +7\pi|\Rs_{\ul{1}}|+7\pi|\Rs_{\ul{2}}|.\label{ANbgamma1}
\eeqa
The growth of $\AN_{\wtq}$ on $\overline{\bgamma}_0^{(0)}$
and $\overline{\bgamma}_0^{(1)}$ indeed
requires (\ref{relaxeddualR}) and (\ref{wtq12dualR}), respectively.
For $-22\pi<\theta^{\vee}<-21\pi$
({\it resp}. $-21\pi<\theta^{\vee}<-20\pi$) 
this is satisfied precisely by the representations in
(\ref{dGRRA}) ({\it resp}. (\ref{dGRRD})).
This holds for any value of $\zeta^{\vee}$ as
$\AN_{\wtq}$ is $\zeta^{\vee}$-independent thanks to
$\Is_{\ul{0}}=0$. Thus, the hemisphere integral on the contour $\bgamma_0$
is convergent for the representations in the set (\ref{dGRRA})
({\it resp}. (\ref{dGRRD}))
everywhere on the strip with $-22\pi<\theta^{\vee}<-21\pi$
({\it resp}. $-21\pi<\theta^{\vee}<-20\pi$).
Also, deep inside the phase with $\pm\zeta^{\vee}\gg 0$,
the contour $\bgamma_0$ can be deformed to $\bgamma_{\pm}$ 
while the hemisphere integral stays convergent.
Thus, the representations in (\ref{dGRRA}) ({\it resp}. (\ref{dGRRD})) are
grade restricted with respect to the window {\bf A} ({\it resp}. {\bf D})
that includes straight paths between $\zeta^{\vee}\gg 0$
and $\zeta^{\vee}\ll 0$ inside the strip.

Let us look at the behaviour of
$\AN_{\wtq_*}={\rm Re}(\im\wt{W}_{\!{\it eff},\wtq_*})$ 
at $t^{\vee}=t_*^{\vee}$ in a neighborhood of the Coulomb branch.
The Hessian of $\wt{W}_{\!{\it eff},\wtq_*}$ at the Coulomb branch
has one zero and two non-zero eigenvalues,
and hence we can find a coordinate system $(\usigma_0,z_1,z_2)$
of $\ttt_{\C}$ such that the Coulomb branch is at $z_1=z_2=0$ around which
\beq
\im \wt{W}_{\!{\it eff},\wtq_*}|^{}_{t^{\vee}=t_*^{\vee}}=z_1^2+z_2^2.
\eeq
Denoting $z_m=x_m+\im y_m$ and
putting $\vec{x}=(x_1,x_2)$ and $\vec{y}=(y_1,y_2)$,
we have
\beq
\AN_{\wtq_*}|_{t^{\vee}=t_*^{\vee}}={\rm Re}(z_1^2+z_2^2)
=\|\vec{x}\|^2-\|\vec{y}\|^2.
\label{ANdualR0}
\eeq
When $t^{\vee}$ is moved off from $t_*^{\vee}$, this deforms to
\beq
\AN_{\wtq_*}|_{t^{\vee}=t_*^{\vee}+\sDelta t^{\vee}}
=\|\vec{x}\|^2-\|\vec{y}\|^2-{\rm Re}(\im\sDelta t^{\vee}\usigma_0).
\label{ANdualRloc}
\eeq
For a fixed $\usigma_0$, the region where this is positive depends
very much on the sign of $c:={\rm Re}(\im\sDelta t^{\vee}\usigma_0)$.
The two-dimensional slices of this region,
one for $\vec{x}$ and one for $\vec{y}$, for
$c<0$, $c=0$ and $c>0$ are as in left, middle, and right of
Fig.~\ref{fig:shaded} respectively.
In particular, when $c>0$, there is a hole $\|\vec{x}\|^2<\|\vec{y}\|^2+c$
where (\ref{ANdualRloc}) becomes negative.

Let us locate the asymptotic region $\overline{\bgamma}_0$ of $\bgamma_0$
in the coordinate system $(\usigma_0,\vec{x},\vec{y})$.
The component $\overline{\bgamma}_0^{(1)}$ of $\overline{\bgamma}_0$
is away from the Coulomb branch
except at the origin $\usigma=0$.\footnote{The hemisphere integral
  on $\overline{\bgamma}_0^{(1)}$ is convergent as long as the representation
  is $USp(4)$ grade restricted, as one can see by (\ref{ANbgamma1}), and
  hence this component can be put aside from the discussion anyway.}
The component $\overline{\bgamma}_0^{(0)}$, 
\beq
\usigma_0=2\Rs_{\ul{0}},\quad
\usigma_1=\Rs_{\ul{0}}+\im \Is_{\ul{1}},\quad
\usigma_2=\Rs_{\ul{0}}+\im\Is_{\ul{2}},
\eeq
intersects the Coulomb branch (\ref{dualRCB}) along the line
$\Is_{\ul{s}}=-\tan\left({\pi n_s\over 7}\right)\Rs_{\ul{0}}$.
Let us reparametrize $\overline{\bgamma}_0^{(0)}$ by setting
$\Is_{\ul{s}}=-\tan\left({\pi n_s\over 7}\right)\Rs_{\ul{0}}+\xi_s$.
Then, a numerical computation shows that $\overline{\bgamma}_0$ is at
$z_1=a\e^{-{\pi\im\over 4}}\xi_1$ and $z_2=b\e^{-{\pi\im\over 4}}\xi_2$
for real positive $a$ and $b$
(after an appropriate $O(2,\C)$ rotation of $(z_1,z_2)$).
That is,
\beq
\usigma_0=2\Rs_{\ul{0}},\quad
\vec{x}=(a\xi_1,b\xi_2),\quad
\vec{y}=(-a\xi_1,-b\xi_2).
\eeq
Exactly at the singular point, $t^{\vee}=t_*^{\vee}$, it is at the edge of the
$\AN_{\wtq_*}>0$ region and can be deformed inside. 
Let us move away from the singular point, $t^{\vee}=t_*^{\vee}+\sDelta t^{\vee}$.
When we can arrange $c={\rm Re}(\im\sDelta t^{\vee}\usigma_0)\leq 0$,
it can still be deformed into the $\AN_{\wtq_*}>0$ region.
However, when $c>0$ is unavoidable,
any deformation of $\overline{\bgamma}_0$
is forced to cross the $\AN_{\wtq_*}<0$
hole $\|\vec{x}\|^2<\|\vec{y}\|^2+c$ on a disk of finite size.

Note that the intersection of $\overline{\gamma}_0$ and the Coulomb branch
is where $\usigma_0$ is real and
is in the chambers I and VI of Fig.~\ref{fig:CBdualR}.
In chamber I ({\it resp}. VI), $\Arg(\usigma_0)$ can move
in the interval $(-{\pi\over 2}+{\pi n_1\over 7},{\pi\over 2})$
({\it resp}. $(-{3\pi\over 2},-{\pi\over 2}-{\pi n_1\over 7})$),
and ${\rm Re}(\im\sDelta t^{\vee}\usigma_0)>0$ is unavoidable when
\beq
R_{n_1}:~-{\pi n_1\over 7}\leq \Arg(\im\sDelta t^{\vee})\leq 0\quad
\left(\,{\it resp}.~L_{n_1}:~\pi\leq \Arg(\im\sDelta t^{\vee})\leq
\pi+{\pi n_1\over 7}\,\right).
\label{dualRodsectors}
\eeq
Thus, the special representation $\C(21)$ ({\it resp}. $\C$)
at $t_3^{\vee}$ is not admissible at $t_3^{\vee}+\sDelta t^{\vee}$
if $\sDelta t^{\vee}$ is in the sector $R_2$ ({\it resp}. $L_2$).
Similarly, the special representation
$\Box(21)$ ({\it resp}. $\Box$) at $t_2^{\vee}$ is not admissible at
$t_2^{\vee}+\sDelta t^{\vee}$
if $\sDelta t^{\vee}$ is in the sector $R_3$ ({\it resp}. $L_3$),
and the special representation
$\mydoo(20)$ ({\it resp}. $\mydoo(-1)$) at $t_1^{\vee}$ is not admissible at
$t_1^{\vee}+\sDelta t^{\vee}$
if $\sDelta t^{\vee}$ is in the sector $R_3$ ({\it resp}. $L_3$).

We expect that these forbidden sectors extend to no-entry zones.
Let us see how the sector on the right of a singular point may extend.
Since this sector is associated to the chamber I of the Coulomb branch
which has $W_3$ and $W_a$ as its lower and upper boundaries, it is expected
that entries from below and from above are blocked 
by these walls.
This leads us to examine the singularities of
the regions $\AN_{\wtq_*}|_{W_3}\geq 0$ and $\AN_{\wtq_*}|_{W_a}\geq 0$.
By a computation similar to the one performed in Section~\ref{subsec:noentry},
we find that $\AN_{\wtq}|_{W_3}\geq 0$ is singular on the curve
\beq
\zeta^{\vee}=-7\log\left|2\cos\left({\pi\wtq^2\over 7}\right)\right|
-7\log\left|2\cos\left({\theta^{\vee}+2\pi\wtq^0+\pi\wtq^2\over 7}\right)\right|
=:\zeta^{\vee}(\theta^{\vee};\wtq^0,\wtq^2),
\label{zetaveetheta}
\eeq
while $\AN_{\wtq}|_{W_a}\geq 0$ is singular on the line
\beq
\zeta^{\vee}=-7\sum_{s=1}^2\log\left|2\cos\left({\pi\wtq^2\over 7}\right)\right|.
\label{zetaveeupper}
\eeq
When $\wtq$ is the shifted highest weight of the representations
$\C(21)$, $\Box(21)$ and $\mydoo(20)$, they indeed pass through the singular
points $t^{\vee}_3$, $t^{\vee}_2$ and $t^{\vee}_1$, respectively, and their slopes
match with the slopes of the boundaries of the forbidden sectors
(\ref{dualRodsectors}).
As $\theta^{\vee}$ moves away from the value $\theta_{\bfa}^{\vee}$
of the singular point $t_{\bfa}^{\vee}$,
the formula (\ref{zetaveetheta}) for the lower boundary
will terminate at some point where the singular point of
$\AN_{\wtq_*}|_{W_3}\geq 0$ reaches the intersection with
another wall.\footnote{For $\bfa=3,2$ and $1$,
the singular point reaches the intersections with $W_4$, $W_p$, $W_2$ at
$\sDelta_3\theta^{\vee}=\pi, 2\pi,3\pi$, 
$\sDelta_2\theta^{\vee}=2\pi, 3\pi, 4\pi$, and
$\sDelta_1\theta^{\vee}=\pi, 3\pi,5\pi$, respectively,
where $\sDelta_\bfa\theta^{\vee}:=\theta^{\vee}-\theta^{\vee}_\bfa$.}
Beyond that point, the lower boundary is given by another formula
which we do not know at the moment.
On the other hand, the formula (\ref{zetaveeupper}) for the upper boundary
will be valid for all $\theta^{\vee}\geq \theta_\bfa^{\vee}$,
as the singular point of $\AN_{\wtq_*}|_{W_a}\geq 0$
is always away from the intersection with other walls.
Therefore, the no-entry zone extends to $\theta^{\vee}\to +\infty$
although we do not know the formula for the lower boundary.
By the reflection symmetry (\ref{dualRodrefsym}), we find that
the forbidden sector to the left of the singular point evolves to
a no-entry zone that extends to $\theta^{\vee}\to -\infty$.
Entry from above is blocked 
by the wall $W_a$, while entry from below is blocked 
by the wall $W_1$ at least near the singular point.
In general, the representation $\Yd(i)$ has
no-entry zones ${\mathscr R}_{\Yd(i)}$ and ${\mathscr L}_{\Yd(i)}$
that extend to the right and to the left of the singular points
$(\zeta^{\vee}_{3-|\Yd|},\theta^{\vee +}_{\Yd(i)})$
and $(\zeta^{\vee}_{3-|\Yd|},\theta^{\vee -}_{\Yd(i)})$
where $\theta^{\vee \pm}_{\Yd(i)}=\pi(\pm 21-|\Yd|-2i)$.

Any path in the window {\bf A}, {\bf B}, {\bf C} or {\bf D} goes through
the no entry zones of the representations outside the sets
(\ref{dGRRA}), (\ref{dGRRB}), (\ref{dGRRC}) or (\ref{dGRRD}) respectively.
Thus, a representation is grade restricted only when it is in this set.
On the other hand, the representations in this set are all grade restricted.
We have already seen this for the windows {\bf A} and {\bf D}: transport
via the contour $\bgamma_0$ works.
For the window {\bf C}, $\bgamma_0$ also works for all
but $\C(21)$ and $\mydoo(-1)$. The latter two can be transported via
certain deformations of $\overline{\bgamma}_0$ as can be found by a
computation similar to (\ref{deform26}) and (\ref{deform00}).
The same holds for the window {\bf B}: $\bgamma_0$ works for all
but $\C(0)$ and $\mydoo(20)$, and deformations of $\overline{\bgamma}_0$
work for the latter two.
This confirms the proposed grade restriction rule.

The picture we expect from the rule is summarized in the following diagram:
\beq
\begin{tikzcd}
  &D_{G^{\vee}}(V^{\vee},W^{\vee})\arrow[d,phantom,"\mbox{\large $\cup$}"]&\\[-0.3cm]
  &{\mathcal S}^{\vee}\arrow[ddl,bend right,"\pi^{\vee}_Y"']\arrow[d,phantom,"\mbox{\large $\cup$}"]\arrow[ddr,bend left,"\pi^{\vee}_X"]&\\[-0.3cm]
  &{\mathcal T}^{\vee}_{\bf w}\arrow[dl,"\cong"']\arrow[dr,"\cong"]\\
  D(Y)&&D(X)
\end{tikzcd}
\label{dualdiagram}
\eeq
${\mathcal S}^{\vee}$ is the set of $USp(4)$
grade restricted branes.
$\pi^{\vee}_X$ and $\pi^{\vee}_Y$ are the flows to low energy sigma models
in the $\zeta^{\vee}\gg 0$ and $\zeta^{\vee}\ll 0$ phases.
The former is defined only on ${\mathcal S}^{\vee}$ because the phase is
strongly coupled and so is the latter because the phase is
strongly coupled at the scale where the effect of superpotential is negligible.
${\mathcal T}^{\vee}_{\bf w}$ is the set of grade restricted branes with respect to
the window ${\bf w}$.
We expect that the restrictions of the projections
$\pi^{\vee}_X$ and $\pi^{\vee}_Y$ to this subset are essentially bijective.
Given this diagram, we know the rule of brane transport between the
two phases through the window.

The details are left for future work.
In the present paper, we would like to discuss the properties of
a particular class of branes and their consequences.
We choose to work in the R-charge frame where the $a_{ij}$'s have R-charge $2$
and others have $0$.
Let us consider the brane datum with matrix factorization of $W^{\vee}$
\beq
Q=\sum_{ij}a_{ij}\eta^{ij}
+\sum_{ij}F^{ij}(p,\tx)\bareta_{ij},
\label{QKdualR}
\eeq
which is
realized on the irreducible representation $\bV$ of the Clifford algebra
\beq
\{\eta^{ij},\bareta_{kl}\}=
      {1\over 2}\left(\delta^i_k\delta^j_l-\delta^i_l\delta^j_k\right),\quad
      \{\eta^{ij},\eta^{kl}\}=\{\bareta_{ij},\bareta_{kl}\}=0.
\eeq
The $\eta^{ij}$'s and $\bareta_{ij}$'s are given the gauge and R-charges
$(\C(-1),-1)$ and $(\C(1),1)$, respectively.
We assign the gauge and R-charges $(\C(21),0)$ to the Clifford vacuum
$|0\rangle\in \bV$ annihilated by the $\eta^{ij}$'s.
This may be described by
\beq
\begin{tikzcd}
  \C(21)\arrow[r,yshift=0.5ex, "F(p{,}\tx)"] &
  \arrow[l,yshift=-0.5ex, "a"]\C(20)^{\oplus 21}
  \arrow[r,yshift=0.5ex, "F(p{,}\tx)"]
  & \arrow[l,yshift=-0.5ex, "a"] \C(19)^{\oplus {21\choose 2}}
  \arrow[r,yshift=0.5ex, "F(p{,}\tx)"]
  & \arrow[l,yshift=-0.5ex, "a"]
  \,\cdots\,\arrow[r,yshift=0.5ex, "F(p{,}\tx)"]
  &\arrow[l,yshift=-0.5ex, "a"] \C(1)^{\oplus 21}
  \arrow[r,yshift=0.5ex, "F(p{,}\tx)"]
  &\arrow[l,yshift=-0.5ex, "a"] \C\\[-0.7cm]
  {}_0&{}_1&{}_2&&{}_{20}&{}_{21}
\end{tikzcd}
\label{KdualR}
\eeq
The left arrow $a$ and the right arrow $F(p,\tx)$ are the first and
the second terms of (\ref{QKdualR}) respectively.
We call this the {\it $a$-Koszul brane} and denote it by $\Kos^{\vee}$.
In the Grassmannian phase $\zeta^{\vee}\gg 0$, the boundary potential
$\{Q,Q^{\dag}\}=(\|a\|^2+\|F(p,\tx)\|^2){\rm id}_{\bV}$ is positive definite
as $a\ne 0$, and hence $\Kos^{\vee}$ is empty at low energies,
\beq
\pi^{\vee}_X(\Kos^{\vee})\cong 0.
\eeq
In the Pfaffian phase $\zeta^{\vee}\ll 0$, the boundary potential vanishes
at $a=F(p,\tx)=0$ which is nothing but $Y$ itself.
Under some relationship between the gauge theory theta angle and the
low energy B-field, we expect that it flows to the structure sheaf of $Y$,
\beq
\pi^{\vee}_Y(\Kos^{\vee})\cong {\mathcal O}_Y.
\eeq
The tensor product of $\Kos^{\vee}$ and a $USp(4)$ grade restricted
representation $\rmV$ flows to the vector bundle associated to $\rmV$,
\beq
\pi^{\vee}_Y(\Kos^{\vee}\otimes \rmV)\cong{\mathcal O}_Y(\rmV).
\eeq
Note that ${\mathcal O}_Y(\rmV(i))={\mathcal O}_Y(\rmV)(-i)$.
The $a$-Koszul brane plays an important r\^ole in the monodromy
along loops around the singular points with a base point in
the Pfaffian phase.
For $\bfa=3,2,1$,
let ${\mathcal L}^{\vee}_\bfa$ be a loop that comes from $\zeta^{\vee}\ll 0$,
goes around $t^{\vee}_\bfa$ clockwise and then goes back to $\zeta^{\vee}\ll 0$,
similarly to the loop ${\mathcal L}'_{\bfa}$
in Fig.~\ref{fig:monodromy2} in the R\o dland model.
Each of them can be regarded as an upward path through a window
followed by a downward path through another.
The difference between the grade restriction rules for the two windows is
which of the two special representations at $t^{\vee}_\bfa$ is included.
The two are $\rmV_\bfa$ and $\rmV_\bfa(21)$,
with $\rmV_\bfa=\C$, $\Box$, $\mydoo(-1)$
for $\bfa=3,2,1$ respectively.
The representation $\rmV_\bfa$ is good
on the right of $t^{\vee}_\bfa$ and bad on its left,
while $\rmV_\bfa(21)$ is good on its left and bad on its right.
If the brane is grade restricted with respect to the first window
and is transported along the upward path that passes by
$t^{\vee}_\bfa$ on its left,
before it goes back through the downward path that passes by $t^{\vee}_\bfa$
on its right,
each of the $\rmV_\bfa(21)$ components in the brane must be traded for
$\rmV_\bfa$ while in the Grassmannian phase.
The trading can be done by binding $\Kos^{\vee}\otimes \rmV_\bfa$
followed by brane-antibrane annihilation, since
it is empty in the Grassmannian phase and
has $\rmV_\bfa$ and $\rmV_\bfa(21)$ at the edge components,
with the other components in the overlap of the two grade restriction rules.
Since $\rmV_\bfa(21)$ appears as the head of the
exact $a$-Koszul complex tensored with $\rmV_\bfa$,
the binding is done by a homomorphism
from the given brane to the empty brane.
The effect is the twist by
$\pi^{\vee}_Y(\Kos^{\vee}\otimes \rmV_\bfa)\cong{\mathcal O}_Y(\rmV_\bfa)$.
To summarize,
\beq
\mdm_{{\mathcal L}^{\vee}_3}=\twist_{{\mathcal O}_Y},\quad
\mdm_{{\mathcal L}^{\vee}_2}=\twist_{{\mathcal O}_Y(\Box)},\quad
\mdm_{{\mathcal L}^{\vee}_1}=\twist_{{\mathcal O}_Y(\sydoo(-1))}.
\label{mdmbasePfdualR}
\eeq
In comparison to the monodromy (\ref{mdmbasePf}) obtained from
the original R\o dland model, the one around $t_3^{\vee}$ matches with
the one around $t_3$. To compare the ones around
$t_2^{\vee}$ and $t_1^{\vee}$ with those around $t_2$ and $t_1$, we need
to find what the vector bundles ${\mathcal O}_Y(\Box)$ and
${\mathcal O}_Y(\mydoo(-1))$ are.
The vector bundle ${\mathcal O}_Y(\Box)$ is defined as the set of
triples $(p,\tx ;w)$ where $(p,\tx)$ solves the vacuum equation
including (\ref{FdualRod})
and $w\in \Box$, subject to the relation
\beq
(p,\tx; w)\sim (\lambda^{-2}p, \lambda^{-1}g\tx; \lambda gw)
\eeq
The symplectic pairing $[\tx^i w]$ is gauge invariant.
Thus, ${\mathcal O}_Y(\Box)$ may also be defined as
the vector bundle whose fiber at $p$ is
the space of $[\tx^i w]$'s.
The vacuum equation implies that the matrix
$(\tx^i_a)$ is of rank 4 and satisfies
$\tx^i_av_i=0$ if $A^{ij}(p)v_j=0$, i.e., if $v\in {\rm Ker} A(p)$.
Therefore, $\{(\tx^i_1),\ldots, (\tx^i_4)\}$ can be regarded as a
basis of $(V/{\rm Ker}A(p))^*$.
In particular, there is a perfect pairing
${\mathcal O}_Y(\Box)\times \ck_Y\to {\mathcal O}_Y$,
which means that ${\mathcal O}_Y(\Box)$ and $\ck_Y$ are dual to each other,
\beq
    {\mathcal O}_Y(\Box)\,\,\cong\,\, \ck^*_Y.
\eeq
It is equipped with a non-degenerate
skew-symmetric pairing
${\mathcal O}_Y(\Box)\times {\mathcal O}_Y(\Box)\to{\mathcal O}_Y(-1)$
and that is dual, again via the vacuum equation,
to the pairing $\ck_Y\times\ck_Y\to {\mathcal O}_Y(1)$ defined by $A(p)$.
Then, ${\mathcal O}_Y(\mydoo)$ is the vector bundle
$\textw^{\!2}_0{\mathcal O}_Y(\Box)$ of its second rank
antisymmetric tensors which are traceless with respect to this pairing.
In particular,
\beq
    {\mathcal O}_Y(\ydoo(-1))\,\,\cong\,\,
    \textw^{\!2}_0\ck^*_Y(1).
\eeq
Thus, the monodromies obtained in the dual model (\ref{mdmbasePfdualR})
match precisely with the result (\ref{mdmbasePf}) from the original model.
In fact,
  the result (\ref{mdmbasePf}) was first obtained in the dual model
  in the way just described. We were able to obtain it directly in the
  original model only after identifying the ``right-fit'' empty branes
  $\Lb^0$, $\Lb^1$, $\Lb^2$ in the $\zeta\gg 0$ phase and finding
  their low energy behaviours in the $\zeta\ll 0$ phase.
It would also be an interesting problem to reproduce (\ref{mdmbaseGr})
in the dual model. For this we need to find the ``right-fit'' empty branes
in the $\zeta^{\vee}\ll 0$ phase and find their low energy behaviour in
the $\zeta^{\vee}\gg 0$ phase.
This is left as a problem for future work.

Through this analysis we have found an explicit correspondence between branes
under the duality. If we identify
$t-7\pi \im$ in the R\o dland model and $t^{\vee}-21\pi\im$ in the dual,
the correspondence is
\beqa
\Lb^2(-1)&\longleftrightarrow&\Kos^{\vee}[-1]\\
\Lb^1&\longleftrightarrow&\Kos^{\vee}\otimes\Box\\
\Lb^0&\longleftrightarrow&\Kos^{\vee}\otimes\ydoo(-1)[-1].
\eeqa
Indeed, they are all empty in the Grassmannian phase,
their low energy images in the Pfaffian phases are the same,
and they generate the monodromies along the corresponding loops.
Construction of the ``right-fit'' empty branes in the
$\zeta^{\vee}\ll 0$ phase of the dual model would lead to another
set of dual pairs, where the left-hand sides are
$\Kos, \Kos\otimes \Box$ and $\Kos\otimes \ydt(-1)$.
In \cite{RenSeg}, an equivalence between the categories of branes
in the dual pair is constructed. It would be interesting to see the relation
between the above correspondence and the equivalence in \cite{RenSeg}.

\subsection{Hosono-Takagi Model}
\label{subsec:HT}

We next consider the model introduced in \cite{Hduality} for the pair of
Calabi-Yau manifolds studied by Hosono and Takagi \cite{HoTa,HoTa14}.
The gauge group is
\beq
G={U(1)\times O(2)\over \{(\pm 1,\pm {\bf 1}_2)\}}
\label{GHT}
\eeq
The matter consists of five fields $x_1,\ldots, x_5$ in $\Box(1)$
and five fields $p^1,\ldots, p^5$ in $\C(-1)$. The superpotential
and the twisted superpotential are
\beqa
W&=&\sum_{i,j,k}S^{ij}_kp^k(x_ix_j),\label{WHT}\\
\wt{W}&=&-2t\,\usigma_{\mathfrak{u}(1)}.
\label{twWHT}
\eeqa
In the expression of the superpotential,
$(x_ix_j)$ is the $O(2)$ invariant symmetric pairing.
The coefficients $S^{ij}_k$ are $ij$ symmetric and generic in the sense
described in \cite{Hduality}.

\newcommand{\PPr}{\!\textit{\textbf{P}}}

The model has two phases.
The $\zeta\gg 0$ phase is weakly coupled. A solution to the vacuum equation
breaks the gauge group completely, and the vacuum manifold is
the locus of $(p,x)$ with $p=0$ and $x\ne 0$ solving the F-term equation
\beq
S_k(x):=\sum_{i,j}S^{ij}_k(x_ix_j)=0,
\eeq
modded out by the $G_{\C}$ action.
This is the free quotient $X$ of the complete intersection $\wt{X}$
of five symmetric bilinear hypersurfaces of
$\CP^4\times\CP^4$ by the exchange of the two $\CP^4$'s.
It is a Calabi-Yau threefold with Hodge numbers
$h^{1,1}=1$ and $h^{2,1}=26$ and the Picard group $\Z_2\oplus\Z H$.
We denote by ${\mathcal O}_X(1)$ a line bundle representing the class $H$.
$X$ has the following topological invariants:
\beq
H^3=35,\quad c_2\cdot H=50,\quad c_3=-50.
\eeq
The $\zeta\ll 0$ phase is strongly coupled. The vacuum equation requires
$p\ne 0$ and $x=0$ and the $O(2)$ subgroup is left unbroken.
We have an $O(2)$ SQCD with five flavors
with varying mass matrix
\beq
S^{ij}(p)=\sum_kS^{ij}_kp^k.
\eeq
After understanding its low energy behaviour,
we find that the theory reduces in the end to a non-linear sigma model.
The target $Y$ is a double cover of the locus of $p\in\CP^4$
of rank $S(p)\leq 4$ which is ramified along the curve of rank $S(p)=3$.
It is a Calabi-Yau threefold with Hodge numbers $h^{1,1}=1$ and $h^{2,1}=26$.
The Picard group is generated by a class $H$ represented by a line
bundle ${\mathcal O}_Y(1)$.
The topological characteristics of $Y$ are
\beq
H^3=10,\quad c_2\cdot H=40,\quad c_3=-50.
\eeq
We shall call $\zeta\gg 0$ the {\it Reye congruence phase}
after the name given to $X$ in \cite{HoTa} and
$\zeta\ll 0$ the {\it determinantal phase} as
$Y$ is a double cover of a subvariety defined by $\det S(p)=0$.

\newcommand{\bfb}{{\bf b}}

A single valued parametrization of the identity component of $G$,
which is also the maximal torus $T$, is given by
\beq
\left[\left(h_{\ul{0}},\left(\begin{array}{cc}
    \cos\psi&-\sin\psi\\
    \sin\psi&\cos\psi
  \end{array}\right)\right)\right]\,\longmapsto\,
(h_1,h_2)=(h_{\ul{0}}\e^{\im\psi},h_{\ul{0}}\e^{-\im \psi}).
\eeq
Let $\usigma_1,\usigma_2$ be the scalar components of the corresponding
vector multiplets. $2\usigma_{\mathfrak{u}(1)}$ in (\ref{twWHT}) is
$\usigma_1+\sigma_2$. The Weyl group
$W=N_T/T\cong \Z_2$ exchanges the two,
\beq
\usigma_1\,\longleftrightarrow\,\usigma_2.
\label{HTWeyl}
\eeq
The weights of the matter fields are
$h_{\ul{0}}\e^{\pm\im \psi}$ for the $x^1_i\pm\im x^2_i$,
and $h_{\ul{0}}^{-2}$ for the $p_k$.
The effective twisted superpotential is
\beq
\wt{W}_{\!{\it eff}}=-t(\usigma_1+\usigma_2)
-5\usigma_1(\log\usigma_1-1)-5\usigma_2(\log\usigma_2-1)
+5(\usigma_1+\usigma_2)(\log(-\usigma_1-\usigma_2)-1).
\eeq
The vacuum equation
$\e^t=(-\usigma_1-\usigma_2)^5/\usigma_1^5=(-\usigma_1-\usigma_2)^5/\usigma_2^5$
is solved by
\beq
\usigma_1/\usigma_2=\omega,\quad \omega^5=1.
\label{HTCB}
\eeq
A solution exists only when $\e^t=-(\omega+1)^5$.
The Weyl group generator (\ref{HTWeyl})
inverts $\omega$ and there are three inequivalent cases
$\omega=1$, $\e^{2\pi\im/5}\sim \e^{-2\pi\im/5}$,
$\e^{4\pi\im/5}\sim\e^{-4\pi\im/5}$.
Thus, there are three singular points 
\beq
\e^t=-(\omega_{\bfb}+1)^5=:\beta_{\bfb},\quad
\omega_{\bfb}:=\e^{2\pi\im \bfb/5},\quad
\bfb=0,1,2,
\label{defbfb}
\eeq
with the Coulomb branch (\ref{HTCB}) with $\omega=\omega_{\bfb}^{\pm 1}$.
The Coulomb branch for $\bfb=0$ is at the locus of
Weyl fixed points and is a union of two copies \cite{Hduality}.
The numerical values of
$\beta_{\bfb}=-\omega_{\bfb}^{5\over 2}(\omega_{\bfb}^{1\over 2}+\omega_{\bfb}^{-{1\over 2}})^5=-\e^{\pi\im \bfb}\left(2\cos\left({\pi\bfb\over 5}\right)\right)^5$ are
\beq
\beta_0\simeq -\e^{3.466},\quad
\beta_1\simeq \e^{2.406},\quad
\beta_2\simeq -\e^{-2.406}.
\eeq

The label $\Yd$ of an irreducible representation of $O(2)$ is 
either a Young diagram of height one or
a ``no box''-diagram with a sign $\emptyset_{\pm}$ \cite{Weyl}.
The representations labeled by $\emptyset_{\pm}$ are one-dimensional:
$\emptyset_+$ is the trivial representation and
$\emptyset_-$ receives a $-1$ sign under $G\setminus T$.
The representation labeled by $\Box\cdots\Box$ ($\elln$ boxes)
is the two-dimensional space $\Sym_0^{\elln}\C^2\subset(\C^2)^{\otimes \elln}$
consisting of symmetric and traceless tensors.
An irreducible representation of $G$ is labeled by $\Yd(i)$.
We write $\C_{\pm}(i):=\emptyset_{\pm}(i)$. Note that $\C_+(i)=\C(i)$.
The representation $\C_{\pm}(i)$ has weight $h_{\ul{0}}^{2i}=h_1^ih_2^i$.
The representation $\Yd(i)$ with $|\Yd|=\elln\geq 1$ has weights
$h_{\ul{0}}^{2i+\elln}\e^{\elln\im\psi}=h_1^{i+\elln}h_2^i$
and $h_{\ul{0}}^{2i+\elln}\e^{-\elln\im\psi}=h_1^ih_2^{i+\elln}$.
We regard the former the highest. That is,
the highest weight of $\Yd(i)$ for any $\Yd$ is
\beq
q=\left(i+|\Yd|,i\right).
\label{hwYi}
\eeq
The growth function that controls
the convergence of the hemisphere integral (\ref{ZV}) is
\beqa
\AN_q&=&\zeta(\Is_1+\Is_2)-\theta(\Rs_1+\Rs_2)
-2\pi \langle q,\Rs\rangle\nn\\
&&
+5\Rs_1\Arg(\im\usigma_1)+5\Rs_2\Arg(\im\usigma_2)
-5(\Rs_1+\Rs_2)\Arg(-\im(\usigma_1+\usigma_2))\nn\\
&&
+5\Is_1\log|\usigma_1|+5\Is_2\log|\usigma_2|
-5(\Is_1+\Is_2)\log|\usigma_1+\usigma_2|.
\label{ANHT}
\eeqa
The integration contour $\bgamma\subset\ttt_{\C}$
is a deformation of the real locus
$\im\ttt$ avoiding the poles of the Gamma function factor,
which is assured if it avoids the walls
$W_1$, $W_2$ and $W_p$ given by the same expressions (\ref{W1}), (\ref{W2})
and (\ref{Wp}) as in the R\o dland model.
If we regard $\bgamma$ as a graph of a function $\Rs\mapsto\Is(\Rs)$,
the condition to avoid the walls is also the same as the R\o dland model:
(\ref{Wedge1}), (\ref{Wedge2}) and (\ref{Wedgep}).
The expression (\ref{ANHT}) of the growth function is almost the same as
(\ref{A}) for the R\o dland model:
the two are related just by the exchange $5\leftrightarrow 7$.
Thus, we may partially reuse the results from the R\o dland model
in the analysis of convergent contours.

In the weakly coupled phase $\zeta\gg 0$, we can take the contour
$\bgamma_+$ given by
\beq
\Is_1=(\Rs_1)^2,\quad
\Is_2=(\Rs_2)^2.
\eeq
The hemisphere intergral on $\bgamma_+$ is convergent
for any representation of $G$ as long as $\zeta>5\log 2$.

In the strongly coupled phase $\zeta\ll 0$, we can insead take
$\bgamma_-$ given by
\beq
\Is_1=\Is_2=-(\Rs_1+\Rs_2)^2.
\label{bgammam}
\eeq
Along $l=\{\Is_1=\Is_2=\Rs_1+\Rs_2=0\}$, the growth function is
$\AN_q|_l=-2\pi(q^1-q^2)\Rs_1+5\pi|\Rs_1|$ and its growth at infinity requires
$|q^1-q^2|<5/2$.
For the highest weight (\ref{hwYi}) of $\Yd(i)$, this means $|\Yd|<5/2$,
that is, $|\Yd|=0,1,2$. Only the four series of representations
\beq
\C_{\pm}(i),\quad
\Box(i),\quad
\ydt(i)
\label{O2GRR}
\eeq
are allowed. Conversely, for any of these representations,
the hemisphere integral at $\zeta\ll 0$
converges on the contour (\ref{bgammam}).
Requiring $\AN_q|_l$ to grow at infinity is nothing but the
grade restriction rule for $O(2)$ SQCD with five fundamentals.
This is consistent with the fact that we have this SQCD 
at an intermediate energy scale in the $\zeta\ll 0$ phase.
We may refer to (\ref{O2GRR}) as the $O(2)$ grade restriction rule.
The hemisphere integral on $\bgamma_-$ is convergent for the representation
$\Yd(i)$ as long as $\zeta$ is less than
\beq
\zeta_{|\Yd|}:=5\log\left(2\cos\left({\pi|\Yd|\over 5}\right)\right),
\eeq
which is the value of the FI parameter at the singular point
$\beta_{\bfb}$ with $\bfb=|\Yd|$.

\begin{figure}[thb]
\psfrag{A}{\mbox{\small ${\mathcal P}_{\!\bf A}$}}
\psfrag{B}{\mbox{\small ${\mathcal P}_{\bf B}$}}
\psfrag{C}{\mbox{\small ${\mathcal P}_{\bf C}$}}
\psfrag{D}{\mbox{\small ${\mathcal P}_{\bf D}$}}
\psfrag{a1}{$t_0$}
\psfrag{a2}{$t_1$}
\psfrag{a3}{$t_2$}
\psfrag{z}{$\zeta$}
\psfrag{th}{$\theta$}
\psfrag{-8pi}{$-6\pi$}
\psfrag{-6pi}{$-4\pi$}
\centerline{\includegraphics[page=48]{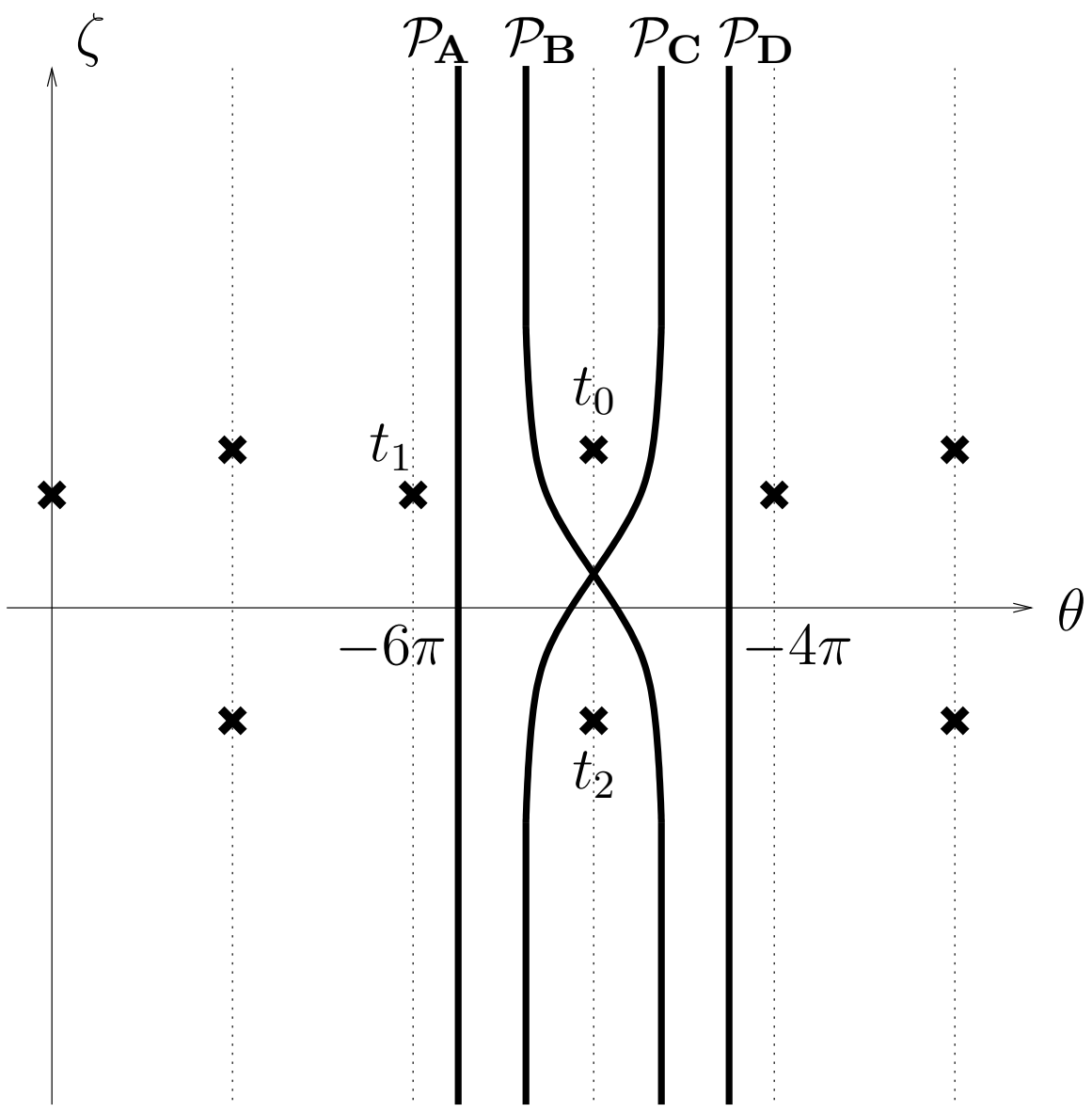}}
\caption{Four windows between the phases.}
\label{fig:HTpaths}
\end{figure}
Next, let us determine the grade restriction rule for paths between
the two phases. Let us consider the four classes of paths
${\mathcal P}_{\bf A},$
${\mathcal P}_{\bf B},$
${\mathcal P}_{\bf C},$
${\mathcal P}_{\bf D}$
that go near the three points
\beqa
t_0&=&5\pi \im +5\log 2,
\label{HTt0}\\
t_1&=&6\pi\im+5\log\left(2\cos\left({\pi\over 5}\right)\right),
\label{HTt1}\\
t_2&=&5\pi\im+5\log\left(2\cos\left({2\pi\over 5}\right)\right),
\label{HTt2}
\eeqa
as depicted in Fig.~\ref{fig:HTpaths}.
We again take the shortcut to find the rule, starting with the
determination of the special representations at the singular points.

The special representation at $t=t_*$
has the highest weight $q_*$ such that
$\dd \wt{W}_{\!{\it eff}, q_*}=0$ in a chamber of the Coulomb branch
(\ref{HTCB}). The equation reads
\beqa
2\pi\im q_*&=&t_*+\sum_iQ_i\log(\im Q_i(\usigma))\nn\\
&=&t_*(1,1)+5(1,0)\log(\im\usigma_1)
+5(0,1)\log(\im\usigma_2)\nn\\
&&+5(-1,-1)\log(-\im(\usigma_1+\usigma_2))
\label{qstHT}
\eeqa
in which we set (\ref{HTCB}) with $\omega=\omega_{\bfb}^{\pm 1}$
for $t_*=t_\bfb$, that is,
\beq
\usigma_1=\e^{\pm \pi\im \bfb}\Cc,\quad
\usigma_2=\e^{\mp \pi\im \bfb}\Cc.
\eeq
The real part of the right-hand side of (\ref{qstHT}) vanishes.
Let us evaluate $\Arg(\im Q_i(\usigma))$ to find the imaginary part.
\begin{figure}[thb]
  \psfrag{w12}{$W_{1,2}$}
  \psfrag{w1}{$W_1$}
  \psfrag{w2}{$W_2$}
  \psfrag{wp}{$W_p$}
  \psfrag{a1}{${2\pi\over 5}$}
  \psfrag{a2}{${4\pi\over 5}$}
  \psfrag{sp}{$\Cc$}
  \psfrag{I0}{${\rm I}_0$}
  \psfrag{II0}{${\rm II}_0$}
  \psfrag{III0}{${\rm III}_0$}
  \psfrag{I1}{${\rm I}_1$}
  \psfrag{II1}{${\rm II}_1$}
  \psfrag{III1}{${\rm III}_1$}
  \psfrag{I2}{${\rm I}_2$}
  \psfrag{II2}{${\rm II}_2$}
  \psfrag{III2}{${\rm III}_2$}
\centerline{\includegraphics[page=49]{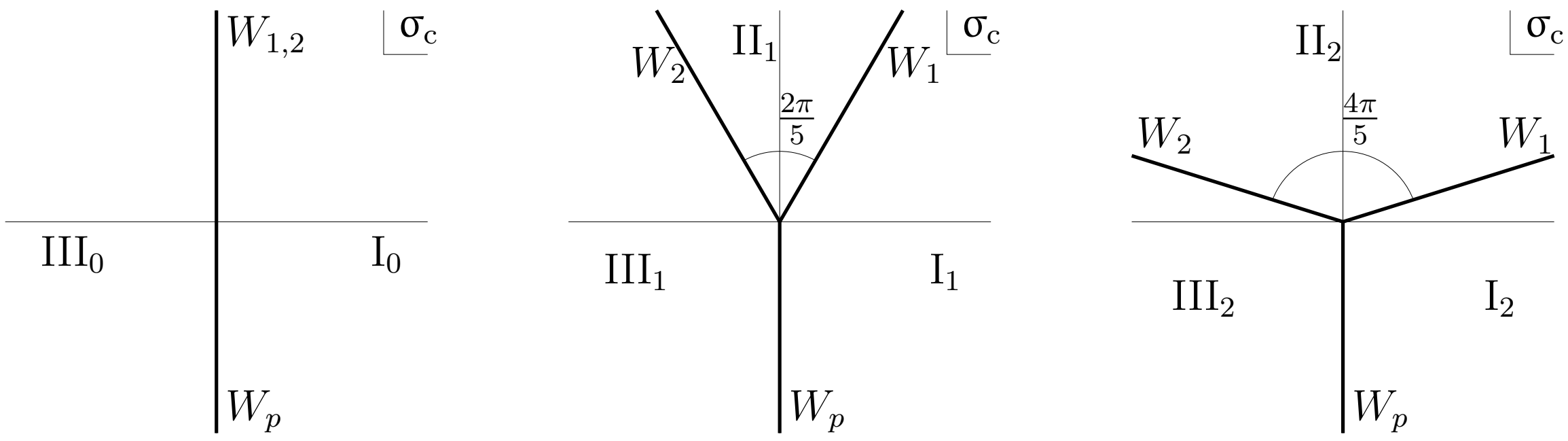}}
\caption{Chambers of the $\omega=\omega_\bfb$ Coulomb branch for $\bfb=0,1,2$
  (left, middle, right).}
\label{fig:CBHT}
\end{figure}
This depends on the chamber of the Coulomb branch separated by the walls
$W_1$, $W_2$ and $W_p$, as depicted in Fig.~\ref{fig:CBHT}
for the branch with $\omega=\omega_{\bfb}$.
If we put $\Cc=|\Cc|\e^{\im\varphi}$ with
$-{\pi\over 2}<\varphi<{3\pi\over 2}$, we find
\beq
\begin{array}{c|ccc|c}
  &\Arg(\im\usigma_1)&\Arg(\im\usigma_2)&\Arg(-\im(\usigma_1+\usigma_2))&
  {1\over 2\pi}\sum_iQ_i\Arg(\im Q_i(\usigma))\\
  \hline
      {\rm I}_{\bfb}&\varphi+{\pi\over 2}+{\pi\bfb\over 5}&
      \varphi+{\pi\over 2}-{\pi\bfb\over 5}&\varphi-{\pi\over 2}&
      \left({5+\bfb\over 2},{5-\bfb\over 2}\right)\\
           {\rm II}_{\bfb}&~~~{}_{''}-2\pi&{}_{''}&{}_{''}&
           \left({-5+\bfb\over 2},{5-\bfb\over 2}\right)\\
                {\rm III}_{\bfb}&{}_{''}&~~~{}_{''}-2\pi&{}_{''}&
                \left({-5+\bfb\over 2},{-5-\bfb\over 2}\right)
\end{array}
\eeq
The row of II${}_0$ should be deleted {\it after} computing $\Arg$'s
in the chamber III${}_0$.
Note that ${1\over 2\pi}\sum_iQ_i\Arg(\im Q_i(\usigma))$
must be equated with $q_*+{\theta_*\over 2\pi}(1,1)$.
The $q_*$ in the chambers
I${}_{\bfb}$ and III${}_{\bfb}$ are highest weights of some
$O(2)$ grade restricted representations
of the gauge group, $(i+|\Yd|,i)$ for $\Yd(i)$:
\beq
\begin{array}{c|c|c}
  t_*&q_*:\Yd(i)~\mbox{on I${}_{\bfb}$}&q_*:\Yd(i)~\mbox{on III${}_{\bfb}$}\\
  \hline
  t_0&(5,5):\C_{\pm}(5)&(0,0):\C_{\pm}\\
  t_1&(6,5):\Box(5)&(1,0):\Box\\
  t_2&(6,4):\ydt(4)&(1,-1):\ydt(-1)
\end{array}
\label{specialrepHT}
\eeq
In the other chambers or on the Weyl image branches,
$q_*$ are not the highest weight of any of the $O(2)$ grade restricted
representations.
Thus, those in the table (\ref{specialrepHT}) are the special representations
at the three points. The special representations
at $t_*+2\pi\im j$ are obtained from those at $t_*$ by $\rmV\mapsto \rmV(j)$.

We anticipate that the two or four representations (\ref{specialrepHT})
at $t_*$ undergo a catastrophe along a small loop around $t_*$.
More precisely,
the representations $\C_{\pm}(5)$ ({\it resp}. $\C_{\pm}$)
cannot stay admissible on the right ({\it resp}. left) of  $t_1$,
and similarly for
$\Box(5)$ ({\it resp}. $\Box$) on the right ({\it resp}. left) of $t_2$,
and
$\ydt(4)$ ({\it resp}. $\ydt(-1)$) on the right ({\it resp}. left) of $t_3$.
If $\rmV(5)$ and $\rmV$ are bad on the right and the left of $t_*$,
so are $\rmV(5+j)$ and $\rmV(j)$ on the right and the left of
$t_*+2\pi\im j$.
Then, the grade restriction rule with respect to the four windows would be
\beqa
\!\!\!\!\!\!\!\!\!\!\!\!
{\rm GRR}_{\bf A}&&
\begin{array}{cccccccc}
&&\C_{\pm}(1)&\C_{\pm}(2)&\C_{\pm}(3)&\C_{\pm}(4)&\C_{\pm}(5)\\
&\Box&\Box(1)&\Box(2)&\Box(3)&\Box(4)&\\
  ~~~~~~~~~~&\ydt&\ydt(1)&\ydt(2)&\ydt(3)&\ydt(4)&
\end{array}
\label{HTGRRA}\\[0.2cm]
\!\!\!\!\!\!\!\!\!\!\!\!
{\rm GRR}_{\bf B}&&
\begin{array}{ccccccc}
&&\C_{\pm}(1)&\C_{\pm}(2)&\C_{\pm}(3)&\C_{\pm}(4)&\C_{\pm}(5)\\
&\Box&\Box(1)&\Box(2)&\Box(3)&\Box(4)&\\
\ydt(-1)&\ydt&\ydt(1)&\ydt(2)&\ydt(3)&&
\end{array}
\label{HTGRRB}\\[0.2cm]
\!\!\!\!\!\!\!\!\!\!\!\!
{\rm GRR}_{\bf C}&&
\begin{array}{ccccccc}
&\C_{\pm}&\C_{\pm}(1)&\C_{\pm}(2)&\C_{\pm}(3)&\C_{\pm}(4)&\\
&\Box&\Box(1)&\Box(2)&\Box(3)&\Box(4)&\\
~~~~~~~~~~&\ydt&\ydt(1)&\ydt(2)&\ydt(3)&\ydt(4)&
\end{array}
\label{HTGRRC}\\[0.2cm]
\!\!\!\!\!\!\!\!\!\!\!\!
{\rm GRR}_{\bf D}&&
\begin{array}{cccccccc}
&\C_{\pm}&\C_{\pm}(1)&\C_{\pm}(2)&\C_{\pm}(3)&\C_{\pm}(4)&\\
&\Box&\Box(1)&\Box(2)&\Box(3)&\Box(4)&\\
\ydt(-1)&\ydt&\ydt(1)&\ydt(2)&\ydt(3)&&
\end{array}
\label{HTGRRD}
\eeqa
We would like to confirm this proposal.

By the similarity of the growth functions, some of the results obtained in
the R\o dland model can be carried over to the Hosono-Takagi model, by
the $7\to 5$ replacement.
For the real contour $\bgamma=\im\ttt$,
the condition for the highest weight of the admissible representation
is the $7\to 5$ version of (\ref{condreal}),
\beq
2\pi |q^1-q^2|<5\pi,\quad
|\theta+2\pi q^1|\,\,\&\,\,
|\theta+2\pi q^2|<5\pi.
\label{condrealHT}
\eeq
For the highest weight (\ref{hwYi}) of $\Yd(i)$, this means
\beq
|\Yd|=0,1,2,\quad -{5\over 2}<i+{\theta\over 2\pi}<{5\over 2}-|\Yd|.
\eeq
For $-5\pi<\theta+2\pi j<-3\pi$, the set of admissible representations
for the real contour is
\beq
\begin{array}{ccccc}
  \C_{\pm}(j)&\C_{\pm}(j+1)&\C_{\pm}(j+2)&\C_{\pm}(j+3)&\C_{\pm}(j+4)\\
  \Box(j)&\Box(j+1)&\Box(j+2)&\Box(j+3)&\\
  \ydt(j)&\ydt(j+1)&\ydt(j+2)&&
\end{array}
\eeq
This set can be enlarged by turning on the imaginary part.
Under the simplifying assumption $\Is_1+\Is_2=0$,
the imaginary part can be chosen as (\ref{sgnP}) and (\ref{sgnN}).
For example, we may take $\bgamma_0$ given by (\ref{bgamma0}).
Then, the conditon is the $7\to 5$ version of (\ref{Cond111}),
\beq
2\pi|q^1-q^2|<5\pi,\quad |\theta+\pi(q^1+q^2)|<5\pi,\quad
|\theta+2\pi q^1|\,\,\&\,\,|\theta+2\pi q^2|
<5{3\pi\over 2}-5\epsilon.
\label{Cond111HT}
\eeq
For the highest weight (\ref{hwYi}) of $\Yd(i)$, this means
\beq
|\Yd|=0,1,2,\quad -{5\over 2}<i+{|\Yd|\over 2}+{\theta\over 2\pi}<{5\over 2}.
\eeq
For $-6\pi<\theta<-5\pi$ ({\it resp}. $-5\pi<\theta<-4\pi$),
this is satisfied precisely by the representations in (\ref{HTGRRA})
({\it resp}. (\ref{HTGRRD})). This is for any value of $\zeta$
as $\AN_q$ is $\zeta$-independent thanks to $\Is_1+\Is_2=0$.
Deep inside the phase with $\pm\zeta\gg 0$,
the contour $\bgamma_0$ can be deformed to $\bgamma_{\pm}$ 
while the hemisphere integral stays convergent.
Thus, the representations in (\ref{HTGRRA}) ({\it resp}. (\ref{HTGRRD}))
are included in the grade restricted set
with respect to the window {\bf A} ({\it resp}. {\bf D}).

The behaviour of $\AN_{q_*}$ near $t=t_*$ in a neighborhood
of the Coulomb branch is almost the same as in the case
of the R\o dland model, except that we need to make the $7\to 5$ replacement.
And we find the similar conclusion concerning the convergence of
the hemisphere integrals for the special representations.
Set
\beq
R_\bfb:~0\leq \Arg(\im\sDelta t)\leq {\pi\bfb\over 5}\quad
\left({\it resp}.~
L_\bfb:~\pi-{\pi\bfb\over 5}\leq\Arg(\im\sDelta t)\leq\pi\right).
\label{HTRandL}
\eeq
Then, the special representations $\C_{\pm}(5)$ ({\it resp}. $\C_{\pm}$)
at $t_0$ are not admissible at $t_0+\sDelta t$ if $\sDelta t$ is in
the direction $R_0$ ({\it resp}. $L_0$). Similarly,
the special representation $\Box(5)$ ({\it resp}. $\Box$)
at $t_1$ is not admissible at $t_1+\sDelta t$ if $\sDelta t$ is in the sector
$R_1$ ({\it resp}. $L_1$), and
the special representation $\ydt(4)$ ({\it resp}. $\ydt(-1)$)
at $t_2$ is not admissible at $t_2+\sDelta t$ if $\sDelta t$ is in the sector
$R_2$ ({\it resp}. $L_2$).

These forbidden directions/sectors extend to no-entry zones.
In general, the representation $\Yd(i)$ has no-entry zones
${\mathscr R}_{\Yd(i)}$ and
${\mathscr L}_{\Yd(i)}$
that extend to the right and the left of the singular points
$(\zeta_{|\Yd|},\theta_{\Yd(i)}^+)$ and
$(\zeta_{|\Yd|},\theta_{\Yd(i)}^-)$, respectively, where
$\theta_{\Yd(i)}^{\pm}=(\pm 5-|\Yd|-2i)\pi$.
As $t$ approaches ${\mathscr R}_{\Yd(i)}$ or ${\mathscr L}_{\Yd(i)}$
from above, any deformation of $\bgamma_+$ is forced to
either go out of the admissible region or
hit the wall $W_1$ or $W_2$ respectively.
As $t$ approaches either from below, any deformation of $\bgamma_-$
is forced to
either go outside the admissible region or hit the wall $W_p$.
The precise locations of the zones are
\beqa
{\mathscr R}_{\Yd(i)}:&&
\zeta_{|\Yd|}\leq \zeta\leq
\left\{\begin{array}{ll}
\zeta(\theta;i)&\,{\rm for}~\,
\theta_{\Yd(i)}^+\leq\theta\leq \theta_{\Yd(i)}^++|\Yd|\pi\\
5\log 2&\,{\rm for}~\, \theta\geq \theta_{\Yd(i)}^++|\Yd|\pi
\end{array}\right.\\
{\mathscr L}_{\Yd(i)}:&&
\zeta_{|\Yd|}\leq\zeta\leq
\left\{\begin{array}{ll}
5\log 2&\,{\rm for}~\,\theta\leq \theta_{\Yd(i)}^--|\Yd|\pi\\
\zeta(\theta;|\Yd|+i)&\,{\rm for}~\,
\theta_{\Yd(i)}^--|\Yd|\pi\leq\theta\leq\theta_{\Yd(i)}^-\\
\end{array}\right.
\eeqa
in which
\beq
\zeta(\theta;q):=5\log\left|2\cos\left({\theta+2\pi q\over 5}\right)\right|.
\eeq
Note that ${\mathscr R}_{\C_{\pm}(i)}$ and ${\mathscr L}_{\C_{\pm}(i)}$
are half lines and might better be called ``no-crossing lines'',
as in the Abelian theories described in Section~\ref{subsec:Abelian}.

Any path in the window {\bf A}, {\bf B}, {\bf C} or {\bf D}
goes through the no-entry zones of the representations outside
(\ref{HTGRRA}), (\ref{HTGRRB}), (\ref{HTGRRC}) or (\ref{HTGRRD}),
respectively. Thus, only the representation
in this set can be grade restricted.
In fact all of these representations are grade restricted.
For the windows {\bf A} and {\bf D}, they can be transported via the contour
$\bgamma_0$ as we have seen above.
For the windows {\bf B} and {\bf C}, transport via the contour
$\bgamma_0$ works for all but the special representations at the singular
points $t_0$ and $t_2$. The special representations can be transported via
an appropriate deformation of $\overline{\bgamma}_0$ which can be found
by a computation similar to (\ref{deform26}) and (\ref{deform00}).
This establishes the grade restriction rules
(\ref{HTGRRA}), (\ref{HTGRRB}), (\ref{HTGRRC}) and (\ref{HTGRRD}).

The picture we expect is the same as (\ref{maindiagram}),
where now ${\mathcal S}$ is the set of $O(2)$ grade restricted branes,
$X$ and $Y$ are the Reye congruence and the determinantal Calabi-Yau,
and ${\mathcal T}_{\bf w}$ is the set of branes satisfying the grade
restriction rule which we have just established.
This picture dictates the rule of D-brane transport between the two phases.
In this paper, we discuss a limited number of important aspects,
postponing more details for future publications.
We shall work in the R-charge frame
where the $p^k$'s have R-charge 2 and the $x_i$'s have 0.
We denote by $\GtoP_{\bf w}$ and $\PtoG_{\bf w}$
the transport through the window ${\bf w}$
from $D(X)$ to $D(Y)$ and from $D(Y)$ to $D(X)$, respectively.
The transport for shifted windows satisfies
\beq
\mdm_Y\circ \GtoP_{\bf w}\circ \mdm_X=\GtoP_{{\bf w}(1)},
\label{YTXHT}
\eeq
in which $\mdm_X$ is the tensor product with ${\mathcal O}_X(1)$
while $\mdm_Y$ is the tensor product with ${\mathcal O}_Y(1)$
combined with the R-charge shift $[-2]$.
The operators $\mdm_X$ and $\mdm_Y$ can be interpreted as the large
volume monodromies at $\e^t=\infty$ and $0$ in the moduli space.

The hemisphere partition function can be evaluated in the two phases
and compared to the large volume central charges.
The situation is similar to the R\o dland model.
In the weakly coupled phase $\zeta\gg 0$, the computation is easy and the
comparison is straightforward as we know well the low energy behaviour of
gauge theory branes.
In the strongly coupled phase $\zeta\ll 0$
on the other hand, the computation and the comparison are both non-trivial.
We need to use Mathematica packages to perform harmonic sums
and to understand the low energy behaviour of gauge theory branes
which involves the boundary dynamics of $O(2)$ supersymmetric QCD.
Comparing the results with the periods
computed in Appendix~\ref{app:periods}, Eqns
(\ref{per-struct1})-(\ref{per-struct4}),
(\ref{periodHTRey}) and (\ref{periodHTDet}), we find
\beq
Z_{D^2}(\brane)=\left\{\begin{array}{cc}
-2(2\pi)^{10}\,Z_X(\pi_X(\brane))&\e^t\to \infty,\\
{1\over 2}(2\pi)^{10}\,\e^t\cdot Z_Y(\pi_Y(\brane))&\e^t\to 0.
\end{array}\right.
\label{ZD2HT}
\eeq
The factor $\e^t$ in the latter case
is consistent with the shift $[-2]$ in the monodromy $\mdm_Y$.

The results can be employed to find the transport of D-brane charges.
Of course, this applies to the charges over $\Q$
that can be captured through the Chern character.
It ignores the torsion information which we know to be non-trivial
in this model.\footnote{The torison part of the topological K-theory
  of a Calabi-Yau 3-fold with $b^1=0$
  is non-trivial when the fundamental group or the torsion part
  of the third integral cohomology group is non-trivial \cite{BruDis}.
  They are $\pi_1(X)=\Z_2$, ${\rm H}^3(X,\Z)_{\rm tor}=0$
  and $\pi_1(Y)=0$, ${\rm H}^3(Y,\Z)_{\rm tor}=\Z_2$
  for $X$ and $Y$ in the Hosono-Takagi model
  \cite{HoTa13,Addington,Addington2}.
  On the other hand, both vanish for $X$ and $Y$
  in the R\o dland model \cite{Addington}.}\label{page:foot}
Let ${\rm B}_X$ and ${\rm B}_Y$ be the basis over $\Q$ of
D-brane charges on $X$ and $Y$ as in (\ref{basis}).
Note that $[{\mathcal O}_{\ell}]=H^2/35+H^3/35$
and $[{\mathcal O}_{\rm pt}]=H^3/35$ in ${\rm B}_X$ while
$[{\mathcal O}_{\ell}]=H^2/10+H^3/10$
and $[{\mathcal O}_{\rm pt}]=H^3/10$ in ${\rm B}_Y$.
The action of the transport $\GtoP_{\bf w}$ on the charges can
be represented as ${\rm B}_X\mapsto {\rm B}_Y\cdot T_{\bf w}$.
Taking the same procedure as in the R\o dland model,
we find the charge transport matrices $T_{\bf w}$
for six of the windows ${\bf w}$
\beqa
&&T_{\bf A}=\left(\begin{array}{cccc}
-3& 55& 3& 4\\
0& -7& -1& -1\\
5& -120& -8& -10\\
-5& 140& 11& 13
\end{array}\right),\quad
T_{\bf B}=\left(\begin{array}{cccc}
-3& 40& 1& 2\\
0& -7& -1& -1\\
5& -120& -8& -10\\
-5& 140& 11& 13
\end{array}\right),
\nn\\
&&T_{\bf C}=\left(\begin{array}{cccc}
-3& -5& -3& -2\\ 
0& -7& -1& -1\\
5& -20& 2& 0\\
-5& 40& 1& 3
\end{array}\right),\quad
T_{\bf D}=\left(\begin{array}{cccc}
-3& -20& -5& -4\\
 0& -7& -1& -1\\
5& -20& 2& 0\\
-5& 40& 1& 3
\end{array}\right),
\label{HTtM}\\
&&T_{{\bf A}(-1)}=\left(\begin{array}{cccc}
-58& 230& -1& 4\\
65& -272& 1& -5\\
55& -120& 2& 0\\
-130& 365& -4& 3
\end{array}\right),\quad
T_{{\bf D}(-1)}=\left(\begin{array}{cccc}
17& -125& -1& -4\\
-10& 83& 1& 3\\
-45& 330& 2& 10\\
70& -535& -4& -17
\end{array}\right).
\nn
\eeqa
The transport matrices for shifted windows are related by the identity
$M_YT_{\bf w}M_X=T_{{\bf w}(1)}$ representing (\ref{YTXHT}) where
\beq
M_X=
\left(\begin{array}{cccc}
1&0&0&0\\
1&1&0&0\\
35&35&1&0\\
-35&-35&1&1
\end{array}\right)
\quad\mbox{and}\quad
M_Y=\left(\begin{array}{cccc}
1&0&0&0\\
1&1&0&0\\
10&10&1&0\\
-10&-10&1&1
\end{array}\right)
\eeq
represent $\mdm_X$ and $\mdm_Y$. Note that the R-charge shift $[-2]$
is invisible at the level of charges.
This identity is indeed satisfied by the above matrices for
${\bf w}={\bf A}(-1)$ and ${\bf D}(-1)$.

Next, we discuss the properties of a class of branes and their consequences.
Let us introduce a brane with matrix factorization
\beq
Q=\sum_{k=1}^5p^k\eta_k+\sum_{k=1}^5S_k(x)\bareta^k,
\eeq
realized on the irreducible representation $\bV$
of the Clifford algebra 
\beq
\{\eta_k,\bareta^l\}=\delta_k^l,~\{\eta_k,\eta_l\}=\{\bareta^k,\bareta^l\}=0,
\quad 1\leq k,l\leq 5.
\eeq
The $\eta_k$'s and the $\bareta^k$'s
are assigned the gauge and R-charges $(\C(-1),-1)$ and $(\C(1),+1)$,
respectively.
We assign trivial gauge and R-charges to the
Clifford vacuum $|0\rangle\in \bV$ annihilated by the $\eta_k$'s.
This brane can also be described by the diagram
\beq
\begin{tikzcd}
  \C\arrow[r,yshift=0.5ex, "S(x)"] &
  \arrow[l,yshift=-0.5ex, "p"]\C(1)^{\oplus 5}\arrow[r,yshift=0.5ex, "S(x)"]
  & \arrow[l,yshift=-0.5ex, "p"] \C(2)^{\oplus {5\choose 2}}
  \arrow[r,yshift=0.5ex, "S(x)"]
  & \arrow[l,yshift=-0.5ex, "p"]
  \C(4)^{\oplus {5\choose 3}}\arrow[r,yshift=0.5ex, "S(x)"]
  &\arrow[l,yshift=-0.5ex, "p"] \C(4)^{\oplus 5}
  \arrow[r,yshift=0.5ex, "S(x)"]
  &\arrow[l,yshift=-0.5ex, "p"] \C(5)\\[-0.7cm]
  {}_0&{}_1&{}_2&{}_3&{}_4&{}_5
\end{tikzcd}
\label{HTKp}
\eeq
We shall call it the $p$-Koszul brane and denote it by $\Kos$.
In the Reye congruence phase $\zeta\gg 0$, we expect that it flows to the
structure sheaf of $X$,
\beq
\pi_X(\Kos)\cong {\mathcal O}_X,
\eeq
under an appropriate relation between the gauge theory theta angle and
the low energy B-field.
The tensor product of $\Kos$ and
a representation $\rmV$ of the gauge group flows to the vector bundle
associated to $\rmV$,
\beq
\pi_X(\Kos\otimes\rmV)\cong {\mathcal O}_X(\rmV).
\eeq
In the determinantal phase $\zeta\ll 0$,
the boundary potential 
$\{Q,Q^{\dag}\}=(\|p\|^2+\|S(x)\|^2){\rm id}_\rmV$
is positive definite as $p\ne 0$, and hence the $p$-Koszul brane is empty
at low energies,
\beq
\pi_Y(\Kos)\cong 0.
\eeq

\newcommand{\vHT}{V}

The $p$-Koszul brane plays an important r\^ole in the monodromy
along loops around the singular points with a base point in
the Reye congruence phase.
For $\bfb=0,1,2$,
let ${\mathcal L}_{\bfb}$ be a loop that comes from $\zeta\gg 0$,
goes around $t_{\bfb}$ clockwise and goes back to $\zeta\gg 0$,
similarly to the loops ${\mathcal L}_{\bfa}$ for
$\bfa=1,2,3$ in Fig.~\ref{fig:monodromy} in the R\o dland model.
Each of them can be regarded as a downward path through a window
followed by an upward path through another.
The difference between the grade restriction rules for the two windows is
which of the two (sets of) special representations at $t_\bfb$ is included.
The two are $\rmV_\bfb$ and $\rmV_\bfb(5)$,
with $\rmV_\bfb=\C_{\pm}$, $\Box$, $\ydt(-1)$
for $\bfb=0,1,2$, respectively.
The representation $\rmV_\bfb$
is good on the right of $t_\bfb$ and bad on its left
while $\rmV_\bfb(5)$ is good on its left and
bad on its right.
If the brane is grade restricted with respect to the first window
and is transported
along the downward path that passes by $t_\bfb$ on its right,
before it goes back through the upward path that passes by $t_\bfb$ on its left,
each of the $\rmV_\bfb$ components of the brane must be traded for
$\rmV_\bfb(5)$ while in the determinantal phase.
The trading can be done by binding $\Kos\otimes \rmV_\bfb$
followed by brane-antibrane annihilation, since
it is empty in the determinantal phase and
has $\rmV_\bfb$ and $\rmV_\bfb(5)$ at the edge components,
with the other components in the overlap of the two grade restriction rules.
Since $\rmV_\bfb$ appears as the head of the
exact $p$-Koszul complex tensored with $\rmV_\bfb$,
the binding is done by a homomorphism
from the given brane to the empty brane.
For $\bfb=0$, this must be done for {\it both} $\rmV_0=\C_+=\C$ {\it and}
$\C_-$.
The effect is the composition of the twists
by $\pi_X(\Kos)={\mathcal O}_X$
and $\pi_X(\Kos\otimes \C_-)={\mathcal O}_X(\C_-)$.
For $\bfb=1,2$, the effect is the single twist by
$\pi_X(\Kos\otimes \rmV_\bfb)\cong{\mathcal O}_X(\rmV_{\bfb})$.
To summarize,
\beq
\mdm_{{\mathcal L}_0}=\twist_{{\mathcal O}_X}\circ\twist_{{\mathcal O}_X(\C_-)},\quad
\mdm_{{\mathcal L}_1}=\twist_{{\mathcal O}_X(\Box)},\quad
\mdm_{{\mathcal L}_2}=\twist_{{\mathcal O}_X(\ydt(-1))}.
\label{mdmHT}
\eeq
Let us describe what the vector bundles ${\mathcal O}_X(\Yd(i))$ are.
Recall that $X$ is the free quotient of $\wt{X}\subset \CP^4\times\CP^4$
by the exchange of the two $\CP^4$'s.
The line bundle
\beq
    {\mathcal O}_X(\C_-)=:{\mathcal O}_X^-
\eeq
is the non-trivial local system associated with this $\Z_2$ quotient.
If we denote by ${\mathcal O}_{\wt{X}}(i,j)$ the restriction to $\wt{X}$
of the line bundle ${\mathcal O}_{\CP^4}(i)\boxtimes {\mathcal O}_{\CP^4}(j)$,
then, ${\mathcal O}_X(i):={\mathcal O}_X(\C(i))$
is the quotient of ${\mathcal O}_{\wt{X}}(i,i)$ by the exchange.
The line bundles ${\mathcal O}_X^-$ and ${\mathcal O}_X(1)$
represent generators of $\Z_2$ and $\Z H$ of the Picard group
(the latter appeared as such already).
We shall write
${\mathcal V}\otimes {\mathcal O}_X(i)=:{\mathcal V}(i)$ for a vector
bundle ${\mathcal V}$ on $X$.
The vector bundle
\beq
{\mathcal O}_X(\Box)=:\vHT_X
\eeq
is the quotient of ${\mathcal O}_{\wt{X}}(1,0)\oplus{\mathcal O}_{\wt{X}}(0,1)$
by the exchange.
It is equipped by a symmetric biliniear form $(\,,\,)$ with values in
${\mathcal O}_X(1)$. Then ${\mathcal O}_X(\ydt)$ is the second symmetric
and $(\,,\,)$-traceless power of $\vHT_X$ which we denote by
$\Sym^2_0 V_X$. In particular,
\beq
{\mathcal O}_X(\ydt(-1))=\Sym^2_0\vHT_X(-1).
\eeq
{\bf Remarks.}
(i) The twists that appear in (\ref{mdmHT}) are all autoequivalences of
$D(X)$ as ${\mathcal O}_X$, ${\mathcal O}_X^-$, $\vHT_X$ and
$\Sym_0^2\vHT_X(-1)$ are spherical objects of $D(X)$.\\
(ii)
The composition of $\twist_{{\mathcal O}_X}$ and
$\twist_{{\mathcal O}_X^-}$ to produce the monodromy along ${\mathcal L}_0$ 
can be in either order. This is consistent only when the two twists
commute with each other.
This is indeed the case as ${\mathcal O}_X$ and ${\mathcal O}_X^-$ are
orthogonal.\\
(iii)
There is another loop $\wt{\mathcal L}_2$ that goes around
$t_2$, similarly to $\wt{\mathcal L}_3$ in Fig.~\ref{fig:monodromy}
in the R\o dland model. The monodromy along $\wt{\mathcal L}_2$
is again the twist by ${\mathcal O}_X(\ydt(-1))$ just as ${\mathcal L}_2$,
even though the two loops are homotopically inequivalent,
obeying
${\mathcal L}_0\circ{\mathcal L}_2\simeq\wt{\mathcal L}_2\circ{\mathcal L}_0$.
This implies that
this monodromy commutes with the monodromy along ${\mathcal L}_0$.
This is indeed the case as ${\mathcal O}_X$ and ${\mathcal O}_X^-$
are orthogonal to $\Sym_0^2\vHT_X(-1)$.\\
A proof of the sphericalness and the orthogonality
is given in Appendix~\ref{subsec:ExtHT}.

Let us compute the effect of the twist operators on the D-brane charges
and compare the result with the monodromies
from the charge transportation rule (\ref{HTtM}).
The D-brane charges of $V_X$ and $\Sym^2_0V_X(-1)$ are
computed in Appendix~\ref{app:classicalReye},
(\ref{AppchVHT}) and (\ref{AppchSym2VHT}),
which we present here in a differernt basis:
\beqa
[V_X]&=&2[{\mathcal O}_X]+[{\mathcal O}_H]+25[{\mathcal O}_{\ell}]
-30[{\mathcal O}_{\rm pt}],
\label{chVHT}\\
{[}\Sym^2_0V_X(-1){]}&=&2[{\mathcal O}_X]-5[{\mathcal O}_{\ell}]
+5[{\mathcal O}_{\rm pt}].
\label{chSym2VHT}
\eeqa
Note also that $[{\mathcal O}_X^-]=[{\mathcal O}_X]$ over $\Q$.
The Euler pairing is
\beq
\begin{array}{c|cccc}
-
&{\mathcal O}_X&{\mathcal O}_H&{\mathcal O}_{\ell}&{\mathcal O}_{\rm pt}\\
\hline
\chi({\mathcal O}_X,-)&0&10&1&1\\
\chi(V_X,-)&-5&45&1&2\\
  \chi(\Sym^2_0V_X(-1),-)&0&15&2&2
\end{array}
\label{EulerReye}
\eeq
Thus, the twists on the charges are represented by
\beqa
\twist_{{\mathcal O}_X},\,\twist_{{\mathcal O}^-_X}:&&\!\!\!{\bf 1}_4
+\left(\begin{array}{c}1\\0\\0\\0\end{array}\right)(0,10,1,1)
  =\left(\begin{array}{cccc}
    1&10&1&1\\
    0&1&0&0\\
    0&0&1&0\\
    0&0&0&1
    \end{array}\right)
\label{ReyeTO}\\
\twist_{V_X}:&&\!\!\!{\bf 1}_4
+\left(\begin{array}{c}2\\1\\25\\-30\end{array}\right)(-5,45,1,2)
  =\left(\begin{array}{cccc}
    -9&90&2&4\\
    -5&45&1&2\\
    -125&1125&26&50\\
    150&-1350&-30&-59
    \end{array}\right)
\label{ReyeTV}
\\
\twist_{\Sym^2_0V_X(-1)}:&&\!\!\!{\bf 1}_4
+\left(\begin{array}{c}2\\0\\-5\\5\end{array}\right)(0,15,2,2)
  =\left(\begin{array}{cccc}
    1&30&4&4\\
    0&1&0&0\\
    0&-75&-9&-10\\
    0&75&10&11
    \end{array}\right)
\label{ReyeTSV}
\eeqa      
One can check that (\ref{ReyeTO}) squared, (\ref{ReyeTV}) and (\ref{ReyeTSV})
match respectively with the monodromy matrices obtained by the 
compositions of the transport matrices in (\ref{HTtM})
\beq
T_{\bf A}^{-1}T_{\bf C}=T_{\bf B}^{-1}T_{\bf D},
\quad
T_{{\bf D}(-1)}^{-1}T_{\bf A},
\quad
T_{\bf C}^{-1}T_{\bf D}=T_{\bf A}^{-1}T_{\bf B}.
\label{MMMeqHT}
\eeq
The last equality of (\ref{MMMeqHT}) is the agreement of
the monodromies along ${\mathcal L}_2$ and $\widetilde{\mathcal L}_2$
at the charge level and is compatible with the commutativity between
${\mathcal O}_X$, ${\mathcal O}_X^-$ and $\Sym^2_0V_X(-1)$.
This is equivalent to the vanishing of the Euler pairing between
${\mathcal O}_X$, ${\mathcal O}_X^-$ and $\Sym^2_0V_X(-1)$
(see the lower-left corner of table (\ref{EulerReye}))
which of course also follows from the orthogonality of these vector bundles.

We may also be interested in the monodromies along loops with the base point
in the determinantal phase. In order to determine them, we need to identify
the ``right-fit'' empty brane in the Reye congruence phase $\zeta\gg 0$
and find their low energy behaviour in the determinantal phase $\zeta\ll 0$.
This needs extra work and will be presented in a future publication.
In the present paper, we shall instead
determine such monodromies using the dual model
in which it is much easier to
identify the ``right-fit'' empty branes in the Reye congruence phase
and to find their low energy behaviour in the determinantal phase,
just as in the R\o dland model versus its dual.

\subsection{Dual Hosono-Takagi Model}
\label{subsec:dualHT}

The dual Hosono-Takagi model is obtained by applying the 2d Seiberg duality
\cite{Hduality} to the $O(2)$ factor of the gauge group (\ref{GHT})
of the Hosono-Takagi model.
The gauge group is
\beq
G^{\vee}={U(1)\times SO(4)\over \{(\pm 1,\pm {\bf 1}_4)\}}
\eeq
The matter consists of five fields $\tx^1,\ldots, \tx^5$ in $\Box(-1)$,
five fields $p^1,\ldots, p^5$ in $\C(-1)$, and fifteen fields $s_{ij}=s_{ji}$
($i,j=1,\ldots, 5$) in $\C(1)$.
The superpotential and the twisted superpotential are
\beqa
W^{\vee}&=&\sum_{i,j}s_{ij}(\tx^i\tx^j)+\sum_{i,j,k}S^{ij}_kp^ks_{ij},
\label{WdualHT}\\
\wt{W}^{\vee}&=&-2 t^{\vee}\usigma_{\mathfrak{u}(1)}.\label{twistedWdualHT}
\eeqa
The expression $(\tx^i\tx^j)$ in $W^{\vee}$ is the $SO(4)$ invariant symmetric
pairing.

The model has two phases. The $\zeta^{\vee}\ll 0$ phase is weakly coupled.
A solution to the vacuum equation breaks the gauge group completely,
and the vacuum manifold is the locus of $(p,\tx,s)$ with $s=0$,
$p\ne 0$ and $(p,\tx)$ solving the F-term equation
\beq
F^{ij}(p,\tx):=S^{ij}(p)+(\tx^i\tx^j)=0,
\label{FdualHT}
\eeq
modded out by the $G^{\vee}_{\C}$ action. This is isomorphic to
the determinantal Calabi-Yau manifold $Y$ of the Hosono-Takagi model.
Indeed, if $(p,\tx)$ solves (\ref{FdualHT}), then $S(p)$ is of rank 4 or less,
and the map $(p,\tx)\mapsto p$ is a double cover which is ramified along the
rank 3 locus.
The $\zeta^{\vee}\gg 0$ phase is strongly coupled.
The D-term equation requires $s\ne 0$ which leaves an $SO(4)$ subgroup unbroken.
After understanding the $SO(4)$ gauge dynamics, we find that
the theory reduces at low energies to
the sigma model on the rank 2 locus of $s$ satisfying the F-term equation
\beq
\sum_{i,j}S^{ij}_ks_{ij}=0.
\label{FdualHTstrong}
\eeq
Regarding $s_{ij}$ as $(x_ix_j)$,
the target space is nothing but the Reye congruence $X$.
In constrast to the original Hosono-Takagi model,
$X$ appears in the strongly coupled phase
and $Y$ appears in the weakly coupled phase in the dual model.

A maximal torus $T^{\vee}\subset G^{\vee}$ can be taken to be the set
of elements of the form
\beq
\left[\left(h_{\ul{0}},\left(\begin{array}{cccc}
    \cos\psi_{\ul{1}}&-\sin\psi_{\ul{1}}&&\\
    \sin\psi_{\ul{1}}&\cos\psi_{\ul{1}}&&\\
    &&\cos\psi_{\ul{2}}&-\sin\psi_{\ul{2}}\\
    &&\sin\psi_{\ul{2}}&\cos\psi_{\ul{2}}
  \end{array}\right)\right)\right].
\eeq
We write $h_{\ul{1}}=\e^{\im\psi_{\ul{1}}}$,
$h_{\ul{2}}=\e^{\im\psi_{\ul{2}}}$. Then, the identification is
$(h_{\ul{0}},h_{\ul{1}},h_{\ul{2}})\sim (-h_{\ul{0}},-h_{\ul{1}},-h_{\ul{2}})$.
A single valued parametrization is
$(h_0,h_1,h_2)=(h_{\ul{0}}^2,h_{\ul{0}}h_{\ul{1}},h_{\ul{0}}h_{\ul{2}})$.
The scalar components of the vector multiplet in the two parametrizations
are related by the same formulae as (\ref{sigmaulsigma}).
Note that $\usigma_{\mathfrak{u}(1)}$ in (\ref{twistedWdualHT})
is $\usigma_{\ul{0}}$.
As in the dual R\o dland model, we shall call
the coordinate systems for $(h_{\ul{0}},h_{\ul{1}},h_{\ul{2}})$
and $(h_0,h_1,h_2)$
the {\it original frame} and 
the {\it single valued frame}, respectively.
The Weyl group is $(\Z_2)^2$ generated by the exchange
$\usigma_{\ul{1}}\leftrightarrow\usigma_{\ul{2}}$ and the simultaneous sign flip
of $\usigma_{\ul{1}}$ and $\usigma_{\ul{2}}$, that is,
$\usigma_1\leftrightarrow \usigma_2$ and
$(\usigma_1,\usigma_2)\to (\usigma_0-\usigma_1,\usigma_0-\usigma_2)$.
Half the sum of positive roots $\rho$ is
\beq
h^{\rho}=h_{\ul{1}}=h_0^{-{1\over 2}}h_1.
\eeq
It is equal to $h_0^{1\over 2}$ modulo a weight.
The weights of the matter fields are
$h_{\ul{0}}^{-2}=h_0^{-1}$ for the $p^k$,
$h_{\ul{0}}^{-1}(h_{\ul{1}}^{\pm 1},h_{\ul{2}}^{\pm 1})$
for the $(\tx_1^i\pm\im\tx_2^i,\tx_3^i\pm\im \tx_4^i)$
(which are $h_0^{-1}h_1,h_1^{-1},h_0^{-1}h_2,h_2^{-1}$)
and $h_{\ul{0}}^2=h_0$ for the $s_{ij}$.
The effective twisted superpotential is
\beqa
\wt{W}^{\vee}_{\it eff}
&=&-t^{\vee}\usigma_0+\pi \im \usigma_0-5(-\usigma_0)(\log(-\usigma_0)-1)
-15\usigma_0(\log\usigma_0-1)\nn\\
&&-5\sum_{s=1}^2(\usigma_s-\usigma_0)(\log(\usigma_s-\usigma_0)-1)
-5\sum_{s=1}^2(-\usigma_s)(\log(-\usigma_s)-1).
\eeqa
The vacuum equation reads
\beq
\e^{t^{\vee}}={(\usigma_1-\usigma_0)^5(\usigma_2-\usigma_0)^5\over
  \usigma_0^{10}},\quad
  {(\usigma_s-\usigma_0)^5
    \over (-\usigma_s)^5}=1~~(s=1,2).
\eeq
A solution is $\usigma_0/\usigma_s-1=\omega_s$ with $\omega_s^5=1$ for $s=1,2$,
that is,
\beq
\usigma_1={\usigma_0\over \omega_1+1},\quad
\usigma_2={\usigma_0\over \omega_2+1},\qquad
\omega_1^5=\omega_2^5=1,
\label{dualHTCB}
\eeq
and it exists only when $\e^{t^{\vee}}=1/((1+\omega_1)^5(1+\omega_2)^5)$.
The Weyl group generators act as
$(\omega_1,\omega_2)\to(\omega_2,\omega_1),(\omega_1^{-1},\omega_2^{-1})$
and the fixed points are excluded as
they have enhanced $SU(2)$ symmetry \cite{HoTo,longflow}.
There are only four inequivalent cases:
$(\omega_1,\omega_2)=(1,\omega)$, $(1,\omega^2)$, $(\omega,\omega^2)$,
$(\omega,\omega^3)$
with $\omega=\e^{2\pi\im/5}$.
As $(1+\omega)(1+\omega^2)=-\omega^4$ and $(1+\omega)(1+\omega^3)=-\omega^2$,
these correspond respectively to
\beq
\e^{t^{\vee}}\,=\,-2^{-5}(1+\omega^2)^5,~\,
-2^{-5}(1+\omega)^5,~\,-1,~\,-1.
\eeq
These are nothing but $2^{-5}$ times
the values
$\beta_2,\beta_1,\beta_0,\beta_0$
of $\e^t$ at the singular points
of the Hosono-Takagi model (\ref{defbfb}).
This means that the FI-theta parameters of
the original and the dual models are related by
\beq
=t^{\vee}+5\log 2\quad\mbox{modulo $2\pi\im$.}
\eeq

For a Young diagram ${\rm Y}$, we have a representation of $SO(4)$,
\beq
\rmV_{\rm Y}=c_{\rm Y}\left[(\C^4)^{\otimes |{\rm Y}|}\right]_0
\eeq
where $\C^4$ is the defining representation of $SO(4)$,
$|{\rm Y}|$ is the number of boxes in ${\rm Y}$,
$c_{\rm Y}$ is the Young symmetrizer for ${\rm Y}$,
and $[-]_0$ is the subspace consisting of vectors that are traceless with
respect to the symmetric bilinear form on $\C^4$.
If the height of ${\rm Y}$ is 0 or 1, it is irreducible.
If the height of ${\rm Y}$ is 2, it splits to a direct sum of two
irreducible representations, $\rmV_{{\rm Y}_+}$ and $\rmV_{{\rm Y}_-}$,
distinguished by the eigenvalues, $1$ or $-1$,
of the Hodge $\ast$ operator on the first column of ${\rm Y}$.
These exhaust all the irreducible representations of $SO(4)$ \cite{Weyl}.
Thus, the label $\Yd$ of an irreducible representation of $SO(4)$ is
a Young diagram of height 0, 1 or a height 2 diagram with a sign.
As described earlier, an irreducible representation of $G^{\vee}$
is labeled by $\Yd(i)$.
If the number of boxes of the first and the second rows of
$\Yd$ are $\Yd_1$ and $\Yd_2$, the highest weight
of $\Yd(i)$ is
$h_{\ul{0}}^{2i+|\Yd|}h_{\ul{1}}^{\Yd_1}h_{\ul{2}}^{\pm\Yd_2}$
where the last factor is $h_{\ul{2}}^{-\Yd_2}$ only for $\Yd={\rm Y}_-$.
Thus the shifted highest weight $\wtq$ for this representation is
\beq
h^{\wtq}=\left\{\begin{array}{ll}
h_{\ul{0}}^{2i+|\Yd|}h_{\ul{1}}^{\Yd_1+1}h_{\ul{2}}^{\Yd_2}
=h_0^{i-{1\over 2}}h_1^{\Yd_1+1}h_2^{\Yd_2}&\Yd\ne {\rm Y}_-\\
h_{\ul{0}}^{2i+|\Yd|}h_{\ul{1}}^{\Yd_1+1}h_{\ul{2}}^{-\Yd_2}
=h_0^{i+\Yd_2-{1\over 2}}h_1^{\Yd_1+1}h_2^{-\Yd_2}&\Yd={\rm Y}_-.
\end{array}\right.
\label{wtqforYiHT}
\eeq
The growth function that controls
the convergence of the hemisphere integral (\ref{ZV}) is
\beqa
\AN_{\wtq}&=&2\zeta^{\vee}\Is_{\ul{0}}-2\theta^{\vee}\Rs_{\ul{0}}
-2\pi \langle\wtq,\Rs\rangle\nn\\
&&+5(-2\Rs_{\ul{0}})\Arg(-2\im\usigma_{\ul{0}})
+15(2\Rs_{\ul{0}})\Arg(2\im\usigma_{\ul{0}})\nn\\
&&+\sum_{s=1}^2\bigl(5(-\Rs_{\ul{0}}+\Rs_{\ul{s}})
\Arg(\im(-\usigma_{\ul{0}}+\usigma_{\ul{s}}))
+5(-\Rs_{\ul{0}}-\Rs_{\ul{s}})
\Arg(\im(-\usigma_{\ul{0}}-\usigma_{\ul{s}}))\bigr)\nn\\
&&+5(-2\Is_{\ul{0}})\log|2\usigma_{\ul{0}}|
+15(2\Is_{\ul{0}})\log|2\usigma_{\ul{0}}|\nn\\
&&+\sum_{s=1}^2\bigl(
5(-\Is_{\ul{0}}+\Is_{\ul{s}})\log|-\usigma_{\ul{0}}+\usigma_{\ul{s}}|
+5(-\Is_{\ul{0}}-\Is_{\ul{s}})\log|-\usigma_{\ul{0}}-\usigma_{\ul{s}}|
\bigr).
\label{ANdualHT}
\eeqa
The integration contour $\bgamma\subset\ttt_{\C}$
is a deformation of the real locus
$\im\ttt$ avoiding the poles of the Gamma function factor,
which is assured if it avoids the walls
$W_p$, $W_s$, $W_{12+}$, $W_{12-}$, $W_{34+}$, $W_{34-}$
whose locations are the same as
$W_p$, $W_a$, $W_1$, $W_3$, $W_2$, $W_4$
given in (\ref{WpdualR})-(\ref{W4dualR}) in the dual R\o dland model.
If we regard $\bgamma$ as a graph of a function $\Rs\mapsto\Is(\Rs)$,
the condition to avoid the walls is
identical to (\ref{Wedgepa}), (\ref{Wedge13}), (\ref{Wedge24}).
In addition to these similarities, note that the growth function
(\ref{ANdualHT}) is the same as the one (\ref{ANdualR}) for the dual
R\o dland model if we make a replacement $7\to 5$.
Thus, many of the computations done for the dual R\o dland model can be reused
in the dual Hosono-Takagi model.

The implication of the condition $\Rs_{\ul{0}}=0\Rightarrow \Is_{\ul{0}}=0$
to avoid the walls $W_p$ and $W_s$ is
\beq
|\wtq^{\ul{1}}|,~|\wtq^{\ul{2}}|\,<\,{5\over 2}.
\label{SO4GRR0}
\eeq
As $\wtq^{\ul{1}}=\Yd_1+1>|\wtq^{\ul{2}}|=Y_2$ by (\ref{wtqforYiHT}),
this is equivalent to $\Yd_1<3/2$ which is solved by $\Yd_1=0,1$.
Only the four series of representations,
\beq
\C(i),\quad
\Box(i),\quad
\ydoopm(i)
\label{SO4GRR}
\eeq
are allowed. We shall refer to this as the $SO(4)$ grade restriction rule,
as it originates from the $SO(4)$ SQCD that appears at some
energy scale. This is imposed of course in the strongly coupled phase
$\zeta^{\vee}\gg 0$ but also in the ``weakly'' coupled phase:
if the Higgsing scale is set higher than the superpotential scale,
the theory is secretly strongly coupled at the intermediate energy scale.

As the good contours deep inside the phases, we can take
$\bgamma_+$ for $\zeta\gg 0$ and $\bgamma_-$ for $\zeta\ll 0$
which are specified by the same expressions as in the dual R\o dland model:
(\ref{bgammapdual}) and (\ref{bgammamdual}).
The hemisphere integrals for the $SO(4)$ grade restricted representations
(\ref{SO4GRR}) are absolutely convergent along the
respective contour in the respective phase.

\begin{figure}[thb]
\psfrag{A}{\mbox{\small ${\mathcal P}_{\!\bf A}$}}
\psfrag{B}{\mbox{\small ${\mathcal P}_{\bf B}$}}
\psfrag{C}{\mbox{\small ${\mathcal P}_{\bf C}$}}
\psfrag{D}{\mbox{\small ${\mathcal P}_{\bf D}$}}
\psfrag{a1}{$t_0^{\vee}$}
\psfrag{a2}{$t_1^{\vee}$}
\psfrag{a3}{$t_2^{\vee}$}
\psfrag{z}{$\zeta^{\vee}$}
\psfrag{th}{$\theta^{\vee}$}
\psfrag{-8pi}{\!\!\!$-16\pi$\,\,\,}
\psfrag{-6pi}{$-14\pi$}
\centerline{\includegraphics[page=50]{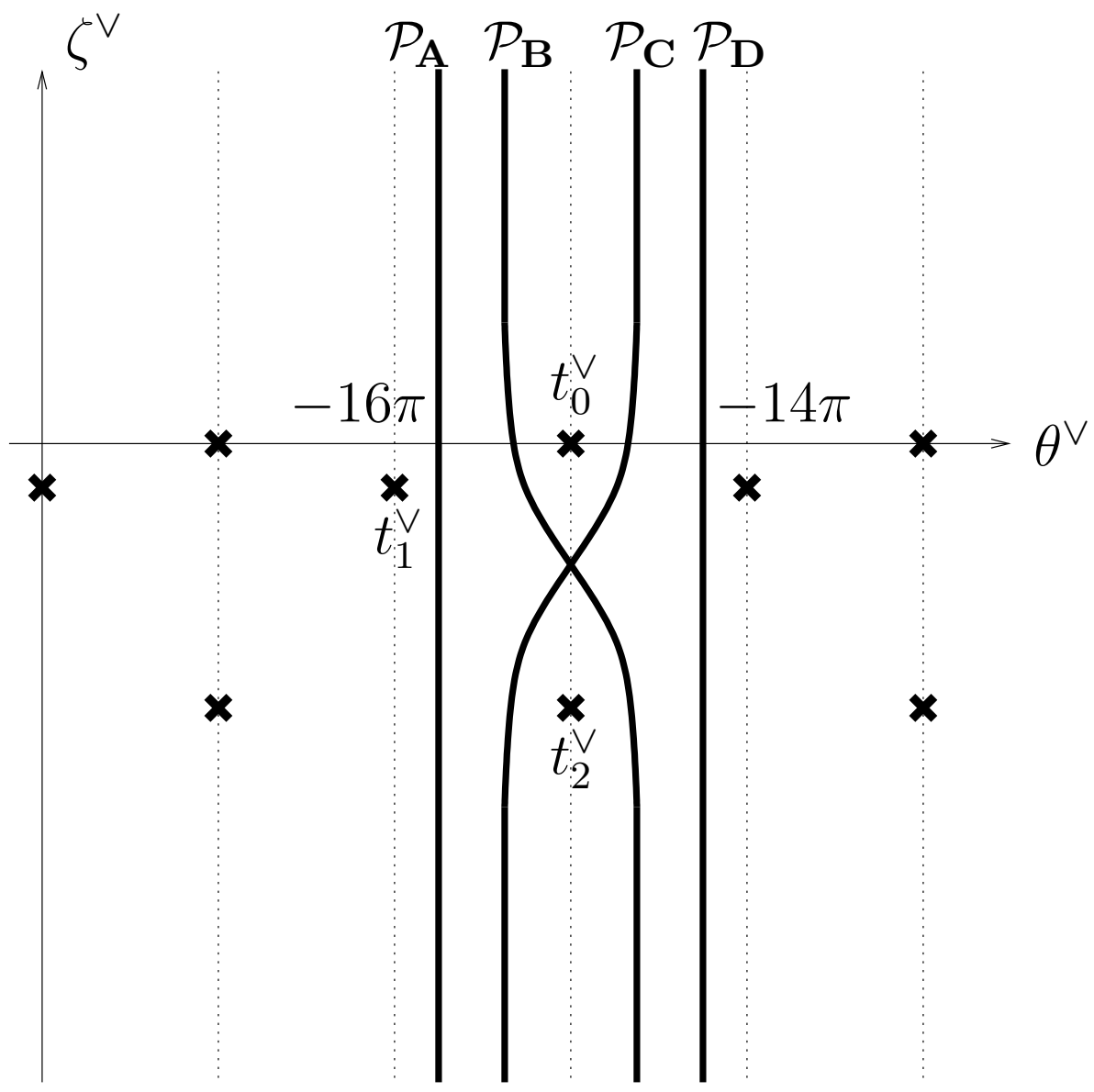}}
\caption{Four windows between the phases}
\label{fig:dualHTpaths}
\end{figure}
Next, let us determine the grade restriction rule for paths between
the two phases. Let us consider the four classes of paths
${\mathcal P}_{\bf A},$
${\mathcal P}_{\bf B},$
${\mathcal P}_{\bf C},$
${\mathcal P}_{\bf D}$
that go near the three points
\beqa
t^{\vee}_0&=&15\pi \im,
\label{dualHTt0}\\
t^{\vee}_1&=&16\pi\im+5\log\left(\cos\left({\pi\over 5}\right)\right),
\label{dualHTt1}\\
t^{\vee}_2&=&15\pi\im+5\log\left(\cos\left({2\pi\over 5}\right)\right),
\label{dualHTt2}
\eeqa
as depicted in Fig.~\ref{fig:dualHTpaths}.
We again take the shortcut to find the rule.

\begin{figure}[thb]
  \psfrag{w1}{$W_{12+}$}
  \psfrag{w2}{$W_{34+}$}
  \psfrag{w3}{$W_{12-}$}
  \psfrag{w4}{$W_{34-}$}
  \psfrag{wp}{$W_p$}
  \psfrag{wa}{$W_s$}
  \psfrag{sp}{$\usigma_0$}
  \psfrag{I}{${\rm I}$}
  \psfrag{II}{${\rm II}$}
  \psfrag{III}{${\rm III}$}
  \psfrag{IV}{${\rm IV}$}
  \psfrag{V}{${\rm V}$}
  \psfrag{VI}{${\rm VI}$}
\centerline{\includegraphics[page=51]{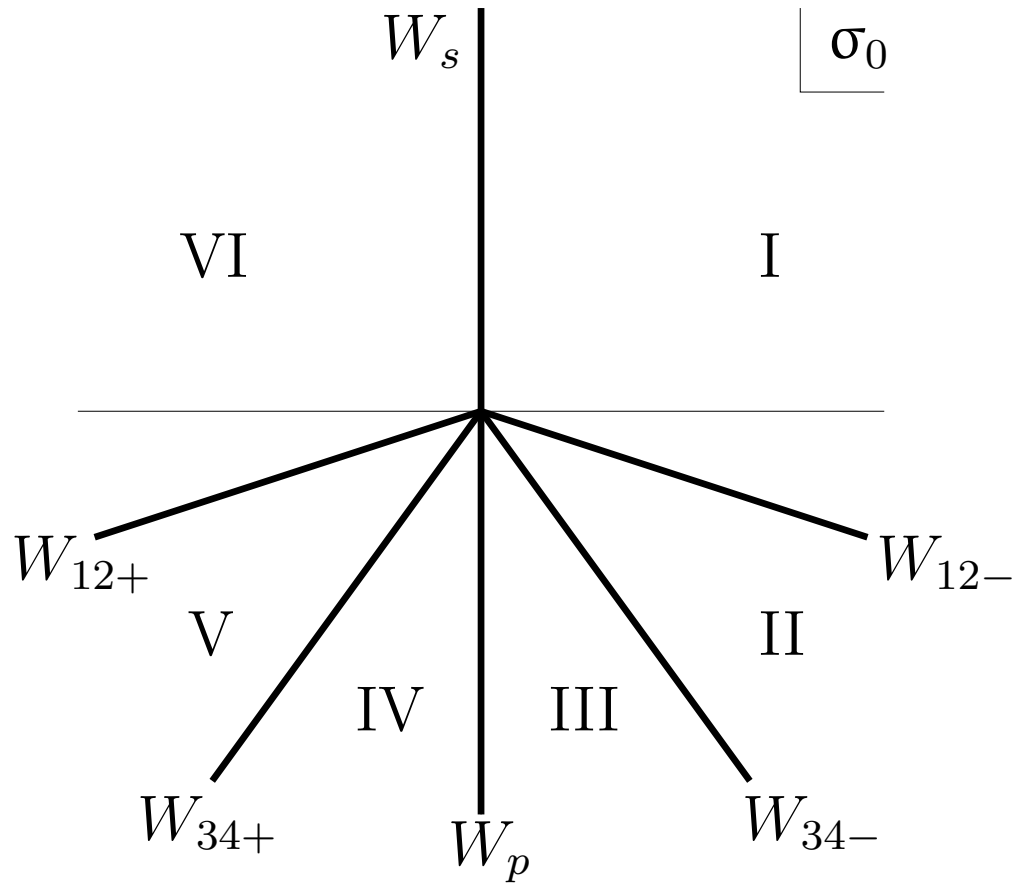}}
\caption{Chambers of the Coulomb branch.}
\label{fig:CBdualHT}
\end{figure}

The shifted highest weight of the
special representation at the singular point $t^{\vee}=t^{\vee}_*$
is, in the single valued frame,
\beqa
2\pi\im\wtq_*
&=&(t^{\vee}_*,0,0)+5(-1,0,0)\log(-\im\usigma_0)
+15(1,0,0)\log(\im\usigma_0)\nn\\
&&+5(-1,1,0)\log(\im(\usigma_1-\usigma_0))
+5(-1,0,1)\log(\im(\usigma_2-\usigma_0))\nn\\
&&+5(0,-1,0)\log(-\im\usigma_1)
+5(0,0,-1)\log(-\im\usigma_2),
\label{1stwtqdualHT}
\eeqa
in a chamber of the Coulomb branch (\ref{dualHTCB}).
For $\omega_s=\e^{2\pi\im n_s/5}$, $s=1,2$, we have
\beq
\usigma_s={\e^{-{\pi\im n_s\over 5}}\usigma_0\over
  2\cos\left({\pi n_s\over 5}\right)},\quad
\usigma_s-\usigma_0=-{\e^{{\pi\im n_s\over 5}}\usigma_0\over
  2\cos\left({\pi n_s\over 5}\right)}.
\eeq
Recall that the Coulomb branches at the singular points
$t_*^{\vee}=t^{\vee}_2$, $t^{\vee}_1$ and $t^{\vee}_0$ (double)
have 
$(\omega_1,\omega_2)=(1,\omega)$, $(1,\omega^2)$,
and ($(\omega,\omega^2)$, $(\omega,\omega^3)$)
with $\omega:=\e^{2\pi \im/5}$ or their Weyl images.
The real part on the right-hand side
of (\ref{1stwtqdualHT}) vanishes under this correspondence.
To find the imaginary part, we need to evaluate
$\Arg(\im Q_i(\usigma))$ which are assumed to take values in
$(-{\pi\over 2},{\pi\over 2})$.
They depend on the chamber of the Coulomb branch separated by the walls.
The chambers for the branch with $(\omega_1,\omega_2)=(\omega^2,\omega)$
are depicted in Fig.~\ref{fig:CBdualHT}.
In general, the walls $W_s$ and $W_p$ are in the positive and negative imaginary
directions in the complex $\usigma_0$-plane, while
the walls $W_{12\pm}$ and $W_{34\pm}$
make angles $\mp {n_1\pi\over 5}$ and $\mp {n_2\pi\over 5}$
with respect to $W_p$,
provided $n_1$ and $n_2$ are chosen from $0,\pm 1,\pm 2$.
Each of the four Coulomb branches has a unique Weyl representative with
$n_1>\pm n_2$ in this range: $(n_1,n_2)=(1,0)$, $(2,0)$ and
($(2,1)$, $(2,-1)$), respectively, at $t^{\vee}_2$, $t^{\vee}_1$
and $t^{\vee}_0$. Fig.~\ref{fig:CBdualHT} is for $(2,1)$, and also
for $(2,-1)$ provided that the labels $W_{34+}$ and $W_{34-}$ are swapped.
For $(1,0)$ and $(2,0)$, the walls $W_{34\pm}$ coincide with $W_p$
and the chambers III and IV are squeezed out.
Let us evaluate $\Arg_i=\Arg(\im Q_i(\usigma))$ for such $(n_1,n_2)$.
For $(n_1,n_2)=(1,0), (2,0), (2,1)$, they are 
\beq
\begin{array}{c|cccccc|c}
  &\Arg_p&\Arg_s&\Arg_{12+}
  &\Arg_{12-}&\Arg_{34+}&\Arg_{34-}
  &{1\over 2\pi}\sum_iQ_i\Arg_i\\
  \hline
              {\rm I}&\varphi&\varphi+\pi
              &\varphi+{\pi n_1\over 5}
              &\varphi-{\pi n_1\over 5}
              &\varphi+{\pi n_2\over 5}
              &\varphi-{\pi n_2\over 5}
              &({15-n_1-n_2\over 2},n_1,n_2)\\
              {\rm II}&{}_{''}&{}_{''}&{}_{''}&{}_{''}+2\pi&{}_{''}&{}_{''}
              &({15-n_1-n_2\over 2},n_1-5,n_2)\\
              {\rm III}&{}_{''}&{}_{''}&{}_{''}&{}_{''}&{}_{''}&{}_{''}+2\pi
              &({15-n_1-n_2\over 2},n_1-5,n_2-5)\\
              {\rm IV}&{}_{''}+2\pi&{}_{''}&{}_{''}&{}_{''}&{}_{''}&{}_{''}
              &({5-n_1-n_2\over 2},n_1-5,n_2-5)\\
              {\rm V}&{}_{''}&{}_{''}&{}_{''}&{}_{''}&{}_{''}+2\pi&{}_{''}
              &({-5-n_1-n_2\over 2},n_1-5,n_2)\\
              {\rm VI}&{}_{''}&{}_{''}&{}_{''}+2\pi&{}_{''}&{}_{''}&{}_{''}
              &({-15-n_1-n_2\over 2},n_1,n_2)
\end{array}
\eeq
where $\usigma_0=\im|\usigma_0|\e^{\im\varphi}$ with $-2\pi<\varphi<0$.
For $(n_1,n_2)=(1,0)$ and $(2,0)$,
the rows of III and IV should be deleted {\it after} computing the $\Arg$'s
in the chambers V and VI.
For $(n_1,n_2)=(2,-1)$, the order of ${}_{''}+2\pi$ for $\Arg_{34+}$
and $\Arg_{34-}$ is exchanged from the above and the expressions for
the sum ${1\over 2\pi}\sum_iQ_i\Arg_i$ are altered
for the middle two chambers
to $({5-n_1-n_2\over 2},n_1-5,n_2+5)$ in III
and to $({-5-n_1-n_2\over 2},n_1-5,n_2+5)$ in IV.
The sum ${1\over 2\pi}\sum_iQ_i\Arg_i$ must be equated with
$\wtq_*+({\theta^{\vee}_*\over 2\pi},0,0)$.
We see that $\wtq_*$ in the chambers I and VI
are shifted highest weights of some $SO(4)$ grade restricted
representations of the gauge group:
\beq
\begin{array}{c|c|c|c}
  t_*^{\vee}&(n_1,n_2) &\mbox{$\wtq_*$:$\Yd(i)$ on I}
  &\mbox{$\wtq_*$:$\Yd(i)$ on VI}\\
  \hline
  t_2^{\vee}&(1,0)&(15-{1\over 2},1,0):\C(15)&(-{1\over 2},1,0):\C\\
  t_1^{\vee}&(2,0)&(15-{1\over 2},2,0):\Box(15)&(-{1\over 2},2,0):\Box\\
  t_0^{\vee}&(2,1)&(14-{1\over 2},2,1):\ydoop(14)&(-1-{1\over 2},2,1):
  \ydoop(-1)\\
  {}_{''}&(2,-1)&(15-{1\over 2},2,-1):\ydoom(14)&(-{1\over 2},2,-1):\ydoom(-1)  
\end{array}
\label{specialrepdualHT}
\eeq
On the other hand, $\wtq_*$ in the other chambers or on the other Weyl
representatives are not the shifted highest weights of any of
the $SO(4)$ grade restricted representations.
Thus, we conclude that those in the table (\ref{specialrepdualHT})
are the special representations at the three points.
The special representations at $t^{\vee}_*+2\pi\im j$
are obtained from those at
$t^{\vee}_*$ by $\rmV\mapsto \rmV(j)$.

We anticipate that the two or four representations (\ref{specialrepdualHT})
at $t^{\vee}_*$ undergo a catastrophe along a small circle around $t^{\vee}_*$:
The representation $\C(15)$ ({\it resp}. $\C$) cannot stay
admissible on the right ({\it resp}. left) of $t^{\vee}_2$,
and similarly for $\Box(5)$ ({\it resp}. $\Box$)
on the right ({\it resp}. left) of $t^{\vee}_2$,
and $\mydoopm(14)$ ({\it resp}. $\mydoopm(-1)$)
on the right ({\it resp}. left) of $t^{\vee}_0$.
If $\rmV(15)$ and $\rmV$ are bad on the right and the left of $t^{\vee}_*$,
so are $\rmV(15+j)$ and $\rmV(j)$ on the right and the left of
$t^{\vee}_*+2\pi\im j$.
Then, the grade restriction rules with respect to the four windows would be
\beqa
\!\!\!\!\!\!\!\!\!\!\!\!
{\rm GRR}_{\bf A}&&
\begin{array}{cccccccc}
&&\C(1)&\cdots&\C(13)&\C(14)&\C(15)\\
&~\Box~&\Box(1)&\cdots&\Box(13)&\Box(14)&\\
  ~~~~~~~~~&~\ydoopm~&\ydoopm(1)&\cdots&\ydoopm(13)&\ydoopm(14)&
\end{array}
\label{dHTGRRA}\\[0.2cm]
\!\!\!\!\!\!\!\!\!\!\!\!
{\rm GRR}_{\bf B}&&
\begin{array}{ccccccc}
&~\C~&\C(1)&\cdots&\C(13)&\C(14)&\\
&~\Box~&\Box(1)&\cdots&\Box(13)&\Box(14)&\\
  ~~~~~~~~~&~\ydoopm~&\ydoopm(1)&\cdots&\ydoopm(13)&\ydoopm(14)&
\end{array}
\label{dHTGRRB}\\[0.2cm]
\!\!\!\!\!\!\!\!\!\!\!\!
{\rm GRR}_{\bf C}&&
\begin{array}{ccccccc}
&&\C(1)&\cdots&\C(13)&\C(14)&\C(15)\\
&~\Box~&\Box(1)&\cdots&\Box(13)&\Box(14)&\\
  \,\,\ydoopm(-1)\!\!&~\ydoopm~&\ydoopm(1)&\cdots&\ydoopm(13)&&
\end{array}
\label{dHTGRRC}\\[0.2cm]
\!\!\!\!\!\!\!\!\!\!\!\!
{\rm GRR}_{\bf D}&&
\begin{array}{cccccccc}
&~\C~&\C(1)&\cdots&\C(13)&\C(14)&\\
&~\Box~&\Box(1)&\cdots&\Box(13)&\Box(14)&\\
  \,\,\ydoopm(-1)\!\!&~\ydoopm~&\ydoopm(1)&\cdots&\ydoopm(13)&&
\end{array}
\label{dHTGRRD}
\eeqa
We next confirm this proposal.

By the similarity of the growth functions, some of the results obtained in
the dual R\o dland model can be carried over to the dual Hosono-Takagi model,
by the $7\to 5$ replacement.
For the real contour $\bgamma=\im\ttt$,
the condition for the highest weight of the admissible representation
is the $7\to 5$ version of (\ref{condrealdualR}),
\beq
\epsilon_0\left(\wtq^{\ul{0}}+{\theta^{\vee}\over \pi}\right)
+\epsilon_1\wtq^{\ul{1}}+\epsilon_2\wtq^{\ul{2}}+15\geq 0,
\quad\epsilon_1\wtq^{\ul{1}}+{5\over 2}\geq 0,
\quad\epsilon_2\wtq^{\ul{2}}+{5\over 2}\geq 0,
\label{condrealdualHT}
\eeq
for $\epsilon_0,\epsilon_1,\epsilon_2\in \{\pm 1\}$.
The latter two are nothing but (\ref{SO4GRR0}), which is
the $SO(4)$ grade restriction rule, and it is satisfied by
(\ref{SO4GRR}): $\C(i)$, $\Box(i)$ and $\mydoopm(i)$ for any $i\in\Z$.
The first inequality imposes a condition on $i$. 
For $-16\pi<\theta<-14\pi$, the set of admissible representations
for the real contour is
\beq
\begin{array}{cccccc}
  \C(1)&\C(2)&\cdots&\C(12)&\C(13)&\C(14)\\
  \Box(1)&\Box(2)&\cdots&\Box(12)&\Box(13)&\\
  \ydoopm(1)&\ydoopm(2)&\cdots&\ydoopm(12)&
\end{array}
\eeq
This set can be enlarged by turning on the imaginary part $\Is$.
Maintaining $\Is_{\ul{0}}=0$ as a simplifying assumption
and taking $|\Is_{\ul{s}}|\gg |\Rs_{\ul{s}}\pm\Rs_{\ul{0}}|$
for $s=1,2$ at most places,
we find that the largest number of representations are allowed
if $\Is_{\ul{s}}$ and $\Rs_{\ul{0}}\Rs_{\ul{s}}$ have opposite signs.
For example, we may take $\bgamma_0$ given by (\ref{bgamma0dual}).
Then, the growth of $\AN_{\wtq}$ at infinity requires,
in addition to $|\wtq^{\ul{1}}|,|\wtq^{\ul{2}}|<5/2$,
\beq
\left|\,\wtq^{\ul{0}}+{\theta^{\vee}\over \pi}\,\right|\,<\, 15.
\eeq
For $-16\pi<\theta<-15\pi$ ({\it resp}. $-15\pi<\theta<-14\pi$),
this is satisfied precisely by the representations in (\ref{dHTGRRA})
({\it resp}. (\ref{dHTGRRD})). This holds for any value of $\zeta$
as $\AN_q$ is $\zeta$-independent thanks to $\Is_{\ul{0}}=0$.
Deep inside the phase with $\pm\zeta\gg 0$,
the contour $\bgamma_0$ can be deformed to $\bgamma_{\pm}$ 
while the hemisphere integral stays convergent.
Thus, the representations in (\ref{dHTGRRA}) ({\it resp}. (\ref{dHTGRRD}))
are grade restricted with respect to the window
{\bf A} ({\it resp}. {\bf D}).

The behaviour of $\AN_{q_*}$ near $t=t_*$ in a neighborhood
of the Coulomb branch is almost the same as in the case
of the dual R\o dland model,
except that we need to make the $7\to 5$ replacement.
And we find the analogous conclusion concerning the convergence of
the hemisphere integrals for the special representations.
Put
\beq
R_{n_1}:~-{\pi n_1\over 5}\leq \Arg(\im\sDelta t^{\vee})\leq 0\quad
\left(\,{\it resp}.~L_{n_1}:~\pi\leq \Arg(\im\sDelta t^{\vee})\leq
     \pi+{\pi n_1\over 5}\,\right).
\eeq
Then, the special representation $\C(15)$ ({\it resp}. $\C$)
at $t_2^{\vee}$ is not admissible at $t_2^{\vee}+\sDelta t^{\vee}$
if $\sDelta t^{\vee}$ is in the sector $R_1$ ({\it resp}. $L_1$).
Similarly, the special representation
$\Box(15)$ ({\it resp}. $\Box$) at $t_1^{\vee}$ is not admissible at
$t_1^{\vee}+\sDelta t^{\vee}$
if $\sDelta t^{\vee}$ is in the sector $R_2$ ({\it resp}. $L_2$),
and the special representations
$\mydoopm(14)$ ({\it resp}. $\mydoopm(-1)$) at $t_0^{\vee}$ are
not admissible at
$t_0^{\vee}+\sDelta t^{\vee}$
if $\sDelta t^{\vee}$ is in the sector $R_2$ ({\it resp}. $L_2$).

These forbidden sectors extend to no-entry zones. In general,
the representation $\Yd(i)$ has no-entry zones ${\mathscr R}_{\Yd(i)}$
and ${\mathscr L}_{\Yd(i)}$ that extend to the right and
to the left of the singular points
$(\zeta^{\vee}_{2-|\Yd(i)|},\theta^{\vee +}_{\Yd(i)})$
and $(\zeta^{\vee}_{2-|\Yd(i)|},\theta^{\vee -}_{\Yd(i)})$
where $\theta^{\vee \pm}_{\Yd(i)}=\pi(\pm 15-|\Yd|-2i)$.
Entry to these zones from above is blocked 
by the wall $W_s$.
Entry to ${\mathcal R}_{\Yd(i)}$ ({\it resp}. ${\mathcal L}_{\Yd(i)}$)
from below is blocked 
by another wall which is
$W_{12-}$ ({\it resp}. $W_{12+}$) near the singular point.
The upper boundaries of the zones are the horizontal half-lines.
The lower boundary of ${\mathscr R}_{\Yd(i)}$ is given by
the $7\to 5$ version of (\ref{zetaveetheta}) near the singular point,
in which $\wtq$ is set to be the shifted highest weight of $\Yd(i)$.
The lower boundary of ${\mathscr L}_{\Yd(i)}$ is obtained from this
by the reflection symmetry.

Any path in the window {\bf A}, {\bf B}, {\bf C} or {\bf D} goes through
the no entry zones of the representations outside the set
(\ref{dHTGRRA}), (\ref{dHTGRRB}), (\ref{dHTGRRC}) or (\ref{dHTGRRD}),
respectively.
Thus, only the representations in this set can be grade restricted.
In fact, they are all grade restricted.
We have already seen this for the windows {\bf A} and {\bf D}: transport
via the contour $\bgamma_0$ works.
For the windows {\bf B} and {\bf C}, $\bgamma_0$ also works for all
but the special representations at the singular points
$t_0^{\vee}$ and $t_2^{\vee}$.
For the latter, deformations of $\overline{\bgamma}_0$ work.
This confirms the proposed grade restriction rule.

The picture we expect is the same as (\ref{dualdiagram}),
where now ${\mathcal S}^{\vee}$ is the set of $SO(4)$ grade restricted branes,
$X$ and $Y$ are the Reye congruence and the determinantal Calabi-Yau,
and ${\mathcal T}_{\bf w}$ is the set of branes satisfying the grade
restriction rule which we have just established.
This picture dictates the rule of D-brane transport between the two phases.

\newcommand{\wHT}{W}

The details are left for future work.
In the present paper, we would like to discuss the properties of
a particular class of branes and their consequences.
We choose to work in the R-charge frame where the $s_{ij}$ have R-charge $2$
and others have $0$.
Let us consider the brane datum with matrix factorization of $W^{\vee}$
\beq
Q=\sum_{ij}s_{ij}\eta^{ij}
+\sum_{ij}F^{ij}(p,\tx)\bareta_{ij},
\label{QKdualHT}
\eeq
which is
realized on the irreducible representation $\bV$ of the Clifford algebra
\beq
\{\eta^{ij},\bareta_{kl}\}=
      {1\over 2}\left(\delta^i_k\delta^j_l-\delta^i_l\delta^j_k\right),\quad
      \{\eta^{ij},\eta^{kl}\}=\{\bareta_{ij},\bareta_{kl}\}=0.
\eeq
The $\eta^{ij}$'s and $\bareta_{ij}$'s are given the gauge and R-charges
$(\C(-1),-1)$ and $(\C(1),1)$, respectively.
We assign the gauge and R-charges $(\C(15),0)$ to the Clifford vacuum
$|0\rangle\in \bV$ annihilated by the $\eta^{ij}$'s.
This may be described by
\beq
\begin{tikzcd}
  \C(15)\arrow[r,yshift=0.5ex, "F(p{,}\tx)"] &
  \arrow[l,yshift=-0.5ex, "s"]\C(14)^{\oplus 15}
  \arrow[r,yshift=0.5ex, "F(p{,}\tx)"]
  & \arrow[l,yshift=-0.5ex, "s"] \C(13)^{\oplus {15\choose 2}}
  \arrow[r,yshift=0.5ex, "F(p{,}\tx)"]
  & \arrow[l,yshift=-0.5ex, "s"]
  \,\cdots\,\arrow[r,yshift=0.5ex, "F(p{,}\tx)"]
  &\arrow[l,yshift=-0.5ex, "s"] \C(1)^{\oplus 15}
  \arrow[r,yshift=0.5ex, "F(p{,}\tx)"]
  &\arrow[l,yshift=-0.5ex, "s"] \C\\[-0.7cm]
  {}_0&{}_1&{}_2&&{}_{14}&{}_{15}
\end{tikzcd}
\label{KdualHT}
\eeq
The left arrow $s$ and the right arrow $F(p,\tx)$ are the first and
the second terms of (\ref{QKdualHT}). We call this the
{\it $s$-Koszul brane} and denote it by $\Kos^{\vee}$.
In the Reye congruence phase $\zeta^{\vee}\gg 0$, the boundary potential
$\{Q,Q^{\dag}\}=(\|s\|^2+\|F(p,\tx)\|^2){\rm id}_{\bV}$ is positive definite
as $s\ne 0$, and hence $\Kos^{\vee}$ is empty at low energies,
\beq
\pi^{\vee}_X(\Kos^{\vee})\cong 0.
\eeq
In the determinantal phase $\zeta^{\vee}\ll 0$, the boundary potential vanishes
at $s=F(p,\tx)=0$ which is nothing but $Y$ itself.
Under some relation between the gauge theory theta angle and the
low energy B-field, we expect that it flows to the structure sheaf of $Y$,
\beq
\pi^{\vee}_Y(\Kos^{\vee})\cong {\mathcal O}_Y.
\eeq
The tensor product of $\Kos^{\vee}$ and an $SO(4)$ grade restricted
representation $\rmV$ flows to the vector bundle associated to $\rmV$,
\beq
\pi^{\vee}_Y(\Kos^{\vee}\otimes \rmV)\cong{\mathcal O}_Y(\rmV).
\eeq
Note that ${\mathcal O}_Y(\rmV(i))$
is the tensor product
${\mathcal O}_Y(\rmV)\otimes {\mathcal O}_Y(1)^{\otimes (-i)}=:{\mathcal O}_Y(\rmV)(-i)$
where ${\mathcal O}_Y(1)$ is a line bundle representing
the Picard group generator $H$.
The $s$-Koszul brane plays an important r\^ole in the monodromy
along loops around the singular points with a base point in
the determinantal phase.
For $\bfb=2,1,0$,
let ${\mathcal L}^{\vee}_\bfb$ be a loop that comes from $\zeta^{\vee}\ll 0$,
goes around $t^{\vee}_\bfb$ clockwise and goes back to $\zeta^{\vee}\ll 0$,
similarly to the loop ${\mathcal L}'_{\bfa}$ for $\bfa=3,2,1$
in Fig.~\ref{fig:monodromy2} in the R\o dland model.
Each of them can be regarded as an upward path through a window
followed by a downward path through another.
The difference between the grade restriction rules for the two windows is
which of the two (sets of) special representations at $t^{\vee}_\bfb$
is included. The two are $\rmV_\bfb$ and $\rmV_\bfb(15)$,
with $\rmV_\bfb=\C$, $\Box$, $\mydoopm(-1)$
for $\bfb=2,1,0$, respectively.
The representation $\rmV_\bfb$ is good
on the right of $t^{\vee}_\bfb$ and bad on its left
while $\rmV_\bfb(15)$ is good on its left and bad on its right.
If the brane is grade restricted with respect to the first window
and is transported along the upward path that passes by
$t^{\vee}_\bfb$ on its left,
before it goes back through the downward path that passes by $t^{\vee}_\bfb$
on its right,
each of the $\rmV_\bfb(15)$ components in the brane must be traded for
$\rmV_\bfb$ while in the Reye congruence phase.
The trading can be done by binding $\Kos^{\vee}\otimes \rmV_\bfb$
followed by brane-antibrane annihilation, since
it is empty in the Reye congruence phase and
has $\rmV_\bfb$ and $\rmV_\bfb(15)$ at the edge components,
with the other components in the overlap of the two grade restriction rules.
Since $\rmV_\bfb(15)$ appears as the head of the
exact $s$-Koszul complex tensored with $\rmV_\bfb$,
the binding is done by a homomorphism
from the given brane to the empty brane.
For $\bfb=2,1$, the effect is the twist by
$\pi^{\vee}_Y(\Kos^{\vee}\otimes \rmV_\bfb)\cong{\mathcal O}_Y(\rmV_{\bfb})$.
For $\bfb=0$, the bindings must be done for {\it both}
$\rmV_0=\mydoop(-1)$ {\it and}
$\mydoom(-1)$.
The effect is the composition of the twists
by $\pi^{\vee}_Y(\Kos^{\vee}\otimes \mydoop(-1))={\mathcal O}_Y(\mydoop(-1))$
and $\pi^{\vee}_Y(\Kos^{\vee}\otimes \mydoom(-1))={\mathcal O}_Y(\mydoom(-1))$.
To summarize,
\beq
\mdm_{{\mathcal L}^{\vee}_2}=\twist_{{\mathcal O}_Y},\quad
\mdm_{{\mathcal L}^{\vee}_1}=\twist_{{\mathcal O}_Y(\Box)},\quad
\mdm_{{\mathcal L}^{\vee}_0}=\twist_{{\mathcal O}_Y(\sydoop(-1))}
\circ\twist_{{\mathcal O}_Y(\sydoom(-1))}.
\label{mdmdHT}
\eeq
The rank four vector bundle
\beq
\wHT_Y\,:=\,{\mathcal O}_Y(\Box)
\eeq
is equipped with a
non-degenerate symmetric pairing
$\wHT_Y\times \wHT_Y\to {\mathcal O}_Y(-1)$, which
defines the Hodge star involution
$\ast:\textw^{\!2}\wHT_Y\to \textw^{\!2}\wHT_Y$.
Then, ${\mathcal O}_Y(\mydoopm)$ is the rank three subbundle 
$\textw^{\!2}_{\pm}\wHT_Y\subset \textw^{\!2}\wHT_Y$
of $\ast=\pm 1$ eigenvectors, that is,
(anti-)self-dual antisymmetric tensors of $W_Y$.
In particular,
\beq
{\mathcal O}_Y(\ydoopm(-1))\,\,\cong\,\,\textw^{\!2}_{\pm}\wHT_Y(1).
\eeq
{\bf Remarks.}
(i) The twists that appear in (\ref{mdmdHT}) are
all autoequivalences of $D(Y)$ if ${\mathcal O}_Y$,
$\wHT_Y$ and $\textw^{\!2}_{\pm}\wHT_Y$ are
spherical objects of $D(Y)$. This is indeed the case for
${\mathcal O}_Y$ as $Y$ is a Calabi-Yau manifold, but needs some computation
for the others.\\
(ii)
The composition of the twists by $\textw^{\!2}_+\wHT_Y(1)$ and
$\textw^{\!2}_-\wHT_Y(1)$ to produce the monodromy along
${\mathcal L}^{\vee}_0$  can be in either order.
This is consistent only when the two twists
commute with each other.
This is the case if $\textw^{\!2}_+\wHT_Y(1)$ and
$\textw^{\!2}_-\wHT_Y(1)$ are orthogonal.\\
(iii)
There is another loop $\wt{\mathcal L}^{\vee}_0$ that goes around
$t_0$, similarly to $\wt{\mathcal L}'_1$ in Fig.~\ref{fig:monodromy2}
in the R\o dland model. The monodromy along $\wt{\mathcal L}^{\vee}_0$
is again the composition of the twists by 
$\textw^{\!2}_+\wHT_Y(1)$ and $\textw^{\!2}_-\wHT_Y(1)$ 
just as ${\mathcal L}^{\vee}_0$,
even though the two loops are homotopically inequivalent,
obeying
${\mathcal L}^{\vee}_2\circ{\mathcal L}^{\vee}_0\simeq\wt{\mathcal L}^{\vee}_0\circ{\mathcal L}^{\vee}_2$.
This implies that
this monodromy commutes with the monodromy along ${\mathcal L}^{\vee}_2$.
This is the case if
$\textw^{\!2}_+\wHT_Y(1)$ and $\textw^{\!2}_-\wHT_Y(1)$ 
are orthogonal to ${\mathcal O}_Y$.\\
We leave the proofs of the sphericalness
and the orthogonality as open problems.

We wish to compute the effect of the twist operators on the D-brane charges
and compare the result with the monodromies from the charge transportation
rule (\ref{HTtM}).
For this, we need to know the Chern characters of $W_Y$ and
$\textw^{\!2}_{\pm}W_Y(1)$.
It turns out that the Chern characters of $\textw^{\!2}_+W_Y(1)$
and $\textw^{\!2}_-W_Y(1)$ are equal to each other and are determined
by the one of $W_Y$ (See Appendix~\ref{app:classicalDet}).
In this paper, we do not attempt to compute ${\rm ch}(W_Y)$ directly but
instead we make an ansatz for it
and see if we find a consistent result with the charge transportation rule.
Let us put
\beqa
[W_Y]&=&
4[{\mathcal O}_Y]+a[{\mathcal O}_H]+b[{\mathcal O}_{\ell}]+c[{\mathcal O}_{\rm pt}],\label{chAnsatz1}\\
{[}\textw^{\!2}_{\pm}W_Y(1)]&=&3[{\mathcal O}_Y]+a'[{\mathcal O}_H]+b'[{\mathcal O}_{\ell}]+c'[{\mathcal O}_{\rm pt}].
\label{chAnsatz2}
\eeqa
The Euler pairing is
\beq
\begin{array}{c|cccc}
-
&{\mathcal O}_Y&{\mathcal O}_H&{\mathcal O}_{\ell}&{\mathcal O}_{\rm pt}\\
\hline
\chi({\mathcal O}_Y,-)&0&5&1&1\\
\chi(W_Y,-)&-5a-b-c&b+20&-a+4&4\\
  \chi(\textw^{\!2}_{\pm}W_Y(1),-)&-5a'-b'-c'&b'+15&-a'+3&3
\end{array}
\label{EulerDet}
\eeq
This would lead to the following representation of the
twists on the charges
\beqa
\twist_{{\mathcal O}_Y}:&&\!\!\!{\bf 1}_4
+\left(\begin{array}{c}1\\0\\0\\0\end{array}\right)(0,5,1,1)
  =\left(\begin{array}{cccc}
    1&5&1&1\\
    0&1&0&0\\
    0&0&1&0\\
    0&0&0&1
    \end{array}\right)
\label{DetTO}\\
\twist_{W_Y}:&&\!\!\!{\bf 1}_4
+\left(\begin{array}{c}4\\a\\b\\c\end{array}\right)(-5a-b-c,b+20,-a+4,4)
  \nn\\
 &=&\left(\begin{array}{cccc}
    1-20a-4b-4c&4b+80&-4a+16&16\\
    -a(5a+b+c)&1+a(b+20)&a(-a+4)&4a\\
    -b(5a+b+c)&b(b+20)&1+b(-a+4)&4b\\
    -c(5a+b+c)&c(b+20)&c(-a+4)&1+4c
  \end{array}\right)
\label{DetTW}
\\
\twist_{\wedge^2_{\pm}W_Y(1)}:&&\!\!\!{\bf 1}_4
+\left(\begin{array}{c}3\\a'\\b'\\c'\end{array}\right)(-5a'-b'-c',b'+15,-a'+3,3)
  \nn\\
 &=&\left(\begin{array}{cccc}
    1-15a'-3b'-3c'&3b'+45&-3a'+9&9\\
    -a'(5a'+b'+c')&1+a'(b'+15)&a'(-a'+3)&3a'\\
    -b'(5a'+b'+c')&b'(b'+15)&1+b'(-a'+3)&3b'\\
    -c'(5a'+b'+c')&c'(b'+15)&c'(-a'+3)&1+3c'
    \end{array}\right)\nn\\
&&\label{Detw2W}
\eeqa      
The matrices (\ref{DetTO}), (\ref{DetTW}) and the square of (\ref{Detw2W})
should match respectively with
the monodromy matrices obtained by the compositions
\beq
T_{\bf B}T_{\bf A}^{-1}=T_{\bf D}T_{\bf C}^{-1},
\quad
T_{\bf A}T_{{\bf D}(-1)}^{-1}
\quad
T_{\bf C}T_{\bf A}^{-1}=T_{\bf D}T_{\bf B}^{-1}
\label{monoDet}
\eeq
of the transport matrices in (\ref{HTtM}).
The first of (\ref{monoDet}) matches with (\ref{DetTO}) indeed.
The second of (\ref{monoDet}) reads
\beq
T_{\bf A}T_{{\bf D}(-1)}^{-1}=\left(\begin{array}{cccc}
  21&40&24&16\\
  -10&-19&-12&-8\\
  -50&-100&-59&-40\\
  75&150&90&61
\end{array}\right).
\eeq
The match with (\ref{DetTW}) holds in the last column when
\beq
a=-2,\quad b=-10,\quad c=15,
\label{abcfixed}
\eeq
under which the rest of the entries all match as well.
This is already a non-trivial consistency check.
Given this ${\rm ch}(W_Y)$, we can compute
${\rm ch}(\textw^{\!2}_{\pm}W_Y(1))$ (see Appendix~\ref{app:classicalDet}):
\beq
a'=0,\quad
b'=-5,\quad
c'=5.
\label{abcfixedprime}
\eeq
The square of the matrix (\ref{Detw2W}) for this $a',b',c'$
exactly matches with
\beq
T_{\bf C}T_{\bf A}^{-1}=T_{\bf D}T_{\bf B}^{-1}
=\left(\begin{array}{cccc}
  1&60&18&18\\
  0&1&0&0\\
  0&-100&-29&-30\\
  0&100&30&31
\end{array}\right).
\label{TATCi}
\eeq
Note also that the bottom-left corner of (\ref{EulerDet}) vanishes
under (\ref{abcfixedprime}), $-5a'-b'-c'=0$,
which is consistent with the expected orthogonality of
${\mathcal O}_Y$ and $\textw^{\!2}_{\pm}W_Y(1)$
and is also related to $T_{\bf C}T_{\bf A}^{-1}=T_{\bf D}T_{\bf B}^{-1}$
in (\ref{TATCi}).
We confirmed that the expected relations among monodromies and twists,
(\ref{mdmdHT}) in particular,
indeed hold at the charge level provided that the Chern character of
$W_Y$ is given by (\ref{chAnsatz1}) with (\ref{abcfixed}).
It would be nice to derive this by a direct computation.

Finding the monodromy along loops with the base point in the
Reye congruence phase $\zeta^{\vee}\gg 0$,
reproducing the result (\ref{mdmHT}) in the original model,
is a more difficult problem.
We need to identify the ``right-fit'' empty branes in the
determinantal phase $\zeta^{\vee}\ll 0$ and find their low energy behaviour
in the Reye congruence phase $\zeta^{\vee}\gg 0$.
This is left as a problem for future work.
Construction of such empty branes, also in the $\zeta\gg 0$ phase of the
original model, would lead to an explicit correspondence between branes
under the duality.

\subsection{The $N$-R\o dland Model And Its Variants}

The R\o dland model has a generalization where
the gauge group is still $U(2)$ but the number of the
fields $x_i$ in $\fund$ and $p^k$ in $\det^{-1}\fund$
is an odd integer $N\geq 5$ other than $7$.
The superpotential remains of the form $W=\sum_{i,j,k}A^{ij}_kp^k[x_ix_j]$.
We shall call this model the ``$N$-R\o dland model''.
The $\zeta\gg 0$ phase is weakly coupled and
reduces to the sigma model whose target space is the Calabi-Yau manifold
$X$ of dimension $(N-4)$, defined by $N$ equations
$A_k(x):=\sum_{i,j}A^{ij}_k[x_ix_j]=0$ in the Grassmannian $G(2,N)$.
The $\zeta\ll 0$ phase is strongly coupled
by an unbroken $SU(2)$ gauge symmetry.
The low energy theory is supported on the Pfaffian subvariety
$Y$ defined as the locus of $p\in \CP^{N-1}$ at which the rank of
$A^{ij}(p):=\sum_kA^{ij}_kp^k$ is non-maximal.
For $N=5$ or $7$, $Y$ is smooth and the theory is indeed a sigma model
with target $Y$. For $N\geq 9$, $Y$ is singular and ``the sigma model''
does not make sense, at least na\"ively. However, there is no sign of
noncompactness and a sound theory is expected.

Most of the results are obtained from the R\o dland model
by $7\to N$ replacement. For example, the effective twisted superpotential
is the $7\to N$ version of (\ref{Weff}).
The singular points are
\beq
\e^t=(1+\omega_{\bfa})^N,\quad\omega_{\bfa}:=\e^{2\pi\im\bfa/N},
\quad \bfa=1,\ldots, {N-1\over 2}
\label{spNRod}
\eeq
at which there is a Coulomb branch $\usigma_1=\omega_{\bfa}^{\pm 1}\usigma_2$.

The growth function is the $7\to N$ replacement of (\ref{A}).
The location of the walls $W_p$, $W_1$, $W_2$ and the condition to avoid them
are identical to the R\o dland model.
The grade restriction rule in the strongly coupled phase $\zeta\ll 0$
is found from the result (\ref{SU2QCDGRR}) for the $SU(2)$ SQCD:
the allowed representations are
\beq
\Sym^h\fund(i),\quad h=0,1,\ldots, {N-3\over 2}.
\label{SU2GRRN}
\eeq
The convergent contours $\bgamma_{\pm}$ in the two phases $\pm\zeta\gg 0$
are identical to that of the R\o dland model.
The hemisphere integral is convergent on $\bgamma_+$ for any representation
as long as $\zeta>N\log 2$, while it is convergent on $\bgamma_-$
for $\Sym^h\fund(i)$ as long as $\zeta<\zeta_{h+1}$ where
$\zeta_{\bfa}:=N\log(2\cos({\pi\bfa\over N}))$ is the value of the $\bfa$-th
singular point.
The convergence condition for the real contour $\bgamma=\im\ttt$ is the
$7\to N$ version of (\ref{condreal}). The set can be enlarged
by turning on the imaginary part maintining $\Is_1+\Is_2=0$
as (\ref{sgnP})-(\ref{sgnN}) such as $\bgamma_0$ given in (\ref{bgamma0}).
The convergence condition is the $7\to N$ version of
(\ref{Cond111}). That is, $i$ of (\ref{SU2GRRN}) is bounded by
\beq
-{N\over 2}<i+{h\over 2}+{\theta\over 2\pi}<{N\over 2}.
\label{NRodstraight}
\eeq
The set of representations satisfying this condition changes as
$\theta$ crosses $\pi\Z$. They are
\beqa
-(N+1)\pi<\theta<-N\pi:&&\!\!\!\!\!
\Sym^h\fund(i)~\,\,
  \left\{\begin{array}{ll}
  h=0,1,\ldots, {N-3\over 2}&\\
~~1\leq i+{h\over 2}\leq N &h\,\,{\rm even}\\
~~{1\over 2}\leq i+{h\over 2}\leq N-{1\over 2}&h\,\,{\rm odd}
  \end{array}\right.
  \label{GRRS}
\\
-N\pi<\theta<-(N-1)\pi:&&\!\!\!\!\!
\Sym^h\fund(i)~\,\,
\left\{\begin{array}{ll}
h=0,1,\ldots, {N-3\over 2}&\\
~~0\leq i+{h\over 2}\leq N-1&h\,\,{\rm even}\\
~~{1\over 2}\leq i+{h\over 2}\leq N-{1\over 2}&h\,\,{\rm odd}
\end{array}\right.\label{GRRShalf}
\eeqa
This changes by $\otimes \C(-1)$ for a $2\pi$ shift of $\theta$.
These representations can be transported along the straight paths
between the two phases in the respective range of $\theta$.
As the number of members, ${N-1\over 2}\times N$, matches with
the Euler number ${N\choose 2}$ of the Grassmannian $G(2,N)$,
we expect that these are the largest possible sets and agree with the
grade restricted subsets for the windows of the straight paths.
If we denote the window of the straight paths  in the range
$-(N+1)\pi<\theta<-N\pi$ by ${\bf S}$ and
its $\theta\mapsto\theta+2\pi j$ shift by ${\bf S}(j)$,
the expectation is that (\ref{GRRS}) and (\ref{GRRShalf}) agree with
$\rmGRR_{\bf S}$ and $\rmGRR_{{\bf S}({1\over 2})}$.

We can also reuse the computation in the R\o dland model to
determine the special representations at the singular points
as well as the no-entry zones.
Let us look at the singular points
\beq
t_*=N\log\left(2\cos\left({\pi\bfa\over N}\right)\right)+\pi\im \bfa
+2\pi \im \left({N-1\over 2}-\left[{\bfa-1\over 2}\right]\right)=:t_\bfa
\eeq
which are aligned at
$\theta=-N\pi$ ($\bfa$ odd) and $\theta=-(N+1)\pi$ ($\bfa$ even).
The Coulomb branch with
$(\usigma_1,\usigma_2)=(\omega_\bfa^{1\over 2}\Cc,\omega_\bfa^{-{1\over 2}}\Cc)$
is separated by the walls to three chambers
${\rm I}_\bfa$, ${\rm II}_\bfa$, ${\rm III}_\bfa$
where $\Arg(\Cc)$ takes values in
$\left(-{\pi\over 2},{\pi\over 2}-{\pi\bfa\over N}\right)$,
$\left({\pi\over 2}-{\pi\bfa\over N},{\pi\over 2}+{\pi\bfa\over N}\right)$,
$\left({\pi\over 2}+{\pi\bfa\over N},{3\pi\over 2}\right)$,
respectively.
The special weight $\wtq_*$ is obtained by
the $7\to N$ version of (\ref{findwtq})-(\ref{tableArg}).
We observe that $\wtq_*$ in the chambers ${\rm I}_\bfa$
and ${\rm III}_\bfa$ are the shifted highest weights
of the $SU(2)$ grade restricted representations
\beq
\rmV_{{\rm I}_\bfa}:=\Sym^{\bfa-1}\fund\left(N-\left[{\bfa-1\over 2}\right]\right)
~\mbox{and}~
\rmV_{{\rm III}_\bfa}:=\Sym^{\bfa-1}\fund\left(-\left[{\bfa-1\over 2}\right]\right).
\eeq
In chamber ${\rm II}_\bfa$ and on the Weyl image
branch with
$(\usigma_1,\usigma_2)=(\omega_\bfa^{-{1\over 2}}\Cc,\omega_\bfa^{1\over 2}\Cc)$,
the special weight $\wtq_*$
is not a shifted highest weight of an $SU(2)$ grade restricted representation.
By a closer look at the behaviour of the growth function, we find that
$\rmV_{{\rm I}_\bfa}$ ({\it resp}. $\rmV_{{\rm III}_\bfa}$) is not admissible
on the right ({\it resp}. left) of $t_\bfa$ --- more precisely, when
$\sDelta t=t-t_\bfa$ is in the sector
\beq
R_\bfa:~0\leq \Arg(\im\sDelta t)\leq {\pi\bfa \over N}\quad
\left({\it resp}.~
L_\bfa:~\pi-{\pi\bfa\over N}\leq\Arg(\im\sDelta t)\leq\pi\right).
\eeq
The representations are $\rmV_{{\rm I}_\bfa}(j)$ and $\rmV_{{\rm III}_\bfa}(j)$
at $t_\bfa+2\pi\im j$ ($j\in \Z$).
These sectors extend to no-entry zones.
The representation $\Sym^h\fund(i)$ has no-entry zones
${\mathscr R}_{h,i}$ and ${\mathscr L}_{h,i}$ which extend to the right of
$(\zeta_{h+1},\theta_{h,i}^+)$ and
to the left of $(\zeta_{h+1},\theta_{h,i}^-)$,
with $\theta_{h,i}^{\pm}=(\pm N-h-2i)\pi$.
Approach from above ({\it resp}. below) is blocked 
by the wall $W_1$ or $W_2$ 
({\it resp}. $W_p$). Their locations are 
given by the $7\to N$ versions of (\ref{Rhi}) and (\ref{Lhi})
with $\zeta(\theta;q):=N\log|2\cos({\theta+2\pi q\over N})|$.

\begin{figure}[htb]
\centerline{\includegraphics[page=52]{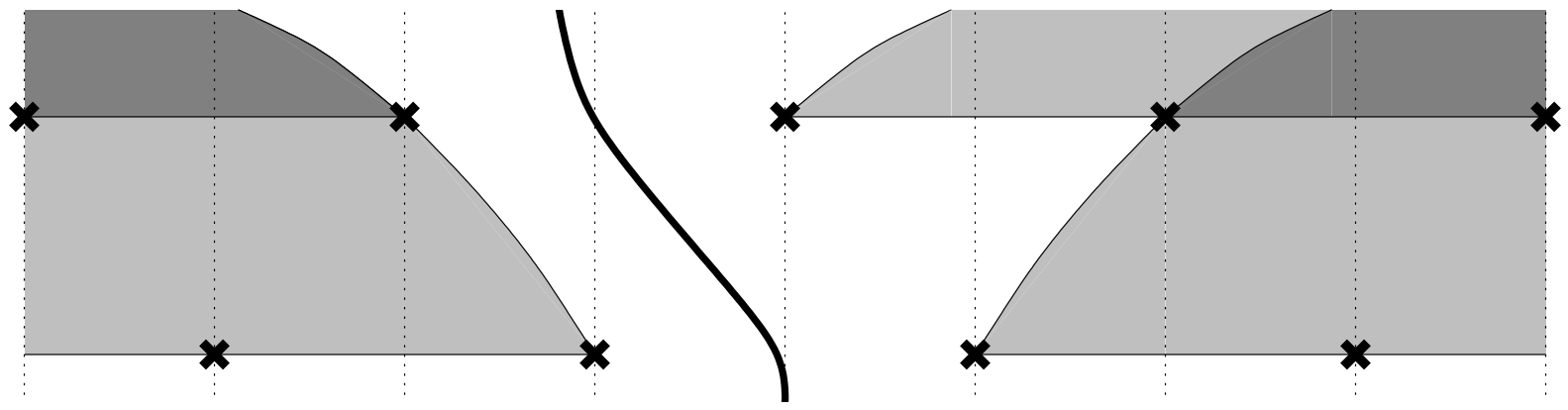}}
\caption{There is a path that avoids no-entry zones.}
\label{fig:NoEntryNRod1}
\end{figure}

Let us discuss the grade restriction rule.
As the singular points are at $\zeta=\zeta_\bfa$
and $\theta\equiv \pi\bfa$ (mod $2\pi$), the paths connecting
$\zeta\gg 0$ and $\zeta\ll 0$ must go through
an interval on $\zeta=\zeta_\bfa$ with
$\theta\in\pi(c_\bfa-1,c_\bfa+1)$ for some
$c_\bfa\in \Z$ which is even for odd $\bfa$ and odd for even $\bfa$.
An example is the broken line through the mid points $(\zeta_\bfa, \pi c_\bfa)$
of these intervals, with the straight lines extending to
$\zeta=\pm\infty$ attached.
Let ${\bf w}_{c_1,\ldots, c_{N-1\over 2}}$ be the window represented by such a path.
For example, the one with
$c_{\bfa}=-(N+1)$ ({\it resp}. $-N$) for odd ({\it resp}. even) $\bfa$ is
the window ${\bf S}$ of straight paths in $-(N+1)\pi<\theta<-N\pi$.
Any path in the window ${\bf w}_{c_1,\ldots, c_{N-1\over 2}}$
must go through the no-entry zone
${\mathscr R}_{h,i}$ or ${\mathscr L}_{h,i}$ if
$\theta_{h,i}^+< \pi(c_{h+1}+1)$ or $\pi(c_{h+1}-1)< \theta_{h,i}^-$,
respectively.
Equivalently, a path in this window can avoid
the no-entry zones of $\Sym^h\fund(i)$ 
only if $\theta_{h,i}^+\geq \pi(c_{h+1}+1)$ and
$\theta_{h,i}^-\leq \pi(c_{h+1}-1)$, that is,
\beq
-{N-1\over 2}\leq i+{h+c_{h+1}\over 2}\leq {N-1\over 2}.
\label{NRodGRRc}
\eeq
If $|c_\bfa-c_{\bfa+1}|$, which is always an odd integer, is equal to $1$
for all $\bfa$,
there is a path in this window which does avoid
the no-entry zones of all $\Sym^h\fund(i)$'s obeying (\ref{NRodGRRc}).
See Fig.~\ref{fig:NoEntryNRod1}.
\begin{figure}[htb]
\centerline{\includegraphics[page=53]{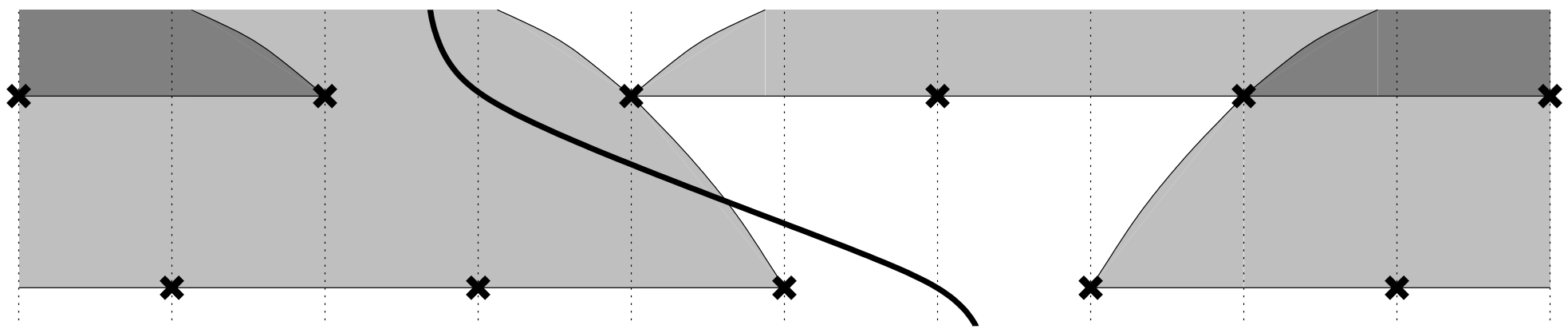}}
\caption{No path can avoid no-entry zones.}
\label{fig:NoEntryNRod2}
\end{figure}
However, if $|c_\bfa-c_{\bfa+1}|\geq 3$,
any path in this window cannot avoid
${\mathscr R}_{\bfa,i}$ with $\theta_{\bfa,i}^+=\pi(c_{\bfa+1}+1)$
or ${\mathscr L}_{\bfa,i}$ with $\theta_{\bfa,i}^-=\pi(c_{\bfa+1}-1)$,
and hence some of
$\Sym^h\fund(i)$ obeying (\ref{NRodGRRc}) cannot be transported.
See Fig.~\ref{fig:NoEntryNRod2}. 
Note that the number of $(h,i)$'s obeying (\ref{NRodGRRc})
is ${N-1\over 2}\times N$ which is the same as the Euler number
of $G(2,N)$.
Thus, if $|c_\bfa-c_{\bfa+1}|\geq 3$ for some $\bfa$,
the number of representations that can be transported through the
window ${\bf w}(c_1,\ldots, c_{N-1\over 2})$
is strictly less than this number, and the set is too small
to produce all branes in the low energy sigma model in the phase
$\zeta\gg 0$.
This motivates us to consider only the windows obeying
\beq
|c_\bfa-c_{\bfa+1}|=1\quad\mbox{for all $\bfa$.}
\label{cccond}
\eeq
The fact that such a window includes a path avoiding all the no-entry
zones of $\Sym^h\fund(i)$ obeying (\ref{NRodGRRc})
and any path in it must go through the no-entry zones of other representations
suggests that these representations constitute the grade restricted subset.
This is confirmed by constructing
the family of contours with which the transport can be done.
For the window of the straight paths, the set precisely matches with
the one determined by (\ref{NRodstraight}).
For example, the sets for the windows ${\bf S}$ and ${\bf S}({1\over 2})$
match with (\ref{GRRS}) and (\ref{GRRShalf}).
For these representations, the contour $\bgamma_0$ works.
For a path that goes mostly along a stright line on $\theta\in\pi\Z$
but circumvents the singular points on the way by small detours,
such as the path ${\mathcal P}$ in Fig.~\ref{fig:thepathP},
the contour $\bgamma_0$ or its deformation can be used, where
the necessary deformation can be found by a computations similar
to (\ref{deform26}) and (\ref{deform00}).
In general, numerical analysis can be used to find the family of contours.
Thus, we conclude that the grade restriction rule is given by
\beq
\rmGRR_{{\bf w}_{c_1,\ldots, c_{N-1\over 2}}}\,=\,\left\{\,\Sym^h\fund(i)\,\Biggl|\,
h=0,1,\ldots, {N-3\over 2};\,\,
(\ref{NRodGRRc})\,\right\}.
\eeq

\begin{figure}[htb]
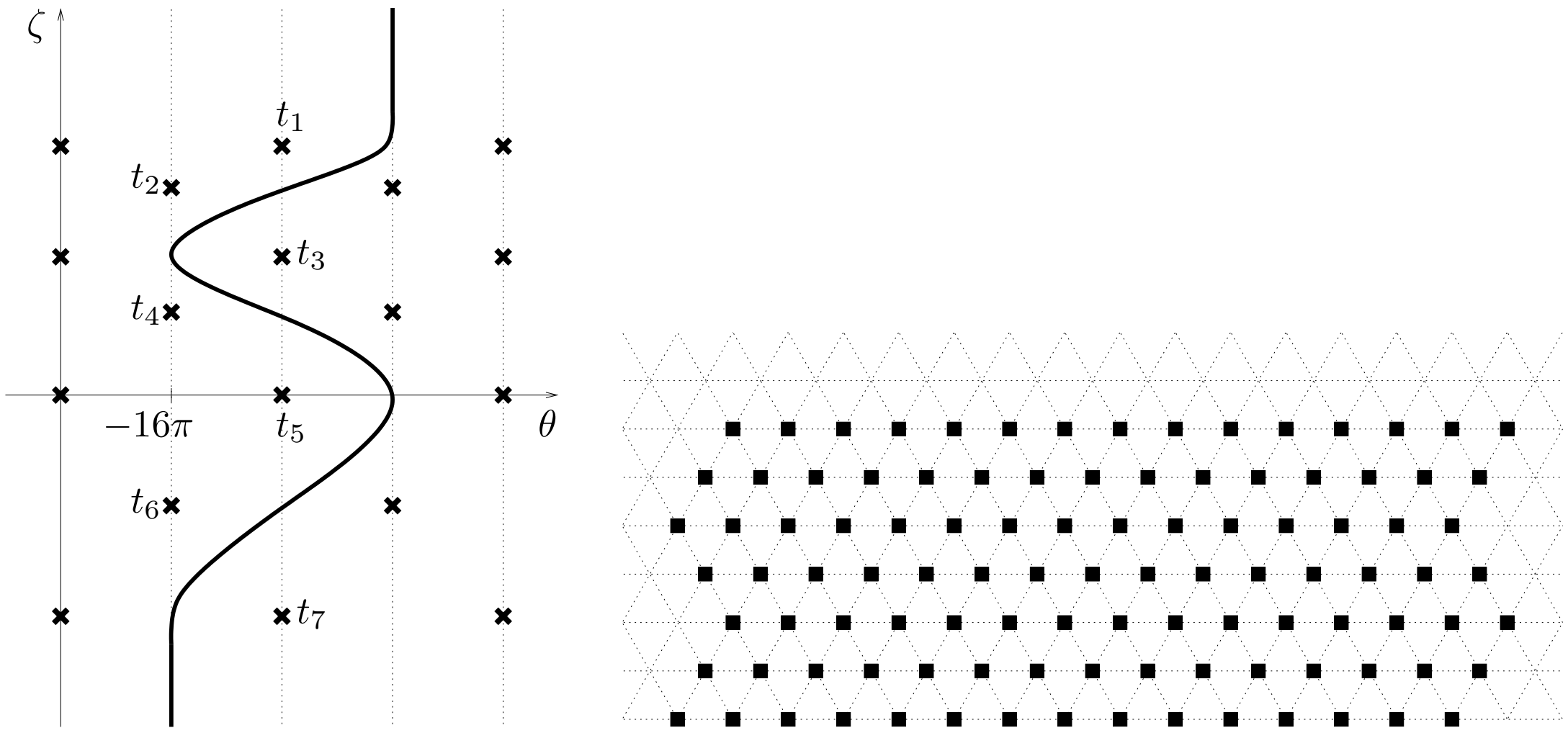

  \psfrag{a1}{\footnotesize $t_1$}
  \psfrag{a2}{\footnotesize $t_2$}
  \psfrag{a3}{\footnotesize $t_3$}
  \psfrag{a4}{\footnotesize $t_4$}
  \psfrag{a5}{\footnotesize $t_5$}
  \psfrag{a6}{\footnotesize $t_6$}
  \psfrag{a7}{\footnotesize $t_7$}
  \psfrag{z}{\footnotesize $\zeta$}
  \psfrag{th}{\footnotesize $\theta$}
  \psfrag{-16pi}{\footnotesize $-16\pi$}
  \centerline{\includegraphics[page=54]{main-pics.pdf}
    ~~~~\includegraphics[page=55]{main-pics.pdf}}
\caption{The window ${\bf w}_{-14,-15,-16,-15,-14,-15,-16}$
  and the corresponding grade restricted set in the $N=15$ model.
The lower-left corner of the grid is $\C(-1)$.}
\label{fig:GRR15Rod}
\end{figure}

\begin{figure}[htb]
\centerline{\includegraphics[page=56]{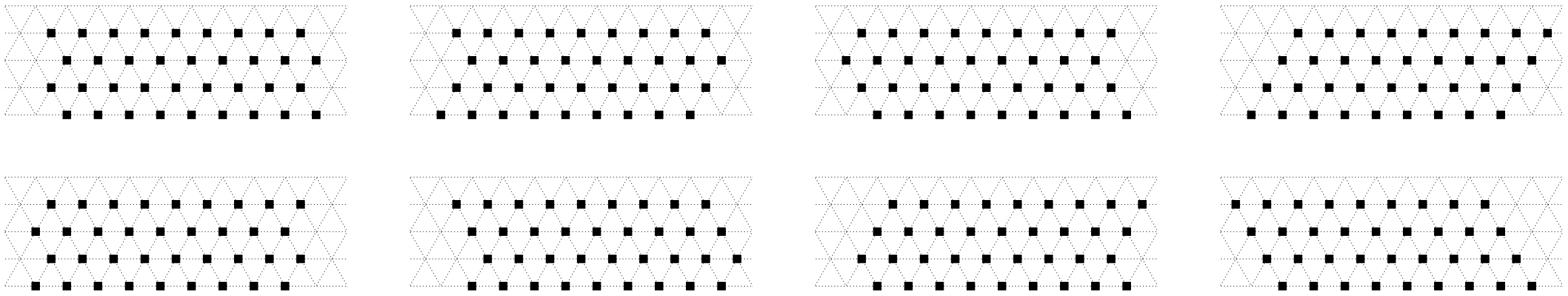}}
\caption{The grade restricted subsets for the windows
${\bf w}_{-10,-9,-10,-9}={\bf S}$,
  ${\bf w}_{-8,-9,-10,-9}$,
  ${\bf w}_{-10,-9,-8,-9}$,
  ${\bf w}_{-8,-9,-10,-11}$ (from left to right in the upper row) and
  ${\bf w}_{-8,-9,-8,-9}={\bf S}({1\over 2})$,
  ${\bf w}_{-10,-11,-10,-9}$,
  ${\bf w}_{-10,-9,-10,-11}$,
  ${\bf w}_{-10,-9,-8,-7}$ (from left to right in the lower row), in the
$N=9$ model.}
\label{fig:GRR9new}
\end{figure}

The grade restricted subset can be plotted in the triangular grid of
$U(2)$ representations.
See Fig.~\ref{fig:GRR15Rod} for an example.
It consists of chains of $N$ consecutive horizontal dots in the $U(2)$ grid,
from height 0 to height ${N-3\over 2}$.
As $i+{h\over 2}$ labels the horizontal position in the grid,
the numbers $-c_1$, $-c_2$, \ldots, $-c_{N-1\over 2}$ show the horizontal
positions of the chains at heights $0$, $1$, \ldots, ${N-3\over 2}$,
respectively.
The condition (\ref{cccond}) means that
the chains at adjacent heights are related by a move along the grid.
Fig.~\ref{fig:GRR9new} shows eight different shapes of grade
restricted subsets that appear in the $N=9$ model.

We expect the same picture as (\ref{maindiagram})
where ${\mathcal S}$ is the set of $SU(2)$ grade restricted branes
(\ref{SU2GRRN}),
$X$ is the Calabi-Yau $(N-4)$-fold in $\Gr=G(2,N)$,
and $D(Y)$ is the set of branes in the low energy theory
supported on $Y$.
Part of this can be understood in the same way as in the R\o dland model.
In particular, the projection ${\mathcal T}_{\bf w}\to D(X)$
is essentially bijective.
This boils down to the fact that
${\mathcal O}_{\Gr}({\rm R})$ for the grade restricted subset
${\rm R}=\rmGRR_{\bf w}$
forms a full and strong exceptional collection of $D(\Gr)$,
which we show next.

The fullness can be seen by the fact
that the Kapranov's collection \cite{KapranovGr,KapranovHomog}
corresponds to a pyramid of height $N-2$ in the $U(2)$ grid
and that ${\rm R}$ can be obtained by cutting down such a pyramid
using the twisted Lascoux complexes.
\begin{figure}[htb]
\centerline{\includegraphics[page=57]{main-pics.pdf}}
\caption{}
\label{fig:Kap9GR}
\end{figure}
See Fig.~\ref{fig:Kap9GR} for the processes in the $N=9$ case to obtain
the sets in Fig.~\ref{fig:GRR9new} from the Kapranov's collection.
The strongness can be checked by the Borel-Weil-Bott theorem.
Let us show that members in ${\rm R}$ have Property EV${}_m$ for $m\geq 0$.
As in the $N=7$ case,
the easiest is the case when ${\rm R}$ is of the rectangular shape,
such as the two right-most sets in Fig.~\ref{fig:GRR9new}.
One rectangle consists of
$\Sym^h\fund(i)$ for $h=0,1,\ldots, {N-3\over 2}$ and $i=1,\ldots, N$.
When $(h,i)$ and $(h',i')$ are in such a range, we would like to show that
the entries of
\beqa
    {\bf x}\!&=&\!(\underbrace{N+h'+i'-i+m-j}_{\displaystyle ~~~=:a},
    \underbrace{N-1-h+i'-i+m+j}_{\displaystyle ~~~=:b},
    N-2,\ldots,2,1),~~~\label{x762}\\
    && j=0,\ldots, {\rm min}\{h.h'\}\nn
\eeqa
are either coincident or strictly decreasing. That is,
$1\leq a\leq N-2$ and/or $b\geq 1$.
Note that $i'-i+m\geq -(N-1)$. If $i'-i+m\geq -{N-1\over 2}$,
then,
\beq
b\geq {N-1\over 2}-h+j\geq 1,
\eeq
where the latter inequality is by $h\leq {N-3\over 2}$ and $j\geq 0$.
If $i'-i+m\leq -{N-1\over 2}-1$,
then
\beq
1\leq h'+1-j\leq a\leq {N-1\over 2}+h'-j\leq N-2,
\eeq
where the first, second and the fourth inequalities are by $j\leq h'$,
$i'-i+m\geq -(N-1)$, and $h'\leq {N-3\over 2}$ with $j\geq 0$.
The other rectangle consists of
$\Sym^h\fund(i-h)$ for $(h,i)$ in the same range.
In such a case, ${\bf x}$ is the same as (\ref{x762}) except that
$h$ and $h'$ are exchanged, and the proof is exactly the same.
The sets of other shapes are obtained from rectangles by a
sequence of shifts of horizontal chains by $\pm 1$.
Property EV${}_m$ for $m\geq 0$ for the shape after each shift
can be proven by the ones before, just as in the $N=7$ case.

We can also show that ${\mathcal O}_{\Gr}({\rm R})$
is full and strong only when ${\rm R}$ is of the above type,
if the height is bounded by the $SU(2)$ rule (\ref{SU2GRRN}).
First of all, the number of members of ${\rm R}$ must be
the Euler number of $\Gr=G(2,N)$ which is ${N\choose 2}=N\times {N-1\over 2}$.
Second, the dots cannot be separated by $N$ or more in the horizontal
direction. Indeed, the Ext group (\ref{ExtComp})
for $(h_1,i_1)=(h,i)$ and $(h_2,i_2)=(h,i')$ with $i-i'\geq N$
has the ${\bf x}$ vector
whose first entry is $a=N+h+i'-i-j\leq 0$ for $j=h$,
meaning that $\Ext^{2(N-2)}\ne 0$.
In particular, in view of the $SU(2)$ bound $0\leq h\leq {N-1\over 2}-1$,
the set ${\rm R}$ must be the union of chains
of $N$ consecutive horizontal dots, from height $0$ to height ${N-3\over 2}$.
Third, chains at adjacent heights must be related by a move along the grid.
Suppose the chains at heights $h$ and $h+1$ voilate
this condition. This means that the set would include
$\Sym^{h}\fund(i)$ and $\Sym^{h+1}\fund(i')$
such that (i) $i'-i\geq N$ or (ii) $i'-i\leq -N-1$.
We would like to show that the ${\bf x}$ vector for
the Ext group of the pair (in one order)
has the first entry $a\leq 0$ for some $j$, so that $\Ext^{2(N-2)}\ne 0$.
In case (i),
for $(h_1,i_1)=(h+1,i')$ and $(h_2,i_2)=(h,i)$,
$a=N+h+i-i'-j\leq h-j=0$ for $j=h$.
In case (ii), for $(h_1,i_1)=(h,i)$ and $(h_2,i_2)=(h+1,i')$,
$a=N+h+1+i'-i-j\leq h-j=0$ for $j=h$.

Monodromy along loops with the base point in the Grassmannian phase
can be obtained using the $p$-Koszul brane (the $7\to N$ version of
(\ref{Kp})) and its tensor product with special representations.
Let ${\mathcal L}_\bfa$ be a loop that comes from $\zeta\gg 0$,
goes around $t_\bfa$ clockwise and goes back to $\zeta\gg 0$.
As in the R\o dland model, we find that the monodromy along ${\mathcal L}_\bfa$
is the twist $\twist$ by
\beq
{\mathcal V}_\bfa
:={\mathcal O}_X(\rmV_{{\rm III}_\bfa})
=\Sym^{\bfa-1}S_X^*\left(-\left[{\bfa-1\over 2}\right]\right).
\eeq
This is indeed an autoequivalence of $D(X)$ as ${\mathcal V}_\bfa$
is a spherical object of $D(X)$.
Also, there is an ambiguity in the choice of ${\mathcal L}_\bfa$ ---
which side of $t_{\bfa'}$ for $\bfa'=\bfa-2, \bfa-4,\ldots$
does it pass through---
but the monodromy is always the twist $\twist_{{\mathcal V}_\bfa}$
independent of the choice.
This is possible only when the twist
$\twist_{{\mathcal V}_\bfa}$ commutes wit
$\twist_{{\mathcal V}_{\bfa'}}$ for such $\bfa'$'s.
This is indeed the case as ${\mathcal V}_\bfa$
is orthogonal to ${\mathcal V}_{\bfa'}$'s.
A proof of the sphericalness and the orthogonality is given in
Appendix~\ref{subsec:ExtRGr}.

\bigskip
\noindent
{\bf Variants}

The $N$-R\o dland model has a variant with the same gauge group $U(2)$ and
$N$ fields $x_i$ in the fundamental representation $\fund$,
but with a field $p^k$ in the representation
$\det^{-d_k}\fund$  ($d_k\geq 1$) for $k=1,\ldots, M$.
The superpotential is of the form $W=\sum_kp^kA_k(x)$ where $A_k(x)$ has degree
$d_k$ in $[x_ix_j]$.
We continue to assume the Calabi-Yau condition
$\sum_kd_k=N$ and that $N$ is odd.
The $\zeta\gg 0$ phase is weakly coupled and reduces
to the sigma model whose target space is the Calabi-Yau manifold $X$
defined by $A_1(x)=\cdots=A_M(x)=0$ in
the Grassmannian $G(2,N)$. It has dimension $2N-4-M$.
The $\zeta\ll 0$ phase is strongly coupled by an unbroken $SU(2)$
gauge symmetry.
Its detail depends on $d_1,\ldots d_M$ and the form of the
superpotential.

The singular points are
\beq
\e^t=\prod_kd_k^{d_k}\cdot(1+\omega_{\bfa})^N,
\quad\omega_{\bfa}:=\e^{2\pi\im\bfa/N},
\quad \bfa=1,\ldots, {N-1\over 2}.
\eeq
Note the similarity with (\ref{spNRod}).
In fact, concerning the effective twisted superpotential and the
growth function, the only diffefrence from the $N$-R\o dland model
is the shift $\zeta\to \zeta-\sum_kd_k\log d_k$.
In particular, the grade restriction rule is exactly the same. 
The $p$-Koszul brane $\Kos$ is
\beq
\begin{tikzcd}
  \C\arrow[r,yshift=0.5ex, "A(x)"] &
  \arrow[l,yshift=-0.5ex, "p"]\rmV\arrow[r,yshift=0.5ex, "A(x)"]
  & \arrow[l,yshift=-0.5ex, "p"] \textw^{\!2}\rmV
  \arrow[r,yshift=0.5ex, "A(x)"]
  & \arrow[l,yshift=-0.5ex, "p"]
  ~~\cdots~~\arrow[r,yshift=0.5ex, "A(x)"]
  &\arrow[l,yshift=-0.5ex, "p"] \textw^{\!M-1}\rmV
  \arrow[r,yshift=0.5ex, "A(x)"]
  &\arrow[l,yshift=-0.5ex, "p"] \textw^{\!M}\rmV=\C(N)
\end{tikzcd}
\eeq
where $\rmV=\oplus_k\C(d_k)$.
The monodromy along the loop around
\beq
t_\bfa=\sum_kd_k\log d_k
+N\log\left(2\cos\left({\pi\bfa\over N}\right)\right)+\pi\im \bfa
+2\pi \im \left({N-1\over 2}-\left[{\bfa-1\over 2}\right]\right)
\eeq
with the base point in the Grassmannian phase is the twist
by ${\mathcal V}_\bfa=\Sym^{\bfa-1}S_X^*(-[{\bfa-1\over 2}])$.
These vector bundles are spherical objects of $D(X)$
and the twists $\twist_{{\mathcal V}_\bfa}$ are indeed autoequivalences of $D(X)$.
And ${\mathcal V}_\bfa$ is orthogonal to
${\mathcal V}_{\bfa'}$ if $\bfa$ and $\bfa'$ are both even or both odd,
so that their twists commute with each other.
See Appendix~\ref{subsec:ExtRGr} for a proof.

\section*{Acknowledgement}

We would like to thank Alexey Bondal, Will Donovan, David Erkinger, D. Faenzi,
Jirui Guo, Wahei Hara, Shinobu Hosono, Misha Kapranov, Ban Lin,
Joseph McGovern, Yoshiki Oshima, Robert Pryor, J\o rgen Rennemo,
Emanuel Scheidegger, Thorsten Schimannek, Ed Segal, Lucy Smith,
Yusuke Suetake, Yukinobu Toda for useful discussions.
We also thank the Simons Center for Geometry and Physics,
MATRIX Institute,
Mainz Institute for Theoretical Physics (MITP) of the Cluster of Excellence PRISMA+ (Project ID 390831469) and
Erwin Schr\"odinger Institute,
for hospitality during several periods where parts of this work were done.
Research of RE was supported in part by
the Deutsche Forschungsgemeinschaft, within the
framework of the Exzellenzinitiative an der Universit\"at Heidelberg.
KH was supported in part by
JSPS Grant-in-Aid for Scientific Research No. 17K05567 and No. 20K03762.
During the course of this project,
JK was supported by the following research grants:
Austrian Science Fund (FWF) Project P30904-N27,
Australian Research Council (ARC) Discovery Projects
DP210101502 and DP250101828,
Australian Research Council (ARC) Future Fellowship FT210100514,
and University of Melbourne Establishment Grant.
MR ackowledges support from
the Research Fund for International Scientists, NSFC grant No. W2632002,
in addition to the Institute for Advanced Study, Princeton and
Yau Mathematical Sciences Center at Tsinghua University.
All the authors are also supported by 
World Premier International Research Center Initiative, MEXT,
Japan at Kavli IPMU, the University of Tokyo, for various periods of time.

\bigskip
\bigskip

\begin{appendix}

\noindent
{\large\bf Appendix}

\section{Supersymmetric Boundary Interactions/Conditions}
\label{app:SUSY}

\mathchardef\mdash="2D
\newcommand{\bfQ}{{\bf Q}}
\newcommand{\obfQ}{\overline{\bf Q}}

We discuss the boundary interactions and boundary conditions
that preserve B-type supersymmetry in
$(2,2)$ supersymmetric field theories in $1+1$ dimensions.
We first consider gauge theories and then
non-linear sigma models possibly with a superpotential.
In this Appendix and Appendix~\ref{app:Cat},
we do not need to assume Calabi-Yau condition on the theories.

\subsection{Gauge Theory}
\label{app:GLSM}

Let us first write down our convention on fields and transformations of
$2d$ $(2,2)$ supersymmetry,
which is the dimensional reduction of the one in \cite{WB}
for 4d ${\mathcal N}=1$ theories, as in \cite{Wittenphases,WittenGrass},
with a rescaling of fields and parameters
that removes $\sqrt{2}$ factors \cite{cmibook}.
A gauge theory has a vector multiplet consisting of
a gauge potential $v_{\mu}$, a complex scalar $\sigma$,
a Dirac fermion $\lambda_{\pm}$ and a real auxiliary field $D$
with values in the adjoint representation $\im\mathfrak{g}$ of the gauge group
$G$ or its complexification $\mathfrak{g}_{\C}$,
and a chiral multiplet consisting of a scalar $\phi$,
a Dirac fermion $\psi_{\pm}$ and an auxiliary field $F$
with values in the matter representation $V$.
We consider a general superpotential $W(\phi)$
and a linear twisted superpotential $\wt{W}(\sigma)=-t(\sigma)$
with FI-theta parameter $t=\zeta-\im\theta$.
We formulate the system on the flat Minkowski spacetime
with time and space coordinates $t=x^0$ and $x=x^1$, or
the light cone coordinates $x^{\pm}=x^0\pm x^1$ with derivatives
$\partial_{x^{\pm}}={1\over 2}(\partial_0\pm\partial_1)$.
The Lagrangian is the sum of the gauge kinetic term,
the matter kinetic term, the superpotential term and
the twisted superpotential term, which are respectively
\beqa
{\mathcal L}_{\rm gauge}
&=&{1\over 2\pi}\left[\,
  \int\dd^4\theta\,\left(-{1\over 2e^2}\bSigma\Sigma\right)
  \,+\,\partial_{x^+}\partial_{x^-}\left({1\over 2e^2}\bsigma\sigma\right)
  \,\right]\nn\\
&=&
{1\over 4\pi e^2}\Bigl[\,\,|D_0\sigma|^2-|D_1\sigma|^2
+\im\blambda_-(\lrD_{\!\! 0}+\lrD_{\!\! 1})\lambda_-
+\im\blambda_+(\lrD_{\!\! 0}-\lrD_{\!\! 1})\lambda_+\nn\\
&&~~~~~~~~
+(v_{01})^2+D^2-{1\over 4}[\sigma,\bsigma]^2
-\blambda_-[\sigma,\lambda_+]
-\blambda_+[\bsigma,\lambda_-]
\,\,\Bigr],\label{Lgauge22}\\[0.2cm]
{\mathcal L}_{\rm matter}
&=&{1\over 2\pi}\left[\,\int\dd^4\theta\,\,\overline{\Phi}\e^V\Phi
\,+\,\partial_{x^+}\partial_{x^-}|\phi|^2\,\right]\nn\\
&=&{1\over 2\pi}\Bigl[\,\,|D_0\phi|^2-|D_1\phi|^2
+\im\bpsi_-(\lrD_{\!\! 0}+\lrD_{\!\! 1})\psi_-
+\im\bpsi_+(\lrD_{\!\! 0}-\lrD_{\!\! 1})\psi_+\nn\\
&&~~~~~~~~+|F|^2+\bphi D\phi
-\half|\sigma\phi|^2-\half|\bsigma\phi|^2
-\bpsi_-\sigma\psi_+-\bpsi_+\bsigma\psi_-\nn\\
&&~~~~~~~~-\im\bphi\lambda_-\psi_++\im\bphi\lambda_+\psi_-
+\im\bpsi_+\blambda_-\phi-\im\bpsi_-\blambda_+\phi\,\,\Bigr],
\label{Lmatter22}\\[0.2cm]
{\mathcal L}_W&=&{\rm Re}\int\dd^2\theta\,\,\im W(\Phi)
~=~-{\rm Im}\left[F^i{\partial W\over \partial \phi^i}
-\psi^i_+\psi^j_-{\partial^2 W\over\partial \phi^i\partial \phi^j}
\right],\label{LW22}\\[0.2cm]
{\mathcal L}_{{\rm FI}\mdash\theta}&=&
{1\over 2\pi}{\rm Re}\int\dd^2\widetilde{\theta}\,
(-t(\Sigma))
~=~{1\over 2\pi}\Bigl[\,-\zeta(D)+\theta(v_{01})\,\Bigr].
\label{LFI22}
\eeqa
The middle expressions are in the superfield formalism (in the convention
of \cite{cmibook}):
$V$ and $\Phi$ are the superfields for the vector and the chiral multiplets
and $\Sigma=\overline{D}_+(\e^{-V}D_-\e^V)$ is the fieldstrength superfield.
$v_{01}=\partial_0v_1-\partial_1v_0+\im[v_0,v_1]$ is the ordinary fieldstrength,
$D_{\mu}$ are the gauge covariant derivatives,
$D_{\mu}\phi=\partial_{\mu}\phi+\im v_{\mu}\phi$ etc.
We use the notation
$a\lrD_{\!\!\!\mu\,}b={1\over 2}a(D_{\mu}b)-{1\over 2}(D_{\mu}a) b$.
We have chosen $G$-invariant and positive definite (hermitian) inner products,
$(a,b)\in \im\mathfrak{g}\times\im\mathfrak{g}\mapsto {1\over e^2}ab\in \R$ 
and $(c,d)\in V\times V\mapsto \bar{c}d\in \C$.
The parameter $e$ is the gauge coupling constant.\footnote{An invariant
inner product on $\im\mathfrak{g}$
  may have multiple parameters. For example, for $G=U(1)^k$, it has
$k(k+1)/2$ parameters.
For $G=U(2)$, we may use
$(a,b)\mapsto {1\over e^2}{\rm tr}(ab)$ for a single $e$
or we may separate the gauge coupling constants for the
$\mathfrak{u}(1)$ and the $\mathfrak{su}(2)$ factors, $e_1$ and $e_2$, respectively, for which the product is
$(a,b)\mapsto {1\over 2e_1^2}{\rm tr}(a){\rm tr}(b)+{1\over e_2^2}\left({\rm tr}(ab)-{1\over 2}{\rm tr}(a){\rm tr}(b)\right)$.
For readability, we denote the product in the general theory
simply as $(a,b)\mapsto {1\over e^2}ab$ and speak as if there is a single
gauge coupling $e$, even if there are several gauge coupling constants.}
The transformation of the fields under the $(2,2)$ supersymmetry
$\delta=\epsilon_+\bfQ_--\epsilon_-\bfQ_+-\bepsilon_+\obfQ_-
+\bepsilon_-\obfQ_+$ is
\beqa
\delta v_{x^{\pm}}\!\!\!&=&\!{\im\over 2}\bepsilon_{\pm}\lambda_{\pm}
+{\im\over 2}\epsilon_{\pm}\blambda_{\pm},
\nonumber\\
\delta\sigma\!&=&\!\!-\im\bepsilon_+\lambda_-
-\im\epsilon_-\blambda_+,
\nonumber\\
\delta\lambda_+\!\!&=&\!
\im\epsilon_+\Bigl(\,D+\im v_{01}+{1\over 2}[\sigma,\bsigma]\,\Bigr)
+2\epsilon_-D_{x^+}\bsigma,
\nonumber\\
\delta\lambda_-\!\!&=&\!
\im\epsilon_-\Bigl(\,D-\im v_{01}-{1\over 2}[\sigma,\bsigma]\,\Bigr)
+2\epsilon_+D_{x^-}\sigma,\nn\\
\delta D\!&=\!&
\epsilon_+\Bigl(\,D_{x^-}\blambda_++{\im\over 2}[\sigma,\blambda_-]\,\Bigr)
+\epsilon_-\Bigl(\,D_{x^+}\blambda_-+{\im\over 2}[\bsigma,\blambda_+]\,\Bigr))
\nn\\
&&~~~~
-\bepsilon_+\Bigl(\,D_{x^-}\lambda_++{\im\over 2}[\bsigma,\lambda_-]\,\Bigr)
-\bepsilon_-\Bigl(\,D_{x^+}\lambda_-+{\im\over 2}[\sigma,\lambda_+]\,\Bigr),
\label{vct22SUSY}\\[0.2cm]
\delta\phi\!&=&\!\epsilon_+\psi_--\epsilon_-\psi_+,
\nonumber\\
\delta\psi_+\!\!&=&\!2\im\bepsilon_-D_{x^+}\phi
-\bepsilon_+\bsigma\phi+\epsilon_+F,
\nonumber\\
\delta\psi_-\!\!&=&\!\!-2\im\bepsilon_+D_{x^-}\phi
+\bepsilon_-\sigma\phi+\epsilon_-F,
\nonumber\\
\delta F\!&=&
\!\!-2\im\bepsilon_+D_{x^-}\psi_+-2\im\bepsilon_-D_{x^+}\psi_-
+\bepsilon_+\bsigma\psi_-+\bepsilon_-\sigma\psi_+
+\im(\bepsilon_-\blambda_+-\bepsilon_+\blambda_-)\phi.
\label{chi22SUSY}
\eeqa
The transformation of the barred fields is obtained from the above by
$\overline{ab}=\overline{b}\overline{a}$.
Under this transformation,
each of ${\mathcal L}_{\rm gauge}$, ${\mathcal L}_{\rm matter}$,
${\mathcal L}_W$ and ${\mathcal L}_{{\rm FI}\mdash\theta}$
is invariant modulo a total derivative term.

\newcommand{\deltaB}{\delta^{}_{\scriptscriptstyle\rm B}}
\newcommand{\deltaBcov}{\delta^{\rm cov}_{\scriptscriptstyle\rm B}}
\newcommand{\bfQA}{\bfQ_{\scriptscriptstyle\rm A}}
\newcommand{\bfQB}{\bfQ_{\scriptscriptstyle\rm B}}
\newcommand{\obfQA}{\obfQ_{\scriptscriptstyle\rm A}}
\newcommand{\obfQB}{\obfQ_{\scriptscriptstyle\rm B}}

When we formulate the system on the left-half $x^1\leq 0$
of the Minkowski spacetime with a timelike boundary at $x^1=0$,
at most a half of the $(2,2)$ supersymmetry can be preserved.
Up to R-symmetry rotations, there are two options,
either A-type 
$\bfQA^{}=\obfQ_++\bfQ_-$ and
$\obfQA^{}=\bfQ_++\obfQ_-$,
or B-type
\beq
\bfQB^{}=\obfQ_++\obfQ_-\quad\mbox{and}\quad
\obfQB^{}=\bfQ_++\bfQ_-
\label{defQB}
\eeq
The latter is our focus in this paper.
The supersymmetry variation of the bulk action has a boundary term
that comes from the total derivative terms mentioned above.
For the gauge and matter kinetic terms and the FI-theta term,
we can find simple boundary counter terms that fix the problem,
\beqa
L^{\rm bdry}_{\rm gauge}&=&
{1\over 4\pi e^2}\left[\,{1\over 2}\partial_1(\bsigma\sigma)
  +D\,{\rm Im}(\sigma)+v_{01}{\rm Re}(\sigma)\,\right],
\label{Lgaugebdry}\\
L^{\rm bdry}_{\rm matter}&=&{1\over 2\pi}\left[\,
  {\im\over 2}(\bpsi_-\psi_+-\bpsi_+\psi_-)
  +\bphi \,{\rm Im}(\sigma) \phi\,\right],
\label{Lmatterbdry}\\
L^{\rm bdry}_{{\rm FI}\mdash\theta}&=&
{1\over 2\pi}\Bigl[\,-\zeta({\rm Im}(\sigma))+\theta({\rm Re}(\sigma))\,\Bigr].
\eeqa
Indeed, the sums
\beq
S_{\alpha}
:=\int_{x^1\leq 0}\dd^2x\,{\mathcal L}_{\alpha}+\int_{x^1=0}\dd t\,
L_{\alpha}^{\rm bdry},\qquad
\alpha={\rm gauge},\,{\rm matter},\,\,{\rm FI}\mdash\theta,
\eeq
have manifestly B-type supersymmetric expressions
\beqa
S_{\rm gauge}&=&\int_{x^1\leq 0}\dd^2x\,{1\over 4\pi}
\obfQB^{}\bfQB^{}[\bfQ_+,\obfQ_+]
\left(\,-{1\over 2e^2}\bsigma\sigma\,\right),\\
S_{\rm matter}&=&\int_{x^1\leq 0}\dd^2x\,{1\over 4\pi}
\obfQB^{}\bfQB^{}[\bfQ_+,\obfQ_+]\bphi\phi,\\
S_{{\rm FI}\mdash\theta}&=&
\int_{x^1\leq 0}\dd^2 x\,{1\over 2\pi}{\rm Re}\left(
\obfQB^{}\bfQB^{}\bigl(\,-t(\sigma)\,\bigr)\right).
\eeqa
These are found from the superspace expressions for
${\mathcal L}_{\alpha}$ and are canonical in that sense. 
On the other hand, the superpotential term
\beq
S_W:=\int_{x^1\leq 0}\dd^2x\,{\mathcal L}_W
=\int_{x^1\leq 0}\dd^2 x\,{\rm Re}\bigl(\bfQ_-\bfQ_+ \im W(\phi)\bigr)
\eeq
does not have such a canonical manifestly B-type supersymmetric expression.
Its B-type variation
$\deltaB=-\bepsilon\bfQB^{}+\epsilon \obfQB^{}$ is
\beq
\deltaB S_W
=-\int_{x^1=0}\dd t \,{\rm Re}\bigl(\,\bepsilon\psi^i\partial_i W(\phi)\,\bigr),
\label{Warner}
\eeq
with $\psi:=\psi_++\psi_-$, which is known as the Warner term \cite{Warner}.
This is cancelled by the B-type variation of the
Chan-Paton factor (\ref{CPfactor})-(\ref{At})
for the matrix factorization $Q$ of $W$, that is, the product
\beq
\e^{\im S_W}\cdot {\rm Pexp}\left(-\im\int_{x^1=0}\dd t\,{\mathcal A}_t\right)
\eeq
is $\deltaB$ invariant. This is because
\beq
\deltaB{\mathcal A}_t
=-{\rm Re}\bigl(\,\bepsilon\psi^i\partial_iW(\phi)\,\bigr)
+\im{\mathcal D}_t\bigl(\,\bepsilon Q+\epsilon Q^{\dag}\,\bigr)
\label{deltaBA}
\eeq
holds provided $Q^2= W{\rm id}_{\bV}$,
where ${\mathcal D}_tX:=\partial_tX+i[{\mathcal A}_t,X]$.

\newcommand{\ckDz}{D_0\hspace{-0.11cm}\check{\vrule height1.2ex width0pt}\,\,}
\newcommand{\ckDt}{D_t\hspace{-0.05cm}\check{\vrule height1.2ex width0pt}\,}

We also need to specify the boundary condition on the fields.
As the action is made B-type supersymmetric
with the inclusion of an appropriate boundary interaction,
the main concern is the supersymmetry of the boundary condition itself.
Let us write down the B-type supersymmetry transformation $\deltaB$,
obtained by setting $\epsilon_{\pm}=\pm\epsilon$ and
$\bepsilon_{\pm}=\pm\bepsilon$ in  (\ref{vct22SUSY}) and (\ref{chi22SUSY}).
We shall use a different set of field variables.
Let us put
\beqa
\chi=\lambda_+-\lambda_-,\quad
\eta=\lambda_++\lambda_-,\label{defchieta}\\
\check{v}_{01}=v_{01}+D_1{\rm Re}(\sigma),\quad\check{D}=D+D_1{\rm Im}(\sigma),\label{defckD}\\
\psi=\psi_++\psi_-~(\mbox{already appeared}),\quad
\wt{\psi}=\psi_+-\psi_-,\\
\wh{D_1\phi}=D_1\phi-{\rm Im}(\sigma)\phi,\quad
\wh{D_1\psi}=D_1\psi-{\rm Im}(\sigma)\psi-\bareta\phi.
\eeqa
We also introduce
\beq
\check{v}_0=v_0-{\rm Re}(\sigma),\quad
\eeq
and denote by $\ckDz$ its covariant time derivative,
$\ckDz\phi=\partial_0\phi+\im\check{v}_0\phi$, etc.
Note that $(\check{v}_0,v_1)$ has
the above $\check{v}_{01}$ as its fieldstrength.
Then, the transformation is
\beq
\begin{array}{l}
  \displaystyle
  \deltaB v_0={\im\over 2}\bepsilon\chi+{\im\over 2}\epsilon\bchi,\\[0.2cm]
  \deltaB{\rm Re}(\sigma)=~~{}_{''}~~,\\[0.2cm]
\displaystyle
\deltaB\chi=2\epsilon\left(\im\check{D}-D_0{\rm Re}(\sigma)\right),\\[0.2cm]
\displaystyle
\deltaB\check{D}=-{1\over 2}\bepsilon \ckDz\chi
          +{1\over 2}\epsilon\ckDz\bchi,
\end{array}
\qquad
\begin{array}{l}
  \displaystyle
  \deltaB v_1={\im\over 2}\bepsilon\eta+{\im\over 2}\epsilon\bareta,\\[0.2cm]
  \deltaB{\rm Im}(\sigma)=-{1\over 2}\bepsilon\eta+{1\over 2}\epsilon\bareta,\quad
\\[0.2cm]
\displaystyle
\deltaB\eta=2\epsilon\left(-\check{v}_{01}+\im \ckDz{\rm Im}(\sigma)\right),\\[0.2cm]
\displaystyle
\deltaB\check{v}_{01}={\im\over 2}\bepsilon\ckDz\eta
+{\im\over 2}\epsilon\ckDz\bareta,
\end{array}
\label{vctB}
\eeq
and
\beq
\begin{array}{l}
  \deltaB\phi=\epsilon\psi,\\[0.2cm]
  \deltaB\psi=-2\im\bepsilon \ckDz\phi,
\end{array}
\qquad
\begin{array}{l}
  \deltaB\wt{\psi}=-2\bepsilon \wh{D_1\phi}
  +2\epsilon F,\\[0.2cm]  
  \deltaB\wh{D_1\phi}=\epsilon\wh{D_1\psi},~
  \deltaB F=\im\bepsilon\left(\wh{D_1\psi}-\ckDz\wt{\psi}\right),\\[0.2cm]
  \deltaB\wh{D_1\psi}=-2\im\bepsilon\ckDz\wh{D_1\phi}.
\end{array}
\label{chiB}
\eeq
The first thing to notice is that $\check{v}_0=v_0-{\rm Re}(\sigma)$ is
$\deltaB$-invariant and hence so is its covariant derivative $\ckDz$.
We also see that the other part of the
vector multiplet splits into two,
$(v_0, {\rm Re}(\sigma), \chi, \check{D})$
and $(v_1,{\rm Im}(\sigma), \eta, \check{v}_{01})$,
and so does the chiral multiplet,
$(\phi,\psi)$ and $(\wt{\psi},\wh{D_1\phi},F,\wh{D_1\psi})$.
Then, as a B-type supersymmetric boundary condition, we can impose that
the boundary values of the second multiplet in each of them to vanish.
In the original variables, this reads
\beqa
&&\begin{array}{l}
v_1=0,\\
{\rm Im}(\sigma)=0,\\
\lambda_++\lambda_-=0,\\
D_1{\rm Re}(\sigma)+v_{01}=0,
  \end{array}
\label{app:vctBC}
\\
\mbox{and}~~~~~~~~&&\nn\\
&&\begin{array}{l}
D_1\phi-{\rm Im}(\sigma)\phi=0,\\
\psi_+=\psi_-,\\
F=0,\\
D_1(\psi_++\psi_-)-{\rm Im}(\sigma)(\psi_++\psi_-)
-(\blambda_++\blambda_-)\phi=0,
  \end{array}~~~~~~~~~~~~~~~~~
\label{app:chiBC}
\eeqa
at the boundary.
If we use the first three of (\ref{app:vctBC}), then the last of
(\ref{app:vctBC}) and the first and the last of (\ref{app:chiBC})
simplify to the usual Neumann boundary condition
on ${\rm Re}(\sigma)-v_0$, $\phi$ and $\psi_++\psi_-$.
This simplifed version is shown in (\ref{vctBC}) and (\ref{chiBC}).
Note also that the
boundary counter terms $L_{\rm gauge}^{\rm bdry}$ and $L_{\rm matter}^{\rm bdry}$
given in (\ref{Lgaugebdry}) and (\ref{Lmatterbdry}) vanish
under this boundary condition.

Note that (\ref{app:vctBC}) requires that $\sigma$ takes the boundary values
in the real locus $\im\mathfrak{g}$ of $\mathfrak{g}_{\C}$.
We wonder if we can deform this condition so that the real locus
is replaced by another Lagrangian submanifold $L$ of
$\mathfrak{g}_{\C}$ with respect to the symplectic structure
$\omega(v,w)={\rm Re}(v)\cdot {\rm Im}(w)-{\rm Im}(v)\cdot {\rm Re}(w)$
for a $G$-invariant inner product $\cdot$ on $\im\mathfrak{g}.$
Since we would like the boundary condition to preserve the gauge symmetry,
this $L$ must be $G$-invariant,  $GL=L$, that is,
each $g\in G$ sends $L$ to itself.
To look for such a possibility, let us write down
the $\deltaB$ variation in yet another set of variables.
Let us put
\beq
\lambda=\im(\blambda_++\lambda_-),\quad
\wt{\lambda}=\im(\blambda_+-\lambda_-),\quad
N=\check{v}_{01}+\im \check{D}.
\eeq
Then, the supersymmetry variation with
$\epsilon=\im(\epsilon_1+\im\epsilon_2)$
and $\bepsilon=-\im(\epsilon_1-\im\epsilon_2)$
takes the form
\beq
\deltaB v_0=\im\epsilon_1{\rm Re}(\lambda)-\im\epsilon_2{\rm Im}(\wt{\lambda}),
\quad
\deltaB v_1=\im\epsilon_1{\rm Re}(\wt{\lambda})-\im\epsilon_2{\rm Im}(\lambda),
\label{deltaBv}
\eeq
and
\beq
\begin{array}{l}
\deltaB\sigma=\im\epsilon_1\lambda+\im\epsilon_2J\wt{\lambda},\\[0.2cm]
\deltaB\lambda=-2\epsilon_1\ckDz\sigma+2\epsilon_2JN,\\[0.2cm]
\deltaB\wt{\lambda}=-2\epsilon_1 N+2\epsilon_2 J\ckDz\sigma,\\[0.2cm]
\deltaB N=\im\epsilon_1\ckDz\wt{\lambda}+\im\epsilon_2J\ckDz\lambda,
\end{array}
\label{deltaBsigma}
\eeq
where $J$ is the complex structure of $\mathfrak{g}_{\C}$ regarded as
a real manifold.
The transformation (\ref{deltaBsigma})
is of the same form as the A-type supersymmery variation
of a chiral multiplet, see Eqn (7.2) in \cite{HHP}. As in such a case,
there is a supersymmetric boundary condition associated to a
$G$-invariant Lagrangian submanifold $L\subset \mathfrak{g}_{\C}$:
\beq
\sigma\in L,~~\lambda\in {\rm T}_{\sigma}L,~~
\wt{\lambda}~\&~N\in J{\rm T}_{\sigma}L,
\label{BCforL}
\eeq
at the boundary. Note that $\ckDz\sigma$ is tangent to $L$ at $\sigma$
on any point of the boundary, as long as $\sigma$ takes boundary values in $L$.
This is because $[X,\sigma]$ for $X\in \mathfrak{g}$ and $\sigma\in L$
is tangent to $L$ at $\sigma$, as $L$ is $G$ invariant.
The condition (\ref{BCforL}) reduces to the latter three of
(\ref{app:vctBC}) in the case of $L=\im\mathfrak{g}$
and hence can be regarded as its generalization to the
case of $L\ne\im\mathfrak{g}$.
What is then the generalization of the first of (\ref{app:vctBC}),
$v_1=0$?
Note that
the right-hand sides of $\deltaB v_0$ and $\deltaB v_1$ in (\ref{deltaBv})
are in ${\rm Re}({\rm T}_{\sigma}L)$ and ${\rm Im}({\rm T}_{\sigma}L)$
respectively under (\ref{BCforL}).
In particular, $\deltaB v_1$ can be non-zero at a point $\sigma\in L$ at which
${\rm Im}({\rm T}_{\sigma}L)\ne \{0\}$, and hence
we cannot impose $v_1=0$ at the boundary if $L\ne\im\mathfrak{g}$.
On the other hand, as we want to preserve the gauge symmetry at the boundary,
we do not want to impose any condition on the boundary values of $v_0$.
We leave the boundary condition on $v_{\mu}$ undertermined at this moment.

At this point, we note a mathematical fact.\\
{\bf Theorem.} \emph{There is a one to one correspondence between
$G$-invariant Lagrangian submanifolds $L$ of $\mathfrak{g}_{\C}$ and
Weyl invariant Lagrangian submanifolds $\bgamma$ of a Cartan subalgebra
$\ttt_{\C}$, via}
\beq
\gamma=\ttt_{\C}\cap L\quad\mbox{and}\quad L=G\bgamma.
\label{LGgam}
\eeq
An example of a Lagrangian submanifold of $\mathfrak{g}_{\C}$
is the graph of the differential
of a function $F$ on $\im\mathfrak{g}$ where we regard
$\mathfrak{g}_{\C}=\im\mathfrak{g}\oplus\mathfrak{g}$
as the total space of the contangent bundle of $\im\mathfrak{g}$
by the identification $\mathfrak{g}\cong\im\mathfrak{g}^*$ using the
invariant inner product. It is $G$-invariant when $F$ is $G$-invariant.
Similarly, the graph
of the differential of a Weyl invariant function $f$ on $\im\ttt$ is a
Weyl invariant Lagrangian submanifold of $\ttt_{\C}$.
Note that the restriction from $\im\mathfrak{g}$
to $\im\ttt$ defines a one to one
correspondence between $G$-invariant functions on $\im\mathfrak{g}$
and Weyl invariant functions on $\im\ttt$ (Chevalley restriction theorem).
When $F$ and $f$ are related by the restriction map, $F\mapsto f=F|_{\im\ttt}$,
then $L={\rm graph}(\dd F)$ and $\bgamma={\rm graph}(\dd f)$ are related by
(\ref{LGgam}). This is a particular example of the correspondence
of the theorem but is also useful for the proof, given 
the fact that any Lagrangian submanifold of the cotangent bundle $T^*M$
is locally the graph of the differential of a function on $M$
as long as it is transversal to the fibers.
\\
{\bf Proof.} 
Suppose $L\subset\mathfrak{g}_{\C}$ is a $G$-invariant
Lagrangian submanifold. By $G$-invariance
$[\mathfrak{g},\sigma]\subset {\rm T}_{\sigma}L$, both
$[\im{\rm Re}(\sigma),\sigma]=-[{\rm Re}(\sigma),{\rm Im}(\sigma)]$
and $[\im{\rm Im}(\sigma),\sigma]=\im[{\rm Im}(\sigma),{\rm Re}(\sigma)]$
are tangent to $L$ at $\sigma$.
Since ${\rm T}_{\sigma}L\cap J{\rm T}_{\sigma}L=\{0\}$ as $L$ is Lagrangian,
this implies $[{\rm Re}(\sigma),{\rm Im}(\sigma)]=0$.\footnote{In \cite{HR},
  this commutativity was a part of the requirement from supersymmetry of the
  boundary condition. We have seen now that this automatically follows
  from the fact that $L$ is $G$-invariant and Lagrangian.}
Therefore, a maximal toral subalgebra of $\mathfrak{g}$ passes through
$\im{\rm Re}(\sigma)$ and $\im{\rm Im}(\sigma)$.
In other words, there is some $g\in G$ that sends $\sigma$ to an
arbitrarily chosen Cartan subalgebra $\ttt_{\C}$ of $\mathfrak{g}_{\C}$.
This means $L=G\bgamma$ for $\bgamma:=L\cap \ttt_{\C}$.
Note that $\bgamma$ is a Weyl invariant subset of $\ttt_{\C}$.
We would like to show that it is a Lagrangian submanifold of $\ttt_{\C}$.
Take a point $\sigma\in\bgamma$. By the Lemma stated below,
there is a phase $u\in U(1)$ such that the tangent space to $L$
at $\sigma$ regarded as a subspace to $\mathfrak{g}_{\C}$ is transversal to
$\mathfrak{g}_u:=u\mathfrak{g}\subset \mathfrak{g}_{\C}$.
We may regard $\mathfrak{g}_{\C}=\im\mathfrak{g}_u\oplus\mathfrak{g}_u$
as the total space of the contangent bundle of $\im\mathfrak{g}_u$.
Then, in a neighborhood $U$ of $\sigma$, $L$ is the graph of the differential
of a function $F$ on $\im\mathfrak{g}_u$
which is invariant under a neighborhood $O$ in $G$ of
the isotropy subgroup $G_{\sigma}\subset G$ at $\sigma$.
Note that $O$ includes a neighborhood of the identity element $1\in G$.
Then, the intersection of $L$ with $\ttt_{\C}$ in $U$ is the graph of the
differential of the restriction of $F$ to
$\im\ttt_u:=\im u\ttt\subset \im\mathfrak{g}_u$
which is invariant under the isotropy subgroup of the Weyl group at $\sigma$.
Thus, $\bgamma$ is a Lagrangian submanifold of
$\ttt_{\C}=\im\ttt_u\oplus\ttt_u$ in a neighborhood of $\sigma$.
Conversely, suppose $\bgamma\subset\ttt_{\C}$ is a Weyl invariant Lagrangian
submanifold and let $L=G\bgamma$ be its adjoint orbit.
Take a point $\sigma\in \bgamma$.
Again by the Lemma below, we can find a phase $u\in U(1)$
such that the tangent space of $\bgamma$ at $\sigma$ is transversal to
$\ttt_u:=u\ttt$ in $\ttt_{\C}$.
Then, in a neighborhood $\mathscr{U}$ in $\ttt_{\C}$ of
$\sigma$ and its Weyl images, $\bgamma$ is the graph of the differential of
a Weyl invariant function $f$ on $\im \ttt_u$.
Then, in the open subset $G\mathscr{U}$ of $\mathfrak{g}_{\C}$,
$L=G\bgamma$ is the graph of the differential of the $G$-invariant function
$F$ on $\im\mathfrak{g}_u:=\im u\mathfrak{g}$ that corresponds to
$f$. In particular it is a Lagrangian submanifold of $G\mathscr{U}$.
It remains to show
\\
  {\bf Lemma.} {\it Let $V$ be a real vector space
    and let $Q\subset V_{\C}$ be a real subspace such that
    $Q\cap \im Q=\{0\}$. Then, the set of phases $u\in U(1)$ such that
    $Q\cap uV\ne \{0\}$ is finite. In particular, $Q$ is transversal to
    $uV$ for some phase $u\in U(1)$, $Q\cap uV=\{0\}$.}\\
{\bf Proof.}
We shall call an element of $V_{\C}$ of the form $uv$
for $u\in U(1)$ and $v\in V$ a \emph{pure phase element}.
Let $P$ be the subspace of $Q$ generated over the reals by the
pure phase elements in $Q$.
It has a basis $\{v_1,\ldots, v_m\}$ consisting
of pure phase elements $v_i\in Q\cap u_iV$.
Note that $P\cap\im P=\{0\}$ since $Q\cap \im Q=\{0\}$.
This implies that $v_1,\ldots, v_m$ regarded as elements of $V_\C$
are $\C$-linearly independent. Indeed, if $\sum_ic_iv_i=0$,
we have $\sum_i{\rm Re}(c_i)v_i=-\im\sum_i{\rm Im}(c_i)v_i$ which vanishes
by $P\cap \im P=\{0\}$ and hence $c_i=0$ for all $i$ by the $\R$-linear
independence of $v_1,\ldots, v_m$. Let us put $w_i:=u_i^{-1}v_i\in V$.
Then, $\{w_1,\ldots, w_m\}\subset V$ form a basis of $P_{\C}$.
Any element of $P$ can be written as $\sum_ir_iv_i$ for some real numbers
$r_i$ and it is expressed as $(r_1u_1,\ldots, r_mu_m)$ with respect to the
basis $\{w_1,\ldots, w_m\}$ of $P_\C$. This shows that the phases of
pure phase elements of $Q$ are $\{\pm u_1,\ldots, \pm u_m\}$.
{\bf Q.E.D.}

\medskip
\noindent
{\bf Remark.}
The condition on $L\subset\mathfrak{g}_{\C}$
that it must be Lagrangian may be relaxed \cite{HHP}
to \emph{totally real}, that is,
${\rm T}_{\sigma}L\cap J{\rm T}_{\sigma}L=\{0\}$
and ${\rm T}_{\sigma}L+J{\rm T}_{\sigma}L=\mathfrak{g}_{\C}$ at each point
$\sigma$ of $L$.
Then, most or possibly all of what is said above can be repeated.
In particular, there can be a one to one correspondence between
$G$-invariant totally real submanifolds of $\mathfrak{g}_{\C}$
and Weyl invariant totally real submanifolds of $\ttt_{\C}$,
by the same relation (\ref{LGgam}).
It holds provided the Chevalley restriction theorem for covariants holds:
the restriction from $\im\mathfrak{g}$ to $\im\ttt$ defines
a one to one correspondence between
$G$-equivariant maps $\im\mathfrak{g}\to \im\mathfrak{g}$
and Weyl equivariant maps $\im\ttt\to\im\ttt$.
Proofs for complex algebraic maps can be
found in \cite{Broer,KNV} for example, which immediately imply the
theorem for real polynomial maps.
The proof in \cite{KNV} can be used to prove the theorem for
real analytic maps.
We do not have a proof for more general smooth maps at this moment.

\subsection{Non-Linear Sigma Model}
\label{app:NLSM}

A $(2,2)$ supersymmetric non-linear sigma model
possibly with a superpotential
is specified by the choice of a K\"ahler manifold $X=(X,J,g)$ (target space),
a closed $(1,1)$-form $B$ on $X$ (B-field)
and a holomorphic function $W$ on $X$ (superpotential).
The system with nonlinear $X$ and non-zero $W$ is sometimes called
``hybrid model'' as it is a hybrid of the non-linear sigma model
in which $X$ is nonlinear and $W=0$
and the Landau-Ginzburg model in which $X$ is flat $\C^n$ and non-zero $W$.
In this Appendix, we refer to the system with general $(X,W)$
a sigma model with superpotential.

The variables are the map $\phi$ from the worldsheet to $X$,
the Dirac fermion $\psi_{\pm}$ with values in $\phi^*T_X$
and an auxiliary field $F$ which is valued in $\phi^*T_X$ but non-covariantly
(see below).
The bulk Lagrangian on the Minkowski spacetime is the sum of three terms
\beqa
{\mathcal L}_g&=&
{\rm Re}\int\dd^2\widetilde{\theta}\,g(\bD_+\bPhi,D_-\Phi)\nn\\
&=&g_{i\bj}(\phi)\left(\partial_0\phi^i\partial_0\bphi^{\bj}
-\partial_1\phi^i\partial_1\bphi^{\bj}
+i\bpsi_-^{\bj}(\lrD_{\!\!\! 0}+\lrD_{\!\!\! 1})\psi_-^i
+i\bpsi_+^{\bj}(\lrD_{\!\!\! 0}-\lrD_{\!\!\! 1})\psi_+^i
\right)
\nn\\
&&+R_{i\bj k\bl}\psi_+^i\psi_-^k\bpsi_-^{\bj}\bpsi_+^{\bl}
+g_{i\bj}\widetilde{F}^i\overline{\widetilde{F}}^{\bj},
\label{Lg22}\\[0.2cm]
{\mathcal L}_B
&=&-{\rm Re}\int\dd^2\widetilde{\theta}\,
B(\overline{D}_+\bPhi,D_-\Phi)\,\,
+\,i\partial_{x^+}B(\bpsi_-,\psi_-)\,
-\,i\partial_{x^-}B(\bpsi_+,\psi_+)
\nn\\
&=&B_{i\bj}(\phi)\left(\partial_1\phi^i\partial_0\bphi^{\bj}
-\partial_0\phi^i\partial_1\bphi^{\bj}\right),
\label{LB22}\\[0.2cm]
  {\mathcal L}_W&=&{\rm Re}\int\dd^2\theta\,\im W(\Phi)
\,=\,-{\rm Im}\left[\,
\widetilde{F}^i\partial_iW(\phi)-D_i\partial_jW(\phi)\psi_+^i\psi_-^j\,\right].
\label{LW22cov}
\eeqa
The middle expressions are again in the superfield formalism
in the convention of \cite{cmibook}.
\beq
\widetilde{F}^i=F^i-\Gamma^i_{jk}\psi_+^j\psi_-^k,
\eeq
are the covariant form of the auxiliary field where $\Gamma^i_{jk}$ is
the Levi-Civita connection of $(X,g)$.
Under the change of local holomorphic coordinates $z^i\to z^{i'}=f^{i'}(z)$,
the chiral superfields $\Phi^i$ change to
$\Phi^{i'}=f^{i'}(\Phi)$, which means $\phi^{i'}=f^{i'}(\phi)$,
$\psi_{\pm}^{i'}=\psi_{\pm}^i\partial_if^{i'}(\phi)$
and $F^{i'}=F^i\partial_if^{i'}(\phi)-\psi_+^j\psi_-^k\partial_j\partial_kf^{i'}(\phi)$.
Indeed $F$ is not covariant but $\widetilde{F}$ is,
$\widetilde{F}^{i'}=\widetilde{F}^i\partial_if^{i'}(\phi)$.
Also, $D_{\mu}$ and $D_i$ are the covariant derivatives of $\phi^*T_X$
and $T_X^*$. The expression for ${\mathcal L}_W$ in (\ref{LW22cov})
is a rewriting of (\ref{LW22}) in terms of $\widetilde{F}$.
The supersymmetry transformation is given by (\ref{chi22SUSY})
in which we set $G=\{1\}$.
Under this, the above Lagrangians are invariant modulo total derivative terms.

Let us formulate the system on the left-half Minkowski spacetime,
$\{(x^0,x^1)|x^1\leq 0\}$.
For ${\mathcal L}_{g,B}={\mathcal L}_g+{\mathcal L}_B$,
there is a manifestly B-type supersymmetric action
\beqa
S_{g,B}&=&\int_{x^1\leq 0}\dd^2 x\,\,{\rm Re}\,\obfQB^{}\bfQB^{}
  \,\Bigl[\,g(\bpsi_+,\psi_-)-B(\bpsi_+,\psi_-)\,\Bigr]\nn\\
&=&\int_{x^1\leq 0}\dd^2x\,{\mathcal L}_{g,B}
  +\int_{x^1=0}\dd t\,L^{\rm bdry}_{g,B}
\eeqa
with the boundary counter term
\beq
L^{\rm bdry}_{g,B}={\im\over 2}\left(g(\bpsi_-,\psi_+)-g(\bpsi_+,\psi_-)\right)
-{\im\over 2}B(\bpsi,\psi),
\label{NLSMctt}
\eeq
where $\psi=\psi_++\psi_-$ and $\bpsi=\bpsi_++\bpsi_-$.
The $\deltaB$-transform of the
superpotential term $S_W=\int_{x^1\leq 0}\dd^2x\,{\mathcal L}_W$
is given by the Warner term (\ref{Warner}).
In order to cancel it, we introduce the
Chan-Paton factor as (\ref{CPfactor})-(\ref{At}).
For this purpose, we introduce a $\Z_2$-graded hermitian vector bundle
$E=E^{\rm ev}\oplus E^{\rm od}$ and an ${\rm End}(E)$ valued boundary
interaction
\beq
    {\mathcal A}_t
    =-\im\dot{\phi}^{\bi}{\partial\over\partial\bpsi^{\bi}}{\mathcal Q}
    +\im\dot{\phi}^i{\partial\over\partial\psi^i}{\mathcal Q}^{\dag}
    -{1\over 2}\psi^i{\partial\over\partial \phi^i}{\mathcal Q}
    +{1\over 2}\bpsi^{\bi}{\partial\over\partial \phi^{\bi}}{\mathcal Q}^{\dag}
    +{1\over 2}\{{\mathcal Q},\mathcal{Q}^{\dag}\}
\label{NLSMcalA1}
\eeq
with
\beq
    {\mathcal Q}=Q^{(0)}(\phi)+\bpsi^{\bi}Q^{(1)}_{\bi}(\phi)
    +{1\over 2}\bpsi^{\bi}\bpsi^{\bj}Q^{(2)}_{\bi\bj}(\phi)+\cdots
    +{1\over n!}\bpsi^{\bi_1}\cdots\bpsi^{\bi_n}Q^{(n)}_{\bi_1\cdots\bi_n}(\phi),
\label{defcQ}
\eeq
in which $Q^{(p)}_{\ldots}$ for even ({\it resp} odd) $p$
takes values in odd ({\it resp}. even) endomorphisms of $E$.
Provided ${\mathcal Q}$ satisfies
\beq
\bpsi^{\bi}{\partial\over\partial\phi^{\bi}}{\mathcal Q}+{\mathcal Q}^2
  = (W(\phi)-c)\,{\rm id}_E
\label{condSC}
\eeq
for a constant $c$,
the $\deltaB$ variation of ${\mathcal A}_t$ takes exactly the same form as
(\ref{deltaBA}) with the replacement $Q\to {\mathcal Q}$.
It cancels the Warner term so that
$\e^{\im S_W}\cdot {\rm Pexp}\left(-\im\int_{x^1=0}\dd t\,{\mathcal A}_t\right)$
is $\deltaB$ invariant.
We would like ${\mathcal A}_t$ to transform like a gauge potential
 ${\mathcal A}_t\to h_\phi^{-1}{\mathcal A}_th_\phi+\im^{-1}h^{-1}_\phi\partial_t h_\phi$ under a gauge transformation
$h_{\phi}$ of $\phi^*E$.
This is the case when ${\mathcal Q}$ transforms under
a local unitary frame change $h$ of $E$ also like a gauge potential
\beq
    {\mathcal Q}\to h^{-1}{\mathcal Q}h+h^{-1}\bpsi^{\bi}\partial_{\bi}h.
\eeq
Under the replacement $\bpsi^{\bi}\mapsto \dd\bz^{\bi}$ in (\ref{defcQ}),
it reads ${\mathcal Q}\to h^{-1}{\mathcal Q} h+h^{-1}\bartial h$.
It defines a differential
$\bartial_{\!\mathcal Q}:\Omega^{0,*}(X,E)\to \Omega^{0,*}(X,E)$
obeying $\bartial_{\!\mathcal Q}(s\cdot\eta)=\bartial_{\!\mathcal Q}s\cdot \eta+(-1)^{|s|}s\cdot\bartial \eta$ for $s\in \Omega^{0,*}(X,E)$ and $\eta\in \Omega^{0,*}(X)$, in which $|s|$ is the total $\Z_2$ degree of $s$, that is,
the sum of mod 2 form degree and the degree from the $\Z_2$-grading of $E$.
Such a structure is called a ($\Z_2$-graded) $\bartial$-\emph{superconnection}
of $E$. The condition (\ref{condSC}) means that its curvature
$\bartial_{\!\mathcal Q}^2$ is $(W-c){\rm id}_E$.
In the case when ${\mathcal Q}$ is first order in $\bpsi$,
where we write
\beq
{\mathcal Q}=Q+\im\bpsi^{\bi}A_{\bi},
\label{1storder}
\eeq    
the boundary interaction (\ref{NLSMcalA1}) takes the form
\beq
{\mathcal A}_t=(\phi^*A)_t-{\im\over 2}F_A(\bpsi,\psi)
-{1\over 2}\psi^iD_iQ+{1\over 2}D_{\bi}Q^{\dag}+{1\over 2}\{Q,Q^{\dag}\},
\label{NLSMcalA2}
\eeq
in which $A=\dd\bz^{\bi}A_{\bi}+\dd z^i(A_{\bi})^{\dag}$ is a unitary
connection of the hermitian vector bundle $E$, $F_A$ is its curvature
and $D_i$ and $D_{\bi}$ are its covariant derivatives.
The curvature condition splits to
\beq
Q^2=(W-c)\cdot{\rm id}_E,\quad
\bartial_{\!A}Q=0,\quad
\bartial_{\!A}^2=0.
\eeq
The last condition means that $\bartial_{\!A}$ defines
a holomorphic structure on $E$ and we often denote the resulting
holomorphic vector bundle by ${\mathcal E}$.
The second condition means that
$Q$ is a holomorphic endomorphism of ${\mathcal E}$.
Thus, we have a $\Z_2$-graded holomorphic vector bundle ${\mathcal E}$
with an odd endomorphism $Q$ that squares to $(W-c){\rm id}_E$.
Such a structure $({\mathcal E},Q)$
is called a \emph{matrix factorization} of $W-c$ on $X$.

The fields split into two multiplets, $(\phi,\psi)$ and
$(\wt{\psi}=\psi_+-\psi_-,\partial_1\phi,F,\partial_1\psi)$,
under the B-type supersymmetry as in (\ref{chiB}).
An invariant boundary condition is to set the latter to vanish
\beq
\partial_1\phi=0, \quad
\psi_+=\psi_-,\quad
F=0,\quad
\partial_1(\psi_++\psi_-)=0
\label{BCNLSM}
\eeq
at the boundary.

The system we encounter in this paper always has a vector $U(1)$ R-symmetry
with charge integrality.
It is asociated with an isometric action of $U(1)$ on $(X,g)$,
$\lambda:x\mapsto R_{\lambda}(x)$, with $R_{-1}={\rm id}_X$, such that
the superpotential $W$ has charge $2$, $R_{\lambda}^*W=\lambda^2 W$.
We suppose that it extends to a holomorphic action of $\C^{\times}$ on $(X,J)$
under which $W$ has charge $2$.
The action of $\lambda\in U(1)$ on the fields is
$\phi\to R_{\lambda}\circ\phi$,
$\psi_{\pm}\to \lambda^{-1}R_{\lambda}^*\psi_{\pm}$ and
$\wt{F}\to \lambda^{-2}R_{\lambda}^*\wt{F}$.
Note that the boundary condition (\ref{BCNLSM}) is invariant under this.
We would also like to consider only boundary interactions that are
$U(1)$ R-symmetric.
This is the case when $E$ is a $U(1)$ equivariant vector bundle and
${\mathcal Q}$ has $U(1)$ charge $1$, that is, for each $\lambda\in\C^{\times}$
there is a linear isomorphism $r_{\lambda}:E\to E$ over $R_{\lambda}:X\to X$,
unitary for $\lambda\in U(1)$,
such that
\beq
r_{\lambda}{\mathcal Q}(R_{\lambda}(\phi,\bpsi))r_{\lambda}^{-1}
=\lambda\,{\mathcal Q}(\phi,\bpsi).
\eeq
Note that the constant $c$ in (\ref{condSC}) necessarily vanishes.
We also require that the charge integrality is preseved, that is,
$r_{-1}$ agress with the $\Z_2$-grading of $E$.
This ${\mathcal Q}$ defines a $\Z$-graded superconnection
whose curvature is $W{\rm id}_E$.
In the first order case (\ref{1storder}),
$({\mathcal E},Q)$ is a $\Z$-graded matrix factorization of $W$.
In the special case where $W=0$ and R-symmetry on $X$ is trivial,
this is a $\Z$-graded complex of holomorphic vector bundles.
The degree $i$ term ${\mathcal E}^i$ is the component of ${\mathcal E}$
of R-charge $i$, on which $r_{\lambda}=\lambda^i\cdot{\rm id}$,
and $Q$ is the differential ${\mathcal E}^i\to {\mathcal E}^{i+1}$.

Let $a$ be a unitary connection of a hermitian line bundle $L$ on $X$ and
let us change the branes as $E\to E\otimes L$ and
${\mathcal Q}\to {\mathcal Q}+\im\bpsi^{\bi}a_{\bi}$.
Then, the condition (\ref{condSC}) continues to hold provided
the curvature $F_a$ is a $(1,1)$-form,
so that $\bartial_a$ defines a holomorphic line bundle ${\mathcal L}_a$.
This changes the boundary interaction as
${\mathcal A}_t\to {\mathcal A}_t+(\phi^*a)_t-{\im\over 2}F_a(\bpsi,\psi)$.
The change in turn is the same as the shift of the B-field that appears
in (\ref{LB22}) and (\ref{NLSMctt}) as $B\to B-F_a$.
Note that the closed two form $-F_a$ represents the class $2\pi c_1(L)$.
Thus, the change of the brane is equivalent to the shift of the B-field.
In the first order case,
\beq
[B]\,\to\, [B]+2\pi c_1({\mathcal L}_a)~\,\mbox{with $({\mathcal E},Q)$ fixed}
~~\Longleftrightarrow~~
({\mathcal E},Q)\,\to\, ({\mathcal E}\otimes {\mathcal L}_a,Q)
~\,\mbox{with $[B]$ fixed}.
\label{Bshift}
\eeq
This is the sigma model version of (\ref{thetashift}).

Suppose the model has a compact symmetry group $G$.
In addition, we assume that $G_{\C}$ acts holomorphically on $(X,J)$ leaving
$W$ invariant and that the $G$ action has a moment map with respect
to the K\"ahler form.
Then, we can gauge the symmetry while preserving the $(2,2)$ supersymmetry.
The vector $U(1)$ R-symmetry is also preserved when it commutes with $G$.
When formulated on a surface with boundary,
the gauged system has B-type supersymmetry
with a canonical boundary counter term and an appropriate Chan-Paton factor
${\rm Pexp}\left(-\im\int \dd t{\mathcal A}_t\right)$.
The latter is associated with a $\bartial$-superconnection
$(E,{\mathcal Q})$ on $X$ with curvature $W$ which is $G$-\emph{equivariant},
that is, for each $g\in G$,
there is a unitary transformation $h(g):E\to E$ over $g:X\to X$ such that
\beq
h(g)^{-1}g^*{\mathcal Q}\,h(g)+h(g)^{-1}\bartial \,h(g)={\mathcal Q}.
\eeq
The Chan-Paton interaction is
\beqa
\lefteqn{{\mathcal A}_t=h(v_t-{\rm Re}(\sigma))}\nn\\
&&-\im\ckDt\phi^{\bi}{\partial\over\partial\bpsi^{\bi}}{\mathcal Q}
    +\im\ckDt\phi^i{\partial\over\partial\psi^i}{\mathcal Q}^{\dag}
    -{1\over 2}\psi^i{\partial\over\partial \phi^i}{\mathcal Q}
    +{1\over 2}\bpsi^{\bi}{\partial\over\partial \phi^{\bi}}{\mathcal Q}^{\dag}
    +{1\over 2}\{{\mathcal Q},\mathcal{Q}^{\dag}\}
\label{gaugedNLSMcalA}
\eeqa
in which $\ckDt\phi^i=\dot{\phi}^i+\im (v_t-{\rm Re}(\sigma))^i(\phi)$
where $\im v^i(\phi)\partial_i-\im v^{\bi}(\phi)\partial_{\bi}$
is the vector field on $X$ generated by $\im v\in\mathfrak{g}$.
When $X$ is the affine space $V$, this reduces to the boudary interaction
in the main text, (\ref{At}), for the trivial
bundle $E=V\times\bV$ with $h(g)=(g,\rho(g))$
and the zero-th order ${\mathcal Q}=Q$.

\medskip
\noindent
{\bf Remark.}
The boundary interactions ${\mathcal A}_t$ of the forms
(\ref{NLSMcalA1}), (\ref{NLSMcalA2}), (\ref{At}) and (\ref{gaugedNLSMcalA}),
which play the central r\^oles in the study of B-branes,
are constructed building on
the ${\mathcal N}=1$ supersymmetric versions obtained earlier
in \cite{Hblin,KrausLarsen,TTU}.

\section{Categories Of D-Branes}
\label{app:Cat}

\newcommand{\Bbrane}{{\mathcal B}}
\newcommand{\Op}{\mathcal{O}p}
\newcommand{\dt}{d_t}
\newcommand{\Th}{{\bf T}}

We describe the category of D-branes of type B in the class of
$(2,2)$ supersymmetric field theories which we encounter in this paper
(though we remove here the Calabi-Yau condition which is assumed
in the bulk of the paper).
We shall also introduce some categories in mathematics
which are known or expected to be equivalent to these categories of D-branes.

The category $D(\Th)$ in a theory $\Th$ is described as follows.
For a pair of branes, $\Bbrane_1$ and $\Bbrane_2$,
let $\Op(\Bbrane_1,\Bbrane_2)$ be the space of
local operators inserted on the boundary on which the branes are
$\Bbrane_1$ and $\Bbrane_2$ before and after the insertion point.
The $U(1)$ R-symmetry and the supercharge $\bfQB$ as in (\ref{defQB})
introduce a $\Z$-grading and  a degree $1$ differential on
$\Op(\Bbrane_1,\Bbrane_2)$ which hence becomes
a $\Z$-graded complex of vector spaces.
The space of morphisms $\Hom^{}_{D(\Th)}(\Bbrane_1,\Bbrane_2)$
is the cohomology of this complex $(\Op(\Bbrane_1,\Bbrane_2),\bfQB)$,
which is a $\Z$-graded vector space.
As usual in the chiral ring argument, there is a composition
$\Hom^{}_{D(\Th)}(\Bbrane_1,\Bbrane_2)\times \Hom^{}_{D(\Th)}(\Bbrane_2,\Bbrane_3)\to \Hom^{}_{D(\Th)}(\Bbrane_1,\Bbrane_3)$
induced from the product of local operators.
In short, $D(\Th)$ is the homotopy category of the  differential graded
(dg) category of branes and local operators.
To be precise, the composition is ill-defined before
taking the cohomology because of the short distance singularity of the
operator product.
However, there are simpler models for $(\Op(\Bbrane_1,\Bbrane_2),\bfQB)$
with the same cohomologies but with well-defined compositions,
so that $D(\Th)$ is equivalent to the homotopy categories of such simpler
and genuine dg-categories.

\subsection{Non-Linear Sigma Model}
\label{app:CatNLSM}

We first consider the non-linear sigma model, possibly with superpotential,
specified by $(X,g,J,B,W)$ with an $U(1)$ action $R_{\lambda}$
as in Appendix~\ref{app:NLSM}, which we simply label by $(X,W)$.
Under the boundary condition (\ref{BCNLSM}), the field variables remaining
on the boundary are $\phi$, $\psi$ and $\bpsi$ which transform under
B-type supersymmetry as
\beq
\deltaB\phi^i=\epsilon\psi^i,\quad
\deltaB\phi^{\bi}=-\bepsilon\bpsi^{\bi},\quad
\deltaB\psi^i=-2\im\bepsilon\dot{\phi}^i,\quad
\deltaB\bpsi^{\bi}=2\im\epsilon\dot{\phi}^{\bi}.
\label{Bvar}
\eeq
As local operators, we consider functions of these variables
and their time derivatives.
If they are inserted between $\Bbrane_1=(E_1,{\mathcal Q}_1)$
and $\Bbrane_2=(E_2,{\mathcal Q}_2)$, we suppose that
the values of these functions at $(\phi,\psi,\bpsi,\dot{\phi},\ldots)$
are linear maps from $E_1|_{\phi}$ to  $E_2|_{\phi}$.
More precisely, we propose that the space $\Op(\Bbrane_1,\Bbrane_2)$ consists of
power series in $\psi,\bpsi,\dot{\phi}$ and their time deriveatives
whose coefficients are sections of ${\rm Hom}(E_1,E_2)$.
There is a $\Z$-grading in $\Op(\Bbrane_1,\Bbrane_2)$ given by the
$U(1)$ R-charge of local operators, specified by
$R_{\lambda}$ on $X$ and $r_{i\lambda}$ on $E_i$ ($i=1,2$).
The supersymmetry acts on the variables as (\ref{Bvar}) but there is also an
action on the Chan-Paton factor as we now find.
When an operator $O\in \Op(\Bbrane_1,\Bbrane_2)$ is inserted at a time $t$,
the path integral measure has a factor of the form
$U_2(t_2,t)O(t)U_1(t,t_1)$ for $t_1<t<t_2$
where
$U(t_f,t_i):={\rm Pexp}\left(-\im\int_{t_i}^{t_f}\dd t\,{\mathcal A}_t\right)$
for each segment $[t_i,t_f]$ with boundary interaction ${\mathcal A}_t$.
As $\deltaB{\mathcal A}_t$ is as in (\ref{deltaBA}),
denoting by $S$ the action $S_{g,B}+S_W$ on the half-strip
$[t_1,t_2]\times \{x^1\leq 0\}$,
we have
\beqa
\lefteqn{\deltaB\left(\e^{\im S}\,U_2(t_2,t)O(t)U_1(t,t_1)\right)}\nn\\
&=&\e^{\im S}\cdot\Biggl\{
\int_t^{t_2}\dd t'{\dd\over \dd t'}\Bigl[\,U_2(t_2,t')(\bepsilon{\mathcal Q}_2+\epsilon{\mathcal Q}^{\dag}_2)(t')
U_2(t',t)\Bigr]O(t)U_1(t,t_1)\nn\\
&&~~~~~~~~~~+U_2(t_2,t)\deltaB O(t) U_1(t,t_1)\nn\\
&&~~~~~~~~~~+U_2(t_2,t)O(t)\int_{t_1}^t\dd t'{\dd\over \dd t'}\Bigl[U_1(t,t')(\bepsilon{\mathcal Q}_1+\epsilon{\mathcal Q}^{\dag}_1)(t')U_1(t',t_1)\,\Bigr]\,\Biggr\}\nn\\
&=&\e^{\im S}\cdot\Biggl\{
(\bepsilon{\mathcal Q}_2+\epsilon{\mathcal Q}^{\dag}_2)(t_2)U_2(t_2,t)O(t)U_1(t,t_1)\nn\\
&&~~~~~~~~~~+U_2(t_2,t)\deltaBcov O(t)U_1(t,t_1)\nn\\
&&~~~~~~~~~~-U_2(t_2,t)O(t)U_1(t,t_1)(\bepsilon{\mathcal Q}_1+\epsilon{\mathcal Q}^{\dag}_1)(t_1)\,\Biggr\}
\label{B2}
\eeqa
with
\beq
\deltaBcov O:=\deltaB O-(\bepsilon{\mathcal Q}_2+\epsilon{\mathcal Q}^{\dag}_2)O
+O(\bepsilon{\mathcal Q}_1+\epsilon{\mathcal Q}^{\dag}_1).
\eeq
The first and the last terms of (\ref{B2}) would contribute to
$\deltaBcov$ on operators inserted at $t_2$ and $t_1$. Therefore,
$\deltaBcov$ is the full action of B-type supersymmetry on local operators.
In particular, for the part $\bfQB$ of
$\deltaB=-\bepsilon\bfQB+\epsilon\obfQB$ which acts on the variables as
\beq
\bfQB\phi^i=0,\quad
\bfQB\phi^{\bi}=\bpsi^{\bi},\quad
\bfQB\psi^i=2\im \dot{\phi}^i,\quad
\bfQB\bpsi^{\bi}=0,
\label{Bvar2}
\eeq
the full action is
\beq
\bfQB^{\rm cov} O=\bfQB O+{\mathcal Q}_2O-(-1)^{|O|}O{\mathcal Q}_1.
\eeq
The space of morphisms in the category $D(X,W)$ of the mdoel
is the cohomology of the complex with this differential,
\beq
\Hom^{}_{D(X,W)}(\Bbrane_1,\Bbrane_2)={\rm H}(\Op(\Bbrane_1,\Bbrane_2),\bfQB^{\rm cov}).
\eeq

There is a simpler complex with the same cohomology. To see this, we introduce
the ``\emph{dimension}'' of operators, where $\phi$ and $\bpsi$ have dimension 0
and $\psi$ has dimension 1 while
each time derivative $\dt:={\dd\over \dd t}$ increases the dimension by $1$.
Let $\bfD$ be the dimension counting operator,
\beq
\bfD\phi=0,\quad \bfD\bpsi=0,\quad
\bfD\psi=\psi,\quad [\bfD,\dt]=\dt. 
\eeq
We see from (\ref{Bvar2}) that $\bfQB$ commutes with $\bfD$.
As the function ${\mathcal Q}$ of $\phi$ and $\bpsi$ has dimension 0,
$\bfQB^{\rm cov}$ also commutes with $\bfD$.
Then, the complex $(\Op(\Bbrane_1,\Bbrane_2),\bfQB^{\rm cov})$
decomposes into the sum of subcomplexes of definite dimensions $d$,
denoted by $(\Op_d(\Bbrane_1,\Bbrane_2),\bfQB^{\rm cov})$,
and so does its cohomology,
\beq
{\rm H}(\Op(\Bbrane_1,\Bbrane_2),\bfQB^{\rm cov})
=\bigoplus_{d=0}^{\infty}
{\rm H}(\Op_d(\Bbrane_1,\Bbrane_2),\bfQB^{\rm cov})
\eeq
In fact, the cohomologies of dimension $d>0$ all vanish.
To show this, we define a fermionic operator $\bfS$ by
\beq
\bfS\,\dt^n\phi^i={n\over 2\im}\dt^{n-1}\psi^i,\quad
\bfS\,\dt^n\phi^{\bi}=0,\quad
\bfS\,\dt^n\psi^i=0,\quad
\bfS\,\dt^n\bpsi^{\bi}=n\cdot \dt^n\phi^{\bi},
\eeq
for $n\geq 0$.
It obviously preserves the dimensions of variables and 
annihilates dimension 0 variables. In particular, it anticommutes with
${\mathcal Q}$.
Moreover, one can show
$\{\bfS,\bfQB\}=\bfD$ by a direct computation.
Thus, we find
\beq
[\bfD,\bfS]=0,\quad
\{\bfQB^{\rm cov},\bfS\}=\bfD.
\eeq
Suppose a $\bfQB^{\rm cov}$-closed operator $O$ has dimension $d>0$.
Then, we have
\beq
O={1\over d}\,\bfD\, O={1\over d}\Bigl(\,\bfQB^{\rm cov}\bfS \,O
+\bfS\underbrace{\bfQB^{\rm cov}O}_0\,\Bigr)=
\bfQB^{\rm cov}\left(\,{1\over d}\,\bfS\, O\,\right).
\eeq
That is, it is $\bfQB^{\rm cov}$-exact.
This is what we wanted to show.
Thus, the dimension 0 component $(\Op_0(\Bbrane_1,\Bbrane_2),\bfQB^{\rm cov})$
is a simpler model with the same cohomology.
The replacement $\bpsi^{\bi}\to\dd\bz^{\bi}$
provides a way to describe it geometrically.
Under this, $\Op_0(\Bbrane_1,\Bbrane_2)$ is identified as the space
of differential forms $\Omega^{0,*}(X,\Hom(E_1,E_2))$
and $\bfQB^{\rm cov}$ is identified as the differential
\beq
\bartial^{}_{{\mathcal Q}_1,{\mathcal Q}_2}O
=\bartial O+{\mathcal Q}_2 O-(-1)^{|O|}{\mathcal Q}_1.
\eeq
Thus, we found that the space of morphisms can also be described as
\beq
\Hom^{}_{D(X,W)}((E_1,{\mathcal Q}_1),(E_2,{\mathcal Q}_2))\,\,\cong\,\,
{\rm H}\left(\Omega^{0,*}(X,\Hom(E_1,E_2)), \bartial_{{\mathcal Q}_1,{\mathcal Q}_2}\right).
\label{morphSC}    
\eeq
Note that the product of dimension 0 operators is free of short distance
singularity and corresponds to the standard product
$\Omega^{0,*}(X,\Hom(E_1,E_2))\times\Omega^{0,*}(X,\Hom(E_2,E_3))\to\Omega^{0,*}(X,\Hom(E_1,E_3))$.
Thus, $D(X,W)$ is equivalent to the homotopy category of the (genuine)
dg-category of $\bartial$-superconnections on $X$ with curvature $W$
whose morphisms are the complexes
$(\Omega^{0,*}(X,\Hom(E_1,E_2)), \bartial_{{\mathcal Q}_1,{\mathcal Q}_2})$
and the composition is the standard product.
In the first order case (\ref{1storder}), we may choose local holomorphic
frames of ${\mathcal E}_1$ and ${\mathcal E}_2$ in which $A_{1\bi}=0$ and
$A_{2\bi}=0$, so that the differential takes the form
\beq
\bartial_{Q_1,Q_2}O=\bartial O+Q_2O-(-1)^{|O|}OQ_1.
\label{bartialQ1Q2}
\eeq

\newcommand{\Ohol}{O_{\scriptscriptstyle \rm hol}}

When $(X,J)$ is a Stein manifold, that is, when
${\rm H}^{0,p}(X,{\mathcal E})=0$ for $p>0$ for any holomorphic vector bundle
${\mathcal E}$, one can find
a further simplification of the dimension 0 complex in the first order
case (\ref{1storder}).
Suppose $O\in \Omega^{0,*}(X,\Hom({\mathcal E}_1,{\mathcal E}_2))$
is $\bartial_{Q_1,Q_2}$-closed and has total R-charge $|O|$. Let us expand it as
$O=O^{(0)}+O^{(1)}+\cdots+O^{(n)}$ where $O^{(i)}$ is a $(0,i)$-form.
Then, the equation $\bartial_{Q_1,Q_2}O=0$ splits to
\beq
\bartial O^{(i-1)}+Q_2O^{(i)}-(-1)^{|O|}O^{(i)}Q_1=0,\quad i=0,1,\ldots,n+1,
\eeq
with $O^{(-1)}=O^{(n+1)}=0$.
Using the Stein property of $X$, we can solve these equations
recursively, starting from the last one $\bartial O^{(n)}=0$.
We find a sequence of forms $\Lambda^{(n-1)}$, $\Lambda^{(n-2)}$,
$\ldots$, $\Lambda^{(0)}$ of total R-charge $|O|-1$
such that
\beqa
O^{(n)}&=&\bartial \Lambda^{(n-1)},\nn\\
O^{(n-1)}&=&\bartial\Lambda^{(n-2)}+Q_2\Lambda^{(n-1)}+(-1)^{|O|}\Lambda^{(n-1)}Q_1,
\nn\\&&\cdots\nn\\
O^{(1)}&=&\bartial\Lambda^{(0)}+Q_2\Lambda^{(1)}+(-1)^{|O|}\Lambda^{(1)}Q_1,\nn\\
O^{(0)}&=&\Ohol+Q_2\Lambda^{(0)}+(-1)^{|O|}\Lambda^{(0)}Q_1
\eeqa
where $\Ohol$ is holomorphic, $\bartial \Ohol=0$,
and satisfies $d_{Q_1,Q_2}\Ohol=0$ for
\beq
d_{Q_1,Q_2}\Ohol\,:=\,Q_2 \Ohol-(-1)^{|O|}\Ohol Q_1.
\label{holdiff}
\eeq
Then, we have $O=\Ohol+\bartial_{Q_1,Q_2}\Lambda$ for
$\Lambda=\Lambda^{(0)}+\cdots \Lambda^{(n-1)}$.
This means that the dimension 0 complex has the same cohomology as
the complex over the space of holomorphic sections of
$\Hom({\mathcal E}_1,{\mathcal E}_2)$ with the differential (\ref{holdiff})
\beq
\Hom^{}_{D(X,W)}(({\mathcal E}_1,Q_1),({\mathcal E}_2,Q_2))~\cong~
{\rm H}(\Gamma_{\rm hol}(X,\Hom({\mathcal E}_1,{\mathcal E}_2)),d_{Q_1,Q_2}).
\label{Homhol}
\eeq

\subsection{Gauge Theory}
\label{app:CatGT}

Let us next discuss the category of D-branes in
the gauge theory, labelled by $(G,V,W)$ as in Appendix~\ref{app:GLSM}.
Let us first put aside the question of regularity.
To simplify the analysis we choose the basic boundary condition
(\ref{app:vctBC})-(\ref{app:chiBC}) so that the remaining variables
on the boundary are $(v_0, {\rm Re}(\sigma),\chi,\check{D})$ from the
vector multiplet and $(\phi,\psi)$ from the chiral multiplet.
See (\ref{defchieta}) and (\ref{defckD}) for the definition of
$\chi$, $\check{D}$ and $\psi$.
The $\deltaB$ variation is given in (\ref{vctB}) and (\ref{chiB}).
As the local operators inserted between
$\brane_1=(\bV_1,Q_1)$ and $\brane_2=(\bV_2,Q_2)$, we consider
functions of these variables and their time derivatives
with values in $\Hom(\bV_1,\bV_2)$ which are gauge invariant.
By gauge invariance, dependence on the variables is only through
${\rm Re}(\sigma)$, $\chi$, $\check{D}$, $\phi$, $\psi$ and their
covariant time derivatives, $D_0X$ or $\ckDz X=D_0X-\im {\rm Re}(\sigma) X$.
The space of local operators $\Op(\brane_1,\brane_2)$
is $\Z$-graded by the $U(1)$ R-symmetry determined by the bulk $R$
and the boundary ${\bf r}_{*1}$, ${\bf r}_{*2}$.
The supersymmetry acts also on the Chan-Paton factor.
By the same computation as in the sigma model, we find that the full action
of $\bfQB$ is
\beq
\bfQB^{\rm cov} O=\bfQB O+Q_2O-(-1)^{|O|}OQ_1
\label{QBcovinGLSM}
\eeq
where the action on the varibales is $[\bfQB,\ckDz]=0$ and
\beqa
&\bfQB{\rm Re}(\sigma)=-{\im\over 2}\chi,\quad
\bfQB\chi=0,\quad
\bfQB\bchi=2D_0{\rm Re}(\sigma)+2\im\check{D},\quad
\bfQB\check{D}={1\over 2}\ckDz\lambda,
\nn\\
&\bfQB\phi=0,\quad
\bfQB\bphi=\bpsi,\quad
\bfQB\psi=2\im \ckDz\phi,\quad\bfQB\bpsi=0.
\label{bfQBgt}
\eeqa
The space of morphisms in the category of branes in the theory,
which we denote by $D_G(V,W)$,
is the cohomology of the complex with differential $\bfQB^{\rm cov}$
in (\ref{QBcovinGLSM}).
We can find a simpler model of the complex with the same cohomology.
First, we introduce the dimensions of variables:
\beqa
\mbox{dimension 0}:&&\phi,~~ \bpsi,\nn\\
\mbox{dimension 1}:&&\psi,~~{\rm Re}(\sigma),~~ \chi,~~\ckDz,\\
\mbox{dimension 2}:&&\bchi,~~ \check{D}.\nn
\eeqa
We see that $\bfQB$ and $\bfQB^{\rm cov}$ preserve the dimension.
Hence the complex $(\Op(\brane_1,\brane_2),\bfQB^{\rm cov})$
decomposes into the sum of complexes of definite dimensions,
and so does its cohomology.
The cohomologies of positive dimensions all vanish. 
Indeed, we define a fermionic operator $\bfS$ by
\beqa
&\bfS{\rm Re}(\sigma)^{(n)}=\bfS\bchi^{(n)}=0,\quad
\bfS\chi^{(n)}=2\im (n+1){\rm Re}(\sigma)^{(n)},\quad
\bfS\check{D}^{(n)}={n+2\over 2\im}\bchi^{(n)},\nn\\
&\bfS\phi^{(n)}={n\over 2\im}\psi^{(n-1)},\quad
\bfS\bphi^{(n)}=\bfS\psi^{(n)}=0,\quad
\bfS\bpsi^{(n)}=n\bphi^{(n)},
\eeqa
for $n\geq 0$ where $X^{(n)}$ stands for $\ckDz^nX$. It preserves the
dimension of variables and annihilates dimension 0 variables.
One can directly check that $\{\bfS,\bfQB\}$ is the dimension counting
operator which we denote by $\bfD$.
Thus, we find $[\bfS,\bfD]=0$ and $\{\bfQB^{\rm cov},\bfS\}=\bfD$.
As in the sigma model, this implies that the cohomology of the 
positive dimension part of the complex $(\Op(\brane_1,\brane_2),\bfQB^{\rm cov})$
vanishes. Therefore, we are left with the dimension zero part.
Via the replacement $\bpsi^{\bi}\to\dd \bphi^{\bi}$, the space
$\Op^{}_0(\brane_1,\brane_2)$ can be identified as
the $G$-invariant part of $\Omega^{0,*}(V,\Hom(\bV_1,\bV_2))$.
Thus, the space of morphisms can also be described as
\beq
\Hom^{}_{D_G(V,W)}(\brane_1,\brane_2)
\,\,\cong\,\,
{\rm H}\left(\Omega^{0,*}(V,\Hom(\bV_1,\bV_2))^G,\bartial_{Q_1,Q_2}\right),
\label{Homgt0}    
\eeq
where $\bartial_{Q_1,Q_2}$ is given by the same expression
as (\ref{bartialQ1Q2}).
As the vector space $V$ is a Stein manifold, a further simplification is
possible. The computation in the sigma model can be repeated with
the $G$-invariance condition taken into account
--- we may employ the averaging over $G$ by the invariant Haar measure
to find $G$-invariant $\Lambda^{(i)}$'s.
Thus, we have a formula like (\ref{Homhol}) in which
$\Gamma_{\rm hol}^{}(-)$ is replaced by $\Gamma_{\rm hol}^{}(-)^G$.
As $G$-invariance and holomorphy imply $G_{\C}$-invariance, we have
\beq
\Hom^{}_{D_G(V,W)}(\brane_1,\brane_2)
\,\,\cong\,\,
{\rm H}\left(\Gamma_{\rm hol}(V,\Hom(\bV_1,\bV_2))^{G_{\C}},d_{Q_1,Q_2}\right),
\label{Homgt1}    
\eeq
with the same expression as (\ref{holdiff}) for $d_{Q_1,Q_2}$.
A further simplification is possible when one can assign positive R-charges
to all the components of the variable $\phi$.
This is the case in the R\o dland model (as well as
in ``other models'' of Section~\ref{sec:other}):
we can choose $0<\epsilon<1$ in (\ref{Rassignment}).
In such a case, a holomorphic function of a definite R-charge
must be a polynomial function.
Therefore, each entry of a matrix factorization
$Q(\phi)$ is a polynomial of $\phi$.
Also, $\Gamma_{\rm hol}(V,-)$ in (\ref{Homgt1}) can be replaced by the space
$\Gamma_{\rm pol}(V,-)$ of polynomial functions.
Denoting $M_i=\bV_i\otimes_{\C} \C[\phi]$
for the polynomial ring $\C[\phi]$ as in (\ref{MvsV}),
$\Gamma_{\rm pol}(V,\Hom(\bV_1,\bV_2))$ is the space of
homomorphisms of $\C[\phi]$-modules from $M_1$ to $M_2$ 
which is denoted by $\Hom_{\C[\phi]}(M_1,M_2)$.
Thus, we can also say
\beq
\Hom^{}_{D_G(V,W)}((M_1,Q_1),(M_2,Q_2))
\,\,\cong\,\,
{\rm H}\left(\Hom_{\C[\phi]}(M_1,M_2)^{G_\C},d_{Q_1,Q_2}\right).
\label{Homgt2}    
\eeq
The category $D_G(V,W)$ is thus isomorphic to the homotopy category of the
dg-category of $G_{\C}$-equivariant matrix factorizations
whose morphisms are $(\Hom_{\C[\phi]}(M_1,M_2)^{G_{\C}},d_{Q_1,Q_2})$
and the composition is the composition of homomorphisms.

We have seen that the vector multiplet does not participate in the
description of the category.
The gauge theory aspect is present only as the gauge invariance requirement,
just as in the orbifold theory where the gauge group is finite.
This is what we expect in the ultraviolet limit $e\to 0$
where the gauge degees of freedom are frozen.
However, we have arrived at the same result without the $e\to 0$ assumption,
that is, even for $e\ne 0$ where the gauge degrees of freedom are present.
By the scale invariance of the category of branes, this would mean
that the above morphism space must agree with that of
the corresponding branes in the low energy theory.
Can it be correct?
Here the question of regularity, which we had put aside, becomes relevant.

If the FI parameter $\zeta$ is deep inside a weakly coupled phase
where the gauge symmetry is broken to finite subgroups by
the solutions to the D-term equation $\mu(\phi)=\zeta$,
regularity should not be an issue.
In particular any $\brane\in D_G(V,W)$ can be acceptable as a brane datum.
However, (\ref{Homgt0}), (\ref{Homgt1}) or (\ref{Homgt2}) cannot be acceptable
as the space of morphisms:
it may not vanish even if $\brane_1$ or $\brane_2$ is
empty in the low energy theory.
Let $V^{\rm ss}(\zeta)\subset V$ be the set of points whose
$G_{\C}$ orbits go through $\mu^{-1}(\zeta)$ so that
$\mu^{-1}(\zeta)/G$ is isomorphic to $V^{\rm ss}(\zeta)/G_{\C}$.
If the boundary potential $\{Q,Q^{\dag}\}$ of a brane $\brane=(\bV,Q)$
is positive definite on $V^{\rm ss}(\zeta)$, 
the brane must be empty at energies below the Higgs scale $e\sqrt{|\zeta|}$.
In particular the brane $\brane$ must be a zero object in the category
in the sense that $\Hom(\brane,\brane')=\Hom(\brane',\brane)=0$
for any brane $\brane'$.
However, if $\{Q,Q^{\dag}\}$ fails to be positive definite on the complement
$V-V^{\rm ss}(\zeta)$, then $\brane$ may fail to be a zero object
if we take (\ref{Homgt0}), (\ref{Homgt1}) or (\ref{Homgt2}) as the definition
of the morphism space.
The problem is that (\ref{Homgt0})
ignores the gauge symmetry breaking that is responsible for the regularity.
There is a simple solution: replace $V$ by $V^{\rm ss}(\zeta)$
in the formula (\ref{Homgt0}).
Indeed, one can easily find $\brane=(\bV,Q)$ as above
where ${\rm id}_{\bV}$ is $\bartial_{Q,Q}$-exact over $V^{\rm ss}(\zeta)$
while it fails to be so on the entire $V$.
We propose that the space of morphisms at
energies below the Higgs scale is
\beq
{\rm H}\left(\Omega^{0,*}(V^{\rm ss}(\zeta),\Hom(\bV_1,\bV_2))^G,\bartial_{Q_1,Q_2}\right).
\label{Homgtphase}
\eeq
$V^{\rm ss}(\zeta)$ may or may not be Stein.
For example, $(\C^n-\{0\})\times \C^f$ is Stein for $n=1$ but not for $n\geq 2$.
In the Stein case, the morphism space
may be simplified to (\ref{Homgt1}) 
in which $V$ is replaced by $V^{\rm ss}(\zeta)$, or even further to
(\ref{Homgt2}) in which $G$, $\C[\phi]$, $Q_i$ and $M_i$
are replaced by their reduced versions.
The space (\ref{Homgtphase}) may be regarded as the morphism space
in the category $D_G(V^{\rm ss}(\zeta),W)$ of
$G$-equivariant matrix factorizations of $W$ on $V^{\rm ss}(\zeta)$,
or more generally, $G$-equivariant $\bartial$-superconnections
on $V^{\rm ss}(\zeta)$ with curvature $W$.
And the flow from the ultra-violet limit to
energies below the Higgs scale may be identified as the functor
$i_{\zeta}^*:D_G(V,W)\to D_G(V^{\rm ss}(\zeta),W)$ associated to the inclusion
$i_{\zeta}:V^{\rm ss}(\zeta)\hookrightarrow V$.

The grade restriction rule, which is the central interest of the present paper,
is another way to circumvent the regularity issue, by restricting the allowed
set of representations that can be included into the Chan-Paton factor.
In this proposal, we can still use (\ref{Homgt2}) as the space of morphisms
in the category at finite energies.

\medskip
\noindent
{\bf Remark.}
The same can be said for branes in a gauged \emph{nonlinear} sigma model
$(G,X,W)$.
For a pair of $G$-equivariant $\bartial$-superconnections on $X$ with
curvature $W$, the morphism space in the ultra-violet limit $e\to 0$ is
\beq
\Hom^{}_{D_G(X,W)}\left((E_1,{\mathcal Q}_1), (E_2,{\mathcal Q}_2)\right)
\,\,\cong\,\,
{\rm H}\left(\Omega^{0,*}(X,\Hom(E_1,E_2))^G,\bartial_{{\mathcal Q}_1,{\mathcal Q}_2}\right).
\label{HomgaugedNLSM}
\eeq
Deep inside a weakly coupled phase, the morphism space below the Higgs scale
is the one in which $X$ is replaced by the subset $X^{\rm ss}(\zeta)\subset X$
consisting of points whose $G_{\C}$ orbits
go through the solutions to the D-term equation.

\subsection{Mathematical Descriptions}
\label{app:Catmath}

The categories of D-branes which we encountered above are
known or expected to be equivalent to categories introduced in areas of
mathematics.
In many of them, key players are the sheaf of regular functions on
an algebraic variety or of holomorphic functions on a complex mainfold,
i.e. the structure sheaf ${\mathcal O}_X$ of the variety/manifold $X$,
and sheaves of ${\mathcal O}_X$-modules of various kinds.

The most classical one is the derived category of coherent sheaves on $X$,
denoted by $\Db(X)$, defined as the full subcategory of
the bounded derived category \cite{Verdier,Hartshorne,GM} of
the category of sheaves of ${\mathcal O}_X$-modules consisting of
complexes with coherent cohomology sheaves.
For a smooth algebraic variety $X^{\rm alg}$, let $X^{\rm an}$ be the associated
complex manifold.\footnote{We thank Alexey Bondal for
  instruction on what is said in this paragraph.}
When $X^{\rm alg}$ is proper ($X^{\rm an}$ compact), for both
$X=X^{\rm alg}$ and $X^{\rm an}$,
$\Db(X)$ is equivalent to the bounded derived
category of the category of coherent sheaves of ${\mathcal O}_X$-modules.
This can be shown using Lemme 2.1.3 of \cite{Illusie}.
In this case, the categories of coherent sheaves on $X^{\rm an}$ and
on $X^{\rm alg}$ are equivalent. (This is Serre's theorem \cite{SerreGAGA} when
$X^{\rm alg}$ is projective. Otherwise, it reduces to Serre's theorem
using a birational morphism from a projective variety to $X^{\rm alg}$
which exists by Chow's lemma.)
Therefore,
the corresponding derived categories are equivalent,
$\Db(X^{\rm an})\cong \Db(X^{\rm alg})$.
These categories are relevant for the sigma model with vanishing
superpotential $W=0$ and with trivial $U(1)$ action on the target $X$.
Recall that branes in the model are
$\Z$-graded $\bartial$-superconnections $(E,{\mathcal Q})$
with vanishing curvature $\bartial_{\mathcal Q}^2=0$ which form a dg-category.
It was proven in \cite{BondalRosly,Block}
that, when $X$ is a compact complex manifold,
its homotopy category, with the morphism spaces (\ref{morphSC}),
is equivalent to the derived category $\Db(X)$.
Thus, the category of B-branes in the sigma model with a compact target space
$X$ is equivalent to $\Db(X)$
and hence also to $\Db(X^{\rm alg})$ when $X$ is algebraic.

Matrix factorizations, relevant for models with non-vanishing superpotential,
also have a relatively long history of study, starting with \cite{Eisenbud}.
The structure of their categories was studied
in \cite{Buchweitz} and revisited more recently in \cite{OrlovLG}.
These are relevant in models 
where $X$ is an affine space and $W$ is a polynomial function,
not necessarily with $U(1)$ R-symmetry so that the category
is $\Z_2$-graded.
Morphism spaces in such a category are of the form (\ref{Homgt2})
where we set $G=\{1\}$.
More recently, a formulation of the category has been developed for
a more general $(X,W)$ without or with $U(1)$ R-symmetry
\cite{Positselski,LinPomerleano,Orlov11,EfimovPositselski}\cite{Shipman}.
Let us consider the case with $U(1)$ R-symmetry, i.e.,
a $\C^{\times}$ action on $X$ under which $W$ has weight $2$.
A \emph{curved dg-sheaf} on $(X,W)$ is a $\C^{\times}$-equivariant
sheaf ${\mathcal E}$ of ${\mathcal O}_X$-modules equipped with an
endomorphism $d_{\mathcal E}$ of $\C^{\times}$-weight $1$ that squares to
$W\cdot {\rm id}_{\mathcal E}$.
It is of course a $\Z$-graded matrix factorization of $W$ on $X$
when ${\mathcal E}$ is locally free of finite rank.
Note that $d_{\mathcal E}=0$ is allowed when ${\mathcal E}$ is supported
in $W^{-1}(0)$.
When $(X,W)$ is algebraic, that is,
when $X$ is an algebraic variety and $W$ is a regular function thereon,
the formulation goes as follows.
First, a definition of the derived category of quasi-coherent
curved dg-sheaves is given,
and then $\Db(X,W)$ is defined as its full subcategory consisting of
objects isomorphic to matrix factorizations.
When $W=0$ and the $\C^{\times}$ action on $X$ is trivial,
this category $\Db(X,0)$ is equivalent to the derived category $\Db(X)$.
The spaces of morphisms in $\Db(X,W)$ are shown in \cite{LinPomerleano,Shipman}
to be isomorphic to the dimension 0 cohomology (\ref{morphSC}).
It would be an interesting problem to extend this definition
to the case when $(X,W)$ is complex analytic, so that
$\Db(X,0)$ with the trivial $\C^{\times}$ action is equivalent to $\Db(X)$
and $\Db(X^{\rm an},W)\cong \Db(X^{\rm alg},W)$ when an analytic $(X^{\rm an},W)$
is associated to an algebraic $(X^{\rm alg},W)$,
and to extend \cite{BondalRosly,Block} to prove
that $\Db(X,W)$ is equivalent to the homotopy category of
the dg-category of $\bartial$-superconnections with curvature $W$,
which is the category of B-branes in the sigma model $(X,W)$.

Let us consider the setup of gauged (non-)linear sigma model $(G,X,W)$.
Then, the complex reductive group $G_{\C}$ acts holomorphically
on a complex manifold $X$
or algebraically on an algebraic variety $X$, leaves $W$ invariant,
and commutes with the $\C^{\times}$-action.
Then, we can consider $G_{\C}$-equivariant curved dg-sheaves on $(X,W)$.
For a pair of such objects, there is an induced $G_{\C}$-action on
the maps between them and only $G_{\C}$-invariants are taken as morphisms.
When $W=0$ (with the trivial $\C^{\times}$-action on $X$)
or when $X$ is algebraic,
the defintion of $\Db(X,W)$ can be extended to define the corresponding
category $\Db_{G_{\C}}(X,W)$ of $G_{\C}$-equivariant curved dg-sheaves.
In the literature, it is also denoted as $\Db([X/G_{\C}],W_{\rm ind})$
where $[X/G_{\C}]$ is a kind of quotient of $X$ by $G_{\C}$
called the ``quotient stack'' and $W_{\rm ind}$ is a function induced from $W$.
In the present paper, ``quotient stacks'' are used only as the notation
to label $G_{\C}$-equivariant things on $X$. For example, the category of
$G_{\C}$-equivariant coherent sheaves on $X$ is denoted by
${\it Coh}([X/G_{\C}])$.
For a vector space $V$ on which $G_{\C}$ acts linearly,
the morphism space in $\Db_{G_{\C}}(V,W)$ for a pair of matrix factorizations is
the same as (\ref{Homgt2}).
In general, we expect that the category
$\Db_{G_{\C}}(X,W)$ is equivalent to the category $D_G(X,W)$
of the ultra-violet limit $e\to 0$ of the gauge theory $(G,X,W)$,
with the morphism space (\ref{HomgaugedNLSM}), that is,
the  homotopy category of the dg-category of
$G$-equivariant $\bartial$-superconnections on $X$ with curvature $W$.
It would be an interesting problem to prove this
by extending the works of \cite{BondalRosly,Block} and
\cite{LinPomerleano,Shipman}.

\noindent
{\bf Remark.}
When $X$ is a smooth algebraic variety, any object of $\Db(X)$ is isomorphic
to a bounded complex of finite rank locally free sheaves.
(The corresponding statement does not necessarily hold for complex manifolds.)
This is the reason for $\Db(X,0)\cong \Db(X)$ in the current definition
of $\Db(X,W)$.
Therefore, in the sigma model with an algebraic target space
and vanishing superpotential,
any B-brane is isomorphic to a B-brane with first order superconnection.
We expect that the same holds also when the superpotential is nonzero
and/or a non-trivial gauge symmetry is present, as long as they are algebraic.
This is the reason why we do not consider higher order superconnections
for branes in gauged linear sigma models.
The picture we are obtaining, such as (\ref{HHPdiagram})
and (\ref{maindiagram}), strongly supports this expectation.

\noindent
{\bf Remark.} The identification of $\Db(X)$ as the category of B-branes in
the sigma model with target $X$ was first suggested by Kontsevich in
\cite{Kontsevich-HMS}.
The relation was further discussed in
\cite{Sharpe,Douglas,Aspinwall-Douglas,Aspinwall-Lawrence,Diaconescu,Lazaroiu}
for example.
The decisive argument for this identification, detailed above,
was originally presented in \cite{HHP} based on \cite{BondalRosly,Block}
(and later revisited in \cite{Bergman,KapRozSau}).
The new contribution of the present Appendix
is the use of the operators $\bfD$ and $\bfS$ to show (\ref{morphSC})
which was previously argued using the decoupling of K\"ahler moduli.
Use of matrix factorizations to describe B-branes in models
with non-zero superpotential $W$ was suggested also by Kontsevich
according to \cite{KapustinLi}.
Earlier than that,
D-branes wrapped on complex submanifolds in level sets $W^{-1}(c)$
were considered \cite{HIV,Hblin}.
In the above mathematical language,
they are curved dg-sheaves ${\mathcal E}$ with $d_{\mathcal E}=0$
supported in $W^{-1}(c)$.

The dg-models for the categories of branes for the class of theories $\Th$
which we have described in Appendix~\ref{app:CatNLSM} and \ref{app:CatGT}
have cones for closed morphisms as described in Section~\ref{subsub:HA}.
This means that these dg-models are \emph{pretriangulated categories}
as introduced in \cite{BK1991}
and are \emph{dg-enhancements} of their homotopy categories
$D(\Th)$ which have the structure of triangulated cateories.
For a proper algebraic variety $X^{\rm alg}$,
the category of superconnections on its analytification $X^{\rm an}$
is equivalent to a class of dg-enhancements of $\Db(X^{\rm alg})$
called the \emph{standard dg-enhancements}.
Any functor between such dg-models that respects the (B-type)
supersymmetry and the $U(1)$ (vector) R-symmetry,
such as the RG flow from gauge theory
to the low energy sigma model, is a dg-functor by definition.
And it is expected to be \emph{pre-exact} \cite{BK1991},
that is, it maps the cone of a closed morphism to the cone
of the image morphism.

\section{Some Facts On Weyl (Anti-)Invariants}
\label{app:Weyl}

Let $G$ be a compact Lie group, $T\subset G$ a maximal torus,
$\ell=\dim T$ the rank of $G$, and $N_T\subset G$ the normalizer of $T$.
The Weyl group $W=N_T/T$ of $G$ acts linearly
on $\ttt={\rm Lie}(T)$ and hence on $\ttt_{\C}=\ttt\otimes_{\R}\C$,
$\im \ttt$, and their duals. Let $W_0=N_T^0/T\subset W$ be the Weyl group
of the identity component $G_0$ of $G$.
A root of ${\rm Lie}(G)$ is regarded as a linear function on $\im\ttt$, i.e.
an element of $\im\ttt^*$.
$W_0$ is generated by reflections with respect to the roots. 
A choice of Weyl chamber in $\im\ttt$
determines the notion of positivity of roots.
We say that a function $f$ on $\im\ttt$ is $W_0$-{odd}
when it is anti-invariant, $f\circ w=(-1)^{\ell(w)}f$, under all $w\in W_0$.
We quote some basic facts about the polynomial functions on
$\im\ttt$ which are $W_0$-invariant or anti-invariant.
See, for example, Chapter 3 of \cite{Humphreys} for a simple proof.
\begin{enumerate}
\item[\!\!(i)]
The product of positive roots $\prod_{\alpha>0}\alpha$ is $W_0$-odd,
and any $W_0$-odd polynomial is $\prod_{\alpha>0}\alpha$ times a
$W_0$-invariant.\\[-0.3cm]
\item[\!\!(ii)]
The algebra of $W_0$-invariant polynomials
has a set of homogeneous and algebraically independent generators
$f_1,\ldots, f_{\ell}$ (the basic Casimir invariants).
The set of their degrees is unique.\\[-0.3cm]
\item[\!\!(iii)]
For affine coordinates $x_1,\ldots, x_{\ell}$ of $\im\ttt$,
the determinant of the Jacobi matrix $\partial f_i/\partial x_j$
is equal to $\prod_{\alpha>0}\alpha$ up to
a scalar multiplication.
\end{enumerate}
In particular, the measure $\prod_{\alpha>0}\alpha(\usigma)\,
\dd^{\ell}\usigma$ of the hemisphere partition function can be written
in terms of the basic Casimir invariants as
\beq
\prod_{\alpha>0}\alpha(\usigma)\,
\dd^{\ell}\usigma
~\propto~
\dd f_1(\usigma)\wedge\cdots\wedge \dd f_{\ell}(\usigma).
\eeq

The fact (i) has the following consequence.
Let $f(x,n)$ be a function of $(x,n)\in \ttt_{\C}\times \ttt_{\C}$
which is invariant under the diagonal $W_0$-action, $f(wx,wn)=f(x,n)$,
$w\in W_0$.
Then
$\Phi(x,n):=\sum_{w\in W_0}f(x,wn)\prod_{\alpha>0}\alpha(x+wn)$
is $W_0$-odd in the $x$-variable,
\beqa
\Phi(wx,n)&=&\sum_{w'\in W_0}f(wx,w'n)\prod_{\alpha}\alpha(wx+w'n)
\nn\\
&=&\sum_{w'\in W_0}f(x,w^{-1}w'n)\prod_{\alpha}\alpha(w(x+w^{-1}w'n))
\nn\\
&=&\sum_{w''\in W_0}f(x,w''n)(-1)^{\ell(w)}\prod_{\alpha>0}\alpha(x+w''n)
=(-1)^{\ell(w)}\Phi(x,n).
\eeqa
If $f(x,wn)$ is a polynomial in
$x$ for an element $n\in \ttt_{\C}$ and for each $w\in W_0$, 
then $\Phi(x,n)$ is a polynomial in $x$ and hence is divisible by
$\prod_{\alpha>0}\alpha(x)$.
In this statement, ``polynomial'' can be replaced by ``power series''
by looking at terms of each degree.

This observation can be used to clarify a point in the
large volume expression of
the hemisphere partition function given in \cite{HR}.
It involves a sum
\beq
\sum_{n\in {\rm Q}^{\vee}_+}\underbrace{\e^{-t'_R(n)}
{\prod_{\beta}\Gamma(1+d_{\beta}(x+n))\over
  \prod_i\Gamma(1+Q_i(x+n))}}_{\displaystyle ~~~~~~~~~~~~~  =:\psi(x,n)}
\cdot\underbrace{
\prod_{\alpha>0}\Gamma(1+\alpha(x+n))
\Gamma(1-\alpha(x+n))
}_{\displaystyle ~~~~~~~~~~~~~  =:\varphi(n,x)}
\label{sumQ}
\eeq
where ${\rm Q}^{\vee}_+$ is a subset of the coroot lattice of $T$,
$t'_R\in \ttt_{\C}^*$ is a Weyl invariant element
and $\{d_{\beta}\}$ and $\{Q_i\}$ are Weyl invariant sets of weights.
It is required that (\ref{sumQ}) is a power series in $x$
so that a cohomology class can be inserted in the place of $x$.
It is supposed that $d_{\beta}(n)$ and $Q_i(n)$  for
$n\in {\rm Q}_+^{\vee}$ are all non-negative integers, and the
factor $\psi(x,n)$ is indeed a power series in $x$.
However, the factor $\varphi(x,n)$ is singular at $x=0$
when there is a positive root $\alpha$ such that
$\alpha(n)$ is nonzero. Indeed, using a well-known Gamma function identity,
it can be written as
\beq
\varphi(x,n)=\prod_{\alpha>0}{\pi\alpha\left(x+n\right)
  \over
  \sin\left(\pi\alpha(x+n)\right)}
=\prod_{\alpha>0}(-1)^{\alpha(n)}{\pi\alpha(x+n)
  \over
  \sin(\pi\alpha(x))}.
\eeq
The denominator $\sin(\pi\alpha(x))$ causes trouble if $\alpha(n)\ne 0$.
A solution to this problem is that
$\varphi(x,n)$ appears under the sum (\ref{sumQ}).
We suppose that the subset ${\rm Q}^{\vee}_+$ is Weyl invariant.
Then, the sum over $n\in {\rm Q}^{\vee}_+$ factors through the sum over
the Weyl orbits. Since $\psi(x,n)\prod_{\alpha>0}(-1)^{\alpha(n)}$
is invariant under the diagonal $W_0$-action $(x,n)\mapsto (wx,wn)$,
the sum
$\sum_{n\in {\rm Q}^{\vee}_+}\psi(x,n)\prod_{\alpha>0}(-1)^{\alpha(n)}\pi\alpha(x+n)$
is divisible by $\prod_{\alpha>0}\alpha(x)$ and hence by the problematic
denominator $\prod_{\alpha>0}\sin(\pi\alpha(x))$.\footnote{We thank
  Yoshiki Oshima for a discussion (September 2014)
  which eventually led us to this solution.}

This clarification is reflected in
the presentation (\ref{LVZ})-(\ref{whGamma12}).
Indeed, the function inside the bracket in (\ref{whGamma12})
is a power series in $x_1$ and $x_2$ after the sum over $(n_1,n_2)$,
and hence the replacement $x_1\to {H_1\over 2\pi\im}$,
$x_2\to {H_2\over 2\pi\im}$ makes sense.
Another minor correction of \cite{HR} is that
the constant $C$ that relates the hemisphere partition function and
the sigma model central charge in a weakly coupled geometric phase is
\beq
C=(-1)^{d_F}(-2\pi\im)^{d_E}(4\pi)^{-|\Delta_+|}\im^{d_G}\cdot |W|,
\eeq
where $d_G$ is the dimension of the gauge group,
$d_E$ is the number of fields for the ambient space,
and $d_F$ is the number of equations.
In the R\o dland model, $d_F=7$, $d_E=14$, $|\Delta_+|=1$,
$d_G=4$, $|W|=2$, and hence
\beq
C=(-1)^7(-2\pi\im)^{14}(4\pi)^{-1}\im^4\cdot 2=(2\pi)^{13}.
\eeq
In the Hosono-Takagi model, $d_F=5$, $d_E=10$, $|\Delta_+|=0$,
$d_G=2$, $|W|=2$, and hence
\beq
C=(-1)^5(-2\pi\im)^{10}\im^2\cdot 2=-2(2\pi)^{10}.
\eeq
These are the constants that appear in (\ref{LVZ})
and the upper case of (\ref{ZD2HT}) respectively.

\section{Representations Of A Compact Disconnected Lie Group}
\label{app:WCF}

\newcommand{\rmI}{{\rm I}}

Let $G$ be a compact Lie group which is not connected,
with the connected component decomposition
$G=\coprod_{i\in \rmI}G_i$, with $G_0$ being the identity component.
Let $T$ be a maximal torus of $G_0$, $N_T^0\subset G_0$
the normalizer of $T$, and $W_0=N_T^0/T$ the Weyl group of $G_0$.
For any element $g\in G$, $gTg^{-1}$ is also a maximal torus of
$G_0$ and hence there is $h_0\in G_0$ such that $h_0gTg^{-1}h_0^{-1}=T$.
This shows that one can find $g_i\in G_i$ for each $i\in \rmI$
so that the normalizer of $T$ in $G_i$ is $N_T^0g_i$. (We may set $g_0=1$.)
In particular, the normalizer of $T$ in $G$ is $N_T=\coprod_{i\in \rmI}N_T^0g_i$,
and the Weyl group $W=N_T/T$ of $G$ can be decomposed as a set as
\beq
W=\bigcup_{i\in \rmI}W_0w_i,
\eeq
where $w_i$ is represented by $g_i$. It fits into
$1\to W_0\to W\to G/G_0\to 1$ as a group.

Let $\rmV$ be a finite dimensional irreducible representation of $G$.
Let $\rmV_0\subset \rmV$ be one of the irreducible components with
respect to $G_0$. Then, each $g_i\rmV_0$ is also an irreducible representation
of $G_0$. The sum $\sum_{i\in\rmI}g_i\rmV_0$ is $G$-invariant and hence
must be $\rmV$ itself.
Since $g_i\rmV_0\cap g_j\rmV_0$ is $G_0$-invariant, either
it is zero or $g_i\rmV_0=g_j\rmV_0$.
Thus, $\rmV$ can be decomposed as
\beq
\rmV=\bigoplus_{i\in {\rm I}'}g_i\rmV_0,
\eeq
for a section ${\rm I}'\subset {\rm I}$ of a quotient $\rmI\to\rmI/\sim$.
Note that
\beq
{\rm tr}^{}_{g_iV_0}\e^{2\pi\usigma}
={\rm tr}^{}_{V_0}(g_i^{-1}\e^{2\pi\usigma}g_i)
={\rm tr}^{}_{V_0}\e^{2\pi w_i^{-1}\usigma}
\eeq
for $\usigma\in \ttt_{\C}$. If $\rmV_0$ is the representation of highest weight
$\lambda$, then the character of $\rmV$ is
\beq
{\rm tr}^{}_V\e^{2\pi\usigma}
=\sum_{i\in {\rm I}'}
{\displaystyle\sum_{w_0\in W_0}(-1)^{\ell(w_0)}
\e^{2\pi w_0(\lambda+\rho)
(w_i^{-1}\usigma)}\over
\displaystyle \prod_{\alpha>0}\left(\e^{\pi\alpha(w_i^{-1}\usigma)}
-\e^{-\pi\alpha(w_i^{-1}\usigma)}\right)}.
\eeq

Note that an automorphism of $\mathfrak{g}$ that stabilizes $\ttt$,
such as ${\rm Ad}_{g_i}$, induces a permutation of the roots.
Since $\alpha(\usigma)\sinh(\pi\alpha(\usigma))$ is invariant
under $\alpha\to -\alpha$, we find that 
$\prod_{\alpha>0}\alpha(\usigma)\sinh(\pi\alpha(\usigma))$ is invariant under
any of the $w_i$'s.
Therefore,
\beqa
\lefteqn{\prod_{\alpha>0}\alpha(\usigma)\sinh(\pi\alpha(\usigma))
\cdot{\rm tr}^{}_V\e^{2\pi\usigma}}\nn\\
&=&\sum_{i\in {\rm I}'}
\prod_{\alpha>0}\alpha(w_i^{-1}\usigma)\sinh(\pi\alpha(w_i^{-1}\usigma))
{\displaystyle\sum_{w_0\in W_0}(-1)^{\ell(w_0)}
\e^{2\pi w_0(\lambda+\rho)
(w_i^{-1}\usigma)}\over
\displaystyle \prod_{\alpha>0}\left(\e^{\pi\alpha(w_i^{-1}\usigma)}
-\e^{-\pi\alpha(w_i^{-1}\usigma)}\right)}\nn\\
&=&\sum_{i\in {\rm I}'}
\prod_{\alpha>0}{\alpha(w_i^{-1}\usigma)\over 2}
\sum_{w_0\in W_0}(-1)^{\ell(w_0)}
\e^{2\pi w_0(\lambda+\rho)
(w_i^{-1}\usigma)}\nn\\
&=&\sum_{i\in {\rm I}'}\sum_{w_0\in W_0}
\prod_{\alpha>0}{w_0\alpha(w_i^{-1}\usigma)\over 2}
\e^{2\pi w_0(\lambda+\rho)
(w_i^{-1}\usigma)}\nn\\
&=&\sum_{i\in {\rm I}'}\sum_{w_0\in W_0}
\prod_{\alpha>0}{\alpha(w_0^{-1}w_i^{-1}\usigma)\over 2}
\e^{2\pi (\lambda+\rho)
(w_0^{-1}w_i^{-1}\usigma)}.
\eeqa
It then follows that the hemisphere integral (\ref{ZVdef})
can be expressed as (\ref{ZV}), where
``Const'' is  $|{\rm I}'||W_0|/2^{|\Delta_+|}$ 
and $\lambda$ is the highest weight of $\rmV_0$.

\section{Outline Of Numerical Analysis Of Contours}
\label{app:numerical}

In this Appendix, we present an outline of the numerical analysis of
convergent contours which is used in Section~\ref{subsec:numerical}.
Of central interest is the region on the three-dimensional sphere $S^3_R$
given in (\ref{S3def}) where the growth function $\AN_{\wtq}$ is positive.
For the purpose of visualization, we consider
a family of two-dimensional slices of $S^3_R$
by the level sets of $\Is_1+\Is_2$, and plot
the positive region on each slice using \texttt{RegionPlot} of Mathematica.
We introduce the coordinates
$\Rs_{\pm}=(\Rs_1\pm\Rs_2)/\sqrt{2}$ and
$\Is_{\pm}=(\Is_1\pm\Is_2)/\sqrt{2}$,
and refer to the value of $\Is_+$ the {\it level}.

The slice at level $\Is_+\in (-R,R)$ is a two-dimensional sphere
of radius $\sqrt{R^2-\Is_+^2}$.
Fig.~\ref{fig:slices} shows the slices at levels $0.6R,\, 0,-0.6R$
from left to right.
\begin{figure}[thb]
  \centerline{\includegraphics[page=58]{main-pics.pdf}~~~~
    \includegraphics[page=59]{main-pics.pdf}~~~~
    \includegraphics[page=60]{main-pics.pdf}}
\caption{}  \label{fig:slices}
\end{figure}
In each of them, the left and the right components are
the projections to the $(\Rs_1,\Rs_2)$-plane of
the hemispheres with $\Is_-\geq 0$ and $\Is_-\leq 0$ respectively,
which we refer to as northern and southern.
The walls $W_1$, $W_2$ and $W_p$ are shown as the vertical, horizontal and
$-45^{\circ}$ lines.
At level $0<\Is_+<R/\sqrt{2}\simeq 0.707R$ (e.g. the left figure),
the complement of the walls is connected.
At level $R/\sqrt{2}< \Is_+ <R$
(not shown in Fig.~\ref{fig:slices} but see e.g. Fig.~\ref{fig:Cplus} below),
the walls $W_1$ and $W_2$ separate the slice into four components.
At level $-R<\Is_+<0$ (e.g. the right figure),
the wall $W_p$ separates the slice into two components.
The boundary of $W_p$ is at level 0 and is shown
as the dashed line in the middle figure.

\begin{figure}[thb]
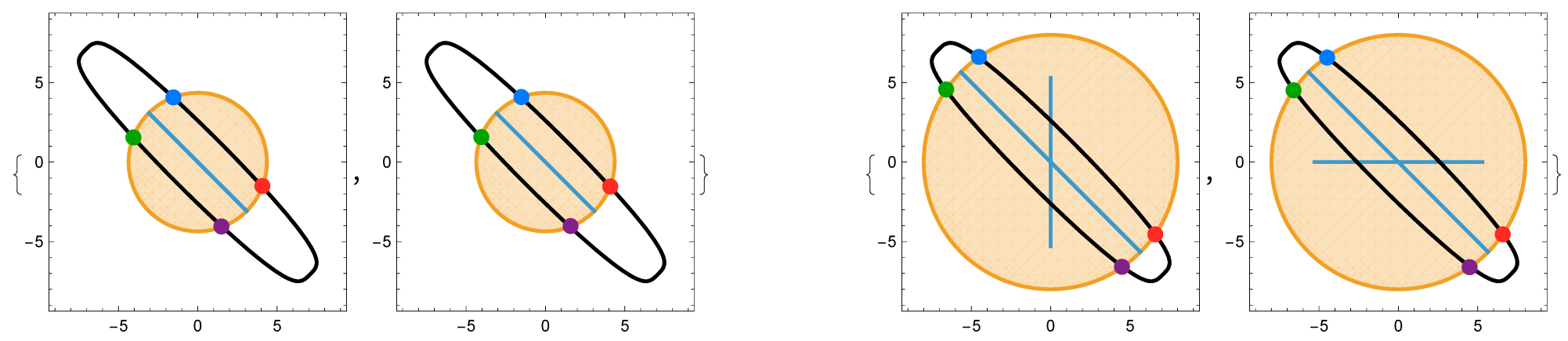

  \centerline{\includegraphics[page=64]{main-pics.pdf}~~~~~~~~~
    \includegraphics[page=65]{main-pics.pdf}}
  \caption{The slices at levels $-0.9R$ and $-0.6R$ (for $R=10$).
    The circle is the projection of $C_-$ to the $\Rs$-plane.
    Its intersection with the equator is where $C_-$ and the slice meet.
  }  \label{fig:Cminus}
\end{figure}

\begin{figure}[thb]
  \centerline{\includegraphics[page=66]{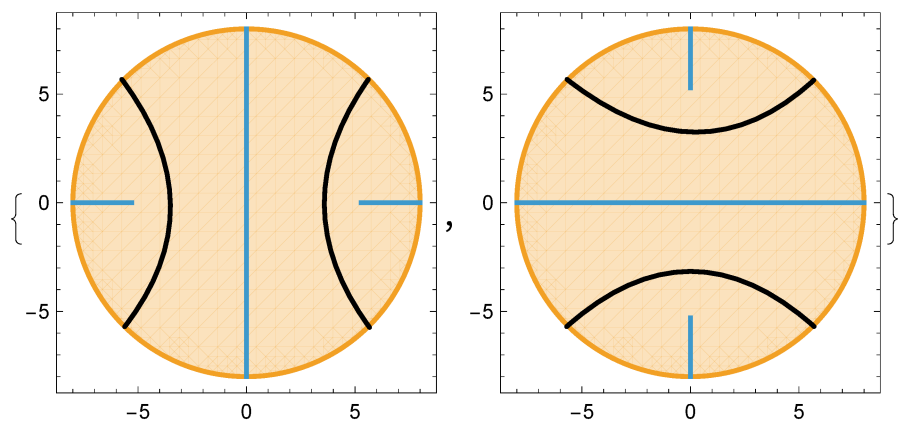}}
  \caption{The level $0.6R$ slice and a circle on it.} \label{fig:Sphe6wC}
\end{figure}

The intersection $C_{\rm real}$ of the real locus $\im\ttt$ and $S^3_R$
is the equator of the level 0 slice.
The intersection $C_0=\bgamma_0\cap S^3_R$
is also on the level 0 slice and is depicted in Fig.~\ref{fig:GG0}
for $R=10$ (left) and $R=100$ (right).
\begin{figure}[thb]
  \centerline{\includegraphics[page=61]{main-pics.pdf}~~~~~~~~~
    \includegraphics[page=62]{main-pics.pdf}}
\caption{}  \label{fig:GG0}
\end{figure}
In the limit $R\to\infty$ it tends to make right angle turns at the north
and south poles and becomes the union of four meridians which
are equally separated in longitude.
The intersection $C_{\pm}=\bgamma_{\pm}\cap S^3_R$ is not on a single slice.
The level of $C_+$ ranges over an interval $[a_+,b_+]$ with $0<a_+<b_+<R$,
where $a_+\nearrow R/\sqrt{2}$ and $b_+\nearrow R$ as $R\to \infty$,
while the level of $C_-$ ranges over an interval $[a_-,0]$
where $a_-\searrow -R$ as $R\to\infty$.
\footnote{The bounds are
$a_+=\sqrt{{R^2\over 2}+{1\over 8}}-{1\over 2\sqrt{2}}$,
  $b_+=\sqrt{R^2+{1\over 2}}-{1\over \sqrt{2}}$ 
  and
$a_-=-\sqrt{R^2-{1\over 32}}+{1\over 4\sqrt{2}}$.}
$C_+$ intersects the slice in the range $a_+<\Is_+<b_+$
at eight points, four each in the northern and southern hemispheres,
as shown in colors in Fig.~\ref{fig:Cplus}.
\begin{figure}[thb]
  \centerline{\includegraphics[page=63]{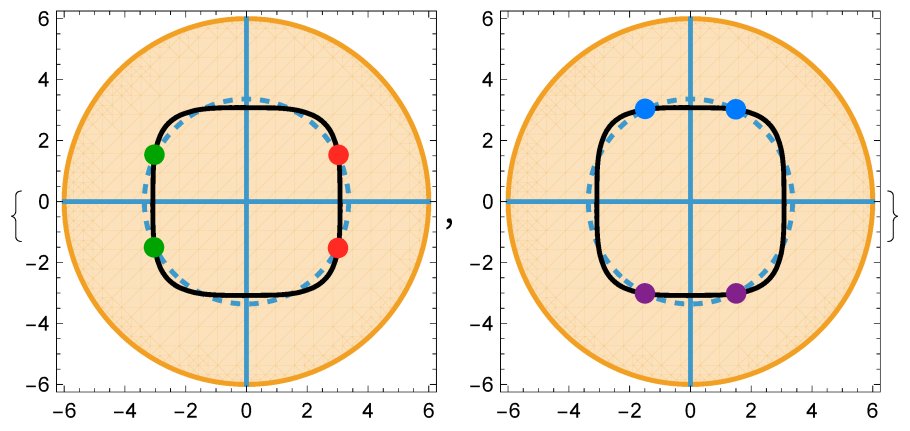}}
  \caption{The level $0.8R$ slice (for $R=10$). The circles
    are the projections to the $\Rs$-plane of $C_+$ (black)
    and the intersection of $\bgamma_+$ and the level set (dashed).
    The intersection of the two is where $C_+$ and the slice meet.}
  \label{fig:Cplus}
\end{figure}
As $\Is_+\searrow a_+$, pairs of points of the same colors
coalesce on the $\Rs_1$ or $\Rs_2$ axis
(the walls $W_2$ and $W_1$ open for $C_+$ exactly at the points of coalescence).
As $\Is_+\nearrow b_+$, pairs of points coalesce on the equator
(red-blue and green-purple on the 45${}^{\circ}$ line
and red-purple and green-blue on the $-45{}^{\circ}$ line).
The circle $C_-$ intersects the slice in the range $a_-<\Is_+<0$
at four points on the equator,
as shown in colors in Fig.~\ref{fig:Cminus}.
$C_-$ can be described as the motion of these points as the level
is varied.
As $\Is_+\searrow a_-$, pairs of points (red-blue and green-purple)
coalesce on the 45$^{\circ}$ line.
As $\Is_+\nearrow 0$, different pairs of points (red-purple and green-blue)
coalesce on the $-45^{\circ}$ line.
Note that the complement of the walls in the slice at $0<\Is_+<R/\sqrt{2}$
retracts to a circle as shown in Fig.~\ref{fig:Sphe6wC}.
Lifted to higher levels or pushed down to level $0$,
this circle can be deformed to $C_+$ or to $C_0$.
And $C_0$ can be flattened inside the level 0 slice to $C_{\rm real}$
which in turn can be pushed down in levels to $C_-$.
This visualizes the fact that all these circles are isotopic to one another
in the complement of the walls in $S^3_R$.

\subsection{Transport Along The Curved Paths}
\label{subapp:curved}

Let us now examine whether there is a family of circles along a curved path,
interpolating between $C_+$ at $\zeta\gg 0$ and $C_-$ at $\zeta\ll 0$,
on which the function $\AN_{\wtq}$ is always positive,
\begin{figure}[htb]
  \psfrag{z}{\scriptsize $\zeta$}
  \psfrag{th}{\scriptsize $\theta$}
  \psfrag{a1}{\scriptsize $t_1$}
  \psfrag{a3}{\scriptsize $t_3$}
  \psfrag{M}{\scriptsize $\mathtt{M}_0$}
  \psfrag{R}{\scriptsize \texttt{R}}
  \psfrag{R1}{\scriptsize $\mathtt{R}_{\zeta_1}$}
  \psfrag{L}{\scriptsize \texttt{L}}
  \psfrag{L3}{\scriptsize $\mathtt{L}_{\zeta_3}$}
  \psfrag{-7Pi}{\scriptsize $-7\pi$}
  \vspace{-0.2cm}
  \centerline{\includegraphics[page=67]{main-pics.pdf}}
  \caption{}\label{fig:RodpathC}
\end{figure}
for the shifted highest weights $\wtq=\wtq_{0,0}$, $\wtq_{0,7}$,
$\wtq_{2,-1}$ and $\wtq_{2,6}$ for the representations
in question: $\C(0)$, $\C(7)$, $\Sym^2\fund(-1)$ and $\Sym^2\fund(7)$.
We first consider the window {\bf C}.
To be concrete, we consider the path as depicted in Fig.~\ref{fig:RodpathC}.
It is a broken line with three segments and is specified by
$\Delta_*\in (0,\pi)$:
the right segment is at $\theta=-7\pi+\Delta_*$ with $\zeta\geq 0$,
the middle segment is at $\zeta=0$ with $-\Delta_*<\theta+7\pi<\Delta_*$,
and the left segment is at $\theta=-7\pi-\Delta_*$ with $\zeta\leq 0$.
We refer to the point on the right ({\it resp}. left) segment with vertical
position $\zeta$ as $\mathtt{R}_{\zeta}$ ({\it resp}. $\mathtt{L}_{\zeta}$) 
and to the point on the middle segment with $\theta=-7\pi+\sDelta\theta$
as $\mathtt{M}_{\sDelta\theta}$.
For example, the two corners are $\mathtt{R}_0=\mathtt{M}_{\Delta_*}$
and $\mathtt{L}_0=\mathtt{M}_{-\Delta_*}$.
Three typical points are shown in the figure.
In the analysis below, we show the results mostly for $\Delta_*=0.1\pi$.

Let us first consider the representation $\C(0)$.
Recall that $\AN_{\wtq_{0,0}}$ is positive on the contour $\bgamma_0$
everywhere on the strip $-7\pi<\theta<-6\pi$.
Thus, after deforming
$\bgamma_+$ to $\bgamma_0$ while in $\zeta\gg 0$ we can continue to take
$\bgamma_0$ just before the point $\mathtt{M}_0$ at the boundary of the strip.
Recall also from Section~\ref{subsec:SU2GRRRodland}
that $\AN_{\wtq_{0,i}}$ (for any $i$) is positive on the contour $\bgamma_-$
as long as $\zeta<f_{0\,{\rm min}}=\zeta_1$.
We see that the valid regions of $\bgamma_0$ and $\bgamma_-$ for
the representation $\C(0)$ have an overlap along the path.
Thus, we aim at connecting the two contours inside the 
$\AN_{\wtq_{0,0}}>0$ region along the overlap. In fact, we can do it
at one point, say, at the corner point $\mathtt{R}_0$.
The $\AN_{\wt{q}_{0,0}}>0$ region in three slices at $\mathtt{R}_0$
are shown as shaded in Fig.~\ref{fig:00R0contours},
\begin{figure}[thb]
  \centerline{\includegraphics[page=68]{main-pics.pdf}~~~~
    \includegraphics[page=69]{main-pics.pdf}~~~~
    \includegraphics[page=70]{main-pics.pdf}}
\caption{}  \label{fig:00R0contours}
\end{figure}
with the levels $0$, $-0.3R$, $-0.6R$ from left to right.
In the figure for the level $0$ slice, the circle $C_0$ is also included.
We see that $C_0$ is indeed inside the shaded region.
In the figures for the other slices, parts of the
deformations of $C_0$ to those levels are included. 
As the circle must avoid the walls, these parts end before reaching
the wall $W_p$. The end points, as shown in colors, continue to higher levels
and meet pairwise at level $0$ where the wall $W_p$ disappears.
It is clear that the circle can be further deformed to $C_-$
inside the positive region.
Altogether, we have found a family of circles along the path
that stays in the $\AN_{\wtq_{0,0}}>0$ region,
interpolating between $C_+$ at $\zeta\gg 0$ and $C_-$ at $\zeta\ll 0$.
Furthermore, each circle can be chosen to be Weyl invariant
as will be explained in Appendix~\ref{subapp:Weylinv}.

Let us next consider the representation $\Sym^2\fund(6)$.
Recall that $\AN_{\wtq_{2,6}}$ is positive on the contour $\bgamma_0$
everywhere on the strip $-8\pi<\theta<-7\pi$.
Thus, we move backward on the path, starting from $\zeta\ll 0$.
After deforming $\bgamma_-$ to $\bgamma_0$ while in $\zeta\ll 0$,
we can continue to take $\bgamma_0$ just before the point $\mathtt{M}_0$
at the boundary of the strip.
The $\AN_{\wt{q}_{2,6}}>0$ region at $\mathtt{M}_0$ in the level 0 slice is
shown as shaded in Fig.~\ref{fig:26atMa},
with an enlarged view in Fig.~\ref{fig:26atMb}-left.
\begin{figure}[thb]
  \begin{subfigure}[b]{.44\linewidth}
    \centering
    \includegraphics[page=71]{main-pics.pdf}
    \caption{}\label{fig:26atMa}
  \end{subfigure}
  \begin{subfigure}[b]{.44\linewidth}
    \centering
    \includegraphics[page=72]{main-pics.pdf}
    \caption{}\label{fig:26atMb}
  \end{subfigure}
    \vspace{-0.4cm}
  \caption{}  \label{fig:26atM}
\end{figure}
The circle $C_0$ in the limit $R\to \infty$ is indeed at the edge
of the region. It goes through the vertex of a checkerboard-like pattern.
The pattern changes as the level is moved.
The shaded region broadens if the level is raised
(Fig.~\ref{fig:26atMb}-middle for level $0.001R$)
and is detached if the level is lowered
(Fig.~\ref{fig:26atMb}-right for level $-0.001R$).
Thus, before reaching $\mathtt{M}_0$ from the left, we can lift a
part of $C_0$ to positive level, so that the resulting circle $C$ can
stay in the positive region as the point moves past $\mathtt{M}_0$.
We may even lift the entire circle so that $C$ is in a single slice.
If the level is high enough, the circle $C$ remains good
as the point reaches the corner $\mathtt{R}_0$ and
gets even better at $\texttt{R}_{\zeta}$ with higher and higher $\zeta$.
Fig.~\ref{fig:26contours} shows
the level $0.1R$ slices at $\mathtt{M}_0$, $\mathtt{R}_0$ and
$\mathtt{R}_{15}$ from left to right,
\begin{figure}[thb]
  \centerline{\includegraphics[page=73]{main-pics.pdf}~~~~
    \includegraphics[page=74]{main-pics.pdf}~~~~
    \includegraphics[page=75]{main-pics.pdf}}
\caption{}  \label{fig:26contours}
\end{figure}
along with examples of the circle $C$.
Thus, at the points beyond $\texttt{M}_0$ in the path,
we can find a circle $C$ in $S^3_R$ inside the $\AN_{\wtq_{0,0}}>0$ region.
While in $\zeta\gg 0$, the circle can be deformed to $C_+$ inside the
$\AN_{\wtq_{2,6}}>0$ region.
We have found a family of circles that stays
in the $\AN_{\wtq_{0,0}}>0$ region everywhere along the path,
interpolating between $C_-$ at $\zeta\ll 0$ and $C_+$ at $\zeta\gg 0$.
However, unlike in the case of $\C(0)$, one cannot choose
the circles to be Weyl invariant for a part of the path,
as will be explained in Appendix~\ref{subapp:Weylinv}.

\begin{figure}[thb]
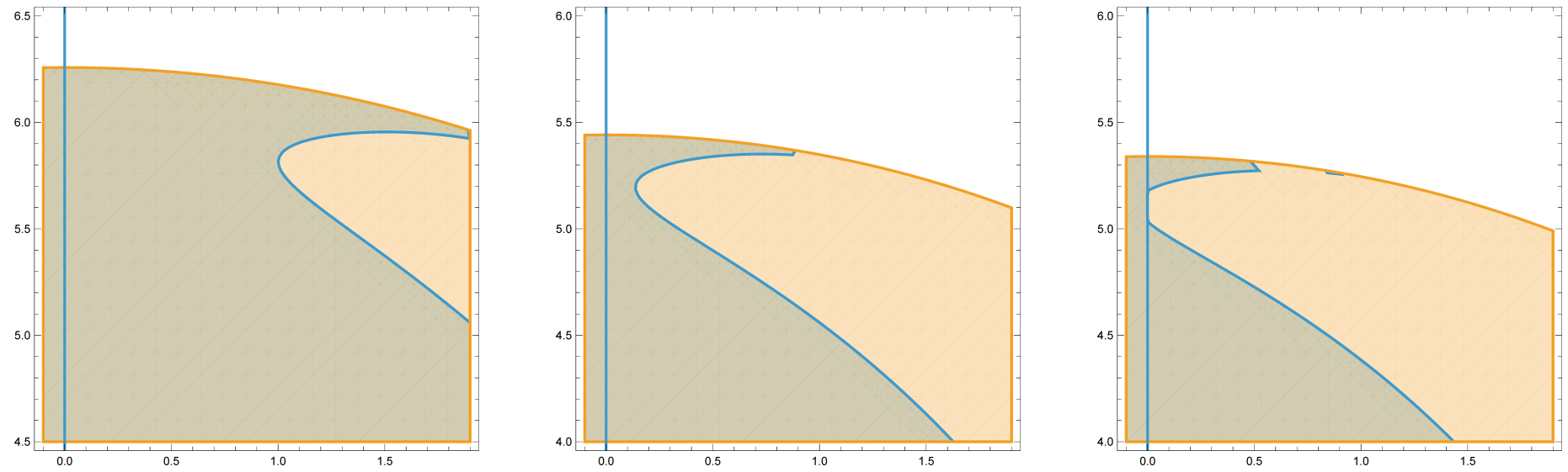

  \centerline{\includegraphics[page=79]{main-pics.pdf}~~~~
    \includegraphics[page=80]{main-pics.pdf}~~~~
    \includegraphics[page=81]{main-pics.pdf}}
  \caption{Enlarged view of some slices at $\mathtt{R}_{\zeta}$ with
    $\sDelta_1\zeta=0.2$ (left), $0.15$ (middle) and $0.14$ (right).
    The levels are fine tuned at $0.78R$, $0.839R$, $0.8455R$
    respectively, so that the road is as wide as possible.}
  \label{fig:coasts}
\end{figure}

\begin{figure}[thb]
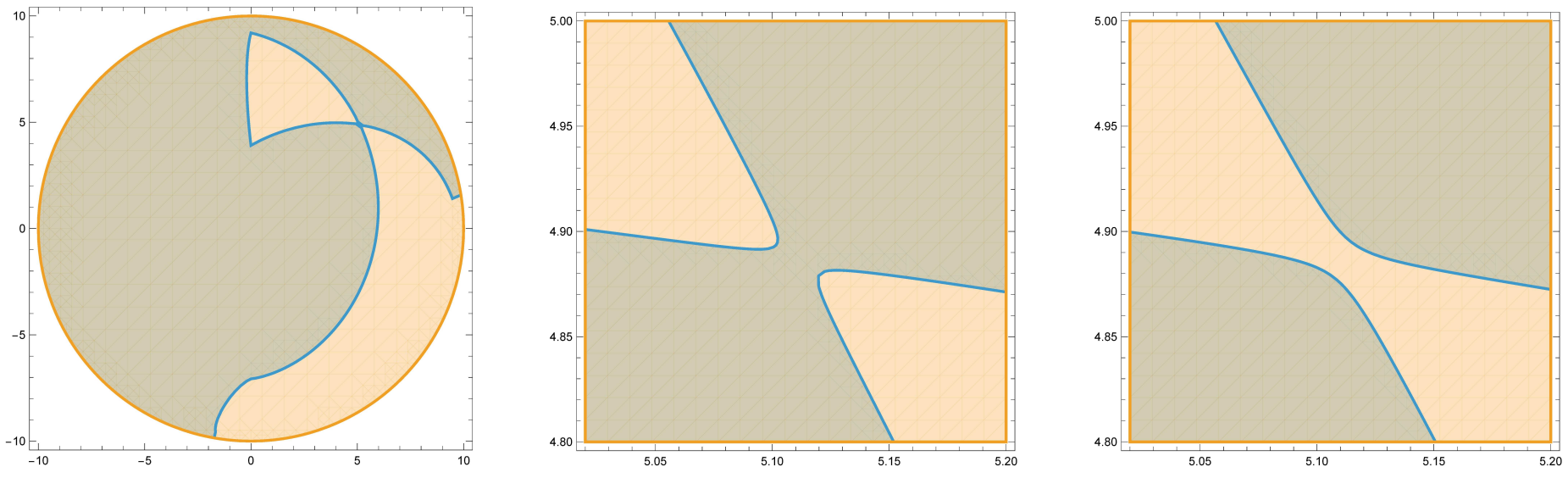

  \centerline{\includegraphics[page=82]{main-pics.pdf}~~~~
    \includegraphics[page=83]{main-pics.pdf}~~~~
    \includegraphics[page=84]{main-pics.pdf}}
  \caption{The wall $W_1$ at $\mathtt{R}_{\zeta}$ with
    $\sDelta_1\zeta=0.143$ (left),
  $0.14273$ (middle) and $0.14272$ (right).}
  \label{fig:W1}
\end{figure}

We next consider the representation $\C(7)$.
Let us first consider the downward direction. 
While in $\zeta\gg 0$, say around $\zeta\simeq 50$,
the function $\AN_{\wtq_{0,7}}$ is positive for $\Is_+\gtrsim 0.1R$,
and $C_+$ can be deformed to a circle
in a slice in the range $0.1R\lesssim \Is_+<R/\sqrt{2}$.
(The upper bound is for a circle
avoiding the walls $W_1$ and $W_2$ to exist in the slice.)
As $\zeta$ is decreased, the $\AN_{\wtq_{0,7}}<0$ region appears and increases,
but a circle can still be drawn in a single slice as long as
$\zeta$ is not too close to $\zeta_1$.
For example, at the point $\mathtt{R}_{\zeta}$ with
$\sDelta_1\zeta:=\zeta-\zeta_1=0.5$,
a circle in the shaded region can be drawn on the $\Is_+=0.7R$ slice,
as shown in Fig.~\ref{fig:07contours1}-left.
\begin{figure}[thb]
  \centerline{\includegraphics[page=76]{main-pics.pdf}~~~~
    \includegraphics[page=77]{main-pics.pdf}~~~~
    \includegraphics[page=78]{main-pics.pdf}}
\caption{}  \label{fig:07contours1}
\end{figure}
As $\zeta$ comes closer to $\zeta_1$, the unshaded region grows
and a circle cannot be drawn in a single slice.
Let us consider the point $\mathtt{R}_{\zeta}$ with $\sDelta_1\zeta=0.2$.
At the level $0<\Is_+<R/\sqrt{2}\simeq 0.707R$ where a circle can be drawn,
the unshaded region reaches the wall $W_1$
and we can only draw an open curve
which ends before entering there (Fig.~\ref{fig:07contours1}-middle
at $\Is_+=0.7R$).
On the slice at some level above $R/\sqrt{2}$,
a coastal road of positive region appears
(Fig.~\ref{fig:07contours1}-right at $\Is_+=0.78R$, with an enlarged view
in Fig.~\ref{fig:coasts}-left).
Lifting the end points to that level and connecting them through the road,
the curve is completed to a circle.
Thus, at the point $\mathtt{R}_{\zeta}$
with $\sDelta_1\zeta=0.2$, we can still find a circle in the
$\AN_{\wtq_{0,7}}>0$ region, though not on a single slice.
The road narrows down and disappears
as $\zeta$ comes much closer to $\zeta_1$.
See Fig.~\ref{fig:coasts}-middle and right.
The coastal road still exists at $\sDelta_1\zeta=0.15$ (middle) but
the coastline reaches the wall on any slice at $\sDelta_1\zeta=0.14$ (right).
The circle in the positive region  is forced to
hit the wall $W_1$ somewhere in $0.14<\sDelta_1\zeta<0.15$.
Below that point, a circle in the $\AN_{\wtq_{0,7}}>0$ region does not exist.
Let us locate the threshold more precisely.
For this purpose, instead of the level by level computation,
we look at the wall $W_1$ itself.
Fig.~\ref{fig:W1} shows the wall $W_1=\{\Rs_1=0,\Is_1>0\}$
with coordinates $(\Rs_2,\Is_2)$ at three points on the path.
The left figure is the entire wall while the middle and right figures show
enlarged views around $(\Rs_2,\Is_2)\simeq (5.1,4.9)$.
Shaded and unshaded are the $W_1$-boundary of the regions
where $\AN_{\wtq_{0,7}}$ is positive and negative.
We see that the $W_1$-boundary of the positive road barely exists at
$\sDelta_1\zeta=0.14273$ but disappears at $\sDelta_1\zeta=0.14272$.
This shows that the threshold is somewhere between these two values.
By looking at the enlarged view with higher resolution,
we can locate the threshold with higher precision.
Let us also find the thresholds on lines closer to $t_1$.
The result on the lines at three values of
$\sDelta_1\theta:=\theta-(-7\pi)$ are
\beq
\begin{array}{c|c|c}
  \sDelta_1\theta & \sDelta_1\zeta & {\rm angle}\\
  \hline
  0.1\pi & 0.1427268 & 0.135739\pi\\
  0.01\pi & 0.01504239 & 0.142143\pi\\
  0.0001\pi & 0.0001512824 & 0.142850\pi
\end{array}
\label{slopematch1}
\eeq
where the ``angle'' in the last column is the arctangent of
$\sDelta_1\zeta/\sDelta_1\theta$. Note that it approaches
$\pi/7\simeq 0.142857 \pi$.

\begin{figure}[thb]
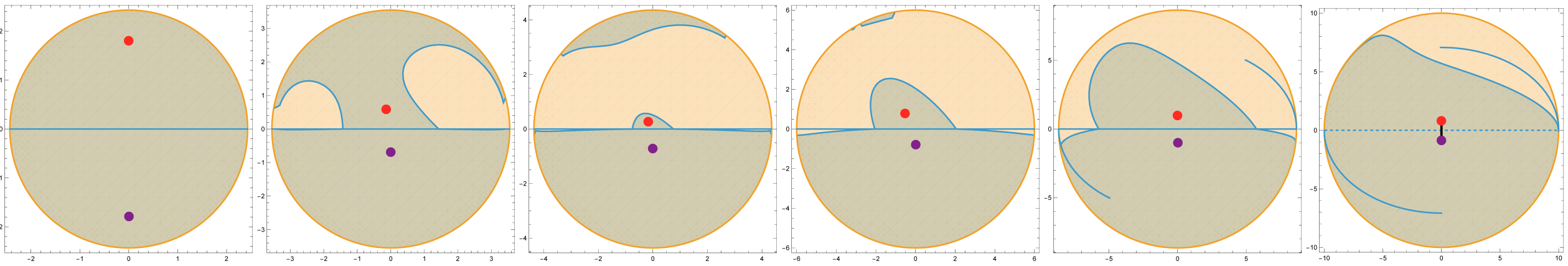

  \centerline{
    \includegraphics[page=85]{main-pics.pdf}
    \includegraphics[page=86]{main-pics.pdf}
    \includegraphics[page=87]{main-pics.pdf}
    \includegraphics[page=88]{main-pics.pdf}
    \includegraphics[page=89]{main-pics.pdf}
         \includegraphics[page=90]{main-pics.pdf}}
  \caption{Projections to the $(-\Is_-,\Rs_+)$ plane of the $\Rs_->0$ half of
    the slices. The wall $W_p$ is shown as the horizontal line.
    (As the relative sizes are very different, we have rescaled the slices
    for the sake of visibility.)}
      \label{fig:0702}
\end{figure}

Next, we again consider the representation $\C(7)$,
now in the upward direction starting from $\zeta\ll 0$.
We again use the fact that $\AN_{\wtq_{0,i}}$ (for any $i$) is positive
on the contour $\bgamma_-$ as long as $\zeta<\zeta_1$.
Thus, we can continue to use $\bgamma_-$ along the path
until arriving at $\mathtt{R}_{\zeta_1}$ from below.
For the representation $\C(7)$,
we find a problem for using $\bgamma_-$ or any of its deformations
as $\zeta$ goes past $\zeta_1$.
To see the problem, it is better to look at the slices from another
angle, in which the intersection in $\Rs_->0$ of the wall $W_p$
and the equator appears at the center.
Fig.~\ref{fig:0702} shows the slices from such an angle, with levels
$-0.97R$, $-0.935R$, $-0.9R$, $-0.8R$, $-0.5R$, $0$ from left to right.
Shaded are the regions where
$\AN_{\wtq_{0,7}}>0$ at the point $\mathtt{R}_{\zeta}$ with
$\sDelta_1\zeta=-0.02$.
Recall that the circle $C_-$ is described as the motion of four colored
points on the equator, as in Fig.~\ref{fig:Cminus}.
We would like to find whether any deformation of such motion can stay
in the $\AN_{\wtq_{0,7}}>0$ region.
Fig.~\ref{fig:0702} also includes a possible motion of two of the four points,
red and purple.
The two are separated by the wall $W_p$ at negative levels but
they can be connected at level 0 where $W_p$ disappears.
We see that, at levels $-0.9R$, $-0.8R$ and $-0.5R$,
the red point is confined in a peninsula that is bounded by the wall $W_p$.
At the point $\sDelta_1\zeta=-0.02$ we are now considering,
the peninsula never vanishes and the red point has a place to stay
at any level from $-R$ to $0$.
This ensures that we have a closed circle in the positive region.
However, as $\sDelta_1\zeta$ approaches 0, the peninsula becomes smaller
and smaller at levels around $\simeq -0.901R$,
and it disappears as $\sDelta_1\zeta$ goes past $0$.
\begin{figure}[thb]
  \centerline{
    \includegraphics[page=91]{main-pics.pdf}
    \includegraphics[page=92]{main-pics.pdf}
    \includegraphics[page=93]{main-pics.pdf}
    \includegraphics[page=94]{main-pics.pdf}
    \includegraphics[page=95]{main-pics.pdf}
    \includegraphics[page=96]{main-pics.pdf}}
  \caption{}
      \label{fig:z1pp}
\end{figure}
Fig.~\ref{fig:z1pp} shows the slices of the same levels as in
Fig.~\ref{fig:0702} but at the point with $\sDelta_1\zeta=+0.005$.
We see that the peninsula that was always present when $\sDelta_1\zeta=-0.02$
is now absent at level $-0.9R$, and a continuous motion of
the red point in the shaded region is impossible.
Thus,  at the point with $\sDelta_1\zeta=+0.005$
we no longer have a closed circle in the positive region.
In order to locate the threshold where the peninsula disappears on some slice,
we look at the wall $W_p$.
\begin{figure}[thb]
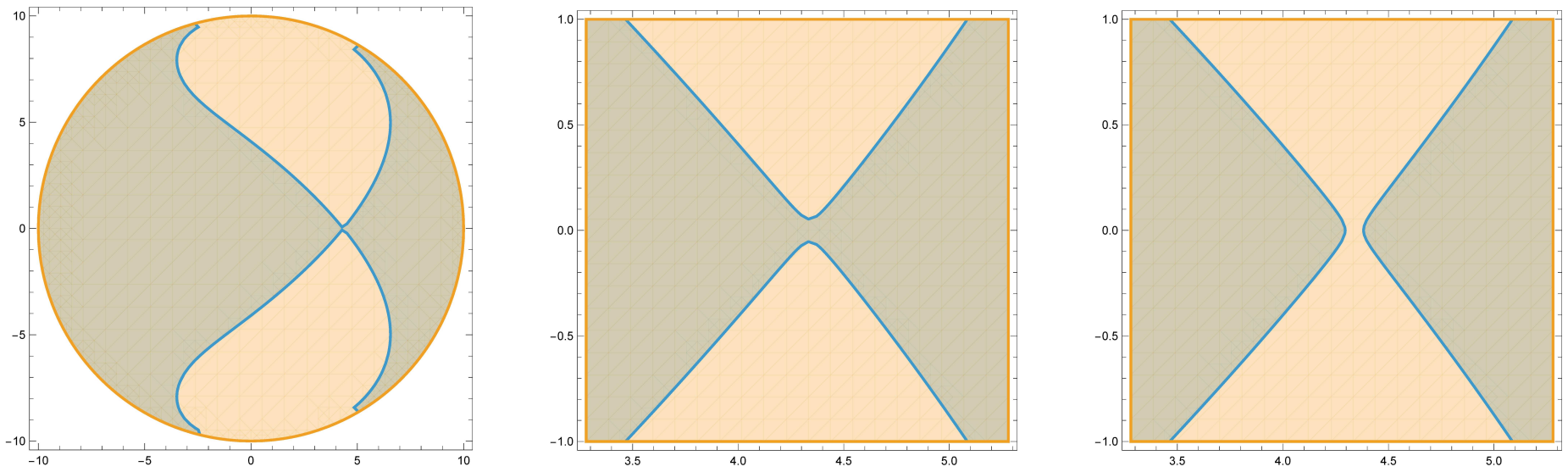

  \centerline{\includegraphics[page=97]{main-pics.pdf}~~~~
    \includegraphics[page=98]{main-pics.pdf}~~~~
    \includegraphics[page=99]{main-pics.pdf}}
  \caption{The wall $W_p$ at $\mathtt{R}_{\zeta}$ with
    $\sDelta_1\zeta=0$ (left),
  $-0.0001$ (middle) and $+0.0001$ (right).}
  \label{fig:07Wp}
\end{figure}
Fig.~\ref{fig:07Wp} shows the wall $W_p=\{\Rs_+=0,\Is_+<0\}$
with coordinates $(\Rs_-,\Is_-)$ at $\mathtt{R}_{\zeta_1}$ and two
nearby points.
Shaded and unshaded are the $W_p$-boundaries of the regions where
$\AN_{\wtq_{0,7}}$ is positive and negative. We see that the
boundary of the peninsula barely exists at $\sDelta_1\zeta=-0.0001$
but disappears at $\sDelta_1\zeta=+0.0001$. This shows that the threshold is
somewhere between these values. Looking at the wall $W_p$ with higher
resolution, it seems that the threshold is exactly at $\sDelta_1\zeta=0$,
that is, at the point $\mathtt{R}_{\zeta_1}$.
Thus, as the point reaches $\mathtt{R}_{\zeta_1}$ from below,
the circle in the $\AN_{\wtq_{0,7}}>0$ region
is forced to step out of the region or hit the wall $W_p$.

\begin{figure}[thb]
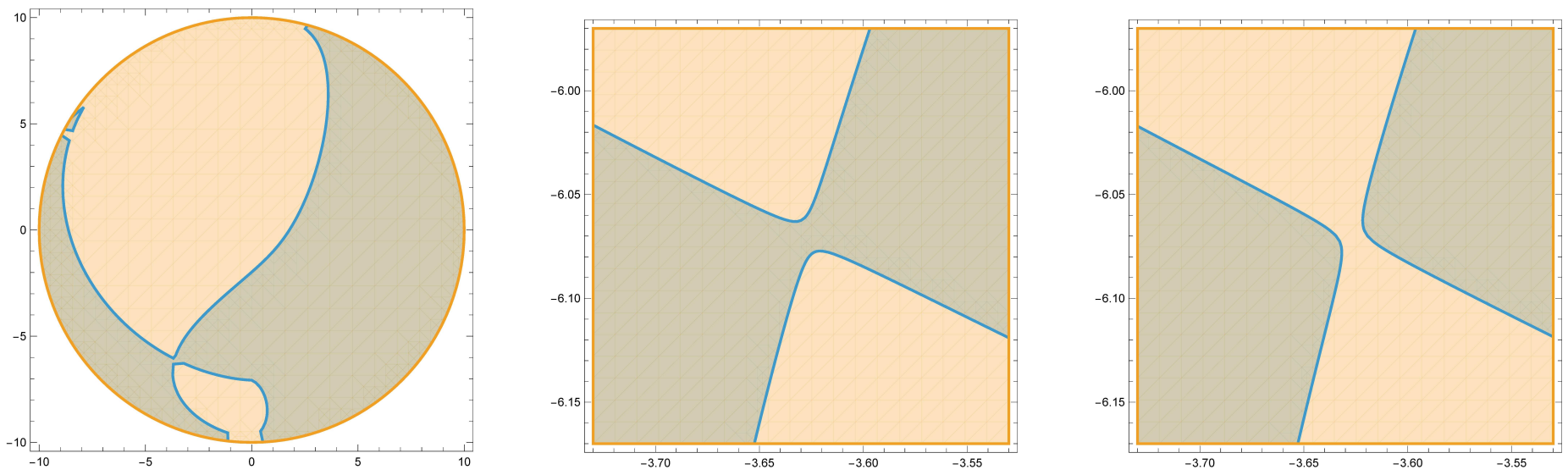

  \centerline{\includegraphics[page=108]{main-pics.pdf}~~~~
    \includegraphics[page=109]{main-pics.pdf}~~~~
    \includegraphics[page=110]{main-pics.pdf}}
  \caption{The wall $W_2$ at $\mathtt{L}_{\zeta}$ with
    $\sDelta_3\zeta=1.251$ (left),
  $1.2504$ (middle) and $1.2502$ (right).}
  \label{fig:W2}
\end{figure}

Let us consider the representation $\Sym^2\fund(-1)$, first in the downward
direction. As $\AN_{\wtq_{2,-1}}$ is positive on the contour $\bgamma_0$
everywere on the strip $-7\pi<\theta<-6\pi$, after deforming $\bgamma_+$
to $\bgamma_0$ at $\zeta\gg 0$, we can contniue to take $\bgamma_0$
until just before the point $\mathtt{M}_0$ at the boundary of the strip.
The $\AN_{\wt{q}_{2,6}}>0$ region at $\mathtt{M}_0$ on the level 0 slice is
shown as shaded in Fig.~\ref{fig:2m1atMa}, with an enlarged view in
Fig.~\ref{fig:2m1atMb}-left.
\begin{figure}[thb]
  \begin{subfigure}[b]{.44\linewidth}
    \centering
    \includegraphics[page=100]{main-pics.pdf}
    \caption{}\label{fig:2m1atMa}
  \end{subfigure}
  \begin{subfigure}[b]{.44\linewidth}
    \centering
    \includegraphics[page=101]{main-pics.pdf}
    \caption{}\label{fig:2m1atMb}
  \end{subfigure}
    \vspace{-0.4cm}
  \caption{}  \label{fig:2m1atM}
\end{figure}
The circle $C_0$ in the limit $R\to \infty$ 
goes through the vertex of a checkerboard-like pattern.
The pattern changes as the level is moved.
The shaded region broadens if the level is raised
(Fig.~\ref{fig:2m1atMb}-middle at $\Is_+=0.001R$)
and is detached if the level is lowered
(Fig.~\ref{fig:2m1atMb}-right at $\Is_+=-0.001R$).
Thus, before reaching $\mathtt{M}_0$ from the right, we can lift
a part of $C_0$ to positive level, so that the resulting circle $C$ can
stay in the positive region as the point on the path moves past $\mathtt{M}_0$.
We may even lift the entire circle so that $C$ is in a single slice.
If the level is high enough, the circle $C$ remains good
until the point goes around the corner $\mathtt{L}_0$.
However, as the point goes down and comes close to
$\mathtt{L}_{\zeta_3}$, it becomes harder for the curve $C$ to stay in the
positive region.
Fig.~\ref{fig:2m1contours} shows the slices at $\mathtt{L}_{\zeta}$
with $\zeta=0$, $-2$, $\zeta_3+2\simeq-3.7$
from left to right, at levels $0.1R$, $0.1R$, $0.061R$ respectively.
\begin{figure}[thb]
  \centerline{\includegraphics[page=102]{main-pics.pdf}~~~~
    \includegraphics[page=103]{main-pics.pdf}~~~~
    \includegraphics[page=104]{main-pics.pdf}}
\caption{}  \label{fig:2m1contours}
\end{figure}
\begin{figure}[thb]
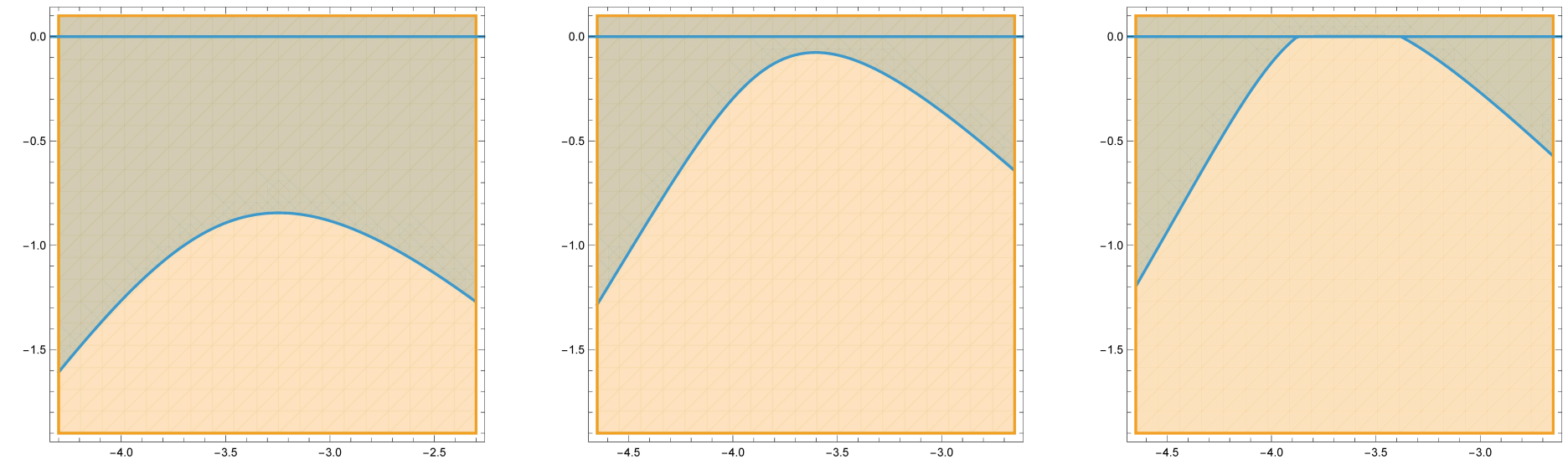

  \centerline{\includegraphics[page=105]{main-pics.pdf}~~~~
    \includegraphics[page=106]{main-pics.pdf}~~~~
    \includegraphics[page=107]{main-pics.pdf}}
  \caption{Enlarged view of some slices at $\mathtt{L}_{\zeta}$ with
    $\sDelta_3\zeta=2$ (left), $1.3$ (middle) and $1.2$ (right).
    The levels are fine tuned at $0.061R$, $0.069R$, $0.071R$
    respectively, so that the road is as wide as possible.}
  \label{fig:2m1coasts}
\end{figure}
We see that a part of the unshaded region comes close to the wall $W_2$,
and only a narrow road is left for the circle to go through.
Enlarged views are shown in Fig.~\ref{fig:2m1coasts}.
The road narrows down
as the distance $\sDelta_3\zeta:=\zeta-\zeta_3$ decreases,
and it disappears somewhere in $1.2<\sDelta_3\zeta<1.3$.
There, the circle in the positive region is forced to hit the wall $W_2$.
Below that point, a circle in the $\AN_{\wtq_{2,-1}}>0$ region does not exist.
We can locate the threshold more precisely by looking at the wall $W_2$ itself.
Fig.~\ref{fig:W2} shows the wall $W_2=\{\Rs_2=0,\Is_2>0\}$
with coordinates $(\Rs_1,\Is_1)$ at three points on the path.
Shaded and unshaded are the $W_2$-boundaries of the regions
where $\AN_{\wtq_{2,-1}}$ is positive and negative.
We see that the $W_2$-boundary of the positive road barely exists at
$\sDelta_3\zeta=1.2504$ but disappears at $\sDelta_3\zeta=1.2502$.
This shows that the threshold is somewhere between these two values.
We can locate the threshold with higher precision by increasing the resolution.
Let us also find the thresholds on lines closer to $t_3$.
The result on the lines at three values of
$\sDelta_3\theta:=\theta-(-7\pi)$ are
\beq
\begin{array}{c|c|c}
  \sDelta_3\theta & \sDelta_3\zeta & {\rm angle}\\
  \hline
  -0.1\pi &      1.250295  & 0.421641\pi\\
  -0.01\pi &     0.1362368 &  0.427859\pi\\
  -0.0001\pi & 0.001376279 & 0.428564\pi
\end{array}
\eeq
where the ``angle'' in the last column is the arctangent of
$\sDelta_3\zeta/|\sDelta_3\theta|$. We see that it approaches
$3\pi/7\simeq 0.428571\pi$.

\begin{figure}[thb]
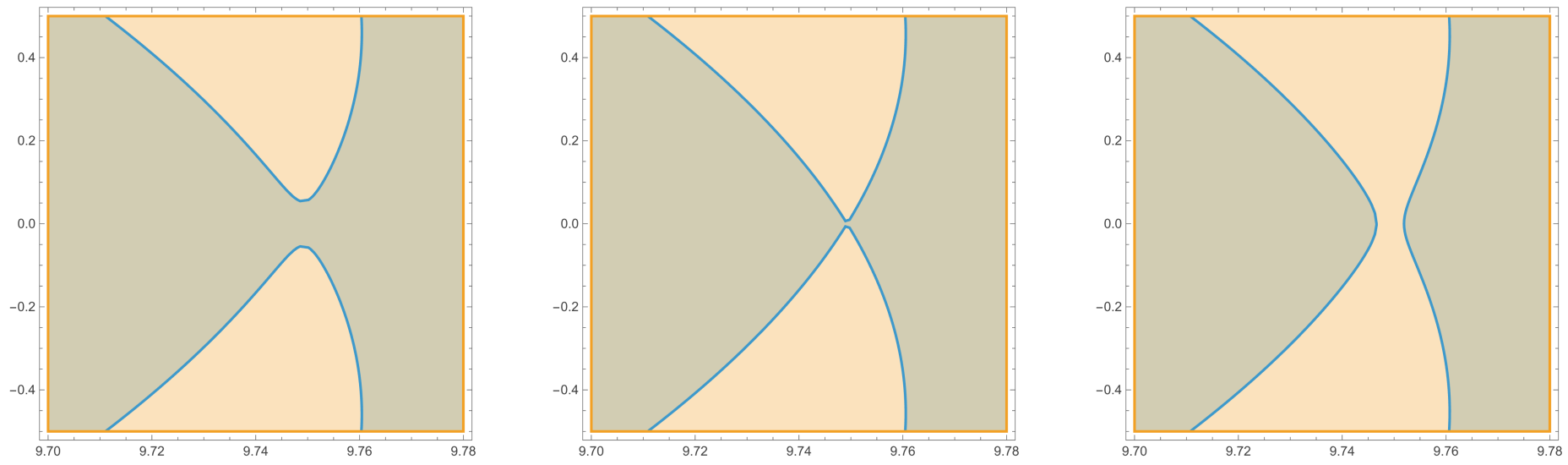

  \centerline{\includegraphics[page=120]{main-pics.pdf}~~~~
    \includegraphics[page=121]{main-pics.pdf}~~~~
    \includegraphics[page=122]{main-pics.pdf}}
  \caption{The wall $W_p$ at $\mathtt{L}_{\zeta}$ with
    $\sDelta_3\zeta=-0.0001$ (left),
    $0$ (middle) and $+0.0001$ (right).}
  \label{fig:2m1Wp}
\end{figure}

Let us again consider the representation $\Sym^2\fund(-1)$,
now in the upward direction starting from $\zeta\ll 0$.
We know from Section~\ref{subsec:SU2GRRRodland} that
$\AN_{\wtq_{2,i}}$ (for any $i$) is positive on the contour $\bgamma_-$
as long as $\zeta<\zeta_3$.
Thus, we can continue to use $\bgamma_-$ along the path
until approaching $\mathtt{L}_{\zeta_3}$ from below.
For the representation $\Sym^2\fund(-1)$,
we find a problem in using $\bgamma_-$ or any of its deformations
as $\zeta$ goes past $\zeta_3$, which is similar to the problem
for the representation $\C(7)$ at $\mathtt{R}_{\zeta}$
as $\zeta$ goes past $\zeta_1$ from below.
To see the problem, we again look at the slices from another angle
as in Fig.~\ref{fig:0702}.
\begin{figure}[thb]
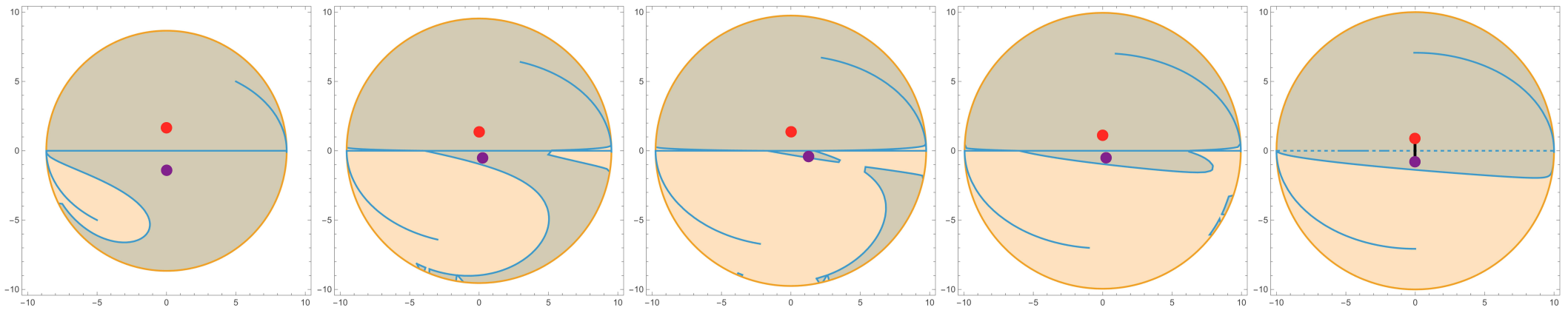

  \centerline{
    \includegraphics[page=111]{main-pics.pdf}
    \includegraphics[page=112]{main-pics.pdf}
    \includegraphics[page=113]{main-pics.pdf}
    \includegraphics[page=114]{main-pics.pdf}
     \includegraphics[page=115]{main-pics.pdf}}
  \caption{Projections to the $(-\Is_-,\Rs_+)$ plane of the $\Rs_->0$ half
    of the slices. (As the relative sizes are not much different,
    we do not make a rescaling this time.)}
  \label{fig:2m1pen}
\end{figure}
\begin{figure}[thb]
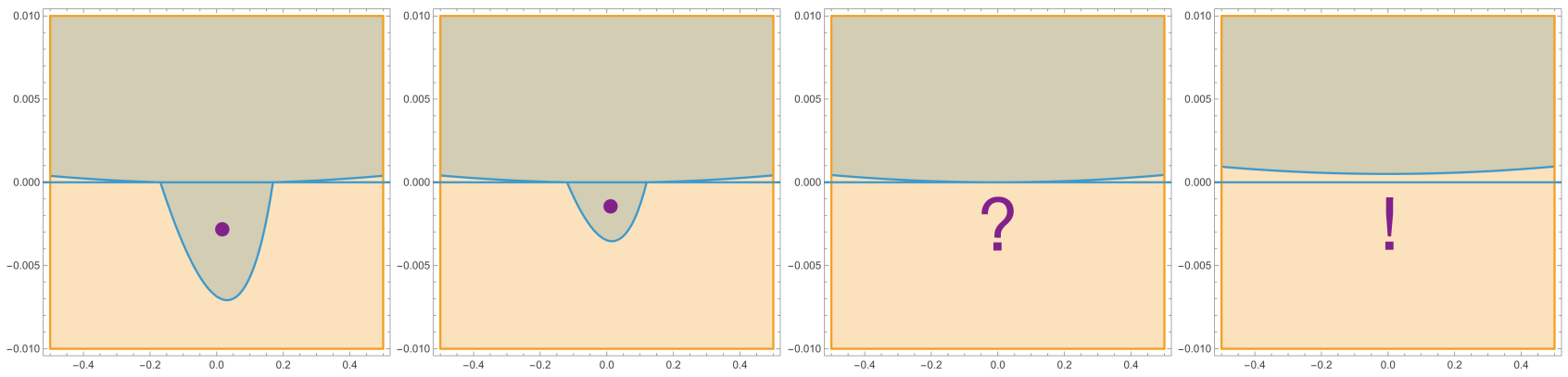

  \centerline{
    \includegraphics[page=116]{main-pics.pdf}
    \includegraphics[page=117]{main-pics.pdf}
    \includegraphics[page=118]{main-pics.pdf}
    \includegraphics[page=119]{main-pics.pdf}}
  \caption{The enlarged views of the center of the level
$-0.2225R$ slice for
    $\sDelta_3\zeta=-0.001$, $-0.0005$, $0$, $+0.01$ from left to right.
    (The horizontal and vertical directions are scaled differently for
    visibility.) For $\sDelta_3\zeta=0$, it is not clear with this resolution
    whether the peninsula exists, but we find with a higher resolution
    ($10^4$ more) that it does exist.}
  \label{fig:2m1EL}
\end{figure}
Figs.~\ref{fig:2m1pen} shows the slices from such an angle, with
levels $-0.5R$, $-0.3R$, $-0.2225R$, $-0.1R$, $0$ from left to right,
where shaded are the $\AN_{\wtq_{2,-1}}>0$ regions at the point
$\mathtt{L}_{\zeta}$ with $\sDelta_3\zeta=-0.1$.
The behaviour of one of the four points, the purple one this time,
is important.
When $\zeta$ is below $\zeta_3$, the purple point at some levels
is confined in a peninsula bounded by the wall $W_p$.
As $\zeta$ approaches $\zeta_3$,
the peninsula at levels around $-0.2225R$
becomes smaller and smaller and it disappears as $\zeta$ goes past $\zeta_3$.
See Fig.~\ref{fig:2m1EL} for the behaviour exactly at level $-0.2225R$.
When $\zeta$ is slightly above $\zeta_3$,
the purple point has no place to stay and a closed circle in
the positive region does not exist.
Looking at the wall $W_p$ itself (Fig.~\ref{fig:2m1Wp}), we find that
the threshold where the peninsula disappears on some slice
seems exactly at $\zeta=\zeta_3$.
Thus, as the point reaches $\mathtt{L}_{\zeta_3}$ from below,
the circle in the $\AN_{\wtq_{2,-1}}>0$ region
is forced to step out of the region or hit the wall $W_p$.

Let us next consider window {\bf B}.
As it is related to window {\bf C} by the reflection across the
line with $\theta=-7\pi$, by the reflection symmetry (\ref{reflectionsym})
at $-7\pi$, we do not need to do any new computation.\footnote{In fact,
  we initially did some computation and observed a complete agreement with
  the result for window {\bf C} modulo change of coordinates.
  This is how we noticed the reflection symmetry (\ref{reflectionsym}).}
We can transfer the result for window {\bf C} to this window
by the exchange of $\C\leftrightarrow \C(7)$ and
$\Sym^2\fund(-1)\leftrightarrow \Sym^2\fund(6)$.

This completes the description of the numerical analysis quoted
in Section~\ref{subsec:numerical}. In particular, we found that
the four representations have ``no-entry zones''
in the FI-theta parameter space.
For the representation $\C(7)$ ({\it resp}. $\C$),
it is a region ${\mathscr R}_{0,7}$ to the right
({\it resp}. ${\mathscr L}_{0,0}$ to the left) of $t_1$,
and for $\Sym^2\fund(-1)$ ({\it resp}. $\Sym^2\fund(6)$),
it is a region ${\mathscr L}_{2,-1}$ to the left
({\it resp}. ${\mathscr R}_{2,6}$ to the right) of $t_1$.
As $t$ enters there from above, 
any deformation of $C_+$ goes out of
the admissible region or is forced to hit the wall $W_1$ or $W_2$.
As $t$ enters there from below,
any deformation of $C_-$ goes out of
the admissible region or is forced to hit the wall $W_p$.
The regions ${\mathscr R}_{0,7}$ and ${\mathscr L}_{0,0}$
are tangent at $t_1$ to $R_1$ and $L_1$
while
${\mathscr R}_{2,6}$ and ${\mathscr L}_{2,-1}$
are tangent at $t_3$ to $R_3$ and $L_3$,
where $R_{\bfa}$ and $L_{\bfa}$ are the forbidden sectors
given in (\ref{RandL}).
This is consistent with the analytical result obtained in
Section~\ref{subsec:catastroph} on the local behaviour of the region
$\AN_{\wtq_*}>0$ around the singular point $t_*$.
The numerical analysis presented above,
which examines all possible finite deformations of $C_{\pm}$,
complements the local analysis in Section~\ref{subsec:catastroph} where
only small deformations of the contour $\bgamma_0$ are examined.

For completeness, let us look at the behaviour of $\AN_{\wtq}$
near the other singular point, $t_2$.
We consider the straight path on its right,
say at $\theta=-8\pi+0.1\pi$, and look at the function
$\AN_{\wtq_{1,7}}$.
As $\zeta$ approaches $\zeta_2$ from above, the negative region grows,
and as $\sDelta_2\zeta=\zeta-\zeta_2$ goes below $0.37$,
any deformation of $C_+$ goes out of the admissible region or
is forced to hit the wall $W_1$. The threshold can be determined
more precisely by looking at the wall $W_1$ itself.
The result for the paths of various distances from $t_2$ is 
\beq
\begin{array}{c|c|c}
  \sDelta_2\theta & \sDelta_2\zeta & {\rm angle}\\
  \hline
  0.1\pi &      0.3764558  & 0.278635\pi\\
  0.01\pi &     0.03921365 &  0.285001\pi\\
  0.0001\pi & 0.0003939251 & 0.285707\pi
\end{array}
\eeq
where the ``angle'' in the last column is the arctangent of
$\sDelta_2\zeta/\sDelta_2\theta$. We see that it approaches
$2\pi/7\simeq 0.285714\pi$.
As $\zeta$ approaches $\zeta_2$ from below and goes beyond it,
the peninsula for the red point becomes smaller and disappears on some slice.
Thus, any deformation of $C_-$ goes out of the admissible region
or is forced to hit the wall $W_p$.
By the reflection symmetry at $-8\pi$,
the same can be said of the function $\AN_{\wtq_{1,0}}$
along the straight path to the left of $t_2$.
Thus, $\fund(7)$ ({\it resp}. $\fund$) has a no-entry zone
${\mathscr R}_{1,7}$ to the right (${\mathscr L}_{1,0}$ to the left)
of $t_2$ which is tangent at $t_2$ to the forbidden
sector $R_2$ ({\it resp}. $L_2$)
in (\ref{RandL}).
This numerical result supports and complements
the analytical result obtained in Section~\ref{subsec:catastroph}.

\subsection{Analysis Away From Singular Points}
\label{subapp:genpaths}

We have so far considered only the paths that go near the singular points,
within $|\sDelta_{\bfa}\theta|\leq 0.1\pi$.
Here we would like to examine more general paths.
In particular, we would like to see whether and how the no-entry zones
${\mathscr R}_{0,7}$ and ${\mathscr L}_{2,-1}$ etc extend away
from the singular points.

We first consider the representation $\C(7)$.
The blocking 
of deformations of $C_+$ by the wall $W_1$ as $\zeta$ goes down,
as depicted in Fig.~\ref{fig:07contours1} and Fig.~\ref{fig:coasts},
continues to hold equally for larger $\sDelta_1\theta$, that is,
at all $\theta\geq -7\pi$.
For $0<\sDelta_1\theta<\pi$, the threshold can be
identified by looking at $\AN_{\wtq_{0,7}}$ on the wall $W_1$
as in Fig.~\ref{fig:W1} for the case $\sDelta_1\theta=0.1\pi$ ---
it is where the positive region $\AN_{\wtq_{0,7}}|_{W_1}\geq 0$ is singular.
Once that is realized, one may try to find the threshold analytically.
Indeed we managed to find it exactly, as presented
in Section~\ref{subsec:noentry}
--- it is given by the formula (\ref{threshold1}).
Given this, the ``angle'' found numerically in (\ref{slopematch1})
approaches exactly $\pi/7$ as $\sDelta_1\theta\searrow 0$,
matching with the angle of the sector $R_1$.
As $\sDelta_1\theta\nearrow \pi$,
the negative region of the triangular shape
as in Fig.~\ref{fig:W1}-left shrinks to zero size.
The shrinking to zero happens at $\Rs_2=0$ which means that
the coast line reaches the wall $W_1$ at the top level $\Is_+=R$
where the slice is a point $p_{\rm top}$ with
$(\Rs_1,\Is_1,\Rs_2,\Is_2)=(0,{R\over\sqrt{2}},0,{R\over\sqrt{2}})$.

The threshold value at $\sDelta_1\theta=\pi$ is $\zeta=7\log 2$
which happens to be the lower bound for the contour
$\bgamma_+$ to be admissible.
Beyond $\sDelta_1\theta=\pi$, that is, for all $\theta\geq -6\pi$,
the threshold continues to be at $\zeta=7\log 2$
as we now explain.
The situation before reaching $\zeta=7\log 2$ from above
is very similar to the case of $\sDelta_1\theta=0.1$,
as shown in Fig.~\ref{fig:07contours1},
except that the road of positive region is narrowest on the equator
$\Is_-=0$, in the direction $\Rs_2>0$.
Let us see what happens as
\beq
\sDelta_+\zeta=:\zeta-7\log 2
\eeq
goes from positive to negative along a line at $\theta> -6\pi$,
say, at $\theta=0$.
It is better to look at
the slice from an angle in which the intersection of the wall $W_1$
and the equator appears at the center.
\begin{figure}[thb]
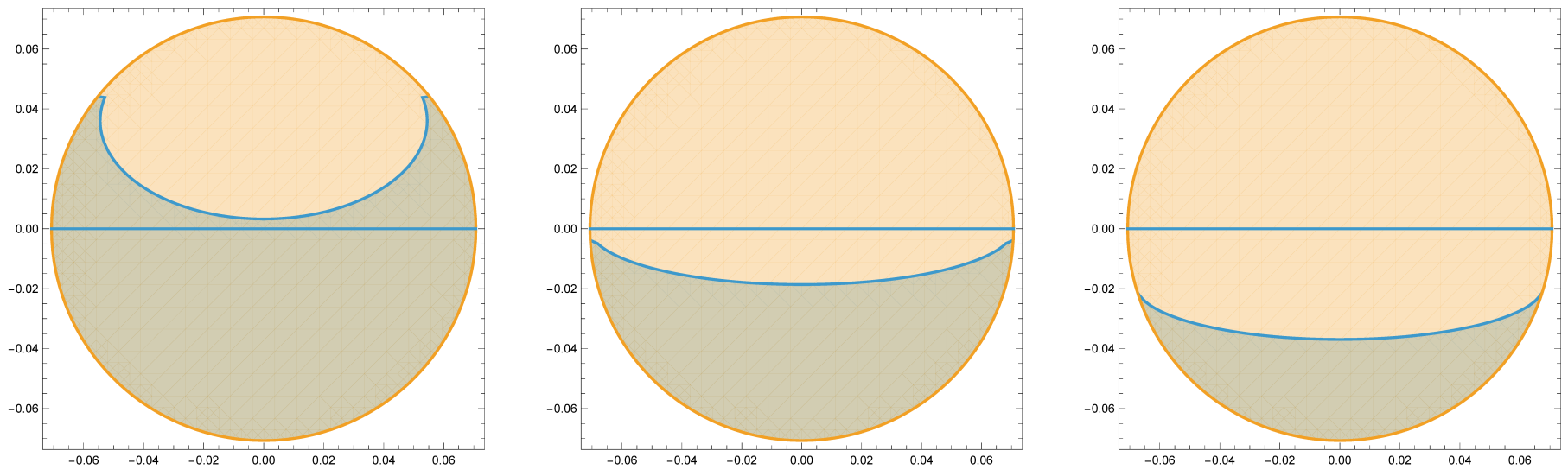

  \centerline{
    \includegraphics[page=123]{main-pics.pdf}~~~
    \includegraphics[page=124]{main-pics.pdf}~~~
\includegraphics[page=125]{main-pics.pdf}}
  \caption{Projection to the $(\Is_-,\Rs_1)$ plane of the $\Rs_2>0$ half
    of the slice at level $0.999975R$.
    The wall $W_1$ is shown as the horizontal line.} \label{fig:07across}
\end{figure}
Fig.~\ref{fig:07across} shows such a view of the slices,
with level $0.999975R$ and at the points with
$\sDelta_+\zeta=0.1$, $0$, $-0.1$ from left to right.
In the left figure, we see a narrow road of shaded region
through which a part of an admissible circle $C$ can go.
Such a road exists if and only if the slice
is above a critical level $\Is_+^{\rm crit}$,
and any admissible deformation of $C_+$ must go through the union of such roads.
If $|\sDelta_+\zeta|$ and $(R-\Is_+)/R$ are much smaller than $1$,
the pictures are invariant under the scaling,
$\sDelta_+\zeta\mapsto \lambda \sDelta_+\zeta$
and $R-\Is_+\mapsto \lambda^2(R-\Is_+)$ for $0<\lambda<1$.
For example, at the points with $\sDelta_+\zeta=0.01, 0, -0.01$,
the pictures at the slice $0.99999975$ are the same as
in Fig.~\ref{fig:07across}, except that the whole slice must be scaled up
by a factor of $10$.
In particular, as $\sDelta_+\zeta>0$ is scaled down by a factor
of $1/10$, the difference $|R-\Is_+^{\rm crit}|$
is scaled down by $1/100$.
As $\sDelta_+\zeta\searrow 0$, the critical level $\Is_+^{\rm crit}$
reaches the top value $\Is_+=R$ where the slice is a point $p_{\rm top}$
on the wall $W_1$.
Then, any deformation of $C_+$, not just $C_+$ itself,
is forced to either go out of the admissible region or
hit the wall $W_1$ as $\zeta\searrow 7\log 2$.

Along the line of threshold, $\zeta=7\log 2$ and $\theta\geq -6\pi$,
the region $\AN_{\wtq_{0,7}}|_{W_1}\geq 0$ is non-singular.
Thus, we have ``no-entry without singularity''.
We would like to understand the difference from
``no-entry with singularity'' for $-7\pi <\theta <-6\pi$.
For this purpose, we first revisit the case of $\sDelta_1\theta=0.1\pi$.
In this case, the positive road exists only on the slices with levels
in a certain interval.
Let us consider the point with $\sDelta_1\zeta=0.1435$ slightly above
the threshold.
Fig.~\ref{fig:0701s}-(a), (b), (c) are parts of the slices
at levels $8.6$, $8.452$, $8.3$, respectively.
\begin{figure}[thb]
  \begin{subfigure}[b]{.24\linewidth}
    \centering
    \includegraphics[page=126]{main-pics.pdf}
    \caption{}\label{fig:0701C}
  \end{subfigure}
  \begin{subfigure}[b]{.24\linewidth}
    \centering
    \includegraphics[page=127]{main-pics.pdf}
    \caption{}\label{fig:0701B}
  \end{subfigure}
  \begin{subfigure}[b]{.24\linewidth}
    \centering
    \includegraphics[page=128]{main-pics.pdf}
    \caption{}\label{fig:0701A}
  \end{subfigure}
    \begin{subfigure}[b]{.24\linewidth}
    \centering
    \includegraphics[page=129]{main-pics.pdf}
    \caption{}\label{fig:0701W1three}
  \end{subfigure}
    \vspace{-0.4cm}
  \caption{}  \label{fig:0701s}
\end{figure}
We see that the road exists only in the case (b).
The picture on the wall $W_1$ is shown in
Fig.~\ref{fig:0701s}-(d), where the lines are the intersection with the slices.
The structure of the shaded region on $W_1$ reflects the need to
fine tune the levels to find the positive road.
As $\zeta$ reaches the threshold, the width of the interval shrinks to
zero size and the region $\AN_{\wtq_{0,7}}|_{W_1}\geq 0$ develops a singularity.
\begin{figure}[thb]
  \begin{subfigure}[b]{.24\linewidth}
    \centering
    \includegraphics[page=130]{main-pics.pdf}
    \caption{}\label{fig:07aboveC}
  \end{subfigure}
  \begin{subfigure}[b]{.24\linewidth}
    \centering
    \includegraphics[page=131]{main-pics.pdf}
    \caption{}\label{fig:07above}
  \end{subfigure}
  \begin{subfigure}[b]{.24\linewidth}
    \centering
    \includegraphics[page=132]{main-pics.pdf}
    \caption{}\label{fig:07aboveB}
  \end{subfigure}
    \begin{subfigure}[b]{.24\linewidth}
    \centering
    \includegraphics[page=133]{main-pics.pdf}
    \caption{}\label{fig:07W1three}
  \end{subfigure}
    \vspace{-0.4cm}
  \caption{}  \label{fig:07s}
\end{figure}scale=0.4->1.0
On the other hand, for $\theta\geq -6\pi$,
the positive road exists as along as the slice is above the critical level
$\Is_+^{\rm crit}$.
For example, let us look at the slices at the point with
$\theta=0$ and $\sDelta_+\zeta=0.1$.
Fig.~\ref{fig:07s}-(a), (b), (c) are parts of
the slices at levels $0.999985R$, $0.999975R$, $0.999965R$ respectively.
We see that the positive road exists in (a) and (b) and not in (c).
The critical level $\Is_+^{\rm crit}$ is somewhwere between
the levels for (b) and (c).
This is reflected in
the picture on the wall $W_1$ shown in Fig.~\ref{fig:07s}-(d).
The scale-down of $\sDelta_+\zeta$ only rescales this picture,
and the limit $\sDelta_+\zeta\searrow 0$ does not lead to a singularity of
the region $\AN_{\wtq_{0,7}}|_{W_1}\geq 0$.

Let us next consider moving $\C(7)$
upward in $\zeta$ on a straight line at $\theta>-7\pi+0.1\pi$.
We look at what happens to the circle $C_-$ or its deformations.
The picture below $\zeta_1$ is similar to the case of
$\sDelta_1\theta=0.1\pi$ (depicted in Fig.~\ref{fig:0702})
in that the red point is confined in a peninsula bounded by $W_p$
which becomes smaller as $\zeta$ comes closer to $\zeta_1$.
The only dependence on $\theta$ is in the details of
how the peninsula is formed:
for low values of $\sDelta_1\theta$ it is formed after an isthmus is pinched off
as in Fig.~\ref{fig:0702},
but for larger $\sDelta_1\theta$ it is formed by expansion of an inland sea.
In any case, the peninsula disappears as $\zeta$ goes past $\zeta_1$.
The threshold can be found again
by looking for the singularity of the region $\AN_{\wtq_{0,7}}\geq 0$
on the wall $W_p$, as in Fig.~\ref{fig:07Wp}.
Numerical analysis suggests that it is exactly $\zeta_1$.
Indeed, this is proven analytically in Section~\ref{subsec:noentry}.
Thus, for all $\theta>-7\pi$, the circle
$C_-$ and any of its deformation are forced to either
go out of the admissible region or
hit the wall $W_p$ as $\zeta$ reaches $\zeta_1$.

To summarize, the no-entry zone ${\mathscr R}_{0,7}$ extends
indefinitely to the right of $t_1$. The upper boundary is a curve
for $-7\pi\leq \theta\leq -6\pi$ and continues to $\zeta=7\log 2$ beyond.
The lower boundary is $\zeta=\zeta_1$ for all $\theta\geq -7\pi$.
By a similar analysis, we find that the no-entry zone
${\mathcal R}_{1,7}$ ({\it resp}. ${\mathcal R}_{2,6}$)
extends indefinitely to the right of $t_2$ ({\it resp}. $t_3$):
the upper boundary is a curve for $-8\pi\leq \theta\leq -6\pi$
({\it resp}. $-7\pi\leq\theta\leq -4\pi$) and continues to
$\zeta=7\log 2$ beyond, while the lower boundary is $\zeta=\zeta_2$
({\it resp}. $\zeta_3$) for all $\theta\geq -8\pi$
({\it resp}. $\theta\geq -7\pi$).
The no-entry zones ${\mathcal L}_{0,0}$,
${\mathcal L}_{1,0}$ and ${\mathcal L}_{2,-1}$ extend indefinitely to
the left, as can be found from ${\mathcal R}_{0,7}$,
${\mathcal R}_{1,7}$ and ${\mathcal R}_{2,6}$
by the reflection symmetry.

During this analysis, we observed that the upper boundary curve of
${\mathcal R}_{1,7}$ goes through $t_1$. In fact, it
matches exactly the upper boundary curve
of ${\mathcal R}_{0,7}$ for $-7\pi\leq \theta\leq -6\pi$.
Also, the $\theta\to\theta-2\pi$ shift of the upper boundary curve of
${\mathcal R}_{2,6}$ goes through both $t_2$ and $t_1$,
and matches exactly the upper boundary curve of ${\mathcal R}_{1,7}$
for $-8\pi\leq \theta\leq -6\pi$.
Indeed, the curves are all described by the same single equation
$\zeta=\zeta(\theta,7-1/2)$ given in (\ref{threshold1}).
This is because the restriction of $\AN_{\wtq_{h,i}}$
on the wall $W_1\subset \{\Rs_1=0\}$ depends only on $i$ but not on $h$.

This means that $\AN_{\wtq_{0,7}}|_{W_1}\geq 0$ is singular on the curve
also for $\theta<-7\pi$. But we know that the representation $\C(7)$
can be safely transported along the straight path in
$-21\pi<\theta<-7\pi$ via the contour $\bgamma_0$.
This part of the curve is a ``harmless singularity''.
We would like to understand the difference from
``harmful singularity'' which would obstruct the transport.
To see what is going on, let us look at the admissible
regions at points around the curve, say, along the line $\theta=-7.5\pi$.
The region $\AN_{\wtq_{0,7}}|_{W_1}\geq 0$ is singular at
$\zeta=\zeta_c:=\zeta(-7.5\pi,7-1/2)\simeq 3.12922$
 and the singular point is
$(\Rs_2,\Is_2)=R(\sin{3\pi\over 7},\cos{3\pi\over 7})/\sqrt{2}$
which is at $\Is_2\simeq 0.689R$ on the level $\simeq 0.611R$ slice.
\begin{figure}[thb]
  \centerline{
    \includegraphics[page=134]{main-pics.pdf}}
  \caption{} \label{fig:fakepglobal}
\end{figure}
Fig.~\ref{fig:fakepglobal} shows the level $0.611R$ slice at the point
$(\zeta,\theta)=(\zeta_c+0.0002,-7.5\pi)$ just above the curve.
At first sight, nothing is special.
We can even draw a circle $C$ in the shaded region as shown in black,
and it can be continued as $\zeta$ is moved below $\zeta_c$.
Let us take a closer look at the red square around the singular point
$(\Rs_1,\Rs_2)=(0,0.689R)$.
Fig.~\ref{fig:fakep}-(a), (b), (c) show the enlarged views of the slices
at levels $0.63R$, $0.611R$, $0.59R$ respectively.
\begin{figure}[thb]
  \begin{subfigure}[b]{.24\linewidth}
    \centering
    \includegraphics[page=135]{main-pics.pdf}
    \caption{}\label{fig:fakepA}
  \end{subfigure}
  \begin{subfigure}[b]{.24\linewidth}
    \centering
    \includegraphics[page=136]{main-pics.pdf}
    \caption{}\label{fig:fakepB}
  \end{subfigure}
  \begin{subfigure}[b]{.24\linewidth}
    \centering
    \includegraphics[page=137]{main-pics.pdf}
    \caption{}\label{fig:fakepC}
  \end{subfigure}
    \begin{subfigure}[b]{.24\linewidth}
    \centering
    \includegraphics[page=138]{main-pics.pdf}
    \caption{}\label{fig:fakepthree}
  \end{subfigure}
    \vspace{-0.4cm}
  \caption{}  \label{fig:fakep}
\end{figure}
We see nothing in (b) but find in (a) and (c)
a small inland sea of negative region with $W_1$ on one side. 
This is reflected in Fig.~\ref{fig:fakep}-(d) of the wall $W_1$
where the curves marked (a), (b), (c) are the intersections of $W_1$ and the
corresponding slices.
As $\zeta$ goes below $\zeta_c$, the shaded region on $W_1$ is disconnected
as in Fig.~\ref{fig:W1}-right.
Accordingly, the inland sea exists at all levels.
Exactly at $\zeta=\zeta_c$, 
the region $\AN_{\wtq_{0,7}}|_{W_1}\geq 0$ develops a singularity.
However, the presence or absence of the inland sea
has no effect on the presence of the admissible circle $C$.
This is the difference from the case of $-7\pi<\theta<-6\pi$
(cf. Fig.~\ref{fig:0701s}) where
the presence or absence of positive road,
which is responsible for the singularity,
has a dramatic consequence on the existence of the admissible circle.

Similarly, the restriction of $\AN_{\wtq_{h,i}}$ on the wall
$W_p\subset\{\Rs_1+\Rs_2=0\}$ is independent of $\theta$,
and hence so is the singularity of $\AN_{\wtq_{h,i}}|_{W_p}\geq 0$.
This means that $\AN_{\wtq_{h,i}}|_{W_p}\geq 0$ is singular on the line
$\zeta=\zeta_{h+1}$ for any value of $\theta$ including the values
at which the transport of $\Sym^h\fund(i)$ across the line
from below is perfectly fine.
This is another ``harmless singularity''.
Let us look at the admissible region for the representation
$\C(7)$ at points near the line $\zeta=\zeta_1$ of singularity but
in the safe range of $\theta$ for $\C(7)$.
\begin{figure}[thb]
  \centerline{
    \includegraphics[page=139]{main-pics.pdf}
    \includegraphics[page=140]{main-pics.pdf}
    \includegraphics[page=141]{main-pics.pdf}
    \includegraphics[page=142]{main-pics.pdf}
    \includegraphics[page=143]{main-pics.pdf}
    \includegraphics[page=144]{main-pics.pdf}}
  \caption{}
      \label{fig:z1pm}
\end{figure}
\begin{figure}[thb]
  \centerline{
    \includegraphics[page=145]{main-pics.pdf}
    \includegraphics[page=146]{main-pics.pdf}
    \includegraphics[page=147]{main-pics.pdf}
    \includegraphics[page=148]{main-pics.pdf}
    \includegraphics[page=149]{main-pics.pdf}
    \includegraphics[page=150]{main-pics.pdf}}
  \caption{}
      \label{fig:z1mm}
\end{figure}
Fig.~\ref{fig:z1pm} and Fig.~\ref{fig:z1mm}
show the regions at the points above and below
the line respectively, with $\sDelta_1\zeta=+0.01$ and $-0.01$,
both with $\theta=-7.1\pi$ in the safe region.
The levels are
$-0.97R$, $-0.935R$, $-0.9R$, $-0.8R$, $-0.5R$, $0$ from left to right,
just as in Fig.~\ref{fig:0702} and \ref{fig:z1pp}.
We see that the only visible difference appears on the third from left
(on level $-0.9R$).
In Fig.~\ref{fig:z1pm}, there is a gap between the upper part
and the lower part near the center but the gap is absent in
Fig.~\ref{fig:z1mm}.
This is reflected in 
\begin{figure}[thb]
  \centerline{
    \includegraphics[page=151]{main-pics.pdf}~~~~
    \includegraphics[page=152]{main-pics.pdf}}
    \caption{}
      \label{fig:z1various}
\end{figure}
Fig.~\ref{fig:z1various} which shows the wall $W_p$ for
$\sDelta_1\zeta=+0.01$ (left) and $\sDelta_1\zeta=-0.01$ (right).
The concentric circles are the intersections with the slices:
the smallest circle at level $-0.97R$
and the largest one, the boundary of $W_p$, at level 0.
As $\zeta\to \zeta_1$, the upper and the lower parts
touch each other at one point on the wall $W_p$,
and the region $\AN_{\wtq_{0,7}}|_{W_p}\geq 0$ develops a singularity.
For both $\sDelta_1\zeta=+0.01$ and $-0.01$,
the colored points can go through the shaded region continuously from
level $\sim -R$ to level $0$, as shown in the figures,
and a closed circle can be found in the admissible region.
The presence or absence of the gap, which is responsible for
the singularity, has no effect on the existence of the admissible circle.
This is in contrast to the difference between
 Fig.~\ref{fig:z1pp} and Fig.~\ref{fig:0702}:
the absence or presence of the peninsula, which is responsible for
the singularity, has a dramatic consequence on the existence of the
admissible circle.

\subsection{Weyl Invariance}
\label{subapp:Weylinv}

Let us examine Weyl invariance of admissible circles.
The fixed point set of the Weyl transformation $w$ of $S^3_R$
is the diagonal $D=\{\usigma_1=\usigma_2\}$ which is a great circle.
It intersects the walls at the top and the bottom level points,
with  $W_1\cap W_2$ and $W_p$ respectively,
and the complement is the union of open intervals $D_+$ and $D_-$
consisting of points with positive and negative real parts $\Rs_1=\Rs_2$.
A circle $C$ in $S^3_R\setminus{\it walls}$ can be regarded as
a continuous map from $S^1$ to $S^3_R\setminus{\it walls}$.
It is Weyl invariant when $w\circ C=C\circ g$ for an isomorphism
$g$ of $S^1$. $g$ is orienation reversing since $C$
is a deformation of the real circle $C_{\rm real}$ for which
it is orientation reversing.  
It has exactly two fixed points and their images under $C$ must be
in $D_+\sqcup D_-$.
When $C$ represents a generator of $\pi_1(S^3_R\setminus{\it walls})\cong\Z$,
which is the case we consider, the two image points are in the
different components, since the winding number would be even otherwise.
Thus, any Weyl invariant circle $C$ of the type we consider
must pass through $D_+$ and $D_-$.

The Weyl reflection $w$ fixes
$\Is_+$ and hence acts on each slice of $S^3_R$.
The action is the reflection with respect to the real diagonal line
$\Rs_1=\Rs_2$ combined with
the exchange of the northern and the southern hemispheres.
It has two fixed points ${\bf p}_{\pm}\in D_{\pm}$
which are the intersection of the equator and the real diagonal line,
with sign $\pm$.
Note that a circle in $S^3_R$ is Weyl invariant if and only if
the intersection with each slice is invariant.

Now, let us examine the circles for the transport of the representation
$\C(0)$ through the window {\bf C}
which we have considered in Appendix~\ref{subapp:curved}.
The only non-trivial part is the interpolation between
$C_0$ and $C_-$ at the corner point $\mathtt{R}_0$
as depicted in Fig.~\ref{fig:00R0contours}.
It is clear that each circle can be chosen to be
invariant under the Weyl reflection.

We next discuss our main target: the Weyl invariance of the circles for
the transport of the representation $\Sym^2\fund(6)$ through the window {\bf C}.
Let us first examine the family of circles which we have considered in
Appendix~\ref{subapp:curved}.
The non-trivial part is the family on the level $0.1R$ slice
as depicted in Fig.~\ref{fig:26contours}.
The circle in the right figure (at $\mathtt{R}_{15}$) looks Weyl invariant
while the ones in the left and middle figures
(at $\mathtt{M}_0$ and $\mathtt{R}_0$) do not.
This is related to the general fact that a Weyl invariant circle
must go through $D_+$ and $D_-$, and hence a circle in a single slice must go
through the diagonal points ${\bf p}_+$ and ${\bf p}_-$ on the equator.
The point ${\bf p}_+$ appears to be in the shaded region in the right figure
but does not in the left and the middle figures.
This is consistent with the formula (\ref{ANdiag})
of the growth function on the diagonal $D$ with
$\Rs_1=\Rs_2=:\Rs$ and $\Is_1=\Is_2=:\Is$,
which reads for $D_+$ with $\Rs>0$ as
\beq
\AN_{\wtq_{2,6}}|_{D_+}=2(\sDelta_+\zeta\,\Is-\sDelta_1\theta\,\Rs).
\label{AN26posdiag}
\eeq
As $15>7\log 2\simeq 4.85>0$,
its positivity requires the level $\Is_+=\sqrt{2}\Is$
to be positive at $\mathtt{R}_{15}$
and negative at $\mathtt{M}_0$ and $\mathtt{R}_0$.
Thus, Weyl invariance of an admissible circle at the positive level $0.1R$
is possible at $\mathtt{R}_{15}$ but is impossible at
$\mathtt{M}_0$ and $\mathtt{R}_0$.

Let us discuss all possible circles at points on the open segment
$S_1=\{\zeta=\zeta_1,\,0<\sDelta_1\theta<2\pi\}$
through which any member of the window {\bf C} must go.
As noted above, any Weyl invariant circle that generates
$\pi_1(S^3_R\setminus {\it walls})$ must go through the diagonal
$D_+\sqcup D_-$. Our focus will be where in $D_+$ it goes.
As $\sDelta_+\zeta_1=\zeta_1-7\log 2$ is negative and
$\sDelta_1\theta$ is positive, positivity of $\AN_{\wtq_{2,6}}|_{D_+}$
requires $\Is_1=\Is_2$ to be negative.
The upper bound of the level $\Is_+$ is found by setting
it equal to zero. It is
\beq
\Is_+^{\rm c}
=-{R\,\sDelta_1\theta\over\sqrt{(\sDelta_+\zeta_1)^2+(\sDelta_1\theta)^2}},
\eeq
and is monotonically decreasing as a function of $\sDelta_1\theta\in [0,2\pi]$,
with the value from $0$ to $\simeq -0.993R$.
Thus, any Weyl invariant circle in the admissible region must go through
the $\Rs_+>0$ side of $W_p$ with levels $\Is_+^{\rm c}\leq \Is_+\leq 0$.
Let us look at the slice at
level $\Is_+^{\rm c}$ for various $\sDelta_1\theta$'s.
\begin{figure}[thb]
  \centerline{\includegraphics[page=153]{main-pics.pdf}~~~~
    \includegraphics[page=154]{main-pics.pdf}~~~~
    \includegraphics[page=155]{main-pics.pdf}}
\caption{}  \label{fig:Weylvc}
\end{figure}
Fig.~\ref{fig:Weylvc} shows the slice for
$\sDelta_1\theta=0.001\pi$, $0.2\pi$, $\pi$ from left to right,
with $\Is_+^{\rm c}=-0.0043R$, $-0.652R$, $-0.974R$ respectively.
The figures also include the meridian at $\Rs_1=\Rs_2\geq 0$, shown in red,
which divides the $\Rs_+>0$ side into halves.
We see that the point ${\bf p}_+$,
at the intersection of the meridian and the equator,
is indeed on the border between the shaded and the unshaded regions.
The picture for a general $\sDelta_1\theta$ falls into one of
these three patterns, depending on where it belongs to among
the three parts of $S_1$ separated by two values $s_*\simeq 0.014\pi$ and
$s_{**}\simeq 0.518\pi$ of $\sDelta_1\theta$ which have
$\Is^{\rm c}_{+*}\simeq -0.066R$ and
$\Is^{\rm c}_{+**}\simeq -0.912R$.
We shall examine whether there is a continuous family of
Weyl invariant configurations of points in the shaded regions of the slices,
going through $D_+$, and thus focusing on the $\Rs_+>0$ side of the wall $W_p$.
Note that a fundamental region of this side is given by
the quarter of the slice, sandwiched between the red meridian
and the $\Rs_1=-\Rs_2>0$ part of the wall $W_p$,
which we refer to as \emph{the quarter $Q$}.

If $s_*<\sDelta_1\theta<s_{**}$, the picture at the level $\Is_+^{\rm c}$
is as in Fig.~\ref{fig:Weylvc}-middle.
We see that the quarter $Q$ is in the unshaded region
except the diagonal point ${\bf p}_+$ on the border.
Thus, at this level,
the only possible Weyl invariant configuration of points in the (barely)
admissible region is at ${\bf p}_+$.
However, at a level just above $\Is_+^{\rm c}$,
the shaded region recedes from the red meridian,
and the point ${\bf p}_+$ belongs to the strictly unshaded region.
\begin{figure}[thb]
  \centerline{\includegraphics[page=156]{main-pics.pdf}}
\caption{}  \label{fig:WeylenlB}
\end{figure}
See Fig.~\ref{fig:WeylenlB} for the enlarged view near ${\bf p}_+$
of the slice at level $-0.651R$ just above $\Is_+^{\rm c}\simeq -0.652R$
of Fig.~\ref{fig:Weylvc}-middle.
Thus, any Weyl invariant configuration of points at the level $\Is_+^{\rm c}$
cannot be continued to a higher level.

If $s_{**}< \sDelta_1\theta<2\pi$, the picture at the level $\Is_+^{\rm c}$
is as in Fig.~\ref{fig:Weylvc}-right.
We see that some part of the quarter $Q$ is in the shaded region,
and it remains so for some levels above $\Is_+^{\rm c}$.
However, at even higher levels, the quarter $Q$ 
is entirely in the unshaded region,
\begin{figure}[thb]
  \centerline{\includegraphics[page=157]{main-pics.pdf}}
\caption{}  \label{fig:Weyl9C}
\end{figure}
as shown in Fig.~\ref{fig:Weyl9C}
(for $\sDelta_1\theta=\pi$ and $\Is_+=-0.900R$).
Thus, there is no Weyl invariant configuration of points in the shaded region.
As $\sDelta_1\searrow s_{**}$, the part of $Q$ in the shaded region at
$\Is_+^{\rm c}$ shrinks to the north and the south poles on the border.
At a level just above $\Is_+^{\rm c}$, the shaded region recedes
and the poles are in the unshaded region.
Thus, the conclusion at the point $\sDelta_1\theta=s_{**}$
is the same as in the case of $s_{**}< \sDelta_1\theta<2\pi$.

If $0<\sDelta_1< s_*$,
the picture at level $\Is_+^{\rm c}$ is as in Fig.~\ref{fig:Weylvc}-left.
In addition to the point ${\bf p}_+$ on the border,
we see that a part of $Q$ near the wall $W_p$ is in the shaded region.
At a level just above $\Is_+^{\rm c}$,
the shaded region recedes from the diagonal point ${\bf p}_+$
as in the case of $s_{*}<\sDelta_1<s_{**}$.
See Fig.~\ref{fig:WeylA}-left for the enlarged view near ${\bf p}_+$
at level $-0.0040R$ just above $\Is_+^{\rm c}\simeq -0.0043R$
of Fig.~\ref{fig:Weylvc}-left.
\begin{figure}[thb]
  \centerline{\includegraphics[page=158]{main-pics.pdf}~~~~
    \includegraphics[page=159]{main-pics.pdf}}
\caption{}  \label{fig:WeylA}
\end{figure}
Thus, a configuration of points at ${\bf p}_+$ cannot be continued above
$\Is_+^{\rm c}$.
On the other hand, at all higher levels up to 0,
a part of $Q$ near $W_p$ remains in the shaded region,
and its Weyl image is well inside the shaded region.
Thus, we do have a family of Weyl invariant configurations
of points in the shaded regions for $\Is_+^{\rm c}\leq \Is_+\leq 0$.
However, it must be connected to an admissible point in the diagonal $D_+$,
which exists at levels below $\Is_+^{\rm c}$.
Unfortunately, the shaded part of $Q$ near $W_p$ disappears
if we go lower in the level,
as shown in Fig.~\ref{fig:WeylA}-right
(for $\sDelta_1\theta=0.001\pi$ and $\Is_+=-0.3R$).
Thus, there is no continious family of Weyl invariant 
configurations of points in the admissible regions that includes
a point in $D_+$.
Things are mostly the same at $\sDelta_1\theta=s_*$.
The only difference at level $\Is_+^{\rm c}$ is that the non-negative part of
$Q$ in addition to ${\bf p}_+$ is the intersection point of the wall $W_p$
and the equator, which is on the border.
It goes into the unshaded region just below
the level $\Is_+^{\rm c}$ and hence does not connect to a point in $D_+$.
Thus, the conclusion is the same as in the case of
$0<\sDelta_1\theta<s_*$.

Thus, in all the three cases where $\sDelta_1\theta$ is in
$(0,s_*]$, $(s_*,s_{**})$ and $[s_{**},2\pi)$, i.e. at any point of $S_1$,
a Weyl invariant circle does not exist in the admissible region.
This proves that there is no family of Weyl invariant admissible circles
for $\Sym^2\fund(6)$ along any path in the window {\bf C}.

Finally, we note that the families of circles that will eventually be
obstructed, such as those in Fig.~\ref{fig:07contours1}
and Fig.~\ref{fig:2m1contours},
are obviously not Weyl invariant. If we required the invariance,
the obstruction would have occured much earlier.

\section{The Periods}
\label{app:periods}
For the readers' convenience we write down the Picard-Fuchs operators and
the series expansions of the solutions (periods) up to order $3$ in the
two limits of the R\o dland and Hosono-Takagi models. The periods are written 
 in the form
\begin{eqnarray}
\Pi_0&=&a_0(z)\label{per-struct1}\\
\Pi_1&=&\frac{1}{2\pi i}\left(a_0(z)\log z+a_1(z)\right)\label{per-struct2}\\
\Pi_2&=&\frac{1}{(2\pi i)^2}\left(a_0(z)\log^2z+2a_1(z)\log z+a_2(z)
\right)\label{per-struct3}\\
\Pi_3&=&\frac{1}{(2\pi i)^3}\left(a_0(z)\log^3 z+3a_1(z)\log^2z
+3a_2(z)\log z+a_3(z)\right),\label{per-struct4}
\end{eqnarray}
where $a_i(z)$ are polynomials in $z$. We shall write $\theta$ for
$z\frac{d}{dz}$.
\subsection{R\o dland Model}

On the mirror of the large volume limit of $X$,
the Picard-Fuchs operator is
\begin{eqnarray}
\mathcal{L}_{\zeta\gg0}&=&9 \theta^4-z\left(519 \theta^4
+1020 \theta^3+816 \theta^2+306\theta+45\right)\nonumber\\
&&-z^2\left(2258 \theta^4+10064
   \theta^3+15194 \theta^2+9546 \theta+2166\right)\nonumber\\
&&+z^3\left(1686
   \theta^4+5256 \theta^3+4706 \theta^2+1350
   \theta+12\right)\nonumber\\
&&-z^4\left(295 \theta^4+608 \theta^3+478
   \theta^2+174 \theta+26\right)\nonumber\\
&&+z^5(\theta+1)^4,
\label{periodRodGra}
\end{eqnarray}
and the periods have the structure (\ref{per-struct1})-(\ref{per-struct4})
with
\begin{eqnarray}
a_0^{\zeta\gg0}(z)&=&1+5 z+109 z^2+3317 z^3+O\left(z^4\right)\nonumber\\
a_1^{\zeta\gg0}(z)&=&14 z+357 z^2+\frac{35105 z^3}{3}
+O\left(z^4\right) \nonumber\\
a_2^{\zeta\gg0}(z)&=&\frac{28 z}{3}+\frac{1477 z^2}{3}+\frac{540757
   z^3}{27}+O\left(z^4\right) \nonumber\\
a_3^{\zeta\gg0}(z)&=&-56 z-1029 z^2-\frac{642089 z^3}{27}+O\left(z^4\right).
\end{eqnarray}
On the mirror of the large volume limit of $Y$,
the Picard-Fuchs operator is
\begin{eqnarray}
\mathcal{L}_{\zeta\ll0}&=&\theta^4-z\left(295 \theta^4
+572 \theta^3+424 \theta^2+138\theta+17\right)\nonumber\\
&&+z^2\left(1686 \theta^4+1488
   \theta^3-946 \theta^2-962 \theta-202\right)\nonumber\\
&&-z^3\left(2258
   \theta^4-1032 \theta^3-1450 \theta^2-318
   \theta+8\right)\nonumber\\
&&-z^4\left(519 \theta^4+1056 \theta^3+870
   \theta^2+342 \theta+54\right)\nonumber\\
&& +9z^5 (\theta+1)^4,
\end{eqnarray}
and the periods are (\ref{per-struct1})-(\ref{per-struct4}) with
\begin{eqnarray}
a_0^{\zeta\ll0}(z)&=&1+17 z+1549 z^2+215585 z^3
+O\left(z^4\right) \nonumber\\
a_1^{\zeta\ll0}(z)&=&70 z+7413 z^2+\frac{3268573 z^3}{3}
+O\left(z^4\right) \nonumber\\
a_2^{\zeta\ll0}(z)&=&84z+15687z^2+{8128729\over 3}z^2
+O\left(z^4\right) \nonumber\\
a_3^{\zeta\ll0}(z)&=& -504 z-36645 z^2-\frac{11142845 z^3}{3}
+O\left(z^4\right).
\label{periodRodPfa}
\end{eqnarray}

\subsection{Hosono-Takagi Model}

On the mirror of the large volume limit of $X$,
 the Picard-Fuchs operator is
\begin{eqnarray}
\mathcal{L}_{\zeta\gg0}&=&49 \theta^4 - z (1085 \theta^4 + 2002 \theta^3
+ 1638 \theta^2 + 637 \theta + 98) \nonumber\\
&&-   z^2 (16105 \theta^4 + 68044 \theta^3 + 102261 \theta^2
+ 66094 \theta + 15736) \nonumber\\
&&+ z^3 (21000 \theta^4 + 68712 \theta^3 + 72568 \theta^2
+ 30072 \theta + 3808) \nonumber\\
&&- z^4 (7440 \theta^4 + 20256 \theta^3 + 23024 \theta^2
+ 12896 \theta + 2944) \nonumber\\
&& + z^5 (512 (\theta + 1)^4),
\end{eqnarray}
and the periods are (\ref{per-struct1})-(\ref{per-struct4}) with
\begin{eqnarray}
a_0^{\zeta\gg0}(z)&=&1+2 z+34 z^2+488 z^3+O\left(z^4\right)\nonumber\\
a_1^{\zeta\gg0}(z)&=& 5 z+\frac{175 z^2}{2}+\frac{4280 z^3}{3}
+O\left(z^4\right)\nonumber\\
a_2^{\zeta\gg0}(z)&=& \frac{20 z}{7}+\frac{580 z^2}{7}+\frac{120665
   z^3}{63}+O\left(z^4\right)\nonumber\\
a_3^{\zeta\gg0}(z)&=&-\frac{120 z}{7}-\frac{1335 z^2}{7}-\frac{158080
   z^3}{63}+O\left(z^4\right).
\label{periodHTRey}
\end{eqnarray}
In the mirror of the large volume limit of $Y$,
 the Picard-Fuchs operator is
\begin{eqnarray}
\mathcal{L}_{\zeta\ll0}&=&\theta^4 - z (465 \theta^4 + 594 \theta^3
+ 431 \theta^2 + 134 \theta + 16) \nonumber\\
&&+ z^2 (42000 \theta^4 + 30576 \theta^3 - 15136 \theta^2
- 14144 \theta - 2816) \nonumber\\
&&- z^3 (1030720 \theta^4 - 231936 \theta^3 - 335424 \theta^2
- 82176 \theta - 2304) \nonumber\\
&&- z^4 (2222080 \theta^4 + 4788224 \theta^3 + 4386816 \theta^2
+ 1992704 \theta + 372736)\nonumber\\
&& + z^5 (3211264 (\theta + 1)^4),
\end{eqnarray}
and the periods are (\ref{per-struct1})-(\ref{per-struct4}) with
\begin{eqnarray}
a_0^{\zeta\ll0}(z)&=&1+16 z+1816 z^2+310336 z^3
+O\left(z^4\right) \nonumber\\
a_1^{\zeta\ll0}(z)&=&70 z+9065 z^2+\frac{4895800 z^3}{3}
+O\left(z^4\right) \nonumber\\
a_2^{\zeta\ll0}(z)&=& 110 z+\frac{44095 z^2}{2}+\frac{41001320 z^3}{9}
+O\left(z^4\right)\nonumber\\
a_3^{\zeta\ll0}(z)&=& -660 z-\frac{113445 z^2}{2}-\frac{65061625 z^3}{9}
+O\left(z^4\right).
\label{periodHTDet}
\end{eqnarray}

\section{Construction Of The Branes $\mathfrak{L}^h$}
\label{app:Lbconstruction}

\newcommand{\lrarrow}{~\substack{\displaystyle\to\\[-0.05cm]\displaystyle\leftarrow}~}
\newcommand{\plrarrow}{~\substack{\displaystyle\nearrow\\[-0.1cm]\displaystyle\swarrow}~}
\newcommand{\nlrarrow}{~\substack{\displaystyle\searrow\\[-0.1cm]\displaystyle\nwarrow}~}

In this appendix, we describe the construction of the height 2
matrix factorizations $\Lb^h$ by binding Lascoux matrix factorizations.
In the main text, the construction is described as
a two step process ---first bind the Lascoux complexes and then
find the matrix factorization completion. That is conceptually simple
and worked in a simple case such as $\Lb^2$.
However, in the actual construction of the complicated cases as $\Lb^1$
and $\Lb^0$, it is more practical to directly
bind the Lascoux matrix factorizations.

When constructing the complexes $\Lc^h$ and the matrix factorizations $\Lb^h$,
we shall take a slightly nonstandard convention
which does not flip the sign of the constituent differentials.
This is achieved by conjugation with ${\rm diag}(\sigma_E,{\rm id}_F)$
for the cone (\ref{cone1}) where $\sigma_E$ is the $\Z_2$ grading of $E$
up to an overall sign.
For the case of the double cone (\ref{cone2}),
the conjugation is by ${\rm diag}(\sigma_E,{\rm id}_F,\sigma_G)$.
The new differentials are respectively
\beq
\left(\begin{array}{cc}
      Q_E&0\\
      f\sigma_E&Q_F
\end{array}\right),~~~~
\left(\begin{array}{ccc}
  Q_E&0&0\\
  f\sigma_E&Q_F&0\\
  \sigma_Gh\sigma_E&\sigma_Gg&Q_G
\end{array}\right).
\label{newdiff}
\eeq

\subsection{Warm-up: $\Lb^2$ Revisited}

We start with reconstructing $\Lb^2$ as a direct binding of
matrix factorizations as a warm-up,
introducing some useful notation.

The brane is constrcuted by binding together
the Lascoux branes of the type (iv) in Fig.~\ref{fig:Lmf}
and of the type (iii)' in the sense of (\ref{otherLcLb}).
Let $B_{\rm iv}$ and $B_{\rm iii}$ be such branes whose height 3 ends
are at the same position in the $U(2)$-grid,
say $\sym^3\fund^*$.
We find a homomorphism $\chi:B_{\rm iii}\to B_{\rm iv}$
that extends the identity map of $\sym^3\fund^*$.
\begin{figure}[htb]
  \centerline{\includegraphics[page=160]{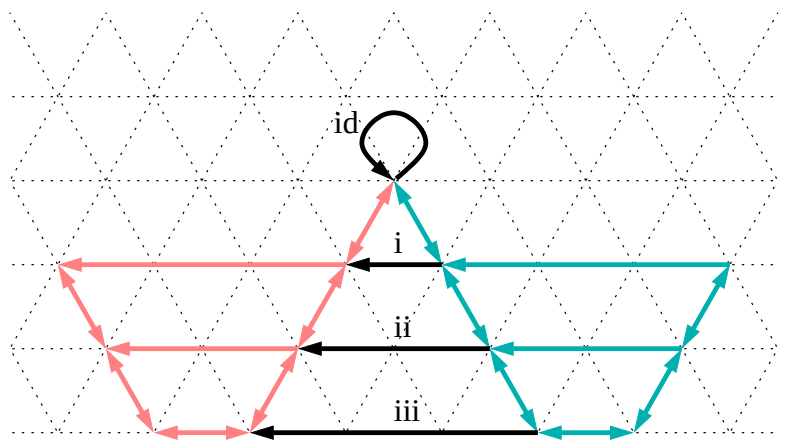}}
  \caption{The homomorphism $\chi$.}
  \label{fig:L2glue}
\end{figure}
Fig.~\ref{fig:L2glue} describes this homomorphism. The arrows with labels
i, ii, iii show the components of $\chi$ other than the identity of
$\sym^3\fund^*$. Their explicit forms are
\beq
\begin{array}{l}
  \chi_{\rm i}=-{2\over 3}\mathring{A},\\
  \chi_{\rm ii}={1\over 18}\bar{A}A+{1\over 3}\mathring{A}^2,\\
  \chi_{\rm iii}=-{1\over 6}\bar{A}\mathring{A}A-{2\over 9}\mathring{A}^3.
\end{array}
\eeq
In the above expressions, we use
\beq
\mathring{A}:=A^{ij}(p)\bareta_i\eta_j,\quad
\bar{A}:=A^{ij}(p)\bareta_i\bareta_j,\quad
A:=A^{ij}(p)\eta_i\eta_j.
\label{defA}
\eeq
These along with
\beq
\bar{\Box}:=\bareta_i\bareta^i,\quad
\Box:=\eta_i\eta^i
\label{defBox}
\eeq
will be used in the construction of $\Lb^1$ and $\Lb^0$.

\begin{figure}[thb]
\centerline{\includegraphics[page=161]{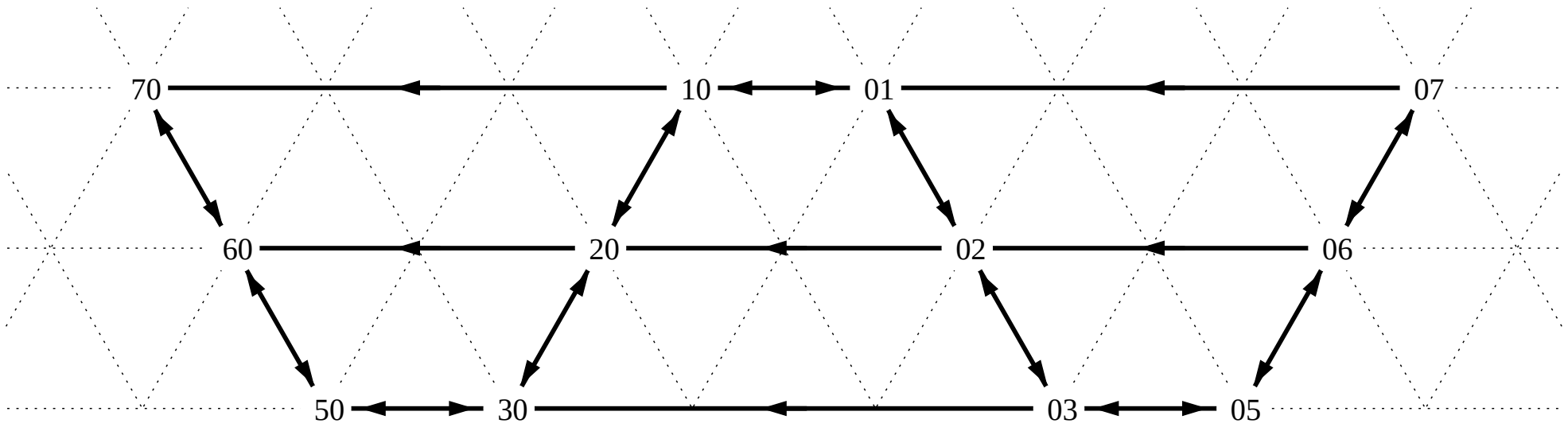}}
\caption{The brane $\mathfrak{L}^2$.}
\label{fig:L2}
\end{figure}

We take the cone of $\chi$
and annihilate the height 3 pair by the tachyon condensation of the identity
map. This induces a zigzag map. 
The resulting matrix factorization is as shown in Fig.~\ref{fig:L2}.
The vertex with label $pq$ has multiplicity
$\textw^{\!p,q}=\textw^{\!p}V^*\otimes \textw^{\!q}V$.
Some of the arrows are
\beqa
&&{\scriptstyle 10}\lrarrow{\scriptstyle 01}
    =\left\{\begin{array}{l}
    -(\searrow)(\nearrow),\\
    -\chi_{\rm i}
    \end{array}\right.\nn\\
&&{\scriptstyle 20}\leftarrow{\scriptstyle 02}=\chi_{\rm ii},\nn\\
&&{\scriptstyle 30}\leftarrow{\scriptstyle 03}=-\chi_{\rm iii},
\label{L2new}
\eeqa    
while the other arrows are the ones induced from the Lascoux
matrix factorizations.
The first line of (\ref{L2new}) is the zigzag map, in which
$(\nearrow)$ is (\ref{Lne}),
and $(\searrow)$ is (\ref{Lse}) with the replacement
$\eta^i\to \bareta^i$.
The alternating signs in front of $\chi_{\bullet}$'s in (\ref{L2new})
come from the conjugation by the $\Z_2$ grading as in ``$f\to f\sigma_E$'' in
(\ref{newdiff})-left.
Note that (\ref{L2new}) agrees with
(\ref{zigzag1}), (\ref{D6mf2}), (\ref{D6mf1}) and (\ref{D6mf0}).

\subsection{Construction Of $\Lb^1$}
\label{sub:L1}

The construction of $\Lb^1$ has two steps.

The first step is to bind
matrix factorizations of the type (v) and (ii)'. 
Let $B_{\rm v}$ and $B_{\rm ii}$ be such branes
whose height 4 ends are at the same position
in the $U(2)$-grid, say $\sym^4\fund^*$.
We find a homomorphism $\chi:B_{\rm ii}\to B_{\rm v}$
that extends the identity map of $\sym^4\fund^*$.
\begin{figure}[htb]
  \centerline{\includegraphics[page=162]{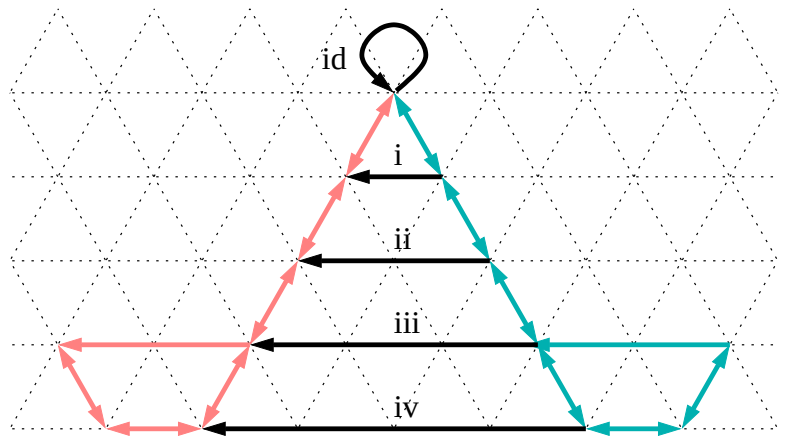}}
  \caption{The homomorphism $\chi$.}
  \label{fig:L1glue1}
\end{figure}
It is described in Fig.~\ref{fig:L1glue1} where the arrows with label i,
ii, iii, iv are given by
\beq
\begin{array}{l}
  \chi_{\rm i}=-{1\over 2}\mathring{A},\\
  \chi_{\rm ii}={1\over 48}\bar{A}A+{1\over 6}\mathring{A}^2,
\end{array}~~
\begin{array}{l}
  \chi_{\rm iii}=-{1\over 36}\bar{A}\mathring{A}A
  -{1\over 18}\mathring{A}^3,\\
  \chi_{\rm iv}={1\over 2^5\cdot 3^2}\bar{A}^2 A^2
  +{1\over 24}\bar{A}\mathring{A}^2A
  +{1\over 36}\mathring{A}^4
\end{array}
\eeq
We take its cone
and annihilate the height 4 pair by the tachyon condensation
of the identity map, which induces a zigzag map.
Let ${\rm C}(\chi)$ be the resulting matrix factorization.
It has two components of
height 3, $\sym^3\fund^*\otimes V$ and $\sym^3\fund^*(-1)\otimes V^*$.

In order to remove the height 3 components, we bind to ${\rm C}(\chi)$
the Lascoux branes of the type (iv) and (iii)' tensored with
$V$ and $V^*$, respectively.
Let $B_{\rm iv}$ and $B_{\rm iii}$ be such branes
which have $\sym^3\fund^*\otimes V$ and $\sym^3\fund^*(-1)\otimes V^*$
respectively at the height 3 ends.
We find homomorphisms $\varphi:{\rm C}(\chi)\to B_{\rm iv}$
and $\bar\varphi:B_{\rm iii}\to {\rm C}(\chi)$
that extend the identity maps at the height 3 positions.
We also find that
$\varphi\circ\bar\varphi:B_{\rm iii}\to B_{\rm iv}$
is null homotopic with a homotopy
$h:B_{\rm iii}\to B_{\rm iv}[-1]$.
\begin{figure}[htb]
  \centerline{\includegraphics[page=163]{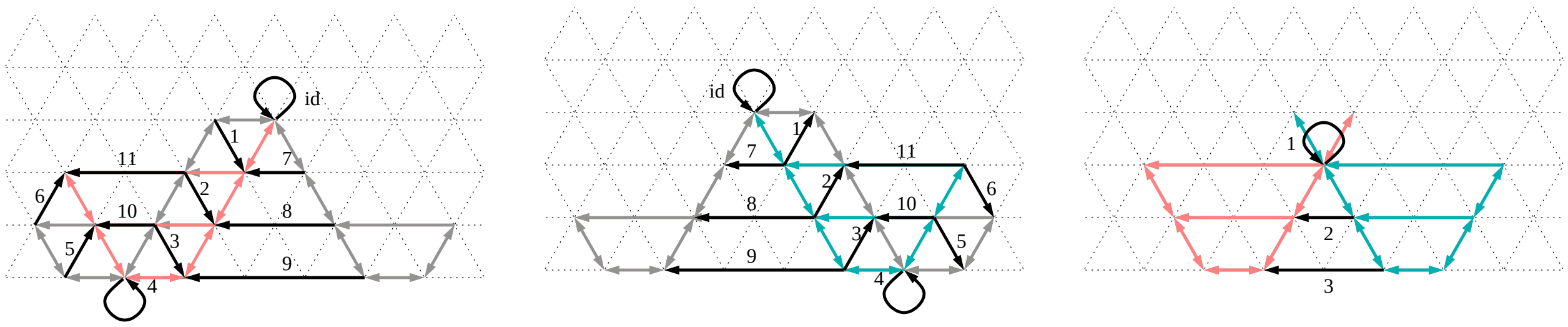}}
  \caption{The homomorphisms $\varphi$ (left),
    $\bar\varphi$ (middle)
    and the homotopy $h$ (right).}
  \label{fig:L1glue2}
\end{figure}
Fig.~\ref{fig:L1glue2} describes these maps. ${\rm C}(\chi)$, $B_{\rm iv}$
and $B_{\rm iii}$ are shown in grey, pink and light-blue.
The arrows with labels $1,2,\ldots$ show the components other
than the identity maps, with the explicit forms
\beq
\begin{array}{l}
\varphi_1={4\over 3}x_i^ae_a\bareta^i+{1\over 3}x_i^ae_a\bar\Box\eta^i,\\
\varphi_2=-2x_i^ae_a\bareta^i-{2\over 3}x_i^ae_a\bar{\Box}\eta^i,\\
\varphi_3=4x_i^ae_a\bareta^i+2x_i^ae_a\bar\Box\eta^i,\\
\varphi_4=4\bar\Box,\\
\varphi_5=-2x_{ia}e^a\bareta^i+2x_{ia}e^a\bar\Box\eta^i,\\
\varphi_6=2x_{ia}e^a\bareta^i,
\end{array}~~
\begin{array}{l}
\varphi_7=-{2\over 3}\mathring{A}+{1\over 12}\bar\Box A,\\
\varphi_8={1\over 18}\bar{A}A+{1\over 3}\mathring{A}^2
-{1\over 9}\bar\Box \mathring{A}A,\\
\varphi_9=-{1\over 6}\bar{A}\mathring{A}A-{2\over 9}\mathring{A}^3
+{1\over 36}\bar\Box\bar{A}A^2+{1\over 6}\bar\Box\mathring{A}^2A,\\
\varphi_{10}=4\bar{A}\bar\Box,\\
\varphi_{11}=\bar{A}^2\bar{\Box},
\end{array}
\label{varphiL1}
\eeq
\beq
\begin{array}{l}
\bar\varphi_1=x_{ia}e^a\eta^i-{1\over 3}x_{ia}e^a\Box\bareta^i,\\
\bar\varphi_2=-{4\over 3}x_{ia}e^a\eta^i+{2\over 3}x_{ia}e^a\Box\bareta^i,\\
\bar\varphi_3=2x_{ia}e^a\eta^i-2x_{ia}e^a\Box\bareta^i,\\
\bar\varphi_4=-4\Box,\\
\bar\varphi_5=-4x_i^ae_a\eta^i-2x_i^ae_a\Box\bareta^i,\\
\bar\varphi_6=2x_i^ae_a\eta^i,
\end{array}~~
\begin{array}{l}
\bar\varphi_7=-{2\over 3}\mathring{A}-{1\over 12}\bar{A}\Box,\\
\bar\varphi_8={1\over 18}\bar{A}A+{1\over 3}\mathring{A}^2
+{1\over 9}\bar{A}\mathring{A}\Box,\\
\bar\varphi_9=-{1\over 6}\bar{A}\mathring{A}^2A-{2\over 9}\mathring{A}^3
-{1\over 36}\bar{A}^2A\Box-{1\over 6}\bar{A}\mathring{A}^2\Box,\\
\bar\varphi_{10}=-4\Box A,\\
\bar\varphi_{11}=-A^2\Box,
\end{array}
\label{barvarphiL1}
\eeq
\beq
\begin{array}{l}
  h_1={4\over 3}-{1\over 3}\bar\Box\Box,\\
  h_2={4\over 3}\mathring{A}+{2\over 9}\bar{A}\Box-{2\over 9}\bar\Box A
  -{4\over 9}\bar\Box\mathring{A}\Box,\\
  h_3={1\over 3}\bar{A}A+{4\over 3}\mathring{A}^2
  -{2\over 3}\bar\Box\mathring{A}A
  +{2\over 3}\bar{A}\mathring{A}\Box
  -{2\over 9}\bar\Box\bar{A} A\Box
  -{2\over 3}\bar\Box\mathring{A}^2\Box
\end{array}
\eeq
We can take the double cone as in (\ref{cone2}), or
(\ref{newdiff})-right. We annihilate the height 3 pairs
by the tachyon condensation of the identity maps.
This yields a matrix factorization of height 2.

It can be simplified by additional brane-antibrane annihilations
at three invertible components of the differentials.
Two of them originate from $\varphi_4$ and $\bar\varphi_4$
acting only on the flavor factors as
\beq
\textw^{\!4,0}\to\textw^{\!5,1}
\quad\mbox{and}\quad
\textw^{\!1,5}\to\textw^{\!0,4}
\label{vp4fla}
\eeq
which are injective and surjective, respectively.
The tachyon condensation leaves us with the cokernel
of $\varphi_4$ and the kernel of $\bar\varphi_4$ along with zigzag maps.
The remaining one is the invertible component $h_1$ of the homotopy.
Its tachyon condensation eliminates the pair and induces zigzag maps.

\begin{figure}[thb]
\centerline{\includegraphics[page=164]{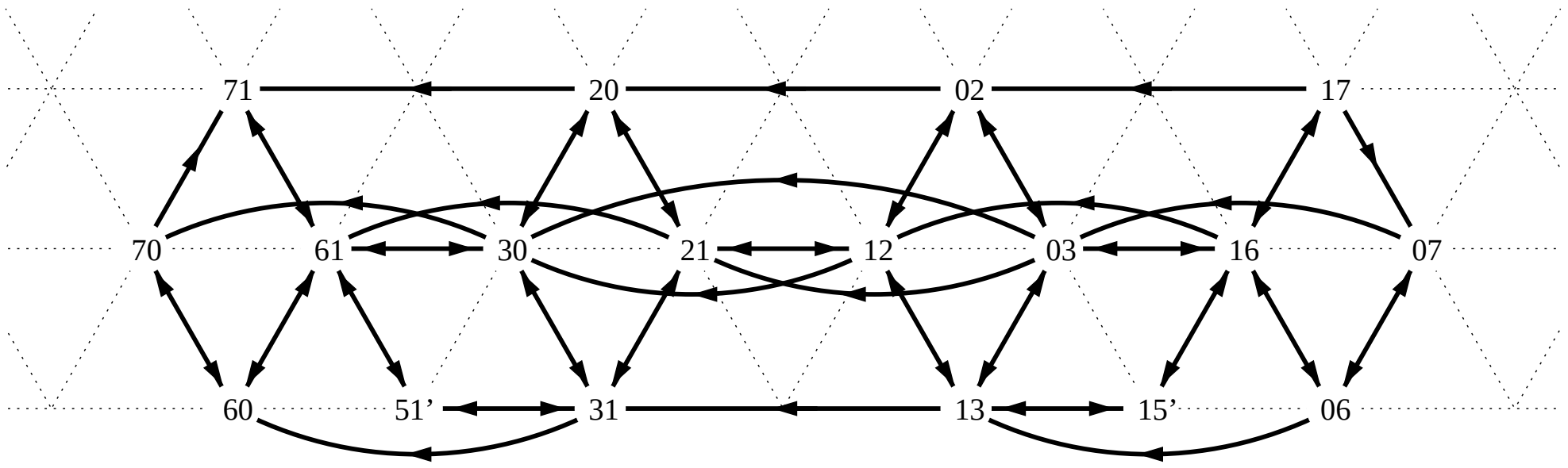}}
\caption{The brane $\mathfrak{L}^1$.}
\label{fig:L1}
\end{figure}

The resulting matrix factorization is as shown in Fig.~\ref{fig:L1}.
The vertex with label {\small $pq$} has multiplicity
$\textw^{\!p,q}=\textw^{\!p}V^*\otimes \textw^{\!q} V$,
with the following exceptions:
the vertex with label {\small 51'} is ${\rm Coker}(\varphi_4)$
and the one with {\small 15'} is ${\rm Ker}(\bar\varphi_4)$.
Some of the arrows are
\beqa
&{\scriptstyle 71}\leftarrow {\scriptstyle 20}=\varphi_{11},~~~
    {\scriptstyle 20}\leftarrow {\scriptstyle 02}
    =\chi_{\rm ii}-\bar\varphi_7h_1^{-1}\varphi_7,~~~
    {\scriptstyle 02}\leftarrow {\scriptstyle 17}=\bar\varphi_{11},
    \nn\\
&
{\scriptstyle 70}\nearrow {\scriptstyle 71}=\varphi_6,~~~
{\scriptstyle 20}\nlrarrow {\scriptstyle 21}=\left\{
\begin{array}{l}
  \varphi_2\\
    \bar\varphi_7 h_1^{-1}(\nearrow)
 \end{array}\right.\!\!\!,~~~
            {\scriptstyle 12}\plrarrow {\scriptstyle 02}=\left\{
            \begin{array}{l}\bar\varphi_2\\
                          (\searrow)h_1^{-1}\varphi_7
\end{array}\right.\!\!\!,~~~
{\scriptstyle 17}\searrow {\scriptstyle 07}=\bar\varphi_6,
\nn\\
&
    {\scriptstyle 61}\lrarrow {\scriptstyle 30}=\left\{
    \begin{array}{l}
      -(\nearrow)\varphi_4^{-1}(\searrow)\\
      -\varphi_{10}
    \end{array}\right.\!\!\!,~~~
    {\scriptstyle 21}\lrarrow {\scriptstyle 12}=\left\{
           \begin{array}{l}
             -(\searrow)h_1^{-1}(\nearrow)\\
             h_2
           \end{array}\right.\!\!\!,~~~
   {\scriptstyle 03}\lrarrow {\scriptstyle 16}=\left\{
   \begin{array}{l}
     -(\nearrow)\bar\varphi_4^{-1}(\searrow)\\
     -\bar\varphi_{10}
   \end{array}\right.\nn\\
&{\scriptstyle 30}\leftarrow {\scriptstyle 12}=\bar\varphi_8,~~~
           {\scriptstyle 30}\leftarrow {\scriptstyle 03}=-\chi_{\rm iii},~~~
           {\scriptstyle 21}\leftarrow {\scriptstyle 03}=\varphi_8,\nn\\
&{\scriptstyle 60}\plrarrow {\scriptstyle 61}=\left\{
           \begin{array}{l}-\varphi_5\\
            -(\leftarrow)\varphi_4^{-1}(\searrow)
           \end{array}\right.\!\!\!,~~~~~~~
{\scriptstyle 16}\nlrarrow {\scriptstyle 06}=\left\{
            \begin{array}{l}
             -\bar\varphi_5\\
           -(\nearrow)\bar\varphi_4^{-1}(\leftarrow)
           \end{array}\right.\!\!\!, 
\nn\\
&
{\scriptstyle 30}\nlrarrow {\scriptstyle 31}=\left\{
            \begin{array}{l}
             -\varphi_3\\
           -(\nearrow)\varphi_4^{-1}(\leftarrow)
           \end{array}\right.\!\!\!,~~~~
{\scriptstyle 13}\plrarrow {\scriptstyle 03}=\left\{
           \begin{array}{l}-\bar\varphi_3\\
            -(\leftarrow)\bar\varphi_4^{-1}(\searrow)
           \end{array}\right.\!\!\!,
\nn\\
&{\scriptstyle 60}\leftarrow {\scriptstyle 31}
           =-(\leftarrow)\varphi_4^{-1}(\leftarrow),~~~
{\scriptstyle 31}\leftarrow{\scriptstyle 13}=h_3,~~~
{\scriptstyle 13}\leftarrow{\scriptstyle 06}
=-(\leftarrow)\bar\varphi_4^{-1}(\leftarrow).
\label{L1arrows}
\eeqa
The other arrows are the ones
induced from the Lascoux matrix factorizations.
$\varphi_4^{-1}$ and $\bar\varphi_4^{-1}$ are left and right inverses
of $\varphi_4$ and $\bar\varphi_4$ respectively,
and $h_1^{-1}$ is the inverse of $h_1$,
whose explicit forms are
\beqa
&\displaystyle \varphi_4^{-1}={1\over 12}\Box, ~~~~
\bar\varphi_4^{-1}=-{1\over 12}\bar\Box\\
&\displaystyle h_1^{-1}={3\over 4}-{1\over 4}\bar\Box \Box.
\label{h1inv}
\eeqa
We use the indentification
${\rm Coker}(\varphi_4)\cong {\rm Ker}(\varphi_4^{-1})$
  and ${\rm Ker}(\bar\varphi_4)\cong {\rm Coker}(\bar\varphi_4^{-1})$
  to define the arrows out of 51' and the arrows into 15'.
Also, $(\searrow)$, $(\nearrow)$ and $(\leftarrow)$ in (\ref{L1arrows})
are the arrows of the Lascoux matrix factorizations, (\ref{Lse}), (\ref{Lne})
and (\ref{Lw}), or the second version
with the replacement $(\eta^i,\bar\eta_i)\to (\bareta^i,\eta_i)$.

\subsection{Construction Of $\Lb^0$}
\label{sub:L0}

The construction of $\Lb^0$ has three steps.

\begin{figure}[htb]
  \centerline{\includegraphics[page=165]{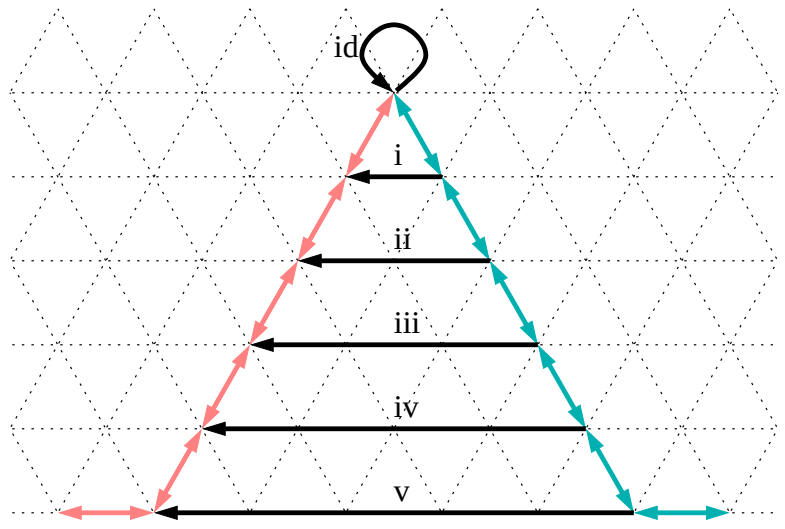}}
  \caption{The homomorphism $\chi$.}
  \label{fig:L0glue1}
\end{figure}

The first step is to bind
matrix factorizations of the type (vi) and (i)'.
Let $B_{\rm vi}$ and $B_{\rm i}$ be such branes
whose height 5 ends are at the same position
in the $U(2)$-grid, say $\sym^5\fund^*$.
We find a homomorphism $\chi:B_{\rm i}\to B_{\rm vi}$
that extends the identity map of $\sym^5\fund^*$,
as described in Fig.~\ref{fig:L0glue1}.
The explicit forms of the arrows are
\beq
\begin{array}{l}
  \chi_{\rm i}=-{2\over 5}\mathring{A},\\
  \chi_{\rm ii}={1\over 10}\mathring{A}^2+{1\over 100}\bar{A}A,\\
  \chi_{\rm iii}=-{1\over 3^2\cdot 5}\mathring{A}^3
  -{1\over 2^3\cdot 3\cdot 5}\bar{A}\mathring{A}A,
\end{array}~~
\begin{array}{l}
  \chi_{\rm iv}={1\over 2^2\cdot 3^2\cdot 5}\left(\mathring{A}^4
  +\bar{A}\mathring{A}^2 A
  +{1\over 2^4}\bar{A}^2A^2\right),\\
  \chi_{\rm v}={-1\over 2\cdot 3^2\cdot 5}\left(
      {1\over 5}\mathring{A}^5+{1\over 2}\bar{A}\mathring{A}^3 A
      +{1\over 8}\bar{A}^2\mathring{A} A^2\right)
\end{array}
\eeq
We take its cone
and annihilate the height 5 pair by the tachyon condensation
of the identity map.
Let ${\rm C}(\chi)$ be the resulting matrix factorization.
It has two components of
height 4, $\sym^4\fund^*\otimes V$ and $\sym^4\fund^*(-1)\otimes V^*$.

\begin{figure}[htb]
  \psfrag{k1}{$k_1$}
  \centerline{\includegraphics[page=167]{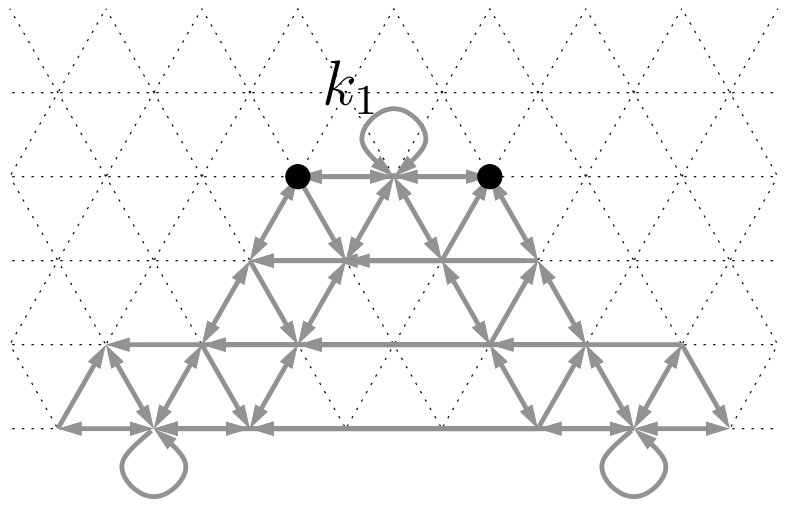}}
  \caption{The matrix factorization ${\rm C}(\psi,k,\bar\psi)$.}
  \label{fig:Cpsi}
\end{figure}

In order to remove the height 4 components, we bind to ${\rm C}(\chi)$
the Lascoux branes of the type (v) and (ii)' tensored with
$V$ and $V^*$, respectively.
Let $B_{\rm v}$ and $B_{\rm ii}$ be such branes
which have $\sym^4\fund^*\otimes V$ and $\sym^4\fund^*(-1)\otimes V^*$,
respectively, at the height 4 ends.
We find homomorphisms $\psi:{\rm C}(\chi)\to B_{\rm v}$
and $\bar\psi:B_{\rm ii}\to {\rm C}(\chi)$
that extend the identity maps at the height 4 positions.
We also find that
$\psi\circ\bar\psi:B_{\rm ii}\to B_{\rm v}$
is null homotopic with a homotopy
$k:B_{\rm ii}\to B_{\rm v}[-1]$.
\begin{figure}[htb]
  \centerline{\includegraphics[page=166]{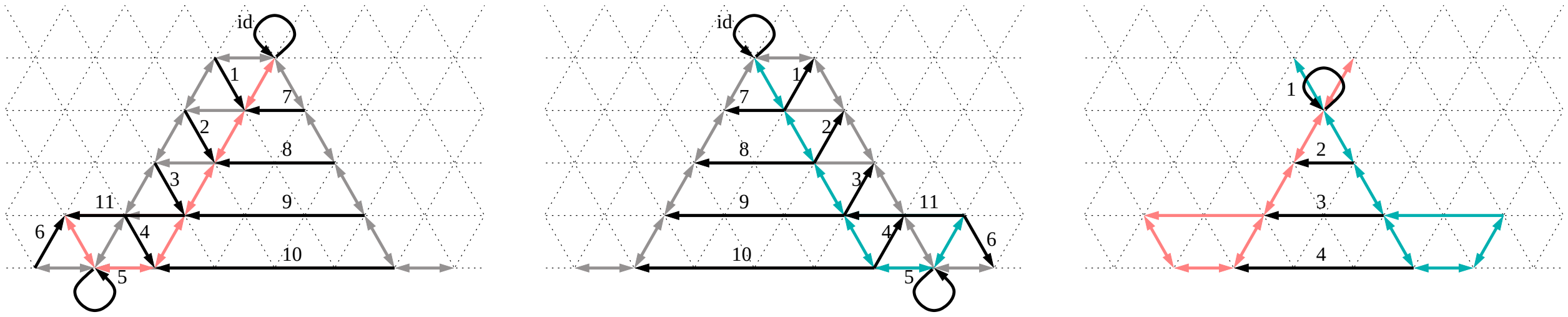}}
  \caption{The homomorphisms $\psi$ (left),
    $\bar\psi$ (middle)
    and the homotopy $k$ (right).}
  \label{fig:L0glue2}
\end{figure}
See Fig.~\ref{fig:L0glue2}. ${\rm C}(\chi)$, $B_{\rm v}$ and
$B_{\rm ii}$ are shown in grey, pink and light-blue.
The explicit forms of the arrows are
\beq
\begin{array}{l}
  \psi_1={5\over 4}x_i^ae_a\bareta^i+{1\over 4}x_i^ae_a\bar\Box\eta^i,\\
  \psi_2=-{5\over 3}x_i^ae_a\bareta^i-{5\over 12}x_i^ae_a\bar\Box\eta^i,\\
  \psi_3={5\over 2}x_i^ae_a\bareta^i+{5\over 6}x_i^ae_a\bar\Box\eta^i,\\
  \psi_4=-5x_i^ae_a\bareta^i-{5\over 2}x_i^ae_a\bar\Box\eta^i,\\
  \psi_5=-5\bar\Box,\\
  \psi_6=5x_{ia}e^a\bareta^i,
\end{array}~~
\begin{array}{l}
  \psi_7=-{1\over 2}\mathring{A}+{1\over 20}\bar\Box A,\\
  \psi_8={1\over 6}\mathring{A}^2+{1\over 2^4\cdot 3}\bar\Box\mathring{A} A
  -{1\over 2^3\cdot 3}\bar\Box\mathring{A}A,\\
  \psi_9={-1\over 2^2\cdot 3^2}\left(2\mathring{A}^3+\bar{A}\mathring{A} A
  -\bar\Box\mathring{A}^2A-{1\over 8}\bar\Box\bar{A}A^2\right),\\
  \psi_{10}={1\over 2^2\cdot 3^2}\left(
  \mathring{A}^4+{3\over 2}\bar{A}\mathring{A}^2 A
  +{1\over 8}\bar{A}^2A^2
  -\bar\Box\mathring{A}^3A
  -{1\over 2}\bar\Box\bar{A}\mathring{A}A^2\right),\\
  \psi_{11}=-5\bar\Box\bar{A},
\end{array}
\label{psiL0}
\eeq
\beq
\begin{array}{l}
  \bar\psi_1={5\over 4}x_{ia}e^a\eta^i-{1\over 4}x_{ia}e^a\bareta^i\Box,\\
  \bar\psi_2=-{5\over 3}x_{ia}e^a\eta^i+{5\over 12}x_{ia}e^a\bareta^i\Box,\\
\bar\psi_3={5\over 2}x_{ia}e^a\eta^i-{5\over 6}x_{ia}e^a\bareta^i\Box,\\
\bar\psi_4=-5x_{ia}e^a\eta^i+{5\over 2}x_{ia}e^a\bareta^i\Box,\\
\bar\psi_5=5\Box,\\
\bar\psi_6=5x_i^ae_a\eta^i,
\end{array}~~
\begin{array}{l}
\bar\psi_7=-{1\over 2}\mathring{A}-{1\over 20}\bar{A}\Box,\\
\bar\psi_8={1\over 6}\mathring{A}^2+{1\over 2^4\cdot 3}\bar{A}A
+{1\over 2^3\cdot 3}\bar{A}\mathring{A}\Box,\\
\bar\psi_9={-1\over 2^2\cdot 3^2}\left(2\mathring{A}^3+\bar{A}\mathring{A}A
+\bar{A}\mathring{A}^2\Box
+{1\over 8}\bar{A}^2A\Box\right),\\
\bar\psi_{10}={1\over 2^2\cdot 3^2}\left(\mathring{A}^4
+{3\over 2}\bar{A}\mathring{A}^2A
+{1\over 8}\bar{A}^2A^2
+\bar{A}\mathring{A}^3\Box
+{1\over 2}\bar{A}^2\mathring{A}A\Box\right),\\
\bar\psi_{11}=5A\Box,
\end{array}
\label{barpsiL0}
\eeq
\beq
\begin{array}{l}
  k_1={5\over 4}-{1\over 4}\bar\Box\Box,\\
  k_2={5\over 6}\left(\mathring{A}-{1\over 8}\bar\Box A+{1\over 8}\bar{A}\Box
  -{1\over 4}\bar\Box\mathring{A}\Box\right),\\
  k_3={5\over 2^2\cdot 3^2}\left(3\mathring{A}^2+{1\over 2}\bar{A}A
  -\bar\Box\mathring{A}A+\bar{A}\mathring{A}\Box
  -{1\over 4}\bar\Box\bar{A}A\Box
  -\bar\Box\mathring{A}^2\Box\right),\\
  k_4={5\over 2^3\cdot 3^2}\left(4\mathring{A}^3+3\bar{A}\mathring{A}A
  +3\bar{A}\mathring{A}^2\Box-3\bar\Box\mathring{A}^2A
  +{1\over 2}\bar{A}^2A\Box
  -{1\over 2}\bar\Box\bar{A}A^2
  -2\bar\Box\mathring{A}^3\Box
  -2\bar\Box\bar{A}\mathring{A}A\Box\right).
\end{array}
\eeq
Then, we can take the double cone as in (\ref{cone2}), or
(\ref{newdiff})-right. We annihilate the height 4 pairs
by the tachyon condensation of the identity maps.
Let us denote by ${\rm C}(\psi,k,\bar\psi)$ the resulting
matrix factorization. See Fig.~\ref{fig:Cpsi}.
It has height 3 components with multiplicities
$\textw^{\!2} V^*$, $(V^*\otimes V)\oplus (V\otimes V^*)$, $\textw^{\!2} V$
from the left to the right of the grid.
The middle two are connected by an invertible component 
$k_1$ of the homotopy
and can be eliminated by its tachyon condensation.
But the height 3 components at the edges, shown as the
bullets in Fig.~\ref{fig:Cpsi}, must be eliminated by additional processes.

\begin{figure}[htb]
  \centerline{\includegraphics[page=168]{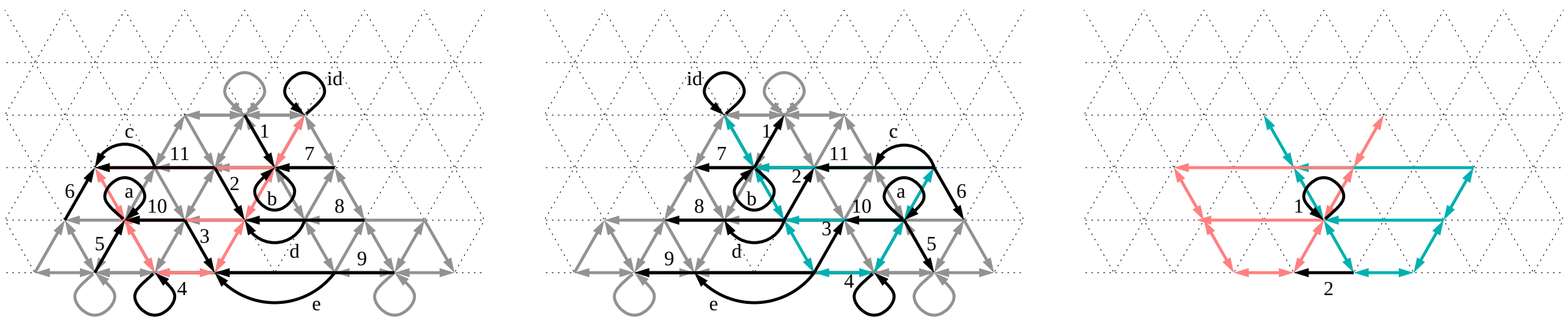}}
  \caption{The homomorphisms $\varphi$ (left),
    $\bar\varphi$ (middle)
    and the homotopy $h$ (right).}
  \label{fig:L0glue3}
\end{figure}

For this purpose,
we bind to ${\rm C}(\psi,k,\bar\psi)$ the Lascoux branes of the type (iv) and
(iii)' tensored with $\textw^{\!2} V$ and $\textw^{\!2} V^*$, respectively.
Let $B_{\rm iv}$ and $B_{\rm iii}$ be such branes which have
$\sym^3\fund^*\otimes \textw^{\!2} V$ and
$\sym^3\fund^*(-2)\otimes\textw^{\!2}V^*$,
respectively, at the height 3 ends.
We find homomorphisms $\varphi:{\rm C}(\psi,k,\bar\psi)\to B_{\rm iv}$
and $\bar\varphi:B_{\rm iii}\to {\rm C}(\psi,k,\bar\psi)$
that extend the identity maps at the height 3 positions.
We also find that
$\varphi\circ\bar\varphi:B_{\rm iii}\to B_{\rm iv}$
is null homotopic with a homotopy
$h:B_{\rm iii}\to B_{\rm iv}[-1]$.
Fig.~\ref{fig:L0glue3} shows these maps.
${\rm C}(\psi,k,\bar\psi)$, $B_{\rm iv}$ and $B_{\rm iii}$ are shown in grey,
pink and light-blue.
The explicit forms of the arrows are
\beqa
&&~\varphi_1,\varphi_2,\varphi_3,\varphi_4,\varphi_5,
\varphi_6,\varphi_7,\varphi_{10},\varphi_{11}=\mbox{the same expressions
  as in (\ref{varphiL1})},\nn\\[0.2cm]
&&\begin{array}{l}
   \varphi_8={1\over 3}\mathring{A}^2+{1\over 18}\bar{A}A
   -{1\over 9}\bar\Box\mathring{A}A+{1\over 2^4\cdot 3^2}\bar\Box^2 A^2,\\
   \varphi_9=-{2\over 9}\mathring{A}^3-{1\over 6}\bar{A}\mathring{A}A
   +{1\over 6}\bar\Box\mathring{A}^2A+{1\over 36}\bar\Box\bar{A}A^2
   -{1\over 36}\bar\Box^2\mathring{A}A^2,\\
   \varphi_{\rm a}=-10\bar\Box^2,\\
   \varphi_{\rm b}=-{5\over 3}+{5\over 12}\bar\Box\Box,\\
   \varphi_{\rm c}=-5\bar\Box^2\bar{A},\\
   \varphi_{\rm d}={5\over 3^2}\left(3\mathring{A}-{1\over 2}\bar\Box A
   +{1\over 2}\bar{A}\Box-\bar\Box\mathring{A}\Box
   +{1\over 8}\bar\Box^2 A\Box\right),\\
   \varphi_{\rm e}={5\over 6}\left(
   -2\mathring{A}^2-{1\over 2}\bar{A}A+\bar\Box \mathring{A}A
   -\bar{A}\mathring{A}\Box+\bar\Box\mathring{A}^2\Box
   +{1\over 3}\bar\Box\bar{A}A\Box
   -{1\over 3}\bar\Box^2\mathring{A}A\Box
   -{1\over 12}\bar\Box^2A^2\right),
  \end{array}
\label{varphiL0}
\eeqa
\beqa
&&~\bar\varphi_1,\bar\varphi_2,\bar\varphi_3,\bar\varphi_4,\bar\varphi_5,
\bar\varphi_6,\bar\varphi_7,\bar\varphi_{10},\bar\varphi_{11}
=\mbox{the same expressions as
  in (\ref{barvarphiL1})},\nn\\[0.2cm]
&&\begin{array}{l}
   \bar\varphi_8={1\over 3}\mathring{A}^2+{1\over 18}\bar{A}A
   +{1\over 9}\bar{A}\mathring{A}\Box+{1\over 2^4\cdot 3^2}\bar{A}^2 \Box^2,\\
   \bar\varphi_9=-{2\over 9}\mathring{A}^3-{1\over 6}\bar{A}\mathring{A}A
   -{1\over 6}\bar{A}\mathring{A}^2\Box
   -{1\over 36}\bar{A}^2A\Box
   -{1\over 36}\bar{A}^2\mathring{A}\Box^2,\\
   \bar\varphi_{\rm a}=-10\Box^2,\\
   \bar\varphi_{\rm b}=-{5\over 3}+{5\over 12}\bar\Box\Box,\\
   \bar\varphi_{\rm c}=-5A\Box^2,\\
   \bar\varphi_{\rm d}={5\over 3^2}\left(3\mathring{A}
   +{1\over 2}\bar{A}\Box
   -{1\over 2}\bar\Box A
   -\bar\Box\mathring{A}\Box
   -{1\over 8}\bar\Box \bar{A}\Box^2\right),\\
   \bar\varphi_{\rm e}={5\over 6}\left(
   -2\mathring{A}^2-{1\over 2}\bar{A}A
   -\bar{A}\mathring{A}\Box
   +\bar\Box \mathring{A}A
   +\bar\Box\mathring{A}^2\Box
   +{1\over 3}\bar\Box\bar{A}A\Box
   +{1\over 3}\bar\Box\bar{A}\mathring{A}\Box^2
   -{1\over 12}\bar{A}^2\Box^2\right),
  \end{array}
\label{barvarphiL0}
\eeqa
and
\beq
\begin{array}{l}
  h_1={10\over 3}\left(1-{1\over 3}\bar\Box\Box
  +{1\over 24}\bar\Box^2\Box^2\right),\\
  h_2={5\over 3}\left(4\mathring{A}-\bar\Box A+\bar{A}\Box
  -2\bar\Box\mathring{A}\Box
  +{1\over 3}\bar\Box^2A\Box
  -{1\over 3}\bar\Box\bar{A}\Box^2
  +{1\over 3}\bar\Box^2\mathring{A}\Box^2\right).
\end{array}
\eeq
We can take the double cone as in (\ref{cone2}), or
(\ref{newdiff})-right. We annihilate the height 3 pairs
by the tachyon condensation of the identity maps as well as the
invertible component $k_1$ of the homotopy.
This yields a matrix factorization of height 2.

It can be simplified by additional brane-antibrane annihilations
at nine invertible components of the differentials.
Three of them come from injective components
$\psi_5$, $\varphi_4$, $\varphi_{\rm a}$ acting only on the flavor factors as
\beq
\textw^{\!5,0}\to\textw^{\!6,1},\qquad
\textw^{\!4,1}\to\textw^{\!5,2},\qquad
\textw^{\!4,0}\to\textw^{\!6,2},
\label{vp54afla}
\eeq
and their tachyon condensation leaves us with the cokernels of these maps.
Another three come from surjective components
$\bar\psi_5$, $\bar\varphi_4$, $\bar\varphi_{\rm a}$
acting only on the flavor factors as
\beq
\textw^{\!1,6}\to\textw^{\!0,5},\qquad
\textw^{\!2,5}\to\textw^{\!1,4},\qquad
\textw^{\!2,6}\to\textw^{\!0,4},
\label{barvp54afla}
\eeq
and their tachyon condensation leaves us with the kernels of these maps.
The remaining three come from isomorphic components
$\varphi_{\rm b}$, $\bar\varphi_{\rm b}$ and $h_1$
of the homomorphisms and the homotopy.
Their tachyon condensation removes the pairs completely.
Of course, all these processes induce zigzag maps.

\begin{figure}[thb]
\centerline{\includegraphics[page=169]{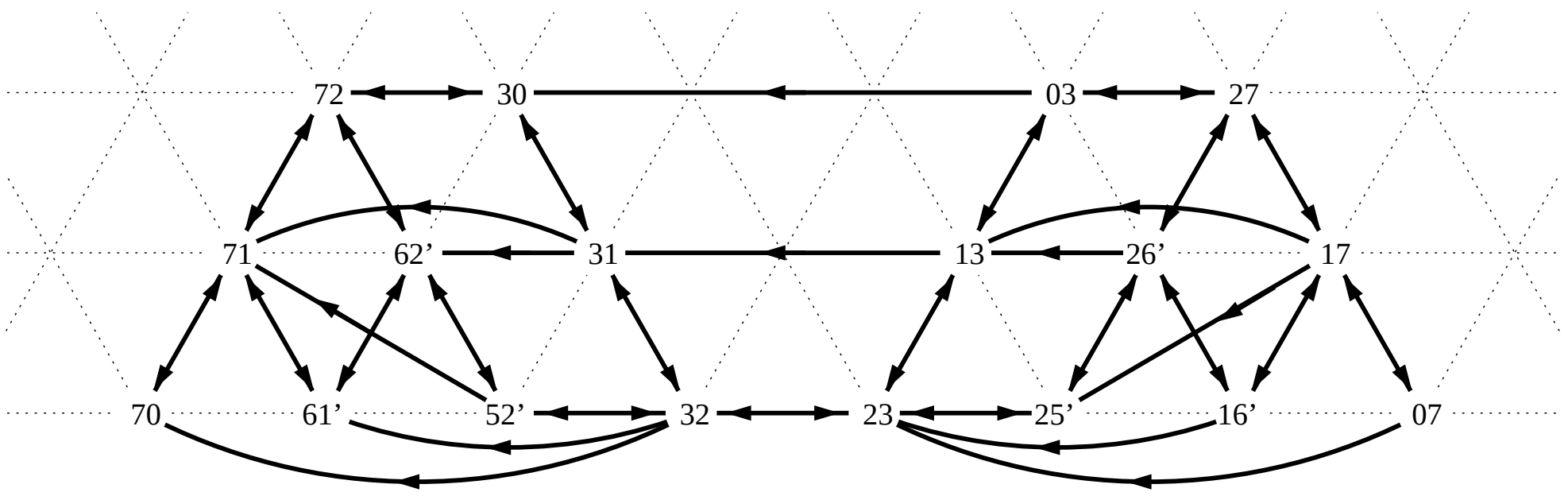}}
\caption{The brane $\mathfrak{L}^0$.}
\label{fig:L0}
\end{figure}

The resulting matrix factorization is as shown in Fig.~\ref{fig:L0}.
The vertex with label {\small $pq$} has multiplicity
$\textw^{\!p,q}$, with the following exceptions:
   the vertices with {\small 61'}, {\small 52'}, {\small 62'}
    and {\small 16'}, {\small 25'}, {\small 26'} stand for the cokernels of
    $\psi_5$, $\varphi_4$, $\varphi_{\rm a}$ and the kernels of
    $\bar\psi_5$, $\bar\varphi_4$, $\bar\varphi_{\rm a}$, respectively.
Some of the arrows are
\beqa
&{\scriptstyle 72}\lrarrow
    {\scriptstyle 30}=\left\{
    \begin{array}{l}
    (\nearrow)\varphi_{\rm a}^{-1}(\searrow)\\
            \varphi_{\rm c}
 \end{array}\right.\!\!\!,~~~~~~~~~
            {\scriptstyle 03}\lrarrow{\scriptstyle 27}=\left\{
            \begin{array}{l}
  (\nearrow)\bar\varphi_{\rm a}^{-1}(\searrow)\\
                            \bar\varphi_{\rm c}
 \end{array}\right.\!\!\!,
\nn\\
&{\scriptstyle 30}\leftarrow{\scriptstyle 03}
    =-\chi_{\rm iii}-\bar\varphi_7\bar\varphi_{\rm b}^{-1}\psi_8
    -\bar\psi_8\varphi_{\rm b}^{-1}\varphi_7+\bar\varphi_7\bar\varphi_{\rm b}^{-1}
    k_2\varphi_{\rm b}^{-1}\varphi_7,\nn\\
&{\scriptstyle 71}\plrarrow{\scriptstyle 72}=\left\{
       \begin{array}{l}\varphi_6\\
       \psi_{11}\varphi_{\rm a}^{-1}(\searrow)
     \end{array}\right.\!\!\!,~~~~~~~
      {\scriptstyle 27}\nlrarrow {\scriptstyle 17}=\left\{
      \begin{array}{l}
      \bar\varphi_6\\
      (\nearrow)\bar\varphi_{\rm a}^{-1}\bar\psi_{11}
    \end{array}\right.\!\!\!,\nn\\
&{\scriptstyle 30}\nlrarrow{\scriptstyle 31}=\left\{
      \begin{array}{l}
        -\psi_3\\
      -(\nearrow)\varphi_{\rm a}^{-1}\varphi_{10}
      -\bar\varphi_7\bar\varphi_{\rm b}^{-1}(\nearrow)
      \end{array}\right.\!\!\!,~~~~
   {\scriptstyle 13}\plrarrow{\scriptstyle 03}=\left\{
   \begin{array}{l}
     -\bar\psi_3\\
     -(\searrow)\varphi_{\rm b}^{-1}\varphi_7
      -\bar\varphi_{10}\bar\varphi_{\rm a}^{-1}(\searrow)
     \end{array}\right.\!\!\!,
\nn\\
&{\scriptstyle 71}\leftarrow{\scriptstyle 31}
=(\leftarrow)-\psi_{11}\varphi_{\rm a}^{-1}\varphi_{10},~~
{\scriptstyle 13}\leftarrow{\scriptstyle 17}
=(\leftarrow)-\bar\varphi_{10}\bar\varphi_{\rm a}^{-1}\bar\psi_{11},
\nn\\
&{\scriptstyle 62'}\leftarrow{\scriptstyle 31}=-\varphi_{10},~~
{\scriptstyle 31}\leftarrow{\scriptstyle 13}
=k_3-\bar\varphi_dh_1^{-1}\varphi_d,~~
{\scriptstyle 13}\leftarrow{\scriptstyle 26'}=-\bar\varphi_{10},\nn\\
&{\scriptstyle 70}\plrarrow{\scriptstyle 71}=\left\{
\begin{array}{l}
  \psi_6\\
  (\leftarrow)\psi_5^{-1}(\searrow)
\end{array}\right.\!\!\!,~~~
{\scriptstyle 17}\nlrarrow{\scriptstyle 07}=\left\{
\begin{array}{l}
  \bar\psi_6\\
  (\nearrow)\bar\psi_5^{-1}(\leftarrow)
\end{array}\right.\!\!\!,
\nn\\
&{\scriptstyle 61'}\plrarrow{\scriptstyle 62'}=\left\{
\begin{array}{l}
  -\varphi_5\\
  -(\leftarrow)\varphi_{\rm a}^{-1}(\searrow)
\end{array}\right.\!\!\!,~~~
{\scriptstyle 26'}\nlrarrow{\scriptstyle 16'}=\left\{
\begin{array}{l}
  -\bar\varphi_5\\
  -(\nearrow)\bar\varphi_{\rm a}^{-1}(\leftarrow)
\end{array}\right.\!\!\!,\nn\\
&{\scriptstyle 71}\nwarrow{\scriptstyle 52'}
=\psi_{11}\varphi_{\rm a}^{-1}(\nwarrow),~~~~
{\scriptstyle 25'}\nwarrow{\scriptstyle 17}
=(\swarrow)\bar\varphi_{\rm a}^{-1}\bar\psi_{11},
\nn\\
&{\scriptstyle 31}\nlrarrow{\scriptstyle 32}=\left\{
\begin{array}{l}
  -\varphi_3\\
  -(\nearrow)\varphi_4^{-1}(\leftarrow)-\bar\varphi_dh_1^{-1}(\nearrow)
\end{array}\right.\!\!\!,~~~
{\scriptstyle 31}\plrarrow{\scriptstyle 32}=\left\{
\begin{array}{l}
  -\bar\varphi_3\\
  -(\searrow)h_1^{-1}\varphi_d-(\leftarrow)\bar\varphi_4^{-1}(\searrow)
\end{array}\right.\!\!\!,
\nn\\
&{\scriptstyle 70}\leftarrow{\scriptstyle 32}
    =-(\leftarrow)\psi_5^{-1}(\leftarrow)\varphi_4^{-1}(\leftarrow),~~~~
    {\scriptstyle 23}\leftarrow{\scriptstyle 07}
    =-(\leftarrow)\bar\varphi_4^{-1}(\leftarrow)\bar\psi_5^{-1}(\leftarrow),
\nn\\
&{\scriptstyle 61'}\leftarrow{\scriptstyle 32}
=-(\leftarrow)\varphi_4^{-1}(\leftarrow),~~~
{\scriptstyle 23}\leftarrow{\scriptstyle 16'}
=-(\leftarrow)\bar\varphi_4^{-1}(\leftarrow),
\nn\\
&{\scriptstyle 32}\lrarrow{\scriptstyle 23}
    =\left\{\begin{array}{l}
    -(\searrow)h_1^{-1}(\nearrow),\\
    h_2
    \end{array}\right.
\label{L0arrows}
\eeqa
while the rest are the ones induced from the Lascoux matrix factorizations.
$\psi_5^{-1}$, $\varphi_4^{-1}$, $\varphi_{\rm a}^{-1}$ {\it resp}.
$\bar\psi_5^{-1}$, $\bar\varphi_4^{-1}$, $\bar\varphi_{\rm a}^{-1}$
and 
$\varphi_{\rm b}^{-1}$, $\bar\varphi_{\rm b}^{-1}$, $h_1^{-1}$
are left inverses {\it resp}. right inverses
and the inverses of the corresponding maps.
Their explicit forms are
\beqa
&\begin{array}{l}
   \psi_5^{-1}=-{1\over 10}\Box,\\
   \varphi_4^{-1}={1\over 6}\left(\Box-{1\over 8}\Box\bar\Box\Box\right),\\
   \varphi_{\rm a}^{-1}=-{1\over 120}\bar\Box^2,
 \end{array}~~
\begin{array}{l}
   \bar\psi_5^{-1}={1\over 10}\bar\Box,\\
   \bar\varphi_4^{-1}=
   -{1\over 6}\left(\bar\Box-{1\over 8}\bar\Box\Box\bar\Box\right),\\
   \bar\varphi_{\rm a}^{-1}=-{1\over 120}\Box^2,
 \end{array}
\\
&\begin{array}{l}
   \varphi_{\rm b}^{-1}=-{3\over 5}\left(1-{1\over 2}\bar\Box\Box\right),\\
   \bar\varphi_{\rm b}^{-1}=-{3\over 5}\left(1-{1\over 2}\bar\Box\Box\right),\\
   h_1^{-1}={3\over 10}\left(1-{1\over 2}\bar\Box\Box
   +{1\over 12}\bar\Box^2\Box^2\right).
 \end{array}
\eeqa
The same remark as the one after (\ref{h1inv}) applies concerning the
definition of the arrows out of the cokernels 61', 52', 62'
and the arrows into the kernels 16', 25', 26' as well as the meaning of
$(\searrow)$, $(\nearrow)$ and $(\leftarrow)$ in (\ref{L0arrows}).

\section{Some Details On The Grade Restriction Process}
\label{app:GRdetails}

We fulfill a promise made in Section~\ref{subsec:GRprocess},
around page~\pageref{page:x}: we show that
there is a homomorphism from $\Complex$ to $N$
that extends the identity map of $\fund(7)$, and also that
there is a homomorphism from $\Complex$ to $N'$
that extends the identity map of $\C(7)$.

We note that
$N=(C_{\rm iii}\otimes V^*\rightarrow C_{\rm ii})$
is quasi-isomorphic to $M_{\rm iii}\otimes V^*\to M_{\rm ii}$
where $M_{\alpha}$ is the $(\GL_2(\C),\C[x])$-module resolved by
the twisted Lascoux complex $C_{\alpha}$ of type $(\alpha)$.
The description of $M_{\alpha}$ is given by \cite{DS} from which
we could learn about this complex. Instead of doing that, we directly study
the complex $N$. The only thing we need from this quasi-isomorphism
is that $N$ is exact except possibly at the last two terms.
Its explicit form is
\beq
\begin{tikzcd}
  \cdots\arrow[r]\arrow[dr]&
  \C(6)\!\otimes\! \Lambda^{\!1,5}
  \ar[draw=none]{d}[sloped,auto=false]{\bigoplus}
  \arrow[r]\arrow[dr,"\!\!\!\bar\varphi_4"]&
  \fund^*(7)\!\otimes\! \Lambda^{\!1,6}
  \ar[draw=none]{d}[sloped,auto=false]{\bigoplus}
  \arrow[r]\arrow[dr,"\!\!\!\!-\bar\varphi_5"]&
  \sym^2\fund^*(8)\!\otimes\! \Lambda^{\!1,7}
  \ar[draw=none]{d}[sloped,auto=false]{\bigoplus}
  \arrow[dr,"\!\!\!\bar\varphi_6"]
  &\\
  \cdots\arrow[r]&
  \fund^*(6)\!\otimes\! \Lambda^{\!0,3}
  \arrow[r]&
  \C(6)\!\otimes\!\Lambda^{\!0,4}\arrow[r]&
  \C(7)\!\otimes\!\Lambda^{\!0,6}\arrow[r]&
  \fund^*(8)\otimes\Lambda^{\!0,7}
\end{tikzcd}
\label{Ndetail}
\eeq
where $\Lambda^{p,q}=\textw^{\!p}V^*\otimes \textw^{\!q}V$ and
$\bar\varphi_i$ are given in (\ref{barvarphiL1}).
The rightward arrows are those from the twisted Lascoux complexes.
Note that $\fund^*(8)\otimes\Lambda^{\!0,7}\cong\fund(7)$.
The problem to find a homomorphism $\Complex\to N$ is
to find the dashed arrows in
\beq
\begin{tikzcd}
  \cdots\arrow[r]&
  M^{-3}\arrow[d,dashed]\arrow[r]&
  M^{-2}\arrow[d,dashed]\arrow[r]&
  M^{-1}\arrow[d,dashed]\arrow[r]&
  M^0\arrow[dd,"f^0"]\arrow[r]&
  M^1\arrow[r]&\cdots\\
  \cdots\arrow[r]\arrow[dr]&
  N_{\uparrow}^{-3}\ar[draw=none]{d}[sloped,auto=false]{\bigoplus}
  \arrow[r]\arrow[dr]&
  N_{\uparrow}^{-2}\ar[draw=none]{d}[sloped,auto=false]{\bigoplus}
  \arrow[r]\arrow[dr]&
  N_{\uparrow}^{-1}\ar[draw=none]{d}[sloped,auto=false]{\bigoplus}
  \arrow[dr]
  &&&\\
  \cdots\arrow[r]&
  N_{\downarrow}^{-3}\arrow[r]&
  N_{\downarrow}^{-2}\arrow[r]&
  N_{\downarrow}^{-1}\arrow[r]&
  N_{\downarrow}^0&&
\end{tikzcd}
\label{theproblemapp}
\eeq
so that the squares commute.
The first line $(M,d_M)$ is $\Complex$
and the last two lines form a copy of (\ref{Ndetail}).
$M^0$ includes $\fund(7)$ as a direct summand and $f^0$ is the projection
to that summand.
We have reset the degree of $N$ so that the last term has degree $0$.
Thus, $N$ is exact at degree $-2$ and below.
Note also that the second line $(N_{\uparrow},d_{\uparrow})$ and
the third line $(N_{\downarrow}, d_{\downarrow})$ are both acyclic.
The first question is whether the image of $f^0\circ d_M^{-1}$ is included
in the image of $d_N^{-1}$.
The latter map is the sum of
$\bar\varphi_6\propto x^a_ie_a\eta^i$ and
$d_{\downarrow}^{-1}=\nearrow=x_{ia}e^a\bareta^i$, while the former
splits to the sum of maps of the form $\varphi$ in (\ref{varphi})
for $(h,i)=(1,8)$ and various $(h',i')$ with $h'=0,1,2$.
The matrix elements $\varphi_a^{\,\,b_1\cdots b_{h'}}(x)$ of $\varphi$
are sums of products
of $x_{ia}$, $x_j^{b_t}$, $\delta_a^{b_t}$, $[x_ix_j]$ as in (\ref{formofmael}).
Each term is divisible by $x_{ia}$ or $x^{b_t}_j$,
thanks to the relation (\ref{fundrel}) for the case $h'=1$
where we recall that $d^{{}^{\,\,\cdot\!\!}}_M$ is at least linear in $x$.
Thus, the image of $f^0\circ d_M^{-1}$ is included in the image of $d_N^{-1}$.
As $M^{-1}$ is a free $\C[x]$-module, there is
a map $f^{-1}:M^{-1}\to N^{-1}$ so that the right-most square commutes.
(It is chosen to be $\GL_2(\C)$ equivariant,
by averaging (\ref{averaging}) over $U(2)$ if necessary.
We will not repeat this kind of remark below.)
Let us write $f^{-1}=(f_{\uparrow}^{-1},f_{\downarrow}^{-1})$.
Note that, as $N_{\uparrow}^{-1}$ is at height 2,
the image of $f_{\uparrow}^{-1}\circ d_M^{-2}$ is included in the image of
$d_{\uparrow}^{-2}=\nearrow$ by the height argument.
Thus, there is a map $f_{\uparrow}^{-2}:M^{-2}\to N_{\uparrow}^{-2}$
so that the small upper-square in the middle commutes,
$f_{\uparrow}^{-1}d_M^{-2}=d_{\uparrow}^{-2}f_{\uparrow}^{-2}$.
Let us consider the map $\psi:M^{-2}\to N_{\downarrow}^{-1}$
which is the sum of $f_{\downarrow}^{-1}\circ d_M^{-2}$ through $M^{-1}$
and $\bar\varphi_5\circ f_{\uparrow}^{-2}$ through $N_{\uparrow}^{-2}$.
Note that
\beqa
d_{\downarrow}^{-1}\circ \psi
&=&d_{\downarrow}^{-1}(f_{\downarrow}^{-1}d_M^{-2}
+\bar\varphi_5f_{\uparrow}^{-2})\nn\\
&=&(d_{\downarrow}^{-1}f_{\downarrow}^{-1}+\bar\varphi_6d_{\uparrow}^{-2})d_M^{-2}
=d_M^{-1}\circ d_M^{-2}=0.
\eeqa
where we used
$d_{\downarrow}^{-1}\bar\varphi_5f_{\uparrow}^{-2}=\bar\varphi_6d_{\uparrow}^{-2}f_{\uparrow}^{-2}=\bar\varphi_6f_{\uparrow}^{-1}d_M^{-2}$ in the second equality.
As $(N_{\downarrow},d_{\downarrow})$ is exact at degree $-1$,
the image of $\psi$ is in the image of $d_{\downarrow}^{-2}$.
As $M^{-2}$ is free, there is a map
$f_{\downarrow}^{-2}:M^{-2}\to N_{\downarrow}^{-2}$ such that
$d_{\downarrow}^{-2}\circ f_{\downarrow}^{-2}=\psi$, that is,
$f_{\downarrow}^{-1}d_M^{-2}=-\bar\varphi_5f_{\uparrow}^{-2}+d_{\downarrow}^{-2}f_{\downarrow}^{-2}$.
We see that $f^{-2}=(f_{\uparrow}^{-2},f_{\downarrow}^{-2})$ is the wanted arrow,
\beq
d_N^{-2}f^{-2}=\left(\begin{array}{cc}
  d_{\uparrow}^{-2}&0\\
  -\bar\varphi_5&d_{\downarrow}^{-2}
\end{array}\right)\left(\begin{array}{c}
  f_{\uparrow}^{-2}\\
  f_{\downarrow}^{-2}
\end{array}\right)
=\left(\begin{array}{c}
  f_{\uparrow}^{-1}d_M^{-2}\\
  f_{\downarrow}^{-1}d_M^{-2}
\end{array}\right)
=f^{-1}d_M^{-1}.
\label{commute111}
\eeq
The rest $f^{-3},f^{-4},\ldots$ can be found
as $M^{-i}$ are free and $(N,d_N)$ is exact at degree $-2$ and below.

The complex
$N'=(C_{\rm iii}\otimes\textw^{\!2}V^*\to (C_{\rm ii}\otimes V^*\to C_{\rm i}))$
is quasi-isomorphic to
$M_{\rm iii}\otimes\textw^{\!2}V^*\to M_{\rm ii}\otimes V^*\to M_{\rm i}$
and hence is exact except possibly at the last three terms.
Its explicit form is
\beq
\begin{tikzcd}
  \cdots\arrow[r]\arrow[dr]&
  \fund^*(6)\!\otimes\!\Lambda^{\!2,6}
  \ar[draw=none]{d}[sloped,auto=false]{\bigoplus}
  \arrow[r]\arrow[dr,"\!\!\!\!-\bar\varphi_5"]
  \arrow[ddr,"~~~~~\,-\bar\varphi_a"{yshift=-23pt}]&
  \sym^2\fund^*(7)\!\otimes\!\Lambda^{\!2,7}
  \ar[draw=none]{d}[sloped,auto=false]{\bigoplus}
  \arrow[dr,"\!\!\!\bar\varphi_6"]&  &\\
  \cdots\arrow[r]\arrow[dr]&
  \C(5)\!\otimes\!\Lambda^{\!1,4}
  \ar[draw=none]{d}[sloped,auto=false]{\bigoplus}
  \arrow[r]\arrow[dr,"\!\!\!\!\!\bar\psi_4"]&
  \C(6)\!\otimes\!\Lambda^{\!1,6}
  \ar[draw=none]{d}[sloped,auto=false]{\bigoplus}
  \arrow[r]\arrow[dr,"\!\!\!\!\!-\bar\psi_5"]&
  \fund^*(7)\!\otimes\!\Lambda^{\!1,7}
  \ar[draw=none]{d}[sloped,auto=false]{\bigoplus}
  \arrow[dr,"\!\!\!\!\bar\psi_6"]
  &\\
  \cdots\arrow[r]&
  \sym^2\fund^*(6)\!\otimes\!\Lambda^{\!0,3}
  \arrow[r]&
  \fund^*(6)\!\otimes\!\Lambda^{\!0,4}
  \arrow[r]&
  \C(6)\!\otimes\!\Lambda^{\!0,5}
  \arrow[r]&
  \C(7)\!\otimes\!\Lambda^{\!0,7}
\end{tikzcd}
\label{Npdetail}
\eeq
where $\Lambda^{\!p,q}$ are as above while $\bar\varphi_i$ and
$\bar\psi_i$ are given in (\ref{barvarphiL0}) and (\ref{barpsiL0}).
Note that $\C(7)\otimes\Lambda^{\!0,7}\cong \C(7)$.
The problem to find a homomorphism $\Complex\to N'$
can also be described by the diagram (\ref{theproblemapp}):
The bottom line and the composite of
the first two lines of (\ref{Npdetail}),
corresponding to $C_{\rm i}$ and
$(C_{\rm iii}\otimes \textw^{\!2}V^*\to C_{\rm ii}\otimes V^*)$
respectively,
are represented by the bottom line $(N_{\downarrow},d_{\downarrow})$
and the middle line $(N_{\uparrow},d_{\uparrow})$ of (\ref{theproblemapp}).
$M^0$ includes $\C(7)$ as a direct summand and $f^0$ is the projection
to that summand.
We have reset the degree of $N'$ so that the last term has degree $0$.
Thus, $N'$ is exact at degree $-3$ and below.
Note also that, in this interpretation of (\ref{theproblemapp}),
the bottom line $(N_{\downarrow}, d_{\downarrow})$ is acyclic.
The first question is whether the image of $f^0\circ d_M^{-1}$ is included
in the image of $d_{N'}^{-1}$.
The latter map is the sum of
$\bar\psi_6\propto x^a_ie_a\eta^i$ and
$d_{\downarrow}^{-1}=\,\to\,={1\over 2}[x_ix_j]\bareta^i\bareta^j$,
while the former
splits to the sum of maps of the form $\varphi$ in (\ref{varphi})
for $(h,i)=(0,7)$ and various $(h',i')$ with $h'=0,1,2$.
The matrix elements $\varphi^{\,\,b_1\cdots b_{h'}}(x)$ of $\varphi$
are sums of products of $x_j^{b_t}$ and $[x_ix_j]$.
Thus, the image of $f^0\circ d_M^{-1}$ is included in the image of $d_{N'}^{-1}$.
As $M^{-1}$ is free, there is a map
$f^{-1}:M\to N^{\prime-1}$ so that the right-most square commutes.
Let us write $f^{-1}=(f_{\uparrow}^{-1},f_{\downarrow}^{-1})$.
The image of $f_{\uparrow}^{-1}\circ d_M^{-2}$ is included in the image of
$d_{\uparrow}^{-2}=\nearrow$.
This was already seen in the previous problem to find $\Complex\to N$,
as ${\rm Im}f^0d_M^{-1}\subset {\rm Im}d_N^{-1}$.
Thus, there is a map $f_{\uparrow}^{-2}:M^{-2}\to N_{\uparrow}^{-2}$
so that the small upper-square in the middle commutes,
$f_{\uparrow}^{-1}d_M^{-2}=d_{\uparrow}^{-2}f_{\uparrow}^{-2}$.
Since $(N_{\downarrow},d_{\downarrow})$ is acyclic,
just as in the previous problem, we also find
$f_{\downarrow}^{-2}:M^{-2}\to N_{\downarrow}^{-2}$
such that 
$f_{\downarrow}^{-1}d_M^{-2}=-\bar\psi_5f_{\uparrow}^{-2}+d_{\downarrow}^{-2}f_{\downarrow}^{-2}$.
Thus, $(f_{\uparrow}^{-2},f_{\downarrow}^{-2})$ is the wanted arrow
$f^{-2}:M^{-2}\to N^{\prime -2}$ just as in (\ref{commute111}).
Now we are facing the problem to find the dashed arrows in
\beq
\begin{tikzcd}
  \cdots\arrow[r]&
  M^{-4}\arrow[d,dashed]\arrow[r]&
  M^{-3}\arrow[d,dashed]\arrow[r]&
  M^{-2}\arrow[d,"f^{-2}"]\arrow[r]&
  M^{-1}\arrow[dd,"f^{-1}"]\arrow[r]&
  M^0\arrow[dd,"f^0"]\arrow[r]&\cdots\\
  \cdots\arrow[r]\arrow[dr]&
  N_{\uparrow}^{\prime-4}\ar[draw=none]{d}[sloped,auto=false]{\bigoplus}
  \arrow[r]\arrow[dr]&
  N_{\uparrow}^{\prime-3}\ar[draw=none]{d}[sloped,auto=false]{\bigoplus}
  \arrow[r]\arrow[dr]&
  N_{\uparrow}^{\prime-2}\ar[draw=none]{d}[sloped,auto=false]{\bigoplus}
  \arrow[dr]
  &&&\\
  \cdots\arrow[r]&
  N_{\downarrow}^{\prime-4}\arrow[r]&
  N_{\downarrow}^{\prime-3}\arrow[r]&
  N_{\downarrow}^{\prime-2}\arrow[r]&
  N_{\downarrow}^{\prime-1}\arrow[r]&
  N_{\downarrow}^{\prime0}&
\end{tikzcd}
\label{theproblemapp2}
\eeq
with the following interpretation:
The first line and the composite of
the last two lines of (\ref{Npdetail}),
corresponding to $C_{\rm iii}\otimes \textw^{\!2}V^*$
and $(C_{\rm ii}\otimes V^*\to C_{\rm i})$, respectively,
are represented by
the middle line $(N_{\uparrow}',d_{\uparrow}')$ and
the bottom line $(N_{\downarrow}',d_{\downarrow}')$
of (\ref{theproblemapp2}).
Note that $N^{\prime -2}_{\uparrow}$ is at height 2
and that the bottom line $(N_{\downarrow}',d_{\downarrow}')$
is exact at degree $-2$ and below.
Then, the stiuation is identical to the problem to find
$f^{-2}:M^{-2}\to N^{-2}$ after finding
$f^{-1}:M^{-1}\to N^{-1}$ in the problem to find $\Complex\to N$.
Therefore, a map $f^{-3}:M^{-3}\to N^{\prime -3}$
can be found.
The rest, $f^{-4},f^{-5},\ldots$, can be found
as $M^{-i}$ are free and $(N',d_{N'})$ is exact at degree $-3$ and below.

\section{Reduction Of $\Lb^1$ And $\Lb^0$ In The Pfaffian Phase}
\label{app:redPf}

\subsection{Reduction Of $\Lb^1$}

\begin{figure}[thb]
  \psfrag{-1}{{\footnotesize $\,-\!1$}}
  \psfrag{-2}{{\footnotesize $\,-\!2$}}
  \psfrag{-3}{{\footnotesize $\,-\!3$}}
  \psfrag{-4}{{\footnotesize $\,-\!4$}}
  \centerline{\includegraphics[page=170]{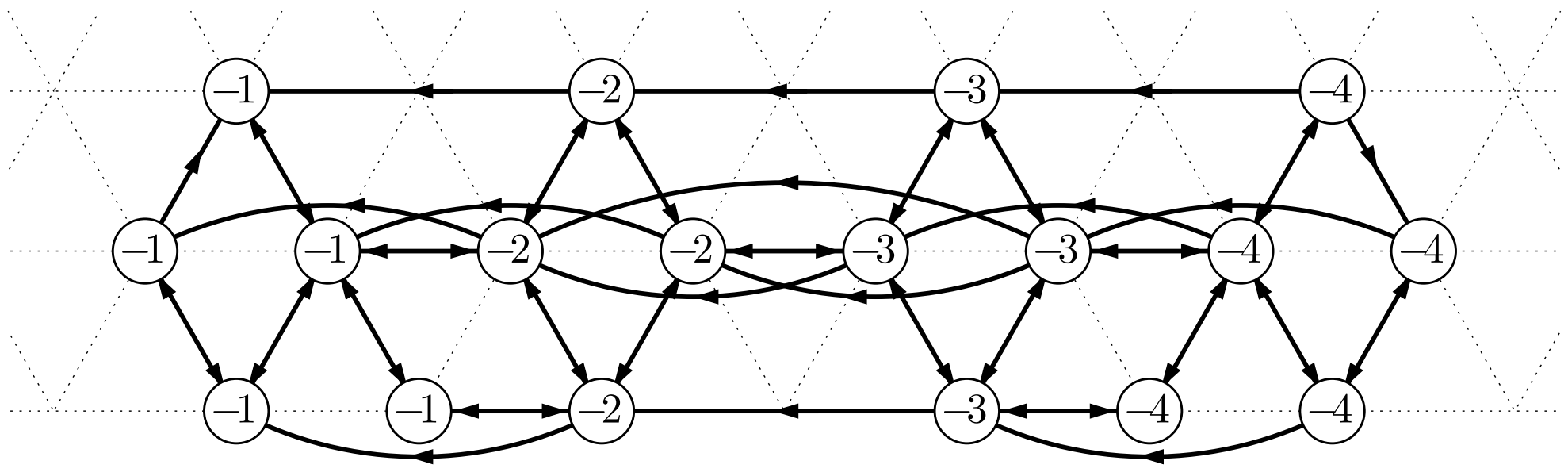}}
  \caption{The R-charge of $\Lb^1$ in the Pfaffian phase.}
  \label{fig:RL1}
\end{figure}

$\Lb^1$ is the matrix factorization Fig.~\ref{fig:L1} 
whose left-most component is $\fund=\fund^*(1)$ with R-charge $0$. 
Let us first compute the R-charge in the Pfaffian phase.
The component $S^h\fund^*(i)\otimes$ (flavor) of original R-charge $r$
has R-charge $r^{\rm Pf}=r+h-2i$ in the Pfaffian frame (\ref{RPf}).
Let us write down $(r,h,i):r^{\rm Pf}$ of the components that appear in
the two-step construction of $\Lb^1$ in Appendix~\ref{sub:L1}.
As the left-most component has $(0,1,1):-1$, the matrix factorization
at the end of the first step has
\beq
\begin{array}{cccc}
  &(5,3,5):-2\!\!&\!\!(6,3,6):-3&\\
  &(4,2,4):-2\,&\,(7,2,6):-3&\\
  (0,1,1):-1~~&(3,1,3):-2\,\,\,&\,\,\,(8,1,6):-3&~~(11,1,8):-4\\
  (1,0,1):-1&\!\!(2,0,2):-2\,\,\,\,\,&
  \,\,\,\,\,(9,0,6):-3\!\!&(10,0,7):-4
\end{array}
\eeq
The Lascoux matrix factorizations that are glued in the
second step have
\beq
\begin{array}{cc}
  &\,~~~(7,3,6):-2\\
  (1,2,2):-1~~\!&~~(6,2,5):-2\\
  (2,1,2):-1~\!\!&\!~(5,1,4):-2\\
  (3,0,2):-1\!\!\!&\!\!(4,0,3):-2
\end{array}\qquad\mbox{and}\qquad
\begin{array}{cc}
  (4,3,5):-3~~~\,&\\
  (5,2,5):-3~~&\!~~(10,2,8):-4\\
  (6,1,5):-3~\!&\!\!~(9,1,7):-4\\
  (7,0,5):-3\!\!&\!\!\!(8,0,6):-4
\end{array}
\eeq
After the tachyon condensation, we are left with the
components with the R-charges as described in Fig.~\ref{fig:RL1}.

Next, let us choose a point $p_*$ in $Y\subset \CP^6$,
take a neighborhood $p_*\in \nhU\subset\CP^6$
on which $A(p)$ can be decomposed as (\ref{AinU}), and
study the part of the field space where $p$ takes values in $\nhU$,
in which we make a change of variables,
(\ref{replacex}), (\ref{etacv}) and (\ref{baretacv}).
A key r\^ole is played by the part $d_*$ of the matrix factorization $Q(p,x)$
obtained by setting $x=0$ and $\widetilde{a}(p)=0$ in (\ref{AinU}).
It includes the components of $Q(p,x)$ that are invertible in this part of
the field space.
It defines a complex at each height,
\beqa
& \mbox{height 2:}~~~
\textw^{\!7,1}~\longleftarrow~
  \textw^{\!2,0}~\longleftarrow~
  \textw^{\!0,2}~\longleftarrow~
  \textw^{\!1,7}\nn\\
  & \mbox{height 1:}~~~
  \textw^{\!7,0}\oplus\textw^{\!6,1}~\longleftarrow~
  \textw^{\!3,0}\oplus\textw^{\!2,1}~\longleftarrow~
  \textw^{\!1,2}\oplus\textw^{\!0,3}~\longleftarrow~
  \textw^{\!1,6}\oplus \textw^{\!0,7}\label{dstarL1}\\
  & \mbox{height 0:}~~~
  \textw^{\!6,0}\oplus\textw^{\!5,1}\mbox{\Large '}~\longleftarrow~
  \textw^{\!3,1}~\longleftarrow~
  \textw^{\!1,3}~\longleftarrow~
  \textw^{\!1,5}\mbox{\Large '}\oplus\textw^{0,6}\nn
\eeqa
where $\textw^{\!p,q}=\textw^{\!p}V^*\otimes\textw^{\!q}V$ while
$\textw^{\!5,1}\mbox{\Large '}$ and
$\textw^{\!1,5}\mbox{\Large '}$ are the cokernel and the kernel of the flavor
part of $\varphi_4$ and $\bar\varphi_4$ given in (\ref{vp4fla}).
Here and in what follows,
we omit the color factors $S^h\sfund^*$ on which $d_*$ acts trivially.
The arrows of (\ref{dstarL1}) are those of (\ref{L1arrows})
in which $\mathring{A}$, $\bar{A}$, $A$
defined in (\ref{defA}) that enter there are replaced by
\beq
\mathring{A}_*={1\over 2}
\left(\bareta_4\eta^{}_5-\bareta_5\eta^{}_4
+\bareta_6\eta^{}_7-\bareta_7\eta^{}_6\right),
\quad
\bar{A}_*=\bareta_{45}+\bareta_{67},\quad
A_*=\eta^{}_{45}+\eta^{}_{67}.
\label{dAstar}
\eeq

The tachyon condensation of the invertible parts of $d_*$
leaves us with a matrix factorization in the reduced form.
The underlying vector space is the cohomology of $d_*$
and the arrows are those induced from the original matrix factorization
and the new arrows coming from the zigzag maps.
After a hard and tedious computation, we find that the cohomology is
\beq
\begin{array}{c|cccc}
  {\rm height}
  &{\rm H}^{-1}_{d_*}&{\rm H}^{-2}_{d_*}&{\rm H}^{-3}_{d_*}&{\rm H}^{-4}_{d_*}\\
  \hline
2&  \textw^{\!3\,}\Ttt^*\otimes \textw^{\!4\,}\cktt^*\otimes \cktt
  &\Ttt^*\otimes \cktt^*,&\Ttt\otimes \cktt
  &\textw^{\!3\,}\Ttt\otimes\textw^{\!4\,} \cktt\otimes \cktt^*\\
1&  \textw^{\!3\,}\Ttt^*\!\otimes\! \textw^{\!3\,}\cktt^*\otimes \cktt&
  \Ttt^*\!\otimes \cktt^*\otimes \cktt&
  \Ttt\otimes (\mbox{16 dim})&
  \textw^{\!3\,}\Ttt\otimes \textw^{\!3\,}\cktt\otimes \cktt^*\\
0&  (\mbox{20 dim})&
  \Ttt^*\!\otimes \textw_0^{\!2\,}\cktt^*\otimes \cktt&
  \Ttt\otimes (\mbox{20 dim})&
  \textw^{\!3\,}\Ttt\otimes (\mbox{20 dim})
\end{array}
\label{redFF1}
\eeq
The computation of the zigzag maps is even harder and has not been
done yet. However, the result of the cohomology
suggests that the reduced matrix factorization is isomorphic to the
tensor product of three factors, one is
$\brane_{\Ttt}$ given by the horizontal diagram (\ref{mf11}),
another is $\brane^{SU(2)}_{3,7}$ which is
either one of the four vertical lines in (\ref{iso4v}),
and the other is a copy of $\cktt\cong \cktt^*$.
This is nothing but $\brane^{\rm D6}_{\nhU}\otimes \cktt$.

\subsection{Reduction Of $\Lb^0$}

\begin{figure}[thb]
  \psfrag{0}{{\footnotesize $0$}}
  \psfrag{-1}{{\footnotesize $\,-\!1$}}
  \psfrag{-2}{{\footnotesize $\,-\!2$}}
  \psfrag{-3}{{\footnotesize $\,-\!3$}}
  \centerline{\includegraphics[page=171]{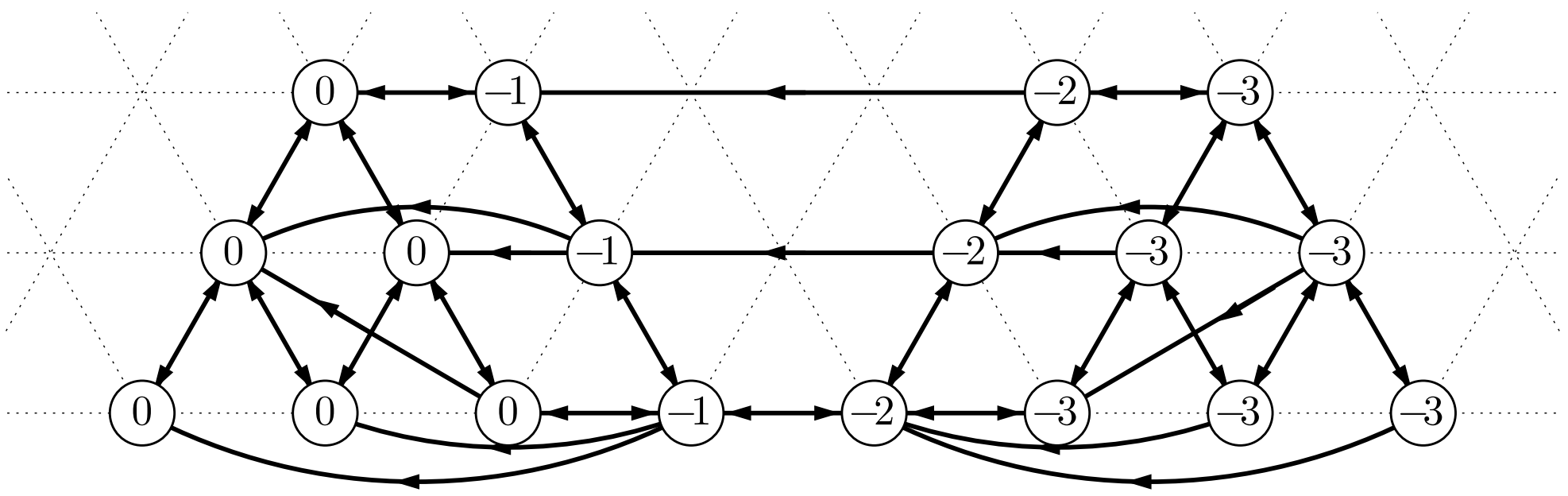}}
    \caption{The R-charge of $\Lb^0$ in the Pfaffian phase.}
  \label{fig:RL0}
\end{figure}

$\Lb^0$ is the matrix factorization Fig.~\ref{fig:L0} 
whose left-most component is $\C$ with R-charge $0$. 
Let us write down $(r,h,i):r^{\rm Pf}$ of the components that appear in
the three-step construction of $\Lb^0$ in Appendix~\ref{sub:L0}.
As the left-most component has $(0,0,0):0$, the matrix factorization
at the end of the first step has
\beq
\begin{array}{cccc}
  &(5,4,5):-1\!\!&\!\!(6,4,6):-2&\\
  &(4,3,4):-1\,&\,(7,3,6):-2&\\
  &\!\!(3,2,3):-1~&~(8,2,6):-2\!\!&\\
  &\!\!\!\!(2,1,2):-1~~&~~(9,1,6):-2\!\!\!\!&\\
  (0,0,0):0~~~&\!\!\!\!\!\!(1,0,1):-1~~~&~~~
  (10,0,6):-2\!\!\!\!\!\!&~~~(11,0,7):-3
\end{array}
\eeq
The Lascoux matrix factorizations that are glued in the
second step have
\beq
\begin{array}{cc}
  &\,~~~(7,4,6):-1\!\!\!\\
    &\,~~~(6,3,5):-1\\
  &~~(5,2,4):-1\\
  (1,1,1):0~\!\!&\!~(4,1,3):-1\\
  (2,0,1):0\!\!\!&\!\!(3,0,2):-1
\end{array}\qquad\mbox{and}\qquad
\begin{array}{cc}
  \!\!\!(4,4,5):-2~~~\,&\\
    (5,3,5):-2~~~\,&\\
  (6,2,5):-2~~&\\
  (7,1,5):-2~\!&\!\!~(10,1,7):-3\\
  (8,0,5):-2\!\!&\!\!\!(9,0,6):-3
\end{array}
\eeq
The Lascoux matrix factorizations that are glued in the
third step have
\beq
\begin{array}{cc}
  &\,~~~(8,3,6):-1\\
  (2,2,2):0~~\!&~~(7,2,5):-1\\
  (3,1,2):0~\!\!&\!~(6,1,4):-1\\
  (4,0,2):0\!\!\!&\!\!(5,0,3):-1
\end{array}\qquad\mbox{and}\qquad
\begin{array}{cc}
  (3,3,4):-2~~~\,&\\
  (4,2,4):-2~~&\!~~(9,2,7):-3\\
  (5,1,4):-2~\!&\!\!~(8,1,6):-3\\
  (6,0,4):-2\!\!&\!\!\!(7,0,5):-3
\end{array}
\eeq
After the tachyon condensation, we are left with the
components with the R-charges as described in Fig.~\ref{fig:RL0}.

Next, we work on the field space where $p$ takes values in the
neighborhood $\nhU\subset \CP^6$ of $p_*\in Y\subset\CP^6$
as in Section~\ref{subsec:theproof}.
The part $d_*$ of the matrix factorization defines a complex
\beqa
& \mbox{height 2:}~~~
\textw^{\!7,2}~\longleftarrow~
  \textw^{\!3,0}~\longleftarrow~
  \textw^{\!0,3}~\longleftarrow~
  \textw^{\!2,7}\nn\\
  & \mbox{height 1:}~~~
  \textw^{\!7,1}\oplus\textw^{\!6,2}\mbox{\Large '}~\longleftarrow~
  \textw^{\!3,1}~\longleftarrow~
  \textw^{\!1,3}~\longleftarrow~
  \textw^{\!2,6}\mbox{\Large '}\oplus \textw^{\!1,7}\label{dstarL0}\\
  & \mbox{height 0:}~~~
  \textw^{\!7,0}\oplus\textw^{\!6,1}\mbox{\Large '}
  \oplus\textw^{\!5,2}\mbox{\Large '}~\longleftarrow~
  \textw^{\!3,2}~\longleftarrow~
  \textw^{\!2,3}~\longleftarrow~
  \textw^{\!2,5}\mbox{\Large '}\oplus\textw^{\!1,6}\mbox{\Large '}
  \oplus\textw^{0,6}\nn
\eeqa
where $\textw^{\!6,1}\mbox{\Large '}$, $\textw^{\!5,2}\mbox{\Large '}$,
$\textw^{\!6,2}\mbox{\Large '}$ and
$\textw^{\!1,6}\mbox{\Large '}$, $\textw^{\!2,5}\mbox{\Large '}$,
$\textw^{\!2,6}\mbox{\Large '}$ 
are the cokernels and the kernels of
the flavor parts of $\psi_5,\varphi_4,\varphi_a$
and $\bar\psi_5,\bar\varphi_4,\bar\varphi_a$ given in
(\ref{vp54afla}) and (\ref{barvp54afla}).
Again we omit the color factors $S^h\sfund^*$ on which $d_*$ acts trivially.
The arrows of (\ref{dstarL0}) are those of (\ref{L0arrows})
in which $\mathring{A}$, $\bar{A}$, $A$
are replaced by (\ref{dAstar}).

The tachyon condensation of the invertible parts of $d_*$
leaves us with a matrix factorization in the reduced form.
The underlying vector space is the $d_*$-cohomology, and we find that it is
\beq
\begin{array}{c|cccc}
    {\rm height}
  &{\rm H}^0_{d_*}&{\rm H}^{-1}_{d_*}&{\rm H}^{-2}_{d_*}&{\rm H}^{-3}_{d_*}\\
  \hline
2&  \textw^{\!3\,}\Ttt^*\otimes \textw^{\!4\,}\cktt^*\otimes \textw^{\!2\,}_0\cktt
  &\Ttt^*\otimes \textw^{\!2\,}_0\cktt^*,&\Ttt\otimes \textw^{\!2\,}_0\cktt
  &\textw^{\!3\,}\Ttt\otimes\textw^{\!4\,} \cktt\otimes \textw^{\!2\,}_0\cktt^*\\
1&  \textw^{\!3\,}\Ttt^*\!\otimes\! (\mbox{20 dim})&
  \Ttt^*\!\otimes \cktt\otimes \textw^{\!2\,}_0 \cktt^*&
  (\mbox{$3\times 20$ dim})&
  \textw^{\!3\,}\Ttt\otimes (\mbox{20 dim})\\
0&  (\mbox{25 dim})&
  \Ttt^*\!\otimes \textw_0^{\!2\,}\cktt\otimes \textw^{\!2\,}_0\cktt^*&
 (\mbox{$3\times 25$ dim})&
\textw^{\!3\,}\Ttt\otimes (\mbox{25 dim})
\end{array}
\label{redFF0}
\eeq
The computation of the zigzag maps has not been
done yet. However, the result of the cohomology
suggests that the reduced matrix factorization is isomorphic to the
tensor product of the horizontal $\brane_{\Ttt}$,
the vertical $\brane^{SU(2)}_{3,7}$, and a copy of
$\textw_0^{\!2}\cktt\cong \textw_0^{\!2}\cktt^*$.
This is nothing but $\brane^{\rm D6}_{\nhU}\otimes \textw_0^{\!2}\cktt$.

\section{Computation Of Chern Characters In The Pfaffian Phase}
\label{app:chPf}

We would like to check the match of the Chern characters
\beq
\lfloor \Psi_{\brane_{A,-1}}(\ck^*)\rfloor
  \,=\,\lfloor i_-^*\Lb^1\rfloor \quad\mbox{and}\quad
\lfloor \Psi_{\brane_{A,-1}}(\textw^{\!2}_0\ck^*)\rfloor
  \,=\,\lfloor i_-^*\Lb^0[1]\rfloor.\label{chL1L0}
\eeq
Inserting the formulae (\ref{chckst}) for the Chern characters of $\ck^*$
and $\textw^{\!2}_0\ck^*$ into the expression (\ref{CHPF}),
we find that the left-hand sides are
\beqa
\lfloor \Psi_{\brane_{A,-1}}(\ck^*)\rfloor
&=&-\Sym^2\fund^*\,\chdot\, 14H^3\cdot 4\left(1-5H+13H^2-{70H^3\over 3}\right)\nn\\
&&-\fund^*\,\chdot\, 14H^3\cdot 4\left(4-18H+42H^2-{135H^3\over 2}\right)\nn\\
&&-\C\,\chdot\, 14H^3\cdot 4\left(5-20H+{83H^2\over 2}-{178H^3\over 3}\right)
\label{LHSL1}
\eeqa
and
\beqa
\lfloor \Psi_{\brane_{A,-1}}(\textw^{\!2}_0\ck^*)\rfloor
&=&-\Sym^2\fund^*\,\chdot\, 14H^3\cdot 5\left(1-{9H\over 2}+{211H^2\over 20}-{171H^3\over 10}\right)\nn\\
&&-\fund^*\,\chdot\, 14H^3\cdot 5\left(4-16H+{166H^2\over 5}-{712H^3\over 15}\right)\nn\\
&&-\C\,\chdot\, 14H^3\cdot 5\left(5-{35H\over 2}+{127H^2\over 4}-{119H^3\over 3}\right).
\label{LHSL0}
\eeqa
On the other hand, in view of the Pfaffian frame R-charges of
$\Lb^1$ and $\Lb^0$ as shown in Fig.~\ref{fig:RL1} and Fig.~\ref{fig:RL0},
we have
\beqa
\lfloor i_-^*\Lb^1\rfloor
\!\!&=&\!\!
\Bigl\lfloor\,-\Bigl(\,\sym^2\fund^*(2)\textw^{\!7,1}+\fund^*(1)\textw^{\!7,0}
+\fund^*(2)\textw^{\!6,1}+\C(1)\textw^{\!6,0}+\C(2)\textw^{\!5,1}\mbox{\Large '}
\,\Bigr)\nn\\
&&+\Bigl(\,\sym^2\fund^*(4)\textw^{\!2,0}+\fund^*(3)\textw^{\!3,0}+\fund^*(4)\textw^{\!2,1}+\C(3)\textw^{\!3,1}\,\Bigr)\nn\\
&&-\Bigl(\,\sym^2\fund^*(6)\textw^{\!0,2}+\fund^*(5)\textw^{1,2}+\fund^*(6)\textw^{\!0,3}+\C(5)\textw^{\!1,3}\,\Bigr)\nn\\
&&+\Bigl(\sym^2\fund^*(8)\textw^{1,7}+\fund^*(7)\textw^{\!1,6}+\fund^*(8)\textw^{\!0,7}+\C(6)\textw^{\!1,5}\mbox{\Large '}+\C(7)\textw^{\!0,6}\,\Bigr)\Bigr\rfloor
\eeqa
and
\beqa
\lfloor i_-^*\Lb^0\rfloor
\!\!&=&\!\!
\Bigl\lfloor\Bigl(\sym^2\fund^*(2)\textw^{\!7,2}+\fund^*(1)\textw^{\!7,1}
+\fund^*(2)\textw^{\!6,2}\mbox{\Large '}+\C\textw^{7,0}+\C(1)\textw^{\!6,1}\mbox{\Large '}+\C(2)\textw^{\!5,2}\mbox{\Large '}\Bigr)\nn\\
&&-\Bigl(\sym^2\fund^*(3)\textw^{\!3,0}+\fund^*(3)\textw^{\!3,1}+\C(3)\textw^{\!3,2}\Bigr)\nn\\
&&+\Bigl(\sym^2\fund^*(6)\textw^{0,3}+\fund^*(5)\textw^{1,3}+\C(4)\textw^{\!2,3}\Bigr)\\
&&-\Bigr(\sym^2\fund^*(7)\textw^{\!2,7}+\fund^*(6)\textw^{2,6}\mbox{\Large '}+\fund^*(7)\textw^{\!1,7}+\C(5)\textw^{\!2,5}\mbox{\Large '}+\C(6)\textw^{\!1,6}\mbox{\Large '}+\C(7)\textw^{\!0,7}\Bigr)\Bigr\rfloor,\nn
\eeqa
in which we omit the $\otimes$ sign to shorten the expressions.
The dimension of
$\textw^{\!p,q}=\textw^{\!p}V^*\otimes\textw^{\!q}V$
is ${7\choose p}\cdot{7\choose q}$
while those of $\textw^{\!p,q}\mbox{\Large '}$, which are
the cokernels or the kernels of the maps
(\ref{vp4fla}), (\ref{vp54afla}), (\ref{barvp54afla}),
are
\beqa
&{\rm dim}\,\textw^{\!5,1}\mbox{\Large '}=
    {\rm dim}\,\textw^{\!1,5}\mbox{\Large '}={7\choose 5}\cdot 7-{7\choose 4}=16\cdot 7,\\
    &{\rm dim}\,\textw^{\!6,1}\mbox{\Large '}=
    {\rm dim}\,\textw^{\!1,6}\mbox{\Large '}={7\choose 6}\cdot 7-{7\choose 5}=28,\\
    &{\rm dim}\,\textw^{\!5,2}\mbox{\Large '}=
    {\rm dim}\,\textw^{2,5}\mbox{\Large '}={7 \choose 5}\cdot{7\choose 2}-{7\choose 4}\cdot 7=49\cdot 4,\\
    &{\rm dim}\,\textw^{\!6,2}\mbox{\Large '}=
    {\rm dim}\,\textw^{\!2,6}\mbox{\Large '}={7\choose 6}\cdot {7\choose 2}-{7\choose 4}=16\cdot 7.
\eeqa
Thus, we find
\beqa
\lfloor i_-^*\Lb^1\rfloor
\!\!&=&\!\!
\sym^2\fund^*\,\chdot\left(\,-7\e^{-2H}+21\e^{-4H}-21\e^{-6H}+7\e^{-8H}\,\right)\nn\\
&&\!\!\!\!\!\!\!\!\!\!\!\!\!\!\!\!
+\fund^*\,\chdot\left(\,-\e^{-H}-49\e^{-2H}+35\e^{-3H}+21\cdot 7\e^{-4H}\right.\nn\\
&&~~~~~~~~~~~~~~~~~~~
\left.-21\cdot 7\e^{-5H}-35\e^{-6H}+49\e^{-7H}+\e^{-8H}\,\right)\\
&&\!\!\!\!\!\!\!\!\!\!\!\!\!\!\!\!
+\C\,\chdot\left(\,-7\e^{-H}-16\cdot 7\e^{-2H}+35\cdot 7\e^{-3H}-35\cdot 7\e^{-5H}+16\cdot 7\e^{-6H}+7\e^{-7H}\,\right)\nn
\eeqa
and
\beqa
\lfloor i_-^*\Lb^0[1]\rfloor
\!\!&=&\!\!
-\sym^2\fund^*\,\chdot\left(\,21\e^{-2H}-35\e^{-3H}+35\e^{-6H}-21\e^{-7H}\,\right)\nn\\
&&\!\!\!\!\!\!\!\!\!\!\!\!\!\!\!\!
-\fund^*\,\chdot\left(\,7\e^{-H}+16\cdot 7\e^{-2H}-35\cdot 7\e^{-3H}+35\cdot 7\e^{-5H}-16\cdot 7\e^{-6H}-7\e^{-7H}\,\right)\nn\\
&&\!\!\!\!\!\!\!\!\!\!\!\!\!\!\!\!
-\C\,\chdot\left(\,1+28\e^{-H}+49\cdot 4\e^{-2H}-35\cdot 21\e^{-3H}\right.\nn\\
&&~~~~~~~~~~~~~~~~~~~
\left.+35\cdot 21\e^{-4H}-49\cdot 4\e^{-5H}-28\e^{-6H}-\e^{-7H}\,\right).
\eeqa
These precisely match with (\ref{LHSL1}) and (\ref{LHSL0}) respectively,
confirming (\ref{chL1L0}).

\section{More On Homological Algebra}
\label{app:moreHA}

\subsection{Some Facts On Complexes Of Vector Bundles}
\label{subsec:locfree}

\noindent
{\bf Theorem.}
Let $({\mathcal E}^*,d^*)$ be a complex of coherent sheaves on a complex
manifold $M$ of dimension $n$.
Suppose $n$ consecutive terms, say
${\mathcal E}^0,\ldots, {\mathcal E}^{n-1}$, are locally free
and the complex is exact at the latter $(n-1)$ terms,
${\mathcal E}^1,\ldots, {\mathcal E}^{n-1}$.
Then, the kernel of $d^0:{\mathcal E}^0\to {\mathcal E}^1$
is locally free.\footnote{We thank Alexey Bondal  (February 2025)
  for the statement and the proof of this theorem.}

\noindent
{\bf Proof.}
We shall denote $\Ext^p$ for $\Ext^p_M$.
Note that a coherent sheaf ${\mathcal E}$ on $M$ is locally free
if and only if $\Ext^{>0}({\mathcal E},{\mathcal O}_x)=0$ for any point
$x$ of $M$.
The short exact sequence
\beq
0\,\longrightarrow\, {\rm Ker}\,d^i\,\longrightarrow\,
{\mathcal E}^i\,\longrightarrow\,
{\rm Im}\,d^i\,\longrightarrow\,0
\eeq
induces the long exact sequence
\beq
\cdots\to \Ext^p({\mathcal E}^i,{\mathcal O}_x)
\to \Ext^p({\rm Ker}\,d^i,{\mathcal O}_x)
\to \Ext^{p+1}({\rm Im}\,d^i,{\mathcal O}_x)
\to \Ext^{p+1}({\mathcal E}^i,{\mathcal O}_x)
\to \cdots
\eeq
As ${\mathcal E}^i$ is locally free for $0\leq i\leq n-1$, we find
\beq
\Ext^p({\rm Ker}\,d^i,{\mathcal O}_x)\,\cong\,
\Ext^{p+1}({\rm Im}\,d^i,{\mathcal O}_x),\qquad
\forall p>0,~0\leq i\leq n-1.
\eeq
Note that
${\rm Im}\,d^i={\rm Ker}\, d^{i+1}$ for $0\leq i\leq n-2$
by the exactness assumption.
Thus, we find
\beqa
\lefteqn{\Ext^p({\rm Ker}\,d^0,{\mathcal O}_x)\cong
\Ext^{p+1}({\rm Ker}\,d^1,{\mathcal O}_x)\cong
\cdots}\nn\\
&&\cdots \cong\Ext^{p+n-1}({\rm Ker}\, d^{n-1},{\mathcal O}_x)\cong
\Ext^{p+n}({\rm Im}\, d^{n-1},{\mathcal O}_x)
\eeqa
This vanishes for $p>0$ as $\Ext^j({\mathcal E},{\mathcal F})=0$
unless $0\leq j\leq n$ for any coherent sheaves ${\mathcal E}$
and ${\mathcal F}$ on $M$ of dimension $n$. {\bf Q.E.D.}

This theorem can be used to see that
a coherent sheaf on a complex manifold
$M$ has a locally free resolution of length $\dim M+1$ or less.
More important for us is the application to
the description of images of the Kn\"orrer map,
such as those that appear in \cite{HHP}
and in Section~\ref{sec:Grassmannian} of the present paper.
Let ${\mathcal C}=({\mathcal C}^*,d^*)$ be a complex of vector bundles
which is infinitely long to the right and is exact
at degrees $i\geq i_*$ for some $i_*$. Then, for any $j\geq i_*-1$,
the truncation ${\mathcal C}^{\leq j}$ of ${\mathcal C}$ obtained by
replacing the degree $j$ term by ${\rm Ker}\, d^j$ and removing
the higher degree terms is quasi-isomorphic to ${\mathcal C}$.
The theorem says that this truncation is also a complex of vector bundles.
Applying this, we find that the low energy image
$\pi_X(\brane)$ of a gauge theory brane $\brane$
in the Grassmannian phase
is quasi-isomorphic to a complex of vector bundles of finite length.
Also, for an infinitely long exact complex of vector bundles,
such as (\ref{longcomplex}) on $X$ and (\ref{longcomplexXY}) on $X\times Y$,
the kernel of each differential is a vector bundle,
and hence so is the cokernel.

\subsection{Computation Of Some Ext Groups: R\o dland Model}
\label{subsec:ExtRGr}

Let $X\subset \Gr=G(2,7)$ be the Calabi-Yau threefold
that appears in the R\o dland model.
We shall show\footnote{We thank
  Will Donovan (January 2015) and Wahei Hara (April 2024) for instructions
  for the proof of (\ref{Sphe}) and (\ref{OSym2orth}), respectively.}
that ${\mathcal O}_X$, $S_X^*$,
$\Sym^2 S_X^*$, or their tensor products with line bundles
are spherical objects of $D(X)$, i.e. if ${\mathcal E}$ is one
of these,
\beq
\Ext^p_X({\mathcal E},{\mathcal E})\,\,\cong\,\,
\left\{\begin{array}{ll}
\C&p=0,3,\\
0&\mbox{else}
\end{array}\right.
\label{Sphe}
\eeq
and that ${\mathcal O}_X$ and $\Sym^2S_X^*(-1)$ are orthogonal
\beq
\Ext_X^p({\mathcal O}_X,\Sym^2 S_X^*(-1))\,\,=\,\,0\quad
\forall p.
\label{OSym2orth}
\eeq

We use the adjunction relation
for the inclusion map $i:=i_X:X\hookrightarrow \Gr$
\beq
\Ext^p_X(i^*E,i^*F)\cong \Ext^p_{\Gr}(E,i_*i^*F),
\label{adjunction}
\eeq
for
\beqa
\mbox{Case I}:&&\!\!\!
E=F={\mathcal O}_{\Gr},\,\, S^*,\,\,\Sym^2 S^*,\\
\mbox{Case II}:&&\!\!\!
E={\mathcal O}_{\Gr},~\,F=\Sym^2 S^*(-1).\nn
\eeqa
The projection formula
\beq
i_*i^*F=i_*({\mathcal O}_X\otimes i^*F)
=i_*{\mathcal O}_X\otimes F
\label{projection}
\eeq
and the Koszul resolution (\ref{KosOX}) of $i_*{\mathcal O}_X$
yields a resolution of $i_*i^*F$. That is, we have
an exact sequence of sheaves on $\Gr$,
\beq
0\longrightarrow F(-7)\stackrel{d^{-7}}{\longrightarrow}
F(-6)^{\oplus 7}\stackrel{d^{-6}}{\longrightarrow}
\cdots \stackrel{d^{-2}}{\longrightarrow}
F(-1)^{\oplus 7}\stackrel{d^{-1}}{\longrightarrow}
F\stackrel{d^0}{\longrightarrow}
i_*i^*F\stackrel{d^1}{\longrightarrow}
0.
\label{theES}
\eeq
For $0\leq i\leq 7$, we have a short exact sequence
\beq
0\,\longrightarrow\,
{\rm Ker}\,d^{-i}\,\longrightarrow\,
F(-i)^{\oplus {7\choose i}}\,\longrightarrow\,
{\rm Ker}\,d^{-i+1}\,\longrightarrow\,
0
\eeq
which induces a long exact sequence
\beqa
\cdots\!\!\!\!\!&&\to
\Ext^p_{\Gr}(E,F(-i)^{\oplus {7\choose i}})\to
\Ext^p_{\Gr}(E,{\rm Ker}\,d^{-i+1})\to \Ext^{p+1}_{\Gr}(E,{\rm Ker}\,d^{-i})\nn\\
&&\to\Ext^{p+1}_{\Gr}(E,F(-i)^{\oplus {7\choose i}})\to~\cdots
\label{longSphe}
\eeqa
At this point, we use
\beqa
\Ext^p_{\Gr}(E,F(-i))\!\!\!&\stackrel{\rm Case\,\, I}{=}&
\!\!\!\left\{\begin{array}{cl}
\delta^p_0\C&i=0,\\
0&i=1,\ldots, 6,\\
\delta^p_{10}\C&i=7
\end{array}\right.
\label{BWBI}\\
&\stackrel{\rm Case\,\, II}{=}&
~~~\,0\qquad i=0,\ldots, 7.
\label{BWBII}
\eeqa
This will be proven below as a special case of a more general one.
Inserting this in (\ref{longSphe}), we find for $p\geq -1$
\beqa
\lefteqn{\Ext^{p+1}_{\Gr}(E,{\rm Ker}\, d^0)\cong
  \Ext^{p+2}_{\Gr}(E,{\rm Ker}\, d^{-1})\cong\cdots}\nn\\
&&\cdots\cong\Ext^{p+7}_{\Gr}(E,{\rm Ker}\,d^{-6})\cong\Ext^{p+7}_{\Gr}(E,F(-7))
=\left\{\begin{array}{cl}
\delta^p_3\C&\mbox{Case I},\\
0&\mbox{Case II}.
\end{array}\right.
\eeqa
The $i=0$ case of (\ref{longSphe}) reads
\beq
\Ext_{\Gr}^p(E,{\rm Ker}\, d^0)\to
\Ext_{\Gr}^p(E,F)\to
\Ext_{\Gr}^p(E,i_*i^*F)\to
\Ext_{\Gr}^{p+1}(E,{\rm Ker}\, d^0)\to
\Ext_{\Gr}^{p+1}(E,F)
\eeq
Combining these, we find that $\Ext_{\Gr}^p(E,i_*i^*F)=\C$ for $p=0,3$
and $0$ for other $p$ in Case I
and $\Ext_{\Gr}^p(E,i_*i^*F)=0$ for all $p$ in Case II.
This proves (\ref{Sphe}) and (\ref{OSym2orth}).

We may repeat this computation in the $N$-R\o dland model, for
$X\subset\Gr=G(2,N)$ the Calabi-Yau manifold of dimension $(N-4)$
that appears in the Grassmannian phase. We show that $\Sym^h S_X^*$
for $h=0,1,\ldots, {N-3\over 2}$ (and their tensor products with line bundles)
are spherical and that
$\Sym^hS_X^*$ is orthogonal to $\Sym^{h-2k}S_X^*(k)$ for $k=1,2,\ldots$.
We use the ajdunction and the projection formulae
(\ref{adjunction})-(\ref{projection})
for $i:X\hookrightarrow \Gr$ and the $7\to N$ version of
the exact sequence (\ref{theES}).
What we need is 
\beqa
\Ext^p_{\Gr}(E,F(-i))\!\!\!&\stackrel{\rm Case\,\, I}{=}&
\!\!\!\left\{\begin{array}{cl}
\delta^p_0\C&i=0,\\
0&i=1,\ldots, N-1,\\
\delta^p_{2(N-2)}\C&i=N
\end{array}\right.
\label{BWBNI}\\
&\stackrel{\rm Case\,\, II}{=}&
~~~\,0\qquad i=0,\ldots, N
\label{BWBNII}
\eeqa
where
\beqa
\mbox{Case I}:&&\!\!\!
E=F=\Sym^h S^*,\quad h=0,1,\ldots, {N-3\over 2}\label{CaseIN}\\
\mbox{Case II}:&&\!\!\!
E=\Sym^h S^*,~\,F=\Sym^{h-2k}S^*(k),\quad k=1,2,\ldots ,\label{CaseIIN}
\eeqa
which generalizes (\ref{BWBI})-(\ref{BWBII}).\\
{\bf Proof of (\ref{BWBNI}) and (\ref{BWBNII}).}
We use the Borel-Weil-Bott theorem as described in Section~\ref{subsub:BWB}.
The most important player is the vector ${\bf x}$ as in
(\ref{vectorx}).
For Case I, 
\beq
{\bf x}
=(\underbrace{N+h-i-j}_{\displaystyle ~~~~ =:a},
\underbrace{N-1-h-i+j}_{\displaystyle ~~~=:b},
N-2,\ldots, 2,1),\quad j=0,1,\ldots, h,
\eeq
$b$ is largest when $i=h-j=0$, in which case ${\bf x}=\rho$,
and this contributes to $\C\subset \Ext^0$ for $i=0$.
When $1\leq (h-j)+i\leq N-2$, $b$ coincides with one of $N-2,\ldots, 2,1$
and this has no contribution to $\Ext^p$.
Thus, we focus on the case $(h-j)+i\geq N-1$ for which $b\leq 0$.
$a\geq N-1$ when $(h-j)-i\geq -1$ which requires
$h-j\geq {N-2\over 2}$, but this is impossible as $h-j\leq {N-3\over 2}$.
Thus, coincident entries can be avoided only when $a\leq 0$ which means
$-(h-j)+i\geq N$. Then, we have $i\geq {2N-1\over 2}$.
In the range $0\leq i\leq N$, this is possible only for
$i=N$ which in turn requires $j\geq h$, i.e., $j=h$.
Then, ${\bf x}=(0,-1,N-2,\ldots, 2,1)$ and a permutation $\sigma$
of $2\times (N-2)$ exchanges is needed to bring it in order.
Since $\Sigma^{\sigma{\bf x}-\rho}V^*=\det^{-2}V^*$, this contributes to
$\C\subset \Ext^{2(N-2)}$ for $i=N$. (\ref{BWBNI}) is proven.
For Case II,
\beqa
    {\bf x}&=&(\underbrace{N+h-i-k-j}_{\displaystyle ~~~=:a},
    \underbrace{N-1-(h-2k)-i-k+j}_{\displaystyle ~~~=:b},N-2,\ldots, 2,1),\\
&&    j=0,1,\ldots, h-2k.\nn
\eeqa
For $0\leq i\leq N$, $a$ is at least $h-k-j\geq k\geq 1$.
$b$ is at most $N-2$ and coincident entries can be avoinded only when
$b\leq 0$ which requires $i\geq N-1-h+k+j$.
In this case,
$a$ is at most $N+h-k-j-(N-1-h+k+j)=2h-2k-2j+1\leq 2h-1$.
As $h\leq {N-3\over 2}$, the upper bound is at most $N-4$.
Thus, $1\leq a\leq N-4$ and coincident entries cannot be avoided.
This proves (\ref{BWBNII}). {\bf Q.E.D.}

The computation can be repeated also in the variants of the $N$-R\o dland model,
for the Calabi-Yau manifold $X\subset \Gr=G(2,N)$ which is the
complete intersection of hypersurfaces
$A_1(x)=\cdots =A_M(x)=0$ of degree $d_1,\ldots,d_M$ in the
Pl\"ucker coordinates. The r\^ole of (\ref{theES}) is played by
\beq
0\longrightarrow \underbrace{F\otimes\textw^MV_{\Gr}^*}_{\displaystyle F(-N)}
\stackrel{d^{-M}}{\longrightarrow}
F\otimes\textw^{\!M-1}V_{\Gr}^*\stackrel{d^{-M+1}}{\longrightarrow}
\cdots \stackrel{d^{-2}}{\longrightarrow}
F\otimes V_{\Gr}^*\stackrel{d^{-1}}{\longrightarrow}
F\stackrel{d^0}{\longrightarrow}
i_*i^*F\stackrel{d^1}{\longrightarrow}
0,
\eeq
where $V_{\Gr}^*=\oplus_k{\mathcal O}_{\Gr}(-d_k)$.
What we need is
\beqa
\Ext^p_{\Gr}(E,F\otimes\textw^{\!i}V_{\Gr}^*)\!\!\!&\stackrel{\rm Case\,\, I}{=}&
\!\!\!\left\{\begin{array}{cl}
\delta^p_0\C&i=0,\\
0&i=1,\ldots, M-1,\\
\delta^p_{2(N-2)}\C&i=M
\end{array}\right.
\\
&\stackrel{\rm Case\,\, II}{=}&
~~~\,0\qquad i=0,\ldots, M,
\eeqa
where Case I and Case II are (\ref{CaseIN}) and (\ref{CaseIIN}).
This reduces to (\ref{BWBNI}) and (\ref{BWBNII})
since $\textw^{\!i}V_{\Gr}^*$ is the sum of
${\mathcal O}_{\Gr}(-d_{k_1}-\cdots -d_{k_i})$ for
$1\leq k_1<\cdots <k_i\leq M$, in which $0\leq d_{k_1}+\cdots+d_{k_i}\leq N$
where the lower bound is attained for $i=0$
and the upper bound is attained for $i=M$.

\subsection{Computation Of Some Ext Groups: Hosono-Takagi Model}
\label{subsec:ExtHT}

Let $X$ be the Reye congruence that appears in the Hosono-Takagi model.
We shall show that ${\mathcal O}_X$, $\vHT_X$,
$\Sym_0^2 \vHT_X$, or their tensor product with line bundles
are spherical objects of $D(X)$, i.e. if ${\mathcal E}$ is one
of these,
\beq
\Ext^p_X({\mathcal E},{\mathcal E})\,\,\cong\,\,
\left\{\begin{array}{ll}
\C&p=0,3,\\
0&\mbox{else}
\end{array}\right.
\label{SpheHT}
\eeq
and that ${\mathcal O}_X$, ${\mathcal O}_X^-$ and $\Sym_0^2\vHT_X(-1)$
are orthogonal to one another, i.e.,
if ${\mathcal E}\ne{\mathcal F}$ are two of these,
\beq
\Ext_X^p({\mathcal E},{\mathcal F})\,\,=\,\,0\quad
\forall p.
\label{orthHT}
\eeq

We recall that $X$ is the free quotient of the
complete intersection $\wt{X}$ of five symmetric bilinear hypersurfaces
in $\wt{\PPr}=\CP^4\times\CP^4$ by the exchange $\tau$
of the two $\CP^4$ factors.
We write $i:=i_{\wt{X}}:\wt{X}\hookrightarrow \wt{\PPr}$,
${\mathcal O}_{\wt{\PPr}}(i,j):={\mathcal O}_{\CP^4}(i)\boxtimes {\mathcal O}_{\CP^4}(j)$ and
${\mathcal O}_{\wt{X}}(i,j)=i^*{\mathcal O}_{\wt{\PPr}}(i,j)$.
A vector bundle on $X$ lifts to a $\Z_2$-equivariant
vector bundle on $\wt{X}$.
The line bundle ${\mathcal O}_X(i)$ lifts to ${\mathcal O}_{\wt{X}}(i,i)$
with the canonical action over $\tau$.
The line bundle ${\mathcal O}_X^-$ lifts to the trivial bundle
${\mathcal O}_{\wt{X}}$ with a sign flip over $\tau$.
The vector bundle $\vHT_X$ lifts to
${\mathcal O}_{\wt{X}}(1,0)\oplus {\mathcal O}_{\wt{X}}(0,1)$
with the canonical action over $\tau$ that exchanges ${\mathcal O}_{\wt{X}}(1,0)$
and ${\mathcal O}_{\wt{X}}(0,1)$. It has a symmetric bilinear form $(\,,\,)$
with values in ${\mathcal O}_X(1)$
that lifts to the pairing
${\mathcal O}_{\wt{X}}(1,0)\times{\mathcal O}_{\wt{X}}(0,1)\to {\mathcal O}_{\wt{X}}(1,1)$.
The symmetric and $(\,,\,)$-traceless $\elln$-times tensor power 
$\Sym^{\elln}_0\vHT_X$ of $\vHT_X$ lifts to
${\mathcal O}_{\wt{X}}(\elln,0)\oplus{\mathcal O}_{\wt{X}}(0,\elln)$ with the canonical action over $\tau$.
Note that the tensor product with ${\mathcal O}_X^-$
changes ${\mathcal O}_X(i)$ to ${\mathcal O}_X^-(i)$
but does not change $\Sym^{\elln}_0V_X(i)$ for $\elln\geq 1$.

If vector bundles ${\mathcal E}$ and ${\mathcal F}$ on $X$ pull back to
$\Z_2$-equivariant vector bundles $\wt{\mathcal E}$ and $\wt{\mathcal F}$
on $\wt{X}$,
\beq
\Ext_X^p({\mathcal E},{\mathcal F})\,\,\cong\,\,
\Ext_{\wt{X}}^p(\wt{\mathcal E},\wt{\mathcal F})^{\Z_2}.
\eeq
For our problem, the vector bundles $\wt{\mathcal E}$ and
$\wt{\mathcal F}$ are all pullbacks of vector bundles on $\wt{\PPr}$.
The computation of $\Ext_{\wt{X}}^p$ for such pairs
can be done in the same way as in the previous subsection.
We employ the adjunction and projection
formulae, like (\ref{adjunction}) and (\ref{projection}),
for $i:\wt{X}\hookrightarrow \wt{\PPr}$ this time,
as well as the exact sequence on $\wt{\PPr}$,
\beq
0\longrightarrow F(-5,-5)\stackrel{d^{-5}}{\longrightarrow}
F(-4,-4)^{\oplus 5}\stackrel{d^{-4}}{\longrightarrow}
\cdots \stackrel{d^{-2}}{\longrightarrow}
F(-1,-1)^{\oplus 5}\stackrel{d^{-1}}{\longrightarrow}
F\stackrel{d^0}{\longrightarrow}
i_*i^*F\stackrel{d^1}{\longrightarrow}
0
\label{eswtP}
\eeq
 for a vector bundle $F$ on $\wt{\PPr}$,
which results from the Koszul resolution of $i_*{\mathcal O}_{\wt{X}}$. 
We shall also use the formula (\ref{BWBProj}) for the case
$V=\C^5$,
\beq
    {\rm H}^p(\CP^4,{\mathcal O}_{\CP^4}(\alpha))\,\,=\,\,\left\{
    \begin{array}{ll}
      \Sym^{\alpha}V^*&\alpha\geq 0,\,\,\,p=0,\\
      \Sym^{-\alpha-5}V\otimes \det V&\alpha\leq -5,\,\,\,p=4,\\
      0&\mbox{otherwise},
    \end{array}\right.
    \label{BWBIII}
\eeq
As (\ref{BWBI})-(\ref{BWBII}) in the R\o dland model, the group 
\beq
\Ext_{\wt{\PPr}}^p({\mathcal O}_{\wt{\PPr}}(k,l),{\mathcal O}_{\wt{\PPr}}(m-i,n-i))
\quad\mbox{for $i=0,1,\ldots,5$}
\label{importantExt}
\eeq
turns out to be important
in the computation of $\Ext_X({\mathcal E},{\mathcal F})$, where
\beq
\begin{array}{rl}
    {\mathcal E}={\mathcal O}_X,\,\,
    {\mathcal F}={\mathcal O}_X,{\mathcal O}_X^-:
    &(k,l)=(m,n)=(0,0),\\
    {\mathcal E}={\mathcal F}=\vHT_X:&
    (k,l),(m,n)=(1,0)\,\,\mbox{and/or}\,\,(0,1)\\
    {\mathcal E}={\mathcal F}=\Sym_0^2\vHT_X:&
    (k,l),(m,n)=(2,0)\,\,\mbox{and/or}\,\,(0,2)\\
    {\mathcal E}={\mathcal O}_X,{\mathcal O}_X^-,\,\,
    {\mathcal F}=\Sym^2_0\vHT_X(-1):&
    (k,l)=(0,0),\,\,(m,n)=(\pm 1,\mp 1).
\end{array}
\label{rangeklmn}
\eeq
For $(k,l)$ and $(m,n)$ in this range,
(\ref{BWBIII}) implies that
the group (\ref{importantExt})
vanishes unless $(k,l)=(m,n)$, in which case it is
\beq
\Ext_{\wt{\PPr}}^p({\mathcal O}_{\wt{\PPr}},{\mathcal O}_{\wt{\PPr}}(-i,-i))
\,\,=\,\,\left\{\begin{array}{ll}
\delta^p_0\C&i=0,\\
0&i=1,\ldots, 4,\\
\delta^p_8\C&i=5
\end{array}\right.
\eeq
Insering this into the long exact sequences that arise from parts of
(\ref{eswtP}), we find
\beq
\Ext^p_{\wt{X}}({\mathcal O}_{\wt{X}}(k,l),{\mathcal O}_{\wt{X}}(m,n))
\,\,=\,\,0,\quad (k,l)\ne (m,n)~\,\mbox{from (\ref{rangeklmn}),}
\label{vanishHT}
\eeq
and
\beq
\Ext^p_{\wt{X}}({\mathcal O}_{\wt{X}}(k,l),{\mathcal O}_{\wt{X}}(k,l))
\,\,=\,\,
\Ext^p_{\wt{X}}({\mathcal O}_{\wt{X}},{\mathcal O}_{\wt{X}})
\,\,=\,\,
\left\{\begin{array}{ll}
\C&p=0,3,\\
0&\mbox{else.}
\end{array}\right.
\label{ExtCY3}
\eeq
$\Ext_X^p({\mathcal O}_X,{\mathcal O}_X)$
and $\Ext_X^p({\mathcal O}_X,{\mathcal O}_X^-)$ are the $\Z_2$ even and odd
parts of
$\Ext^p_{\wt{X}}({\mathcal O}_{\wt{X}},{\mathcal O}_{\wt{X}})$.
The group $\Ext_{\wt{X}}^0({\mathcal O}_{\wt{X}},{\mathcal O}_{\wt{X}})$
consists of constant functions on $\wt{X}$ which are
of course $\Z_2$ invariant.
The group $\Ext_{\wt{X}}^3({\mathcal O}_{\wt{X}},{\mathcal O}_{\wt{X}})$
is generated by a holomorphic three form on $\wt{X}$
which is also $\Z_2$ invariant by the residue formula \cite{HoTa}.
This shows (\ref{SpheHT}) for ${\mathcal E}={\mathcal O}_X$
and (\ref{orthHT}) for ${\mathcal E}={\mathcal O}_X$ and
${\mathcal F}={\mathcal O}_{\wt{X}}^-$.
$\Ext_X^p(\vHT_X,\vHT_X)$ is the $\Z_2$-invariant part of
\beq
\Ext_{\wt{X}}^p({\mathcal O}_{\wt{X}}(1,0)\oplus{\mathcal O}_{\wt{X}}(0,1),
    {\mathcal O}_{\wt{X}}(1,0)\oplus{\mathcal O}_{\wt{X}}(0,1))
\label{thisExt}
\eeq
which is the sum of
$\Ext_{\wt{X}}({\mathcal O}_{\wt{X}}(1,0),{\mathcal O}_{\wt{X}}(1,0))$ and
$\Ext_{\wt{X}}({\mathcal O}_{\wt{X}}(0,1),{\mathcal O}_{\wt{X}}(0,1))$
by (\ref{vanishHT}).
The two are exchanged by the $\Z_2$ and hence
the invariant part is a single copy of (\ref{ExtCY3}).
This shows (\ref{SpheHT}) for ${\mathcal E}=\vHT_X$.
Similarly for ${\mathcal E}=\Sym_0^2\vHT_X$: we just have to replace
$(1,0)$ and $(0,1)$ in this computation by $(2,0)$ and $(0,2)$.
$\Ext_X^p({\mathcal O}_X,\Sym_0^2\vHT_X(-1))$
and $\Ext_X^p({\mathcal O}^-_X,\Sym_0^2\vHT_X(-1))$
are the $\Z_2$ even and odd parts of
\beq
\Ext_{\wt{X}}^p({\mathcal O}_{\wt{X}},{\mathcal O}_{\wt{X}}(1,-1)\oplus{\mathcal O}_{\wt{X}}(-1,1))
\eeq
which vanishes by (\ref{vanishHT}). This shows (\ref{orthHT}) for
${\mathcal E}={\mathcal O}_X,{\mathcal O}_X^-$ and
${\mathcal F}=\Sym_0^2\vHT_X$.

\noindent
{\bf Remark.}
If we use the fact that $X$ and $\wt{X}$ are both Calabi-Yau,
it is obvious that ${\mathcal O}_X$ and ${\mathcal O}_X^-$
are spherical and orthogonal to each other.

\section{Some Topological Invariants}
\label{app:someinv}

We compute or constrain the Chern classes of
vector bundles that appear in the two phases of the R\o dland
and Hosono-Takagi models.
The results are mainly used in the computation of
the action of spherical twists on D-brane charges.
As a preparation, for rank 2 and 4 complex vector bundles $E$ and $F$
on a manifold of real dimension $\leq 6$, let us express
${\rm ch}(\Sym^2 E)$ and ${\rm ch}(\textw^{\!2}F)$
in terms of ${\rm ch}(E)$ and ${\rm ch}(F)$.
If $x_1, x_2$ are Chern roots of $E$,
the total Chern class of $\Sym^2 E$ is $\prod_{i\leq j}(1+x_i+x_j)
=1+3c_1(E)+2c_1(E)^2+4c_2(E)
+4c_1(E)c_2(E)$ which means
\beq
{\rm ch}(\Sym^2 E)=3+3{\rm ch}_1(E)+{1\over 2}{\rm ch}_1(E)^2
+4{\rm ch}_2(E)-{1\over 2}{\rm ch}_1(E)^3+4{\rm ch}_1(E){\rm ch}_2(E).
\label{chS2}
\eeq
If $x_1,x_2,x_3,x_4$
are Chern roots of $F$, the total Chern class
of $\textw^{\!2}F$ is $\prod_{i<j}(1+x_i+x_j)=1+3c_1(F)+3c_1(F)^2+2c_1(F)+c_1(F)^3+4c_1(F)c_2(F)$ which means
\beq
{\rm ch}(\textw^{\!2}F)=6+3{\rm ch}_1(F)+{1\over 2}{\rm ch}_1(F)^2+2{\rm ch}_2(F)+{\rm ch}_1(F){\rm ch}_2(F).
\label{chW2}
\eeq

\subsection{Grassmannian Calabi-Yau}
\label{app:classicalGr}

The cohomology group of the Grassmannian $G(k,N)$ is freely generated by
the Schubert classes $\sigma_{a_1,\ldots, a_k}$ labelled by a
non-increasing sequence of $k$ integers
between $0$ and $N-k$.
$\sigma_{0,\ldots, 0}$ is the identity element $1$ of the cohomology ring, and
we write $\sigma_{1,0,\ldots, 0}=\sigma_1$ etc for simplicity.
$\sigma_1$ is the positive generator of the Picard group.
Also,
\beq
\int_{G(k,N)}\sigma_{a_1,\ldots,a_k}\sigma_{b_1,\ldots, b_k}=\left\{
\begin{array}{ll}
  1&\mbox{if $a_i+b_{k-i+1}=k$ for all $i$,}\\
  0&\mbox{else}
\end{array}\right.
\eeq
See for example \cite{GH} for more.
The Chern classes of the dual $S^*$ of the tautological subbundle $S$
and the tautological quotient bundle $Q$ are
$c_i(S^*)=\sigma_{1,\ldots, 1}$ ($i$ 1's) and $c_i(Q)=\sigma_i$.
The two sets of classes are related by $c(S)c(Q)=1$
and either of them generates the cohomology ring of $G(k,N)$.
The expression for Chern classes of the tangent bundle
is found by ${\rm ch}(T_{G(k,N)})={\rm ch}(S^*){\rm ch}(Q)$.

For $\Gr=G(2,7)$, Pieri's formula yields
\beqa
\sigma_1^2&=&\sigma_2+\sigma_{1,1},\label{Pieri2}\\
&&\cdots\nn\\
\sigma_1^6&=&5\sigma_5+9\sigma_{4,2}+5\sigma_{3,3},\label{Pieri6}\\
\sigma_1^7&=&14\sigma_{5,2}+14\sigma_{4,3},\label{Pieri7}\\
\sigma_1^8&=&28\sigma_{5,3}+14\sigma_{4,4},\label{Pieri8}\\
&&\cdots\nn\\
\sigma_1^{10}&=&42\sigma_{5,5}.\label{Pieri10}
\eeqa
The Calabi-Yau threefold $X$ is the zero locus of a section of
${\mathcal O}_{\Gr}(1)^{\oplus 7}$ on $\Gr=G(2,7)$,
and hence $c(T_X)=i_X^*c(T_{G(2,7)})/c({\mathcal O}_X(1))^7$.
The Poincar\'e dual of $X\subset \Gr$ is 
the Euler class of ${\mathcal O}_{\Gr}(1)^{\oplus 7}$, which is
$c_1({\mathcal O}_{\Gr}(1))^7=\sigma_1^7$.
Therefore, for a cohomology class $\omega$ of $\Gr$,
\beq
\int_Xi_X^*\omega=\int_{G(2,7)}\omega\cdot \sigma_1^7.
\label{XtoGr}
\eeq
$i_X^*\sigma_1$ is the positive generator $H$ of the Picard group of $X$.
Using these and (\ref{Pieri2}), (\ref{Pieri8}), we find
$i_X^*\sigma_{1,1}=H^2/3$ and (\ref{topGra}).
In particular,
\beq
{\rm ch}(S_X^*)=2+H+{1\over 6}H^2.
\label{AppchSv}
\eeq
Inserting this into (\ref{chS2}) with $E=S^*$, we find
${\rm ch}(\Sym^2S_X^*)=3+3H+{7\over 6}H^2+{1\over 6}H^3$
and hence
\beq
    {\rm ch}(\Sym^2S_X^*(-1))=
{\rm ch}(\Sym^2S_X^*)\e^{-H}
=    3-{1\over 3}H^2.
\label{AppchSym2Sv}
\eeq
Written in terms of the Chern characters of
${\mathcal O}_X$, ${\mathcal O}_H$, ${\mathcal O}_{\ell}$,
${\mathcal O}_{\rm pt}$,
(\ref{AppchSv}) and (\ref{AppchSym2Sv}) can be read as
(\ref{chSv}) and (\ref{chSym2Sv}), respectively.
We also note that $\sigma_{1,1}^2=\sigma_{2,2}$. This together with
(\ref{Pieri6}) and (\ref{Pieri10})
yields $\sigma_1^6\sigma_{1,1}^2=5\sigma_{5,5}=5\sigma_1^{10}/42$,
which is the second of the relations in (\ref{cohrelGr}).
The Euler pairing (\ref{EulerG}) is found by the Riemann-Roch formula using 
the expressions (\ref{AppchSv}), (\ref{AppchSym2Sv}) and
\beq
    {\rm td}(X)=1+{1\over 12}c_2(X)=1+{1\over 6}H^2=1+7\ell.
\eeq

\subsection{Pfaffian Calabi-Yau}
\label{app:classicalPf}

In this paper, we do not attempt to directly compute the Chern classes of
the rank 4 and 5 vector bundles $\ck^*_Y$ and $\textw^2_0\ck^*_Y(1)$ on
the Pfaffian Calabi-Yau $Y$.
We take for granted the formulae (\ref{chck}) and (\ref{chw2ck})
that follow from the results (\ref{chL1}) and (\ref{chL0})  
from the hemisphere partition function and the identification
$\pi_Y(\brane^1)\cong \ck^*_Y$ and
$\pi_Y(\brane^0)\cong\textw^{\!2}_0\ck^*_Y(1)[-1]$.
However, we make a consistency check of this claim:
It must be compatible with the relation
bwteeen the Chern classes of $\ck^*_Y$ and 
$\textw^{\!2}_0\ck^*_Y(1)$.
In terms of the basis $(1,H,H^2,H^3)$ of the cohomology group of $Y$,
the formula (\ref{chck}) reads
\beq
{\rm ch}(\ck^*_Y)=4-2H+{1\over 6}H^3.
\label{Appchck}
\eeq
Inserting this into (\ref{chW2}) with $F=\ck^*_Y$, we find
${\rm ch}(\textw^{\!2}\ck^*_Y)=6-6H+2H^2$.
Recall that $\textw^{\!2}_0\ck_Y$ is defined as the kernel of a surjective map
$\textw^{\!2}\ck_Y\to {\mathcal O}_Y(1)$.
Using $\ck_Y\cong \ck^*_Y(1)$, (\ref{ckckdual}),
we have a short exact sequence
$0\to \textw^{\!2}_0\ck^*_Y(1)\to\textw^{\!2}\ck^*_Y(1)\to {\mathcal O}_Y\to 0$
which means
\beq
{\rm ch}(\textw^{\!2}_0\ck^*_Y(1))={\rm ch}(\textw^{\!2}\ck^*_Y)\e^H-1
=5-H^2.
\label{Appchw2ck}
\eeq
In terms of the basis
$([{\mathcal O}_Y], [{\mathcal O}_H],[{\mathcal O}_{\ell}],[{\mathcal O}_{\rm pt}])$, this is nothing but (\ref{chw2ck}). The claim is indeed self-consistent.
The Euler pairing (\ref{EulerP}) is found by the Riemann-Roch formula
using the expressions (\ref{Appchck}), (\ref{Appchw2ck}) and
\beq
{\rm td}(Y)=1+{1\over 12}c_2(Y)=1+{1\over 3}H^2=1+{14\over 3}\ell.
\eeq

\subsection{Reye Congruence}
\label{app:classicalReye}

We consider the Reye congruence $X$.
See the beginning of Appendix~\ref{subsec:ExtHT} for its description along with
$\wt{X}$, $\wt{\PPr}=\CP^4\times\CP^4$
and the vector bundles thereof.
We denote by $H_1$ and $H_2$ the first Chern classes of
${\mathcal O}_{\wt{\PPr}}(1,0)$ and ${\mathcal O}_{\wt{\PPr}}(0,1)$
and their restrictions to $\wt{X}$.
As $\wt{X}$ is the zero locus of five sections
of ${\mathcal O}_{\wt{\PPr}}(1,1)$, a cohomology class $\omega$ of
$\wt{\PPr}$ satisfies
\beq
\int_{\wt{X}}i_{\wt{X}}^*\omega
=\int_{\wt{\PPr}}\omega\cdot (H_1+H_2)^5.
\eeq
This shows the following relations on $\wt{X}$,
\beq
2H_1^2+2H_2^2=3H_1H_2,\quad
H_1^3=H_2^3=5[{\rm pt}],\quad
H_1^2H_2=H_1H_2^2=10[{\rm pt}].
\label{relwtP}
\eeq
The first Chern class $H$ of ${\mathcal O}_X(1)$ pulls back to $H_1+H_2$.
The rank 2 vector bundle $V_X$ pulls back to
${\mathcal O}_{\wt{X}}(1,0)\oplus {\mathcal O}_{\wt{X}}(0,1)$
with Chern character
\beq
\e^{H_1}+\e^{H_2}
=2+H_1+H_2+{1\over 2}(H_1^2+H_2^2)+{1\over 6}(H_1^3+H_2^3).
\label{chV22}
\eeq
The relations (\ref{relwtP}) yield $(H_1+H_2)^2={7\over 3}(H_1^2+H_2^2)$
and $(H_1+H_2)^3=7(H_1^3+H_2^3)$, and hence
\beq
{\rm ch}(V_X)=2+H+{3\over 7\cdot 2}H^2+{1\over 7\cdot 6}H^3.
\label{AppchVHT}
\eeq
Inserting this into (\ref{chS2}) with $E=V_X$ we find
${\rm ch}(\Sym^2V_X)=3+3H+{19\over 14}H^2+{5\over 14}H^3$.
The vector bundle $\Sym^2_0V_X$ is defined as the kernel of a surjective map
$\Sym^2V_X\to {\mathcal O}(1)$ and hence
we have a short exact sequence
$0\to \Sym^2_0V_X(-1)\to\Sym^2V_X(-1)\to {\mathcal O}_X\to 0$ which means
\beq
{\rm ch}(\Sym^2_0V_X(-1))={\rm ch}(\Sym^2V_X)\e^{-H}-1
=2-{H^7\over 7}.
\label{AppchSym2VHT}
\eeq
In the basis
$([{\mathcal O}_Y], [{\mathcal O}_H],[{\mathcal O}_{\ell}],[{\mathcal O}_{\rm pt}])$, these are (\ref{chVHT}) and (\ref{chSym2VHT}). 
The Euler pairing (\ref{EulerReye}) is found by the Riemann-Roch formula
using the expressions (\ref{AppchVHT}), (\ref{AppchSym2VHT}) and
\beq
{\rm td}(X)=1+{1\over 12}c_2(Y)=1+{5\over 7\cdot 6}H^2=1+{25\over 6}\ell.
\eeq

\subsection{Determinantal Calabi-Yau}
\label{app:classicalDet}

We are interested in the Chern characters of
the rank 4 and 3 vector bundles $W_Y$ and $\textw^2_{\pm}W_Y(1)$ on
the determinantal Calabi-Yau $Y$.
It turns out that the Chern characters of
$\textw^2_+W_Y$ and $\textw^2_-W_Y$ are equal to each other.
To explain this, recall that $Y$ is the free quotient of the space $P^{\vee}$
of solutions to the D-term and F-term equations by the gauge group
$G^{\vee}=(U(1)\times SO(4))/\Z_2$ and that the vector bundles
of our interest are realized by
\beq
W_Y=P^{\vee}\mathop{\times}_{\,G^{\vee}\!\!}\Box,\qquad
\textw^{\!2}_{\pm}W_Y=P^{\vee}\mathop{\times}_{\,G^{\vee}\!\!}\ydoopm.
\eeq
The action of $G^{\vee}$ on $P^{\vee}$ extends to
an action of $\bar{G}^{\vee}=(U(1)\times O(4))/\Z_2$ and it induces
a holomorphic involution $\tau$ on $Y$.
The group $\bar{G}^{\vee}$ acts linearly on $\Box$ and its various powers
including $\mydoo$. Elements of $\bar{G}^{\vee}\setminus G^{\vee}$
exchange the subspaces $\mydoop$ and $\mydoom$ of $\mydoo$.
This means that the involution $\tau$ lifts to an action on
$\textw^2W_Y$, exchanging $\textw^2_+W_Y$ and $\textw^2_-W_Y$.
In particular, $\textw^{\!2}_+W_Y\cong \tau^*\textw^{\!2}_-W_Y$.
Therefore we find
\beq
{\rm ch}(\textw^{\!2}_+W_Y)=\tau^*{\rm ch}(\textw^{\!2}W_Y)=
{\rm ch}(\textw^{\!2}_-W_Y).
\eeq
The second equality holds because $\tau$ is a holomorphic involution
and hence $\tau^*$ fixes the generator $H$ of 
the diagonal cohomology ring of $Y$.
As the two bundles sum up to $\textw^{\!2}W_Y$, we find
${\rm ch}(\textw^{\!2}_{\pm}W_Y)={1\over 2}{\rm ch}(\textw^{\!2}W_Y)$.

As stated in the main text, we do not directly compute ${\rm ch}(W_Y)$
but make a consistency check instead.
The ansatz (\ref{chAnsatz1}) and (\ref{chAnsatz2}) read
\beqa
{\rm ch}(W_Y)&=&4+aH+\left(-{a\over 2}+{b\over 10}\right)H^2+\left({a\over 6}+{b+c\over 10}\right)H^3\\
{\rm ch}(\textw^2_{\pm}W_Y(1))&=&3+a'H+\left(-{a'\over 2}+{b'\over 10}\right)H^2+\left({a'\over 6}+{b'+c'\over 10}\right)H^3
\label{AppchAnsatz2}
\eeqa
The Euler pairing (\ref{EulerDet}) is found by the Riemann-Roch formula
using these and
\beq
{\rm td}(X)=1+{1\over 12}c_2(Y)=1+{H^2\over 3}=1+{10\over 3}\ell.
\eeq
Comparison of twist and monodromy matrices fixes
$a,b,c$ as $-2,-10,15$ (\ref{abcfixed}), i.e.,
\beq
{\rm ch}(W_Y)=4-2H+{7\over 6}H^3.
\eeq
Inserting this into (\ref{chW2}) with $F=W_Y$, we find
${\rm ch}(\textw^{\!2}W_Y)=6-6H+2H^2$ and hence
\beq
{\rm ch}(\textw^{\!2}_{\pm}W_Y(1))=(3-3H+H^2)\e^H=3-{1\over 2}H^2.
\eeq
In comparison with (\ref{AppchAnsatz2}), we find
$a'=0$, $b'=-5$ and $c'=5$ as quoted in (\ref{abcfixedprime}).

\end{appendix}

\small\baselineskip=.93\baselineskip

\end{document}